\documentclass[12pt]{elsart.nw}
\usepackage{amssymb,amsmath}
\usepackage{cite}
\usepackage{afterpage}
\usepackage{float,rotate}
\usepackage{pstricks,pst-all}
\usepackage{graphicx,graphics,latexsym,epsfig}
\catcode`\@=11
-0.45cm \evensidemargin -0.45cm \font\cmss=cmss12  \def\1{\hbox{{1}\kern-.25em\hbox{l}}} \def\bfZ{\relax{\hbox{\cmss
Z\kern-.4em Z}}}

\def\nbar      {\ensuremath{\bar{\mathrm{n}}}}

\def\piz      {\ensuremath{\pi^0}}
\def\pip      {\ensuremath{\pi^+}}
\def\pim      {\ensuremath{\pi^-}}

\def\K     {\ensuremath{\mathrm{K}}}

\def\Kp    {\ensuremath{\mathrm{K}^+}}
\def\Km    {\ensuremath{\mathrm{K}^-}}

\def\Kb    {\ensuremath{\overline{\mathrm{K}}}{}}

\def\mevc {\ensuremath{\,{\mathrm{Me\hskip -1pt V}\hskip -2pt/\hskip -1pt c}}}
\def\gevc {\ensuremath{\,{\mathrm{Ge\hskip -1pt V}\hskip -2pt/\hskip -1pt c}}}

\def\ppb              {\ensuremath{\bar{\mathrm{p}}\mathrm{p}}}
\let\pbp=\ppb

\def\NNb              {\ensuremath{\overline{\mathrm{N}}\mathrm{N}}}

\def\qqb        {\ensuremath{\bar{q}q}}
\let\qqbar=\qqb
\def\ssb        {\ensuremath{\bar{{s}}{s}}}
\def\ccb        {\ensuremath{\bar{{c}}{c}}}

\def\uub        {\ensuremath{\bar{{u}}{u}}}
\def\ddb        {\ensuremath{\bar{{d}}{d}}}
\def\nnb        {\ensuremath{\bar{{n}}{n}}}
\def\udb        {\ensuremath{\bar{{d}}{u}}}
\def\dub        {\ensuremath{\bar{{u}}{d}}}

\newcommand{\SLJ}[3]    {\relax\ensuremath{{}^{#1}{{\mathrm #2}}_{#3}}}

\newcommand{\tso}{\SLJ3S1}
\newcommand{\ssz}{\SLJ1S0}
\newcommand{\tpz}{\SLJ3P0}
\newcommand{\tpo}{\SLJ3P1}
\newcommand{\tpt}{\SLJ3P2}
\newcommand{\spo}{\SLJ1P1}

\catcode`\@=11
\def\eqalign#1{\null\,\vcenter{\openup\jot\m@th
\ialign{\strut\hfil$\displaystyle{##}$&$\displaystyle{{}##}$\hfil
     \crcr#1\crcr}}\,}
\catcode`\@=12
\newcommand{\ket}[1]{\ensuremath{\vert\,#1\,\rangle}}
\newcommand{\nonett}[9]
{
\setlength{\unitlength}{0.8mm}
\begin{picture}(150.00,90.00)
\put(10.00,45.00){\vector(1,0){70.00}}
\put(45.00,10.00){\vector(0,1){70.00}}
\put(110.00,45.00){\vector(1,0){30.00}}
\put(125.00,30.00){\vector(0,1){30.00}}
\put(82.50,42.50){\makebox(5.00,5.00){$I_3$}}
\put(142.50,42.50){\makebox(5.00,5.00){$I_3$}}
\put(127.50,60.00){\makebox(5.00,5.00){\quad\ S\quad\ Singlet }}
\put(35.50,78.00){\makebox(5.00,5.00){\quad\ S }}
\put(45.50,78.00){\makebox(5.00,5.00){\quad\ \quad\ Octet }}
\put(45.00,45.00){\circle*{2.00}}
\put(70.00,45.00){\circle*{2.00}}
\put(20.00,45.00){\circle*{2.00}}
\put(32.50,70.00){\circle*{2.00}}
\put(57.50,70.00){\circle*{2.00}}
\put(32.50,20.00){\circle*{2.00}}
\put(57.50,20.00){\circle*{2.00}}
\put(125.00,45.00){\circle*{2.00}}
\put(45.00,45.00){\circle{0.00}}
\put(45.00,45.00){\circle{5.00}}
\put(32.50,70.00){\line(1,0){25.00}}
\put(57.50,70.00){\line(1,-2){12.50}}
\put(70.00,45.00){\line(-1,-2){12.50}}
\put(57.50,20.00){\line(-1,0){25.00}}
\put(32.50,20.00){\line(-1,2){12.50}}
\put(20.00,45.00){\line(1,2){12.50}}
\put(60.00,70.00){\makebox(15.00,5.00)[l]{#2}}
\put(15.00,70.00){\makebox(15.00,5.00)[r]{#1}}
\put(15.00,15.00){\makebox(15.00,5.00)[r]{#3}}
\put(6.75,37.50){\makebox(13.25,5.00)[r]{#5}}
\put(60.00,15.00){\makebox(15.00,5.00)[l]{#4}}
\put(70.00,37.50){\makebox(13.75,5.00)[l]{#6}}
\put(47.50,37.50){\makebox(12.50,5.00)[l]{#7}}
\put(47.50,47.50){\makebox(12.50,5.00)[l]{#8}}
\put(127.50,47.50){\makebox(12.50,5.00)[l]{#9}}
\end{picture}
}
\newcommand{\nonet}[9]
{
\setlength{\unitlength}{0.8mm}
\begin{picture}(150.00,90.00)
\put(10.00,45.00){\vector(1,0){70.00}}
\put(45.00,10.00){\vector(0,1){70.00}}
\put(82.50,42.50){\makebox(5.00,5.00){$I_3$}}
\put(35.50,78.00){\makebox(5.00,5.00){\quad\ S }}
\put(45.00,45.00){\circle*{2.00}}
\put(70.00,45.00){\circle*{2.00}}
\put(20.00,45.00){\circle*{2.00}}
\put(32.50,70.00){\circle*{2.00}}
\put(57.50,70.00){\circle*{2.00}}
\put(32.50,20.00){\circle*{2.00}}
\put(57.50,20.00){\circle*{2.00}}
\put(45.00,45.00){\circle{0.00}}
\put(45.00,45.00){\circle{4.00}}
\put(45.00,45.00){\circle{6.00}}
\put(32.50,70.00){\line(1,0){25.00}}
\put(57.50,70.00){\line(1,-2){12.50}}
\put(70.00,45.00){\line(-1,-2){12.50}}
\put(57.50,20.00){\line(-1,0){25.00}}
\put(32.50,20.00){\line(-1,2){12.50}}
\put(20.00,45.00){\line(1,2){12.50}}
\put(60.00,70.00){\makebox(15.00,5.00)[l]{#2}}
\put(15.00,70.00){\makebox(15.00,5.00)[r]{#1}}
\put(15.00,15.00){\makebox(15.00,5.00)[r]{#3}}
\put(6.75,37.50){\makebox(13.25,5.00)[r]{#5}}
\put(60.00,15.00){\makebox(15.00,5.00)[l]{#4}}
\put(70.00,37.50){\makebox(13.75,5.00)[l]{#6}}
\put(32.00,37.00){\makebox(12.50,5.00)[l]{#7}}
\put(32.00,49.00){\makebox(12.50,5.00)[l]{#8}}
\put(48.00,45.00){\makebox(12.50,5.00)[l]{#9}}
\put(100.00,30.00){\line(1,0){40.00}}
\put(100.00,30.00){\makebox(15.00,5.00)[l]{$\sigma= n\bar nn\bar n$,
\ \ \ 650 MeV}} \put(100.00,50.00){\line(1,0){40.00}}
\put(100.00,50.00){\makebox(15.00,5.00)[l]{$\kappa= n\bar sn\bar n$, \
~$\sim$810 MeV}} \put(100.00,70.00){\line(1,0){48.00}}
\put(100.00,70.00){\makebox(15.00,5.00)[l]{$f_0,a_0=n\bar ns\bar s$,
\ ~980 MeV}}
\end{picture} }

\def\sf{\sin\theta_\eta}%
\def\xybar#1#2{\mbox{$\mbox{#1}\bar{\mbox{#2}}$}}%
\def\xxbar#1{\xybar{#1}{#1}}%
\def\qqbar{\xxbar{q}}%
\def\ppbar{\xxbar{p}}%

\newcommand{\bc}        {\begin{center}}
\newcommand{\ec}        {\end{center}}
\newcommand{\bi}        {\begin{enumerate}}
\newcommand{\ei}        {\end{enumerate}}
\newcommand{\be}        {\begin{equation}}
\newcommand{\ee}        {\end{equation}}

\newcommand{\etg}       {\mbox{$\eta$}}
\newcommand{\etp}       {\mbox{$\eta^{\prime}$}}

\renewcommand{\theequation}{\thesection.\arabic{equation}}

\begin{document}
\begin{frontmatter}
\title{
Glueballs, Hybrids, Multiquarks. \vspace{2mm}\\
\large\bf Experimental facts versus QCD inspired concepts.}

\author[hiskp]{Eberhard Klempt}                                                              %
\author[protvino]{and Alexander Zaitsev}
\address[hiskp]{
Helmholtz-Institut f\"ur Strahlen- und Kernphysik
der Rheinischen Friedrich-Wilhelms Universit\"at, \\
Nu\ss allee 14-16, D--53115 Bonn, Germany
}
\address[protvino]{
Institute for High-Energy Physics,
Moscow Region,
RU-142284 Protvino, Russia
}
\date{\today\  }

\begin{abstract}
\underline{Glueballs}, hybrids and multiquark states are predicted as
bound states in models guided by quantum chromodynamics, by QCD sum
rules or QCD on a lattice. Estimates for the (scalar) glueball ground
state are in the mass range from 1000 to 1800\,MeV, followed by a
tensor and a pseudoscalar glueball at higher mass.
Experiments have reported evidence for an abundance of meson resonances
with $0^{-+}, 0^{++}$ and $2^{++}$ quantum numbers. In particular the
sector of scalar mesons is full of surprises starting from the elusive
$\sigma$~and~$\kappa$ mesons. The $a_0(980)$ and $f_0(980)$, discussed
extensively in the literature, are reviewed with emphasis on their
Janus-like appearance as $K\bar K$ molecules, tetraquark states or
$q\bar q$ mesons. Most exciting is the possibility that~the three
mesons $f_0(1370)$, $f_0(1500)$, and $f_0(1710)$ might reflect the
appearance of a scalar glueball in the world of quarkonia. However, the
existence of $f_0(1370)$ is not beyond doubt and there is evidence that
both $f_0(1500)$ and $f_0(1710)$ are flavour octet states, possibly in a
tetraquark composition. We suggest a scheme in which the scalar
glueball is dissolved into the wide background into which all scalar
flavour singlet mesons collapse.

There is an abundance of meson resonances with the quantum numbers of
the $\eta$. Three states are reported below 1.5\,GeV/c$^2$  whereas
quark models expect only one, perhaps two. One of these states,
$\iota(1440)$, was the prime glueball candidate for a long time. We
show that $\iota(1440)$ is the first radial excitation of the $\eta$
meson.

\underline{Hybrids} may have exotic quantum numbers which are not
accessible by $q\bar q$ mesons. There are several claims for
$J^{PC}=1^{-+}$ exotics, some of them with properties as predicted from
the flux tube model interpreting the quark--antiquark binding by a
gluon string. The evidence for these states depends partly on the
assumption that meson--meson interactions are dominated by s--channel
resonances. Hybrids with non-exotic quantum numbers should appear as
additional states. Light-quark mesons exhibit a spectrum of (squared)
masses which are proportional to the sum of orbital angular momentum
and radial quantum numbers. Two states do not fall under this
classification. They are discussed as hybrid candidates.

The concept of \underline{multiquark} states has received revived
interest due to new resonances in the spectrum of states with open and
hidden charm. The new states are surprisingly narrow and their masses
and their decay modes often do not agree with simple quark-model
expectations.

Lattice gauge theories have made strong claims that glueballs and
hybrids should appear in the meson spectrum. However, the existence of
a scalar glueball, at least with a reasonable width, is highly
questionable. It is possible that hybrids will turn up in complex
multibody final states even though so far, no convincing case has been
made for them by experimental data. Lattice gauge theories fail to
identify the nonet of scalar mesons. Thus, at the present status of
approximations, lattice gauge theories seem not to provide a
trustworthy guide into unknown territory in meson spectroscopy.
\vspace*{2mm}   \end{abstract} \vspace{-6mm} \end{frontmatter}

\clearpage\setcounter{equation}{0}
\markboth{\sl Meson spectroscopy} {\sl Table of contents}
\clearpage \tableofcontents

\markboth{\sl Meson spectroscopy} {\sl Introduction}
\clearpage\setcounter{equation}{0} \section{\label{Introduction}
Introduction}

\subsection{\label{Highlights and perspectives}
Highlights and
perspectives}

\subsubsection{\label{Scope of the review}
Scope of the review}

One of the most challenging topics in contemporary particle physics is
the dynamics of non-Abelian gauge theories. Quantum Chromo Dynamics
(QCD) is the most explored example of a class of theories in which the
bosonic field quanta carry the charge of the fermionic agents. QCD
predicts very new effects like confinement, quark and gluon
condensates, glueballs and hybrids, perhaps instantons. Some of them
have been observed, others not. We start to understand how the
enigmatic $\sigma(485)$ meson is created but we do not know if gluons
add new degrees of freedom to spectroscopy. We know that the structure
of the QCD vacuum plays a decisive $\rm r\hat{o}le$, but we are unable
to link the vacuum structure to the confinement property of QCD.

Spectroscopy is a unique way to access QCD, and substantial progress
has been made in recent years. This review covers topical features
in meson spectroscopy, from light quark mesons to heavy quarkonia, from
glueballs and hybrids to multiquark states. We are convinced that a
global view of the whole field provides better insight than even a
detailed discussion of separate issues would provide. At present, such
a global view cannot be derived from first principles. A general frame
needs to be defined within which detailed observations can be
discussed. For this reason, we begin this report with a summary of
highlights. This will allow us to discuss in the subsequent sections
important results and their $\rm r\hat{o}le$ for the full picture. Meson
spectroscopy is an active field of research which implies that the
frontier is open and that the interpretation of new results is often
controversial. This review tries to give a fair account of the
different views and still to emphasize our own position. We do not claim
evidence for new physics when observations find consistent explanations
with a minimal set of assumptions: `plurality ought never be posed
without necessity' (Occam's razor). As authors we take the privilege to
present, here in the summary, our own view in a series of clear
statements with little reasoning or justifications. We are convinced
that we have identified questions and that we give answers that should
guide further theoretical and experimental work. Readers interested in
other interpretations and in the scientific {\it if's} and {\it why's}
would have to read the full text. Our conclusions and suggestions for
future directions will be presented in this section as well.

\subsubsection{\label{Recent outstanding achievements}
Recent outstanding achievements}

There have been outstanding achievements in recent years:

\begin{enumerate}

\vspace{-1mm}\item The long-sought $h_c(1P)$ state has been found
precisely at the mass given by the centre-of-gravity of the
$\chi_{cJ}(1P)$ states. The $\eta_c$ first radial excitation
$\eta_c(2S)$ is now established. With these two discoveries, all
charmonium states expected by quark models to fall below the $D\bar D$
threshold are found, and no additional state. This is an important
confirmation of the quark model; the precise mass values provide
stringent constraints to model builders on the structure of the
confining interactions.

\vspace{1mm}\item Several unexpectedly narrow states, $X(3872)$,
$X(3940)$, $Y(3940)$, $Z(3930)$, $Y(4230)$, were discovered over the
last years. Their interpretation as $c\bar c$ states, as molecules, or
as hybrids is hotly debated. In the channel $e^+e^-\to J/\psi$ plus
recoil mesons, peaks due to $\eta_c(1S)$, $\eta_c(2S)$, and $X(3940)$
are observed (and a convolution of $\chi_{cJ}(1P)$ states). We suggest
to identify $X(3940)$ with $\eta_c(3S)$. The $Z(3930)$ resonance is
observed in $\gamma\gamma \to D\bar D$ with a helicity distribution
consistent with a $\chi_{c2}(2P)$ interpretation. The $Y(3940)$,
observed in its J/$\psi\,\omega$ decay mode, is tentatively interpreted
by us as the $\chi_{c0}(2P)$ $q\bar q$ state with a strong tetraquark
component, in analogy with the $\phi\omega$ decay mode of $f_0(1760)$.
The $\chi_{c1}(2P)$ quark model state is identified with $X(3872)$. The
state is attracted by the $DD^*$ threshold to which it couples
strongly.  We present reasons why we interpret $Y(4260)$ as $\psi(4S)$
quark-model state and identify $\psi(4415)$ with $\psi(5S)$.

\vspace{1mm}\item New mesons with open charm, those which are found and
those which are not found, and their decays provide new insight into
the $\rm r\hat{o}le$ of tetraquark configurations. The properties of
these mesons are often dominated by their strong affinity to $qq\bar
q\bar q$. In spite of this, also these mesons are compatible with a
$q\bar q$ assignment: there are no `flavour-exotics'. The $q\bar q$
component is essential to form a resonant state.

\vspace{1mm}\item In the light-quark sector, a very large number of
high-mass states was suggested and awaits future confirmation. They
follow a simple mass pattern compatible with a string picture of
mesonic excitations. Contributions from exotic partial waves, with
quantum numbers beyond the quark model, have been identified. Their
resonant nature is controversially discussed. Particular interest has
been focussed on scalar mesons, their properties and abundance. A
generally accepted consistent picture has not yet emerged.

\vspace{1mm}\item J/$\psi$ decays -- studied at the Beijing storage ring
BES -- have yielded unexpected results, in particular on scalar mesons.
Their production and decay pattern when recoiling against an $\omega$ or
$\phi$ meson is a key witness of their anomalous structure.

\vspace{1mm} \item Last not least, the $\sigma$ at $\sim
500$\,MeV/c$^2$, called $\sigma(485)$ in this report, and $\kappa$
mesons at $\sim 700$\,MeV/c$^2$ called $\kappa(700)$ seem to be
firmly established, not only because of new data but also due to
substantial advances in the theoretical description of low-energy
$\pi\pi$ (and $K\pi$) interactions. Both particles are now understood
as consequence of chiral symmetry and unitarity, constraints which can
both be met only when these two particles are introduced.
\end{enumerate}\vspace{-1mm}

\subsubsection{\label{New insights}
New insights}

{\it Glueballs, hybrids, tetraquarks} and their $\rm r\hat{o}le$ in
{\it meson spectroscopy} are the central themes of this review.

\begin{enumerate}\vspace{-1mm} \item A unified picture of {\it meson
spectroscopy} emerges from new (and old) experimental findings
connecting the family of $\Upsilon$ states with light-quark
spectroscopy, a view emphasized often by Nathan Isgur. The first
orbital excitation energy, the mass difference between the first tensor
and the first vector meson (with quantum numbers $J^{PC}=2^{++}$ and
$1^{--}$, respectively), is about 500\,MeV/c$^2$, independent of the
quark flavour. The hyperfine splitting between the two $S$-wave ground
states (with quantum numbers $J^{PC}=1^{--}$ and $0^{-+}$,
respectively) scales with $1/M$ as expected for a magnetic hyperfine
interaction originating from one-gluon exchange. Only the -- predicted
-- $\Upsilon$-$\eta_b$ mass difference breaks this simple rule. The
pion plays a very special $\rm r\hat{o}le$ in meson spectroscopy due to
its dual nature as $q\bar q$ meson and as Goldstone boson of QCD with
massless quarks. It is extremely surprising that the $1/M$ dependence
is obeyed by all states from the J/$\psi$-$\eta_c$ mass difference down
to the $\rho$-$\pi$ splitting. These two characteristic mass gaps can
be identified even in the baryon excitation spectrum (see
Fig.~\ref{fig:forces} in section \ref{Flavour independence of forces}).

 \vspace{1mm}\item We do not discuss individual high orbital and radial
excitations. Rather, we point out regularities in their mass spectrum.
For sufficiently high excitations in orbital angular momentum $l$ and
for small radial quantum numbers $n$, the squared masses of light
mesons are proportional to $l+n$ (see Fig. \ref{fig:allmesons} in
section \ref{High radial and orbital excitations of light mesons}). The
$l+n$ dependence of meson masses can be derived in a string model but
not in constituent quark models which are close to a $l+2n$ behaviour.
Note that $n=0$ is allowed. The ground states have $N=1$, we define
$n=N-1$.

\vspace{1mm}\item The $l+n$ dependence of meson masses leads naturally
to a mass-degeneracy of even- and odd-parity resonances for all but
`stretched' states (with $j=l+s$). Chiral symmetry restoration,
suggested to be responsible for the mass-degeneracy, does not give an
explanation why stretched states have no parity partners. In the string
model, this aspects follows automatically.

\vspace{1mm}\item {\it Hybrids} may have exotic or non-exotic
spin-parity quantum numbers. In the sector of non-exotic states, there
are two outstanding observations which do not fit into any quark-model
assignment and which have the properties predicted for hybrids. These
are the resonances $\pi_2(1880)$ and $\eta_2(1870)$. Comparatively
narrow hybrids are predicted in this partial wave. On the other hand,
the $1^{++}$ sector would be an ideal case to find hybrids as well; but
no candidate has been seen. One important difference between these two
partial waves is that the $2^{-+}$ wave has important $S$-wave
couplings to $2^{++}+0^{-+}$ mesons while there are no strong $S$-wave
decays of the $1^{++}$ wave. Further careful studies of dynamical
effects induced by thresholds are therefore certainly needed before the
claims for $\pi_2(1880)$ and $\eta_2(1870)$ can be accepted as resonant
states of exotic nature.

\vspace{1mm}\item Exotic mesons have quantum numbers which are not
accessible to $q\bar q$ configurations. Spin-parity exotic states can
have contributions from $q\bar qg$, $qq\bar qq$, and meson-meson
configurations; flavour exotics are restricted to $qq\bar qq$ and
meson-meson dynamics. Since there is no established evidence for
flavour exotics, the most likely interpretation of spin-parity exotic
states would be that they are hybrids.

Exotic partial waves are an established phenomenon. However, there is
still a controversy if exotic waves show a resonant behaviour. The
`best case', the $\pi_1(1400)$, has decay modes which rule out a hybrid
interpretation. As $\pi\eta$ system in a $qq\bar qq$
decuplet-antidecuplet configuration, it should be accompanied by flavour
exotic partners like $K^+\pi^+$. The latter configuration does not show
any phase motion. The interpretation of the $\pi\eta$ exotic wave in
terms of non-resonant meson-meson interactions seems thus more likely
than hybrid interpretations even though practical fits based on such
assumptions yield a worse $\chi^2$. The $\pi_1(1600)$ is a much
more viable hybrid candidate. But as long as the experimental evidence
for $\pi_1(1400)$ and $\pi_1(1600)$ is of similar quality, doubts
remain why to reject $\pi_1(1400)$  on general grounds and why to
accept $\pi_1(1600)$.

\vspace{1mm}\item There is a long controversy concerning the $\rm
r\hat{o}le$ of $gg$, $q\bar q$, $q\bar qg$, $qq\bar qq$, and
meson-meson configurations in meson spectroscopy. ($gg$ and $q\bar qg$
are short-hand notations for glueballs and hybrids.)  A meson wave
function of a `normal' meson like $f_2(1270)$ may contain contributions
from all these configurations, at an -- unknown -- momentum-dependent
fraction. The main emphasis is therefore laid on states beyond the
quark model expectations.

So far, there are no forcing arguments for the existence of any
{\it multiquark states} which could not be accommodated as a $q\bar q$
object. For example, there are good arguments to interpret the narrow
$D^{*+}_{s0}(2317)$ state as $D\,K$ bound state. Since $D$ and $K$ both
have isospin 1/2, an isospin singlet and a triplet of states (with
charges $0,1,2$) should be expected. Yet, only the singlet was found.
The latter state is compatible with a $c\bar s$ assignment. Thus we
conjecture that there must be a $c\bar s$ `seed'. Generalising this
concept, the $q\bar q$ seed is essential for all states which are
seemingly of tetraquark or molecular nature. The $q\bar q$ seed may
dress into a tetraquark configuration if this is energetically favoured
(e.g. due to diquark-diquark correlations). These correlations act by
merging the quark-antiquark pairs into two colour-neutral mesons. Thus,
a molecular component develops. The fractional contributions of these
three configurations depend on the vicinity of thresholds for hadronic
decays, and can be functions of the spatial separations or of the
momentum transfer with which the system is probed. At large momentum
transfer, in fragmentation, the $q\bar q$ nature is explored and
seemingly molecular states are produced with the same rate as ordinary
mesons. Soft processes (like the decay chain $\phi\to\gamma f_0(980)$;
$f_0(980)\to K\bar K$ with minimal momentum transfers) reveal the
molecular character of a state.

\vspace{1mm}\item The $q\bar q$, tetraquark, and molecular components
lead to a rich phenomenolgy, in particular for scalar mesons.
Some mesons like $a_0(980)$, $\sigma(485)$, $f_0(980)$, and
$\kappa(700)$ reveal clearly a molecular character: their pole
positions are very close to their respective thresholds and the poles
can be generated from hadronic loops of the decay products. These
arguments exclude tightly bound tetraquark configurations. The
need for hadronic loops does not imply that the forces between the
constituents are of molecular type. Two pseudoscalar mesons in the
$0^{+}$ channel, e.g. in the 1\,GeV region, may interact attractively
because at short distances their constitutive quarks can form diquarks
with energetically favoured correlations.

The mass difference of the ground-state scalar meson to the
threshold defined by the mass of the two related pseudoscalar
mesons (i.e. $M_{\chi_{b0}(1P)}$ - $M_{B\bar B}$, $\cdots$,
$M_{f_0(980)}$ - $M_{K\bar K}$, $M_{\sigma(485)}$ - $M_{\pi\pi}$)
is a smooth function of the mass of the decay products (see Fig.
\ref{scalarview}a in section \ref{Scalar states from the sigma to
chib0(1P)}). The $\chi_{b0}(1P)$ has a mass 700\,MeV/c$^2$ below
the threshold for $B\bar B$ decays. It connects via
$\chi_{c0}(1P)$, via $D^*_{s0}(2317)$, via a postulated
$D^*_0(1980)$, via $f_0(980)$ and $\kappa(700)$ to the
$\sigma(485)$. The $\sigma(485)$ mass is 200\,MeV/c$^2$ above the
$\pi\pi$ threshold. Current views mostly assume the $\sigma(485)$
to be a $\pi\pi$ effect, unrelated to $q\bar q$ spectroscopy. The
continuous transition from $\chi_{b0}(1P)$ to the $\sigma(485)$
does of course not imply similar wave functions for these two
mesons. Instead, their nature is very different as evidenced by
the 1/$N_c$ behaviour of their unitarized chiral amplitude (see
Fig. \ref{fig:nc} in section \ref{Scalar mesons from chiral
symmetry}). The by far predominant part of the $\chi_{b0}(1P)$
wave function is of $b\bar b$ nature. When the quark mass is
slowly reduced from the $b$-quark mass to light-quark masses, the
$\chi_{b0}(1P)$ state changes continuously into the $\sigma(485)$,
a predominantly $\pi$-$\pi$ molecular object. But it has a $q\bar
q$ seed. If a low-mass light-quark $q\bar q$ pair is created with
vacuum quantum numbers, the $\sigma(485)$ is created with unit
probability. It has a complicated wave function but it is still
the $^3P_0$ scalar ground state; there is no additional
light-quark scalar meson with $q\bar q$ structure. It is the
$\sigma(485)$ which occupies the slot reserved for the $1^3P_0$
ground state. QCD provides very different mesonic wave functions
when a quark and an antiquark is created out of the vacuum: the
wave function for a clean $q\bar q$ meson in case of $b$ quarks, a
highly correlated and complex system having the appearance of a
$\pi\pi$ interaction phenomenon when light quarks are involved.

\vspace{1mm}\item Likewise, $\chi_{b0}(2P)$, $\chi_{c0}(2P)$, the
disputed Selex state $D_s^{*+}(2630)$, $D^*_0(2350)$, $f_0(1500)$,
$K^*_0(1430)$, and $f_0(1300)$ may form a group of first radial
excitations. The postulated $D^*_0(1980)$ is below the $D\pi$ threshold
and can decay either via its high-mass tail (if the coupling is strong
enough) or by weak interactions only. Its existence would resolve the
apparent puzzle with the $D^*_{s0}(2317)$ mass which should not be
below the $D^*_0(2350)$ mass if these two mesons just differ by the
exchange of an $s$ quark into a $d$ or $u$ quark. The $f_0(1300)$ is
the wide scalar background which was called $\epsilon(1300)$ for a long
time.

 \vspace{1mm}\item  In our view, the scalar flavour-singlet mesons are
very wide and merge into one continuous scalar background. Their widths
arise from their strong coupling to the QCD vacuum by emission of two
Goldstone bosons, i.e. of two pions. The $\sigma(485)$ is part of this
background. The wide poles at $\sim 500, \sim 1300$, and $\sim
1540$\,MeV/c$^2$ reported in the literature are very far from the real
axis. These states are supposed to be flavour-singlet mesons having
$\frac{1}{\sqrt 3}|\uub + \ddb + \ssb\rangle$ and $\frac{1}{\sqrt
3}|\uub\ddb + \ddb\ssb + \uub\ssb\rangle$ components. One exception is
the $\sigma(485)$ meson. Due to its low mass, flavour SU(3) is broken
and the \ssb\ part of its wave function is small. The wide background
interferes with the flavour-octet scalar mesons, $f_0(980)$,
$f_0(1500)$, $f_0(1760)$, and $f_0(2100)$, which have normal hadronic
widths. At present, the flavour content of these mesons cannot be
deduced experimentally like, e.g., the flavour decomposition of $\eta$
and $\eta^{\prime}$ or of $\omega$ and $\phi$. Dynamical arguments and
indirect evidence need to be invoked. Any data shedding light on the
flavour wave functions of scalar mesons is therefore extremely
welcome.

 \vspace{1mm}\item  We propose that the three observed meson states
called $f_0(1710)$, $f_0(1790)$, and $f_0(1810)$ originate from one
single meson, $f_0(1760)$, which has a (mainly) flavour-octet wave
function with a large tetraquark component. The peculiar $f_0(1760)$
production and decay pattern observed in J/$\psi$ decays into $\omega
f_0(1760)$, $\phi f_0(1760)$, and $\gamma f_0(1760)$ is interpreted as
dressing of the initially formed $q\bar q$ system into $qq\bar q\bar q$
and its interference with a broad flavour-singlet background amplitude.
The glueball candidate $f_0(1500)$ is interpreted as being mainly a
flavour octet meson. All scalar mesons have sizable $q\bar q$ and
$qq\bar q\bar q$ components. Their $q\bar q$ nature is revealed in
their production. Their decays and mass pattern require substantial
$qq\bar q\bar q$ components. There is no 'narrow' $f_0(1370)$.

\vspace{1mm}\item In our view, the $f_0(980)$ resonance is a
flavour-octet meson, at least in its $q\bar q$ component. The vicinity
of the $K\bar K$ threshold breaks SU(3) symmetry and transforms
$f_0(980)$ into state with a $\frac{1}{\sqrt 2}(u\bar u+d\bar d)s\bar
s$ flavour structure of largely molecular decomposition.

\vspace{1mm}\item There is no evidence for a narrow scalar {\it
glueball} below 1.8\,GeV/c$^2$ in mass, or for a scalar decuplet
consisting of a $q\bar q$ nonet and a glueball with mixing of two
isoscalar $^3P_0$ states. The wide flavour singlet background may
contain a glueball fraction. The glueball content can be searched for
in the same way as a glueball fraction was sought for in the wave
function of the $\eta^{\prime}$. The existence of glueballs is an
unproven hypothesis.

\vspace{1mm}\item Glueballs could exist as broad objects with 1\,GeV
widths, or even broader. For scalar, pseudoscalar and tensor quantum
numbers, broad and likely flavour-blind intensities have been
identified. It is, of course, very difficult to establish a phase motion
of such broad states. Even if the existence of such a wide state is
accepted, this does not imply a particular interpretation: in J/$\psi$
decays a very wide object with quantum numbers of the $\rho$ has been
reported which could be a molecular background, generated by
$t$-channel exchange forces, a hybrid, but certainly not a glueball.

\vspace{1mm} \item The $\eta(1295)$ is unlikely to be
the radial excitation of the $\eta$ meson; more likely it is faked by
feedthrough from the mass-degenerate $f_1(1285)$ and the Deck effect in
$\eta$-$\sigma$ $S$-wave interactions. The $\eta(1440)$ is not split
into $\eta(1405)$ and $\eta(1475)$; rather, the node in the
$\eta(1440)$ wave function - expected for a radial excitation - leads
to an apparent splitting of $\eta(1440)$. The $\eta(1440)$, the newly
suggested X(1835) (if it proves to have pseudoscalar quantum numbers),
the $\pi(1375)$ and $K(1450)$ form a consistent nonet of pseudoscalar
radial excitations.  A second nonet of pseudoscalar radial excitations
is proposed but all four states require experimental confirmation. The
excitation energies for the first and second radial excitation of
pseudoscalar mesons decrease monotonously with increasing meson mass,
when going from $\pi$ to $K$, $\eta$, $\eta^{\prime}$, and to $\eta_c$
(see Fig. \ref{pic:m2} in section \ref{The nonet of pseudoscalar radial
excitations}).

\end{enumerate}

\subsubsection{\label{Critique of QCD folklore}
Critique of QCD folklore}

In our view, efforts in this field have too often been driven by the
quest for discoveries instead of the search for a deeper understanding.
In this report, we review critically some QCD folklore. We conclude the
following:

\begin{enumerate}

\vspace{-1mm}\item Central production in a 450\,GeV/c proton beam is
not dominated by Pomeron-Pomeron interactions; Regge exchange is still
important. The so-called `glueball filter' finds a straightforward
interpretation; it does not project out extraordinary states.

\vspace{1mm}\item In radiative J/$\psi$ decays, the dominance of
initial-state radiation is often assumed but experimentally not proven.

\vspace{1mm}\item Proton-antiproton annihilation has no large affinity
to glue-rich states. The process prefers (partial) rearrangement of
quarks.
\end{enumerate}\vspace{-1mm}

There is no experimental evidence that one of these three processes is
a source of glueballs.

\subsubsection{\label{Successes and limitations of lattice QCD}
Successes and limitations of lattice QCD}

 Lattice QCD is growing into a predictive theory of strong
interactions. Its many achievements include: accurate calculations of
meson and baryon ground state masses, reliable calculations of weak
matrix elements and of hadron decays, enlightening studies of chiral
symmetry breaking, instanton induced interactions, and the confinement
problem. The numerical successes of Lattice QCD are most striking when
they can be connected to extrapolations using chiral perturbation
theory. There are, however, questions for which no chiral expansion
exists; the usefulness of Lattice QCD for this kind of applications
needs to be tested with scrutiny. We emphasize that glueballs are
predicted in Yang-Mills theories, hybrids in quenched QCD. It is an
open issue if these concepts find their correspondence in Nature.

Many of the experimental findings and their interpretation
suggested here are at variance with firm predictions of Lattice QCD,
other important phenomena were not predicted. A large number of Lattice
QCD results were preceded by results obtained within `QCD inspired
models'. In this general overview, we restrict ourselves however to a
discussion of Lattice QCD and list a few conclusions which are based on
our personal interpretation of the spectrum of scalar mesons:

\begin{enumerate}

\vspace{-1mm}\item A full spectrum of glueballs is
predicted but not even its ground state exists as identifiable object.

\vspace{1mm}\item Hybrids are predicted to exist. Present data
give only little evidence for hybrid degrees of freedom in meson
spectroscopy but do not exclude their existence.

\vspace{1mm}\item Lattice calculations did not find that scalar mesons
are organised along their flavour structure, into singlet and octet
states.

\vspace{1mm}\item Lattice calculations did not predict the large $\rm
r\hat{o}le$ of tetraquark configurations for scalar mesons.

\vspace{1mm}\item Lattice calculations missed the continuous transition
from $\chi_{b0}(1P)$ to the $\sigma(485)$ when quark masses are changed
from the $b$ mass to $u$ and $d$ quark masses.

\vspace{1mm}\item Lattice calculations seem to fail to reproduce mass
gaps due to excitations with vacuum quantum numbers. This is an
observation outside of meson spectroscopy, but Lattice QCD prefers a
mass of the Roper $N(1440)P_{11}$ above and not below $N(1535)S_{11}$.
In this case, a chiral extrapolation to realistic quark masses exists
but has not cured the problem.

 \end{enumerate}\vspace{-1mm}

Some of the most important questions in meson spectroscopy got answers
from Lattice QCD (with the presently required approximations)
which turned out to be misleading. Other important questions are out of
reach of present-day Lattice QCD. It is certainly premature to question
if QCD as fundamental theory is valid also in the low energy domain.
Intense computing and new ideas are needed to study flavour-singlet
mesons in tetraquark configuration in Lattice QCD and to unquench
glueballs ready to decay into 135\,MeV/c$^2$ pions. It is an irony of
science that answers from Lattice QCD which we believe to be in
conflict with experimental findings led to an enormous stimulation of
the field and to a much improved understanding of strong interactions.

\subsubsection{\label{Future options}
Future options}

We conclude this section by defining open issues, by listing some
specific propositions and by giving some general recommendations. The
aim of experiments is not just to define basic resonance parameters
like masses, widths and decay branching fractions but rather to learn
more about the structure of mesons. Hence it is important to use a
variety of different methods. Production of mesons in hard reactions
depends on the mesonic wave function at the origin of the
centre-of-mass system. Diffractive production of light-quark resonances
off nuclei is sensitive to the spatial extent of the wave function,
colour transparency of charmonium states gives access to its $q^2$
evolution. The amplitude of radiative decays is proportional to the
inverse of the quark mass $m_q^{-1}$. But, how does a constituent
quark mass change if one considers high orbital excitations?

\begin{enumerate}

\vspace{-1mm}\item Physics is an open science with
competing ideas, also in data analysis. After publication, data should
be made public. In baryon physics, data are accessible via internet. In
meson spectroscopy, data are private property of collaborations also
beyond publication. This can be changed; look at web page
\begin{verbatim}http://hadrondb.hiskp.uni-bonn.de\end{verbatim}
where
you find Dalitz plots and full data samples for multibody final states,
both real data and Monte Carlo data. An open-access policy should be
enforced not only for journals but also for data.

\vspace{1mm}\item Quantum numbers and decay properties of the recently
discovered states with open and hidden charm need to be studied. The
$B$ factories and the CLEO experiment have opened a window to a new
spectroscopy but the existence of many of the states is based on poor
statistics. A substantial increase of statistics (and running time) is
required. Since BaBaR -- and SLAC as particle laboratory -- will be
closed, the hope for advances in the immediate future rests on BELLE.
In the time of continuous evaluations, the citation index is an
important aspect of science. Half of the BELLE papers with more than
100 citations report discoveries in spectroscopy. Surprisingly, also
CDF and $D\O$ are shown to be capable to make significant contributions
to this field, and additional results can be hoped for from LHCb.

\vspace{1mm}\item An intense antiproton source is part of the FAIR
project at GSI, Darmstadt. The spectroscopy of open and hidden charm
states is a central issue of PANDA, a universal detector proposed to
be installed at the antiproton facility. Mass and widths of states
below the $c\bar c$ threshold and their decay rates will be  measured
with high precision. Above the $c\bar c$ threshold, there are
candidates for all $N=1$ and $N=2$ states which can be interpreted in
the quark model as $S$- and $P$-states; the $D$ states (except the
$^3D_1$ states which can be observed in $e^+e^-$ annihilation) are
unknown. The completion of the quark model and the search for states
beyond the quark model are central themes at FAIR. In this report we
argue that there is no forcing evidence for states beyond the quark
model. The high-mass charmonium region is likely the best place to
search for mesons with exotic quantum numbers, exotic in spin-parity
like $J^{PC}=1^{-+}$, or exotic in their flavour structure like charged
companions of the $X,Y,Z$ states at 3940\,MeV (with flavour content
$cu\bar d\bar c$, ...).

\vspace{1mm}\item Scalar mesons still remain a most fascinating
research object. In this report, a new view is developed which is in
conflict with generally accepted interpretations. Interpretational
differences manifest themselves in the flavour wave functions of scalar
isoscalar mesons. The best approach to substantiate or to reject the
view are likely coupled channel analyses of J/$\psi$ decays into vector
plus scalar mesons. Urgently needed are high-statistics and
high-quality data on processes like J/$\psi\to\gamma, \omega, \phi$
recoiling against $\pi^0\pi^0$, $\eta\eta$, $\eta\eta^{\prime}$ and
$\eta^{\prime}\eta^{\prime}$. Radiative decays like
$f_0\to\gamma\omega$ and $f_0\to\gamma\phi$  will constrain the flavour
structure. The possibility of a sizeable glueball fraction in the
$\eta^{\prime}$ was studied in the so-called Rosner plot (see Fig.
\ref{plotetap} in section \ref{Mixing}). A similar plot is required for
$f_0(980), f_0(1500), f_0(1760)$, and for the region between the narrow
resonances. Such experiments will become feasible at BESIII.

A coupled-channel partial wave analysis on J/$\psi$ data on
$\gamma, \omega, \phi$ recoiling against $\pi\pi$, $K\bar K$, $4\pi$,
and $\omega\omega$ should be performed on existing data. An amplitude
for the broad isoscalar scalar background should be included in
the fit. If possible, a mass independent phase shift analysis should be
made, not an isobar fit only.

The two scalar mesons $f_0(1500)$ and $f_0(1760)$ are interpreted here
as flavour octet states. At present, experimental data permit
the interpretation of $f_0(2100)$ as a glueball (even though we
believe it to be an octet meson). The most direct way to decide this
alternative is to measure the ratio of its decay modes into
$\eta^{\prime}\eta^{\prime}$ and $\eta\eta^{\prime}$. BESIII should
have good chances to search for both channels in radiative J/$\psi$
decays.

\vspace{1mm}\item The systematic of scalar mesons suggests the
existence of a (nearly) stable scalar charm state, called $D_0^*(1980)$
here, with a mass below its only hadronic decay mode $D\pi$. The
experimental $D\pi$ mass distribution exhibits a -- statistically not
significant -- threshold spike and wider scalar background
contribution (see Fig. \ref{fig:d2p-d3p}b in section
\ref{Heavy--quark spectroscopy}); the partial wave analysis of
$D\pi\pi$ data assigns a fraction of the data to two subthreshold
virtual $D^*$ intermediate states. The possibility that this is a
$D_0^*$ object was not tested. One of these observations might be a
trace of the postulated state.

\vspace{1mm}\item There is the possibility that a narrow $D^*_0$ exists
at about 2450\,MeV/c$^2$, just below the $D_s\,K$ threshold, the
analogue of $a_0(980)$ with open charm. It would be an isodoublet. Both
these states can be searched for at PANDA. If our scheme to organize
scalar mesons is right, it will not be found.

\vspace{1mm}\item The $X(3872)$ is observed in its $\gamma J/\psi$
decay. The $X(3872)\to\gamma\psi(2S)$ decay mode should be searched
for. The $\chi_{bJ=1,2}(2P)$ states have decay branching ratios to
$\Upsilon(2S)$ which exceed those to the $\Upsilon(1S)$ ground state by
a factor 2; for the $\chi_{b0}(2P)$ state, this enhancement factor is
5. The relative yield of $X(3872)\to\gamma\psi(2S)$ should shed light
on the $X(3872)$ structure.

\vspace{1mm}\item The $D^*, D^*_s$ and $B$ decays display the flavour
structure of isoscalar scalar mesons; the statistics is, however, at
the lower limit for far-reaching conclusions. A Super $b$ factory
planned at Frascati will provide $10^{10}$ $B$ and $B_s$ mesons; in
recent years, the discovery potential of $B$ decays for meson
spectroscopy was demonstrated in an impressive way. An immediate remedy
would be to combine the data from BaBaR and BELLE, and for $D^*, D^*_s$
decays those of Focus, E687 and E791 in addition. Such a combined
effort would also help to see if the threshold spike in the $D\pi$ mass
distribution in $B\to D\pi\pi$ decays is fake or real, and if its
quantum numbers are $0^{++}$ or $1^{--}$.

\vspace{1mm}\item In our view, data analysis was often guided by
prejudices. Of course, it is boring (and unnecessary) to test charge
conjugation invariance in every new reaction. However, it is a
legitimate question to ask to what extend isovector mesons are
suppressed in radiative J/$\psi$ decays. For this kind of process, the
Particle Data Group lists one entry only, J/$\psi\to\gamma\pi^0$, which
is small but in a very unfavourable kinematical region. Radiative
yields for $f_2(1270)$ are known, for $a_2(1320)$ they are not.
Likewise, radiative yields for J/$\psi\to\gamma f_4(2050)$ or $\gamma
f_1(1285)$ and $\gamma f_1(1420)$ are large, but no upper limits exist
for their isovector counterparts. Not only J/$\psi\to\gamma\pi^0\pi^0$,
J/$\psi\to\gamma\eta\eta$, J/$\psi\to\gamma\rho\rho$, and
J/$\psi\to\gamma\omega\omega$ should be studied but
J/$\psi\to\gamma\pi^0\eta$ and J/$\psi\to\gamma\rho\omega$ as well. It
is unclear whether these modes have just not been published, if they
escaped detection so far because nobody looked, or if the modes are
suppressed.

In central production, isoscalar final states were studied intensively,
but isovector states were not. The collaborations focussed on detection
of glueballs in Pomeron-Pomeron fusion, hence final states like
$\rho\rho$ and $\omega\omega$ were believed to be interesting,
$\rho\omega$ not. A large fraction of the data had a forward
$\Delta^{++}$ (instead of a fast proton). These data were not analysed.
The $\Delta^{++}$ requires Regge exchanges to be produced; the
$\Delta^{++}$ - proton ratio depends on fraction of Regge exchange in
the data. The Compass experiment at CERN is going to clarify many of
the open issues.

\vspace{1mm}\item There are several experiments which will study
exotic partial waves. New information can be expected from
Pomeron-Regge (Compass) and from Pomeron-photon fusion. First
evidence for the latter process has been reported by the STAR
collaboration at BNL. The search for exotic mesons is also a central
topic of the Hall D project at JLab.

\vspace{1mm}\item The high-mass spectrum of light-quark mesons gives
information on the confinement region. Pioneering new results have
been presented on antiproton-proton annihilation in flight. This
process appears to give the best access to the full meson spectrum. By
their nature, $p\bar p$ formation experiments do not allow the study of
exotic waves; this is however possible in subsystems. At FAIR,
glueballs are hoped to show up at high energy as narrow states above a
continuum background; light meson spectroscopy is not one of the
research goals. Hegels `craft of reason' (List der Vernunft) will
entail the quest for a detailed knowledge of the `background' of $q\bar
q$ mesons. Thus an improved understanding of confinement physics will
emerge from the FAIR research. Experimentally, an extracted low-energy
antiproton beam with momenta between 400\,MeV/c and 2.5 to 3\,GeV/c is
required. The extended energy range (compared to LEAR) gives access to
states at even higher masses. This will provide information on whether
the linear Regge trajectories start to bend, stop, or continue their
linear rise as a function of the squared mass. For part of the data, a
polarised target should be used. A source of polarised antiprotons as
foreseen by the PAX collaboration at FAIR would of course lead to very
significant constraints in the partial wave analysis. The upper part of
the spectrum will be covered by the planned facilities; here, the PANDA
detector is an ideal instrument. For momenta antiproton below 2\,GeV/c,
the options are not yet clear. In case, a second facility would be
required, the asymmetric BaBaR detector would be an ideal detector.

Antiproton-proton annihilation formation experiments give access only
to non-strange mesons; mesons with hidden strangeness are formed only
to the extent to which they mix with $n\bar n$. In production
experiments, planned to be performed with PANDA, the full meson
spectrum including exotic partial waves can be covered. We lack
knowledge of mesons carrying strangeness. With the intense antiproton
beams available at FAIR, a strangeness exchange reaction $p\bar p\to
\Lambda\bar \Lambda$ will produce a source of $\bar \Lambda$.
Exploiting secondary reactions should help to understand the spectrum
of Kaon excitations.

\vspace{1mm}\item Baryons -- which are not discussed here -- have the
potential to provide additional insights into strong interactions which
cannot be gained studying mesons alone.
\end{enumerate}\vspace{-1mm}

\subsubsection{\label{Hints for expert readers}
Hints for expert readers}

Experts in meson spectroscopy may not have time to read
through this lengthy paper but might be interested to see new
developments and interpretations.

Readers interested in the history of meson spectroscopy may consult
subsection \ref{The A,B,C, .. of meson spectroscopy: the naming scheme}.
The subsection \ref{Regge trajectories} contains a brief discussion of
the basics of string models not commonly known to experimental
physicists. Section \ref{Heavy--quark spectroscopy} on heavy quark
spectroscopy is rather compact, and our personal view of the
newly discovered narrow states is presented along with the experimental
data.

For high-mass states, a remarkable mass pattern evolves which we
believe to provide very significant constraints on the dynamics of
quarks at large excitation energies. In subsection \ref{Patterns at
high l, n}, we scrutinise the conjecture of restored chiral symmetry
under these conditions.

The search for hybrids, for mesons with exotic quantum numbers, is
reviewed in section \ref{Mesons with exotic quantum numbers}.

The $\eta(1440)$ is a long-discussed particle and deeply connected with
the search for glueballs among $q\bar q$ mesons. Our view is
given in subsections \ref{Isoscalar resonances revisited} and
\ref{The nonet of pseudoscalar radial excitations}.

An update on the existence of the $\sigma(485)$ and $\kappa(700)$ can
be found in section \ref{Scalar mesons}. There is a full section on
different interpretations of scalar mesons (\ref{Scalar mesons and
their interpretation}). Our interpretation containing new ideas on the
dynamical $\rm r\hat{o}le$ of tetraquark configurations for scalar
mesons is given in section \ref{Global views of scalar mesons and the
scalar glueball}. Comments on dynamically generated resonances and on
multiquark states are presented in subsection  \ref{Dynamical
generation of resonances and flavour exotics}. A global view of scalar
mesons from heavy to light quarks is given at the end of this review,
in section \ref{Scalar states from the sigma to chib0(1P)}.

\subsubsection{\label{Outline}
Outline}

After the introductory section, the review begins with a survey
of expectations based on lattice gauge calculations and
QCD-inspired models (section~\ref{From QCD to strong interactions}).
These expectations provide a `coordinate system' for experiments and
their interpretations. In section \ref{Major experiments} major
experiments in this field are introduced to allow the reader to
appreciate better the significance of results. The spectrum of
meson resonances seen in a particular experiment depends critically on
the method applied. Fusion of $2\gamma$ e.g. couples to quark charges,
while radiative J/$\psi$ decays are a good place to search for
glueballs. The main experimental methods are described in
section~\ref{Experimental methods}. In this part, some experimental
results are discussed which are of general interest for $q\bar q$
spectroscopy but unrelated to the main issues of this report.

The spectroscopy of heavy quarks -- reviewed in
section~\ref{Heavy--quark spectroscopy} -- reveals a very simple pattern
and serves as textbook example providing evidence for Coulomb-like
interactions and a linear confinement potential between a heavy quark
and an antiquark at large separations. In spite of this apparent
simplicity, surprising discoveries were made recently. Some $D_s^*$
resonances are surprisingly narrow and have masses which fall much
below the expected values. The $X(3872)$ resonance has large decay
branching ratios into both, a J/$\psi$ recoiling against a $\rho$ or
against an $\omega$ meson. These two decays are observed with similar
strengths, hence isospin symmetry is broken violently. Higher mass
states, $X(3940)$, $Y(3940)$, $Z(3930)$, and $X(4260)$, are observed
which could need an interpretation beyond the standard $q\bar q$
scenario.

The discussion of light-quark spectroscopy and of searches for
glueballs and hybrids begins in section \ref{High radial and orbital
excitations of light mesons} with a global survey of the meson
spectrum. A statistical approach to the numerous results exhibits
striking regularities pointing at a simple structure of the confining
interactions. The regularities are very helpful in assigning
spectroscopic quantum numbers to the observed states and to identify
mesons which fall outside of the usual nonet classification of meson
resonances. These additional mesons may be glueballs, hybrids or
multiquark states.

In hybrids, the gluonic fields transmitting the forces between quark
and antiquark are themselves excited. The gluon field may acquire an
effective `constituent' mass and thus hybrids are no longer
two-particle bound states; they may have quantum numbers outside of the
$q\bar q$ scheme. Mesons with exotic quantum numbers have been
reported but the evidence for their existence as resonant states is not
beyond doubt; their interpretation as hybrids (instead of $ qq\bar
q\bar q$ tetraquark states) is not yet based on solid experimental
facts.  Evidence for exotic mesons is reviewed in section
\ref{Mesons with exotic quantum numbers}.

Light-quark spectroscopy receives its continued fascination from the
QCD-based prediction that the self-interaction between gluons as
mediators of strong interactions should lead to new forms of hadronic
matter. Glueballs were predicted as mesonic bound states having no
valence quarks. The $\eta(1440)$ was for a long time a prime glueball
candidate; its $\rm r\hat{o}le$ is discussed in section
\ref{Pseudoscalar mesons} in the context of other pseudoscalar mesons.

Most exciting is the possibility that the three mesons $f_0(1370)$,
$f_0(1500)$, and $f_0(1710)$ might reflect the intrusion of a scalar
glueball into the world of quarkonia. However, the existence of
$f_0(1370)$ is not beyond doubt, and there is no unanimous agreement if
and how gluonic degrees of freedom manifest themselves in spectroscopy.
The interpretation of the scalar meson spectrum is a most controversial
topic which is discussed in sections \ref{Scalar mesons} - \ref{Scalar
mesons and their interpretation}.

A short outlook is given in section \ref{Outlook}. A summary of the
report and our conclusions were already presented above.

\subsubsection{\label{A guide to the literature}
A guide to the literature}

There is an enormous number of papers related to meson
spectroscopy covering a wide range of topics and results. It is of
course impossible to discuss them all in depth or even just to mention
them. The selection of results presented in this review may reflect
personal biases and even important contributions may have escaped our
attention. Apologies go to all authors whose results are missing or
not properly represented here.

A prime source of information is of course contained in the Review of
Particle Properties which contains most published results on meson
spectroscopy. The last one which is used for this review appeared in
2006 \cite{Eidelman:2004wy}. The main review is published in even
years; in odd years, additions can be found in the web edition. The
Review was first issued by Gell-Mann and Rosenfeld in 1957
\cite{Gell-Mann:1957wh}. This was the birth of series which is
indispensable for any particle physicists. This year (2007), the
50$^{\rm th}$ anniversary is celebrated.

The development of meson spectroscopy can be followed closely by
reading the Proceedings of specialised Conferences and Workshops. The
series of International Conferences on Meson Spectroscopy was begun in
1969 and continued for 14 years
\cite{Meson69,Meson71,Meson72,Meson74,Meson77,Meson81,Meson83};
after 1985 it was held regularly as biennial conference on Hadron
Spectroscopy \cite{Hadron85,Hadron87,Hadron89,Hadron91,Hadron93,%
Hadron95,Hadron97,Hadron99,Hadron01,Hadron03,Hadron05}. The last
conference of this series took place in Rio de Janeiro in 2005;
Hadron07 will be hosted by Frascati. In some years, additional
workshops were organised which were often devoted to specific topics or
were born out of regional needs: an incomplete list
comprises~\cite{EXTRA89,EXTRA90,EXTRA92,EXTRA99,EXTRA03a,EXTRA03b}.
Confinement and other theoretical aspects were felt to require a deeper
discussion in an own conference series on Quark Confinement and the
Hadron Spectrum \cite{Conf94,Conf97,Conf98,Conf00,Conf02,Conf04,Conf06}.

More pedagogical introductions can be found in the Proceedings of
dedicated Schools. We quote here a few lectures which may still be
useful, from the CERN summer school~\cite{Quinn:2002ti}, the PSI summer
school at Zuoz~\cite{Klempt:2000ud}, The Advanced School on Quantum
Chromodynamics at Benasque in Spain \cite{Page:2000sp}, the Schools at
Erice, Sicily~\cite{Seth:2003pp}, the Scottish Universities Summer
School~\cite{Ioffe:2004nk}, and the Hampton University Graduate School
at Jefferson Lab~\cite{Klempt:2004yz,Jaffe:2004ph}.

Meson spectroscopy is part of the much wider field of strong
interaction physics. Related are workshops and conference series'
devoted to chiral dynamics~\cite{CD03}, lattice QCD~\cite{LATTICE05},
hadron structure \cite{HSQCD04} or to meson and nuclear physics
\cite{MENU04}. Often, interesting results in meson spectroscopy are
discussed. We give here references only to the most recent workshop.

Finally, we refer to some recent reviews on meson spectroscopy. Godfrey
and Napolitano~\cite{Godfrey:1998pd} provide a good introduction to the
field; recent reviews concentrate on gluonic degrees of freedom
\cite{Amsler:1998up,Li:2003xv,Amsler:2004ps,Bugg:2004xu,Zhu:2007wz}.
Due to the non-Abelian structure of QCD, the dynamics of quarks and
gluons cannot be calculated analytically from QCD but effective
theories provide quantitative predictions in restricted areas. At low
energies, strong interactions can be expanded in terms of (low) pion
momenta and the pion mass; reviews on Chiral Perturbation Theories can
be found in \cite{Gasser:1983yg,Gasser:1984ux,Meissner:1993ah,%
Weinberg:1996kr,Bernard:2006gx}. If one of the quarks is very heavy, it
remains stationary in the hadron rest frame and the spin remains
frozen. This gives rise to Heavy Quark Effective Theories
\cite{Korner:1991kf,Neubert:1993mb}. QCD inspired models still play a
decisive $\rm r\hat{o}le$ for understanding the phenomenology. Three
papers by Barnes and colleagues present detailed information on meson
masses and decays \cite{Barnes:1996ff,Barnes:2002mu,Barnes:2005pb}. The
remarkable progress in QCD simulations on a lattice is reviewed
extensively in \cite{McNeile:2003dy,Khan:2005cf}. An detailed
documentation of the physics of heavy quarkonia can be found
in~\cite{Brambilla:2004wf}.

\subsection{\label{Phenomenologyy of qbar q mesons}
Phenomenology of $q\bar q$ mesons}

\subsubsection{\label{The A,B,C, .. of meson spectroscopy: the naming
scheme}
The A,B,C, .. of meson spectroscopy: the naming scheme}

Particles have names which go back into the early history of particle
physics. Today, a rational scheme is used where letter and
subscripts have a precise meaning. When the first mesons were
discovered, no scheme existed, and names were often invented to
memorise the discoverer, his wife or his children.

The naming of mesons started with a lepton. The mesotron had a mass
between proton and electron mass and proved quickly not be the
predicted mediator of nuclear forces. The name changed to $\mu$-meson
and to muon to avoid any confusion with hadronic mesons. The quantum of
nuclear interaction proved to be the \boldmath$\pi$\unboldmath-meson,
discovered \cite{Lattes:1947mw} by Cecil Frank {\bf P}owell who
received the Nobel prize in 1950 for this discovery. It became
convention to use Greek symbols for new particles. The
\boldmath$\eta$\unboldmath\ followed, discovered by {\bf A}hud Pevsner
\cite{Pevsner}, the $\omega$ \cite{Maglic:1961nz}, the
\boldmath$\phi$\unboldmath\ discovered by {\bf P}.L. Connolly
\cite{Connolly} (the name $\phi$ was suggested by Sakurai). The
$\sigma$ (with mass $\sim 500$\,MeV/c$^2$) was introduced to describe
nuclear interactions by meson exchange current and later understood as
scalar $2\pi$ correlation effect. It is separated in mass from the
$\epsilon$ (with mass $\sim 1300$\,MeV/c$^2$) by a deep valley, for some
time called S$^{\ast}$. The strange analogue of the $\sigma$ is the
$\kappa$. The $\theta$ and $\tau$ particles proved to be same particle
which was then named Kaon $K$. Then the habit changed to denote new
meson resonances by roman letters.

{\bf A:} The $a_2(1320)$ meson e.g. was formerly
called \boldmath$A$\unboldmath\ meson. Later it was observed to be split
into a pseudovectorial state A$_1$ and a tensor A$_2$. The couple
Gerson and Sulamith Goldhaber discovered  jointly the resonance complex
at the Bevatron in Berkeley \cite{Goldhaber}, and decided to call it
\boldmath$A$\unboldmath\ because of their son {\bf A}mos. Sulamith died
a year after the discovery, in 1965, at an age of 42 after having made
remarkable contributions to particle physics.

\vspace*{-3mm}{\bf B:} None of us authors knows if A. {\bf B}ondar
played a leading $\rm r\hat{o}le$ in the discovery of the $b_1(1235)$, formerly
called \boldmath$B$\unboldmath\ meson \cite{Bondar}. It used to have
the nickname {\bf B}uddha particle, a buddhist member of the
collaboration suggested this name. In \pbp\ annihilation into
$\omega\pi\pi$, it resides like a {\bf B}uddha on the $\omega\pi$
background.

\vspace*{-3mm}{\bf C:} The \boldmath$C$\unboldmath\ meson, now called $
K_1(1270)$, was discovered by Lucien Montanet and others \cite{Astier}.
Its name reminds of its discovery at {\bf C}ERN.

\vspace*{-3mm}{\bf D:} The \boldmath$D$\unboldmath\ meson, or
$f_1(1285)$, was discovered independently at BNL \cite{Miller} and at CERN
\cite{Andlau}\footnote{\scriptsize \cite{Miller} was submitted in May
1965 and published in June, two weeks after \cite{Andlau} was
submitted.}.  The first author {\bf D}onald H. Miller had a son {\bf
D}ouglas Allen. The letter $D$ is now used to denote mesons with open
charm.

\vspace*{-3mm}{\bf E:} The history of the \boldmath$E$\unboldmath\
or $\eta(1440)$ meson is particularly interesting. It was first reported
in 1965 at the Sienna conference on High Energy Physics. It was found
in $ p\bar p$ annihilation at rest and was the first meson
discovered in {\bf E}urope. People at CERN were obviously very cautious
not to present a wrong result, only two years later the
\boldmath$E$\unboldmath\ meson was reported in a journal publication
\cite{Baillon:1967}. A detailed study was presented which included
several final states and a full partial wave analysis yielding
pseudoscalar quantum numbers. The name \boldmath$E$\unboldmath\ meson
may stand for {\bf E}urope but remembers also Lucien Montanet's
daughter {\bf E}lisabeth. In the same year, a second meson was
discovered which is now called $f_1(1420)$. It has nearly the same mass
as the \boldmath$E$\unboldmath\ meson and was called
\boldmath$E$\unboldmath , too, since the quantum numbers in
\cite{Baillon:1967} were believed to be possibly wrong (see, however,
\cite{Baillon:1982ap}). Now we know that the state produced in $ p\bar
p$ annihilation has pseudoscalar quantum numbers. But this confusion
led the MarkIII collaboration to call a state at 1440\,MeV/c$^2$,
discovered in radiative J/$\psi$ decay \cite{Scharre:1980zh},
$iota(1440)$ or $\iota(1440)$ to underline the claim
\cite{Chanowitz:1980gu} that the first glueball was discovered
($\iota=1$ in Greek). Indeed, the meson proved to have pseudoscalar
quantum numbers \cite{Edwards:1982nc}. The history of this meson is not
yet closed and the discussion will be resumed in this report.

\vspace*{-3mm}{\bf F:} The $f_2(1270)$ was introduced by Walter Selove
as \boldmath$f$\unboldmath\ meson to memorise his wife {\bf F}ay
Ajzenberg \cite{Selove}. Fay herself is a well-known nuclear physicist
awarded with 3 doctorates honoris causa. She wrote the book `A Matter
of Choices: Memoirs of a Female Physicist (Lives of Women in Science)'.
For some time, $ D_s$ mesons were called F mesons.

\vspace*{-3mm}{\bf G,H:} The former \boldmath$G$\unboldmath-meson was
discovered by {\bf G}oldberg and others \cite{Goldberg} (now
$\rho_3(1690)$), the $H$ (claimed at 975\,MeV/c$^2$\,) \cite{Bartsch:1964}
has become the $h_1(1170)$. The letter $G$ was later used to claim the
discovery of a glueball. The $g_s(1470)$ \cite{Etkin:1982se} may have
been an early sighting of the glueball candidate $f_0(1500)$. Certainly
the $G(1590)$ \cite{Binon:1983ny} is closely related to it. A tensor
glueball was suggested to be mixed into 3 tensor resonances $g_T(2010),
g_T^{\prime}(2300)$ and $g_T^{\prime\prime}(2340)$ \cite{Etkin:1987rj}.

\begin{figure}[ph]
\begin{minipage}[c]{0.50\textwidth}
{\bf J:} The {\bf J}\boldmath/$\psi$\unboldmath\ must not be missed.
This narrow resonance revolutionised physics, it was discovered
independently in 1974 at Brookhaven National Laboratory (BNL) in Long
Island, New York, in the process proton + Be $ \to e^+e^-$ + anything
\cite{Aubert:1974js}. The group leader was Samuel Chao Chung Ting, the
Chinese symbol for Ting reads {\bf J}. At Stanford University, the new
resonance was observed in the SPEAR storage ring in $ e^+e^-$
annihilation to $\mu^+\mu^- , e^+e^-$ and into
hadrons~\cite{Augustin:1974xw}. It was seen in a series $\psi$,
$\psi^{\prime}$, $\psi^{\prime\prime}$. Its name is derived from the
$\psi$--shaped pattern in $\psi^{\prime}$ decays into $\pi^+\pi^-$ pair
and the J/$\psi$ (decaying into $e+e-$), see Fig. \ref{psiprime}. The
name J/$\psi$ honours both discoveries. \end{minipage}
\hspace{10mm}\begin{minipage}[r]{0.40\textwidth}
\hspace{10mm}\mbox{\epsfig{file=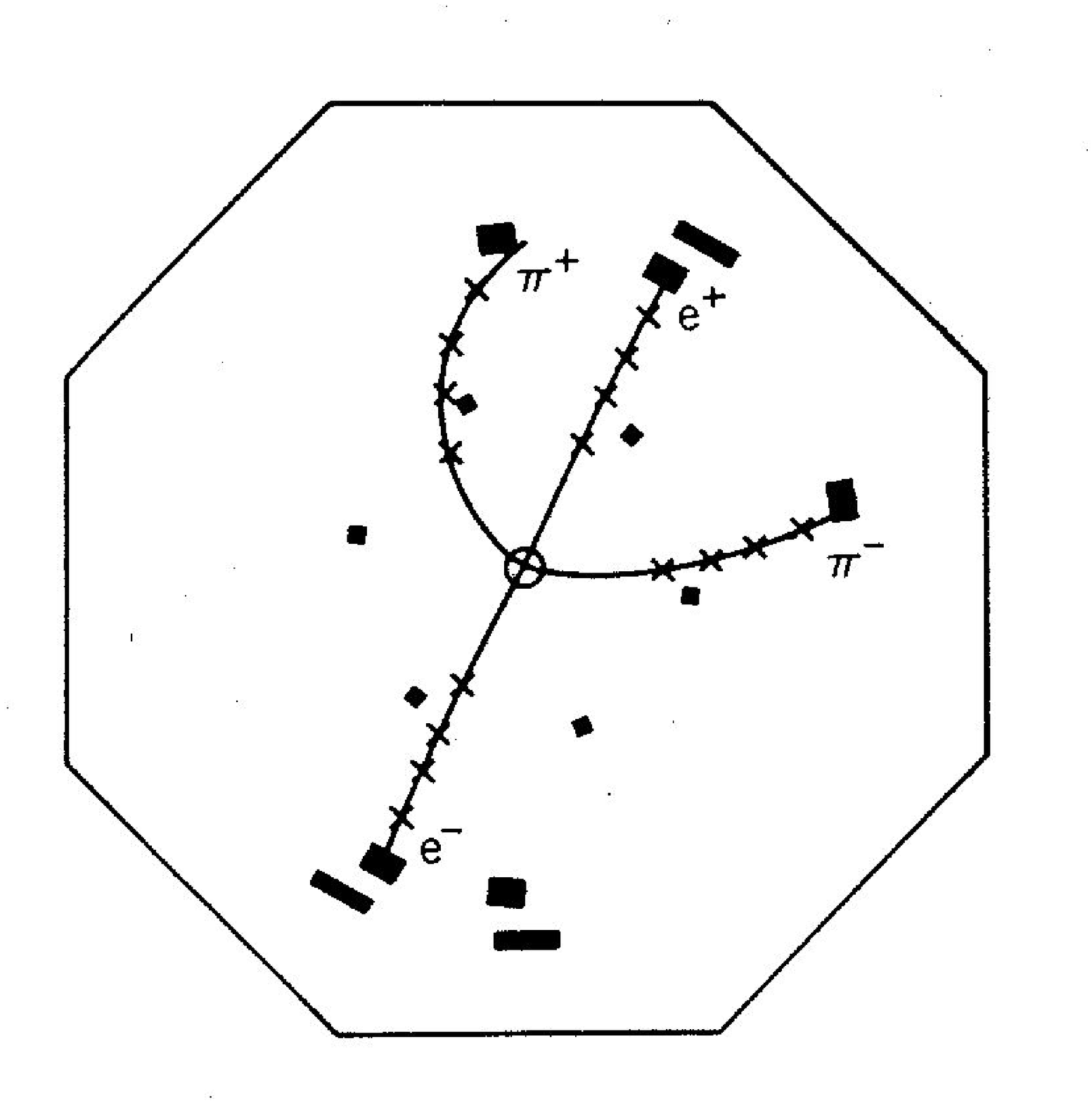,width=.8\textwidth}}
\vspace{-5mm}\hspace{5mm}\caption{\label{psiprime}
Decay of a $\psi^{\prime}$ into $\pi^+\pi^-J/\psi$ with
J/$\psi$ decaying into a high--energy $e^+e^-$ pair in the MARKIII
experiment. }
\end{minipage}
\end{figure}
\vspace*{-1mm}{\bf K,L, ... Z:} The Kaon $K$ was already mentioned; the
$K_2(1770)$ was discovered as $L$(1770) \cite{Bartsch}. The $M(1405)$
was a precursor of the $\pi_1(1400)$ \cite{Alde:1988bv}. The $N$ and
$P$ are reserved for neutron/nucleon and proton, the O is void. There
were two $Q$ mesons (now $K_1(1270)$ and $K_1(1400)$), {\bf R, S, T}
were used to name enhancements in the missing mass in the reaction $
\pi^-p\to p X$ \cite{Focacci}. With the $W$ as charged weak interaction
boson, with $X$ as the unknown, with $\Upsilon$ as bottomonium family
and with $Z$ as neutral weak interaction boson, the alphabet was
exhausted.

With the increasing number of inhabitants in the particle zoo, rules had
to be invented like Carl v. Linn\'e (Linnaeus, 1707 - 1778) did in
biology. The system was introduced by the Particle Data Group in
1986. New particle symbols were chosen which are simple, convey the
quantum numbers, and are close, whenever possible, to the traditional
names listed above. Table \ref{names} translates meson names into
quantum numbers. Mesons are characterised by their quantum numbers,
their total angular momentum $J$ , their parity $P$ and their charge
conjugation parity $C$:
\be
P=(-1)^{L+1}~~~~~~~C=(-1)^{L+S}
\ee
where $L$ denotes the intrinsic orbital angular momentum in the
$q\bar q$ meson and $S$ the total spin of the $q\bar q$ pair. The
notation $J^{PC}$ is used not only for the neutral mesons for which
$C$ parity is a good quantum number but also for the other components
of a SU(3) nonet. For instance, since $C\pi^0=\pi^0$, we define
$C\pi^\pm=\pi^\mp$, and $C\K^0=\Kb{}^0$, and with $C\rho^0=-\rho^0$, we
adopt $C\rho^\pm=-\rho^\mp$ and $C\K^{\star\,\pm}=-\Kb{}^{\star\,\mp}$.
In addition, a sign for each flavour needs to be defined. The
convention is used that the sign of a flavour is the sign of its
charge, $c$ quarks have positive charm, $s$ quarks have negative
strangeness.

The {\it up} and the {\it down} quark form an isospin doublet
We define antiquarks by $\bar u\,=\,C\,u$ and
$\bar d\,=\,C\,d$.  We obtain:

\begin{equation}\label{mes:eq:states}
\renewcommand{\arraystretch}{1.3}
\begin{tabular}{ccccc}
$\ket{I=1, I_3=1 }$&=&$-\ket{\udb}~,$\\
$\ket{I=1, I_3=0 }$&=&$\frac{1}{\sqrt{2}}(\ket{\uub} - \ket{\ddb})~,$\\
$\ket{I=1, I_3=-1}$&=&$\ket{\dub}~,$\\
$\ket{I=0, I_3=0 }$&=&$\frac{1}{\sqrt{2}}(\ket{\uub} + \ket{\ddb})$&=&
$\ket{\nnb}~,$\\
$\ket{I=0, I_3=0 }$&=&$\ket{\ssb}~.$\\
\end{tabular} \renewcommand{\arraystretch}{1.0}
\end{equation}

The \ket{\nnb} and \ket{\ssb} states have the same quantum numbers
and mix to form two physical states. With $n$ we denote the two
lightest quarks, $u$ and $d$, the symbol $q$ stands for any quark,
including heavy quarks.

The $G$-parity is approximately conserved in strong interactions. It is
defined as charge conjugation followed by a rotation in isospin space
about the $y$-axis,
\begin{equation} G =C e^{i\pi I_y} =  (-1)^I C =
(-1)^{L+S+I}\,. \end{equation}
For a system of $n_{\pi}$ pions, $G=(-1)^{n_\pi}$. The $G$-parity plays
a similar  $\rm r\hat{o}le$ as $C$-parity in QED where the relation
$C=(-1)^n$ holds for $e^+e^-\rightarrow n\gamma$.

The $\omega_3$ is a isoscalar meson with a flavour wave function
$\frac{1}{\sqrt 2} (d\bar d+u\bar u)=\nnb$ even though it may have a
small $s\bar s$ component, its total angular momentum is $J=3$, its
intrinsic orbital angular momentum is even (with $L=2$ dominant), and
it is member of the spin triplet $\omega_1,\omega_2,\omega_3$. Only the
$\omega_3(1670)$ is experimentally known. The mass is added to identify
a meson. The index for $J$ and the mass are omitted when no confusion
can occur like for the $\pi, \eta$, or $\phi$. The name J/$\psi$ is
kept.

\begin{table}[pt]
\caption{\label{names}
Symbols of mesons with no open strangeness, charm or
beauty\vspace*{2mm}} \bc \renewcommand{\arraystretch}{1.7}
\begin{tabular}{lccccc}
\hline\hline
$q\bar q$ content & $^{2S+1}L_J=$ &$^1(L\ even)_J$       & $^1(L\
odd)_J$       &  $^3(L\ even)_J$    & $^3(L\ odd)_J$      \\
                  &$ J^{PC}$   &
$0^{-+},2^{-+}\cdots$&$1^{+-},3^{+-}\cdots$
&$1^{--},2^{--},3^{--}\cdots$&$0^{++},1^{++},2^{++}\cdots$\\ $u\bar d,
d\bar d-u\bar u,d\bar u$&&     $\pi$           &      $b$             &
$\rho$ &       $a$           \\ $d\bar d+u\bar u, s\bar s$
&&$\eta,\eta^{\prime}$ &    $h,h^{\prime}$    &   $\omega, \phi$    &
$f,f^{\prime}$    \\ $c\bar c$                         &&     $\eta_c$
&       $h_c$          &         $\psi$      &        $\chi_c$     \\
$b\bar b$                         &&     $\eta_b$        &       $h_b$          &      $\Upsilon$     &        $\chi_b$      \\
\hline\hline
\end{tabular}
\renewcommand{\arraystretch}{1.0}
\ec
\end{table}

Mesons with open strangeness, charm or beauty are no
charge-conjugation eigenstates and the four columns of
Table~\ref{names} collapse into two. The normal series ($^1(L\ odd)_J$,
$^3(L\ odd)_J$) with $J^P=0^+,1^-,\ldots$ is tagged by a star, the
'anomalous' series ($^1(L\ even)_J$, $^3(L\ even)_J$) with
$J^P=0^-,1^+,\ldots$ is untagged. Mesons belonging to the normal series
are, e.g., $K_3^{\ast}$ and $D^{\ast+} _{s0}(2317)$. The latter
describes a $c\bar s$ state with $J=0$ and positive parity. Anomalous
mesons have no star, like the $D_{s1} ^+(2460)$ with $J=1$ and
$P=+1$. Bottom mesons $B^{\pm}$ have a mass of more than 5\,GeV/c$^2$;
their decays cover the full charmonium spectrum and may even contribute
to glueball spectroscopy. The wave functions of the $K_1(1270)$ and
$K_1(1400)$ have $L$ even and can have intrinsic quark spin 1 or 0 (and
would then belong to the $a_1(1260)$ or the $b_1(1235)$ nonet,
respectively). In practice, they are mixtures of the two quark model
configurations.

Quantum numbers like $ J^{PC}=0^{+-},1^{-+},2^{+-}\cdots$  are not
accessible in fermion-antifermion systems. In analogy to the $\pi$
(with  $ J^{PC}=0^{-+}$) and $\pi_2(1670)$ (with $ J^{PC}=2^{-+}$),
these states are called $\pi_1$. Its isoscalar companion would be
$\eta_1$. If a  $ J^{PC}=1^{-+}$ exotic resonance in the charmonium
spectrum would be discovered, its name would be $\eta_{c1}$. Exotic
states $b_0, h_0, b_2, h_2$ with $0^{+-}$ and $2^{+-}$, respectively,
may exist as well. So far, only evidence for states with exotic
quantum numbers $(I^G)J^{PC}=(1^-)1^{-+}$ has been reported.

\subsubsection{\label{Meson nonets}
Meson nonets}

SU(3) symmetry considerations have shown to be extremely useful in
classifying mesons and baryons \cite{Gell-Mann:1962xb} as composed of
quarks and antiquark or of 3 quarks, respectively. Light mesons
with $u, d$ and $s$ quarks have a flavour wave function which is
conveniently decomposed into a flavour basis. For pseudoscalar mesons,
this is given by $\pi^+=-d\bar u$, $\pi^-=\bar ud$, $ K^0=\bar sd$,
$K^+=\bar su$, $ \bar K^0=\bar ds$, $ K^-=\bar us$. The quark model
representation of the neutral pseudoscalar mesons is $\pi^0 =
\frac{1}{\sqrt 2}(\uub - \ddb)$,

\begin{equation}
\label{qcd:mes:eq:neutrals}
\eta_8 = \sqrt{\frac{1}{6}}(\uub +\ddb -2\ssb )~,\qquad {\rm
and}\qquad\eta_1 = \sqrt{\frac{1}{3}}(\uub +\ddb +\ssb )~.
\end{equation}

\begin{figure}[pt]
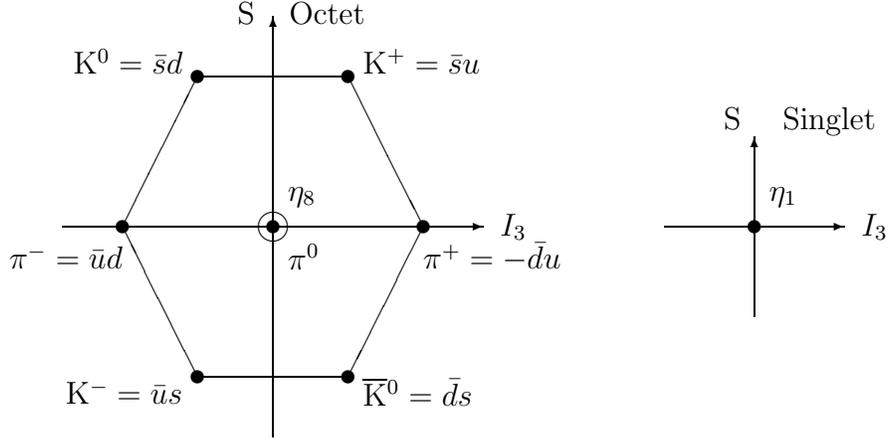

\begin{center}
\nonett{$\K^0=\bar
sd$}{$\Kp=\bar su$}{$\Km=\bar us$} {$\Kb^0=\bar ds$}{$\pi^-=\bar
ud$}{$\pi^+=-\bar du$}{$\pi^0$} {$\eta_8$}{$\eta_1$}
\end{center}
\vspace*{-8mm}\caption{The nonet of pseudoscalar mesons.
\label{mes:fig:pseudo} }
\end{figure}

We thus have families of nine mesons with the same radial and
angular-momentum structure of the wave function. For pseudoscalar
mesons, a nonet is shown in Fig.~\ref{mes:fig:pseudo}.

Other nonets are formed by the ground state vector mesons with $\rm
J^{PC}=1^{--}$,~~\{~$\rho , \phi , \omega$,  $K^*$~\},~~by
tensor mesons with $\rm J^{PC}=2^{++}$,~~\{~$a_2(1320) , f_2'(1525),
f_2(1270),  K_2^*(1430)$~\}, and many others.

\subsubsection{\label{Mixing}
Mixing}

Mesons having identical external quantum numbers can mix even if they
have different internal flavour structures or different angular momentum
configurations. The isoscalar pseudoscalar mesons are not realised as
SU(3) eigenstates; the physical mass eigenstates, using the
abbreviation  $\nnb=\frac{1}{\sqrt 2}(\uub +\ddb)$, can be written as
\begin{eqnarray}
\label{qcd:meson:eq:psmix}
\ket{\eta\phantom{^\prime}} &= \cos\Theta_P\,\ket{\nnb} -
     \sin\Theta_P\,\ket{\ssb} &=  \cos\theta_P\ket{\eta_8} -
                                \sin\theta_P\ket{\eta_1}\cr
\ket{\eta^{\prime}} &=          \sin\Theta_P\,\ket{\nnb} +
     \cos\Theta_P\,\ket{\ssb} &=  \sin\theta_P\ket{\eta_8} +
                                \cos\theta_P\ket{\eta_1}\,.
\end{eqnarray}

Nothing forbids the wave functions to have a $c\bar c$ component. The
huge mass difference between $\eta$ and $\eta_c$ makes, however, the
mixing angle very small. The pseudoscalar mixing angle is defined using
eigenstates of SU(3) symmetry as basis vectors. The two mixing
angles are related by $\Theta_P=\theta_P-\Theta_{\rm id}+\pi/2$. The
angle $\Theta_{\rm id}=\arctan(1/\sqrt{2})$ is defined below.

This mixing scheme is highly simplified: it is assumed that the spatial
wave functions of all ground-state pseudoscalar mesons are identical.
Clearly, it is not to be expected that these assumptions hold to a good
precision, that the wave function of the pion is the same as that of
the $\bar nn$ component of the $\eta$ and the $\eta^{\prime}$. Further,
the $\eta$ and $\eta^{\prime}$ could also mix with other states, in
particular with a pseudoscalar glueball. This has led to speculations
that the $\eta^{\prime}$ (and to a lesser extend also the $\eta$) may
contain a large fraction of glue. There is an abundant literature on
this subject, and we refer the reader to~\cite{Feldmann:1998vh}.

In a simplified form, one can extend the mixing scheme
\ref{qcd:meson:eq:psmix} to include an inert (gluonic) component which
does not couple to charges. The $\eta$ and $\eta^{\prime}$ wave
functions are then written as
\begin{equation}
\label{mixglue}
\begin{array}{rcl}
|\eta\rangle &=&
X_\eta|\eta_q\rangle+Y_\eta|\eta_s\rangle+Z_\eta|G\rangle\ ,\\[1ex]
|\eta^\prime\rangle &=&
X_{\eta^\prime}|\eta_q\rangle+Y_{\eta^\prime}|\eta_s\rangle+Z_{\eta^\prime}|G\rangle\ ,
\end{array}
\end{equation}
with $X_{\eta(\eta^\prime)}^2+Y_{\eta(\eta^\prime)}^2+Z_{\eta(\eta^\prime)}^2=1$
and thus $X_{\eta(\eta^\prime)}^2+Y_{\eta(\eta^\prime)}^2\leq 1$.
A gluonic admixture implies that
$Z_{\eta(\eta^\prime)}^2=1-X_{\eta(\eta^\prime)}^2-Y_{\eta(\eta^\prime)}^2$
must take on a positive value.
The mixing scheme assumes isospin symmetry, \emph{i.e.~}excludes
mixing with $\pi^0$, and neglects other possible admixtures from $c\bar
c$ states and/or radial excitations. In absence of gluonium,
$Z_{\eta(\eta^\prime)}\equiv 0$, the mixing parametrisation
(\ref{mixglue}) is reduced to the scheme (\ref{qcd:meson:eq:psmix}).

\begin{figure}[pt]
\centerline{\includegraphics[width=0.4\textwidth]{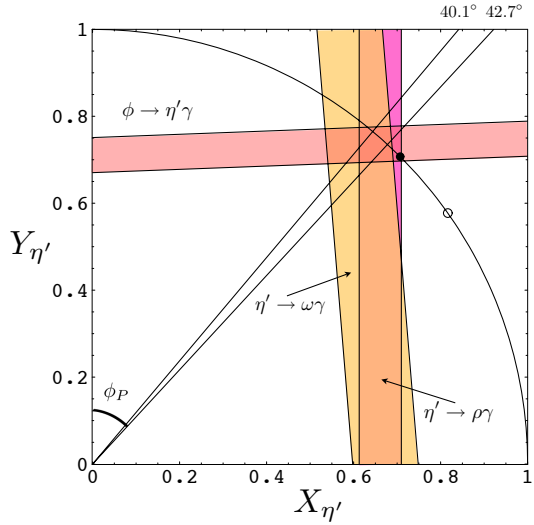}}
\caption{\label{plotetap}Constraints on the flavour wave function of
$\eta^{\prime}$ mesons from radiative decays. A point in the plane
defines the non-strange ($X_{\eta^\prime}$) and strange
($Y_{\eta^\prime}$) component in the $\eta^\prime$. The relation
$X_{\eta(\eta^\prime)}^2+Y_{\eta(\eta^\prime)}^2\leq 1$ must hold. The
two small circles define a pure singlet wave function
(open circle) or a mixing angle of $-11.1^{\circ}$ in the basis of eqs.
(\ref{qcd:mes:eq:neutrals}). The vertical and inclined bands show the
constraints for $X_{\eta^\prime}$ and $Y_{\eta^\prime}$ from
$\eta^\prime\to(\rho,\omega)\gamma$ and $\phi\to\eta^\prime\gamma$
transitions.} \vspace{-4mm}
\end{figure}

The first analysis of this type was carried out by Rosner
\cite{Rosner:1982ey} and diagram \ref{plotetap} is often called Rosner
plot. New data on the ratio $R_{\phi} = BR(\phi \to \eta^{\prime}
\gamma)/BR(\phi \to \eta \gamma)$ led the KLOE collaboration to
estimate the gluonium fractional content of $\eta^{\prime}$ meson as
$Z^2 = 0.14\pm0.04$.

A re-evaluation of this data and data on radiative decays of vector to
pseudoscalar and to pseudoscalar to vector mesons by Escribano and
Nadal was presented in \cite{Escribano:2007cd}. The radiative widths
for $\eta^{\prime}\to\rho\gamma$ and $\eta^{\prime}\to\omega\gamma$
decays are proportional to the $n\bar n$ content of the $\eta^{\prime}$
wave function, and a measurement of the rate constrains
$X_\eta^{\prime}$. The $\omega$ meson has a small $s\bar s$ component
and the area allowed by experiment is tilted (see Fig. \ref{plotetap}).
The $\phi\to\eta\gamma$ and $\phi\to\eta^{\prime}\gamma$  yields
determine the $s\bar s$ components.

The latter analysis is consistent with a negligible gluonic content in
$\eta^\prime$ (and in $\eta$) mesons, $Z_{\eta^\prime}^2=0.04\pm 0.10$;
$Z_\eta^2=0.00\pm 0.07$. For a vanishing gluonic component, the
$\eta$ and $\eta^\prime$ mixing angle is found to be
$\Theta_P=(41.5\pm 1.2)^\circ$ in the quark-basis scheme and
$\theta_P=-(13.2\pm 1.2)^\circ$ in the singlet-octet scheme. A variety
of reactions has been used in the past to determine the pseudoscalar
mixing angles. Recently, the transition form factors $F_{\eta
\gamma}(Q^2)$ and $F_{\eta' \gamma}(Q^2)$ have been used to derive
$\Theta_P= 38.0^{\circ}\pm 1.0^{\circ}\pm 2.0^{\circ}$
\cite{Huang:2006as} where references to earlier work can be found.

The sign in (\ref{qcd:meson:eq:psmix}) is a convention used only for
pseudoscalar mesons; for all other meson nonets, the convention is
chosen as for the vector meson nonet:
 \begin{equation}
\label{meson:eq:vmix}
\eqalign{
&\ket{\omega} = \ \cos\Theta_V\,\ket{\omega_1} +
\sin\Theta_V\,\ket{\omega_8} \cr &\ket{\phi} =  -
\sin\Theta_V\,\ket{\omega_1} + \cos\Theta_V\,\ket{\omega_8}}
\end{equation}
 For $\Theta_V=\Theta_{\rm id}=\arctan(1/\sqrt{2})$, the
wave functions reduce to $|\omega\rangle = |\nnb\rangle$, $|\phi\rangle
= |\ssb\rangle$. This angle is called {\it ideal mixing angle}
$\Theta_{\rm id}$.

A second mixing example was already discussed: there are two nonets
with $J^{PC}=1^{++}$ and $1^{+-}$, respectively, the \{$a_1(1260),
f_1(1285), f_1(1510),  K_{1A}$\} and \{$b_1(1235), h_1(1170),
h_1(1380),  K_{1B}$\}. $ K_{1A}, K_{1B}$ are defined as spin
triplet and spin singlet states. The internal quark spin is not a
measurable quantity, the two states $| 1^1{ P}_1\rangle $ and $|
1^3{ P}_1\rangle $ mix forming $ K_1(1280)$ and $K_1(1400)$.
\begin{equation}
 |K_1(1280)\rangle = +\cos(\theta) |1^1{ P}_1\rangle
+ \sin(\theta) |1^3{ P}_1\rangle \phantom{\ .}
\end{equation}
and
\begin{equation}
 |K_1(1400)\rangle =
-\sin(\theta) |1^1{ P}_1\rangle
+
\cos(\theta) |1^3{ P}_1\rangle \ .
\end{equation}
The charge conjugation operator $\mathcal{C}$ gives opposite phases when
applied to $|^1{ P}_1\rangle$ and $|^3{ P}_1\rangle$ basis states. We
impose $K_1(1280)$ and $\bar K_1(1280)$ to have the same decomposition
into $| 1^1{ P}_1\rangle $ and $| 1^3{ P}_1\rangle $, then $\rm\theta$
must be $\pi/4$. This is compatible with data on $ K_1$ decays.
Likewise, there is evidence for two states $ K_2(1770)$ and $
K_2(1820)$ which are driven by the quark model states  $|^1{
D}_2\rangle$ and $|^3{ D}_2\rangle$. Similar mixing phenomena can be
expected for the $ D$ and $ B$ families.

\subsubsection{\label{mes:sub:Zweig}
The Zweig rule}

\begin{figure}[!htpc]
\begin{center}
\includegraphics*[width=.7\textwidth]{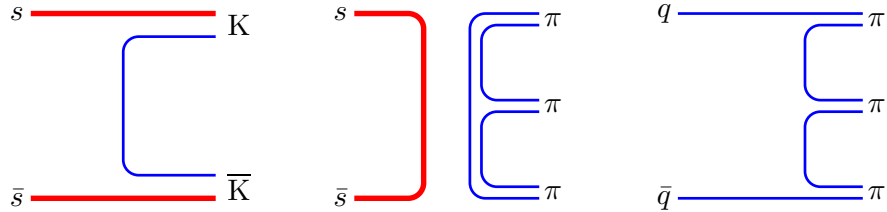}
\end{center}
\caption{\label{mes:fig:OZI1}
Connected (left) and disconnected (centre)  contribution to
$\phi(1020)$ decay. The latter contribution to $\phi(1020)\to\pi\pi\pi$
can be described as an allowed decay from a small impurity in the wave
function (right).}
\end{figure}

The  Zweig or OZI rule, after Okubo, Zweig and Iizuka
 \cite{Okubo:1963fa,Zweig:1964jf,Iizuka:1966fk}, has played an
important $\rm r\hat{o}le$ in the development of the quark model. The $\phi(1020)$
is e.g. an isoscalar vector meson like the $\omega$. But it is much
narrower, in spite of the more favourable phase-space. It decays
preferential into $ K\bar K$ pairs, and rarely into three pions.
The suppression of the $\phi\to 3\pi$ decay was explained by its almost
pure \ssb\ flavour function and the assumption that the decay proceeds
mostly with the strange quark and antiquark flowing from the initial
state to one of the final-state mesons, as seen in
Fig.~\ref{mes:fig:OZI1}, left, while disconnected diagrams (centre) are
suppressed. The decay $\phi\to \pi\pi\pi$  is attributed to the small
$n\bar n$ component which is mixed into the dominantly $\ssb$ wave
function. The $n\bar n$ component decays into pions by a perfectly
allowed process (right).
 Vector mesons have a mixing angle of $\Theta_V
= 39^{\circ}$ which deviates from the ideal mixing angle $\Theta_{\rm
id}$ by $\delta_V =3.7^{\circ}$. The physical $\phi (1020)$ is then
written
\begin{eqnarray}
\label{mes:eq:OZI }
\ket{\phi}=\cos\delta_V\,\ket{s\bar s} + \sin\delta_V\,\ket{n\bar n}, &
\qquad \ket{\omega}= - \sin\delta_V\,\ket{s\bar s} +
\cos\delta_V\,\ket{n\bar n} \nonumber \\
 \ket{f_2'{1525}}=\cos\delta_T\,\ket{s\bar s} +
\sin\delta_T\,\ket{n\bar n}, & \qquad \ket{f_2(1270)}= -
\sin\delta_T\,\ket{s\bar s} + \cos\delta_T\,\ket{n\bar n}\end{eqnarray}
The mixing angle of tensor mesons is about $\delta_T=-6^{\circ}$.

The OZI rule was decisive when the narrow J/$\psi$ was discovered with
its high mass and narrow width of about 90\,KeV/c$^2$. Its narrow width
entailed -- together with the OZI rule -- that a new quark flavour was
discovered, and that J/$\psi$ is a $c\bar c$ state.

\subsubsection{\label{Regge trajectories}
Regge trajectories}

\begin{figure}[h]
\bc
\includegraphics[width=0.5\textwidth,height=0.4\textwidth]{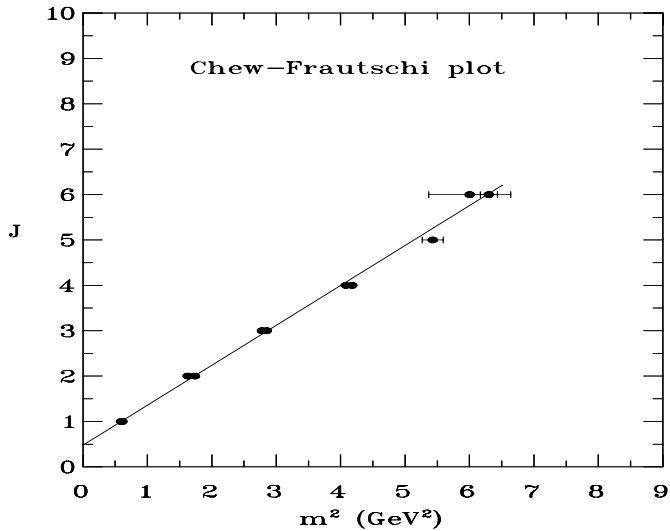}
\ec
\vspace{-4mm}
\caption{Spin of meson resonances versus M$^2$ \cite{Gauron:2000ri}.
The squared masses of $\rho, \omega$, $f_2(1270), a_2(1320)$, $\omega_3,
\rho_3$,  $a_4(2040), f_4(2050)$, $\rho_5(2350)$, and $a_6(2450),
f_6(2510)$ are plotted against their total angular momentum. For all
mesons shown, $j=l+s$. In Regge theory, non-integer spins $\alpha$ are
allowed in exchange processes. Particles have integer spins.}
\label{intro:regge}
\end{figure}

Light mesons have been observed with large angular momenta, up to
$J=6$. Fig.~\ref{intro:regge} shows the meson mass square as a function
of the total angular momentum $J$.
Spin of $q\bar q$ resonances are linearly related to their squared
masses over the full range of M$^2$ and $J$. The width of resonances
increases with their available phase space for decay; the widths are
approximately given by a constant ratio $\Gamma/M\simeq 0.1$.

Regge theory relates the particle spectrum to the forces between the
particles and the high-energy behaviour of the scattering amplitude. If
scattering data on the Regge trajectory are included (for momentum
transfer $q^2=t\leq 0$), Regge trajectories require a nonlinear term. An
important question is related to these observations: do trajectories
continue to arbitrarily high values of $J$ or is there a trajectory
termination point beyond which no resonances
occur?\footnote{\footnotesize Amazingly, our galaxy falls onto the
same Regge trajectory as mesons and baryons as Muradian discovered
\cite{Muradian:1994mk,Muradian:1997np}.} Abundant
literature can be found on this topic. We refer to a few recent papers
\cite{Brisudova:2003dj,deTeramond:2005su,Bigazzi:2004ze}.

An attractive interpretation of linear Regge trajectories was proposed
by Nambu \cite{Nambu:1978bd}. He assumed that the gluon flux between
the two quarks is concentrated in a rotating flux tube or a rotating
string with a homogeneous mass density. The velocity at the ends may be
the velocity of light. Then the total mass of the string is given by
$$
Mc^2 = 2\int_{0}^{r_0}\frac{\sigma dr}{\sqrt{1-v^2/c^2}} = \pi\sigma r_0
$$
and the angular momentum by
$$
l = \frac{2}{\hbar c^2}\int_{0}^{r_0}\frac{\sigma r v
dr}{\sqrt{1-v^2/c^2}} = \frac{\pi\sigma r_{0}^{2}}{2\hbar c} + {\rm
constant} = \frac{M^2}{2\pi\sigma\hbar c} + {\rm constant} $$ where
$\sigma$ is the string tension. We set the constant to $-\frac{1}{2}$
(thus replacing $l$ by $l+\frac{1}{2}$). From the slope in
Fig.~\ref{intro:regge} we find $\sigma =
0.182$\,(GeV)$^2/\hbar c=0.92$\,GeV/fm, and diameters of $$
 r_0(\rho ) = 0.26\, {\rm fm} \qquad\qquad\
 r_0(a_6) = 0.86\, {\rm fm}
$$
\vskip -1mm

Mass and radius $r_0$, and mass and width, are proportional. Hence the
total width of resonances is proportional to $r_0$: the probability of
string breaking per unit length is a constant. The $a_6^{+}(2450)$ is a
meson with a mass of 2450\,MeV/c$^2$ in which a $u$ quark and a $\bar{d}$
quark are bound by the confining forces; both quark spins are parallel,
the orbital angular momentum between $u$ and $\bar d$ is $l=5\hbar$.
Quark and antiquark are separated by 1.7\,fm\,.

This is a mechanical picture of a quantum system which was expanded
by Baker and Steinke \cite{Baker:2002km} and by Baker
\cite{Baker:2003fn} to a field theoretical approach.  The
origin of the string connecting quark and antiquark is an effect,
similar to the Meissner effect in superconductivity, which confines the
colour electric fields to a narrow tube. For quark-antiquark
separations larger than the tube size, a linear potential develops that
confines the quarks in hadrons. Flux tube fluctuations to this
quark-antiquark potential can be determined by path integral methods.
Regge trajectories of light mesons can then be calculated by attaching
massless scalar quarks to the ends of a rotating string. A path
integral is then calculated around the classical rotating straight
string solution. After quantisation of the quark motion at the end of
the string, energy levels of a rotating string at large excitations are
obtained:
\begin{equation} M_n^2(l) =
2\pi\sigma\left(l+n+\frac{1}{2}\right) \,. \label{string}
\end{equation}
Eq. (\ref{string}) contains an important message: for large $l,n$ and
$l\gg n^2$, degeneracy of orbital and radial excitations is expected.
Quark spins are not included in the picture. Spin-spin interactions
vanish with the inverse meson mass forces, see right panel of
Fig.~\ref{fig:forces} and consider $M_1^2-M_2^2 = (M_1-M_2)\cdot
(M_1+M_2)=$\ constant. Spin-orbit interactions are known to be small
even though this was an unexpected observation.

This semiclassical picture can be reproduced within a `5-dimensional
theory holographically dual to strongly coupled QCD (AdS/QCD)'. In
\cite{Karch:2006pv}, the linearity of the mass square spectrum is used
as a constraint for the infrared behaviour of the theory. Two variants
have been proposed how to model confinement: by a `hard
wall'~\cite{Polchinski:2001tt} or by a `soft wall' ~\cite{Karch:2006pv}
which cut off effectively AdS space in the infrared region. The two
variants lead to different relations between mass and orbital and
radial quantum numbers. The hard wall solutions gives  $M_n(l) \sim
l+2n $, the soft wall $M_n^2(l) \sim l+n$.

The duality between the anti-de-Sitter space and conformal field
theories seems to us as one of the most promising directions to clarify
the relation between hadron properties, including the excitation
spectrum, and their structure functions \cite{Brodsky:2006uq}. A very
short introduction can be found in section \ref{Solvable models}.

\subsubsection{\label{Flavour independence of forces}
Flavour independence of forces}

\begin{figure}[ph]
\begin{tabular}{cc}
\epsfig{file=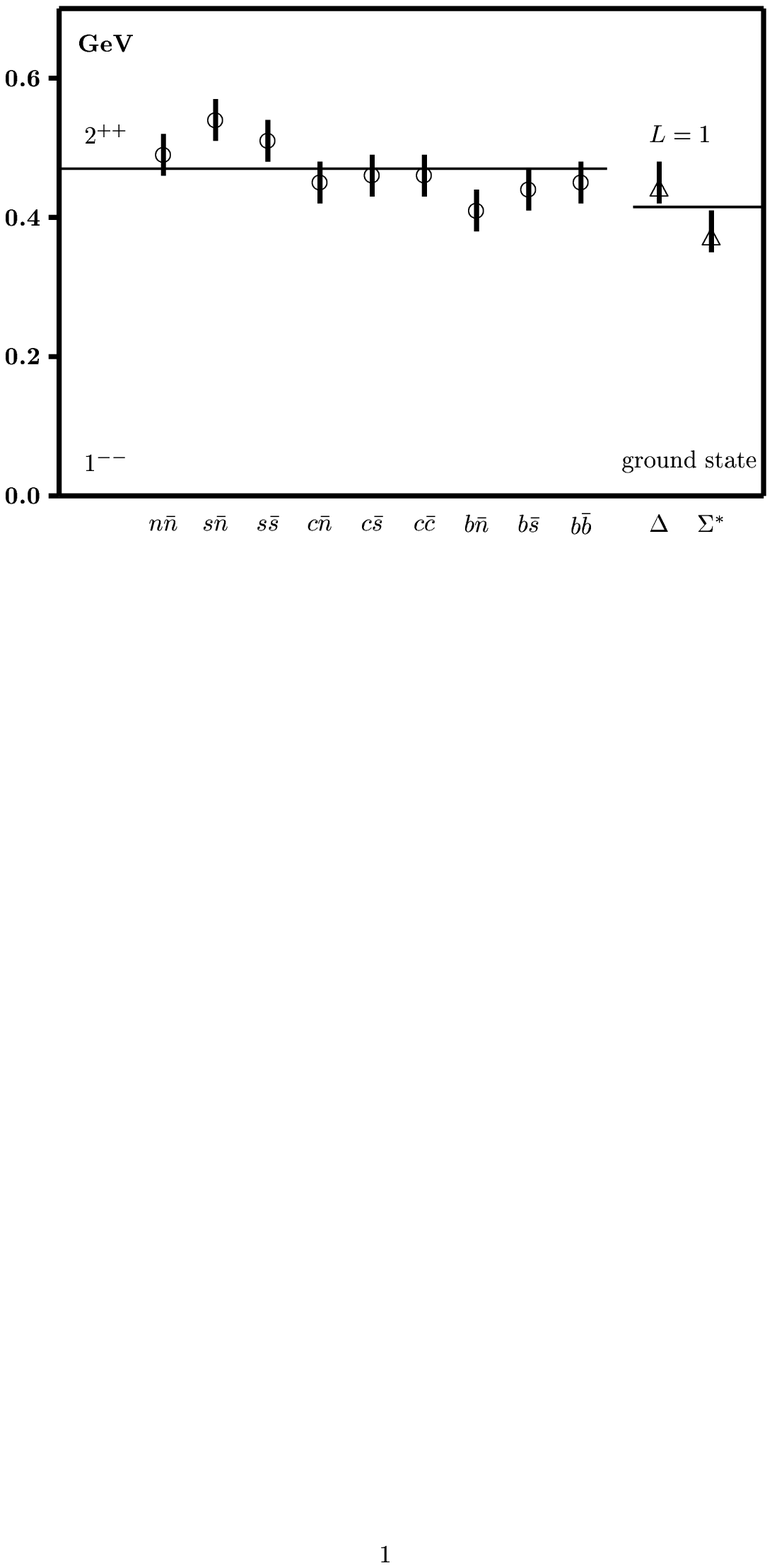,width=0.45\textwidth,clip=}&\hspace*{-2mm}
\epsfig{file=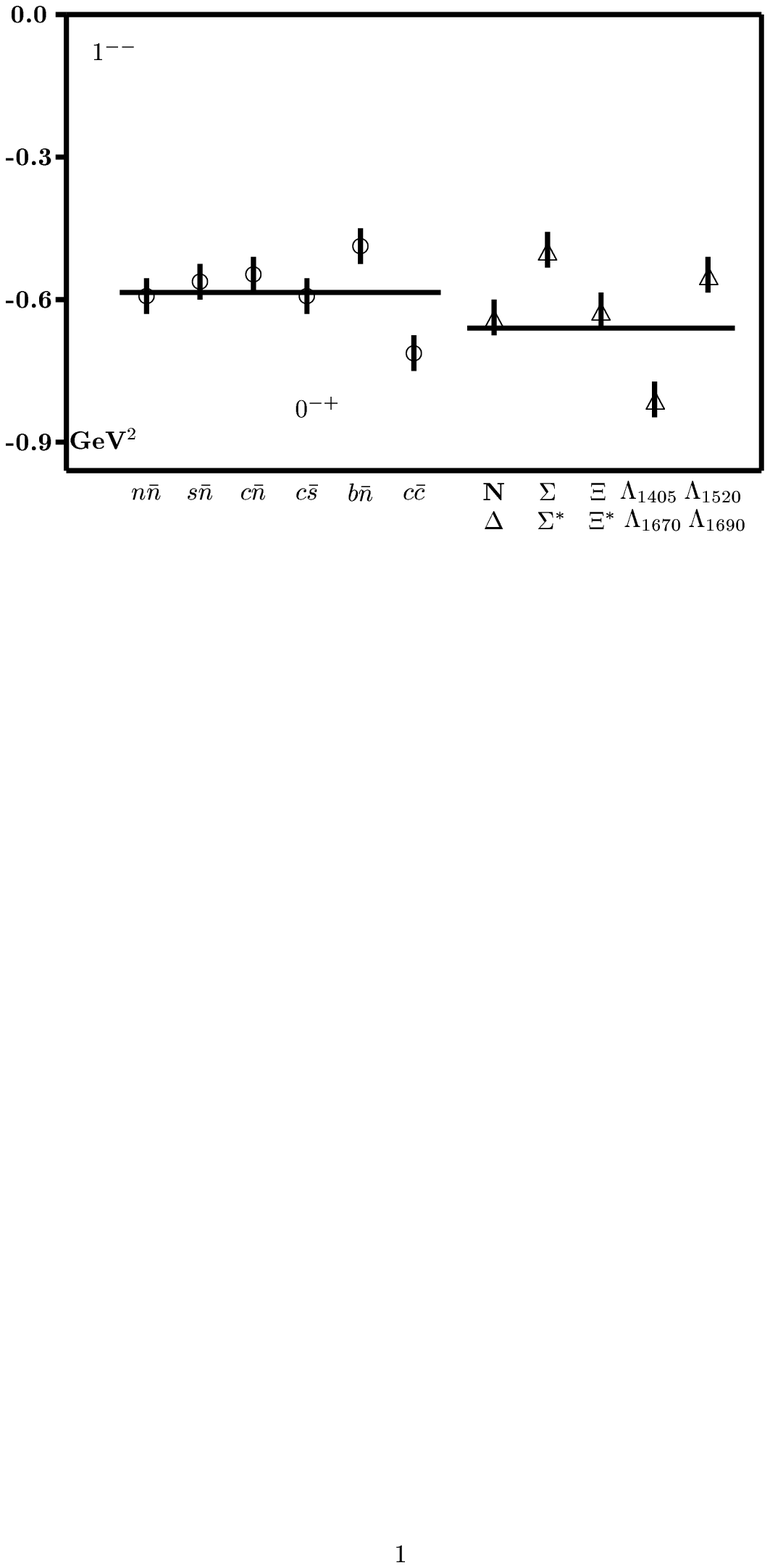,width=0.45\textwidth,clip=}
\end{tabular}
\caption{\label{fig:forces}
Flavour independence of the strong forces: Left panel: Mass differences
for $L=1$ excitations of mesons and baryons (for systems with aligned
spin). Right panel: Mass square differences between pseudoscalar and
vector mesons, for ground states ($L=0$) octet- and decuplet-baryons,
and for $L=1$ excited singlet- and octet-baryons. }
\end{figure}

Meson masses vary over a wide range; e.g. $\omega(782)$ and
$\Upsilon(9460)$ are both ground state vector mesons. The different
masses are due to different quark masses, the interactions are the
same: gluons do not experience any difference when coupling to a {\it
blue} up, down or bottom quark. Quarks having different flavour couple
to gluons with their colour, irrespective of their flavour.
Surprisingly, the flavour independence of quark-antiquark interactions
can be seen in meson mass spectra (Fig.~\ref{fig:forces}). The spectra
show a similarity of mass splittings from very light to very heavy
mesons implying that even the wave functions must be similar.

The left panel of Fig.~\ref{fig:forces} shows the mass gap
for $L=1$ excitations with quark spins aligned for mesons and some
baryons.  An arbitrary (model) error of 30\,MeV/c$^2$ is assigned to
the masses. The mass difference between the lowest-mass tensor meson
and the lowest-mass vector meson are given for different flavours. The
figure starts with the difference of the mean $f_2(1270)$, $a_2(1320)$
mass and the mean $\omega$, $\rho$ mass, and ends with the difference
in mass of $\chi_{b2}(1P)$ and $\Upsilon(1S)$. For baryons, the mass
difference is given for the mean value of the orbital angular momentum
L=1 excitations, $\Delta_{{1/2}^-}(1620)$ and $\Delta_{{3/2}^-}(1700)$,
and the ground state $\Delta_{{3/2}^+}(1232)$, and for the
corresponding $\Sigma$ states $\Sigma_{{3/2}^+}(1385)$,
$\Sigma_{{1/2}^-}(1750)$, and $\Sigma_{{5/2}^-}(1775)$. For the
comparison of baryon excitations, it is important to select states in
which either spin or isospin wave functions are symmetrical with
respect to exchange of two quarks. The first orbital angular momentum
excitation of the nucleon, $N_{{1/2}^-}(1520)$, would be inappropriate
to use for this comparison. It has a larger excitation energy with
respect to the nucleon ground state; both, nucleon mass and
$N_{{1/2}^-}(1520)$ mass are reduced by spin-spin interactions (likely
induced by instantons) which are absent for members of a decuplet.

On the right panel of Fig.~\ref{fig:forces}, the 'magnetic' mass
splitting between states with aligned and not aligned spins is plotted.
In this case, the differences in mass square are plotted. Note that
$M_1^2-M_2^2 = (M_1-M_2)\cdot (M_1+M_2)$. If this is a constant,
$M_1-M_2$ scales with $1/M$. (Thus there is rather an invariance and
not an independence under flavour rotations.) It is extremely
surprising to see that the $\rho$--$\pi$ mass square splitting is very
similar to that for J/$\psi$--$\eta_c$. For the latter we invoke the
magnetic interaction due to one-gluon exchange. For the pion, the full
machinery of chiral symmetry and Goldstone bosons is needed to explain
its low mass. The $\eta_b(1S)$ has so far not been detected
\cite{Mahmood:2002jd}. Using a theoretical prediction from Godfrey and
Rosner \cite{Godfrey:2001eb}, the $\Upsilon(1S)$--$\eta_b(1S)$ mass
square difference is (1.08\,GeV/c$^2$)$^2$, confirming the trend to
larger negative values seen already in the $c\bar c$ family. We note
that the ALEPH collaboration found in their 6-track event sample one
event which was compatible $\gamma\gamma\to\eta_b(1S)$, and with the
expected background rate \cite{Heister:2002if}. In
\cite{Eidelman:2004wy}, mass and decay mode of the $\eta_b(1S)$
candidate event are given.

The $\Delta(1232)$ and nucleon wave functions differ by a flip in spin
(and isospin). The mass splitting is similar to the one observed when
the spin is flipped in mesons (like in $\rho$ and $\pi$). In the
language of instanton-induced forces: there is one $q\bar q$ pair in
the $\pi$ which is affected by instantons, in the $\rho$ there is none.
And their is one $qq$ pair in the nucleon affected by instantons and
none in the $\Delta(1232)$. In both cases, the mass square is reduced
by the same amount. Adding one or two strange quarks does not introduce
large changes in the decuplet-octet mass splittings. More tricky are
the $\Lambda$ states. The mass difference here is between flavour
singlet and octet states instead of octet and decuplet. In the flavour
singlet states, there are 3 quark pairs  which are antisymmetric in
spin and flavour, in octet states there is only one. The fraction of
quark pairs which are antisymmetric in spin and in flavour is however
reduced in odd angular momentum excitations by a factor 2. In total,
we expect and observe
the same splitting as in the case of the $\Delta(1232)$ and nucleon.
The mass square splitting in Fig. \ref{fig:forces} is proportional the
fraction in the wave function which is antiysmmetric in spin and in
flavour \cite{Klempt:2002vp}.

Orbital angular momentum excitations and the spin-spin interactions
lead to very similar mass splittings over a very wide range of meson
masses demanding a unified description. This viewpoint was often
stressed by Isgur \cite{Isgur:1999ic}.

\subsubsection{\label{The light meson spectrum and quark model
assignments}
The light meson spectrum and quark model assignments}

\begin{figure}[pb]
  \begin{center}
\includegraphics[%
bb=15 375 565 725,height=0.40\textheight,clip=on%
 ]{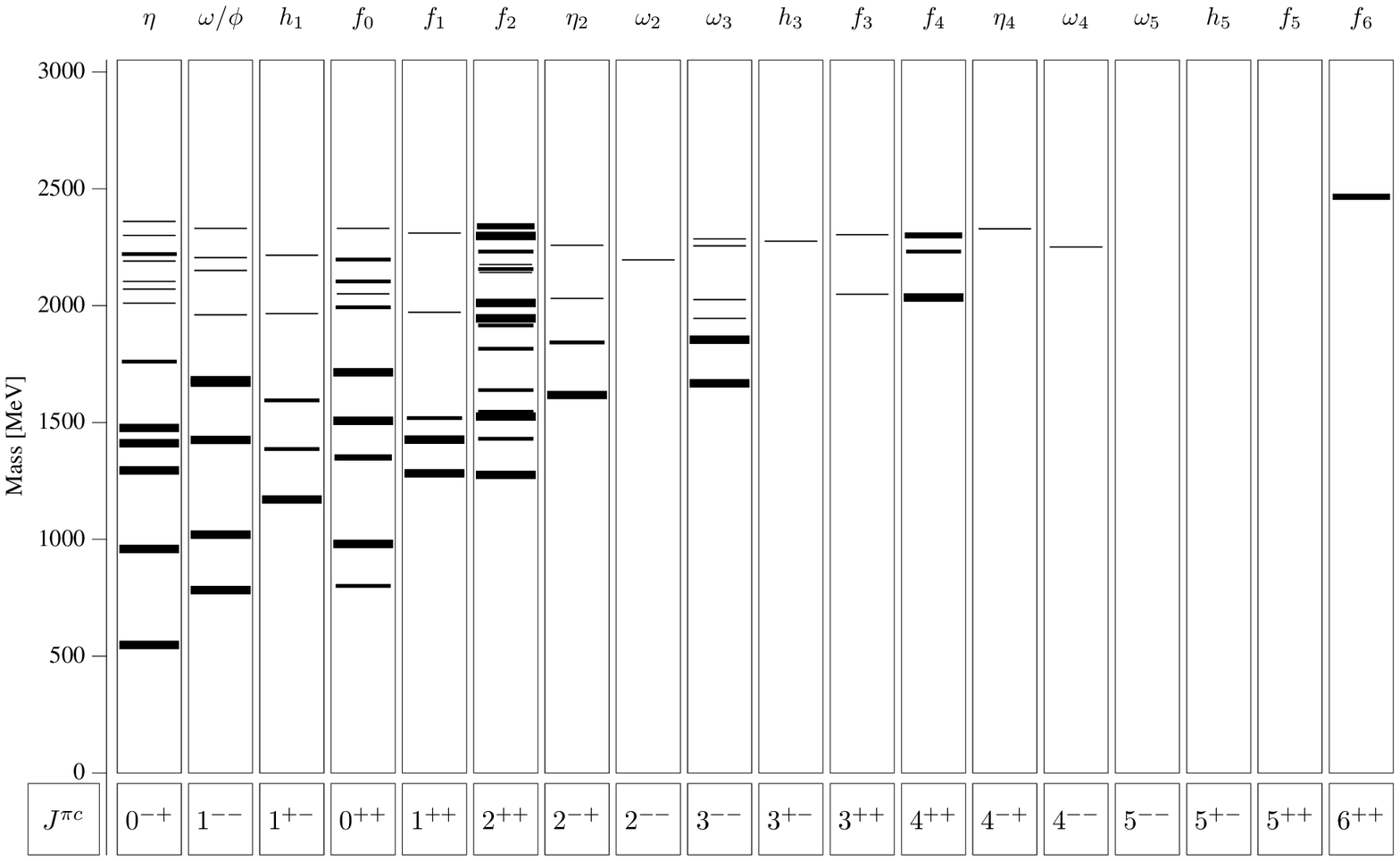}
    \caption{\label{fig:intromodel:isoscalarspectrum}
Experimental light flavoured isoscalar meson spectrum. Data are from
\protect\cite{Eidelman:2004wy}. Mean values of resonance positions are
indicated by thick lines, less established resonances are represented
by medium thick lines, 'further states' by very thin lines.      }
  \end{center}
\vspace{5mm}
\end{figure}

\begin{figure}[pt]
  \begin{center}
    \includegraphics[%
bb=15 375 565 725,height=0.40\textheight,clip=on%
    ]{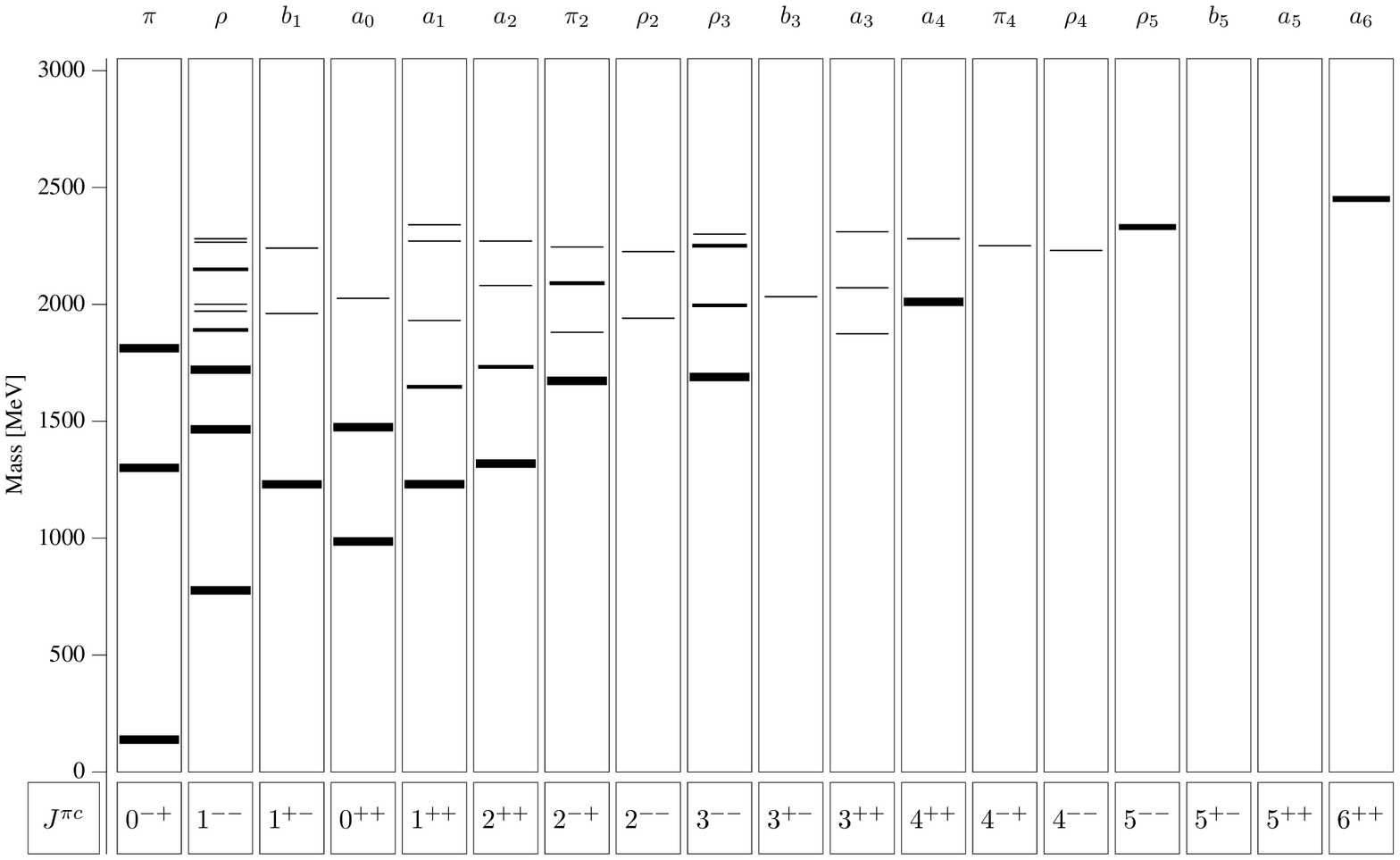}
    \caption{\label{fig:intromodel:isovectorspectrum}
Experimental light flavoured isovector meson spectrum. Data are from
\protect\cite{Eidelman:2004wy}. Mean values of resonance positions are
indicated by thick lines, less established resonances are represented
by medium thick lines, 'further states' by very thin lines.    }
  \end{center}
  \begin{center}
    \includegraphics[%
bb=15 375 565 725,height=0.40\textheight,clip=on%
  ]{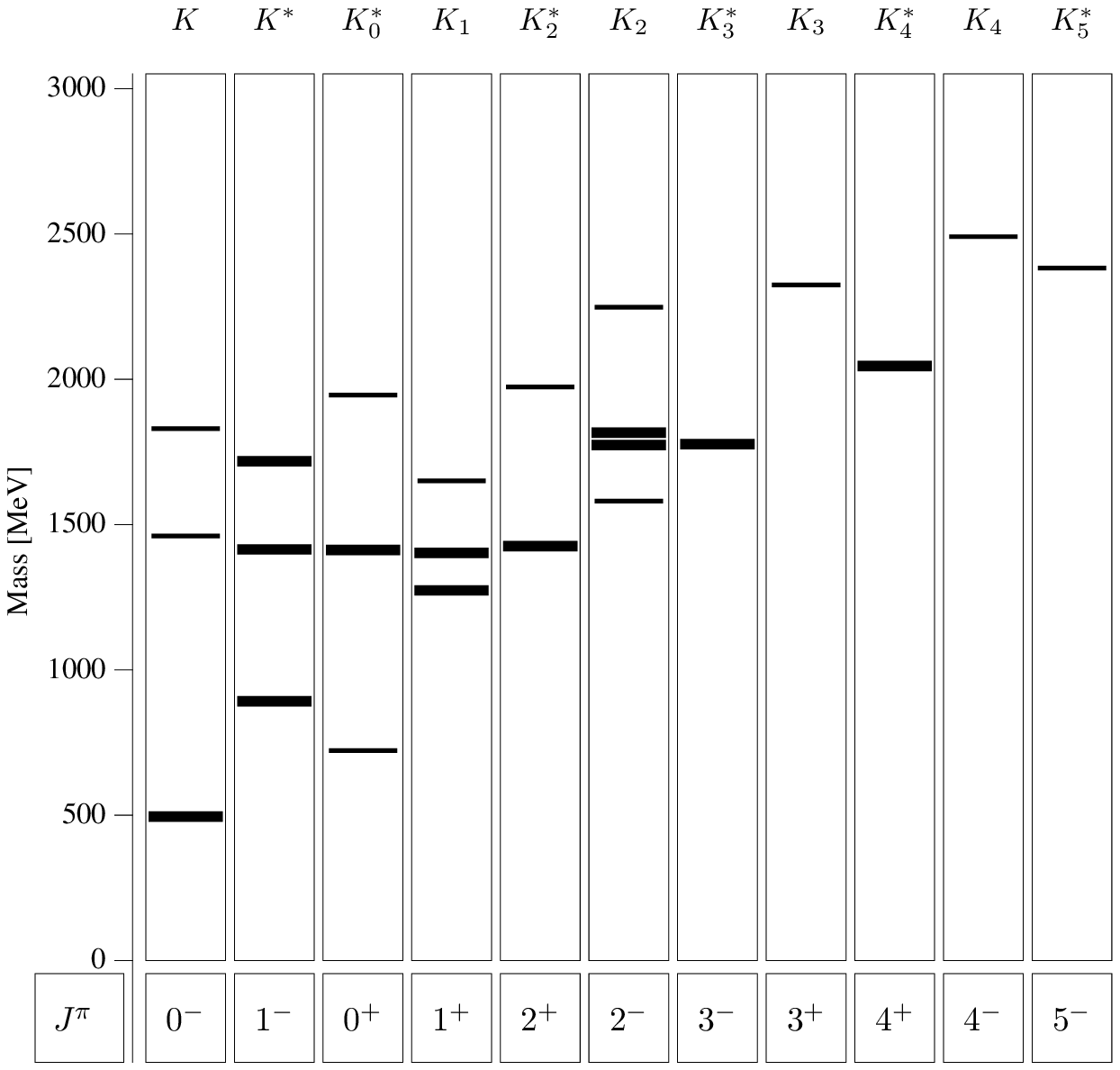}
  \end{center}
  \caption{\label{fig:intromodel:strangespectrum}
Experimental strange-meson spectrum. Data are       from
\protect\cite{Eidelman:2004wy}. Mean values of resonance positions are
indicated by thick lines, less established resonances are represented
by medium thick lines. }
\end{figure}

Apart from ground state mesons and those on the leading Regge
trajectories, a large number of meson resonances is known. In the
following three figures the experimental spectrum of light flavoured
mesons is displayed. Figure~\ref{fig:intromodel:isoscalarspectrum}
shows isoscalar mesons, Fig.~\ref{fig:intromodel:isovectorspectrum}
isovector mesons, and Fig.~\ref{fig:intromodel:strangespectrum} the
experimental Kaon spectrum. 
Included in the figures are all entries in the Review of
Particle Properties \cite{Eidelman:2004wy}, also those which are
less well established and which are {\it omitted from the summary
table}. The Particle Data Group has a third classification, {\it
further states}, a collection of candidates which are not considered to
have a good chance to survive future experiments. In spite of this
warning, we include the results based on analyses of Crystal Barrel
data by the QMC-Rutherford-Gatchina group \cite{Bugg:2004xu}
since these are the only data providing a systematic view of
the high-mass region of meson resonances.

These mesons have known spin and parity, their flavour structure is
however unknown. In some cases, like $\omega$, $\phi$, $f_2(1270)$ and
$f_2^{\prime}(1520)$, the decay pattern reveals their $n\bar n$ and
$s\bar s$ nature. The nature of other mesons is more difficult to
establish. For the isospin quartet $a_0^{\pm,0}(980),f_0(980)$ numerous
interpretations exist. A solid knowledge of the spectrum of $q\bar
q$ mesons is mandatory if other forms of hadronic matter like
glueballs, hybrids, tetraquark states, $ N\bar N$ bound
states and resonances, or meson-meson molecules bound by nuclear
forces are to be discriminated against the `background' of $q\bar q$
mesons.

Of course, these different forms of hadronic matter are not realised in
nature as separate identities. Mesons should rather be
considered as complex objects with a Fock space expansion to which
different configurations make their contribution, even though at an
unknown level. So the question arises if leading configurations can be
identified by comparing simplified models with experimental data.
From the patterns of the well established resonances in these figures
we infer:
\begin{itemize}
\item  With the exception of pseudoscalar and scalar mesons,
  for every isovector resonance there exist an isoscalar partner almost
  degenerate in mass within 50\,MeV/c$^2$. Prominent pairs are:
\bc
\vspace*{2mm}
\begin{tabular}{cll}
$\rho(770)$---$\omega(782)$&$a_1(1260)$---$f_1(1285)$&$a_2(1320)$---$f_2(1270)$\\
$\rho_3(1690)$---$\omega_3(1670)$&$a_4(2040)$---$f_4(2050)$&
$a_6(2450)$---$f_6(2510)$ \,. \\
\end{tabular} \\ \vspace*{2mm} \ec
  From this one can conclude that in these sectors  $n\bar n
\leftrightarrow s\bar s$  flavour mixing does not play a significant
$\rm r\hat{o}le$.

\item Again with the exception of (pseudo)scalars, spin-orbit splittings
seem to be small, note \textit{e.g.} the approximate degeneracy of
\bc
\vspace*{2mm}
\begin{tabular}{cll}
$f_1(1285)$---$f_2(1270)$&$a_1(1260)$---$a_2(1320)$&$K_1(1400)$---$K_2^*(1430)$\\
$K_2(1770)$---$K_3^*(1780)$&$K_4(2500)$---$K_5^*(2380)$\,.&\\
\end{tabular}
\vspace*{2mm}
\ec
\item  Also, again with the exception of (pseudo)scalars, spin-spin
splittings are not very large, see \textit{e.g.} the near degeneracy of
\bc
\vspace*{2mm}
\begin{tabular}{cc}
$b_1(1235)$---$a_1(1260)$---$a_2(1320)$&$h_1(1170)$---$f_1(1285)$---$f_2(1270)$\\
$\pi_2(1670)$---$\rho_3(1690)$&$\eta_2(1645)$---$\omega_3(1670)$\,.\\
\end{tabular}
\vspace*{2mm}
\ec

\item for the (pseudo)scalar mesons distinguished spin-spin splitting
and flavour mixing is observed, the most prominent example being the
$\pi-\eta-\eta'$-pattern.
\end{itemize}

These regularities help to assign spectroscopic quantum numbers to
the states. Of course, these quantum numbers are an interpretation. The
$\rho(1700)$ e.g. could be the second $\rho$ radial excitation with
spectroscopic quantum numbers $3^3S_1$  or the ground state with
intrinsic orbital angular momentum $L=2$, $1^3D_1$. The external
quantum numbers are $J^{PC}=1^{--}$ in both cases. In general,
the physical state can be mixed. An assignment of the dominant part in
an expansion of the physical state into states with spectroscopically
defined quantum numbers can be made by comparison with models, for its
mass and for the decay modes.

Table~\ref{tab:nonets} displays the spectrum of light $q\bar q$ mesons
having small intrinsic orbital angular momenta ($L=0,1,2$). It is
surprising how many states are still unobserved; this deficiency
certainly increases with increasing mass. Even more intriguing are
those states, indicated by question marks in Table~\ref{tab:nonets},
which are expected to have low masses, in the region between 1000 and
1600\,MeV/c$^2$, for which candidates exist but for which our knowledge of
the meson spectrum does not suffice for an unambiguous interpretation.

\begin{table}[ph]
\caption{\label{tab:nonets}
The light meson spectrum. Mesons masses marked $xxxx$ are so
far unobserved, those marked with $????$ will be discussed in detail in
this report. }
\vskip 2mm
\begin{center}
\renewcommand{\arraystretch}{1.4}
\begin{tabular}{cccccccccc}
\hline
\hline
\ $L$\quad &\ $S$\quad &\ $J$\quad &\ $n$\qquad & $I=1$\qquad & $I=1/2$
\qquad & $I=0$\qquad & $I=0$\qquad & $J^{PC}$\qquad & $n^{2s+1}L_J$\qquad  \\
\hline
0 & 0 & 0 & 1 & $\pi$  &  $K$ & $\eta$& $\eta^{\prime}$   & $0^{-+}$ &
$1\ssz$\\ 0 & 1 & 1 & 1 & $\rho$ & $ K^*$ &$\phi$ & $\omega$ & $1^{--}$
& $1\tso$ \\ \hline 0 & 0 & 0 & 2 &$\pi$(1370) &  $ $K$(1460)$
&$\eta(????)$  & $\eta(????)$
 & $0^{-+}$ & $2\ssz$\\
0 & 1 & 1 & 2 &$\rho$(1450)& $K^*(1410)$ &$\phi (1680)$
&$\omega$(1420)
 & $1^{--}$ & $2\tso$\\
\hline
1 & 0 & 1 & 1 &b$_1(1235)$&$ K_{1B}$ &h$_1(1380)$ &h$_1(1170)$ \quad
 & $1^{+-}$ & $1\spo$\\
1 & 1 & 0 & 1 &a$_0(????)$&$ K_{0}^*(1430)$ &f$_0(????)$ &f$_0(????)$
 & $0^{++}$ & $1\tpz$\\
1 & 1 & 1 & 1 &a$_1(1260)$&$ K_{1A}$ &f$_1(1510)$ &f$_1(1285)$
 & $1^{++}$ & $1\tpo$\\
1 & 1 & 2 & 1 &a$_2(1320)$&$ K_{2}^*(1430)$ &f$_2(1525)$ &f$_2(1270)$
 & $2^{++}$ & $1\tpt$\\
\hline
2 & 0 & 2 & 1 &$\pi_2(1670)$&$ K_{2}(1770)$ &$\eta_2(1870)$
&$\eta_2(1645)$
 & $2^{-+}$ & $ 1^1D_2$\\
2 & 1 & 1 & 1 &$\rho(1700)$&$ K^*(1680)$ &$\phi(xxxx)$ &$\omega(1650)$
 & $1^{--}$ & $ 1^3D_1$\\
2 & 1 & 2 & 1 &$\rho_2(xxxx)$&$ K_2(1820)$ &$\phi_2(xxxx)$
&$\omega_2(xxxx)$
 & $2^{--}$ & $ 1^3D_2$\\
2 & 1 & 3 & 1 &$\rho_3(1690)$&$ K^*_3(1780)$
&$\phi_3(1850)$&$\omega_3(1670)$ & $3^{--}$ & $ 1^3D_3$\\ \hline
\hline
\end{tabular}
\renewcommand{\arraystretch}{1.0}
\end{center}
\end{table}

From deep inelastic scattering we know that mesons do not have a simple
$q\bar q$ structure; the coloured quarks distort the QCD vacuum, the
quark and gluon condensates become polarised and the mesonic wave
function contains further virtual $q\bar q$ pairs and violent gluonic
field configurations. From lattice QCD and from QCD-inspired models we
expect states beyond the ordinary spectrum of $q\bar q$ mesons,
glueballs, hybrids, and multiquark states. These additional states may
have unusual production and/or decay properties which may support an
identification with non-$q\bar q$ states. Their identification as
glueballs, hybrids, or multiquark states requires comparison between
model calculations and experiment, and a subtle discussion of the
experimental conditions under which they have been observed. States
which have served as meaningful candidates for glueballs, hybrids and
multiquark states are in the focus of this review.

\markboth{\sl Meson spectroscopy} {\sl From QCD to strong interactions}
\clearpage\setcounter{equation}{0}\section{\label{From QCD to strong interactions}
From QCD to strong interactions}

\subsection{\label{Quantum Chromo Dynamics}
Quantum Chromo Dynamics}

\subsubsection{\label{The QCD Lagrangian}
The QCD Lagrangian}

When quarks were first introduced by Gell-Mann and Zweig in 1964
\cite{Gell-Mann:1964nj,Zweig:1964jf}, the quark model was often
interpreted as a merely useful scheme for classifying hadrons, and the
lack of any experimental evidence for free quarks was not considered to
be a real problem. The absence of isolated quarks became an urgent
issue only with the successes of the quark-parton model and of quantum
chromo dynamics as a fundamental theory of strong interactions
\cite{Fritzsch:1972jv}. QCD provides a frame within which experimental
findings in hadron physics are discussed even though the physical
consequences of the theory are often not worked out. The QCD Lagrangian
\cite{Politzer:1973fx,Gross:1973id} has a structure similar to its
analogue in QED:

\vspace{-5mm}
\begin{equation}
\label{lagrangian}
{\mathcal L}_{QCD} =\bar{q}_i (i\partial_\mu\gamma^\mu \delta_{ij}
 +g {\lambda_{ij}^a\over 2}A_\mu^a\gamma^\mu -m\delta_{ij})
q_j -{1 \over 4} F_{\mu\nu}^a F^{a \mu\nu}
\end{equation}\vspace{-7mm}

with

\vspace{-5mm}\begin{equation}
F_{\mu\nu}^a =\partial_\mu A_\nu ^a -\partial_\nu A_\mu ^a +gf_{abc}
A^b _\mu A^c _\nu\,.
\end{equation}\vspace{-7mm}

where $q_i$ are the quark fields with colour indices $i=1,2,3,\cdots$,
${\lambda^a \over 2}$ are the generators of colour $SU(3)$, $A^\mu_a$
are the gluon fields which transform according to the adjoint
representation of $SU(3)$ with $a=1,...,8$,  $g$ is the bare coupling,
$m$ is the quark mass. As in QED, the Lagrangian can be derived from
the equation for free Dirac particles by requiring invariance under
local gauge transformations, $U(1)$ in case of QED or $SU(3)$ for QCD.

Eq.~(\ref{lagrangian}) shows that quarks couple to gluons
like electrons to photons. The electromagnetic current $e\gamma^\mu$ of
QED is replaced by $g\gamma^\mu {\lambda\over 2}$ of QCD. t'Hooft
showed in the early 1970's \cite{'tHooft:1971fh} that a gauge theory
with a local $SU(3)$ colour symmetry is renormalisable; divergences can
be absorbed by a redefinition of the parameters of the Lagrangian in
all orders of perturbation theory.

\begin{figure}[ph]
\bc
\epsfig{file={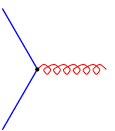},width=0.15\textwidth,origin=c}
\epsfig{file={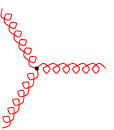},width=0.15\textwidth,origin=c}
\epsfig{file={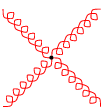},width=0.15\textwidth,origin=c}
\ec
\caption{\label{fig:1}Interactions in QCD at the tree level.}
\end{figure}

In comparison to QED, QCD has important additional components arising
from the special structure of the (non-Abelian) $SU(3)$ group. The
electric charge is a single-valued number while quarks come in coloured
triplets. Photons carry no electric charge, gluons in QCD are colour
octet states. The nonlinear terms in the field strength $F^{\mu\nu}$
give rise to trilinear and quadratic vertices in the theory so that
gluons couple to themselves in addition to interacting with quarks as
depicted in Fig. \ref{fig:1}. This self-interaction leads to important
consequences:

\begin{enumerate}

\vspace{-1mm}\item The self-interaction between gluons makes the theory
nonlinear and very difficult to solve. Nevertheless in some special
cases approximate solutions can be found.\\ At low energies, an
expansion in terms of the (small) pion mass and small pion momenta
provides a frame which connects different hadronic reactions. This
expansion is known as Chiral Perturbation Theory ($\rm\chi PT$). The
dynamics of heavy quarks can be accessed in an expansion in powers of
the inverse heavy quark mass known as Heavy Quark Effective Theory
(HQET). In other cases, a possible solution of QCD can be obtained by
numerical calculations on a lattice. With increasing computer power and
the development of appropriate functions for chiral extrapolations,
lattice QCD has grown to a predictive theory. However, lattice QCD
often predicts numbers but does not provide intuitive understanding;
the gap from QCD to intuition is bridged by so-called `QCD-inspired
models'.

\vspace{1mm}\item Likely, the gluon self-interaction leads to
confinement of colour. Originally, the confinement hypothesis was
invented to `explain' the absence in Nature of free quarks, i.e. of
particles with fractional electric charge. However, no fundamental
principle excludes the existence of a scalar field carrying colour. A
bound state of a quark and the scalar field would still have fractional
charge but could be colour-neutral. Hence a modern version of the
confinement hypothesis denies the existence of observable objects with
non-vanishing colour charge: all observable particle states are colour
singlets.

\vspace{1mm}\item The self-interacting leads to the suggestion of the
existence of new forms of hadronic matter with excited gluonic degrees
of freedom, known as glueballs and hybrids.

\end{enumerate}

\subsubsection{\label{Chiral symmetry}
Chiral symmetry}

In the limit of vanishing quark masses, the QCD Lagrangian is invariant
under chiral transformations

\vspace{-5mm}\begin{eqnarray} q & \to & \exp (i \gamma_5 \alpha^a
t^a)\, q , \nonumber \\ \overline{q} & \to & \overline{q}\,\exp (i
\gamma_5 \alpha^a t^a) ,
\end{eqnarray}\vspace{-7mm}

where $t^a$ are the generators of the flavour symmetry group - $SU(2)$
for isotopic symmetry or $SU(3)$ for symmetry of $u, d, s$ quarks. In
this limit, positive- and negative-parity states are mass-degenerate.
This degeneracy is not seen in the spectrum of low lying hadrons. For
example, the negative-parity companion of the nucleon, $\rm
N(1535)S_{11}$, is separated from the nucleon by about
600\,MeV/c$^2$.This mass shift is substantial and indicates that chiral
symmetry is broken. The spontaneous symmetry breaking is signalled by a
$\overline{q}q$ vacuum expectation value, the chiral condensate
\cite{Gell-Mann:1968rz}. The value of the quark condensate density is
known with high precision \cite{Colangelo:2000dp} from the
Gell-Mann-Oakes-Renner relation:

\vspace{-5mm}\begin{eqnarray} \langle\bar qq\rangle
= -\frac{f^2_{\pi} m^2_{\pi}}{2(m_u+m_d)}=-(240\pm
10\,{\mathrm{MeV}})^3
\end{eqnarray}\vspace{-7mm}

for which higher-order corrections are small.

Spontaneously broken chiral symmetry is associated with the existence
of eight Goldstone bosons, of eight massless pseudoscalar mesons.
Pseudoscalar mesons observed in Nature,
$\pi^{\pm},\pi^0, \eta , K^{\pm}, K^0, \bar K^0$, are not massless; for
small quark masses, chiral symmetry is broken explicitly leading to
finite (but still small) masses. The $\eta^{\prime}$ mass, however, is
not small; the $\gamma_5$-invariance is broken on the quantum
mechanical level due to an anomaly which implies that the
divergence of the flavour singlet axial vector current $A_\mu =
\overline{q}\gamma_\mu\gamma_5 q$ does not vanish:

\vspace{-5mm}\be \partial_\mu
A_\mu(x) = N_f q(x), \label{eq:ano}\qquad\quad{\rm where} \ee \be q(x)
= -\frac{1}{32\pi^2} \epsilon_{\mu\nu\alpha\beta}G^a_{\mu\nu}(x)
G^{a}_{\alpha\beta}(x)
\ee\vspace{-7mm}

is called topological charge density.

\subsubsection{\label{Instantons}
Instantons}

The QCD vacuum has a nontrivial topological structure. There are
trivial vacua having vanishing classical fields ($A^a_{\mu}=0,
G^a_{\mu\nu}=0$) with small perturbative oscillations near this zero.
In addition,  there are infinitely many other solutions with $
G^a_{\mu\nu}=0$ which can not be reduced to $A^a_{\mu}=0$ by a continuous
gauge transformation \cite{Polyakov:1975rs}. These different solutions
are labelled by an integer number, called winding number. The tunnelling
transitions between different vacua leads to nonperturbative field
configurations \cite{Belavin:1975fg,'tHooft:1976fv} called instantons
and anti-instantons. The tunnelling decreases the vacuum energy making
it negative.

The theory of the instanton vacuum is far from being completed.
Its inferences of any of its consequences for the real world requires
the use of models and of experimental data. Nevertheless it can be used
as powerful approximation to evaluate nonperturbative effects. We refer
the reader to reviews by Sch\"afer and Shuryak \cite{Schafer:1996wv},
Forkel \cite{Forkel:2000sq} and Diakonov \cite{Diakonov:2002fq}. As
a qualitative guide we can consider instantons as nonperturbative
fluctuations with a very strong field in a relatively small volume
($r\approx1/3$ fm) with average space-time density $n\approx 1$
fm$^{-4}$ which provide a significant contribution to the gluon
condensate.

Right-handed and left-handed quarks can propagate  in this
highly non-trivial vacuum in zero modes of definite chirality (left
handed for instantons, right handed for anti-instantons). As there is
exactly one zero mode per flavour, only the propagation of quark pairs
of opposite chirality are influenced by instantons. This has important
consequences for meson spectroscopy: direct instanton effects should be
expected for pseudoscalar mesons {\it and} for scalar mesons.

Instantons influence the dynamics of quarks in two ways: i) The quark
and antiquark propagators get dressed and obtain a dynamical mass
\cite{Diakonov:2002fq}. ii) Quark and antiquark may scatter
off instantons leading to an effective quark
interaction. Such processes are described by non-local $2N_{f} $-quark
interactions known as 't Hooft vertices \cite{'tHooft:1976fv}. They
generate the dominant instanton effects in the light-quark sector. The
't Hooft vertices manifestly break the axial $U_A(1)  $ symmetry of the
QCD Lagrangian with $m_{q}=0$ and thereby resolve (at least
qualitatively) the $U_A(1)$ problem, i.e. they explain why the
$\eta^{\prime}$ meson has almost twice the mass of the $\eta$ meson and
why it cannot be considered as a (quasi) Goldstone boson.

\subsection{\label{Chiral Perturbation Theory chi PT}
Chiral Perturbation Theory $\chi$PT}

Chiral Perturbation Theory
\cite{Gasser:1983yg,Gasser:1984ux,Meissner:1993ah,Weinberg:1996kr} is
based on the construction of an effective Lagrangian which respects the
QCD symmetries, and contains $\pi, K$ and $\eta$ mesons as elementary
ingredients of the theory. These particles are the Goldstone bosons of
spontaneous chiral symmetry breaking of massless QCD and are the
low-energy degrees of freedom of the theory.  In the chiral limit, all
amplitudes vanish at zero momentum. This fact allows one to construct
$\chi$PT as an expansion in even powers of momenta, generically denoted as
$O(p^2), O(p^4)$. At each order in $p^2$, $\chi$PT is the sum of all
terms compatible with the symmetries, each multiplied by a free
parameter. Once a set of parameters up to a given order is determined,
it describes, to that order, any other process involving mesons. In the
leading order, the only parameters are masses and decay constants; 10
further chiral parameters are needed to describe meson-meson
interactions in second order ($p^4$). The natural range within which
$\chi$PT can be applied is governed by the parameter
$\Lambda_{\chi}\sim4\pi f_{\pi}$ at about 1\,GeV/c$^2$. The low-mass scalar
interactions can be derived from the leading order Lagrangian and
unitarity, from which the existence of a $\sigma(485)$ and
$\kappa(700)$ was deduced (see section \ref{Scalar mesons}). In
\cite{Oller:1997ng,Oller:1998hw,GomezNicola:2001as}, second order
corrections and coupled channels effects were calculated and
meson-meson cross section data up to 1.2\,GeV/c$^2$ described in a
consistent way. Resonances were constructed by extrapolation of the
$\chi$PT amplitude via the Inverse Amplitude Method
\cite{Truong:1991gv} which avoids conflicts with unitarity.

For sufficiently light quarks, chiral perturbation theory can be used
to estimate quark mass ratios \cite{Weinberg:1977hb}. To first order,
the squared masses of pseudoscalars are proportional to the quark
masses. Electromagnetic interactions add to squared masses as well:

\vspace{-5mm}\begin{eqnarray}
m_{\pi^0}^2 &=& B(m_u + m_d) \nonumber\\
m_{\pi^+}^2 &=& B(m_u + m_d) + \Delta_{em}\nonumber\\
m_{K^0}^2 &=& B(m_d + m_s) \nonumber\\
m_{K^+}^2 &=& B(m_u + m_s) + \Delta_{em}
\end{eqnarray}\vspace{-7mm}

Here, $B$ and $\Delta_{em}$ are unknown constants. From these equations
quark mass ratios can be determined with high precision:

\vspace{-5mm}\be
m_u / m_d = (2 m_{\pi^0}^2 -m_{\pi^+}^2 + m_{K^+}^2 - m_{K^0}^2)/(m_{K^0}^2 - m_{K^+}^2 + m_{\pi^+}^2 ) = 0.56
\ee\be
m_s / m_d = (m_{K^0}^2 - m_{\pi^+}^2 + m_{K^+}^2)/(m_{K^0}^2 - m_{K^+}^2 + m_{\pi^+}^2 ) = 20.1
\ee\vspace{-7mm}

Higher order $\chi$PT corrections change these ratios by $\approx
10\%$. To estimate absolute values of quark masses we have to use
sources from outside of chiral perturbation theory. As a first estimate
we can assume that, as a result of chiral symmetry breaking, the same
mass is added to all bare masses. In this approximation

\vspace{-5mm}\be
m_s = (m_{\phi} - m_{\rho})/2 \approx 120\, {\rm MeV/c^2}.
\ee\vspace{-7mm}

More refined approaches based on sum rules  \cite{Narison:2005wc}
give an $s$-quark mass of about $100$ MeV/c$^2$, with about $10\%$
precision. Thus, $u$ and $d$ quarks are very light on a hadronic scale
of 1\,GeV/c$^2$: $m_u \approx 4$ MeV/c$^2$, $m_d \approx 7$ MeV/c$^2$. This
is a justification of the low mass approximation of $\chi$PT. Masses of
constituent quarks cannot be determined without addressing the model in
which they are used.

\subsection{\label{Lattice QCD}
Lattice QCD}

Lattice QCD aims at simulating full QCD on a discrete Euclidean
space-time lattice and seems to be the most promising technique to
calculate the properties of hadrons from the QCD Lagrangian. The four
dimensions of space and time are discretised with a regular lattice of
points separated by a spacing $a$ in a volume $L^3T$. Anisotropic
lattices use more lattice points in the time direction than in the
space directions. The time is chosen to be Euclidean, i.e. the metric
tensor is definite. The reciprocal spacing $1/a$ should be much smaller
than the masses or momenta involved in the calculation. It acts as a
ultra-violet cut-off parameter while maintaining gauge invariance of
the theory. In the formulation defined by Wilson \cite{Wilson:1974sk},
quarks are placed on the sites of the lattice, while the gluonic fields
are represented by links connecting neighbouring sites. Every field
configuration receives a Boltzmann weight $\exp\{ -S\}$ where $S$ is
the Euclidian action. Expectation values of physical quantities are
estimated from ensemble averages. Lattice QCD provides a
non-perturbative regularisation of QCD and a means by which observables
can be predicted. The advantage of this approach is that it is based on
first principles of QCD and that its approximations are controllable.
The mass of a state is e.g. extracted from the decay of a two-point
correlation function

\vspace{-5mm}\begin{equation}
    C(t) = \langle \Phi(t) \Phi(0) \rangle \propto e^{-M t}
\end{equation}\vspace{-7mm}

where $\Phi(t)$ is a lattice operator that creates or annihilates the
state of mass $M$ at time $t$. After some thermalisation, the
correlation function falls off exponentially, the fall-off is given by
the particle mass. Excited states fall off much more rapidly
and are thus difficult to extract on the lattice. A further problem
originates from the Wilson-Dirac operator which does not respect
chiral symmetry: light quarks of lattice QCD are too massive, $\sim
50$\,MeV/c$^2$. A remedy is the so-called chiral extrapolation: at low
momenta, the interactions e.g.\ in $\pi-\pi$ scattering and $\pi-N$
scattering are weak and can be treated perturbatively. The results
depend on the assumed quark mass which can be made large. Hence the
dependence of an observable as a function of the pion mass can be
determined and a chiral expansion to the true quark masses can be
performed; this limit often reproduces accurately physical quantities.

Still,  the inclusion of dynamical quarks is a major problem. In many
calculations, quark loops are neglected, the calculations are done in
{\it quenched approximation}. Unquenching is very computer time
consuming, additionally it suffers very often from too large quark
masses.

\subsubsection{\label{Confinement}
Confinement}

The non-observation of free quarks has led to the general conviction
that quarks (or colour sources, to be more precise) are confined. It is
assumed but not proven that quantum chromodynamics confines quarks
due to some special class of gauge field configurations like
Abelian monopoles or centre vortices which are supposed to
dominate the QCD vacuum at large distance scales. We refer the reader
to a few reviews for more detailed discussions of the ideas
\cite{Simonov:1987rn,Kronfeld:1987vd} and the present status
\cite{Alkofer:2000wg,Greensite:2003bk}.

\newcommand{\bb}{\hspace{1.25ex}}
\begin{figure}[ph]
\begin{tabular}{cc}
\hspace{-5mm}\psfig{file=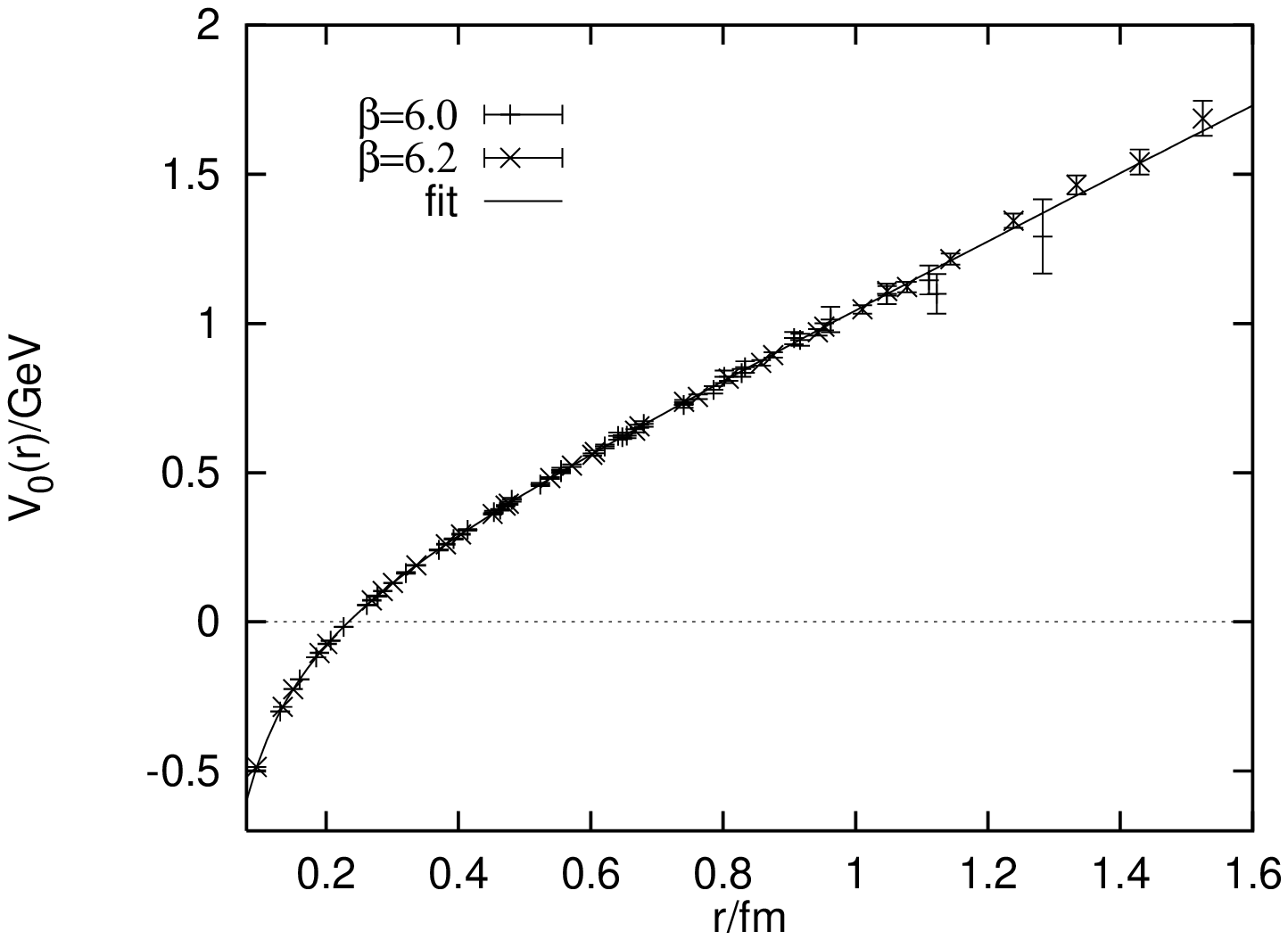,width=0.49\textwidth,height=0.4\textwidth,clip=}&
\hspace{-5mm}\psfig{file=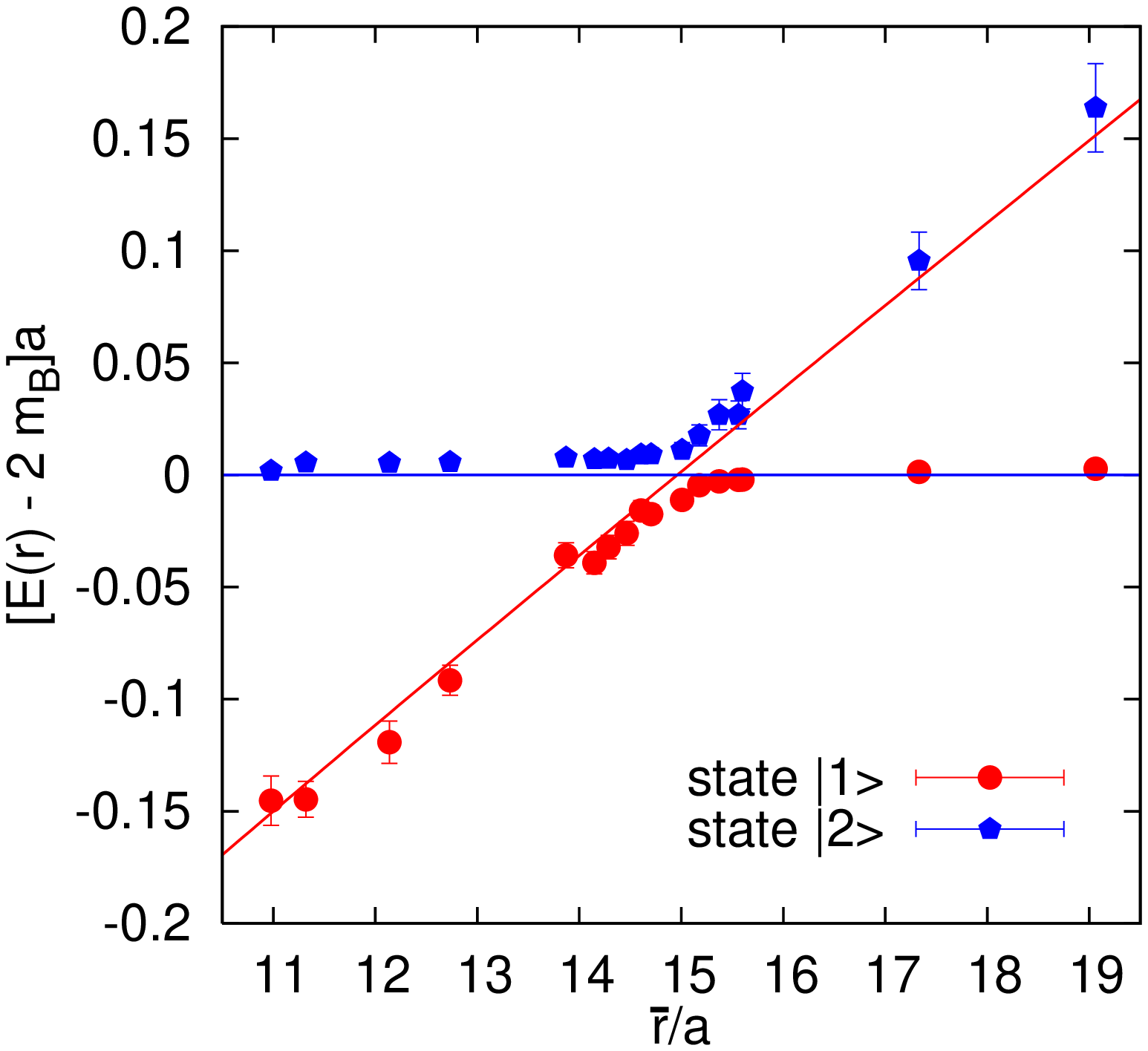,width=0.49\textwidth,height=0.39\textwidth,clip=}
\end{tabular}
\vspace*{-2mm}
\caption{\label{fig:potential}
Left: The potential energy between two heavy quarks (with fixed
positions) as a function of their separation from lattice QCD.
The solid line represents a potential in the form
$V(r) = \frac{4}{3}\frac{\alpha_s}{r} + b\cdot r$;
from ~\protect\cite{Bali:2000gf}.
Right: In full QCD, a sea quark-antiquark can be created from
the vacuum and at large distances, two-meson states are energetically
preferred.  For static quarks, the levels cross at some distance $R$
(with $a\approx 0.083$\,fm), the string breaking introduces mixing of
the energy levels defined by the potential $V(R)$ and the threshold
$2m(B)$
\cite{Pennanen:2000yk,Bernard:2001tz,Michael:2005kw,Michael:2006hf}.}
\begin{minipage}[c]{0.49\textwidth}
\epsfig{file=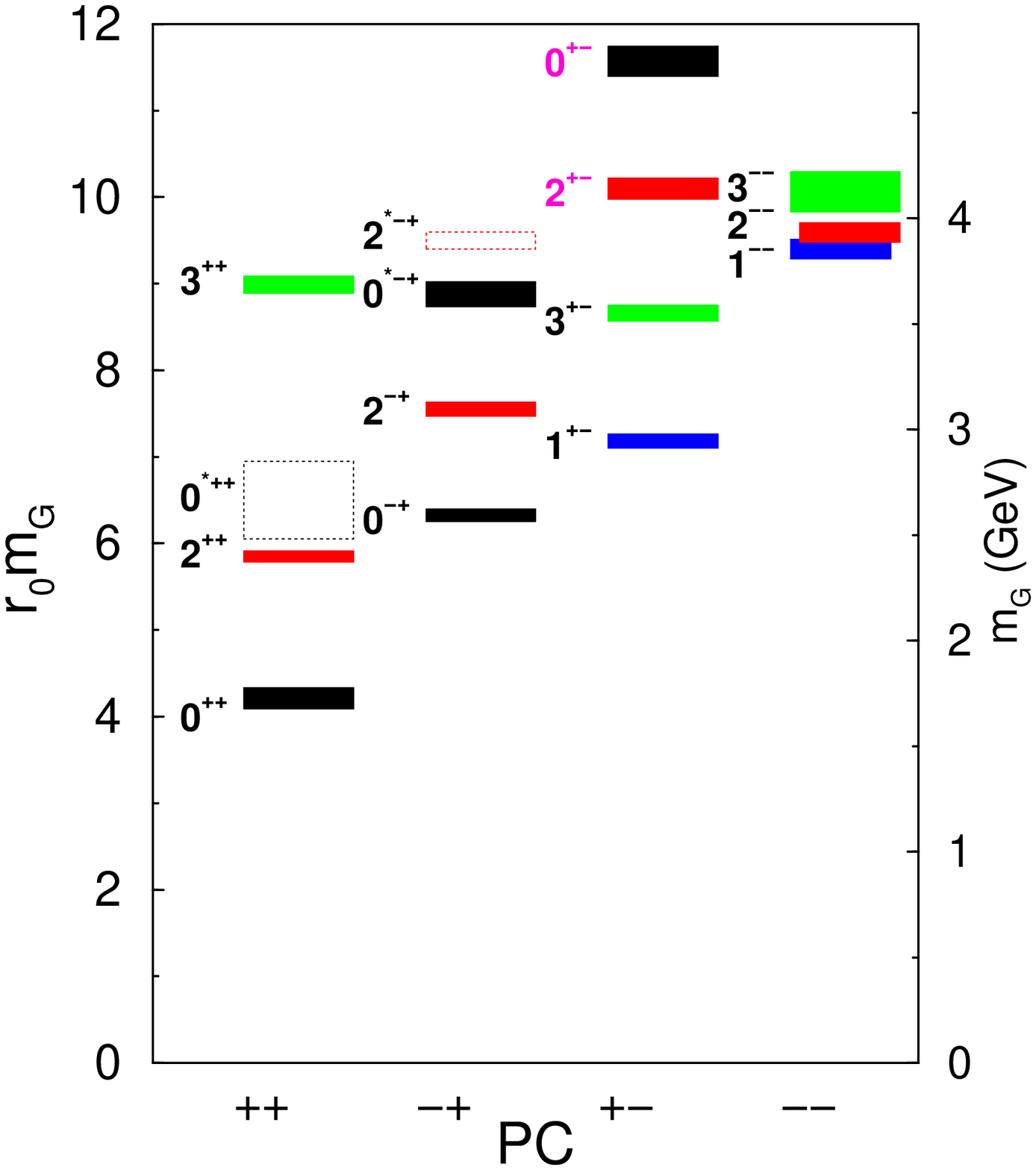,width=\textwidth}
\end{minipage}
\begin{minipage}[c]{0.49\textwidth}
\begin{footnotesize}
\renewcommand{\arraystretch}{1.4}
\begin{tabular}{llll}
 $J^{PC}$ & Other $J$ & \hspace{3ex} $r_0 m_G$ & \hspace{2ex} $m_G$ (MeV/c$^2$) \\
\hline
 $0^{++}$  &                & \bb 4.21 (11)(4)            & 1730 (50)\bb(80)  \\
 $2^{++}$  &                & \bb 5.85 (2)\bb(6)          & 2400 (25)\bb(120) \\
 $0^{-+}$  &                & \bb 6.33 (7)\bb(6)          & 2590 (40)\bb(130) \\
 $0^{*++}$ &                & \bb 6.50 (44)(7)$^\dagger$  & 2670 (180)(130)  \\
 $1^{+-}$  &                & \bb 7.18 (4)\bb(7)          & 2940 (30)\bb(140) \\
 $2^{-+}$  &                & \bb 7.55 (3)\bb(8)          & 3100 (30)\bb(150) \\
 $3^{+-}$  &                & \bb 8.66 (4)\bb(9)          & 3550 (40)\bb(170) \\
 $0^{*-+}$ &                & \bb 8.88 (11)(9)            & 3640 (60)\bb(180) \\
 $3^{++}$  & $6,7,9,\dots$  & \bb 8.99 (4)\bb(9)          & 3690 (40)\bb(180) \\
 $1^{--}$  & $3,5,7,\dots$  & \bb 9.40 (6)\bb(9)          & 3850 (50)\bb(190) \\
 $2^{*-+}$ & $4,5,8,\dots$  & \bb 9.50 (4)\bb(9)$^\dagger$& 3890 (40)\bb(190) \\
 $2^{--}$  & $3,5,7,\dots$  & \bb 9.59 (4)\bb(10)         & 3930 (40)\bb(190) \\
 $3^{--}$  & $6,7,9,\dots$  &    10.06 (21)(10)           & 4130 (90)\bb(200) \\
 $2^{+-}$  & $5,7,11,\dots$ &    10.10 (7)\bb(10)         & 4140 (50)\bb(200) \\
 $0^{+-}$  & $4,6,8,\dots$  &    11.57 (12)(12)           & 4740 (70)\bb(230)
\end{tabular}
\renewcommand{\arraystretch}{1.0}
\end{footnotesize}
 \end{minipage}
\caption{\label{fig:cont}
The mass spectrum of glueballs in pure {$\rm SU_C(3)$} gauge theory
\cite{Morningstar:1999rf}. The masses are given in units of the
hadronic scale $r_0$ along the left vertical axis and in GeV/c$^2$ along the
right vertical axis. The mass uncertainties indicated by the vertical
extents of the boxes do {\em not} include the uncertainty in $r_0$.
Numerical results are listed in tabular form. In some cases, the
spin-parity assignment for a state is not unique. The figure shows the
smallest $J$ value, the other possibilities are indicated in the second
column of the table. In column 3, the first error is the statistical
uncertainty from the continuum-limit extrapolation and the second is
the estimated uncertainty from the anisotropy. In the final column, the
first error comes from the combined uncertainties in $r_0 m_G$, the
second from the uncertainty in $r_0^{-1}=410(20)$ MeV/c$^2$. The dagger
indicates states for which the authors of \cite{Morningstar:1999rf}
suggest further studies before interpreting them.}
\end{figure}

Lattice simulations have also confirmed properties of the static quark
potential and the string-like behaviour of the QCD flux tube. In Fig.
\ref{fig:potential}, the potential energy is shown as a function of
the separation between two static quarks. The potential energy can be
described very well by the superposition of a Coulomb-like potential
and a linearly rising (confinement) potential. At sufficiently large
separations, for $R \approx 0.12$\,fm, the total energy suffices to
allow mesons to decay into two (colour-neutral) mesons: spring
breaking occurs. Spring breaking can be simulated on a lattice
\cite{Pennanen:2000yk,Bernard:2001tz,Michael:2005kw}.
The right panel in Fig.
\ref{fig:potential} displays the energy levels due to a $q\bar q$ and a
two-meson system in an adiabatic approximation. In a hadronic
reaction, the sudden approximation $-$ where the system follows the
straight line $-$ is more realistic, and $q\bar q$ pairs can be excited
to much higher energies.

\subsubsection{\label{Glueballs}
Glueballs}

The calculation of the glueball mass spectrum belongs to the early
achievements of lattice QCD and is still of topical interest. A
convenient way to represent the results was given by
Morningstar and Peardon \cite{Morningstar:1999rf} who calculated a
large number of spin-parity configurations, see Fig.
\ref{fig:cont}. The numerical results are reproduced here as
well; for noisy states, the right $J^{PC}$ assignment is difficult to
extract.

There exists a large number of papers, partly using different
techniques, to extract glueball properties. We refer the reader to
a recent calculation \cite{Loan:2005ff}
where references to previous work and to different methods are given.
Essentially all calculations agree that the lowest mass scalar glueball
should have a mass between 1.6 to 1.8\,GeV/c$^2$, followed by a tensor
glueball and a pseudoscalar glueball at about 2.3 and 2.5\,GeV/c$^2$,
respectively. It can be shown that unquenching the lattice does not
lead to significant changes in the glueball mass spectrum
\cite{Gregory:2005yr}. However, in these calculations the sea quarks
still have sizable masses corresponding to $m_{\pi}/m_{\rho}\sim 0.7$.

The lattice glueballs are compact objects, the scalar glueball has a radius
of about 0.3\,fm  \cite{Loan:2006gm}.

\subsubsection{\label{Ground-state qbar q mesons}
Ground-state $q\bar q$ mesons}

\begin{figure}[pb]
\begin{center}
\epsfig{file=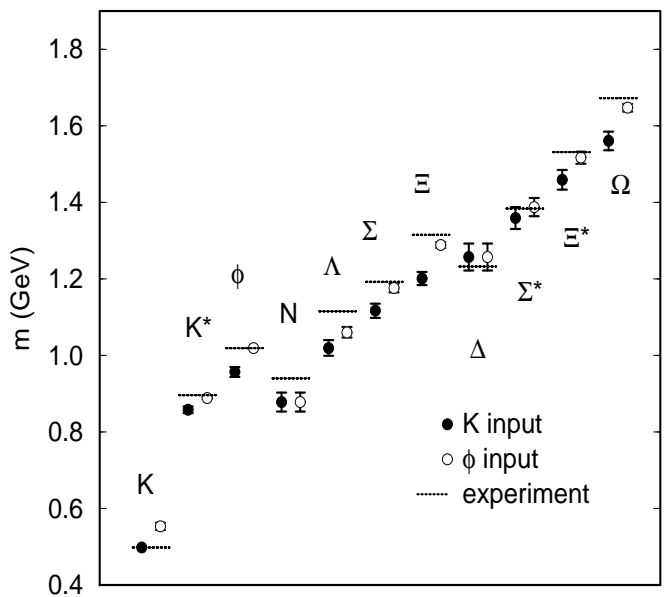,width=0.5\textwidth,height=0.45\textwidth}
\caption{\label{fig:spectrum}
Light hadron spectra in quenched QCD
\protect\cite{Aoki:1999yr}.}
\end{center}
\end{figure}

A precise calculation of the light hadron spectrum in quenched QCD,
i.e. without quark loops, was carried out by the CP-PACS Collaboration.
Their results are summarised in the plot shown in
Fig.~\ref{fig:spectrum}. The agreement is impressive: quenched QCD
describes the light hadron spectrum at the level of~10\,\%. Still,
there remain some significant deviations from the experimentally
observed spectrum, dynamical quarks do play a $\rm r\hat{o}le$ even
though at a surprisingly low level.

The CP-PACS results in Fig. \ref{fig:spectrum} do not yet show any
$SU(3)$ flavour singlet meson. Pseudoscalar and scalar flavour singlet
mesons are deeply affected by the topological structure or quantum
fluctuations of the QCD vacuum. The quenched approximation thus misses
the most important aspects of the $\eta$ and $\eta^{\prime}$ and of
their controversial scalar counterparts. An introduction to objectives
and methods to overcome these difficulties, at least in part, can be
found in some lectures notes  \cite{Schilling:2004kg}. The masses

\vspace{-5mm}\begin{equation}
 M_{\eta} = 292 \pm 31 \; \mbox{MeV/c}^2, \quad M_{\eta'} = 686 \pm 31
\; \mbox{MeV/c}^2. \label{eq:values}
\end{equation}\vspace{-7mm}

are low with respect to experimental values but yield a  mass splitting
that compares well with data. A similar study for scalar isoscalar
mesons has not been made but calculations of the $a_0(980)$ and $
K^*(1430)$ masses have been carried out \cite{McNeile:2006nv}.

The spectrum of scalar mesons is hotly debated. The glueball of lowest
mass has scalar quantum numbers, and every reliable information how to
organise the scalar nonet is extremely welcome. In Table~\ref{tab:a0}
recent results on the mass of the ground-state scalar isovector meson
$a_0$ are collected. The results are spread over a wide range: lattice
QCD can support the identification of the $a_0(980)$, of the Crystal
Barrel $a_0(1475)$, or of the Obelix $a_0(1290)$ state as the ground
state scalar isovector $q\bar q$ state,.

\begin{table}[ph]
\caption{\label{tab:a0}
The mass of the $a_0$ meson from  lattice QCD}.
\renewcommand{\arraystretch}{1.4}
\bc
\begin{tabular}{ccc}
\hline\hline
Group & Method &  $m_{a_0}$  GeV/c$^2$ \\
\hline
Bardeen at al.~\cite{Bardeen:2001jm} & quenched & $1.34(9)$ \\
Hart et al.~\cite{Hart:2002sp}  & $n_f=2$, partially quenched,
                                     & $1.0\pm0.2$ \\
Prelovsek et al.~\cite{Prelovsek:2004jp} & $n_f=2$, unquenched,
                                     & $1.58\pm0.34$ \\
Prelovsek et al.~\cite{Prelovsek:2004jp}  & partially quenched
                                     & $1.51\pm0.19$ \\
Burch et al.~\cite{Burch:2006dg} & quenched & $\sim 1.45$ \\
McNeile and Michael \cite{McNeile:2006nv} & $n_f=2$, unquenched, &
$1.01\pm0.04$  \\
\hline\hline
\end{tabular}
\ec
\renewcommand{\arraystretch}{1.0}
\end{table}

This is a puzzling situation: lattice QCD predicts meson and baryon
masses of undisputed states, in close agreement with experiment. When
open physical questions are involved, there is often no guidance
provided.

Recently, masses of scalar flavour singlet mesons were calculated in
($n_f=2$) unquenched lattice QCD \cite{Hart:2006ps} and it was shown
that the mass of the lightest $0^{++}$ meson is much lower than
the mass of the $0^{++}$ glueball in quenched QCD. The lowest-mass
scalar flavour singlet meson could hence be identified with $f_0(980)$
or even with $\sigma(485)$. It was stressed that for scalar mesons,
there is a strong dependence on lattice spacing. Strong mixing between
scalar glueball and scalar mesons is found, again with a significant
lattice-spacing dependence, demanding future calculations with finer
lattice spacing, dynamical simulations with light quarks (plus a
strange quark in the sea), and large statistics. The main effect of
unquenching the glueball is to drive its mass from 1.6\,GeV to 1\,GeV.
The authors conclude that ``it is not clear that a mixing scheme based
on only the states $f_0(1370)$, $f_0(1500)$, and $f_0(1710)$ is
complete enough to determine the fate of the quenched glueball".

\subsubsection{\label{Hybrids}
Hybrids}

Hybrid mesons are mesons in which the gluonic degrees of freedom are
excited. Most fascinating is the possibility that they may acquire
spin-exotic quantum numbers which cannot be created from  a $q \bar{q}$
state with unexcited glue. However, spin-exotic mesons can also be
formed by $q \bar{q}q \bar{q}$ systems or by meson-meson interactions.

The lowest-lying hybrids are expected to have $J^{PC}=1^{-+}$. Hybrids
in the $\Upsilon$ family are predicted in the range 10.7 to
11.0\,GeV/c$^2$, above the $\Upsilon(4S)$  and
with 20\,MeV/c$^2$ width. Bottomonium hybrids are expected to decay
with a width of about 100\,MeV/c$^2$ into one of the $\chi_b$ states
and an isoscalar scalar meson, via its coupling to its flavour-singlet
component \cite{Juge:1999ie,Manke:2001ft,Michael:2003xg}. In the
charmonium family, hybrids are predicted at 4.34\,GeV/c$^2$
\cite{Bernard:1997ib,Manke:1998qc,Mei:2002ip} with an estimated
uncertainty of 100 to 200\,MeV/c$^2$. The predicted mass is above the
$Y(4260)$ (which we interpret as $\psi(4S)$, see section \ref{A new ccb
vector state Y(4260)}). Light-quark hybrids should have masses at about
2\,GeV/c$^2$ \cite{McNeile:2002az} but even masses as low as
1.5\,GeV/c$^2$ are not excluded \cite{Hedditch:2005zf}. Hybrid decay
widths depend of course on the hybrid mass. For masses around
2.2\,GeV/c$^2$, typical partial widths are $(400\pm 120)$ or $(90\pm
60)$\,MeV/c$^2$ for $\pi b_1$ and $\pi f_1$ decays
\cite{McNeile:2006bz} and about 60\,MeV/c$^2$ for $\pi a_1$ decays
\cite{Cook:2006tz}.

\subsection{\label{QCD sum rules}
QCD sum rules}

In the method of QCD sum rules \cite{Shifman:1978bx}, problems due to
long-distance interactions are avoided by considering only quarks
propagating at short distances. Hadrons are represented by their interpolating
currents taken at large virtualities, i.e. at large, space-like momenta
$Q^{2}\equiv-q^{2}\gg\Lambda_{QCD}^2$. For normal mesons, the
interpolating currents are constructed from $q$ and $\bar q$. In the
case of glueballs, the currents are given by gluonic interpolating
fields, e.g. by  $O_{S}\left(  x\right) =\alpha_{s}G_{\mu\nu}^{a}\left(
x\right) G^{a\mu\nu}\left(  x\right)$ for a scalar glueball. The
correlation function of these currents is introduced as

\vspace{-5mm}\begin{equation}
\Pi_{G}(-q^{2})=i\int d^{4}x\,e^{iqx}\left\langle 0|T\,O_{G}\left(  x\right)
O_{G}\left(  0\right)  |0\right\rangle
\end{equation}\vspace{-7mm}

and treated in the framework of the operator product expansion where
the short distance interactions are calculated using QCD perturbation
theory. Long-distance interactions are parametrised in terms of
universal vacuum condensates. The result of the QCD calculation is then
matched, via dispersion relation, to a sum over hadronic states

\vspace{-5mm}\begin{equation}
\Pi_{G}\left(  Q^{2}\right)  =\frac{1}{\pi}\int_{0}^{\infty}%
ds\frac{{\rm Im} \Pi_{G}\left(  -s\right)  }{s+Q^{2}}\,.
\label{disprel}
\end{equation}\vspace{-7mm}

The spectral function is usually parametrised by one or more resonances
and a step-like continuum

\vspace{-5mm}\begin{equation} {\rm Im} \Pi_{G}^{\left(  ph\right)
}\left(  s\right) = {\rm Im}\Pi_{G}^{\left(  pole\right)  }\left(  s\right) +
{\rm Im} \Pi_{G}^{\left(  cont\right)  }\left(  s\right), \label{phspecdens}%
\end{equation}\vspace{-7mm}

leading to the following representation:

\vspace{-5mm}\begin{equation}
\Pi(q^2)= \frac{q^2f^2}{m^2(m^2-q^2)} +
q^2\int\limits_{s_0^h}^{\infty}ds \frac{\rho^h(s)}{s(s-q^2)} +\Pi(0)\,.
\label{dispSR}
\end{equation}\vspace{-7mm}

After a Borel transformation, a more convenient form of the sum rule is
obtained

\vspace{-5mm}\be \Pi(M^2)= f^2e^{-m^2/M^2}+ \int_{s^h_0}^\infty ds
~\rho^h(s)e^{-s/M^2}, \label{BorelSR}
\ee\vspace{-7mm}

where contributions from high masses are suppressed.
Shifman, Vainshtain and Zakharov \cite{Shifman:1978bx} related
the short distance behaviour of current correlation functions to the
vacuum expectation values of gluon operators and thus deduced
the magnitude \cite{Colangelo:2000dp} of the gluon condensate:

\vspace{-5mm}\begin{eqnarray}
\langle\frac{\alpha_s}{\pi} G^2\rangle =(0.012\, {\rm GeV})^4\pm 30\%\,.
\end{eqnarray}\vspace{-7mm}

QCD sum rules have the advantage that they deal with fundamental
parameters like current quark masses and vacuum condensate densities,
contrary to phenomenological models which use not so well defined
objects like constituent quarks etc. An impressive number of results on
light quark spectroscopy were obtained using sum rules. For a
discussion of perspectives and limitations of sum rules, we refer the
reader to the review of Colangelo and Khodjamirian
\cite{Colangelo:2000dp}.

Narison has calculated glueball masses and widths in the QCD sum rule
approach and low-energy theorems \cite{Narison:1996fm}. In the scalar
sector, he found a gluonium state having a mass $M_G=(1.5\pm 0.2)$ GeV,
which should decay into the flavour octet (!) $\eta\eta'$ channel
and into 4$\pi^0$. In addition, he suggests wide gluonium states at low
masses which he identifies with the $\sigma(485)$. In a more recent
paper, Narison proposed that the two isoscalar resonances $\sigma(485)$
and $f_0(980)$ result from a maximal $q\bar q$--glueball mixing. The
scalar states $f_0(1500)$, $f_0(1710)$, $f_0(1790)$ are suggested to
have a significant glueball component \cite{Narison:2005wc}.

\subsection{\label{Flux-tube model}
Flux-tube model}

The flux tube approach describes mesons in a dynamical nonrelativistic
model motivated by the strong coupling expansion of lattice QCD
\cite{Isgur:1984bm}. In this model quarks are connected by a string of
massive beads, with a linear confining potential between the beads.
By analogy with lattice QCD only locally transverse spatial
fluctuations of the bead positions are possible. For a string of $N$
mass points which connects a quark at site $0$ to an antiquark at site
$N+1$ the  flux tube model Hamiltonian is written as

\vspace{-5mm}\begin{equation}
H = H_{quarks} + H_{flux\ tube} \ ,
\label{tube:hamiltonian}
\end{equation}\vspace{-7mm}

\vspace{-5mm}\begin{equation}
H_{quarks} =
-{1\over 2 m_q} \vec \nabla_q^2
-{1\over 2 m_{\bar q}} \vec \nabla_{\bar q}^2
+ V_{q\bar q} \ , \nonumber
\end{equation}\vspace{-7mm}

\vspace{-5mm}\begin{equation}
H_{flux tube} = b_0 R + \sum_n\left[ {p_n^2\over 2b_0a} +
{b_0\over 2a}(y_n-y_{n+1})^2\right], \nonumber
\end{equation}\vspace{-7mm}

Here $m_q$ and $m_{\bar q}$ are the quark and antiquark masses, $m_b$ is the
bead mass, $y_n$ is the transverse displacement of the $n$th bead,
$p_n$ is its momentum, $b_0$ is a string tension, and $R=(N+1)a$ is the
separation between the static quarks. The potential $V_{q\bar q}$ is
meant to describe the colour Coulomb interaction. When the flux tube is
in its ground state, the excitation of $q \bar q$ yields the
conventional meson spectrum.  Flux-tube excitations lead to a new kind
of hadronic states called hybrids. The Schr\"odinger equation with
Hamiltonian (\ref{tube:hamiltonian}) can be solved in adiabatic
approximation thus giving the masses of ordinary and of hybrid mesons.
The excitations of the flux tube are characterised by $\Lambda$, the
projection of the flux tube orbital angular momentum along the $q\bar
q$ axis. For a single flux tube excitation, $\Lambda = \pm 1$. With
this additional degree of freedom the quantum numbers of hybrid mesons
with the $q \bar q$ spin $S=0$ are $J^{PC} = 1^{++}, 1^{--}$; for $S =
1$ the quantum numbers  $J^{PC} = 2^{+-}, 2^{-+}, 1^{+-}, 1^{-+},
0^{+-}, 0^{-+}$ are accessible. The list contains exotic quantum
numbers, $J^{PC} = 0^{-+}, 1^{-+}, 2^{+-}$, which cannot be generated
by normal $q\bar q$ states. The lightest hybrid states are
predicted to have masses $M \approx 2.0$ GeV/c$^2$
\cite{Swanson:2003kg}.

Mesons can decay via flux-tube breaking at any point along its length,
producing a $\bar{q}q$ pair in $J^{PC}=0^{++}$ state. Hybrid states
have a node along the initial $q \bar q$ axis leading to a suppression of
hybrid decays into two mesons with identical spatial wave functions:
hybrid decays into two pseudoscalar mesons or one pseudoscalar and one
vector meson are forbidden. Instead, hybrids prefer decays into one
$S$-wave and one $P$-wave meson. Typical examples are decays into $\pi
b_1(1235)$ or $\pi f_1(1285)$. Hybrids can therefore be observed only
in complicated final states and at rather high masses. These conditions
make experimental searches for them very difficult.

The decay widths can be calculated using some models for wave functions
of the mesons involved and assuming a universal string breaking constant
for all decays \cite{Close:1994pr}. This universality is of course a
rather restrictive assumption.

\begin{figure}[ph]
\bc
\includegraphics[angle=-90,width=0.6\textwidth]{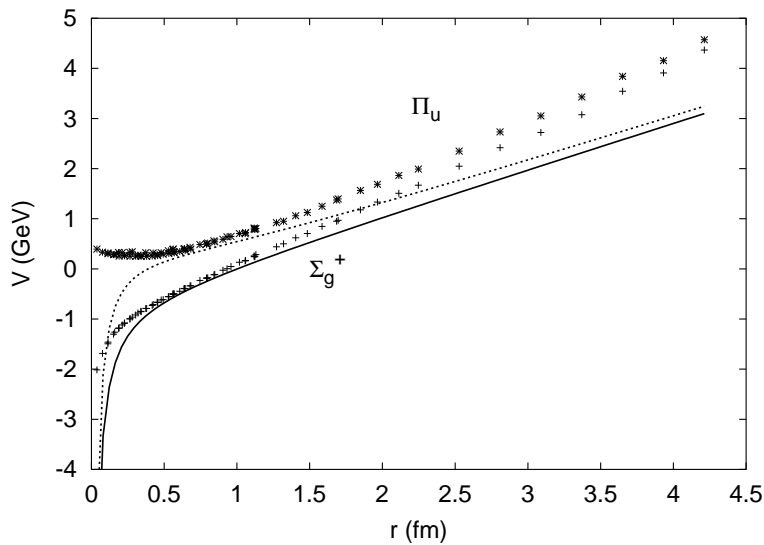}
\ec
\caption{\label{Vplot}
Comparison of the flux-tube potentials represented by a solid line
(dotted line first excited state)  with lattice gauge calculations
given by crosses \protect{\cite{Juge:1997nc}}. \vspace{2mm}}
\end{figure}

To estimate the reliability of the flux-tube model we can compare its
results on hybrids with those of lattice QCD. Figure \ref{Vplot}
shows the flux tube potential for the ground state (solid line) and the
first excited state (dotted line) \cite{Swanson:2003kg}. These are
compared to lattice computations of the same adiabatic potentials
(points) \cite{Juge:1997nc}. At a $q \bar q$ separation of about $1$
fermi, both models nearly coincide while at smaller separations the
flux tube model potential overestimates the strength of the attractive
Coulomb potential, at large separations, it underestimates the string
tension. Hybrid mesons at small $q \bar q$ separations can be treated
as states formed by a $q \bar q$ pair and a gluon. The gluon field
carries colour, the $q \bar q$ form a colour octet and thus their
interaction is repulsive.

Apart from these details, flux-tube-model and lattice potential are
rather similar, and the flux tube provides a viable model and a very
convenient language. It has significant predictive power and a
sufficient number of tuneable parameters for further development.

\subsection{\label{Quark models}
Quark models}

The quark model still plays an important $\rm r\hat{o}le$ in the interpretation of
the mass spectra of mesons. Quark models assume that a meson is
composed of a constituent quark and an antiquark; their masses may
arise from the spontaneous breaking of chiral symmetry, they
are free parameters of the model. Between quark and antiquark a
confining potential is assumed to rise linearly with the interquark
distance. The confinement potential is however not sufficient to
explain details of the spectra and some additional quark-quark
interaction is required to reproduce the known meson masses. The main
differences between different quark models are related to the choice of
the appropriate spin-dependent quark interactions which find their
roots in either One-Gluon-Exchange, Goldstone-Boson-Exchange or
instanton effects.

\subsubsection{\label{Gluon-exchange interactions}
Gluon-exchange interactions}

The first unified constituent quark model for all $q\bar q$-mesons was
developed by Godfrey and Isgur~\cite{Godfrey:1985xj}. It is widely used
as a reference for comparison of new experimental data, and has had a
very significant impact on the development of the field.  The model
starts from a Hamiltonian

\vspace{-5mm}\begin{equation}
        H \Psi = (H_0 + V) \Psi = E \Psi,
\qquad
H_0 = \sqrt{{m_q}^2+|\vec p|^2} +
\sqrt{{m_{\bar q}}^2+|\vec p|^2}\,,
\end{equation}\vspace{-7mm}

with $\vec p$ the relative momentum in the centre-of-mass frame. The
interaction is written as
$$
        V = H^{c} + H^{SS} + H^{LS} + H^{A}
$$
which contains a central potential -- linear confinement $br + c$ and
Coulomb potential --, a spin-spin and a spin-orbit interaction, and an
annihilation contribution for flavour-neutral mesons.

The potential is generated from gluon exchange with a running coupling
constant for which a parametrisation
$$
        G(Q^2) = -\frac{4}{3}\,\alpha_s(Q^2)\frac{4\pi}{Q^2}
$$
is chosen where $\alpha_s(Q^2)=\sum_k\,\alpha_k\,\
e^{-\frac{Q^2}{4\gamma_k^2}}$, with ${\alpha_s(0)}$\,finite, and $\vec
Q = \vec p' -\vec p$. These potentials are ``smeared out'' to avoid
singularities at the origin. Relativistic effects are partly taken into
account, but spin-orbit forces are suppressed; {\it there are no
spin-orbit forces in the Hamiltonian}. The excuse for this suppression
is the experimental observation that these are weak or absent in the
data. From the theoretical side, spin-orbit forces are at least partly
compensated by the so-called Thomas precession, a relativistic
generalisation of Coriolis forces. The $\rho$-$\pi$-splitting is
assigned to the magnetic interaction caused by one-gluon exchange.

Annihilation is taken into account by
parametrising the annihilation amplitude, one for non-pseudoscalar
flavour-neutral mesons and a different one for pseudoscalar mesons. All
mesons are assumed to be ``ideally mixed'', except the pseudoscalar
mesons.

\subsubsection{\label{Exchange of Goldstone particles}
Exchange of Goldstone particles}

The model proposed by Vijande, Fernandez, and Valcarce
\cite{Vijande:2004he} uses a quark-antiquark potential which includes
a (screened) confinement potential in the form

\vspace{-5mm}\begin{equation}
  \label{eq:model:VcGBE}
  V_{\textit{\footnotesize{con}}}(\vec x) =
  \left(\Delta-a_c\left(1-\exp({-\mu_c|\vec
x|})\right)\right)\,\lambda^{(c)}_q\cdot
  \lambda^{(c)}_{\bar q}\,,
\end{equation}\vspace{-7mm}

a short range one-gluon-exchange inspired potential and  Goldstone
Boson exchanges in the form of modified Yukawa potentials

\vspace{-5mm}\begin{equation}
  \label{eq:model:VGBE}
  V(\vec x) = \sum_{\mu=\pi,\sigma,K,\eta} V_{\mu}(\vec x)\,,
\end{equation}\vspace{-7mm}

where the isoscalar ($\sigma$) part contains a central and a spin-orbit
term, and the other potentials ($\pi,K,\eta$) consist of central,
spin-spin and a tensor contribution. Flavour mixing for pseudoscalar
mesons is induced through the flavour dependence of Goldstone boson
exchange. Spectra range from light quark mesons to heavy flavours.

\subsubsection{\label{Instanton-induced interactions}
Instanton-induced interactions}

In the Bonn quark model \cite{Koll:2000ke,Ricken:2000kf},
the residual interactions (in addition to a linear confining potential)
are induced by instantons. Meson mass spectra, wave functions and
transition matrix elements are calculated from a homogeneous,
instantaneous Bethe-Salpeter equation. In this relativistic model,
confinement is described by a linear potential with a suitable Dirac
structure. In the Bonn model, two different Dirac structures were used
to calculate the meson mass spectra. The first one has a scalar and a
time-like vector structure in the form

\vspace{-5mm}\begin{equation}
\frac{1}{2}(\Bbb{I}\cdot\Bbb{I}-\gamma_0\cdot\gamma_0)
\end{equation}\vspace{-7mm}

where $\Bbb{I}$ is the identity operator. Alternatively, a
confinement potential  invariant under $U_A(1)$ was assumed:

\vspace{-5mm}\begin{equation}
\frac{1}{2}(\Bbb{I}\cdot\Bbb{I}-\gamma_5\cdot\gamma_5
                -\gamma^\mu\cdot\gamma_\mu).
\end{equation}\vspace{-7mm}

In the limit of vanishing quark masses, this structure leads to parity
doublets in the meson spectrum. The two variants define two models,
called $\mathcal A$ and $\mathcal B$ in
\cite{Koll:2000ke,Ricken:2000kf}. Instanton-induced interactions lead
to  $u\bar u\to d\bar d$ and $u\bar u\to s\bar s$ transitions with
adjustable strengths; their values were determined to describe the
ground state pseudoscalar mesons.

\begin{figure}[pb]
\begin{minipage}[c]{0.49\textwidth}
\includegraphics[width=0.9\textwidth,height=0.7\textwidth]{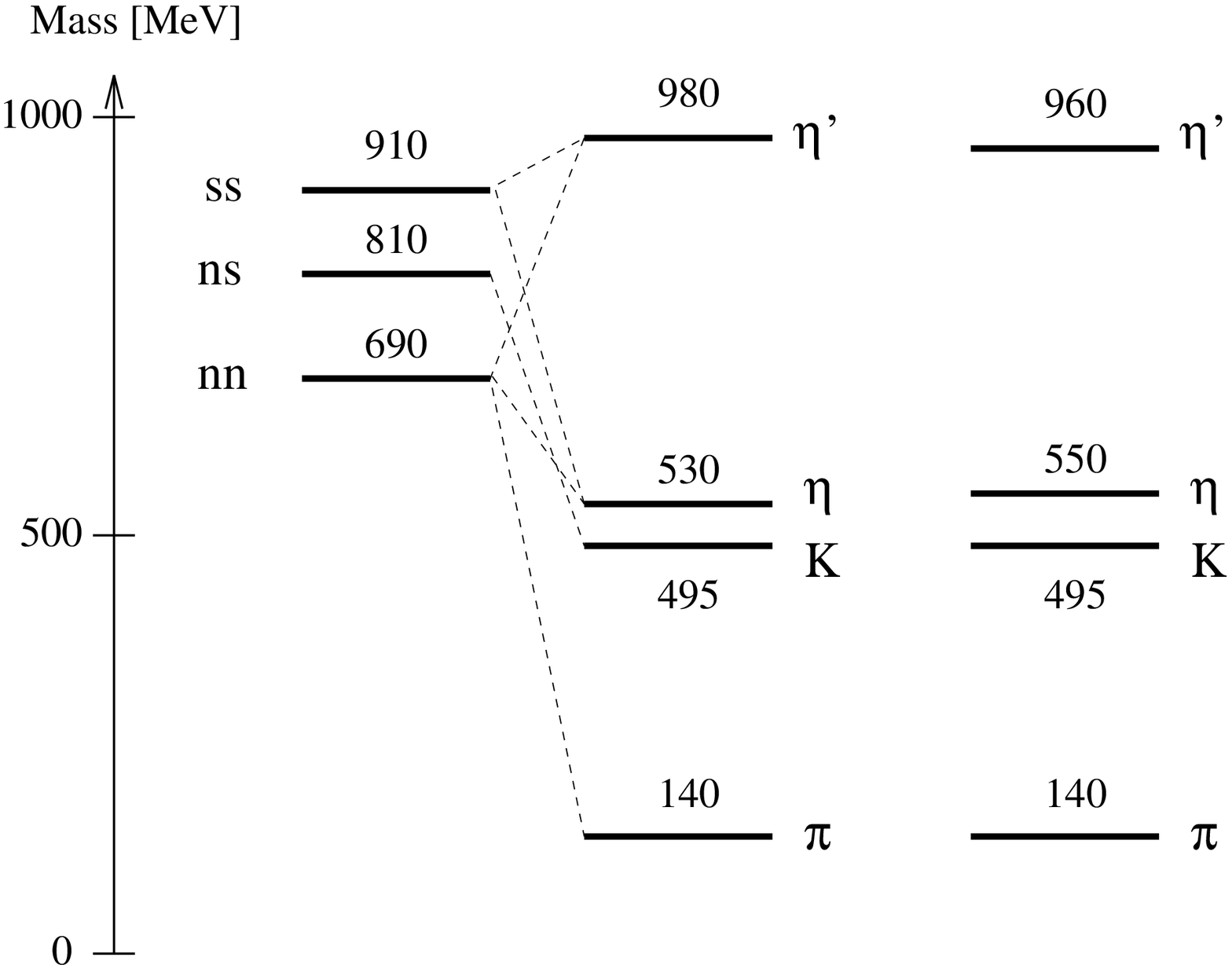}
\end{minipage}
\begin{minipage}[t]{0.49\textwidth}
\includegraphics[width=0.9\textwidth,height=0.5\textwidth]{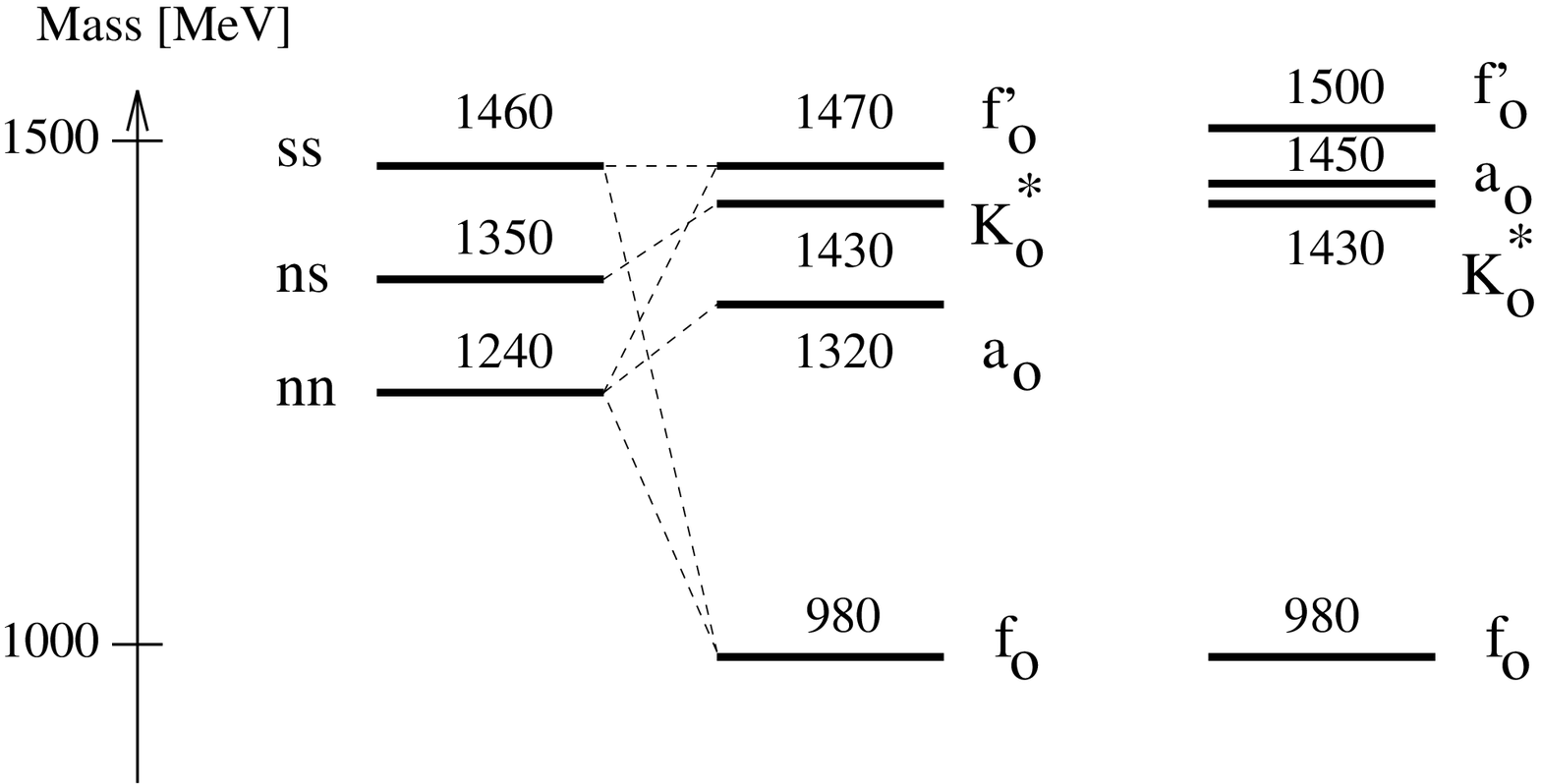}
\end{minipage}
\caption{\label{instanton-for-mesons}
Schematic splitting of pseudoscalar (left panel) and scalar (right
panel) flavour nonets with confinement interaction (left), with
confinement and instanton-induced force (middle) compared to data
(right) \protect\cite{Klempt:1995ku}.}
\end{figure}

Instanton-induced forces have been `invented' to explain the $U_A(1)$
anomaly, the large value of the $\eta^{\prime}$ mass
\cite{'tHooft:1976up}. Hence it is not surprising that a model based on
instanton-induced forces does well in describing pseudoscalar meson
masses. Instantons in mesons induce coherent spin flips for quark and
antiquark; such transitions are possible only in pseudoscalar mesons
where spin $up$ and spin $down$ can be interchanged, or in scalar
mesons where the spins of quark and antiquark may flip and the orbital
angular momentum changes its direction. Fig. \ref{instanton-for-mesons}
shows the influence of instantons for pseudoscalar and scalar
ground-states mesons \cite{Klempt:1995ku}. Instantons lead to
substantial mass splittings; the $\rho\pi$ splitting is now due to
a shift in mass of the pseudoscalar pion which is absent in $\rho$
mesons. Particularly interesting is the fact that scalar mesons organise
themselves into a high-mass flavour-octet and a low-mass flavour
singlet state. The predicted $f_0(1470)$ is easily identified with the
known $f_0(1500)$; the $ K^*_0(1470)$ is correctly reproduced. The
authors in \cite{Klempt:1995ku} identified the predicted $a_0(1320)$
with the $a_0(1450)$, in spite of a larger discrepancy. The predicted
$f_0(980)$ was identified either with the $f_0(980)$; (this
assignment leads to an asymmetric treatment of the twins $a_0(980)$ and
$f_0(980)$); alternatively, it was identified with the wide scalar
`background' called $f_0(1000)$ by Au, Morgan, and Pennington
\cite{Au:1986vs} (and nicknamed 'red dragon' by Minkowski and Ochs
\cite{Minkowski:1998mf}). Later, we will call it $f_0(1300)$.

\subsubsection{\label{Heavy Quark Symmetry}
Heavy Quark Symmetry}

Heavy quark symmetry \cite{Isgur:1989ed,Shifman:1986sm,Caswell:1985ui}
exploits the large separation of mass scales in hadronic bound states of
a heavy quark with light constituents (quarks, antiquarks and gluons) :
the Compton wave length of the heavy quark ($\lambda_Q\sim 1/m_Q$) is
much smaller than the size of the hadron containing the heavy quark
($R_{\rm had}\sim 1/\Lambda_{\rm QCD}$). This symmetry is widely used
in the analysis of semileptonic decays of heavy-light hadrons and in
spectroscopy (see review \cite{Neubert:1996si} by Neubert). In the
rest frame of the heavy quark, relativistic effects such as colour
magnetism vanish as $m_Q\to\infty$ since the heavy-quark spin decouples
from the light-quark degrees of freedom. For $N_h$ heavy-quark
flavours, there is thus an SU$(2 N_h)$ spin-flavour symmetry group,
under which the effective strong interactions are invariant.  (The
flavour symmetry is analogous to the fact that different isotopes have
the same chemistry, since to a good approximation the wave function of
the electrons is independent of the mass of the nucleus; the nuclear
spin decouples in the limit $m_e/m_N\to 0$). Symmetry-breaking
corrections in heavy-light systems can be studied in a systematic way
within the Heavy Quark Effective Theory (HQET) where these effects
depends on few phenomenological parameters. Here we limit ourselves to
spectroscopic applications and do not touch semileptonic decays, most
extensively studied within HQET.

\begin{table}[pb]
\caption{\label{DTab}
Low-lying $Q\bar q$ states in the LS and JJ coupling
scheme.\vspace{2mm}}
\bc
\renewcommand{\arraystretch}{1.4}
\begin{tabular}{l|cc|cccc} \hline
 & $D$ &  $D^*$    & $D_0^*$  &  $D_1$  & $D_1'$  & $D_2^*$ \\
\hline
\hline
$J^P$         & $0^-$ & $1^-$ & $0^+$ & $1^+$ & $1^+$ & $2^+$ \\
$^{(2S+1)}L_J$  & $^1S_0$ & $^3S_1-{}^3D_1$ & $^3P_0$ & $^3P_1-{}^1P_1$ & $^3P_1-{}^1P_1$ & $^3P_2$ \\
$\ell_q$ &  0 & 0 & 1 & 1 & 1 & 1 \\
$j_q^p$  &  $\frac{1}{2}^-$  &  $\frac{1}{2}^-$ & $\frac{1}{2}^+$  &  $\frac{1}{2}^+$ &
$\frac{3}{2}^+$  &  $\frac{3}{2}^+$ \\
decay & --  & -- &  $(DP)_S$  & $(D^*P)_S$  & $(D^*P)_D$ & $(DP)_D$ \\
\hline
\hline
\end{tabular}
\renewcommand{\arraystretch}{1.4}
\ec
\end{table}

In the limit $m_Q\to\infty$, the spin of the heavy quark and the total
angular momentum $j$ of the light quark inside a hadron
are separately conserved by the strong interactions. Because of
heavy-quark symmetry, the dynamics is independent of the spin and mass
of the heavy quark. Hadronic states can thus be classified by the
quantum numbers (flavour, spin, parity, etc.) of the light-quark
degrees of freedom. Instead of the usual LS coupling scheme, $Q\bar q$
mesons are more efficiently characterised in the $jj$
coupling scheme. The total spin of a $Q\bar q$ meson then decomposes
into the spin $\vec s_Q$ and the angular momentum $\vec j_q =
\vec\ell_q + \vec s_q$ where $\vec \ell_q$ and $\vec s_q$ are angular
momentum and spin of the light quark. The total spin-parity is then
given by $J^P = \frac{1}{2}^- \otimes j_q^p$. In this basis, the lowest
states comprise three doublets labelled by $j_q^p = \frac{1}{2}^-$,
$\frac{1}{2}^+$, and $\frac{3}{2}^+$. Table \ref{DTab} lists the
classification of $Q\bar q$ states in the two coupling schemes.

The different bases impose different multiplet structures on the $D$
spectrum. For small spin-orbit interactions, the spectroscopic $LS$
basis has an $S$-wave ($D$,$D^*$) doublet and a $P$-wave quadruplet, the
$(jj)$ scheme has three doublets. The light components of
heavy-quark-symmetry wave functions do not depend on the spin of the
heavy quark. This leads to predictions \cite{Isgur:1991wq} for the
decay-width ratios for hadronic transitions between different $jj$
multiplets like $ D_1 \to D+\pi,~D_1 \to D^* +\pi,~ D_0 \to D +\pi,~
D_2 \to D + \pi$, which are found to be in good agreement with
experiment.

The predictions of heavy quark effective field theory will be used as a
guide to appreciate the significance of recently discovered $D$ and
$D_s$ resonances.

\subsection{\label{Meson decays}
Meson decays}

Lattice gauge calculations have shown how hadronic decays may occur:
when a meson is excited, the string connecting quark and antiquark is
expanded until it reaches an energy excitation from which a transition
to a meson-meson final state is allowed. It is natural to assume that
the $q\bar q$ pair just created carries the quantum numbers of the
vacuum.

The $^3P_0$ model of strong decays was suggested nearly 40 years ago by
Micu \cite{Micu:1968mk} and further developed by the Orsay group
\cite{LeYaouanc:1972ae,LeYaouanc:1973xz}. Extensive calculations were
carried out by Barnes and collaborators
\cite{Barnes:1996ff,Barnes:2002mu,Barnes:2005pb};  references to
earlier work can be found in these papers. The results provide an
important auxiliary tool in the discussion of the nature of new states.

Other approaches were attempted. The discovery of the charmonium system
with its positronium-like level pattern motivated the use of a
one-gluon-exchange potential and of the assumption that hadronic decays
proceed via time-like gluons \cite{Eichten:1978tg,Eichten:1979ms}.
Instantons induce transitions not only from one $q\bar q$ pair by a
four-point interaction but also transitions of type $u\bar u\to
(d\bar ds\bar s)$. Due to their nature, instantons become effective
only for decays of scalars into two pseudoscalar mesons and for
pseudoscalar decays into a pseudoscalar plus a scalar meson. Results of
first calculations were reported in \cite{Ritter:1996xh}. Of course,
instanton-induced decays could interfere with other decay mechanisms;
an assignment of scalar and pseudoscalar mesons to flavour multiplets
or glueballs using arguments based on a particular decay model is
therefore extremely difficult or even impossible.

\subsection{\label{Effective chiral symmetry restoration}
Effective chiral symmetry restoration}

For light-quark mesons and for sufficiently high radial or orbital
excitations both, chiral symmetry and $U_A(1)$ symmetry, could be
effectively restored as suggested by Glozman \cite{Glozman:1999tk}. The
momenta of valence quarks could increase at large hadron excitation
energies and could then decouple from the chiral condensates of the QCD
vacuum. As a consequence, the dynamical quark mass is reduced. For the
physics of low-lying hadrons, the chiral-symmetry-breaking condensates
are crucially important.  The physics of highly-excited states is
suggested to be decoupled from these chiral symmetry breaking effects.
Asymptotically, the states may approach a regime where their properties
are determined by the underlying unbroken chiral symmetry (i.e. by the
symmetry in the Wigner-Weyl mode). In this case hadrons should
gradually decouple from the Goldstone bosons \cite{Glozman:2006xq} and
form new multiplets. If chiral symmetry is restored, the states fall
into approximate multiplets of $SU(2)_L\times SU(2)_R$
\cite{Cohen:2001gb}. In the usual potential model classification
$I,J^{PC}$, chiral symmetry restoration leads to the following
mass relations:
\bc
${\bf J~=~0}~~~~~~~~~~ 1,0^{-+} \leftrightarrow 0,0^{++};~~~
 1,0^{++} \leftrightarrow 0,0^{-+}$

${\bf J~=~2k}~~~~~~~~ 0,J^{--} \leftrightarrow 0,J^{++};~~~
 1,J^{-+} \leftrightarrow 0,J^{++};~~~
 1,J^{++} \leftrightarrow 0,J^{-+};~~~
 1,J^{++} \leftrightarrow 1,J^{--}$

$ {\bf J~=~2k-1}~~~ 0,J^{++} \leftrightarrow 0,J^{--};~~~
 1,J^{+-}\leftrightarrow 1,J^{--};~~~
 1,J^{--}\leftrightarrow 0,J^{+-};~~~
 1,J^{--} \leftrightarrow 1,J^{++}$
\ec
The $U_A(1)$ symmetry connects opposite-parity states of the same
isospin but from different $SU(2)_L \times SU(2)_R$ multiplets, for
example $~~1,0^{-+} \leftrightarrow 1,0^{++}$. Chiral symmetry
restoration requires a doubling of some of the radial and angular Regge
trajectories for $J > 0$.  At large excitations, some of the
$\rho$-mesons have $a_1$ mesons as their chiral partners, while the
other $\rho$-meson excitations are chiral partners of $h_1$ mesons. In
section \ref{High radial and orbital excitations of light mesons}, the
predictions will be confronted with experimental results.

\subsection{\label{Solvable models}
Solvable models}

The relation between the spectroscopy of excited hadrons in the
resonance region and their partonic degrees of freedom observed in
deep inelastic scattering is one of the most challenging tasks in
strong interaction physics. A promising direction has been proposed
by Maldacena \cite{Maldacena:1997re} who proposed a connection between
the strongly coupling limit of a conformal field theory (CFT) defined
on the AdS asymptotic boundary \cite{Klebanov:2000me,Witten:1998qj},
and the propagation of weakly coupled strings in a higher dimensional
Anti-de-Sitter (AdS) space, where physical quantities can be computed.
QCD is not a conformal theory but for an approximately constant
coupling constant and for vanishing quark masses, QCD is similar to a
strongly-coupled conformal theory. Holographic duality requires a
higher dimensional warped space with negative curvature and a
four-dimensional boundary. The equations of motion in the AdS space are
expressed as effective light-front equations which describe holographic
light-front eigenmodes dual to QCD bound states~\cite{Brodsky:2006uq}.
The eigenvalues of the effective light-front equation reproduce the
mass spectra of light-quark mesons and baryons, the eigensolutions
their valence wavefunctions. Confinement can be modelled by a `hard
wall' cutting off AdS space in the infrared region
\cite{Polchinski:2001tt} or spacetimes can be capped off smoothly by a
`soft wall' \cite{Karch:2006pv}.

On the basis of this ADS/CDF correspondence, a number of results on
nonperturbative QCD were obtained: from chiral symmetry breaking
\cite{Erlich:2005qh,Shock:2006gt} to the pattern of highly excited
states \cite{Brodsky:2006uq,Karch:2006pv}, to meson decay constants,
form factors, and to exclusive scattering amplitudes
\cite{Brodsky:2003px}. From these amplitudes, valence, sea-quark and
gluon distributions, and generalized parton distributions measured in
deeply virtual Compton scattering can be determined. A survey of the
achievements can be found in recent papers of Brodsky and de Teramond
\cite{Brodsky:2007xx,deTeramond:2007yy}.

\subsection{\label{Have we gained insight from q bar q mesons?}
Have we gained insight from $q\bar q$ mesons?}

``The purpose of computing is insight not numbers'', a memorable phrase
used repeatedly by Hamming throughout his book on Numerical Methods for
Scientists and Engineers \cite{Hamming:1962ek}. Quark models have the
intent to provide insight: to identify the leading mechanisms which
result in the observed spectrum of hadron resonances and their decays.
However, there are three fully developed quark models; one could try to
identify the model with the best $\chi^2$ for a minimum number of
parameters as matching physics reality best, but a good $\chi^2$ does
not prove that the model is correct.

All calculations reproduce the $\pi-\eta-\eta'$ splitting, although
the underlying dynamics differ widely. In the relativised quark model
of Godfrey and Isgur \cite{Godfrey:1985xj}, the non-degeneracy of $\pi$
and $\eta$ is due to the inclusion of an extra annihilation amplitude,
whereas in the other models it is attributed to an explicit flavour
dependence in the quark-antiquark interaction. In the Bonn quark model
these are based on instanton effects, which simultaneously accounts for
the $\rho/\omega-\pi$, $K^*-K$ and $\pi-\eta-\eta'$ splittings through
non-vanishing $u\bar{u}\leftrightarrow d\bar{d}$,
$u\bar{u}\leftrightarrow s\bar{s}$, and $d\bar{d} \leftrightarrow
s\bar{s}$ amplitudes. In the Goldstone-boson-exchange  model these
splittings are due to the explicit flavour dependence of the various
meson exchanged. The largest differences turn up for the scalar mesons:
Whereas the relativised quark model concurs with the usual folklore to
yield degenerate isoscalar and isovector states around 1\,GeV/c$^2$,
and the strange scalar state around 1.25 GeV/c$^2$, the other models
predict that the scalar spectrum show splittings similar to the
pseudoscalar case but of opposite sign.

All models account very well for the position of the lowest state in
each flavour sector even up to the highest masses and total angular
momenta. Note however, that Vijande, Fernandez and Valcarce
\cite{Vijande:2004he} only quote results for $J \le 3$, related to
their choice of a ``screened'' confining potential. The two models
which include one-gluon exchange account somewhat better for the
observed splittings (and mixing) of radially excited vector meson
states. The Bonn quark model contains only the confining potential in
this sector and has nearly no spin-orbit and/or tensor interactions.
Thus the spin-singlet and spin triplet $1^+$ strange mesons are in fact
degenerate. None of the models produces more than one $\eta$ states
between 1 and 1.5\,GeV/c$^2$ (where the PDG \cite{Eidelman:2004wy}
lists 3), for a discussion see section \ref{Pseudoscalar mesons}.

Obviously, strong interaction dynamics can be based on very different
physical pictures, and still give reasonable descriptions of
experimental data. Lattice QCD can serve here as additional support for
the assumptions underlying the different quark models.
Instantons can be observed in  lattice QCD by cooling the highly
fluctuating field configurations~\cite{Negele:1998ev}. Cooling removes
the perturbative part of the gauge fields and leaves the classical
configurations. Likely, the rapid field changes in the initial phase
are caused by one-gluon exchange dynamics. The $\rho-\pi$ and the
$N-\Delta$ splitting can e.g. originate from one-gluon exchange; in
this case, cooling would be expected to lead to a shrinkage of this
mass splitting. If instantons are responsible, the splittings should
not be effected by cooling. It turned out (Chu et al.
\cite{Chu:1994vi}) that most particle masses, in particular also the
$\rho$ and $\pi$ masses, changed very little when the gauge
configurations were cooled. The observation suggests that the mass
splittings are largely due to instantons. However, the mass splitting
between the nucleon and $\Delta$ was reduced by smoothing; thus
one-gluon exchange dynamics seems to be responsible for the $N-\Delta$
splitting. However, Chu et al. \cite{Chu:1994vi} assigned the latter
observation to a technical problem. We are not aware that this
important question was addressed more recently.

\markboth{\sl Meson spectroscopy} {\sl Major experiments}
\clearpage\setcounter{equation}{0}\section{\label{Major experiments}
Major experiments}

The importance of the development of new scientific instruments for
particle physics cannot be overestimated. New experimental methods have
often opened new directions while new demands and new questions
stimulated important technical achievements. In this section
we give a short description of major detectors with which the results
to be presented and discussed in this report were obtained.
The section is not meant to review the achievements in detector
technology; rather it should serve the reader as short reference
information on experimental aspects when physics results are discussed.

 \subsection{\label{Diffractive and charge exchange experiments}
Diffractive and charge exchange experiments }

\subsubsection{\label{The CERN-Munich experiment}
The CERN-Munich experiment}

The CERN-Munich spectrometer at the CERN Proton Synchrotron is described
in detail in \cite{Grayer:1974cr}. The spectrometer was used to study
final states with two charged particles originating from the
interaction of a $\pi^-$ beam with hydrogen and transversely-polarised
butanol targets. The beam particles were momentum analysed and tagged by
a beam spectrometer and Cherenkov counter. A 1 mm thick scintillation
counter behind the target was used to determine the charged particle
multiplicity. The particle trajectories were measured, using
proportional chambers and magnetorestrictive spark chambers, in front
and behind of a magnetic spectrometer with $50$\,cm$\times 150$\,cm
aperture. Scintillation counters and a hodoscope were used in the
trigger.  Particle identification was provided by two multicell
threshold Cherenkov counters. The target $-$ suspended inside a 2.5\,T
magnetic field $-$ was surrounded by a veto box of
tungsten-scintillation shower counters which vetoed events with
$\pi^0$'s and charged recoil multiplicities above one. A 36-element
hodoscope measured the azimuthal angle of the recoil proton to ensure,
for events with $|t|> 0.08$ (GeV/c)$^2$, coplanarity of the reaction.
Most of the data were recorded at 17.2\,GeV/c momentum and a trigger
requiring two charged particles in the final state. The results of the
CERN-Munich collaboration are published in
\cite{Hyams:1973zf,%
Grayer:1974cr,Hyams:1974wr,Estabrooks:1974pb,Hoogland:1974cv,%
Blum:1975rk,Hyams:1975mc,Hoogland:1977kt}.
For later data, a polarised target was used (CERN-Munich-Krakow
collaboration) \cite{Becker:1978ks,Becker:1978kt}.

\subsubsection{\label{The WA3 experiment}
The WA3 experiment}

The experiment WA3 carried out by the ACCMOR collaboration started data
taking at CERN SPS in 1977. In the initial stage it used secondary
beams of 60\,GeV/c and 93\,GeV/c momenta tagged with Cherenkov counters.
The apparatus
comprised a two-magnet forward spectrometer with 80 magnetorestrictive
wire spark chamber planes and two multicellular Cherenkov detectors.
The 50 cm liquid-hydrogen target was surrounded with scintillator-lead
sandwiches. Proportional chambers and scintillation hodoscopes were used
to select the forward charged-particle multiplicity. Veto sandwiches
around the target and in front of magnets were used to suppress events
with secondaries out of aperture. Most results on meson spectroscopy
are devoted to diffractive production of  meson resonances in pion and
Kaon beams~\cite{Daum:1979iv,Daum:1979sx,%
Daum:1979ix,Alper:1980hi,Daum:1980ay,Daum:1981hb}.
In these studies two triggers were used, either an unbiased trigger
asking only for a preset multiplicity in forward direction, or a trigger
with an additional cut on the momentum squared transfer, $|t|>0.16$
(GeV/c)$^2$. In 1981 this detector was upgraded to study inclusive
production of charmed particles.

\subsubsection{\label{The LASS experiment}
The LASS experiment}

The LASS facility was a general purpose spectrometer at SLAC designed
primarily for meson spectroscopy \cite{Aston:1986jb,Ratcliff:1991pf}.
It had $4\pi$ geometrical acceptance with excellent angular and
momentum resolution, full azimuthal symmetry, particle identification,
and high-rate data acquisition capability. LASS contained two magnets
filled with tracking detectors. The first magnet was a superconducting
solenoid with a 2.24\,T field parallel to the beam direction. This
magnet was followed by a 3\,Tm dipole magnet with a vertical field.
The solenoid was used in measurements of interaction products having
large production angles and relatively low momenta. High-energy
secondaries with small polar angle passed through the dipole for
measurements of their tracks. Particle identification was provided by a
Cherenkov counter, a time-of-flight hodoscope which filled the exit
aperture of the solenoid, and by a Cherenkov counter at the exit of the
dipole spectrometer. In addition, $dE/dx$ ionisation energy loss was
measured in a cylindrical drift chamber surrounding the liquid $H_2$
target to separate wide-angle protons from pions at momenta below 600
MeV/c. LASS was situated in an RF-separated beam line delivering a
5-16\,GeV/c Kaon beam of high purity. Cherenkov counters were used to tag
Kaons. The  poor duty factor of the SLAC beam limited the useful flux
to $\sim 250\div500$ particles per second. The trigger accepted all
interactions in the target except all-neutral final states. The total
LASS Kaon program contained 135 million events collected in 1982. The
results were published in \cite{Aston:1986jb,Aston:1986rm,Aston:1987ir,%
Aston:1987ak,Aston:1987am,Aston:1993qc}.

\subsubsection{\label{The VES experiment}
The VES experiment}

\begin{figure}[pb]
\bc
\includegraphics[width=0.6\textwidth,height=0.5\textwidth]{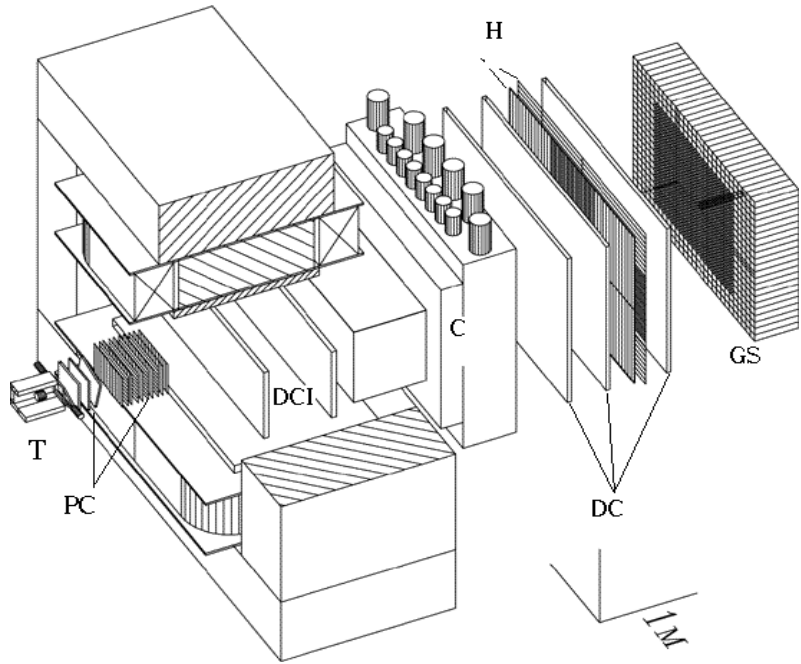}
\ec
\caption{\label{fig:VES}
Experiment VES~\protect\cite{Bityukov:1991zk}: T - target with veto counters
and lead-scintillator sandwiches, PC - proportional chambers, DCI - inner
drift chambers, DC - drift chambers, C - Cherenkov counter, H -
hodoscope, GS - gamma spectrometer.} \end{figure} The detector VES at
Protvino was designed to study multiparticle decays of mesonic
resonances produced in $ \pi^- N$ and $  K^- N$ interactions at beam
momenta of $20\div40$\,GeV/c produced in the 70\,GeV/c proton
synchrotron (Fig. \ref{fig:VES}). The beam particles were tagged with
three threshold Cherenkov counters and their coordinates measured at
the entrance to the spectrometer with proportional chambers. A
beryllium target of 4\,cm in diameter and 4\,cm in length was used. On
both sides, a 3\,cm aluminium shield absorbed low momentum charged
particles; veto scintillation counters were used in the trigger,
lead-scintillator sandwiches to tag events in which $\pi^0$ mesons were
emitted into the backward hemisphere in the centre of mass of the
reaction. In forward direction the required multiplicity was selected
with scintillation counters near the target and two coordinate
scintillation hodoscopes $H$ behind the wide aperture ($100$\,cm
$\times 150$\,cm) magnet. Tracking was provided by 16 planes of
proportional chambers ($PC$) in front of the magnet, four planes of
minidrift chambers ($DCI$) inside the magnet, and nine planes of drift
chambers behind the magnet. Particles were identified with multicell
Cherenkov counters ($C$) which discriminate pions and Kaons in the
$4.4\div 18$\,GeV/c  momentum range. Photons were registered in a
multicell  lead glass spectrometer (in total 1200 cells). Veto
sandwiches in front of the magnet and behind it made the spectrometer
nearly completely hermetical for charged particles and photons.
Most data were collected with a trigger on two or more charged
particles in forward direction and a veto on high-momentum charged
particles in side direction. The results on meson spectroscopy are
published in
\cite{Bityukov:1991ri,Bityukov:1991aw,Berdnikov:1992jd,Beladidze:1993km,%
Amelin:1994ii,Berdnikov:1994kc,Kachaev:1994hs,Amelin:1995fg,Amelin:1995gu,%
Amelin:1995gt,Amelin:1999gk,Amelin:2000nm,Khokhlov:2000tk,Dorofeev:2001xu,%
Kachaev:2001jj,Amelin:2004ns,Nikolaenko:2004tz,Amelin:2005ry,Amelin:2006wg}.

\subsubsection{\label{Experiment E852 at BNL}
Experiment E852 at BNL}

\begin{figure}[pb]
\begin{center}
\includegraphics[width=0.6\textwidth]{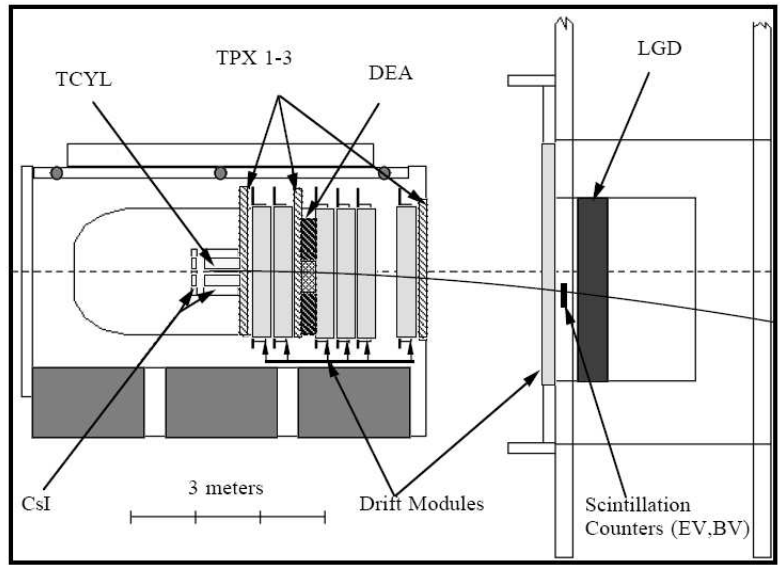}
\end{center}
\caption{\label{fig:bnl}
Experiment E~852 at BNL \protect\cite{Teige:1996fi}:
TCYL: four-layer cylindrical drift chamber,
$CsI$: caesium iodide array,
DEA: lead scintillator sandwich photon veto counter,
TPX1-3: proportional wire chambers,
LGD: lead glass detector.}
\end{figure}
The Multi-Particle spectrometer (MPS) shown in Fig. \ref{fig:bnl} was
located at the Brookhaven Alternating Gradient Synchrotron.  All data
were obtained in a 18\,GeV/c $\pi^-$ beam. The basic element of this
detector was a 5\,m long wide-aperture magnet. The apparatus
\cite{Teige:1996fi} consisted of 3 regions: target,
tracking, and downstream region. The target region was located in the
middle of the magnet and contained a liquid hydrogen target, multilayer
wire-chambers used to trigger on recoil protons, and a 198-element
cylindrical thallium-doped caesium iodide array ($CsI$) capable of
rejecting events with wide-angle photons. Main components of the
tracking system were located at the downstream end of the magnet where
3 proportional wire chambers and 6 drift chamber modules with 7 planes
each were placed. Interspersed among these were three proportional wire
chambers for trigger purposes, a window-frame lead-scintillator photon
veto counter to ensure photon hermeticity, a scintillation counter to
veto forward charged tracks for neutral triggers, and scintillation
counters to identify charged particles entering the photon veto
counter. The downstream region contained a 3045 element lead-glass
calorimeter. For some experiments, multicell Cherenkov counters were
used for particle identification of secondaries. The experiment was
mainly devoted to the study of mesonic states with exotic quantum
numbers. The results can be found in
\cite{Teige:1996fi,Thompson:1997bs,Adams:1998ff,Chung:1999we,Adams:2000zg,%
Manak:2000px,Gunter:2000am,Eugenio:2000rf,Ivanov:2001rv,Adams:2001sk,%
Chung:2002pu,Nozar:2002br,Dzierba:2003fw,Kuhn:2004en,Lu:2004yn,Dzierba:2005jg}.


\subsection{\label{Central production experiments}
Central production experiments}

\subsubsection{\label{The WA76 and WA91 experiments}
The WA76 and WA91 experiments}

The WA76 experiment was designed to study mesons produced centrally
at the largest beam momenta available. First data were taken at
85\,GeV/c; the majority at 300\,GeV/c and some data at 450\,GeV/c.
The experiment was based on the CERN $\Omega$ spectrometer. The hadron
(pions, protons) beam impinged on a 60\,cm long H$_2$ target; two
Cherenkov counters identified the beam particle. Excitation of the
target proton was vetoed in horizontal scintillation slabs surrounding
the target. One (and only one) fast particle had to traverse the
forward particle region, equipped with multi-wire proportional
chambers, but should not hit veto counters defining non-interacting
beam particles. At least two particles were required in the two drift
chambers near the target. Cherenkov counters were used to identify
pions and Kaons.

 The experiment was devoted to the study of meson resonances and the
search for glueballs. The results can be found in
\cite{Armstrong:1984rn,Armstrong:1986ci,%
Armstrong:1986cm,Armstrong:1989hy,%
Armstrong:1989zr,Armstrong:1989rv,%
Armstrong:1989jk,Armstrong:1989es,Armstrong:1990qw,%
Armstrong:1991ch,Armstrong:1991rg,Armstrong:1992uu,French:1999tm}.
A few related results were published by the WA91 experiment
\cite{Abatzis:1994ym,Antinori:1995wz,Barberis:1996tu}
using a similar setup.

\subsubsection{\label{The GAMS experiment}
The GAMS experiment}

\begin{figure*}[pb]
\begin{center}
\includegraphics[angle=-91,width=.7\textwidth]{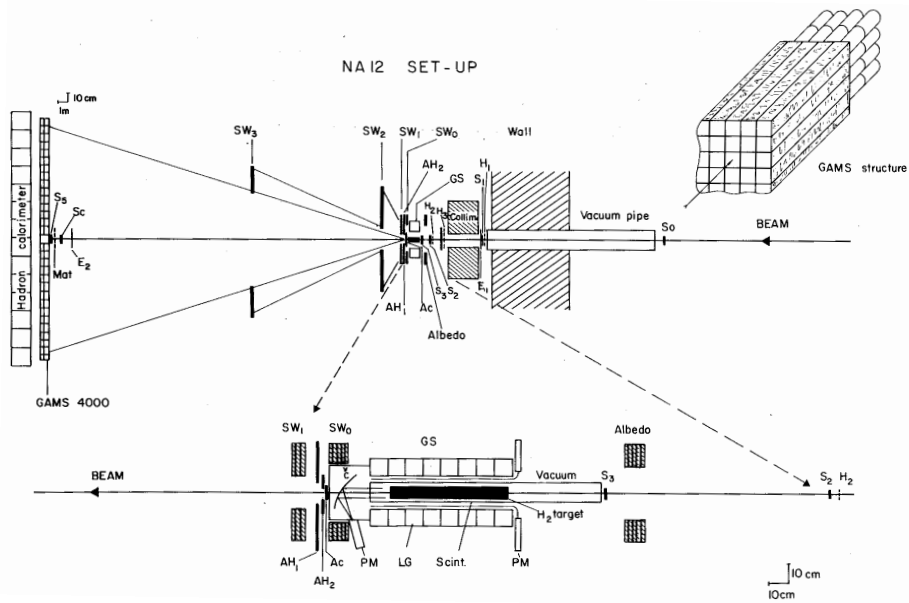}
\end{center}
\caption{\label{exp:gams}
Schematic view of the GAMS-4000 experiment at CERN.
$S$:  beam scintillation counters; $H$: beam hodoscope; $SW$ and albedo:
lead-scintillator sandwich counters; $AC, AH$: scintillation
counters~\protect\cite{Boutemeur:1989gr}.
}
\end{figure*}
The GAMS experiment was designed to detect neutral mesons decaying into
photons in high-energy reactions. Its main component was a wall of lead
glass cells. 2000 cells were combined for experiments at IHEP,
Protvino, in a 40\,GeV/c pion beam;  GAMS-4000 was built for
experiments at the CERN SPS. The latter setup is reproduced in
Fig.~\ref{exp:gams}.
The pion beam traversed the 50\,cm long liquid H$_2$ target and was
detected in scintillation counters and by Cherenkov light produced in
H$_2$. The opening angle of the light cone was used to determine the
longitudinal coordinate of the interaction point. A range of scattering
angles was allowed in forward scintillators. Target fragmentation or
excitation of the target proton was suppressed by scintillators and lead
glass counters surrounding the target.

The lead glass wall consists of 4092 cells and covers an area of
5\,m$^2$. A 100\,GeV photon was measured with a precision of about
1.5\,GeV in energy; the impact point was localised to 1\,mm$^2$.
Results from GAMS can be found in
\cite{Alde:1985kp,Alde:1986nx,Alde:1986xr,Alde:1994kf,Alde:1991ns,%
Alde:1994jj,Alde:1994jm,Alde:1997ri,Alde:1997vq,Alde:1998mc,%
Alde:1999gh,Bellazzini:1999sj}.

\subsubsection{\label{The WA102 experiment}
WA102 experiment}

The WA102 experiment was built as combined effort of the former WA76
and GAMS collaborations. The  WA76  served as charged-particle tracker,
the GAMS-4000 detector was installed in the forward region to detect
photons. This combination provided the possibility to study events with
charged and neutral particles and thus extended the range of the WA76
and GAMS experiments considerably. The results of the experiment were
reported in a series of letter publications \cite{Barberis:1996iq,%
Barberis:1997ve,Barberis:1997vf,Barberis:1998ax,Barberis:1998by,%
Barberis:1998bq,Barberis:1998in,Barberis:1998tv,Barberis:1998sr,%
Barberis:1999am,Barberis:1999an,Barberis:1999ap,Barberis:1999be,%
Barberis:1999cq,Barberis:1999id,Barberis:1999wn,Barberis:1999zh,%
Barberis:2000cx,Barberis:2000cd,Barberis:2000em,Barberis:2000kc,%
Barberis:2000vp,Barberis:2001bs}.

\subsection{\label{Antiproton-proton annihilation}
Antiproton-proton annihilation}

\subsubsection{\label{The Asterix experiment}
The Asterix experiment}

The Asterix experiment studied \ppb\ annihilation from
$S$- and $P$-wave orbitals by stopping antiprotons in H$_2$ gas at room
temperature and pressure and observing the coincident X-ray
spectrum. The detector consisted of
a gas target (45\,cm length and 14\,cm in diameter), a X-ray drift
chamber and  seven multi-wire proportional chambers, partly with
cathode readout to provide spatial resolution along the wires.
Two end-cap detectors with three wire planes and cathode readout
on both sides gave large solid-angle coverage. The assembly was
situated in a homogeneous magnetic field of 0.8\,T.
The energy resolution of the detector for 8 keV X-rays was
about 20\%. The momentum resolution  for $\pbp\to\pip\pim$ events
at 928\mevc\ was 3\%. Pions and Kaons could be separated up to 400\mevc.
The detector is fully described in \cite{Ahmad:1990gx}. Physics results
related to meson spectroscopy were published in
\cite{Ahmad:1984bp,Duch:1989sx,May:1989sw,May:1990ju,May:1990jv,%
Klempt:1990vc,Reifenroether:1991ik}.

\subsubsection{\label{The Crystal Barrel experiment}
The Crystal Barrel experiment}

\begin{figure*}
\begin{center}
\includegraphics[width=0.8\textwidth]{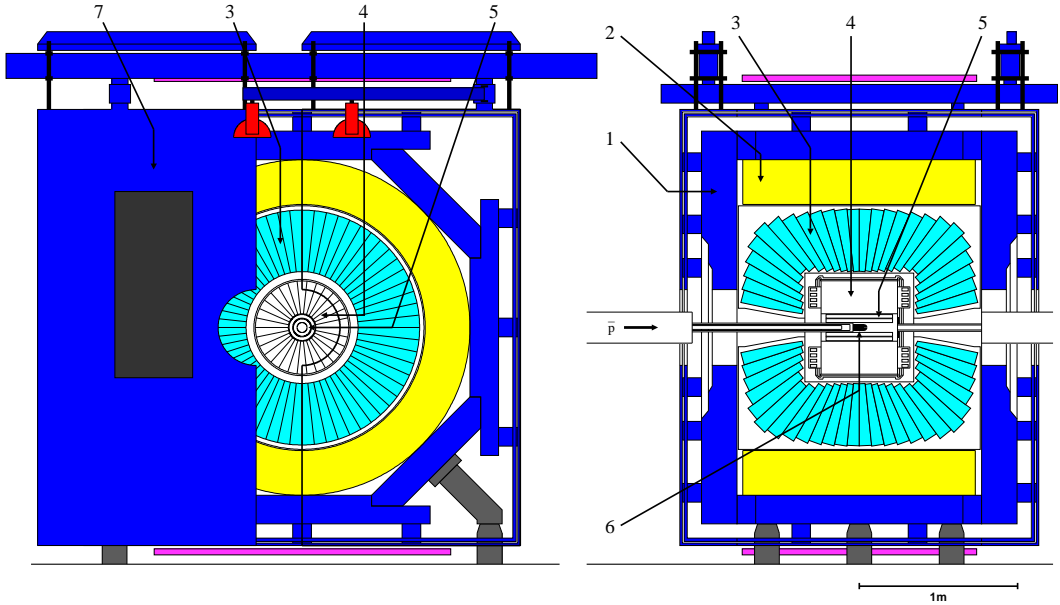}
\end{center}
\caption{\label{exp:fig:CBAR}%
Overall layout of the Crystal Barrel detector showing
(1) magnet yoke, (2) magnet coils, (3) $CsI$ barrel,
(4) jet drift chamber, (5) proportional chamber, (6) liquid hydrogen
target, (7) one half of endplate. Left - longitudinal cross section;
Right- transverse view.}
\end{figure*}

The Crystal Barrel spectrometer is shown in Fig.~\ref{exp:fig:CBAR}.  A
detailed description of the apparatus, as used for early data-taking
(1989 onwards), is given in \cite{Aker:1992ny}. In 1995, a microstrip
vertex detector surrounding the target was added
\cite{Doser:1998ge}. Target and vertex detector were
surrounded by a cylindrical jet drift chamber having 30 sectors with
each sector having 23 sense wires. The coordinate along the wire was
determined by charge division. A momentum resolution for pions of less
than 2\% at 200\mevc\ was obtained, rising to 4.2\% at 1\gevc\ for
those tracks traversing all layers of the JDC. The JDC also provided
$\pi$/K separation below 500\mevc\ by ionisation sampling. The whole
detector was situated in a 1.5\,T solenoidal magnet with the incident
antiproton beam direction along its axis.

Physics results related to meson spectroscopy are published in
\cite{Amsler:1992rx,Amsler:1994ah,Amsler:1994ey,Amsler:1994rv,Amsler:1994pz,%
Amsler:1995bf,Amsler:1995bz,Amsler:1995wz,Adomeit:1996nr,Abele:1996fr,%
Abele:1996nn,Abele:1997qy,Abele:1997dz,Abele:1997wg,Abele:1997vu,%
Abele:1997vv,Abele:1998gn,Abele:1998kv,Abele:1998qd,Baker:1999ac,Abele:1999tf,%
Abele:1999fw,Abele:1999cjb,%
Abele:1999pf,Abele:2000qq,Abele:2001js,Amsler:2001fh,Abele:2001pv,%
Amsler:2002qq,Baker:2003jh,Amsler:2004rd,Amsler:2004kn,Dunnweber:2004vc}.
Most Crystal Barrel data on \pbp\ annihilation in flight were
analyzed by the QMC-Rutherford-Gatchina group \cite{Anisovich:1999af,%
Anisovich:1999ag,Anisovich:1999fe,Anisovich:1999fd,Anisovich:1999jx,%
Anisovich:1999jw,Anisovich:1999pt,Anisovich:1999xm,Anisovich:2000ae,%
Anisovich:2000af,Anisovich:2000bs,Anisovich:2000ix,Anisovich:2000jx,%
Anisovich:2000kx,Anisovich:2000mv,Anisovich:2000us,Anisovich:2000ut,%
Anisovich:2001cr,Anisovich:2001hj,Anisovich:2001jf,Anisovich:2001pn,%
Anisovich:2001pp,Anisovich:2001vt,Anisovich:2002su,Anisovich:2002sv}.
 Meson spectroscopy with the Crystal Barrel is reviewed for
\NNb\ annihilation at rest by Amsler \cite{Amsler:1998up} and Amsler
and Tornqvist \cite{Amsler:2004ps}, and in flight by Bugg
\cite{Bugg:2004xu}.

\subsubsection{\label{The Obelix experiment}
The Obelix experiment}

The Obelix spectrometer is based on the open-axial field magnet which
had previously been used for experiments at the ISR. The magnet
provides a field of 0.5\,T in an open volume of about 3\,m$^3$. A full
description of the detector can be found in \cite{Adamo:1992bb}. \par
The Obelix detector consists of an imaging vertex detector with
three-dimensional readout for charged tracks and X-ray detection,
a jet drift chamber (JDC) for tracking and particle identification
by dE/dx measurement with 3280 wires and flash-analog-to-digital
readout, a system of two coaxial barrels of plastic scintillators
consisting of 30 (84) slabs positioned at a distance of 18\,cm
(136\,cm) from the beam axis for time-of-flight (TOF) measurements, and
a high-angular-resolution gamma detector (HARGD) \cite{Affatato:1993nim}.
The momentum resolution for monoenergetic pions (with 928\mevc) from
the reaction $\pbp\to\pi^+\pi^-$ was determined to  3.5\%, \piz\ were
reconstructed with a mass resolution of $\sigma_{\piz} = 10$\,MeV/c$^2$
and a momentum-dependent efficiency of 15 to 25\%. The detector system
was used with a liquid H$_2$ (D$_2$) target, a gaseous H$_2$ target at
room temperature and pressure, and a target at low pressures (down to
30 mbar). The wide range of target densities provided detailed
information about the influence of the atomic cascade on the
annihilation process. For part of the time, a $\nbar$ beam was produced
by charge exchange in a liquid H$_2$ target (positioned 2\,m upstream
of the centre of the main detector).  The intensity of the collimated
beam was about  $40\,\bar n/10^6\,\bar p$ of which about 30\% interact
in the central target. The $\bar n$ beam intensity was monitored by a
downstream $\bar n$ detector.

The Obelix Collaboration had a broad program of experiments covering
atomic, nuclear and particle physics \cite{Adamo:1992bb}.  Main
results related to meson spectroscopy can be found in
\cite{Bertin:1995fx,Bertin:1997kh,Bertin:1997vf,%
Bertin:1997zu,Bertin:1998sb,Bertin:1998hu,Cicalo:1999sn,%
Nichitiu:2002cj,Bargiotti:2003bv,Bargiotti:2003ev,Salvini:2004gz}.

\subsubsection{\label{Experiment E835 at FNAL}
Experiment E760/E835 at FNAL}

\begin{figure}[!ht]
\bc
\includegraphics[width=0.8\textwidth]{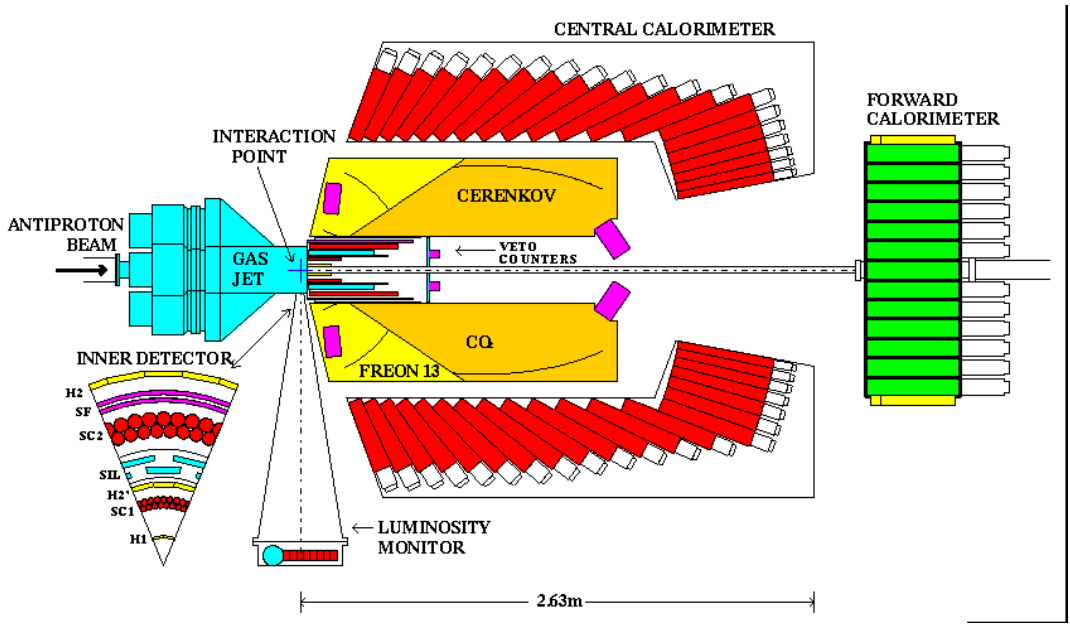}
\ec
\caption{\label{fnal}
Experiment E760/E835 at FNAL~\protect\cite{Ambrogiani:2001jw}.}
\end{figure}

Antiproton-proton annihilation in flight was studied
by the E760/E835 experiment at Fermilab~\cite{Ambrogiani:2001jw}
(see figure~\ref{fnal}).
Antiprotons were produced by bombarding a target with high-energy
protons. The antiprotons were cooled in phase space in a storage ring
and then accelerated to study $\bar pp$ collisions at extremely high
energies. A fraction of the antiprotons were used for medium-energy
physics: $8\cdot 10^{11}$ antiprotons circulated in the Fermilab
accumulator ring with a frequency $f_{rev}=0.63$\,MHz. At each
revolution, antiprotons passed through a hydrogen gas jet target, with
$\rho_{jet}= 3\cdot 10^{14}$H$_2$/cm$^3$, which results in a luminosity
$ L=N_{\bar p}f_{rev}\rho_{jet}$ of $2\cdot 10^{31}$/cm$^2$s. The
energy of the antiproton beam, and thus the invariant mass of the $\bar
pp$ system, could be tuned very precisely according to $\sqrt{s} =
m_p\cdot\sqrt{2(1+E_{\bar p}/m_p)}$.  The observed rate $R=2\cdot
10^{6}$/s due to the hadronic background is related to the luminosity
by $R=\sigma\cdot L$.

Most data taken at this experiment were devoted to a study of the
charmonium system. We give reference to a few of them
\cite{Ambrogiani:2001wg,Bagnasco:2002si,Ambrogiani:2003md,%
Andreotti:2004ru,Andreotti:2005vu,Andreotti:2005zr}. E760 also produced results on
light meson spectroscopy \cite{Armstrong:1993ey,Armstrong:1993fh,Uman:2006xb}.

\subsection{\label{Electron-positron annihilation experiments at
Phi and cbarc factories}
Electron-positron annihilation experiments at \protect{$\Phi$} and
\protect{$c\bar c$} factories}

\subsubsection{\label{VEPP}
VEPP}

The collider VEPP-2M was constructed at BINP, Novosibirsk, in 1974 and
stopped in 2000. Its main parameters were:

\vspace*{4mm}
\begin{tabular}{ccc}
 2E = 0.4-1.4\,GeV&\qquad & $L_{max} = 4 \times 10^{30}$\,cm$^{-2}$s$^{-1}$\\
total integrated luminosity $\approx 80$\,pb$^{-1}$&&$10^9$ recorded
events. \\ \end{tabular}

 For 25 years this facility was one of the main sources of information
on $e^+e^-$ annihilation at relatively low energies. Precision data in
this region are important not only for hadron spectroscopy but for
precision electroweak physics as well. Indeed, this region gives the
main hadronic contribution to the muon anomalous magnetic moment
\cite{Bennett:2006fi}.

One of the well known achievements of this laboratory is the method of
high precision mass measurements by resonant depolarisation. It was
proposed and developed at BINP \cite{Bukin:1975db,Derbenev:1980gp} and
widely used by BINP and by other laboratories. Electrons and positrons
in storage rings can become polarised due to emission of synchrotron
radiation. Spins of polarised electrons (positrons) precess around the
vertical magnetic field with frequency $\Omega$ which is related to the
particle energy $E$ and the revolution frequency $\omega$:
\begin{equation}
\Omega=\omega(1+\gamma\cdot\mu_a/\mu_0)
\vspace*{-3mm}
\end{equation}
where $\gamma=E/m_e$, $m_e$ is the electron mass, $\mu_a$ and $\mu_0$
are anomalous and normal parts of the electron magnetic moment. The
precession frequency can be measured using resonant depolarisation. The
polarised beam is exposed to an external electromagnetic field with the
frequency $\Omega_D$:
\begin{equation}
\Omega\pm\Omega_D=\omega\cdot n
\vspace*{-3mm}
\end{equation}
with any integer $n$. The frequency $\Omega_D$ is scanned. At resonance,
the polarisation disappears. This allows them to measure the value of
$E=\gamma\times m_e$ with very high precision. The relative mass
precision achieved with this method is $3\cdot10^{-5}$ for
$\phi(1020)$, $ 5\cdot 10^{-6}$ for J/$\psi(1S)$ and $ 2\cdot10^{-5}$
for $\Upsilon(1S)$.

\begin{figure}[pt]
\bc
\includegraphics[width=0.6\textwidth]{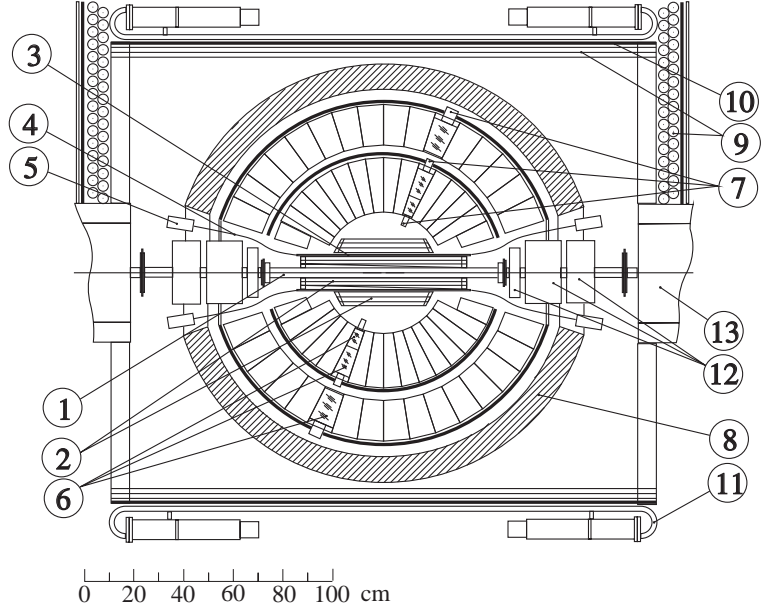}
\ec
\caption{\label{detectors:SND}
Experiment SND~\protect\cite{Achasov:1999ge}. Side view: 1 - beam pipe, 2 - drift chambers,
3 - scintillation counters, 4 - fiber lightguides, 5 - PMTs, 6 -
NaI(Tl) counters, 7 - vacuum phototriodes, 8 - iron absorber, 9 -
streamer tubes, 10 - iron plates, 11 - scintillation counters, 12 -
magnetic lenses, 13 - bending magnets.} \end{figure}

First results at VEPP-2M were obtained with the nonmagnetic detectors
OLYA \cite{Ivanov:1981wf} and ND \cite{Golubev:1985vd}. Later two more
advanced detectors SND \cite{Aulchenko:1998be,Abramov:2001gn} and CMD
\cite{Akhmetshin:1995vz} were commissioned. The main part of the SND
(Spherical Neutral Detector, Fig.~\ref{detectors:SND}) was a three-layer
spherical highly granulated NaI(Tl) calorimeter, consisting of 1632
individual crystals. The calorimeter thickness was 13.5 radiation
lengths, its total mass 3.6 tons. Its energy resolution for
photons as a function of their energy was determined to
$\sigma_E/E=4.2\% /E_{[\rm GeV]}^{\frac{1}{4}}$. The inner part of the
detector was a cylindrical drift chamber system for tracking of charged
particles. The solid angle coverage for the inner chamber was $96\% $
of $4 \pi$. Outside of the calorimeter muon veto detectors were
installed consisting of streamer tubes and scintillation counters.
Results of this experiment related to meson spectroscopy are published
in \cite{Achasov:1998cc,Aulchenko:1998xy,Aulchenko:1999dz,%
Achasov:1999gj,Achasov:1999jc,Achasov:1999ib,Achasov:1999ge,%
Achasov:2000ym,Achasov:2000ku,Achasov:2001hb,Achasov:2001ni,%
Achasov:2001xi,Achasov:2002hs,Achasov:2002ud,Achasov:2003bv,%
Achasov:2003ed,Achasov:2003ir,Aulchenko:2003wv,Achasov:2006bv,%
Achasov:2006dv,Achasov:2006vp}.

The detector CMD-2 (Cryogenic Magnetic Detector) was commissioned at
BINP in 1990. The detector  consisted of a cylindrical drift chamber
surrounding the collision region, with $250 \mu$ resolution in the
plane transverse to the beam axis, and a double-layer multiwire
cylindrical proportional chamber. Both chambers were placed inside of
a very thin (0.38 radiation lengths) superconducting solenoid with a
field of $B=1.5$\,T. The barrel $CsI$ calorimeter $-$ with a thickness
of $8.1 X_0$ $-$ and a muon range detector were placed outside of the
solenoid. The energy resolution for photons in the energy range from
100 to 700 MeV was about $9\%$, the angular resolution of the order of
$0.02$ radians. End-cap BGO calorimeters of $13.4 X_0$ thickness
were placed inside the solenoid. The detector was nearly fully hermetic
for photons. Results of this experiment related to meson spectroscopy
are published in \cite{Akhmetshin:1997tz,Akhmetshin:1999zv,%
Akhmetshin:1999dh,Akhmetshin:1999di,Akhmetshin:1999ty,%
Akhmetshin:1999wd,Akhmetshin:2000ca,Akhmetshin:2000hp,
Akhmetshin:2001hm,Akhmetshin:2001ig,Akhmetshin:2000it,%
Akhmetshin:2000wv,Akhmetshin:2003ag,Akhmetshin:2003zn,%
Akhmetshin:2004dy,Achasov:2004zf,Akhmetshin:2005vy,Akhmetshin:2006bx,%
Aulchenko:2006na,Akhmetshin:2006sc,Akhmetshin:2006wh}.
In the near future, both detectors -- SND and CMD-2 -- will work at the
new high-luminosity $e^+ e^-$ collider VEPP2000 presently being
constructed at BINP.

Recently one more modern facility, KEDR, started data taking at BINP.
It occupies one interaction region at the VEPP-4M - $e^+ e^-$ collider.
VEPP-4M will reach a maximum beam energy of $E=6$\,GeV and a luminosity
of $L=10^{31}$\,cm$^{-2}$\,s$^{-1}$. The main goal of the experiments
at this facility are very high precision measurements of the
$\tau$-meson \cite{Anashin:2006xp} and of the $\psi$
\cite{Aulchenko:2003qq} and $\Upsilon$ families as well as two-photon
physics using a dedicated zero-angle spectrometer for scattered
electrons and positrons. The KEDR detector consists of a vertex
detector, drift chamber, time of flight system of scintillation
counters, a particle identification system based on aerogel Cherenkov
counters, a calorimeter (with liquid krypton in the barrel part and
$CsI$ crystals in the end caps), and a muon tube system inside and
outside of the magnet yoke.

\subsubsection{\label{Daphne}
KLOE at Daphne}

DAPHNE, the Frascati $\phi$ factory, is an $e^+ e^-$ collider
working at $2E\approx m_{\phi}\approx 1.02$\,GeV with a design
luminosity of $5\cdot10^{32}\rm cm^{-2} s^{-1}$. $\phi$ mesons are
produced essentially at rest with a cross section of $\approx 3.2 \mu
b$. The main decay modes are $  K^+ K^-$ and $  K^0_S K^0_L$ pairs so
that pure and monochromatic $  K^0_S,K^0_L,K^+$ and $  K^-$ beams can
be obtained. A survey of tests of CP and CPT invariance is given in
\cite{Ambrosino:2006ek}. This facility is used also to study light quark
spectroscopy.

The KLOE detector consists of a drift chamber\cite{Adinolfi:2002uk}
surrounded by an electromagnetic calorimeter \cite{Adinolfi:2002zx} and
a superconducting solenoid providing a 0.52 T magnetic field. The drift
chamber is of cylindrical shape, 4\,m diameter and 3.3\,m in length.
This unusually big volume is needed to detect $  K^0_L$ decays. The
momentum resolution is $\sigma_p/p \le 0.4\%$. The electromagnetic
calorimeter is a lead-scintillating-fiber calorimeter consisting of a
barrel and two endcaps covering $98\%$ of the solid angle. The
energy resolution is $\sigma_E/E=5.7\%/\sqrt E$ (E in GeV). This
detector has taken data for two years ($2001\div 2002$) with a maximum
luminosity of up to $7.5\cdot 10^{32}$\,cm$^{-2}$s$^{-1}$. An integrated
luminosity of $450$\,pb$^{-1}$, equivalent to $1.4 \cdot10^9~~\phi$
decays, was achieved. In 2004, KLOE resumed data taking with an upgraded
machine. The KLOE results related to meson spectroscopy are published
in \cite{Aloisio:2002ac,Aloisio:2002bs,Aloisio:2002bt,Aloisio:2002vm,%
Aloisio:2003dw,Aloisio:2004bu,Ambrosino:2005wk,Ambrosino:2006gk,unknown:2006hb}.

\subsubsection{\label{Argus, Mark3 and DM2}
Argus, Mark3 and DM2}

Significant progress in meson spectroscopy came from the Argus (DESY),
Mark3 (SLAC) and DM2 (Orsay) detectors. They stopped operation more
than 20 years ago, and often their contributions are now replaced by
new data with higher statistics. Some results are important still
today; these will be discussed in the appropriate sections. Detectors,
their performance and their achievements have been reviewed by
K\"opke and Wermes \cite{Kopke:1988cs}. A report of the physics
achievements of the ARGUS collaboration can be found in
\cite{Albrecht:1996gr}.

\subsubsection{\label{BES}
BES}

\begin{figure}[!ht]
\bc
\includegraphics[width=0.4\textwidth]{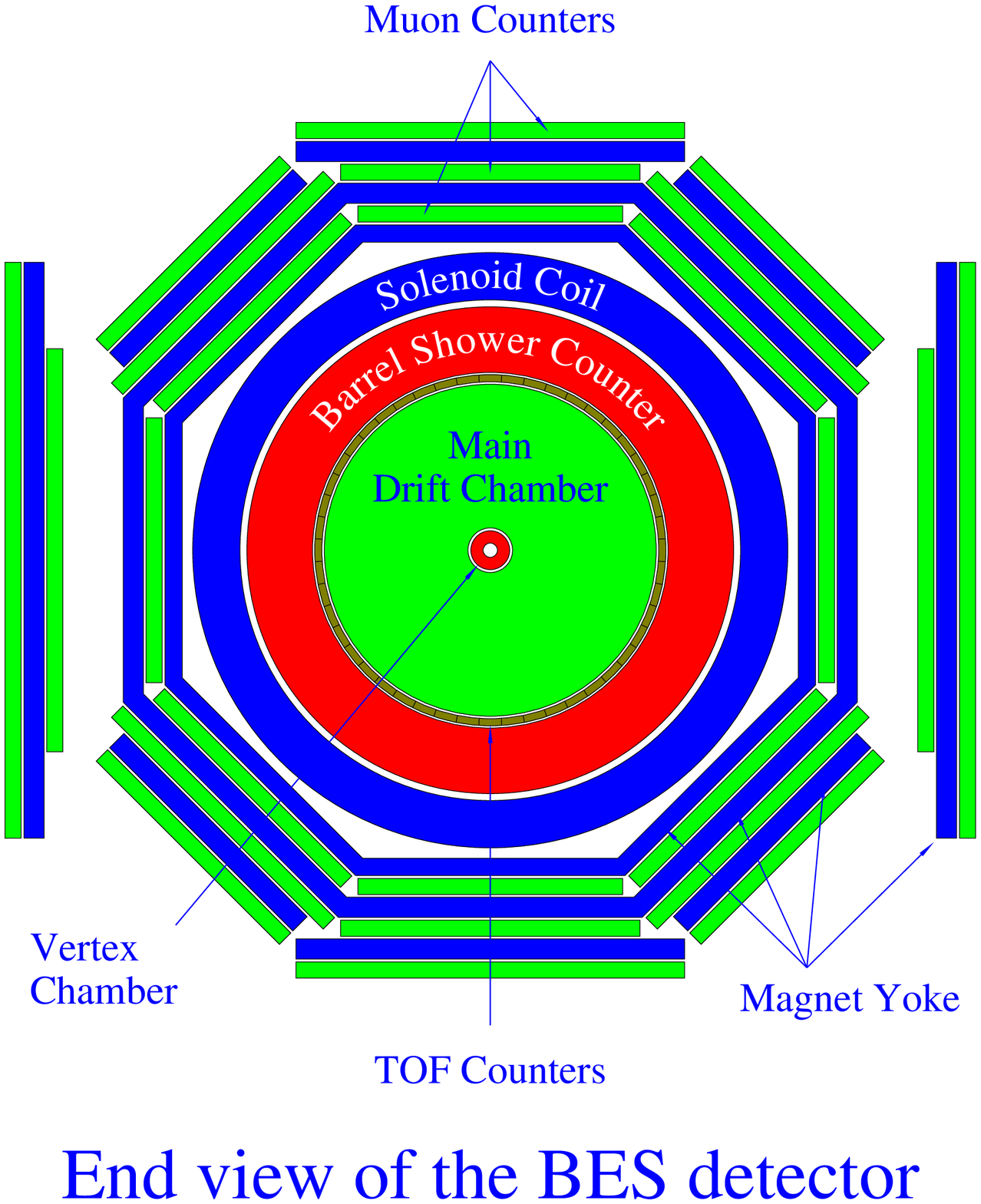}
\ec
\caption{\label{fig:exp:BES}
Experiment BES~\protect\cite{Bai:2001dw}.}
\end{figure}

The detector BES is installed at the $e^+ e^-$ collider BEPC at IHEP,
Peking. The collider has a maximum energy of $2E=4.4$\,GeV and a
luminosity reaching $10^{31}\rm cm^{-2} s^{-1}$. It started to work in
1989. A very significant upgrade is planned for 2007 aiming at
$L=10^{33}\rm cm^{_2} s^{-1}$ luminosity. Up to now this facility
provided record samples of J/$\psi$ ($64\cdot10^6$) and $\psi$'
($18\cdot10^6$). The detector BES is based on a conventional solenoidal
magnet. BES II is an upgraded version of the initial BES I detector. The
layout is shown in Fig. \ref{fig:exp:BES}. Inner tracking is provided by
a 12-layer vertex  chamber and a main drift chamber. The vertex
detector is used in the trigger.  Particle identification relies on
$dE/dx$ measurements in the drift chamber (with $\approx 8\%$
resolution) and an array of 48 time of flight counters ($\approx 200$ps
time resolution). Neutral particle detection is provided by a lead-gas
shower detector with $\sigma_E/E=28\% /\sqrt E$ (E in GeV). The iron
return yoke is instrumented with counters for muon identification.
There is a large number of papers giving precise branching ratios for
J/$\psi$ and $\psi^{\prime}$ decays. The branching ratios and the
comparison of J/$\psi$ and $\psi^{\prime}$ decays is an interesting
subject; here we just mention the 12\% rule. Results on meson
spectroscopy are published in
\cite{Bai:1990hs,Bai:1998eg,Bai:1998gh,Bai:1999mj,Bai:1999mk,Bai:1999mm,Bai:1999mq,Bai:1999tg,Xu:2000sf,%
Bai:2000sr,Bai:2000ss,Bai:2001ha,Bai:2003sw,Bai:2003ww,Bai:2003vf,%
Bai:2004qj,Bai:2004cg,Bai:2004av,Ablikim:2004mv,Ablikim:2004dj,%
Ablikim:2004qn,Ablikim:2004cg,Ablikim:2004st,Ablikim:2004wn,%
Ablikim:2004qe,Ablikim:2005um,Ablikim:2005cd,Ablikim:2005ni,%
Ablikim:2005jy,Ablikim:2005ju,Yang:2005ej,Li:2005ch,Ablikim:2005yd,Ablikim:2006bz,Ablikim:2006ca,%
Ablikim:2006db,Ablikim:2006dw,Jin:2006hd,Ablikim:2006hp,Ablikim:2006md,Ablikim:2007gd}.

\subsection{\label{Heavy-quark spectroscopy}
Heavy-quark spectroscopy}

\subsubsection{\label{Experiments at FNAL}
Experiments at FNAL}

Charmed quarks are copiously produced in high energy hadron beams
owing to the high cross section, reaching a few hundreds $\mu b$ at
beam momenta of a few hundred GeV/c. As a rule, these events are singled
out with precision microvertex detectors. A number of experiments were
performed to study heavy quark physics in hadron and photon
beams. Early studies at CERN \cite{Alvarez:1990sy,Adamovich:1996ih,%
Adamovich:1999df} had only exploratory character and this direction of
physics was given up. At Fermilab, important results were achieved
by E687 \cite{Frabetti:1990au} upgraded to the FOCUS experiment
\cite{Arena:1999bc}, by E691 \cite{Anjos:1988uf}, upgraded to E791, and
the SELEX experiment. Here we describe the SELEX experiment as an
example of this approach. Experiment E760/E835 has been discussed in
section \ref{Experiment E835 at FNAL}.

The SELEX experiment used the Fermilab charged hyperon beam at 600
GeV/c to produce particles in a set of thin target foils of Cu or
diamond. The negative beam was composed of about 50\% $\Sigma^-$
hyperons and 50\% $\pi^-$. The positive beam had $90\%$ protons. The
main goal of the experiment was the study of  particles carrying charm
and strangeness. The beam particles were unambiguously identified with
a transition radiation detector. The trajectories of charged secondaries
were reconstructed in a three-stage magnetic spectrometer providing
momentum resolution of $\sigma_P/P < 1\%$ for a 150\,GeV/c proton. A
10\,m long Ring-Imaging CHerenkov detector separated $\pi$ from
$K$ up to 165\,GeV/c. Three-stage lead glass spectrometers were
used for electromagnetic  calorimetry. A very high precision silicon
vertex detector provided an average proper-time resolution of 20\,fs
for charm decays. A scintillation trigger was used to select events
with the required topology. Charm candidates were selected using an
online secondary vertex algorithm.

A large variety of results was obtained. Those relevant in the context
of this paper were published in
\cite{Aitala:2000xt,Aitala:2000xu,Aitala:2002kr,%
Aitala:2005yh,Frabetti:1994di,Frabetti:1995sg,Frabetti:1997sx,%
Frabetti:2001ah,Frabetti:2003pw,Link:2003bd,Link:2003gb,%
Malvezzi:2003jp,Link:2004mx,Link:2004wx,Link:2004zc,Benussi:2005db,%
Link:2005ge,Smith:1997ud,Molchanov:2001qk,Evdokimov:2004iy}.
It may be a surprise that even the $D\O$ and CDF experiments working
on \pbp\ collisions at $\sqrt 1.96$\,TeV  have made significant
contributions to meson spectroscopy
\cite{Abazov:2004kp,Acosta:2003zx,Abulencia:2005ry,Abulencia:2005zc}.

\subsubsection{\label{exp:Cleo}
CLEO}

The Cornell $e^+ e^-$
storage ring CESR is a symmetric collider which started to work in
1979. Since then, the CLEO collaboration has conducted studies of $b,
c, \tau$ and $\gamma\gamma$ physics in $e^+e^-$ interactions near 10
GeV \cite{Andrews:1982dp}. Successive detector upgrades have been
performed in parallel with luminosity improvements to CESR, which has
delivered over $9\rm fb^{-1}$ integrated luminosity. The CLEO-II
detector \cite{Kubota:1991ww}, operational since 1989, consisted of
drift chambers for tracking and $dE/dx$ measurements, time-of-flight
counters, a 7800-element $CsI$ electromagnetic calorimeter, a 1.5\,T
superconducting solenoid, iron for flux return and muon identification,
and muon chambers. A three-layer silicon vertex detector was added in
1995. The CLEO-II detector was the first to combine a large magnetic
volume with a precision crystal electromagnetic calorimeter. A major
upgrade, the CLEO-III detector \cite{Kopp:1996kg,Viehhauser:2000cu},
was installed in 1999. \begin{figure}[!ht] \bc
\includegraphics[angle=-90,width=0.8\textwidth]{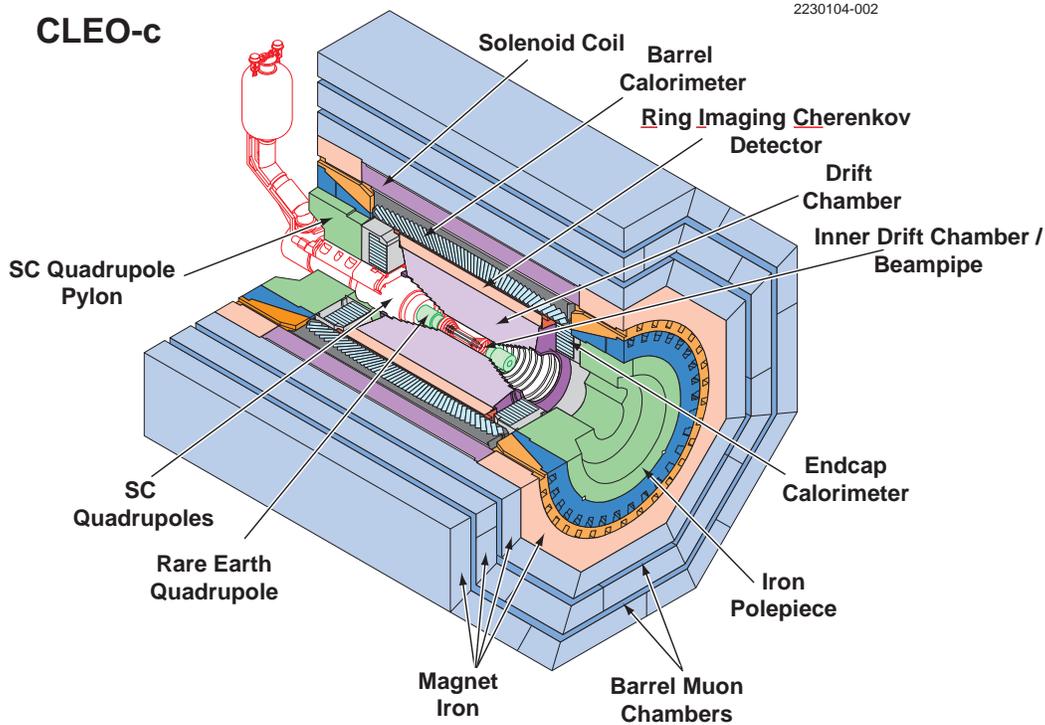}
\ec
\caption{\label{exp:CLEO-c}
Experiment CLEO-c}
\end{figure}
It contained a new four-layer silicon-strip vertex detector, a new
wire drift chamber and a particle identification system based on
Cherenkov ring imaging. The integrated luminosity accumulated by the
CLEO-III detector in 1999-2003 was  $16\rm fb^{-1}$.  In 2003 the
detector was transformed into CLEO-c (Fig.~\ref{exp:CLEO-c}) aiming
for charm physics at $3\div5$\,GeV centre-of-mass energy at very
high luminosity \cite{Briere:2005ci}. The silicon detector was replaced
with a wire chamber that has significantly less material, the magnetic
field was lowered from 1.5\,T to 1\,T. Selected results of the
CLEO-collaboration on meson spectroscopy are published in
\cite{Andrews:1980ha,Andrews:1980bk,Berkelman:1983xe,Chen:1983kr,%
Besson:1984bd,Avery:1989ui,Brock:1990pj,Butler:1993rq,Kubota:1994gn,%
Bergfeld:1994af,Alexander:1998dq,Alam:1998wh,Jessop:2000bv,Kopp:2000gv,%
Muramatsu:2002jp,Besson:2003cp,Frolov:2003jf,Artuso:2004pj,%
Bonvicini:2004yj,Rubin:2004cq,Adam:2004pr,Artuso:2004fp,Briere:2005ci,%
Zweber:2005em,Adam:2005mr,Rubin:2005px,Skwarnicki:2005rv,Rosner:2005ry,%
Dubrovin:2005uc,Besson:2005ud,Ahohe:2005ug,Andreotti:2005vu,Athar:2006gh,%
Cawlfield:2006hm,Adams:2006na,%
He:2006kg,Coan:2006rv,Cawlfield:2007dw,Bonvicini:2007tc}.

\subsubsection{\label{Meson spectroscopy at the barbb factories BaBaR
and BELLE}
Meson spectroscopy at the \protect{$ \bar bb$} factories BaBar and BELLE}

In 1999, two general purpose detectors BaBaR and BELLE started data
taking at the SLAC and KEK $e^+ e^-$ B-factories. Both facilities
operate at the $\Upsilon(4S)$ resonance to produce pairs of $B \bar B$
mesons to study time-dependent $CP$ asymmetries in their decays. Owing
to the very high luminosity of the $e^+ e^-$ colliders and the
universality of the detectors, these facilities have excellent
potential for the study of meson spectroscopy. Five groups of reactions
can be used for this purpose:

- decays of B-mesons produce a broad spectra of final states
with hidden charm, open charm and with no charm;\\
 - $c \bar c$ pairs are produced with a cross section comparable with
$\Upsilon(4S)$ production;\\
- even-$C$ parity states can be produced in $\gamma\gamma$
collisions; \\
- vector mesons can be produced in $e^+ e^-$ interactions with
sufficiently high-energy initial state radiation (radiative return);\\
- flavour tagging reactions like $e^+~e^-\to J/\psi ~\eta_c$ can be used
to study charmonium states.

The SLAC PEP-II B-factory operates at 10.58\,GeV, with energies of the
colliding electron and positron beam of 9\,GeV and 3.1\,GeV,
respectively, resulting in a Lorentz boost of the centre of mass of
$\beta=0.55$. The maximum luminosity is $\sim 9\cdot
10^{33}$\,cm$^{-2}$s$^{-1}$, the peak cross section for formation of
the $\Upsilon(4S)$   $\sim 1$\,nb. Thus about $10^9$ $B$ mesons were
recorded. The detector BaBaR \cite{Aubert:2001tu} is asymmetric along
the beam which reflects the asymmetry in the beam energies. The inner
part of the detector includes tracking, particle identification and
electromagnetic calorimetry. It is surrounded by a superconductive
solenoid providing a magnetic field of 1.5\,T. The tracking system
is composed of a Silicon Vertex Tracker (SVT) and a drift chamber. The
SVT is used for precision measurements of primary and secondary decay
vertices as well as for measurements of low-momentum tracks. A 40-layer
drift chamber is used to measure particle momenta and the ionisation
loss $dE/dx$. Charged particles are identified with momenta up to
$\approx 700$\,MeV/c, the momentum resolution is about
$\sigma_{p_t}\approx 0.5\%$ at $p_t=1.0$\,GeV/c and the resolution for
the track impact-parameter is about 25 and 40\,$\mu$m in the
transverse plane and along the detector axis, respectively. Separation
of pions and Kaons at momenta from 0.5\,GeV/c to 4\,GeV/c is provided
by a novel ring-imaging detector based of fused silica bars. The
electromagnetic calorimeter is a finely segmented array of $CsI(Tl)$
crystals with energy resolution of $\sigma_E/E\approx 2.3\%\cdot
E^{-1/4} + 1.9\%$ (E in GeV). The iron return yoke is instrumented
with resistive plate chambers and limited streamer tubes for detection
of muons and neutral hadrons. Results related to meson spectroscopy are
published in \cite{Aubert:2001tu,%
Bartoldus:2001tu,Aubert:2003fg,Aubert:2003pe,Aubert:2003pt,%
Krokovny:2003zq,Aubert:2004fc,Aubert:2004kj,Aubert:2004ku,%
Aubert:2004ns,Aubert:2004pw,Robutti:2004yz,Aubert:2004zr,Aubert:2005cb,%
Aubert:2005ce,Aubert:2005eg,Aubert:2005gw,Yu:2005hz,Aubert:2005iz,%
Aubert:2005kd,Aubert:2005rm,Robutti:2005ru,Aubert:2005sk,Aubert:2005sm,%
Aubert:2005tj,Aubert:2005vi,Aubert:2005wb,Aubert:2005yj,Aubert:2005zh,%
Aubert:2006aj,Aubert:2006bk,Aubert:2006bm,Aubert:2006bu,Aubert:2006jq,%
Aubert:2006mh,Aubert:2006nu,Aubert:2006qx,Collaboration:2007dc,%
Aubert:2007ij,Aubert:2007ur,:2007rv,:2007wg}.

\begin{figure}[pt]
\bc
\includegraphics[width=0.58\textwidth]{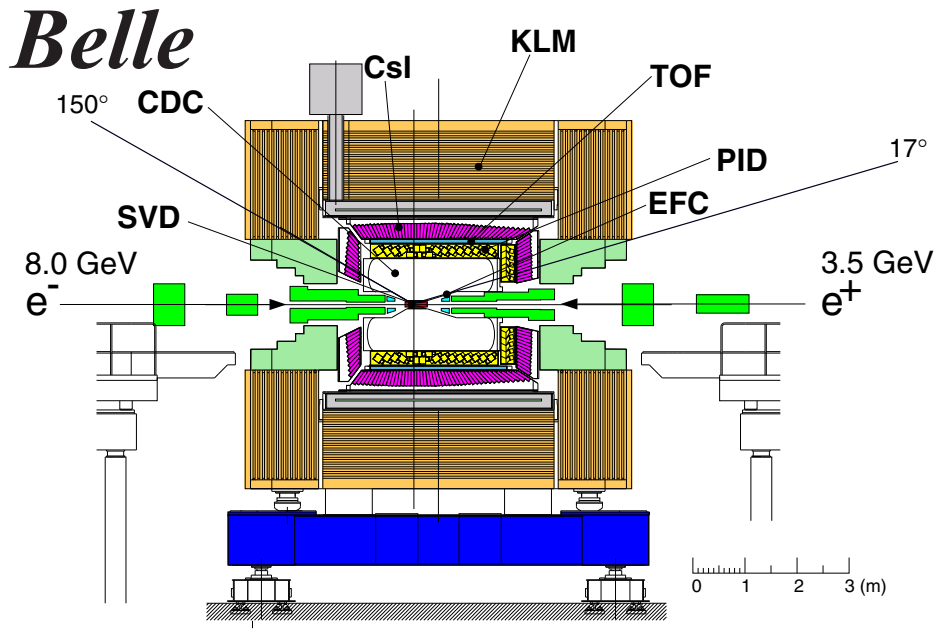}
\ec
\caption{\label{exp:BELLE}
Experiment BELLE}
\end{figure}

The BELLE detector is located at the KEKB $e^+ e^-$ collider at the KEK
laboratory in Tsucuba, Japan. The KEKB storage rings have asymmetric
energies: 8\,GeV for electrons and 3.5\,GeV for positrons, that provides
to the $\Upsilon(4S)$ resonance a Lorentz boost of $\beta\gamma=0.425$.
At KEKB the world record in luminosity of
$1.4\cdot10^{34}$\,cm$^{-2}$s$^{-1}$ was achieved. The side view of the
Belle detector \cite{Iijima:2000cq} is shown at Fig. \ref{exp:BELLE}.
Tracking, identification and calorimetric systems are placed inside a
1.5\,T superconducting solenoid magnet of 1.7\,m radius. Precision
tracking and vertex measurements are provided by a silicon vertex
detector and central drift chamber. The four-layer double-sided silicon
detector with strip pitches of $75\,\mu$m$(z)$ and $50\,\mu $m
surrounds the beryllium beam pipe having 1.5\,cm radius. The central
drift chamber has 50 layers of anode wires for tracking and $dE/dx$
measurements. The momentum resolution of the tracking system is
$\sigma_{p_t}/p_t=(0.30/\beta + 0.19p_t)\%$, where $p_t$ is the
transverse momentum in GeV/c. The Kaon/pion separation is achieved
using the central drift chamber, time of flight counters with rms
resolution of 0.95\,ps and aerogel Cherenkov counters with refractive
indices from 1.01 to 1.03, depending on polar angle. The
electromagnetic calorimeter consists of 8737 $CsI(Tl)$ crystals of
projective geometry with energy resolution for photons of
$\sigma(E)/E\approx 1.8\%$ at $E_{\gamma}$ above 3\,GeV. The flux return
is instrumented with 14 layers of resistive plate chambers for muon
identification and detection of neutral hadrons. Results on meson
spectroscopy can be found in \cite{Kim:2001tv,Abe:2002ds,Abe:2002tw,%
Abe:2002av,Garmash:2003er,Abe:2003jk,Choi:2003ue,Wang:2003iz,%
Abe:2003vn,Abe:2003zm,Wang:2003yi,Lee:2004mg,Garmash:2004wa,Abe:2004zs,%
Wang:2005fc,Abe:2005hd,Abe:2005ix,Abe:2005iy,Kuo:2005nr,Xie:2005tf,%
Ziegler:2005tt,Chistov:2005zb,Drutskoy:2005zr,Drutskoy:2005dr,%
Abe:2006hf,Mori:2006jj,Chen:2006qm,Sokolov:2006sd,unknown:2006xm,Brodzicka:2007aa,%
Wang:2007as,Mori:2007bu,Abe:2007jn,Wang:2007ea}.

\subsection{\label{Meson spectroscopy at LEP}
Meson spectroscopy at LEP}

The main goal of the LEP program was the study of electroweak physics in
$e^+ e^-$ collisions on and above the $Z^0$. The operation of the LEP
collider at CERN started in August 1989 and stopped in November 2000.
LEP had four intersection regions, each surrounded by a particle
detector. The detectors (DELPHI, ALEPH, L3 and OPAL) were optimised
differently to study various aspects of physics. The initial LEP
energy was chosen to be around 91\,GeV as to produce $Z^0$ particles.
Since the end of 1995, LEP has moved away from the $Z^0$ and entered
its second phase. Its energy was doubled to study the production of
$Z^0Z^0$ and $W^+W^-$ pairs, and to search for new particles, in
particular the Higgs boson and/or supersymmetric particles.
Even though the main goal of these experiments was electroweak physics,
very significant results in meson spectroscopy were obtained by all
four detectors. In the first data taking period, about $5 \times 10^6$
of $Z^0$ decays were collected in each of the four detectors. Owing to
the high branching ratios of $Z^0$ decays to heavy quarks
($BR(Z^0\rightarrow c\bar c)=11.8\%,~BR(Z^0\rightarrow b\bar
b)=15.1\%$), considerable statistic was collected with heavy quarks in
the final state. This data is especially fruitful to study heavy-light
mesons. In the second phase, the very high luminosity, up to
$10^{32}$\,cm$^{-2}$s$^{-1}$, in combination with the high energy
provided unique possibilities to study $\gamma \gamma$ interactions.

\subsubsection{\label{The Aleph experiment}
The Aleph experiment}

ALEPH was a $4\pi$ detector designed to give detailed information on
complex events from high-energy $e^+e^-$ collisions
\cite{Decamp:1990jr}. A superconducting coil, 5\,m in diameter and 6\,m
long, produced a uniform 1.5\,T field in beam direction. Closest to
the interaction point, a silicon vertex detector was installed with
$12\rm\,\mu m$ resolution in $r - \phi$ and $r - z$ and capability to
identify secondary vertices in decays of $\tau$-leptons, and of $c$
and $b$ quarks. The vertex detector was surrounded, in order of
increasing radius, by a drift chamber (Inner Tracking Chamber), a
Time Projection Chamber (TPC) of 3.6\,m diameter and 4.4\,m length, and
an electromagnetic calorimeter. A resolution in transverse momentum
of $\sigma(p_t)/p_t=0.0006\cdot p_t + 0.005 (p_t$ in GeV/c) was
reached. The electromagnetic calorimeter, consisting of 2\,mm lead
sheets with proportional wire sampling, has an energy resolution for
electromagnetic showers of $\sigma_E/E=0.18/\sqrt E + 0.0009~ (E$ in
GeV). Outside of the coil, a 1.2\,m thick iron return path was used as
hadron calorimeter, a double layer of drift tubes provided muon
identification. Strong points of the detector are precision of
momentum measurements for charged particles due to a high magnetic
field and a TPC,  good identification of electrons and muons, and good
spatial resolution in electron and $\gamma$ calorimetry. A
silicon-tungsten calorimeter installed in 1992 provided high precision
of the luminosity measurement via Bhabha scattering. Results on meson
spectroscopy are published in \cite{Buskulic:1995mt,Buskulic:1996uk,%
Barate:1998cq,McNeil:1997qa,Muheim:1998sx,Barate:1999ze,Heister:2001nj,%
Bartalini:2002je,Pal:2004yi}.

\subsubsection{\label{The Delphi experiment}
The Delphi experiment}

DELPHI was a general purpose detector offering 3-dimensional information
on curvature and energy deposition with fine spatial granularity, as
well as identification of leptons and hadrons over most of the solid
angle \cite{Aarnio:1990vx}. A superconducting coil provided a 1.2\,T
solenoidal field of high uniformity. Tracking relied on a microvertex
detector, an inner detector (a multiwire proportional chamber), a Time
Projection Chamber (TPC) measuring up to 16 space points per
track, an outer detector with 5 layers of drift tubes, and forward drift
chambers. The 3-layer silicon microvertex detector was used for
precision measurements of the interaction vertex and of decay vertices
of short-lived particles such as bottom and charm hadrons and $\tau$
leptons. The single-hit resolution was found to be $9\,\mu$m in
$z$-direction and $7.6\,\mu $m in $r - \phi$-direction. A Ring
Imaging Cherenkov detector (RICH) used gaseous ($C_5F_{12}$) and liquid
($C_6F_{14}$) radiators. As a result, particle identification was
achieved at nearly all momenta: below 1\,GeV/c with $dE/dx$
measurements in the TPC, from 0.7\,GeV/c to 8\,GeV/c with the liquid
radiator and from 2.5\,GeV/c to 25\,GeV/c with the gaseous radiator.
Electromagnetic showers were measured in the barrel with high
granularity by a High Density Projection Chamber (HPC), and in the
endcaps by 1 x 1 degree projective towers composed of lead glass. A
segmented magnet yoke served for hadron calorimetry and as a filter for
muons which were identified in two drift chamber layers. In addition,
scintillator systems were implemented in the barrel and forward
regions. A small-angle Shashlik-type calorimeter was used to monitor
the luminosity. Results on meson spectroscopy are reported in
\cite{Abreu:1992rv,Abreu:1994hj,Abreu:1996nz,Abreu:1998cn,%
Abreu:1998vk,Abdallah:2003xk}.

\subsubsection{\label{The L3 experiment}
The L3 experiment}

The detector L3 consisted of a large-volume ($\O=11.9$\,m,
$L=12$\,m) low-field (0.5 T) solenoidal magnet, a small central
tracking detector with high spatial resolution, a high-resolution
electromagnetic calorimeter encapsulating the central detector, a
hadron calorimeter acting also as a muon filter, and high-precision
muon tracking chambers \cite{xxx:1989kx}. The detector was designed to
measure energy and position of leptons with the highest obtainable
precision allowing a mass resolution $\delta(m)/m$ of about $2\%$ in
di-lepton final states. Hadronic energy flux was detected by a
fine-grained calorimeter, which also served as a muon filter and a
tracking device. The outer boundary of the detector was given by the
iron return yoke of a conventional magnet.  The muon momentum
measurement was performed with a precision of $\sigma(p)/p\approx 2.5\%$
by three sets of high precision drift chambers with long lever arm in
the central detector region. A forward-backward muon detection system
extended the polar angle coverage to 22 degrees in the forward region.
Radially inwards was a combined hadron calorimeter and muon absorber.
The electromagnetic energy flow was determined by approximately 11000
BGO crystals.  Full electromagnetic shower containment over nearly
$4\pi$ solid angle coverage was achieved. The energy resolution varied
from $5\%$ at 100 MeV to $1.4\%$ at high energy. Surrounding the 10\,cm
diameter beam pipe, a high-precision Silicon Microstrip Detector and a
small drift chamber operating in the time expansion mode acted as
charged particle vertex detectors, providing momentum resolution
$\sigma(p_t)/p_t=0.021\cdot p_t$ ($p_t$ in GeV/c). Results on
meson spectroscopy can be found in \cite{Adriani:1993ta,%
Acciarri:1995rs,Acciarri:1997rb,Acciarri:1997yx,Acciarri:1998xh,%
Acciarri:1999jx,Acciarri:2000ev,Achard:2001uu,Achard:2003qa,%
Achard:2004ux,Achard:2004us,Achard:2005pb,Vorobev:2004zd}.

\subsubsection{\label{The OPAL experiment}
The Opal experiment}

The general purpose detector OPAL  was designed to
study a wide range of unexplored physics at LEP \cite{Ahmet:1990eg}.
The tracking system of the apparatus, in order of increasing distance
from the interaction point, were a silicon microvertex detector
providing single hit resolution of $5\,\mu$m in $r - \phi$ and $13\,\mu
$m in $z$, central detectors consisting of a vertex and a jet chamber,
and a barrel-shaped chambers for precise $z$-measurements. The main
tracking with the jet chamber provided up to 159 space points per
track. It was used for tracking and for particle identification using
$dE/dx$. The momentum resolution was $\sigma_p/p=0.0022\cdot p$ ($p$ in
GeV/c). A warm conductor solenoid provided a uniform magnetic field of
0.4\,T. A TOF scintillator barrel detector, complemented by a
scintillating tile endcap detector allowed particle identification at
momenta up to 2.5\,GeV/c. Electromagnetic showers were measured in a
multi-cell $10\times10$\,cm$^2$ lead glass electromagnetic calorimeter
with resolution of about $\sigma(E)/E=0.05/\sqrt E$ ($E$ in GeV).
Hadrons were measured with streamer tubes and thin-gap wire chambers.
Muons were detected in an external muon identifier composed of four
layers of drift chambers. The luminosity was measured via Bhabha
scattering into a forward small-angle silicon-tungsten calorimeter.
Results on meson spectroscopy are published in
\cite{Alexander:1991su,Acton:1992zq,%
Ackerstaff:1998ec,Thiergen:2000sp,Abbiendi:2001qp}.


\markboth{\sl Meson spectroscopy} {\sl Experimental methods}
\clearpage\setcounter{equation}{0}\section{\label{Experimental methods}
Experimental methods}

A variety of experimental techniques has been developed to search for
new meson resonances and to establish properties of known states. In
{\it production} experiments, the total energy is shared between a
recoil particle and a multi-meson final state. Angular momentum can be
transferred, and the multi-meson system can contain contributions from
several resonant partial waves with different angular momenta. The
quantum numbers of the multi-meson system are restricted only by
conservation laws; production of partial waves with exotic quantum
numbers, i.e. of quantum numbers which cannot be attributed to a $q\bar
q$ system, is allowed. In {\it formation} experiments, their is no
recoil particle, mass and quantum numbers of the final state are given
by the initial state. Formation of mesons with exotic quantum numbers
in $e^+e^-$ or $ p\bar p$ annihilation is thus forbidden.

Consider, i.e., antiproton-proton annihilation in flight. Nucleon and
antinucleon provide their mass and their momentum to a mesonic final
state.  When all final-state mesons are combined, their invariant mass
is determined by the invariant mass $\sqrt s$ of the initial state. By
scanning the total energy by variation of the antiproton momentum, the
mass of the final state can be tuned to cover the range of masses in
which meson resonances are to be studied. The mass resolution is given
by the precision with which the beam energy can be kept stable. The
quantum numbers of $q\bar q$ mesons which can be formed are restricted
to those  of the $p\bar p$ system which contains no exotic partial
wave.

Alternatively, the initial momentum can be kept fixed,
antiproton-proton annihilation can e.g. be studied at rest. Meson
resonances can now be produced recoiling against another meson, often a
pion. In case of three pions in the final state, each pion recoils
against the other two pions possibly forming a resonance. In the
analysis, one has to take into account all three possibilities. Of
course, not their probabilities but their amplitudes have to be added.
The mass of resonances is reconstructed from the measured particle
momenta in the final state, the mass resolution is given by the
accuracy with which the final-state particles are measured.

 \subsection{\label{Meson resonances in production experiments}
Meson resonances in production experiments}

\subsubsection{\label{Charge exchange and strangeness exchange
scattering}
Charge exchange and strangeness exchange scattering}

The model of Reggeised particle exchange was developed in the 60's of
the last century; it provides efficient tools for the analysis of
two-body hadronic reactions. It was named after T.~Regge who proposed
the method of complex angular momentum in nonrelativistic quantum
mechanics \cite{Regge:1959mz,Regge:1960zc}. In field theory this method
links the two-body scattering amplitude at high energy with the
amplitude in the crossed channel. Consider the reaction
\begin{equation}
a+b \rightarrow c+d
\label{a+b}
\end{equation}
at high energy $s\gg m^2$ and fixed transfer momentum squared $|t| \sim
m^2$. The amplitude for $t$-channel exchange of a particle with mass
$m$ and spin $J$ can be expressed as
\begin{equation}
T(s,t)=\frac{1}{k}g_1 g_2 \left( \frac{s}{s_0} \right) ^J (m^2-t)^{-1}
\label{singlepole}
\end{equation}
where $s_0$=1 GeV$^2$ is a scale and $g_1$, $g_2$ are coupling
constants. For exchanges with $J \ge 2$ this amplitude violates the
Froissart limit, $T(s,t)\sim ln^2 (s)$ for $s \rightarrow \infty $, for
binary reactions \cite{Froissart:1961ux}. This problem appears
as (\ref{singlepole}) gives the correct amplitude only near the $m^2$
pole which is very far from the physical region of the reaction
(\ref{a+b}), where $t$ is negative. At very high $s$ and moderate $t$,
the Regge model predicts a simple functional form for the extrapolation
of the amplitude from the $t$-channel resonance region to the physical
region of reaction (\ref{a+b}) :
\begin{equation}
R(s,t) \sim \frac{1}{k}~ \beta (t)~
(\frac{s}{s_0})^{\alpha_i (t)} \eta(\alpha_i(t))
\label{regge}
\end{equation}
where $\alpha_i (t)$ is a smooth function of $t$, called Regge
trajectory. $J$ is the spin of the lightest particle in the $t$-channel
($0$ for the $\pi$ trajectory, $1$ for $\rho$, $\frac{1}{2}$ for $N$
etc.), $\eta(\alpha_i(t)) = -\left(\frac{1+\sigma_i
e^{-i\pi\alpha_i(t)}}{\sin\pi\alpha_i(t)}\right) $ is a signature
factor for Regge pole $i$ with trajectory $\alpha_i(t)$, signature
$\sigma_i = \pm1$. The amplitude (\ref{regge}) represents $t$-channel
exchange of objects having identical flavour quantum numbers (labelled
by $i$) and all possible spins.

\begin{figure}[H]
\begin{center}
\includegraphics[width=0.3\textwidth]{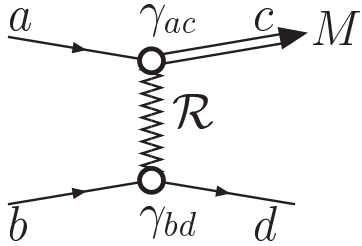}
\end{center}
\caption{\label{pic:twobody}
The diagram for a two-body reaction with Reggeon exchange.
}
\vspace*{-2mm}
\label{twobody}
\end{figure}

The differential cross section of the reaction (\ref{a+b}) can be
obtained by squaring the amplitude (\ref{regge})
\begin{equation}
\frac{d\sigma}{dt} = f(t)~(\frac{s}{s_0})^{2\alpha_i(t)-2}\,.
\label{dsigmadt}
\end{equation}
The trajectories $\alpha_i (t)$ are supposed
to be linear near $t=0$
\begin{equation}
\alpha_i (t) = \alpha_i (0) + \alpha'_i (0) t\,.
\label{trajectory}
\end{equation}
From (\ref{dsigmadt}) we see that $\alpha_i (0)$ determines the
dependence of $\frac{d\sigma}{dt}(t=0)$ on energy while $\alpha'_i (0)$
describes the $t$-dependence on energy. The parameters $\alpha_i (0)$
and $ \alpha'_i (0)$ for different Regge trajectories are known from
experiments. The extrapolation of $\alpha_i (t)$ from small negative
$t$ (reaction \ref{a+b}) to positive $t$ (reaction $a + \bar c
\rightarrow \bar b + d$) shows that $\alpha_i (m^2_{r})=J$ as suggested
by the model ($J$ is the spin of the resonance $r$). Most trajectories
have almost equal slopes, $\alpha'(0)\approx 1\mbox{ GeV}^{-2}$, except
the trajectory with vacuum quantum numbers. The intercept of the Regge
trajectory $\alpha_i (0)$ can be estimated to $\alpha_i (0) = J - m^2_r
/ \alpha'(0)$. It gives $\alpha (0) \approx 0.5$ for the $\rho$
trajectory and $\alpha(0) \approx 0 $ for the $\pi$-trajectory. Using
(\ref{dsigmadt}) we conclude that for reactions with $\rho$-trajectory
exchange, like $ \pi^- p \rightarrow \pi^0 n $, the differential cross
section $\frac{d\sigma}{dt}(t=0)$ falls off with energy as $1/s$, for
$\pi^- p \rightarrow \rho^0 n$ (with $\pi$ and $a_1$-trajectory
exchange) the cross section drops as $1/s^2$.

The trajectory with vacuum quantum numbers plays a very special $\rm r\hat{o}le$ in
Regge-phenomenology. It gives the dominant contribution to elastic
scattering at asymptotically high energies. From the optical theorem
\begin{equation}
\sigma_{tot}=\frac{4\pi}{k} Im R(t=0)
\label{optical}
\end{equation}
and  (\ref{regge}) we derive that $\alpha(0)=1$ leads to a constant
total cross section. This trajectory was called Pomeron after
I.Ya. Pomeranchuk who proposed arguments in favour of asymptotically
constant total cross sections \cite{Pomeranchuk:1}. At the first glance, the
upper limit for any trajectory intercept should be $\alpha(0)=1$
since for $\alpha(0)>1$, the Froissart limit is violated. This
assumption is however not supported by data showing an asymptotically
rising total cross sections of hadron-hadron interactions at high
energy. To describe the total hadronic cross sections, the Pomeron
cannot be represented by a simple $t$-channel Regge pole; it has to be
replaced by a more complicated object \cite{Ter-Martirosyan:1970yh}
with $\alpha(0) \approx 1.1$ and $\alpha'(0) \approx 0.3$, modified by
absorption and inelastic intermediate states. It is a still debated
question if such a complicated object can be linked to any resonances
(possibly glueballs) in the crossed channel. The physics of Pomerons
is, e.g., highlighted in the excellent book by Donnachie, Dosch,
Nachtmann and Landshoff \cite{Donnachie:2002en}.

To get information on the function $\beta(t)$ in (\ref{regge}) we have
to rely on considerations beyond the Regge model. By looking at
Fig. \ref{pic:twobody} it is natural to assume that the function
$\beta(t)$ in (\ref{regge}) can be represented in factorised form
\begin{equation} \beta(t)=(-t)^{n/2} \gamma_{acr}(t) \gamma_{bdr}(t)
\label{factor}
\vspace{2mm}
\end{equation}
The functions $\gamma_{acr}$ and $\gamma_{bdr}$ can be written as
$\gamma_{acr}=g_{acr} exp(r_{acr} t)$and $\gamma_{bdr}=g_{bdr}
exp(r_{bdr} t)$. Here, $g_{acr}$ and $g_{bdr}$ are coupling constants,
$r_{acr}$ and $r_{bdr}$ are slope parameters representing formfactors.
At $t \rightarrow 0$, the amplitude can be suppressed at some power of
$sin \theta \sim \surd(-t)$ to provide a net helicity ($\lambda_i$)
flip, and the extra flip to guarantee parity conservation $n
=|\lambda_c-\lambda_a|+|\lambda_d-\lambda_b|$.

\subsubsection{\label{Two--photon fusion}
Two-photon fusion}

\begin{figure}[H] \begin{center}
\includegraphics[width=.3\textwidth]{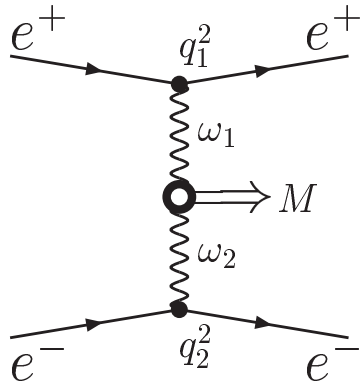}
\vspace*{-2mm}
\end{center}
\caption{\label{pic:gammagamma}
The diagram describing meson production in
$e^+ e^-$ annihilation. Here
 $q_i^2 = (p - p')^2$ and $p(p')$ are the momenta of the initial (final)
electrons, $\omega_i$ are the energies of the virtual photons.
}
\vspace*{-2mm}
\end{figure}

Production of resonances by two-photon fusion, as well as decays of
resonances into two photons, provide important information on hadron
structure. For $q \bar q$ mesons the matrix element for these processes
is proportional to $ \Sigma Q^2_i$. The decay width of a neutral meson
with isospin $I=1~~ (M=\frac{1}{\surd 2}(u \bar u - d \bar d))$ is thus
proportional to $\frac{1}{2}(Q^2_u - Q^2_d)^2$, for $I=0~~ SU_3$-octet
mesons, it is proportional to $\frac{1}{3}(Q^2_u + Q^2_d)^2$ and for
$SU_3$-singlet mesons to $\frac{1}{6}(Q^2_u + Q^2_d - 2 Q^2_s)^2$. From
a comparison with data, singlet-octet mixing angles can be derived. For
pseudoscalar mesons the mixing angle is $\theta_P \approx -20^0$, for
tensor mesons ($f_2, ~f_2'$) it is $\theta_T \approx 28^0$. The latter
mixing angle leads to an accidental suppression of two-photon decays
of $f_2'$ due to an approximate cancellation of the $u \bar u + d \bar
d$ and $s \bar s$ contributions. Two-photon fusion is studied at $e^+
e^-$ colliders (Fig. \ref{pic:gammagamma}).

Common sense, as well as specific models, tell us that $\gamma \gamma$
decay of non-$q \bar q$ mesons like glueballs, hybrids, multiquark
objects, or mesonic molecules is suppressed. The two-photon width can
therefore be used to identify non-$q \bar q$ states. Production of
glueballs can, e.g., be expected in radiative J/$\psi$ decays via two
gluons $ J/\Psi \rightarrow \gamma g g$, but in two-photon fusion it
should be suppressed because of their small $\gamma \gamma$ coupling. A
convenient estimator called stickiness was proposed by Chanowitz
\cite{Chanowitz:1984cb} to discriminate glueballs and $q\bar q$ mesons.
The stickiness of a resonance R with mass $m_{\rm R}$ and two-photon
width $\Gamma _{{\rm R} \to \gamma\gamma}$ is defined as
\begin{equation}
 S_{\rm R} = N_l \left(\frac{m_{\rm R}}{K_{{\rm
J}\to\gamma {\rm R}}}\right)^{2l+1} \frac{\Gamma _{{ J}\to\gamma
{\rm R}}}{\Gamma _{{\rm R} \to \gamma\gamma}} \ ,
\label{sticky}
\vspace{2mm}
\end{equation}
 where $K_{{ J}\to\gamma {\rm R}}$ is the energy of the photon in
the J rest frame, $l$ is the orbital angular momentum of the two
initial photons or gluons ($l=1$ for $0^-$), $\Gamma _{{ J}\to\gamma
{\rm R}}$ is the J radiative decay width for R, and $N_l$ is a
normalization factor chosen to give $S_{\eta} = 1$.

The gluiness ($G$) was introduced~\cite{Close:1996yc,Paar:2000pr} to
quantify the ratio of the two-gluon and two-photon coupling of a
particle and is defined as

\begin{equation}
G = \frac{9\,e^4_q}{2}\,\biggl(\frac{\alpha}{\alpha _s}\biggr)^2
\, \frac{\Gamma _{{\rm R} \to {\rm gg}}}{\Gamma _{{\rm R} \to
\gamma\gamma}} \ ,
\label{gluish}
\vspace{2mm}
\end{equation}
where $e_q$ is the relevant quark charge. $\Gamma _{{\rm R} \to
{\rm gg}}$ is the two-gluon width of the resonance {\rm R},
calculated from equation (3.4) of ref.~\cite{Close:1996yc}.
Stickiness is a relative measure, gluiness is a normalised
quantity and is expected to be near unity for $q\bar{q}$ mesons.

The cross section for $e^+ e^-\rightarrow M$ can be related to the
cross section for $\gamma \gamma \rightarrow Me^+ e^-$ in the
equivalent photon approximation:
 \begin{equation} d\sigma_{e^+ e^-
\rightarrow e^+ e^- M} = dn_1 dn_2 d\sigma_{\gamma \gamma \rightarrow
M} (W^2) \label{equivalent}
\end{equation}
were $n_{1,2}$ is number of photons with energy $\omega_{1,2}$ and
four-momentum squared $q^2_{1,2}$
\begin{equation}
dn_{1,2}=\frac{\alpha}{\pi}[1-\frac{\omega_{1,2}}{E_{1,2}} +
\frac{\omega^2_{1,2}}{2E^2_{1,2}}-\frac{m_e^2\omega^2_{1,2}}
{(-q^2_{1,2})E^2_{1,2}}]\frac{d\omega_{1,2}}{\omega_{1,2}}
\frac{d(-q^2_{1,2})}{-q^2_{1,2}}\,.
\label{fotonspectr}
\end{equation}
The photon spectrum peaks very sharply at $(-q^2)\rightarrow
(-q^2)_{min}=m^2_e\omega^2/E(E-\omega)$. The peak is much narrower
than any hadronic form factor, most photons are therefore nearly real
($q^2 \approx 0$). After integration of (\ref{equivalent}) over
$q^2_{1,2}$ and $\omega_{1,2}$, the cross section for production of
resonance $M$ via two-photon fusion is
\begin{equation} \sigma_{e^+ e^-
\rightarrow e^+ e^- M}=(2J+1)\frac{8\alpha^2\Gamma_{M \rightarrow
\gamma \gamma}}{m^3_M} G(s,m_M,f) \label{tagtag}
\end{equation}
where $G(s,m_M,f)$ is a known function of $s$, $m_M$, and the form
factor slope $f$. The dependence on the form-factor slope is very weak
and can be neglected in most of cases. In the equivalent photon
approximation, two-photon fusion looks like a formation
experiment in a wide-band photon beam with known spectra ($dn
\approx
\frac{\alpha}{\pi}\frac{d\omega}{\omega}\frac{d(-q^2)}{(-q^2)})$.

Not all positive-charge-parity states can couple to two real photons.
Consider a system of two spin-one particles, $s_1=s_2=1$. The total
spin of these two particles is composed of spin $S=0,~1,~2$ and orbital
angular momentum $L=0,~1,~2,~3....$. With these $S$ and $L$, one
positive parity state and one negative parity state exist for $J=0$,
three positive parity states and four negative parity states for
$J=1$, and five positive parity states and four negative parity states
for any $J>1$. Only few of these states can be realised in a system of
two real photons since real photons have only two
helicities ($\lambda=\pm 1$) and obey Bose symmetry.
Neither $J^p=1^-$ nor $J^p=1^+$  states couple to two real photons.
This statement is called Landau-Yang theorem: the cross section for
production of $J^{PC}=1^{++}$ states like $f_1(1285),~f_1(1420)$ or
$a_1(1260)$ in two-photon fusion with nearly real photons is strongly
suppressed. If at least one of the two photons is sufficiently virtual
having $|q^2| \sim m_h^2$ then all the three helicities are
allowed ($\lambda =1,0,-1$) and production of $J=1$ states in
two-photon fusion is no longer suppressed. Events with larger $|q^2|$
can be selected by applying a cut on the scattering angle of the
electron or positron (single tag) or both (double-tag).

Some other reactions can also be used to study two-photon fusion, like
production of mesons in Coulomb field of nuclei $\gamma Z \rightarrow M
Z$ (Primakoff reaction) and production of mesons in peripheral
interactions of nuclei $Z_1 Z_2 \rightarrow Z_1 Z_2 M$. In both cases,
photons are selected by requiring very small $|q^2|$. The
first reaction is used to study light mesons since in fixed-target
experiments, production of heavy mesons is suppressed by the minimal
transfer-momentum squared $-q^2_{min}$ which increases with the mass
$m$ of the produced meson: $-q^2_{min} \approx m^4/4E^2_{beam}$.

\subsubsection{\label{Central production}
Central production}

\begin{figure}[ph]
\begin{center}
\includegraphics[width=.3\textwidth,height=.28\textwidth]{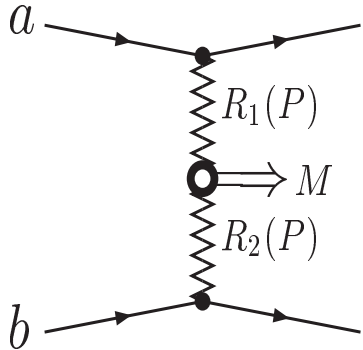}
\vspace*{-2mm}
\end{center}
\caption{\label{cepro}
 Central production of a system X. The momentum transfer is defined by
$q_i=(p_i - p'_i)$ where $p_i, p'_i$ are the in- and outgoing momenta
of the two hadrons.  $\omega_i$ are the energies transferred to $X$.
}
\vspace*{-2mm}
\end{figure}

\begin{figure}[pb]
\begin{center}
\includegraphics[width=.45\textwidth,height=0.35\textwidth]{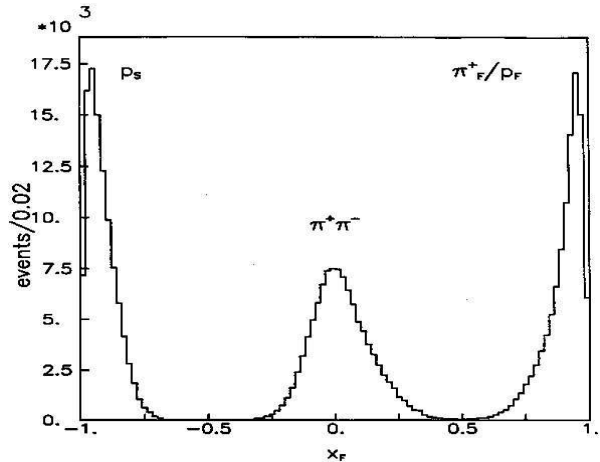}
\vspace*{-3mm}
\end{center}
\caption{\label{pic:centralprod}
The distribution of the Feynman variable $x_F$ in the reaction $p\,p\to
p_s (\pi^+\pi^-) p_f$. In the lab system, the scattered proton goes
forward as fast proton $p_f$. Little momentum is transferred to the
target proton  $p_s$. The $\pi^+\pi^-$ pair is slow in the
centre-of-mass system \protect\cite{Armstrong:1991ch}. }
\end{figure}

Central production is a process similar to two-photon fusion. Two
hadrons at a large energy scatter, keeping their identity and loosing a
small fraction of their energy. In a fixed target experiment, a hadron
$h_{\rm beam}$ (mostly protons were used but also pions or Kaons),
scatters off a target proton $ p_{\rm target}$ and produces a particle
or system of particles $X$ \be \label{centprod}
 h_{\rm beam}\ { p}\ \to\ h_{\rm fast}\ X\ p_{\rm slow}
\ee
where the subscripts 'fast' and 'slow' indicate the fastest and slowest
particles in the laboratory respectively. The fast hadron scatters into
forward direction emitting a Reggeon. The fast (slow) hadron
transfers a squared four-momentum $t_1 (t_2)$ to the central system.
(See Fig.~\ref{cepro} for definitions.)
The centre of mass energy of the
scattered particle is reduced by $x_1$. The target-proton is slow in
the laboratory system; it emits a Reggeon and its centre of
mass energy is reduced by $x_2$. The two virtual particles collide
producing a particle or system of particles X with mass
$M\sim\sqrt{s(1-x_1)(1-x_2})$ where $s$ is the squared centre-of-mass
energy. For $\sqrt s\sim 30$\,GeV/c$^2$  and $1-x_1,1-x_2$ in the range from
0.0 to 0.1 the available phase space is limited to about 3.0\,GeV/c$^2$ .  In
the centre of mass system, particle X carries only a small fraction
$x_F$ of the momenta of the scattered particles; they are produced
centrally, see Fig.~\ref{pic:centralprod}. Scattered hadron and target
proton keep a large fraction $x_F$ of their total energy. The two
scattering particles transfer a four-momenta $t$ to the central
system.

At sufficiently high energy $s_{1,2}>>1$GeV$^2$ and moderate momentum
transfer $-t_{1,2}<1$GeV$^2$ the amplitude for double Regge production
of the resonance can be written as (see \cite{Kaidalov:2003fw} and
references therein):
\vspace*{-6mm}

\begin{eqnarray}
T_{\lambda_1\lambda_2}^{\lambda_3\lambda_h\lambda_4}(s_1,s_2,t_1,t_2,\phi) =
\phantom{rrrrrrrrrrrrrrrrrrrrrrrrrrrrrrrrrrrrrrrrrrrrrr}\nonumber \\
\qquad\qquad\
\sum_{i,k}g_{\lambda_1\lambda_3}(t_1)g_{\lambda_2\lambda_4}(t_2)
\left(\frac{s_1}{s_0}\right)^{\alpha_i(t_1)}\left(\frac{s_2}{s_0}\right)^{\alpha_k(t_2)}
\eta(\alpha_i(t_1))\eta(\alpha_k(t_2))g_{ik}^{\lambda_h}(t_1,t_2,\phi)
\label{central}
\end{eqnarray}
Here, $\phi$ is the angle between the transverse momenta $\vec
p_{3\perp}$ and $\vec p_{4\perp}$ of the outgoing protons and
$\eta(\alpha_i(t))$ is the signature of Regge pole $i$ with
trajectory $\alpha_i(t)$. The vertex couplings for different helicities
are the same as in single-Regge exchange (\ref{factor}). This amplitude
looks as a very natural choice if we compare Fig. \ref{twobody} with Fig.
\ref{pic:centralprod} and amplitude (\ref{regge}) with
(\ref{central}).

 The spin structure of the central vertex
$g_{ik}^{\lambda_h}(t_1,t_2,\phi)$ depends on the product of the
naturalities of particle $h$ and the exchanged Reggeons. With this
amplitude, the cross section for production of resonances with
$J^{PC}=0^{++}, 0^{-+}, 1^{++}$ can be expressed as
\hspace{-6mm}

\begin{eqnarray}
\hspace{-3mm}\frac{d\sigma(0^{++})}{dt_1 dt_2 d \phi}
\sim\ & \ {G^p_E}^{2} (t_1)
{G^{p}_{E}}^2 (t_2)[F_1^2(t_1,t_2,\vec
p_{3\perp}\!\cdot\!\vec p_{4\perp},M^2)
 +\frac{\sqrt{t_1t_2}}{\mu^2}
\cos(\phi) \times F_2^2(t_1,t_2,\vec
p_{3\perp}\!\cdot\!\vec p_{4\perp},M^2) ]^2
\nonumber\\
\hspace{-3mm}\frac{d\sigma(0^{-+})}{dt_1 dt_2 d \phi }
\sim\ & \ t_1 t_2 {G^p_E}^{2} (t_1)
{G^{p}_{E}}^2 (t_2) \sin^2(\phi)\times F^2(t_1, t_2, \vec
p_{3\perp}\!\cdot\!\vec p_{4\perp},M^2)
\qquad\qquad\qquad\qquad\ \\
\hspace{-3mm}\frac{d\sigma(1^{++})}{dt_1dt_2d\phi } \sim\ & \ {G^p_E}^{2}
(t_1) {G^{p}_{E}}^2 (t_2) [(\sqrt{t_1} - \sqrt{t_2})^2 \times
F^2_3(t_1,t_2,\vec p_{3\perp}\!\cdot\!\vec p_{4\perp},M^2 \nonumber
\qquad\qquad\qquad\qquad\ \\
 &\qquad\qquad\qquad\ + \sqrt{t_1t_2}
\sin^2(\phi /2)\times F^2_4(t_1,t_2,\vec
p_{3\perp}\!\cdot\!\vec p_{4\perp},M^2)] \nonumber
\end{eqnarray}

The experimental data agrees with these predictions. It is important to
note that these predictions follow from general rules of the Regge model
and do not tell us anything on the structure of Reggeons or mesons. The
physics of resonances and their interactions with Reggeons is hidden in
unknown formfactors $F^2(t_1,t_2,\vec p_{3\perp}\!\cdot\!\vec
p_{4\perp},M^2)$.

Several processes may compete in central production; scattering of two
Reggeons, Pomeron-Reggeon scattering or Pomeron-Pomeron scattering.
The cross section for Pomeron-Pomeron scattering should  be
independent of $s$ while collisions of two Reggeons in the central
region are predicted to scale with $1/s$ and Pomeron-Reggeon scattering
as $1/\sqrt s$ \cite{Ganguli:1980ec}.  The nature of Pomerons is a
topic of intense discussions. Here it is sufficient to recall that
Pomerons have no valence quarks. It has therefore been argued that
central production is a good place to search for glueballs
\cite{Robson:1977pm,Close:1987er}. Likewise, Pomeron-Reggeon exchange
may be well suited for the study of hybrid mesons.

\begin{figure}[pt]
\begin{center}
\includegraphics[width=0.65\textwidth]{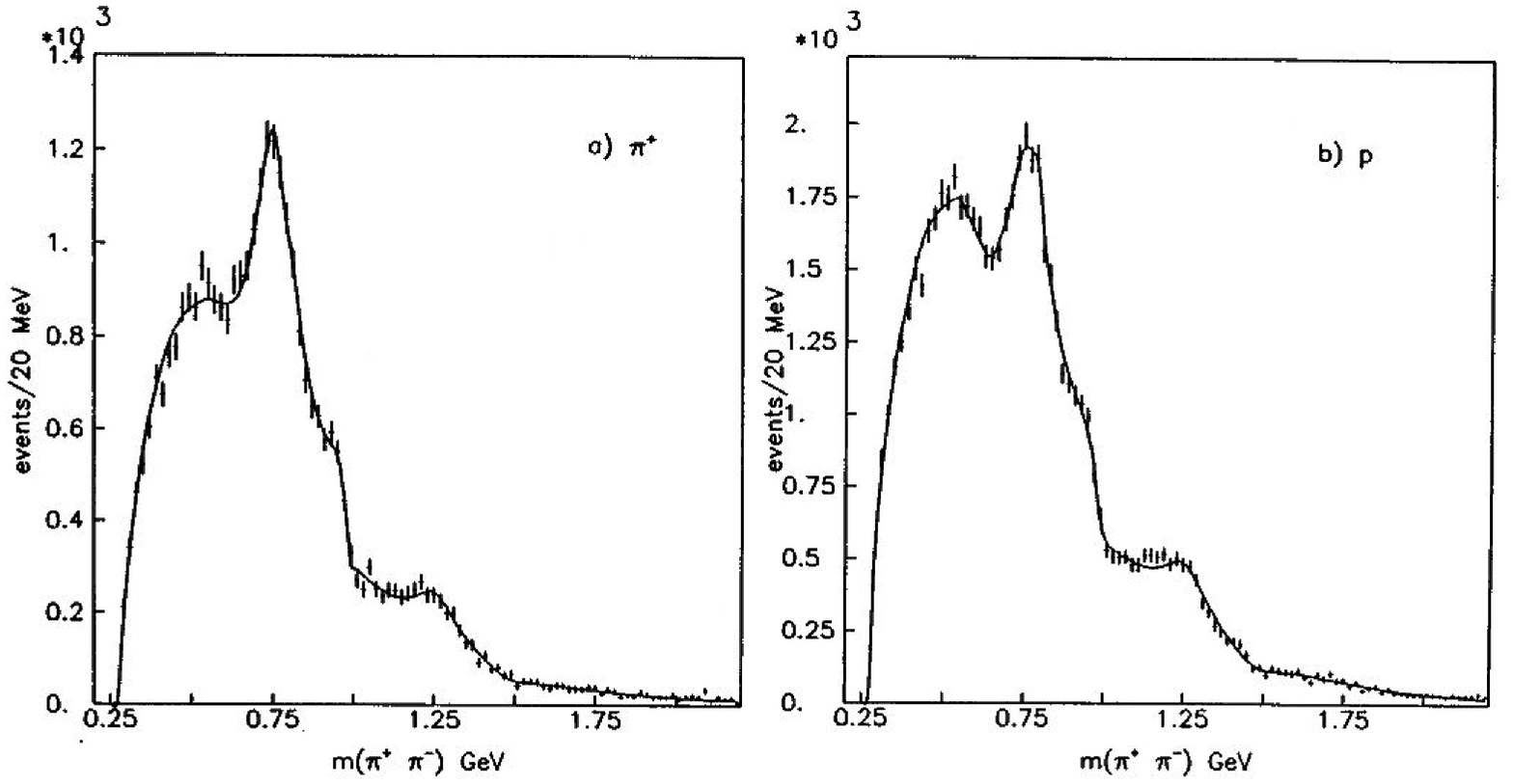}
\includegraphics[width=0.33\textwidth]{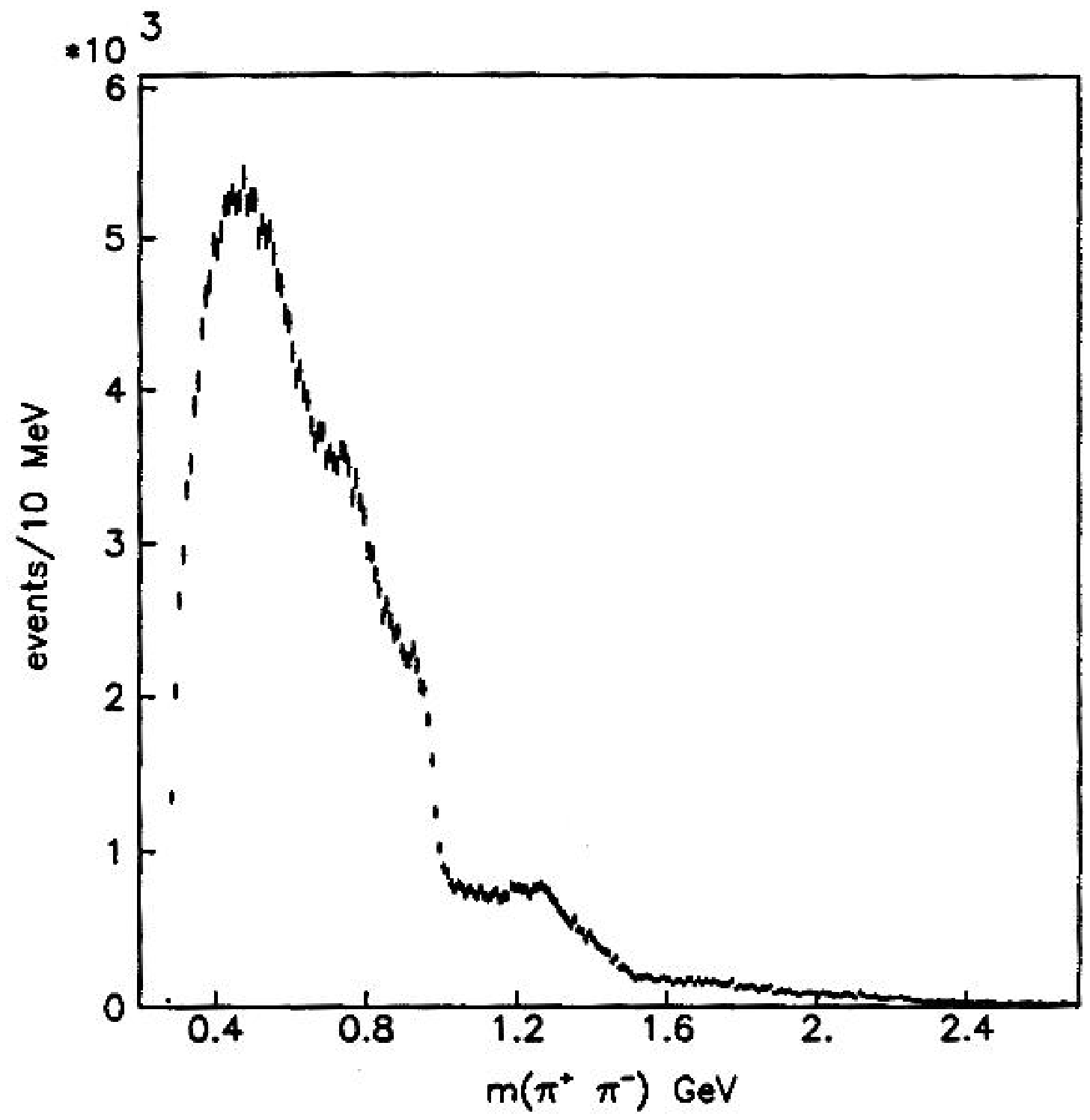}
\vspace{-50mm}

{\tiny
\hspace{135mm}c) p  \\
\hspace{30mm}85\,GeV/c \hspace{40mm} 85\,GeV/c \hspace{45mm}300\,GeV/c
\\ } \vspace{42mm} \end{center} \caption{The $\pi^+\pi^-$ invariant
mass distribution using a 85\,GeV/c pion (a), proton (b), or a
300\,GeV/c proton beam~\protect\cite{Armstrong:1991ch}. }
\label{fig:wa76-pipi}
\end{figure}

The suggested scaling of cross sections as a function of $\sqrt s$  is
evidenced in  Fig.~\ref{fig:wa76-pipi}. It shows the centrally produced
$\pi^+\pi^-$ invariant mass using a 85\,GeV/c proton and $\pi^+$ beam,
and a 300\,GeV/c proton beam \cite{Armstrong:1991ch}. A low-mass
enhancement is seen, followed by $\rho$ production and a sharp drop
of intensity due to $f_0(980)$. As expected, production of $\rho$
disappears with increasing $\sqrt s$. This is also seen when the pion
is replaced by a proton. The very fast decrease of the cross-section
with energy can be explained by a Reggeised pion-pion contribution to
$\rho$ production which decreases with energy as $\approx 1/s^2$. The
low-mass enhancement remains as dominant feature when $\sqrt s$
increases.

The production rate for $\eta^{\prime}$ production is reduced by a
factor $0.72\pm 0.16$ when going from WA92 ($\sqrt s=12.7$) to WA102
($\sqrt s=29.1$) \cite{Barberis:1999zh}. The reduction factor is nearly
compatible with the predicted 1 for Pomeron-Pomeron fusion and
incompatible with the prediction of 0.2 for $\rho\rho$
fusion\footnote{\tiny These numbers should not be stressed to
much.  In \cite{Close:1997nm}, a preliminary value $0.20\pm 0.05$ was
given for $\sigma_{\eta^{\prime}(\sqrt s=29.1\,\rm
GeV)}/\sigma_{\eta^{\prime}(\sqrt s=12.7\,\rm GeV)}$ and used to argue
that pseudoscalar meson production is suppressed in double Pomeron
exchange. The ratio for central production
$\sigma_{\eta}/\sigma_{\omega}=0.09\pm 0.01$ at $\sqrt s=29.1$\,GeV was
given in \cite{Barberis:1998ax} and compared to
$\sigma_{\eta}/\sigma_{\omega}=0.20\pm 0.02$ at $\sqrt s=12.7$\,GeV.
From Table \ref{wacompile}, this ratio is $0.52\pm 0.06$ at $\sqrt
s=29.1$\,GeV. The large discrepancies demonstrate the
difficulty of extracting absolute cross section. The systematic errors
were obviously not fully under control.}.

A large number of centrally produced final states X has been studied by
the WA102 collaboration, like $\pi , \eta ,\eta^{\prime}$
\cite{Barberis:1998ax}, $\pi\pi$ and $ K\bar K$
\cite{Barberis:1999an,Barberis:1999am,Barberis:1999ap,Barberis:1999cq},
$\eta\pi$ \cite{Barberis:2000cx}, $\eta\eta$ \cite{Barberis:2000cd},
$\eta\eta^{\prime}$ and $\eta^{\prime}\eta^{\prime}$
\cite{Barberis:1999id}, $3\pi$ \cite{Barberis:1998in,Barberis:2001bs},
$\pi^+\pi^-\eta$, $\rho\gamma$, and $\phi\gamma$
\cite{Barberis:1998by,Barberis:1999be}, $4\pi$
\cite{Barberis:1997ve,Barberis:1999wn,Barberis:2000em},
$ K\bar K\pi$ \cite{Barberis:1997vf}, $\omega\omega$
\cite{Barberis:2000kc}, $\phi\phi$ \cite{Barberis:1998bq}, $ K^*\bar
K^*$ and $\phi\omega$ \cite{Barberis:1998tv}, baryon-antibaryon
\cite{Barberis:1998sr} and charmonium states \cite{Barberis:2000vp}.
Earlier experiments investigated the same channels but had smaller
statistics. Here we discuss only the latest results and refer to
earlier publications only when needed.

At a first view, a lab momentum of 450\,GeV/c seems to guarantee that we
deal with an asymptotic hadron-hadron scattering situation. However,
this is not automatically the case. Let us consider central
production of events in which both scattered hadrons are protons and
loose the same energy in their centre of mass system. The two protons
with four-momenta $p_1, p_2$ have a total energy $\sqrt s$. After
scattering, the two protons have four momenta $p_3, p_4$, the central
system is described by the mass $M_X^2$ and four-momentum $p_5$. Then
$s=(p_1+p_2)^2,~~ s_1=(p_3+p_5)^2,~~ s_2=(p_4+p_5)^2$ where $s_1,s_2$
are the invariant masses squared of scattered protons with particle
$X$. When both protons scatter at zero angle, $s_1=s_2=M_X \sqrt s +
M_p^2 $ which is, for 450\,GeV/c lab momentum, about (5\,GeV)$^2/c^4$
at $M_X=1$\,GeV/c$^2$, and (2.2\,GeV)$^2/c^4$ at
$M_X=M_{\pi}$\,GeV/c$^2$. This is a regime where not only Pomeron
exchange but also other trajectories should make significant
contributions. Hence it is not surprising that both Regge-Regge and
Pomeron-Pomeron scattering contribute to central production under the
experimental conditions of the WA102 experiment.

A compilation of important cross sections~\cite{Kirk:2000ws} is
presented in Table~\ref{wacompile}. Pomerons are expected to have
positive parity and charge conjugation. Double Pomeron Exchange
should thus lead to production of isoscalar particles X with positive
$G$-parity; isovector particles require the exchange of Reggeons. The
production of mesons by Double Pomeron Exchange is predicted to yield a
fall-off with squared momentum transfer $t$ in the form $e^{bt}$
\cite{Ganguli:1980ec}. Scalar mesons (discussed in chapter \ref{Scalar
resonances from central production}) show compatibility an exponential
$e^{bt}$ distribution  but also axial vector mesons and some tensor
mesons including the $a_2(1320)$.

\begin{table}[!ht]
\caption{\label{wacompile}Compilation of cross sections for resonance
production at $\sqrt{s}$=29.1\,GeV
in the WA102 collaboration~\protect\cite{Kirk:2000ws}. The
error quoted represents the statistical and systematic errors summed in
quadrature.\vspace*{2mm}}
\label{summary}
\renewcommand{\arraystretch}{1.4}
\begin{center}
\begin{tabular}{cccccc}
\hline\hline
$J^{PC}$ & Resonance&$\sigma$ ($\mu$b) & $J^{PC}$ & Resonance&$\sigma$ ($\mu$b)\\
$0^{-+}$ & $\pi^0$     &22.0 $\pm$ 3.3 & $1^{--}$ & $\rho(770)$&3.1 $\pm$ 0.25 \\
         & $\eta$    & 3.9 $\pm$ 0.4  && $\omega(782)$&7.4 $\pm$ 0.6\\
         & $\eta^\prime$& 1.7 $\pm$ 0.2 &&$\phi(1020)$& 0.06 $\pm$ 0.02\\
\hline
$0^{++}$ &$a_{0}(980)$   & 0.64 $\pm$ 0.06 & $2^{++}$ &$a_{2}(1320)$  &1.7 $\pm$ 0.2\\
         & $f_{0}(980)$  &5.7 $\pm$ 0.5 & & $f_{2}(1270)$&3.3 $\pm$ 0.4\\
         & $f_{0}(1370)$  & 1.8 $\pm$ 0.6& &$f_{2}(1525)$& 0.07$\pm$ 0.01 \\
         & $f_{0}(1500)$  & 2.9 $\pm$ 0.3& &$f_{2}(1910)$  &0.53$\pm$ 0.04 \\
         & $f_{0}(1710)$  & 0.25 $\pm$ 0.07&&$f_{2}(1950)$&2.8 $\pm$ 0.18\\
         & $f_{0}(2000)$  &3.1 $\pm$ 0.5 & & $f_{2}(2150)$ & 0.12$\pm$0.2\\
\hline
$1^{++}$ &$a_{1}(1260)$  &10.0 $\pm$ 0.9 & $2^{-+}$ & $\pi_{2}(1670)$  &1.5 $\pm$ 0.15\\
         &$f_{1}(1285)$  &6.9 $\pm$ 1.3& &$\eta_{2}(1645)$ &1.9 $\pm$ 0.2\\
         &$f_{1}(1420)$  &1.1 $\pm$ 0.4 & & $\eta_{2}(1870)$ &1.9$\pm$0.2\\
\hline\hline \end{tabular} \renewcommand{\arraystretch}{1.0}
\end{center}
\end{table}

The largest cross section is given by single $\pi^0$ production. Its
large yield is explained by the large $\pi p$ scattering contribution
at small effective energies ($\sqrt{s_1}\approx \sqrt{s_2}\approx 2.2$
GeV/c$^2$). The strong yield of $f_0$ resonances and the large fraction
of high-mass tensor mesons in vector-vector final states are compatible
with Double Pomeron exchange. Double Pomeron Exchange is often
considered as flavour blind (even though structure functions reveal
that the nucleon contains more $\bar nn$ than $\bar ss$ quarks).
Production of $f_2(1525)$ is however strongly suppressed, by a factor
$\sim 50$ compared to $f_2(1270)$, $\phi$ production by $\sim 100$
compared to $\omega$: flavour symmetry is badly broken. Donnachie and
Landshoff \cite{Donnachie:1998gm} have suggested that there may be two
Pomerons; a so-called soft Pomeron with y axis intercept at
1.08\,GeV$^2$ on the Chew-Frautschi plot and a hard Pomeron with
intercept at $\sim 1.4$\,GeV$^2$. The soft Pomeron should have a weaker
coupling to heavier quarks, and thus SU(3) symmetry can be broken. For
Regge exchange, breaking of flavour symmetry is not a problem. Regge
trajectories of $s\bar s$ mesons have a higher intercept on the
Chew-Frautschi plot, and their exchange falls off with $s$ faster than
the exchange of normal Regge trajectories.

The abundant $a_1(1260)$ production definitely requires at least one
Reggeon. Apart from single $\pi^0$ production, it provides the largest
cross section. The isovector $a_1(1260)$ is observed in central
production with a larger cross section than the isoscalar $f_1(1285)$
though the two mesons have the same $J^{PC}$ and about the same mass.
Similarly, $\pi_2(1670)$ is produced with a similar rate as
$\eta_2(1645)$. The hypothesis that central production at $\sqrt
s=29.1$\,GeV is dominated by Pomeron-Pomeron collisions cannot hold
true for all reactions.

Further information can be obtained from $t$ distributions and from
angular distributions. The cross sections fall off with $t\cdot e^{bt}$
for mesons like $\eta$, $\eta^{\prime}$, $f_2(1270)$, $f_2(1525)$,
$\eta_2(1645)$, and $\eta_2(1870)$ while for other mesons (all
scalar mesons, $\rho$, $\Phi$, $a_1(1260)$, $f_1(1285)$, $a_2(1320)$), a
simple fall-off  with $e^{bt}$ as expected for Double Pomeron Exchange
is observed. Other mesons ($\pi^0$, $\omega$, $\eta_2(1645)$) show a
mixed behaviour, possibly indicating different reaction mechanisms. The
WA91 collaboration noticed that the yield of centrally produced
resonances depends on the vector {\it difference} of the transverse
momentum recoil of the final state protons \cite{Barberis:1996tu}. For
fixed four-momentum transfers in the transverse plane, $t_1 \sim
-k_{T1}^2, \ t_2 \sim -k_{T2}^2$, the quantity $dk_T \equiv
|\vec{k}_{T1} - \vec{k}_{T2}|$ can vary from configurations where two
protons are scattered into the same direction or into opposite
directions. Generalising this concept, an angle $\phi$ (in the plane
perpendicular to the beam) between slow and fast proton can be defined
\cite{Close:1997us}.

The $dk_T$ (or $\phi$) dependence of central production has been
studied intensively in nearly all WA102 publications; its implications
are discussed in a series of papers by Close and collaborators
\cite{Close:1997pj,Close:1997nm,Close:1997us,Close:1999is,Close:1999bi,%
Close:2000dx,Close:2000yk}. The study revealed two surprises. First,
many $q\bar q$ mesons show an angular distribution as those produced in
double-tag events in two-photon fusion. The exchanged particle must
have J~$>$~0, J~=~1 is the simplest assumption. Using $\gamma^*
\gamma^*$ collisions as an analogy, Close and Sch\"uler
\cite{Close:1997nm} calculated the $\phi$ dependencies for the
production of resonances with different quantum numbers. They have
found that for a $J^{PC}$~=~$0^{-+}$ state \begin{equation}
\frac{d^3\sigma}{d\phi dt_1dt_2} \propto t_1 t_2 sin^2 \phi .
\end{equation}
For the  $J^{PC}$~=~$1^{++}$ states the model predicts that
$J_Z$~=~$\pm1$ should dominate, and
\begin{equation}
\frac{d^3\sigma}{d\phi dt_1dt_2}\propto (\sqrt{t_2} - \sqrt{t_1})^2 + \sqrt{t_1
t_2}sin^2 \phi/2 \ .
\label{1pp}
\end{equation}

The WA102 data were found to be consistent with these calculated
angular distributions. If Pomeron-Pomeron scattering were the leading
contribution in central production, the angular distribution would
imply a vectorial interaction of Pomerons \cite{Close:1997us}. Here we
have to stress that these $\phi$ dependencies are of merely kinematical
origin in Regge phenomenology \cite{Kaidalov:2003fw}. In
\cite{Petrov:2004hh}, the concept is generalised to exchanges of mesons
with arbitrary spin with subsequent reggeisation. Absorptive effects
are taken into account by unitarisation of the production amplitudes.

The angular distributions revealed a further unexpected phenomenon. For
some states, the angular distributions were found to be markedly
different from those for most well established $q\bar q$ mesons having
the same quantum numbers. It has been suggested that the angular
distribution identifies intrinsic angular momenta \cite{Close:1997pj}.
Thus scalar and tensor mesons in $ ^3P_0$ or $ ^3P_2$ $q\bar q$
configurations with internal orbital angular momentum $L=1$ behave
differently than glueballs or $ K\overline K$ molecules which need
no intrinsic angular momentum. Thus it is suggested that a cut in
$dk_T$  may act as a {\it glueball filter} separating non-$q\bar q$
objects from ordinary $q\bar q$ mesons. This question will be resumed
in  section \ref{Scalar resonances from central production} on
scalar mesons in central production.

A final word of caution seems appropriate on the assumption that
central production is dominated by Pomeron-Pomeron fusion. This
assumption relies on the evolution of $\rho$ and $\eta^{\prime}$
production with $s$. While $\rho$ production decreases with
increasing $s$ indicating Regge exchange as production mechanism,
$\eta^{\prime}$ production remains approximately constant. This is a
behaviour predicted for Pomeron-Pomeron fusion \cite{Ganguli:1980ec}.
The large yield of isovector mesons in central production makes
it however unlikely that Pomeron-Pomeron fusion is the dominant
feature of central production. Fusion of two Regge trajectories must
play an important $\rm r\hat{o}le$. The scattering of two vector mesons
can create scalar, pseudoscalar or axial vector and tensor mesons, and
also $\eta_{2}$ and $\pi_{2}$ couple to $\omega\omega$ (and $\rho\rho$)
and to $\rho\omega$, respectively. For  $J^{PC}=0^{-+}$, $1^{++}$,
$2^{-+}$ mesons, the high yield of isovector mesons and the angular
distribution make it likely that at $\sqrt s=30$\,GeV, the dominant
mechanism for central production is still fusion of two Regge
trajectories. For scalar mesons we notice an one order of magnitude
reduction for the $a_0(980)$ cross section compared to $f_0(980)$
production; the $a_0(1450)$ remained unobserved \cite{Barberis:2000cx}
while $f_0(1500)$ is seen. The $t$ dependence of scalar mesons shows
the expected $e^{bt}$ behaviour. Hence scalar mesons are compatible
with Pomeron-Pomeron fusion as main source of their production
mechanism. The $\rm r\hat{o}le$ of tensor mesons is less clear. The
$a_2(1320)/f_2(1270)$ production ratio is about 1/2. At higher masses,
little is known about isovector states. Isoscalar tensor mesons from
central production of $\rho\rho$, $\rho\omega$, $\omega\omega$ and
$\phi\omega$ have been reported by WA102
\cite{Barberis:2000kc,Barberis:1998bq,Barberis:1998tv} but data on
$\rho\omega$ are missing. WA76, using a 300\,GeV/c proton beam, studied
the centrally produced $\omega\rho$ and $\omega\omega$ systems. They
found $304\pm50$ $\omega\rho$ events. From their distributions one can
estimate $200\pm 100$ $\omega\omega$ events. Since $\omega\rho$
requires at least one Regge trajectory to participate, there is no real
evidence that two centrally produced vector mesons originate from
Pomeron-Pomeron fusion.  Obviously, at $\sqrt s=29.1$ the regime of
double Pomeron exchange is not yet fully reached and depends on the
quantum numbers of the produced system: scalar meson production is
compatible in all aspects with Pomeron-Pomeron fusion, production of
tensor mesons seems to receive contributions from both, Regge and
double Pomeron exchange while pseudoscalar, vector and axial vector
mesons are still in the regime of Reggeon exchange.

\subsubsection{\label{The Deck effect}
The Deck effect}

In the study of quasi-two body reactions one has to take into account
production of resonances as well as nonresonant production of final
states. A specific model was proposed by Deck ~\cite{Deck:1964hm} to
describe nonresonant effects in quasi-two body reactions at high
energy. Consider e.g. the reaction $\pi^-p\to p\,\rho \pi$ in the
region of the $a_1(1260)$ meson, with the $\rho \pi$ system having $I^G
J^{PC}=1^-1^{++}$ quantum numbers.
\begin{figure}[!ht]
\bc
\includegraphics[width=0.6\textwidth]{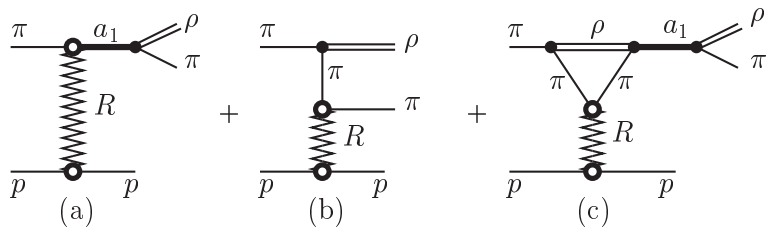}
\ec
\caption{\label{pic:deck}
Direct resonance production (a)and the Deck mechanism via $\pi$ exchange
(b) and via $\pi$ exchange and rescattering (c).}
\end{figure}
This reaction can proceed via direct $a_1$ production (Fig.~\ref{pic:deck} (a)),
via $\pi$ exchange (Fig.~\ref{pic:deck} (b)) and via $\pi$ exchange with
rescattering (Fig.~\ref{pic:deck} (c)). The Deck model amplitude
has the form \cite{Bowler:1978jd}:
\begin{equation}
\frac{A}{k}\cdot e^{i\delta}\cdot sin \delta + B\cdot e^{i\delta}\cdot (cos
\delta + \frac{\alpha}{k}\cdot sin \delta)
\end{equation}
where $A$ represents $a_1(1260)$ production and $B$ the Deck amplitude
with a pion propagator. The elastic $\rho\pi$ phase shift $\delta$ is
given by a Breit-Wigner amplitude,  $k$ is the $\rho \pi$ phase
space. The predictions of this model are far from being trivial. It
gives the intensities and phases of nonresonant contributions in
different final $J^{PC}$ states with different isobars as well as the
$t$ dependence of all these contributions. In particular with
increasing squared momentum transfer $|t|$, the nonresonant
contribution has to decrease since the nonresonant $\rho \pi$ system
has a larger spatial extension than the $a_1(1260)$. In diffractive
reactions, the Deck model predictions are in reasonable agreement with
experimental results \cite{Daum:1979iv}.

\subsubsection{\label{meth:Antiproton-proton annihilation}
Antiproton-proton annihilation}

In $N\bar N$ annihilation,  a nucleon and an antinucleon undergo a
transition from the baryonic world to the quanta of strong
interactions, to mesons. Even for annihilation at rest, mesons of up to
1.7\,GeV/c$^2$  can be produced. The mass of meson resonances carrying
open strangeness is restricted to about 1.4\,GeV/c$^2$.  The OZI rule
disfavours production of mesons with hidden strangeness even though in
some reactions, $\phi$ production was found to surprisingly large.
There is an abundant literature on the question if the large $\phi$
production rate signals an $s\bar s$ component in the proton wave
function \cite{Ellis:1994ww}, or the $\phi$'s are produced via
rescattering of Kaons in the final state
\cite{Locher:1994cc,Anisovich:1996xr}. See \cite{Donnachie:2004qk} for
a review.

A large number of meson resonances has been studied in
$N\bar N$ annihilation at rest \cite{Amsler:1998up,Amsler:2004ps} and
in flight \cite{Bugg:2004xu}. The annihilation process and the
formation of $\bar pp$  atoms preceding annihilation has been reviewed
in some detail in \cite{Klempt:2002ap,Klempt:2005pp}; here we emphasize
a few points relevant for the further discussion.

In $ N\bar N$ annihilation at rest in H$_2$ or D$_2$,
annihilation is preceded by capture of an antiproton by a hydrogen or
deuterium atom. Collisions between the protonium atom and surrounding
molecules induce transitions from high orbital angular momentum states
via Stark mixing; and this mixing is fast enough to ensure dominant
capture from $S$-wave orbitals when antiprotons are stopped in liquid
H$_2$ or D$_2$. In gas, the collision frequency
is reduced and $P$-wave annihilation makes significantly larger
contributions. In particular at very low target pressures the $P$-wave
fractional contribution is very large. Alternatively, rather pure
samples of $P$-wave annihilation can also be studied by coincident
detection of X-rays emitted in the atomic cascade of the $\bar pp$  system
(which feed mostly the 2$P$ level).

Proton and antiproton both carry isospin $ |I,I_3> =
|\frac{1}{2},\pm\frac{1}{2}>. $ The two isospins couple to
$|I=0,I_3 = 0>  \quad {\rm or} \quad
|I=1,I_3 = 0 >  \quad {\rm with} \quad I_3 = 0$.
In the absence of initial state interactions between
proton and antiproton the relation
$$
\bar pp =\sqrt{\frac{1}{2}}\left(|I=1,\,I_3=0> + |I=0,\,I_3=0>\right).
$$
holds. One could expect processes of the type $\bar pp\to \bar nn$ in
the initial state but charge exchange - which would lead to unequal
weight of the two isospin components - seems to be not very important
\cite{Abele:2000xt}.

The quantum numbers of the $\bar pp$  system are the same as those for
$q\bar q$ systems; both are bound states of fermion and anti-fermion.
Since isospin can be $I=0$ or $I=1$, every atomic $\bar pp$  state may
have $G$-parity +1 or -1. Table \ref{ini} lists the quantum numbers of
atomic levels from which annihilation may occur. The $\bar pn$ system
has always $I=1$, and every atomic level has defined $G$-parity. For
annihilation into specific final states, selection rules may restrict
the number of initial states. Annihilation into any number of \piz\ and
\etg\ mesons is e.g. forbidden from all initial states with negative
$C$-parity .

\begin{table}[htb]
\caption{\label{ini}
Quantum numbers of levels of the $\bar pp$  atom
from which annihilation may occur.\vspace{2mm}}
\begin{center}
\renewcommand{\arraystretch}{1.4}
\begin{tabular}[t]{c l l | c l l}
\hline\hline
$^{2S+1}L_J$ &
\multicolumn{2}{c|}{$I^{G} \left(J^{PC}\right)$} &
$^{2S+1}L_J$ &
\multicolumn{2}{c}{$I^{G} \left(J^{PC}\right)$} \\
\hline
$^1 S_0$ & $1^- (0^{-+})$ & $0^+ (0^{-+})$ &
$^1 P_1$ & $1^+ (1^{+-})$ & $0^- (1^{+-})$ \\
$^3 S_1$ & $1^+ (1^{--})$ & $0^- (1^{--})$ &
$^3 P_0$ & $1^- (0^{++})$ & $0^+ (0^{++})$ \\
\hline
&&&$^3 P_1$ & $1^- (1^{++})$ & $0^+ (1^{++})$ \\
&&&$^3 P_2$ & $1^- (2^{++})$ & $0^+ (2^{++})$ \\
\hline\hline
\end{tabular}
\renewcommand{\arraystretch}{1.0}
\end{center}
\end{table}

Unlike other processes, annihilation dynamics has no preference of
producing low-mass mesons compared to mesons having a large mass. In
central production or two-photon fusion, the meson mass spectrum falls
off with $1/M^2$ (with $M$ being the meson mass); there is no similar
suppression of high masses in $p\bar p$ annihilation. Rather, $N\bar N$
annihilation can be described as if the final state would be produced
as `white' spectrum distributed according to the phase space which then
needs to be multiplied  with a dynamical weight accounting for
final-state interactions. Experiment shows \cite{Amsler:1993kg} that
the angular momentum barrier (for $\ell =0,1$ or $2$) does not play a
significant $\rm r\hat{o}le$: the ratio of $\omega\eta^{\prime}$ to $\omega\eta$
annihilation frequencies is about 1/2. The angular momentum in the
reaction is $\ell =1$, the linear momenta differ by a factor 2. Hence a
reduction in the order of $2^{2\ell+1}=8$ due to phase space and
angular momentum barrier should be expected, in addition to a factor 0.5
accounting for the larger $n\bar n$ component of the $\eta$ wave
function (compared to the $\eta^{\prime}$ wave function). The reduction
found experimentally is about 0.5,  compatible with the smaller $n\bar
n$ component of the $\eta^{\prime}$ wave function from which the $\bar
pp$ system decouples. Likewise, production of $a_2(1320)\pi$ --
requiring $\ell=2$ between  $a_2(1320)$ and $\pi$ -- is as frequent as
production of $\rho\pi$ with  $\ell=1$.

The initial $N\bar N$ state contains three constituent quarks and three
antiquarks. In annihilation into two mesons, two $q\bar q$ pairs are
going out. $N\bar N$ annihilation is a superposition of quark-antiquark
annihilation and rearrangement diagrams, with an intermediate state of
quarks and gluons. One may therefore expect $N\bar N$ annihilation to be a
good process to search for multiquark (tetraquark- or baryonium)
states and for hybrids.  Glueballs can be produced by their mixing with
$q\bar q$ mesons.

\subsubsection{\label{Flavour tagging}
Flavour tagging}

\begin{figure}[ph]
\begin{center}
\includegraphics[width=0.7\textwidth]{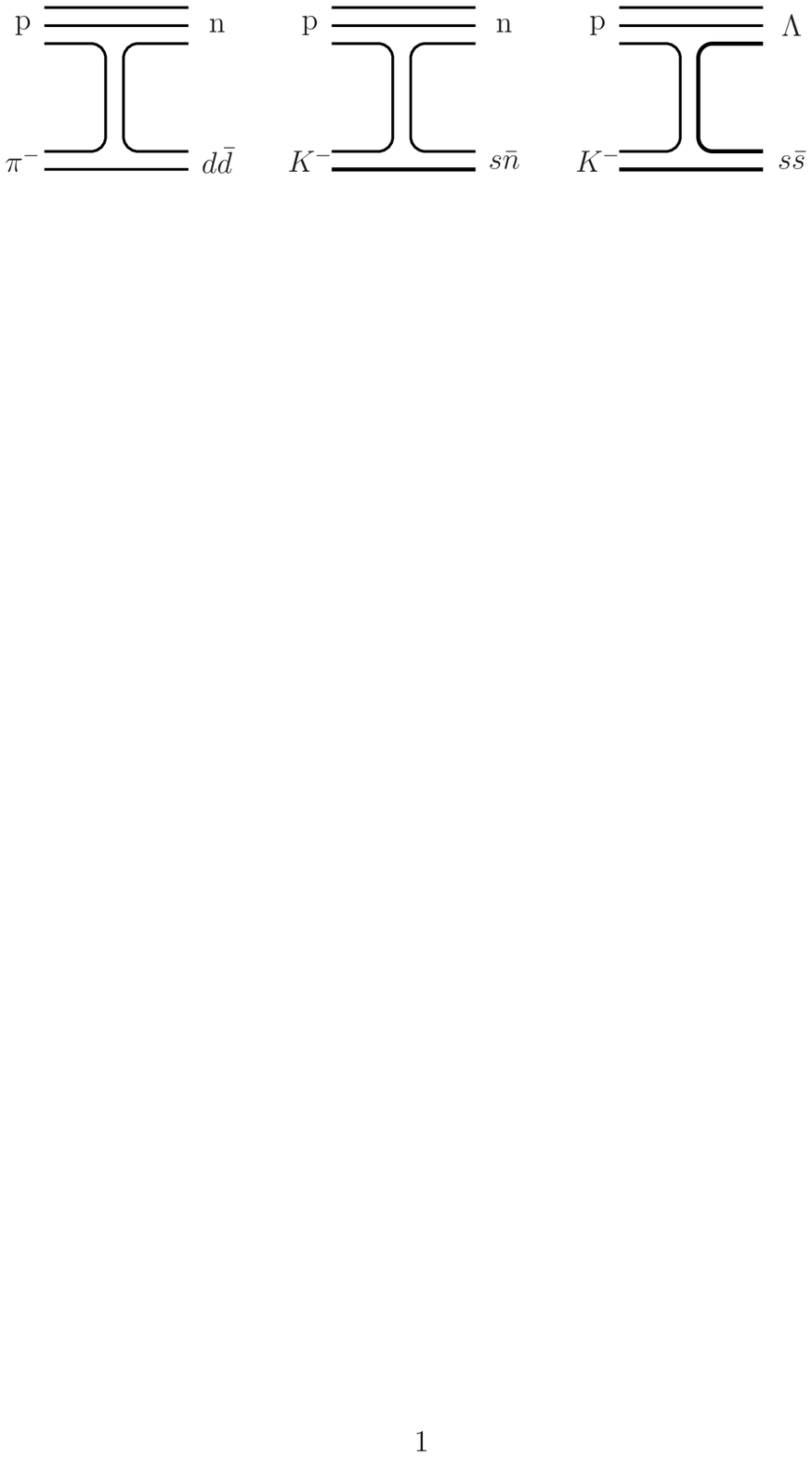}
\vspace*{-2mm}\end{center}
\caption{\label{flavour-scatt}
Flavour tagging in scattering experiments. Up and down quarks are
shown as thin, strange quarks as thick lines. In the first two reactions,
a charged $n\bar n$ meson is exchanged, and the meson in the
final-state meson is neutral and carries the strangeness of the
incoming pion or Kaon. If a hyperon is detected in coincidence, the
produced meson is likely to carry hidden strangeness.
}
\end{figure}

\paragraph{Charge and strangeness exchange scattering:} In the charge
exchange reaction, $ \pi^- + proton  \to neutron + meson$, charged
mesons like $\pi$ or $b_1$ (unnatural parity exchange) or $\rho /a_2$
(natural parity exchange) are exchanged and mesons are produced with a
preferred $n\bar n$ flavour structure. For incident $ K^-$, the
baryon in the final state can be neutron or a $\Lambda$. Under these
experimental conditions, mesons with open ($n\bar s$) and with hidden
($s\bar s$) flavour are produced dominantly, see
Fig.~\ref{flavour-scatt}. This method was extensively exploited by the
LASS collaboration~\cite{Dunwoodie:1990xe,Dunwoodie:1990xd}. At high
energies, the incoming meson can be excited without changing its
flavour structure. Such processes are often called Pomeron exchange
reaction even though $f_2$ exchange leads to the same flavour in the
final state.

\begin{figure}[pb]
\begin{center}
\includegraphics[width=0.6\textwidth]{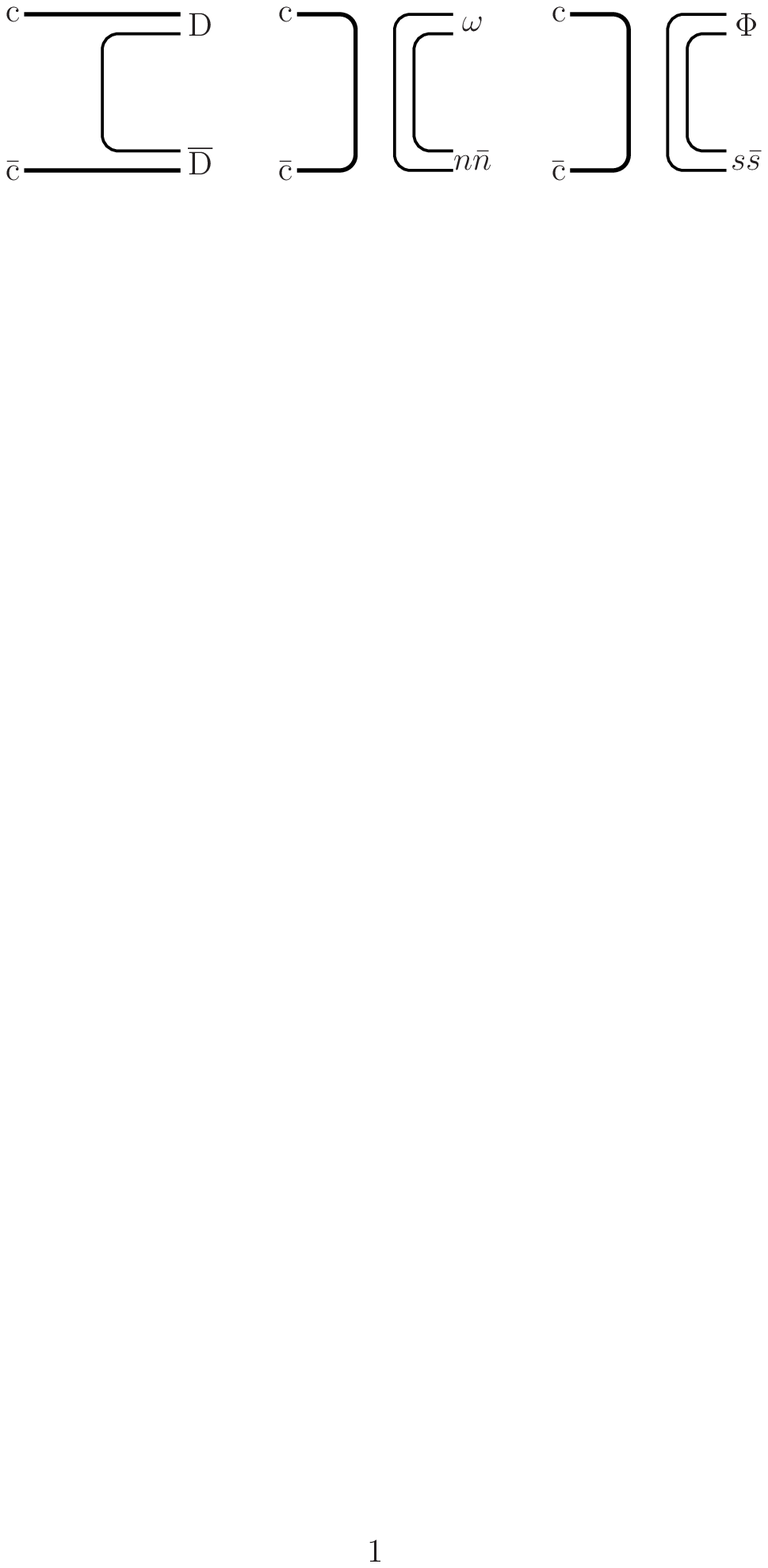}
\vspace*{-2mm}\end{center}
\caption{\label{ddecay}Decays of charmonium states into $\rm\bar DD$
are allowed only above the  $\rm\bar DD$ threshold, the J/$\psi$ (and
the $\eta_c$ and $\chi$ states) can decay only into light quarks. A
$\omega$ or $\Phi$ signal determines the $\bar uu +\bar dd$ and
$\bar ss$ component, respectively, of the recoiling meson. The
thick lines represent charmed quarks. }
\vspace*{-4mm}
\end{figure}
\paragraph{J/$\psi$ decays:}
The mass of the J/$\psi$ is below the threshold for decay into
mesons having {\it open} charm, hence the $c\bar c$ pair must annihilate
and new particles have to be created (see figure~\ref{ddecay}). Such
processes are suppressed. The $c\bar c$ bound state converts into
gluons carrying a large four-momentum, and the coupling is small. Hence
the J/$\psi$ is narrow. The OZI rule can be exploited to tag the
flavour of mesons produced in J/$\psi$ decays in cases where one of the
two mesons has a known flavour content. If it is a $\bar uu +\bar dd$
meson like the $\omega$, the recoiling meson is produced via its $\bar
uu +\bar dd$ component. If a $\phi(1020)$ is produced, the recoiling
meson is produced through its $\bar ss$ component. In this way, the
flavour structure of mesons can be determined. This was done e.g. for
the $\eta$ and $\eta^{\prime}$ mesons
~\cite{Coffman:1988ve,Jousset:1988ni} and led to the pseudoscalar
mixing angle as discussed in section \ref{Mixing}. Flavour tagging of
scalar mesons has led to surprising results which will be discussed in
section \ref{Scalar mesons in hadronic J/psi decays}.

\par
\paragraph{$D$, $D_s$ and $B$ decays:}
Due to their large mass, $D$ and $D_s$ decays are well suited to study
the light meson mass spectrum. A number of different three- and
four-body final states has been analysed and interesting results,
particularly on scalar mesons, were obtained. The results from partial
wave analyses will be discussed in section \ref{Scalar mesons in D, Ds
and B decays}. Here we restrict ourselves to general considerations.

\begin{figure}[H]
\bc
\begin{tabular}{cccc}
\includegraphics[width=0.22\textwidth,height=24mm]{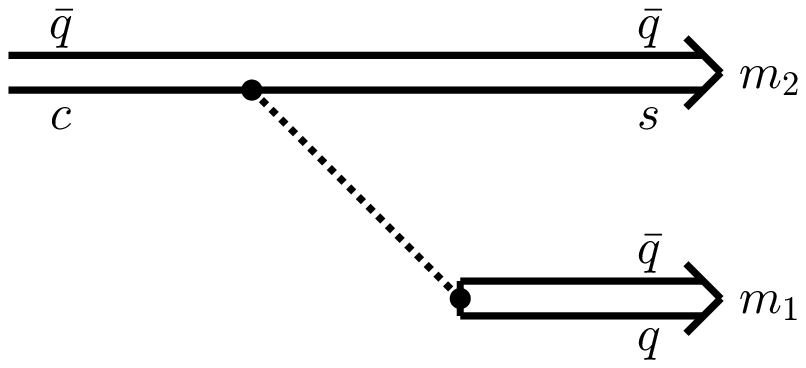}&
\hspace{3mm}\includegraphics[width=0.22\textwidth,height=24mm]{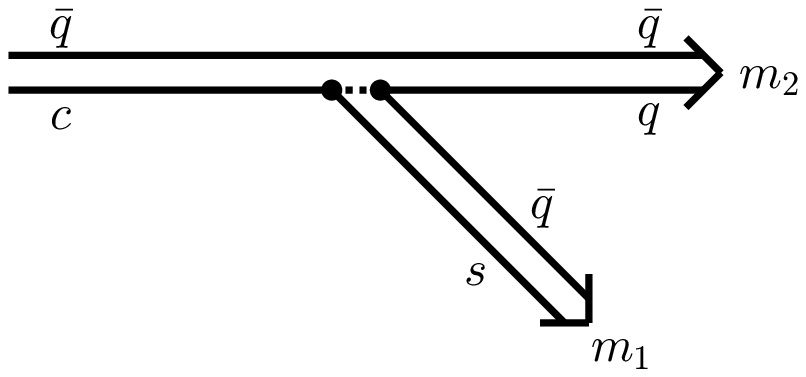}&
\hspace{3mm}\includegraphics[width=0.22\textwidth,height=24mm]{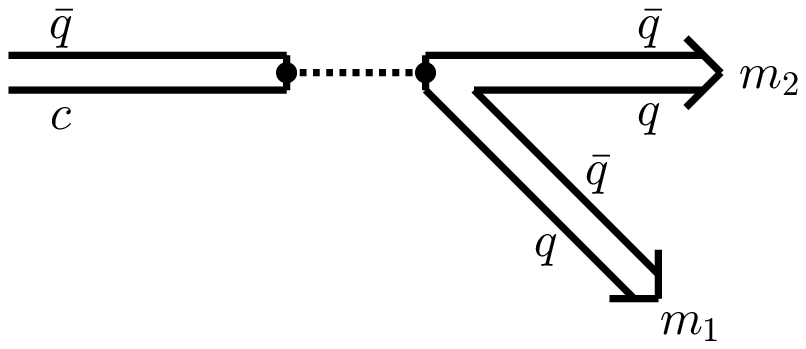}&
\hspace{3mm}\includegraphics[width=0.22\textwidth,height=24mm]{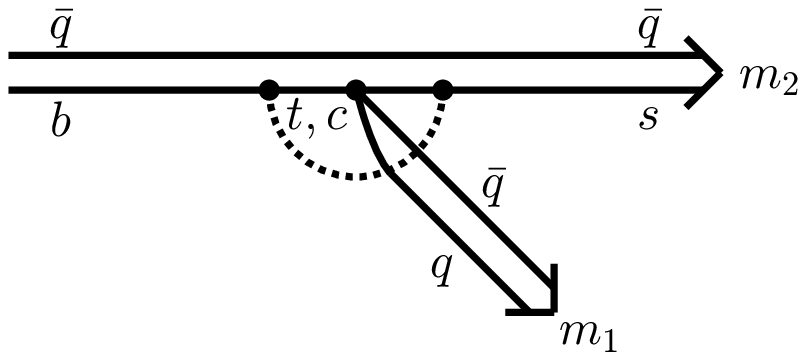}
\end{tabular}

(a)\hspace{40mm}(b)\hspace{40mm}(c)\hspace{40mm}(d)
\ec
\caption{\label{fig:d-decay} Decays of
charmed meson: a) leading mechanism, b) colour suppressed diagram, c)
annihilation diagram. Here, $q$ can be $d$ or $s$ quarks. d) $B$
decays into light quarks prefer penguin diagrams to which $b$ quarks
contribute, too. } \end{figure}

$D$ and $D_s$ mesons may decay via different mechanisms depicted in
Fig.~\ref{fig:d-decay}. A $c$ quark converts into a quark $q$
($q=d,s,b$) by emission of a virtual $W^+$. The $W^+$ decays into a
$q\bar q$ meson. The $q\bar q$ meson may escape; this is the leading
diagram (a). Quark and antiquark may split to find a antiquark-quark
pair. Now the colours have to match, the process is colour suppressed
(b). Finally, the initial $c\bar q$ may annihilate directly into a
virtual $W^+$ (c). The diagrams can be Cabibbo allowed or suppressed.
The $c$ quark prefers to convert into an $s$ quark, conversion into a
$d$ quark is suppressed by the matrix element $|V_{cd}|^2=(0.224\pm
0.012)^2$ of the CKM matrix. Likewise, the $W^+$ can convert into
$u\bar d$ or $u\bar s$ where the latter is Cabibbo suppressed, too. We
estimate the relevance of the contributions by comparing branching
ratios of simple reactions.

$D^0=c\bar u$ mesons cannot decay via the annihilation diagram (c). In
colour allowed decays,  the $W^+$ converts into a positively charged
meson; in an Cabibbo allowed decay mode, this is a $\pi^+$. The $c$
quark converts in a Cabibbo allowed transition into an $s$ quark making
a $K^-$ by picking up the 'spectator' $\bar u$ quark. In colour
suppressed decays, the strange quark picks up the $u$ quark from
internal $W^+$ conversion, the $\bar d$ and the $s$ quark form a $\bar
K^0$ or $\bar K^{0*}$. Table~\ref{tab:d-decay} compares these two types
of decay branching ratios. The colour-suppressed reactions are observed
with branching ratios which are smaller by a factor 2. This factor
depends on the particular reaction, the factor 2 provides an order of
magnitude estimate for further discussions.

\begin{table}[H]
\caption{\label{tab:d-decay} Branching ratios for $D^0$ decays
\protect\cite{Eidelman:2004wy}. \vspace*{2mm} } \bc
\renewcommand{\arraystretch}{1.3}
\begin{tabular}{cccc}
\hline\hline
\multicolumn{2}{c}{colour allowed}& \multicolumn{2}{c}{colour suppressed}\\
$\rm\pi^+ K^-$ & $(3.80\pm 0.09)$\%&$\rm\pi^0\bar K^0$&$(2.30\pm0.22)$\% \\
$\rm\rho^+ K^-$ & $(10.1\pm
0.8)$\%&$\rm\rho^0\bar K^0$&$(1.55^{+0.12}_{-0.16})$\% \\
                                &&$\rm\omega\bar K^0$&$(2.3\pm0.4)$\%\\
$\rm\pi^+ K^{*-}$& $(5.9\pm 0.5)$\%&$\rm\pi^0\bar  K^{*0}$&$(2.8\pm0.4)$\% \\
\hline\hline
\end{tabular}
\ec
\renewcommand{\arraystretch}{1.0}
\end{table}

In a similar spirit we derive an estimate for the fraction of
annihilation in $D_s^+$ decays. The fraction of leptonic and
semileptonic decay modes is about 25\%, the branching ratio for
$D_s^+\to \tau^+\nu_{\tau}$ is ($6.4\pm 1.5$)\%. For $D_s^+\to
\tau^+\nu_{\tau}$, helicity conservation plays no $\rm r\hat{o}le$ since
$\frac{v}{c}=0.09$; decays into lighter particles are suppressed. For
constituent quark masses of 350\,MeV, $\beta\sim 0.85$.  There is
the colour factor 3 but still, we expect an annihilation contribution
to hadronic decay modes of a few percent only.

\begin{figure}[H]
\bc
\includegraphics[width=0.5\textwidth]{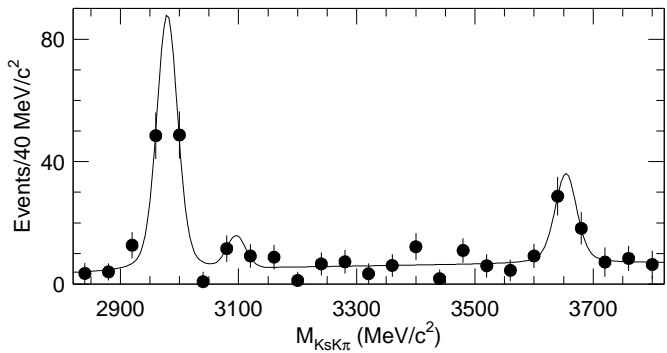}
\ec
\caption{\label{fig:belle-eta_c}
The $ K^0_S K\pi$ mass distribution from $B^{\pm}\to K^{\pm}\,(K^0_S
K\pi)$ events. The two peaks are assigned to $B^{\pm}\to\rm
K^{\pm}\,\eta_C(1S)$ and $B^{\pm}\to K^{\pm}\,\eta_C(2S)$ events
\protect\cite{Choi:2002na}. } \end{figure}

Due to its high mass, $B$ decays offer access to a wide range of
spectroscopic  issues. In $B$ decays, the first three diagrams
in Fig.~\ref{fig:d-decay} lead to charmed mesons. Here we use one
example, the $\eta_{c}(2S)$, to show the power of this method. The BELLE
collaboration searched for the $\eta_{c}(2S)$ in the reaction
$B^{\pm}\to K^{\pm}\eta_{c}(2S)$ and $B^{0}\to K^0_S\eta_{c}(2S)$ with
$\eta_c(2S)$ decaying into  $\rm K^{\pm}\pi^{\mp}K^0_S$
\cite{Choi:2002na}. After cuts to reduce background from non $B$
events, a spectrum shown in Fig.~\ref{fig:belle-eta_c} is obtained. The
fit to the mass spectrum yields $39 \pm 11$ events centred at $3654
\pm 6$\,MeV/c$^2$.  The width was determined to $15^{+24}_{-15}$\,MeV/c$^2$.  The
properties of the resonance, mass, width, production and decay mode,
are compatible with it being the $\eta_{c}(2S)$. The peak was confirmed
by BaBaR \cite{Aubert:2003pt} and CLEO \cite{Asner:2003wv} in
$\gamma\gamma$ fusion; the angular distributions favour $0^{-+}$ so
that the interpretation of the peak as $\eta_{c}(2S)$ seems to be
settled.

Production of light mesons by one of the diagrams in Fig.
\ref{fig:d-decay} proceeds via the matrix element  $|V_{bu}|$ which is
only $(0.00367\pm 0.00047)$. The (rare) hadronic decays into final
states containing no $c$ quarks make use of so-called Penguin diagrams
(Fig. \ref{fig:d-decay}d) in which the $b$ quark converts into a $u,c$
or $t$ quark by emission of a $W^-$ which is reabsorbed to produce a
$d$ or $s$ quark.

Flavour symmetry can be invoked to relate charmless two-body decay
modes \cite{Gronau:2006eb} into two pseudoscalar mesons. The large
branching ratio for reactions like $B\to K\eta^{\prime}$ can be
understood when a phenomenological flavour singlet amplitude is
introduced which takes into account the {\it $\eta^{\prime}$ affinity}
to gluons in the initial state. This observation motivated
Minkowski and Ochs \cite{Minkowski:2004xf} to extend their study to
$B$ to scalar- and pseudoscalar-meson decay modes and to discuss
scalar glueball contributions in $B$ decays.

\subsection{\label{Meson resonances in formation experiments}
Meson resonances in formation experiments}

\subsubsection{\label{e+e- annihilation}
$e^+e^-$ annihilation}

A typical example of a {\it formation} experiment is
annihilation of a $e^+e^-$ pair into hadrons.
The energies of an electron and positron beam
(in $e^+e^-$ storage rings these are often the same, except for the $b$
factories) are tuned to the desired value and the probability to
produce hadronic final states is determined. The cross sections are
then obtained by scanning the $e^+e^-$ total energy $\sqrt s$ over the
desired energy range.

\begin{figure}[H]
\bc
\begin{tabular}{cc}
\hspace{-3mm}\includegraphics[width=0.65\textwidth]{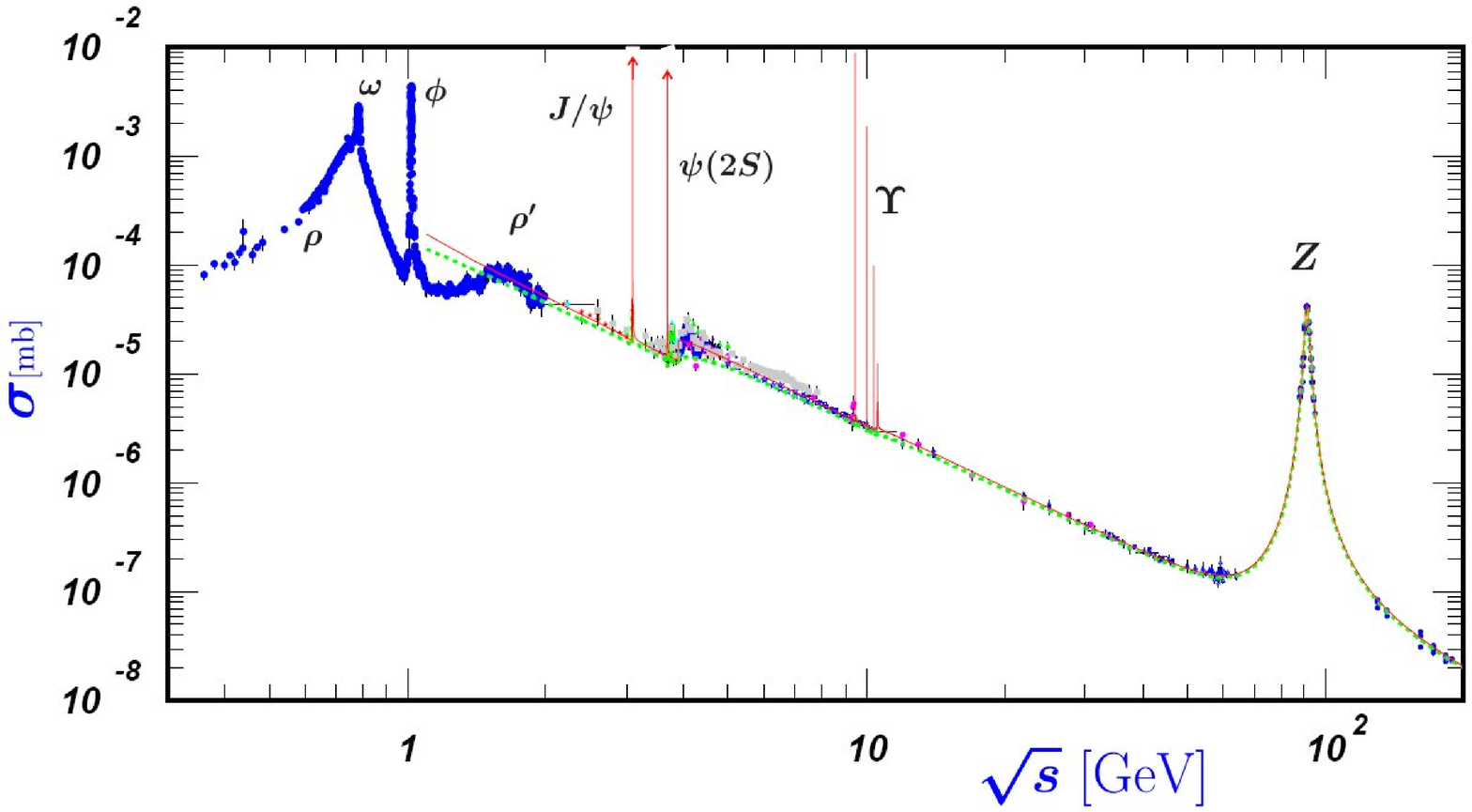}&
\hspace{-1mm}\includegraphics[width=0.34\textwidth,height=0.34\textwidth]{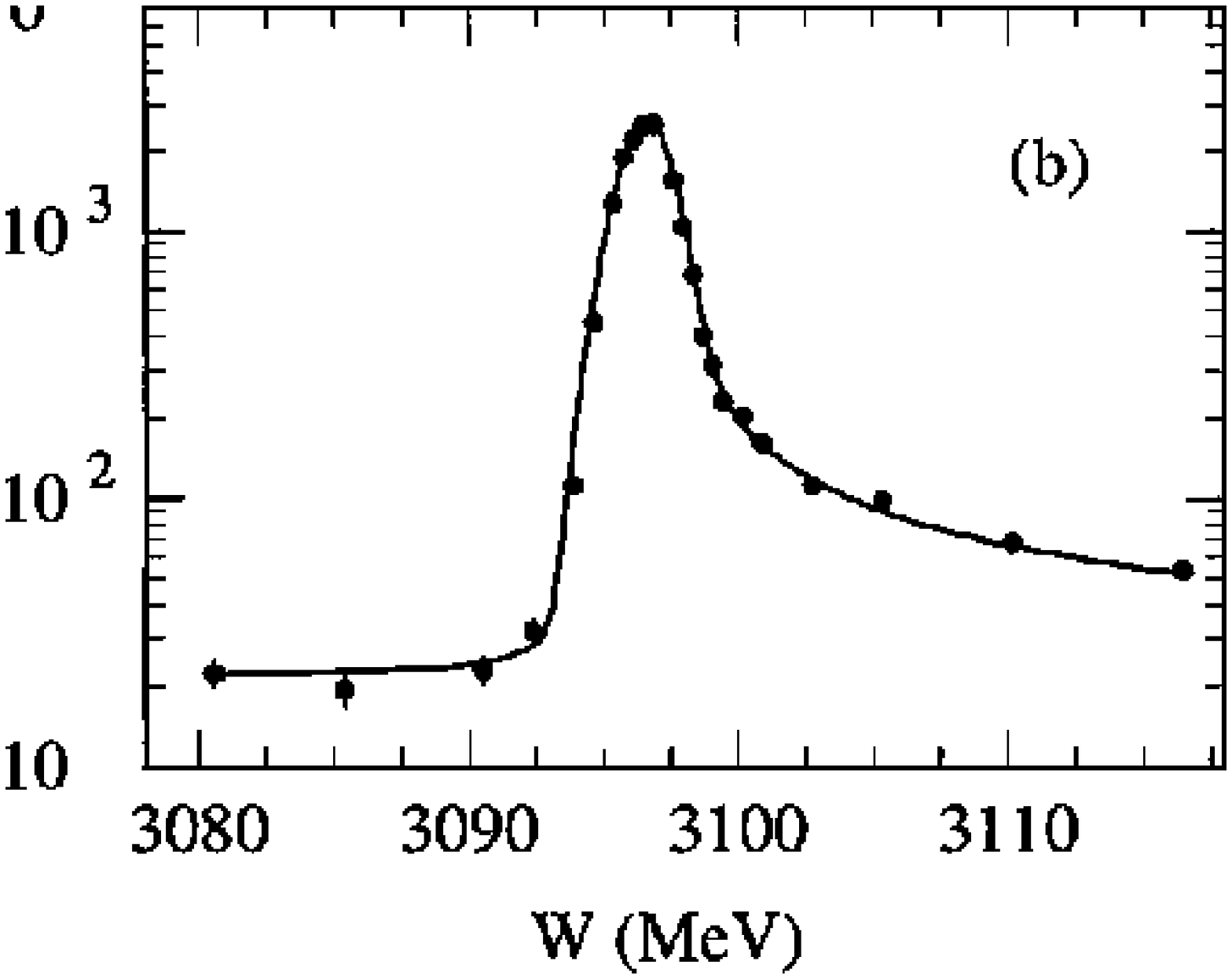}
\end{tabular}
\ec
\caption{\label{pdg40.6top}
a: $e^+e^-$ hadronic cross section (from~\protect\cite{Eidelman:2004wy}).
b: $e^+e^-$ cross section (in $nb$)
for the reaction $e^+e^-\to\rho\pi$~\cite{Bai:1996rd}.}
\end{figure}

Fig.~\ref{pdg40.6top} shows the  total cross
section for $e^+e^-$ annihilation into hadrons. Peaks are seen which
can be identified with $\rho$, $\omega$ and $\phi$. The series of
narrow resonances belongs to the $c\bar c$ and $b\bar b$ families; at
the highest energy, the $Z$, the neutral weak interaction boson, is
seen. The reaction is dominated by a one-photon intermediate state;
hence only resonances with quantum numbers $J^{PC}=1^{--}$ are formed.
Strangeness, charm beauty (and truth) are conserved in electromagnetic
interactions, $ K^*$'s are not seen $e^+e^-$ formation (but can be
produced with a recoiling Kaon).

\subsubsection{\label{Width of the J/psi}
Width of the J/$ \psi$}

Fig.~\ref{pdg40.6top}b shows an expanded view  of the J/$\psi$ region
for the reaction $e^+e^-\to\rho\pi$~\cite{Bai:1996rd}. A clear peak at
the J/$\psi$ mass is observed. The asymmetry of the line shape is due
to the loss of some energy by initial state radiation which is taken
into account in the fit. The width of the curve is completely dominated
by the momentum spread of the beam; the natural width of the J/$ \psi$
is too narrow to be determined directly from the scan. However, the
width is related to the total cross section observed in a specific
reaction. The cross section can be written in the form

\be
\sigma (E) =
(4\pi)(\lambda^2/4\pi^2)
\frac{\Gamma_{e^+e^-}\Gamma_{final}/4}
{\left[\left(E-E_R\right) + \Gamma^2/4\right)]}
\frac{2J+1}{(2s_1+1)(2s_2+1)}
\label{jeq} \ee

with $ \lambda/2\pi = 1/p = 2/E$ being the de Broglie wavelength of $
e^+$ and $ e^-$ in the centre-of-mass system (cms), $E$ the cms
energy, and $ \Gamma$ the total width. The first quotient of the
right-hand side of (\ref{jeq}) is the usual Breit-Wigner function
describing a resonant behaviour. The second quotient sums over the spin
components in the final state and averages over the spin components in
the initial state. $s_1=s_2=1/2$ are electron and positron spin; $J=1$
is the J/$ \psi$ total angular momentum.  $\Gamma_{e^+e^-},
\Gamma_{final}$ are the partial widths for the decay into the
initial and final state, respectively. Integration over the cross
section yields
\be
\int_0^{\infty}
\sigma(E_{e^+e^-\to e^+e^-,\mu^+\mu^-,\mathrm{hadrons}})dE =
\frac{6\pi^2}{E_R^2\Gamma}
\Gamma_{e^+e^-}
\Gamma_{(e^+e^-,\mu^+\mu^-,\mathrm{hadrons})}.
\ee
for the three final states $e^+e^-,\mu^+\mu^-$, and hadrons.
The total width is given by the sum of the partial
decay width
\be
\Gamma = \Gamma_{e^+e^-} + \Gamma_{\mu^+\mu^-} + \Gamma_{hadrons}.
\ee
Imposing $ \Gamma_{e^+e^-} = \Gamma_{\mu^+\mu^-}$ yields
3 equations and thus 3 unknown widths.
\par
The J/$\psi$ has a mass of $3096.87\pm 0.04$\,MeV/c$^2$ and
$87\pm 5$\,keV/c$^2$ width. We may compare this to the
$\rho$ mass, 770\,MeV, and its width 150\,MeV/c$^2$.
Obviously the J/$ \psi$ is extremely narrow. This can be
understood by assuming that the J/$ \psi$ is a bound
state of a new kind of quarks called charmed quarks $c$, and that
\bc
J/$\psi = c\bar c$.
\ec
\par
The Okubo-Zweig-Ishida (OZI) rule then explains why the
J/$ \psi$ is so narrow. Historically, the discovery of the J/$\psi$
was a major breakthrough in particle physics and a final
proof of the quark model.

\subsubsection{\label{ISR}
Initial state radiation}

The shape of J/$\psi$ signal in Fig. \ref{pdg40.6top} differs from a
Breit-Wigner one (\ref{jeq}) for two main reasons - the energy spread
of the $e^+$ and $e^-$ beams and for initial state radiation. In
$e^+e^-$ colliders, the energy spread is driven mainly by quantum
fluctuations of synchrotron radiation. The spread depends on the
magnetic fields, a typical value is a few MeV/c$^2$.  For this reason,
narrow resonances like J/$\psi$ are significantly distorted; the
visible peak cross section is reduced by
$\Gamma(J/\psi)/\delta(E_{cm})$ (where $\delta(E_{cm})$ is the spread
of the total energy) and the visible width is close to
$\delta(E_{cm})$.

The effects of initial state radiation (ISR) can be
illustrated in the reaction $e^+~e^- \to
\mu^+~\mu^-~\gamma$ in the vicinity of J/$\psi$ peak. The Born cross
section for J/$\psi$ production is given by \begin{equation}
\frac{d\sigma^{Born}_{J/\psi(s,x)}}{dx}=W(s,x)\times\sigma_0(s(1-x))
\end{equation}
where $\sqrt s$ is the $e^+~e^-$ invariant mass,
$x=2E_{\gamma}/\sqrt s$,~ $E_{\gamma}$ is the photon energy in the CMS,
and $\sigma_0$ is the Born cross section for $e^+~e^- \to J/\psi \to
\mu^+~\mu^-$.
The function
\begin{equation}
W(s,x)=\frac{2\alpha}{\pi x} \times (2 ln \frac{\sqrt s}{m_e} - 1) \times(1-x+\frac{x^2}{2})
\end{equation}
describes the probability of ISR photon emission. To first
approximation, the Born cross section for $e^+~e^- \to J/\psi \to
\mu^+~\mu^-$ is given by the Breit-Wigner formula (\ref{jeq}).
For a narrow resonance like J/$\psi$ we can replace the
Breit-Wigner function by a $\delta$ function, integrate over the photon
energy and find \begin{equation}
\sigma^{Born}_{J/\psi}(s)=\frac{12\pi^2\Gamma_{ee}B_{\mu\mu}} {m~s}
\times W(s, x_0),~~ x_0=1-\frac{m^2}{s}
\end{equation}
The asymmetry of J/$\psi$ signal (tail to high $s$) is clearly seen.
Interference of J/$\psi$ resonant and QED nonresonant amplitudes has
of course to be accounted for. By detecting the ISR photon, one can
study vector final states with masses well below the energy of
$e^+~e^-$ collider. This effect is called radiative return
\cite{Arbuzov:1998te,Binner:1999bt,Benayoun:1999hm}. The method is a
new tool in meson spectroscopy, in particular for the study of vector
mesons. It has been exploited in Frascati, using the Daphne detector,
to measure the hadronic cross section for $e^+e^-$ annihilation in the
region below the $\phi$ mass \cite{Denig:2005eb}. The BaBaR
collaboration, with a collider energy set to the $\Upsilon (4s)$,
covered the mass range from low-energies up to the bottomonium states.
The dynamics of $e^+ e^-$ annihilation into several final states were
studied from $\pi^+\pi^-\pi^0$ to $2(\pi^+2\pi^-)\,K^+K^-$
\cite{Aubert:2004kj,Aubert:2005eg,Aubert:2006jq}.

\subsubsection{\label{tau}
Decay of $\tau$ mesons}

The vector and axial vector couplings in $\tau$ decays gives access to
studies of the properties of $\rho(1450)$ and $a_1(1260)$  mesons.
Evidence is also observed for a radial excitation at 1700\,MeV/c$^2$.
We refer the reader to \cite{Asner:1999kj} for more detailed
information. In $\pi$ scattering, $a_1(1260)$ is obscured by the Deck
effect and unstable results were obtained for a long time.

\subsubsection{\label{The chi states in pbp annihilation}
The $\chi$ states in $\bar pp$  annihilation}

The charmonium $\chi$ states
can be observed in $\bar pp$  annihilation by a scan of the antiproton
momenta. Of course, $c\bar c$ states couple only weakly to $p\bar p$
since all 3 quarks and antiquarks in the initial state have to
annihilate into gluons before a meson with hidden charm can be
produced. The  $\chi$ states radiate down to the J/$\psi$ ground state
which then decays into an electron-positron or $\mu^+\mu^-$ pair.
Lepton pairs with having a large invariant mass are rarely produced in
$p\bar p$ annihilation; hence high-mass lepton pairs can be used as
very selective trigger.

\begin{figure}[!ht]
\bc
\begin{tabular}{cc}
\includegraphics[width=0.48\textwidth,height=0.35\textwidth]{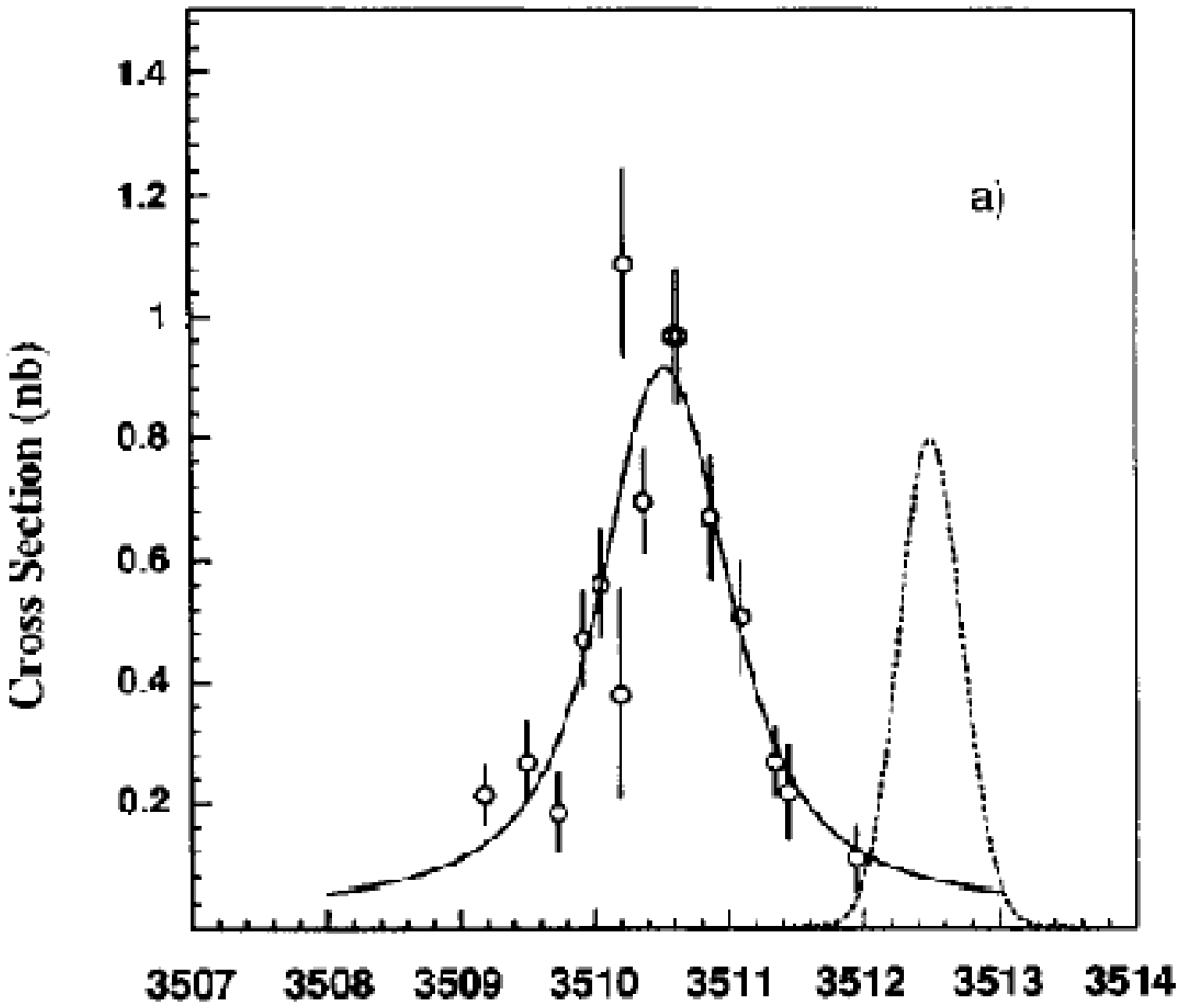}&
\includegraphics[width=0.48\textwidth,height=0.35\textwidth]{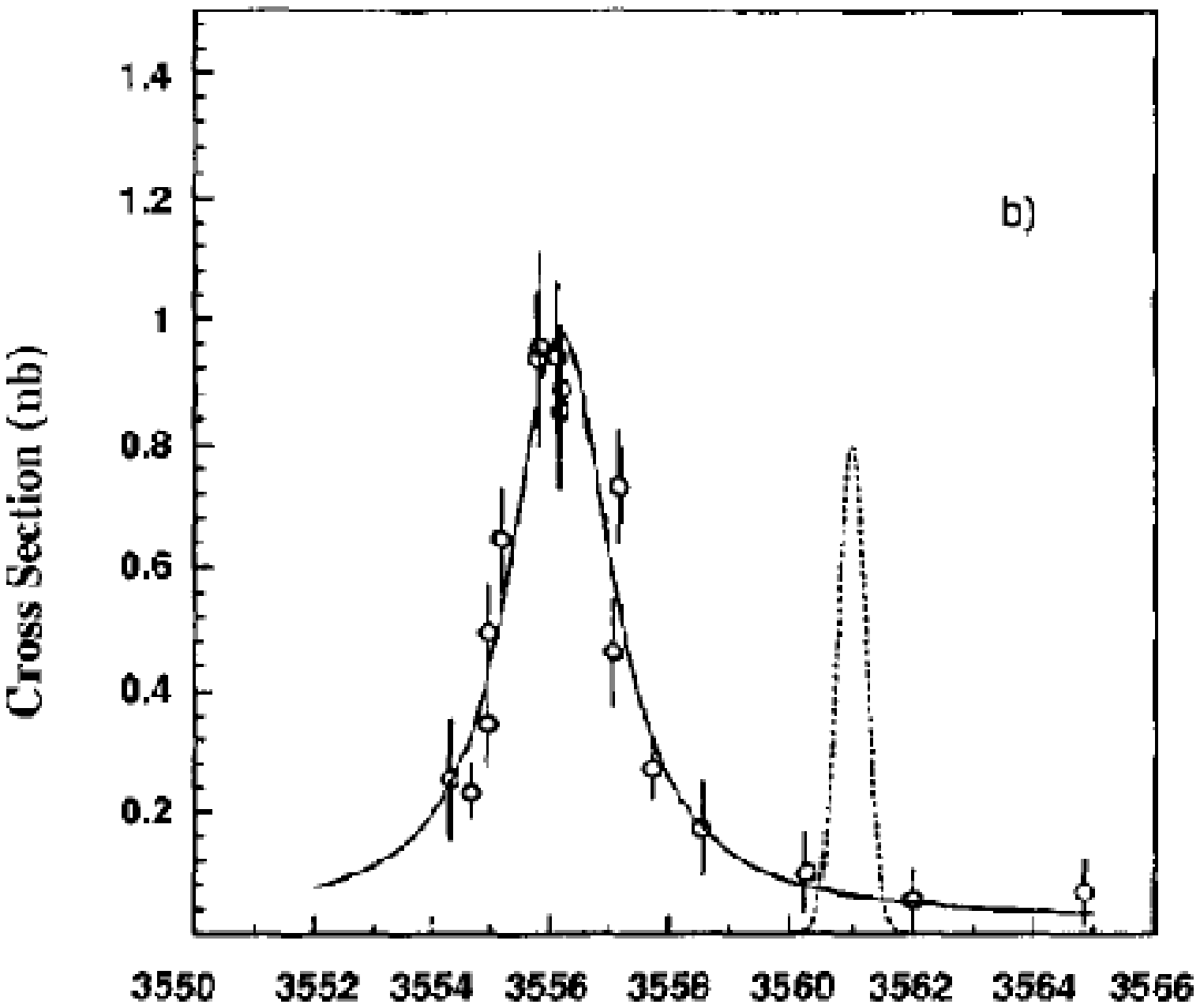}
\end{tabular}
\ec
\vspace*{-5mm}
\caption{\label{cb-scan}
The number of J/$\psi$ as a function of the $\bar pp$ mass in the
$\chi_1$ (a) and $\chi_2$ (b) mass regions
\protect\cite{Ambrogiani:2001jw}.}
\end{figure}

The decisive advantage of this approach is the possibility to study the
$\chi$ states in formation. The observed width is determined by the
spread of beam momenta, and this can be made narrow.
Figure~\ref{cb-scan} shows scans of the $\chi_{c1}(1P)$ and
$\chi_{c1}(2P)$ regions. The experimental resolution, given by the
precision of the beam momentum, is shown as a dashed line. The observed
distributions are broader: the natural widths of the states due to
their finite life time can be directly determined.

\begin{figure}[pt]
\bc
\includegraphics[width=0.6\textwidth,height=0.48\textwidth]{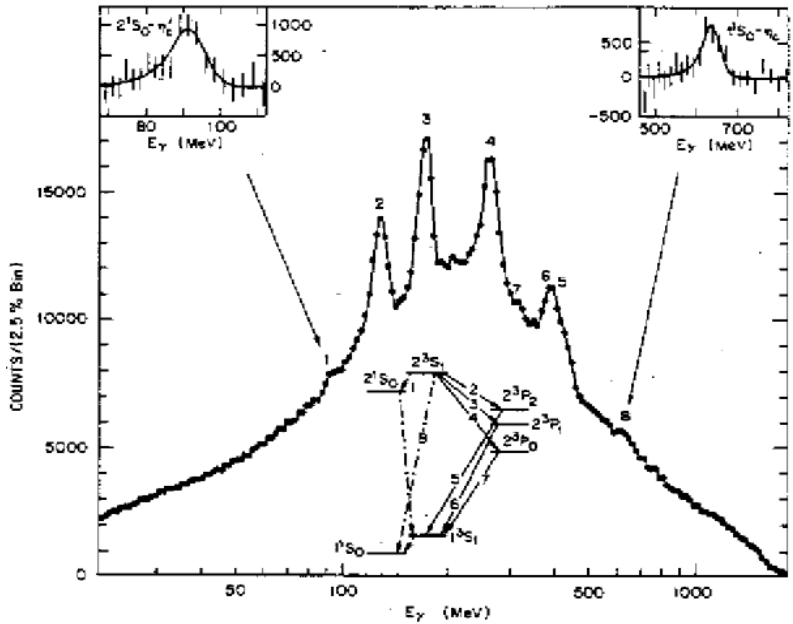}
\ec
\vspace*{-5mm}
\caption{\label{cb-rad}
Radiative transitions between charmonium
levels~\protect\cite{Gaiser:1985ix}).}
\end{figure}

The impressive gain in accuracy achieved in a formation
experiment can be visualised in a comparison with the
$\chi$ states as observed in production.
Figure~\ref{cb-rad} shows
the inclusive photon spectrum from the $\psi(2S)$
states~\cite{Gaiser:1985ix}. A series
of narrow states is seen identifying the masses of intermediate
states. The level scheme assigns the lines to specific transitions
as expected from charmonium models. The width of the lines
is given by the experimental resolution of the detector; the
charmonium states are {\it produced}. The lowest mass state, the
$1^1S_0$ state, is called $\eta_c$. It is the analogue of the
$\eta^{\prime}$-meson.

\subsection{\label{Glueball rich processes}
Glueball rich processes}

\begin{figure}[ph]
\bc
\begin{tabular}{ccc}
\includegraphics[width=0.27\textwidth,height=28mm]{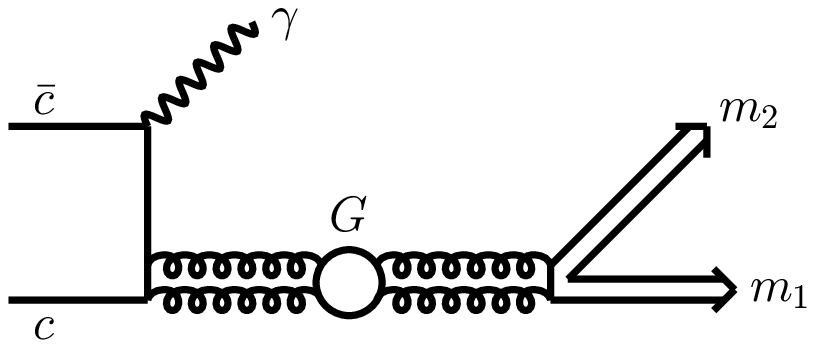}&
\hspace{3mm}\includegraphics[width=0.24\textwidth,height=25mm]{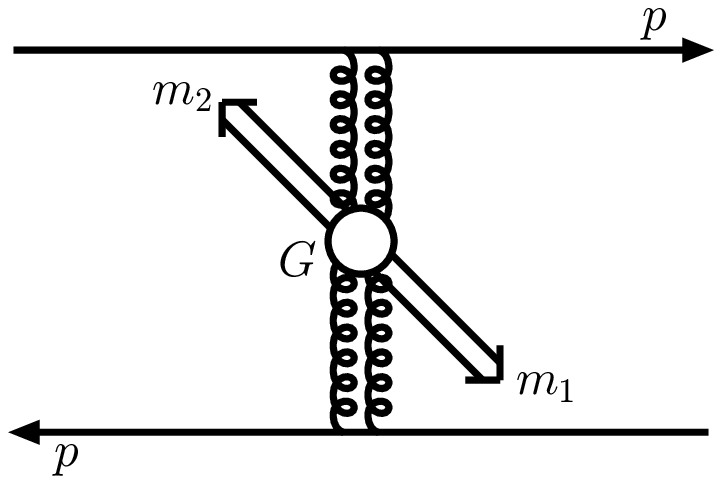}&
\hspace{3mm}\includegraphics[width=0.30\textwidth,height=27mm]{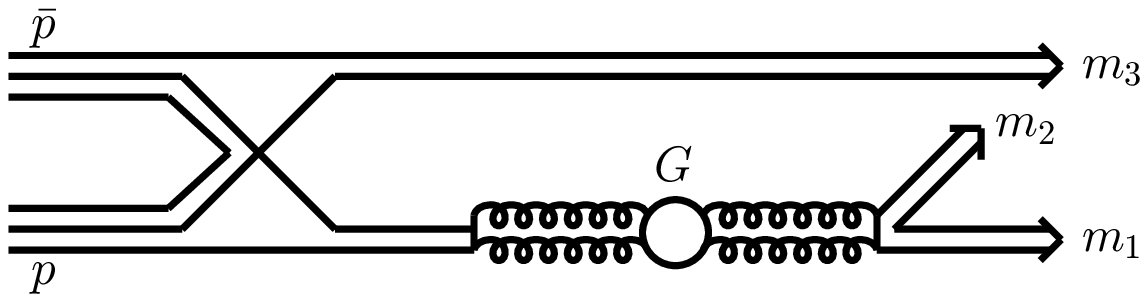}
\end{tabular}
\ec
\vspace*{-3mm}
\caption{\label{gluemod}
Diagrams possibly leading to the formation of glueballs: radiative
J/$\psi$ decays, Pomeron-Pomeron collisions in hadron-hadron
central production, and in $ p\bar p$ annihilation. }
\end{figure}

QCD, at least in its formulation on a lattice, predicts the existence
of glueballs, of bound states of gluons with no constituent quarks. The
search for these states and their possible $\rm r\hat{o}le$ within the
family of $q\bar q$ mesons is one of the topical issues of this review.
There is an extensive folklore on how to hunt for glueballs
\cite{Robson:1977pm} and which distinctive features should identify
them as non-$\bar qq$ mesons. Glueballs should, e.g., be produced
preferentially in so-called gluon-rich processes; some are depicted in
Fig.~\ref{gluemod}.
\par

The most suggestive process is the radiative J/$\psi$ decay. The
J/$\psi$ is narrow; the $D\bar D$ threshold is above the mass of the
J/$\psi$ and the OZI rule suppresses decays of the $c\bar c$ system
into light quarks. In most decays, the J/$\psi$ undergoes a transition
into 3 gluons which then convert into hadrons. But J/$\Psi$ can also
decay into 2 gluons and a photon. The photon can be detected, the two
gluons interact and must form glueballs - if they exist.

Central production is another process in which glueballs should be
abundantly produced. In central production two hadrons pass by each
other `nearly untouched' and are scattered diffractively into forward
direction. No valence quarks are exchanged. The process is often called
Pomeron-Pomeron scattering or double Pomeron exchange. The absence of
valence quarks in the production process makes central production a
good place to search for glueballs.

In $\bar pp$  annihilation, quark-antiquark pairs annihilate into gluons,
they interact and may form glueballs. Glueballs decay into hadrons and
hence hadro-production of glueballs is always possible. The probability
that two gluons escape from the $p\bar p$ system to form a glueball is
a bit remote.

Production of glueballs should be suppressed in $\gamma\gamma$
collisions since photons couple to the intrinsic charges. So we should
expect a glueball to be strongly produced in radiative J/$\psi$ decays
but not in  $\gamma\gamma$ fusion. Radial excitations might be visible
only weakly in J/$\psi$ decays but they should couple to
$\gamma\gamma$.
Further distinctive features can be derived from their decays
(glueballs are flavour singlets). Decays to $\eta\eta^{\prime}$ identify
a flavour octet; radiative decays of glueballs are forbidden. All these
arguments have to be taken with a grain of salt: mixing of a glueball
with mesons having the same quantum numbers can occur and would dilute
any selection rule.


\markboth{\sl Meson spectroscopy}{\sl Heavy--quark spectroscopy}
\clearpage\setcounter{equation}{0}\section{\label{Heavy--quark spectroscopy}
Heavy-quark spectroscopy}

The discovery of a narrow high-mass resonance at Brookhaven National
Laboratory (BNL) in  the process proton + Be $\to e^+e^-$ + anything
\cite{Aubert:1974js} and at the Stanford University in $e^+e^-$
annihilation to $ \mu^+\mu^- , e^+e^-$ and into hadrons
\cite{Augustin:1974xw} initiated the `November revolution of particle
physics'. In the subsequent years, several narrow charmonium states
were observed. The states below the $D\bar D$ threshold have a small
available phase space for decays mode into other charmonium states;
they may also decay into mesons with light quarks only, via gluonic
intermediate states. The strong interaction fine structure constant
$\alpha_s$ is already sufficiently small at these momentum transfers
($\alpha_s\sim 0.3$) resulting in small strong interaction widths.

\begin{figure}[pb]
\bc
\includegraphics[width=0.7\textwidth,height=0.8\textwidth,clip=on]{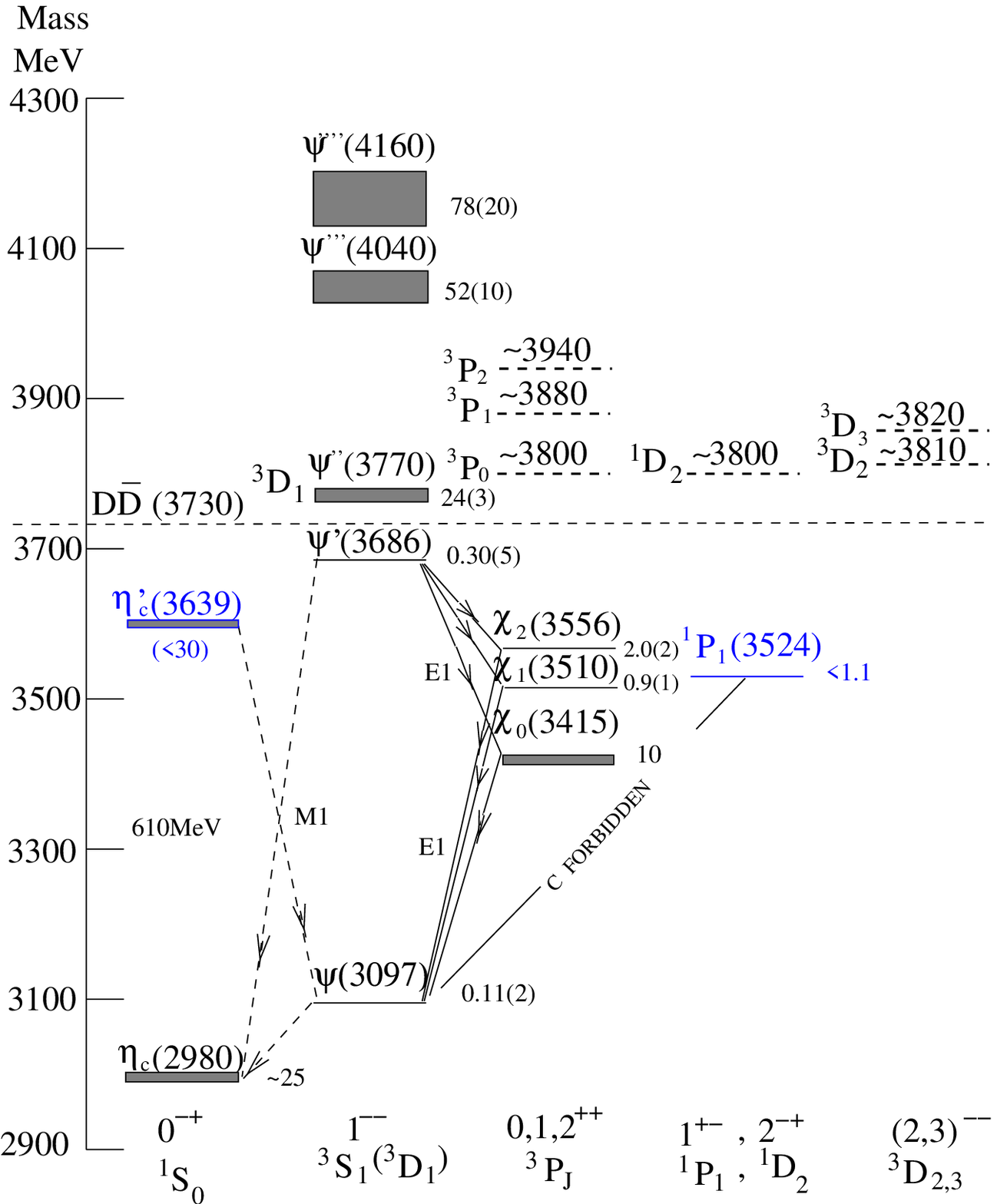}
\ec
\caption{The charmonium family of states and transitions between them
\cite{Seth:2005pj}. Established resonances are shown as solid lines,
predicted states as dashed lines. Widths and their errors are added as
small numbers. In the bottomonium family, the $3S$ and $2P$
states are also stable against decays into $  B\bar B$.
 }
\label{fig:ccb}
\end{figure}

The quark model predicts a rich spectrum of charmonium states. Fig.
\ref{fig:ccb} gives a survey of observed and expected states. Some
transitions are indicated by arrows. The $h_{c}(1P)$ -- shown as
$^1P_1(3254)$ -- is, for example, forbidden to decay into J/$\psi$
radiatively; allowed are decays into $\pi^0J/\psi$  or
$\gamma\eta_c(1S)$ (as E1 transition).

The physics of $c\bar c$ states has found renewed interest due to the
discovery of a series of unexpectedly narrow states, and several
excellent review articles have been written which document the
achievements \cite{Brambilla:2004wf,Seth:2005pj,Quigg:2005tv,%
Swanson:2006st,Eichten:2007qx,Zhu:2007wz}. Here, we give a short survey
of the field. This will allow us to emphasize the intimate relation
between the spectroscopy of heavy quarks and of their light cousins.

\subsection{\label{The J/psi states below the Dbar D threshold}
The  J$/\psi$ states below the $D\bar D$ threshold}

In $e^+e^-$ annihilation, charmonium states with $  J^{PC}=1^{--}$
can be observed and their mass be determined with the (high) precision
with which the frequency of the circulating beam can be determined (see
section \ref{VEPP}). Positive $C$-parity states can be detected via
radiative transitions from the $\psi(2S)$ resonance but only with the
(poorer) accuracy of the photon energy measurement
\cite{Gaiser:1985ix}. This limitation can be overcome by
$ p\bar p$ formation experiments as discussed in section
\ref{Experiment E835 at FNAL}. Exploiting this technique, the E835
experiment added to our knowledge an impressive amount of
high-precision data on the $c\bar c$ system. From BES and earlier
experiments, a large number of branching ratios of the charmonium
states is known. Their discussion will be resumed when they provide
information on light-quark spectroscopy. A comparison of charmonium
decay modes with model calculations is beyond the scope of this review.
The two states  $\eta_c(2S)$ and $h_c(1P)$ were observed only recently.
Their discovery will be briefly discussed.

\subsubsection{\label{The h_c(1P)}
The $h_c(1P)$}

The $h_{c}(1P)$ was searched for intensively by the E760/835
collaboration in the reaction chain $ p\bar p\to h_{c}(1P)\to
 \pi^0J/\psi\to\pi^0e^+e^-$ but, at the end, no conclusive evidence
 was found \cite{Andreotti:2005vu}. Stimulated by an observation of the
 CLEO collaboration \cite{Rosner:2005ry} discussed below, E835 studied
 the reaction $ p\bar p\to h_{c}\to \gamma\eta_c(1S)\to 3\gamma$
 and reported a 3$\sigma$ signal \cite{Andreotti:2005vu} which was
 assigned to $h_c(1P)$. Its mass was determined to $M(h_c)=3525.8 \pm
0.2 \pm 0.2$\,MeV/c$^2$, its width $\Gamma(h_c)<1$\,MeV/c$^2$.

\begin{figure}[pb]
\vspace{-3mm}
\bc\includegraphics[width=0.45\textwidth,height=0.35\textwidth]{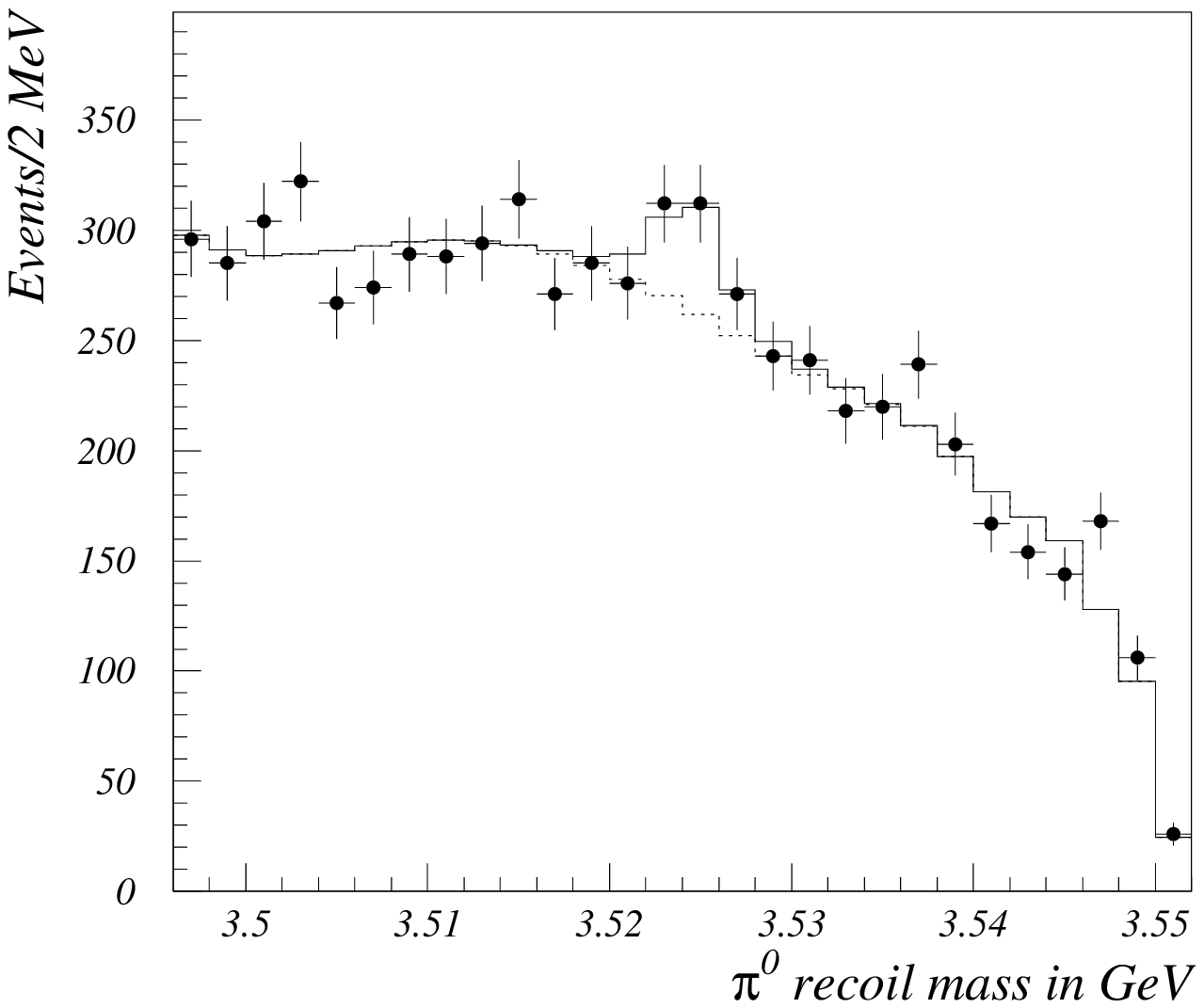}
\raisebox{0.31\textwidth}{\rotatebox{270}{\includegraphics[width=0.30\textwidth]{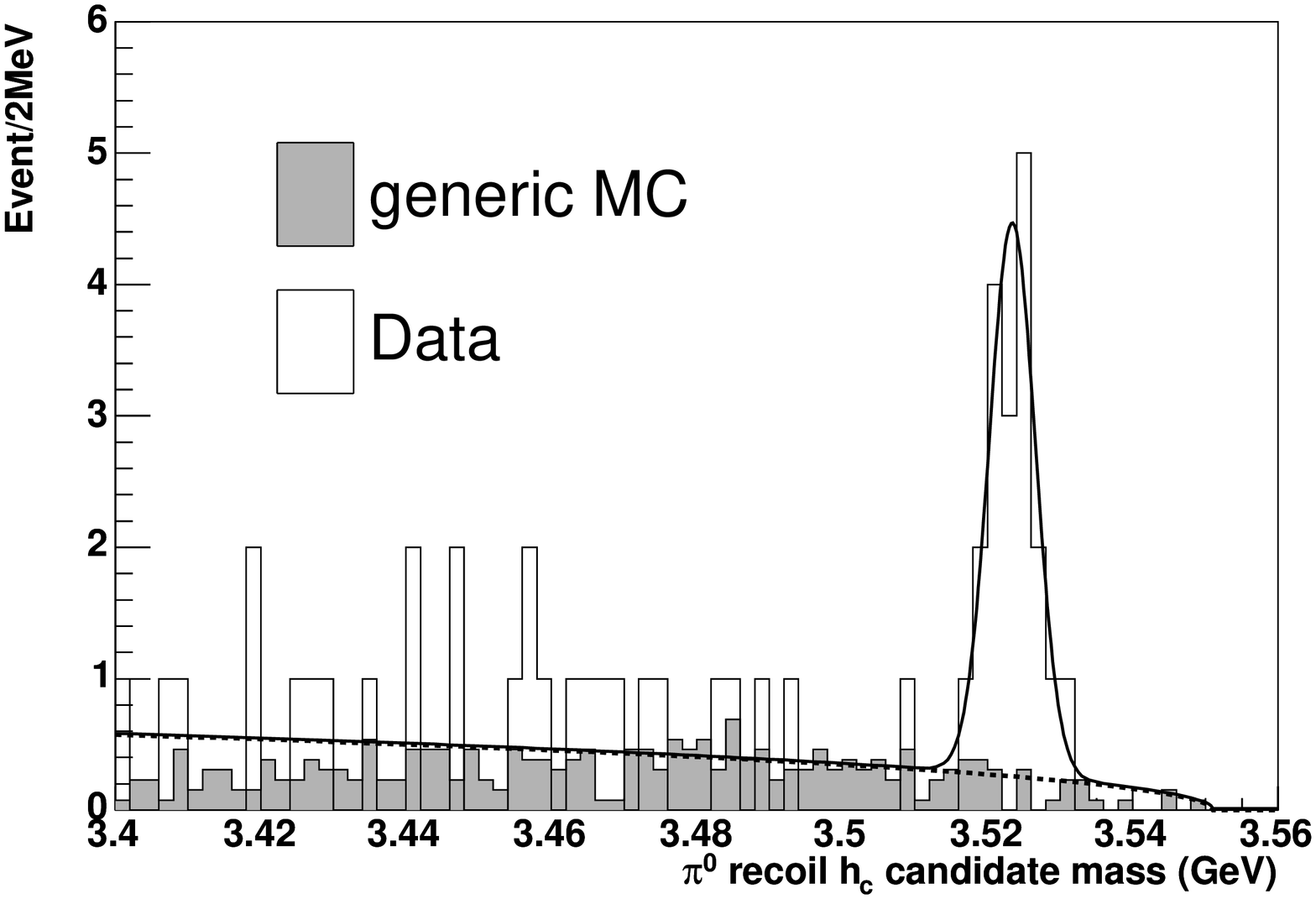}}}
\vspace{-50mm}\\
\hspace{53mm}a\hspace{86mm}b\vspace{42mm}\\
\ec
\caption{\label{cleohc}
 Observation of $h_c(1^1P_1)$ in
 $\psi(2S) \to \pi^0 h_c \to (\gamma\gamma)(\gamma \eta_c$ with
 $\eta_c$ observed inclusively (a) or exclusively (b)
 \cite{Rosner:2005ry}.}
 \end{figure}

The best evidence for the $h_c(1P)$ stems from the CLEO collaboration
searching for the isospin-violating reaction $\psi(2S) \to \pi^0 h_c
\to (\gamma\gamma)(\gamma \eta_c)$. The $\eta_c$ is measured
inclusively giving a larger yield, and exclusively in several hadronic
decays modes.  Both data sets are shown in Fig. \ref{cleohc}.
 The data yield a $h_c(1P)$ with $M(h_c)=3524.4 \pm 0.6 \pm 0.4$\,MeV/c$^2$.
 This mass can be compared with the centre-of-gravity of the triplet
 charmonium states
 \be
 \left< M_{\chi (^{3}P_{J})} \right> - M_{h_{c}(^{1}P_{1})} =
 +1.0\pm0.6\pm0.4\;\mathrm{MeV/c^2}
\label{hfschi}
\vspace{-5mm}
 \ee
$$
{\rm with} \qquad\ \left<M_{\chi (^{3}P_{J})}\right> = \frac{\tiny 1}
{\tiny 9}\left(5\cdot M_{\chi_{c2}}+3\cdot M_{\chi_{c1}}+
M_{\chi_{c0}}\right).
$$
With the new BES mass measurements of $\chi_{cJ}(1P)$ states
 \cite{Ablikim:2005yd}, the centre of gravity $\left< M_{\chi
 (^{3}P_{J})} \right> = 3524.85\pm 0.32 \pm 0.30$\,MeV/c$^2$ is even
 closer to the $h_c(1P)$ mass.

In leading order, the difference in Eq. (\ref{hfschi}) vanishes for a
$c\bar c$ central potential composed of a vector Coulomb ($\sim 1/r$)
and a scalar confining potential ($\sim r$). Spin-spin interactions and
tensor interactions are expected to be small \cite{Godfrey:2002rp}. The
result is thus an important confirmation of the assumptions on which
most quark models rely.

\subsubsection{\label{The eta_c(2S)}
The $\eta_c(2S)$}

The transition to $\eta_c(2S)$ (often called $\eta_c^{\prime}$) was
first reported by the Crystal Ball Collaboration \cite{Edwards:1981mq},
see Fig.~\ref{cb-rad} in section \ref{Experimental methods}, but the
signal was later shown to be a fluctuation. Evidence for the
$\eta_c(2S)$ was deduced from the $b$-factories, where $\eta_c(2S)$ is
now observed in three different reactions.

\begin{figure}[pb]
\bc
\begin{tabular}{ccc}
\includegraphics[width=0.32\textwidth,height=0.29\textwidth,clip=]{eps-hq/belle-etac}&
\includegraphics[width=0.32\textwidth,height=0.3\textwidth,clip=]{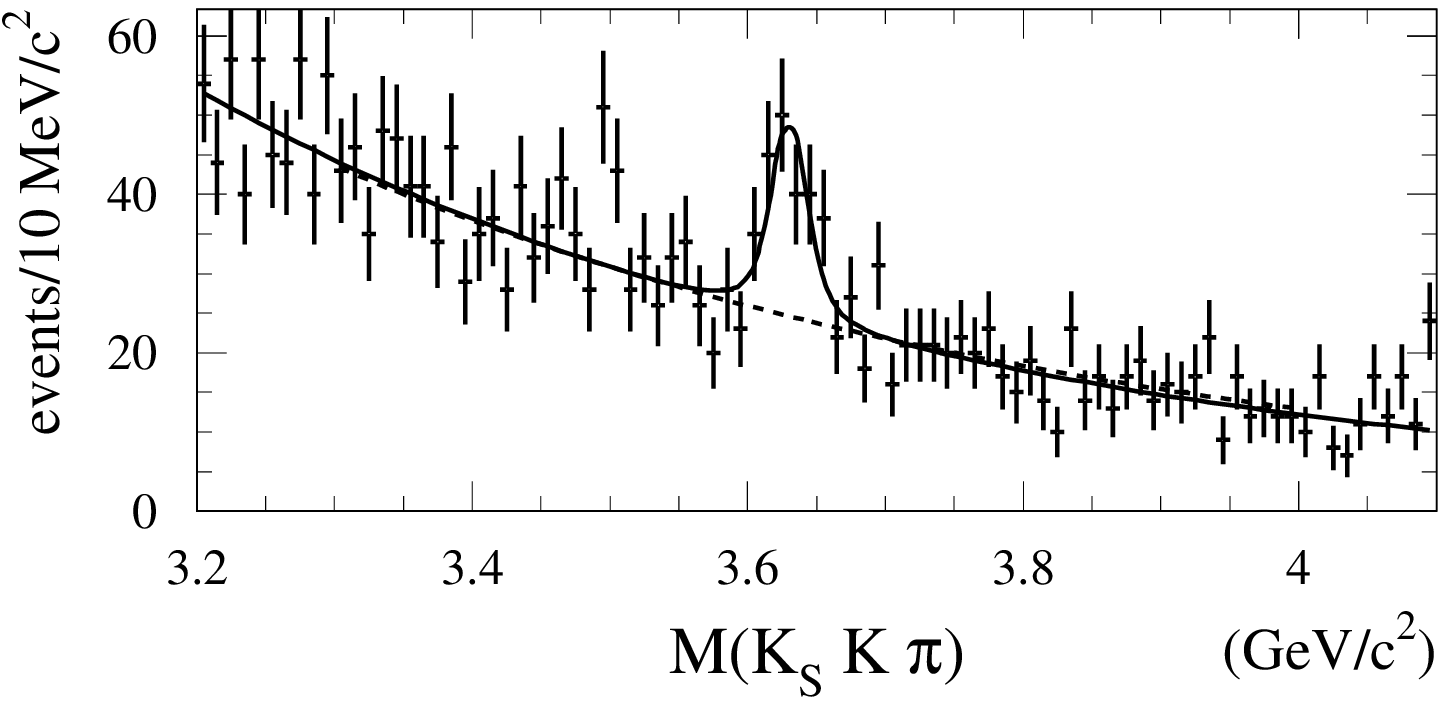}&
\hspace{-3mm}\includegraphics[width=0.32\textwidth,height=0.29\textwidth,clip=]{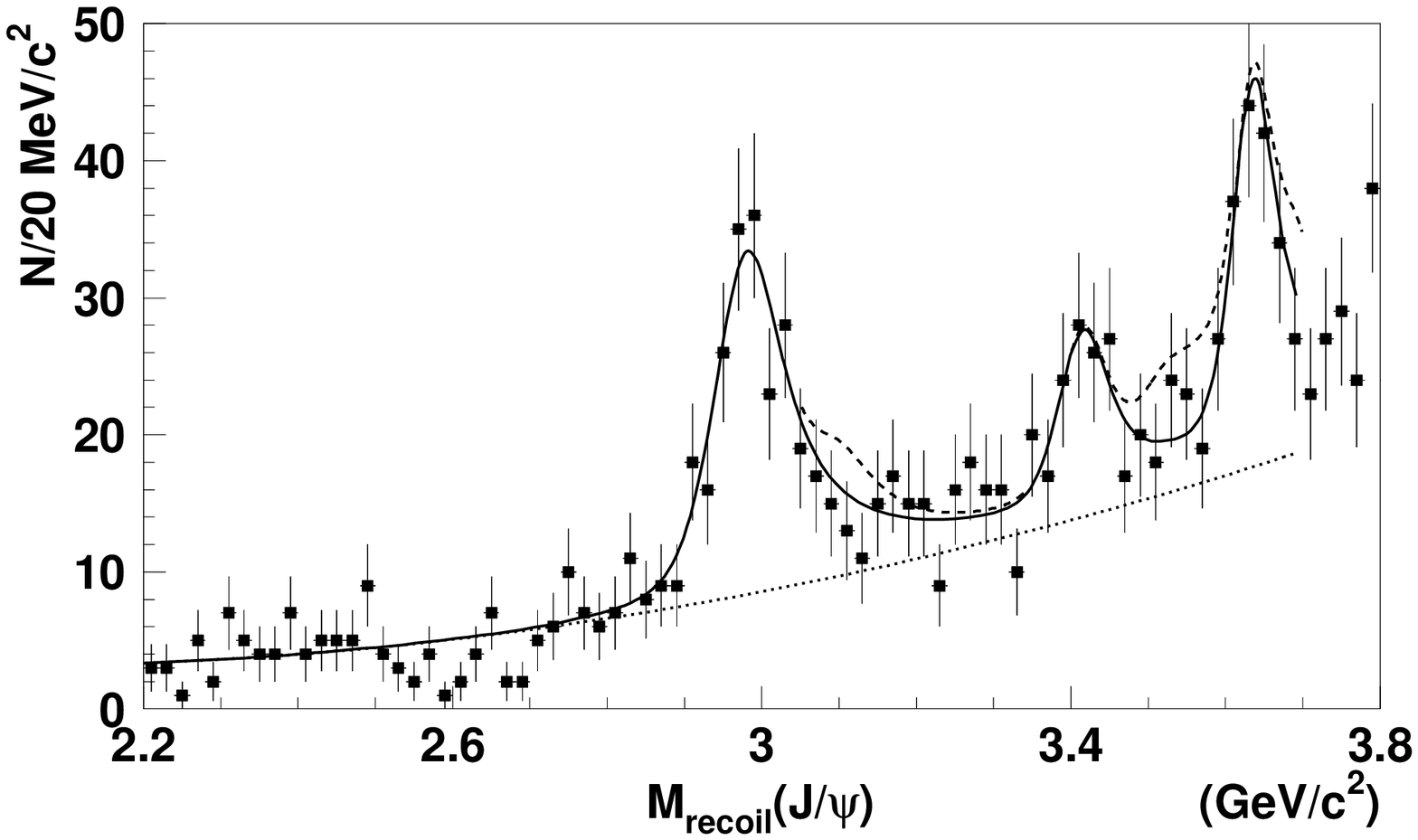}
\end{tabular}\vspace{-50mm}\\
\hspace{10mm}a\hspace{58mm}b\hspace{18mm}c\vspace{43mm}\\
\ec
\caption{\label{fig:etac2s}a: $ K^0_S K^-\pi^+$  mass distribution in
exclusive $ B\to K K^0_S K^-\pi^+$ decays \cite{Choi:2002na}. The
distribution shows a strong peak at the $\eta_c(1S)$ mass, a few events
due to J/$\psi$ production and a third peak assigned to  $\eta_c(2S)$.
b: $ K^0_S K^-\pi^+$ mass distribution from two-photon fusion
\cite{Aubert:2003pt}.  c: The mass of the
$c\bar c$ system recoiling against a reconstructed J/$\psi$ in
inclusive $e^+e^- \to J/\psi\,X$ events. The $\eta_c(1S)$ and
$\eta_c(2S)$ resonances are observed, the small peak above 3.4\,GeV/c$^2$ is
assigned to $\chi_{cJ}$ production \cite{Abe:2004ww}. }
\end{figure}

The $\eta_c(2S)$ resonance was discovered by the BELLE collaboration as
a narrow peak in the $ K^0_S K^-\pi^+$  mass distribution in a sample
of exclusive $ B\to K\,K^0_S K^-\pi^+$ decays \cite{Choi:2002na}. Fig.
\ref{fig:etac2s} (left) shows the $ K^0_S K^-\pi^+$  mass distribution.
There is a significant excess of events at about 3.65\,GeV/c$^2$. A narrow
peak is expected only if it belongs to the charmonium family; $c\bar c$
states can be produced in $B$ decays via the penguin diagram (see Fig.
\ref{fig:d-decay}d in section \ref{Flavour tagging}.) The peak was also
observed by the BaBaR collaboration in the same reaction
\cite{Aubert:2005vi}.

The BaBaR collaboration reported the observation of $\eta_c(2S)$ in
two-photon fusion with two untagged photons (i.e. with unobserved
electrons) \cite{Aubert:2003pt}. It was detected through its $ K^0_S
K^-\pi^+$ decay mode. The relevant part of the spectrum exhibits a
narrow peak shown in Fig. \ref{fig:etac2s} (centre). The full spectrum
(not shown here) contains a large peak due to $\eta_c(1S)$ and a
smaller J/$\psi$ peak. The latter is assigned to initial state
radiation. The production characteristics of the J/$\psi$ and
$\eta_c(1S)$ are different; $\eta_c(2S)$ production is compatible with
$J^{PC}=0^{-+}$ quantum numbers. At Cornell, $\eta_c(2S)$ was found in
CLEOII and CLEOIII data; the two-photon width was estimated, assuming
similar $ K^0_S K^-\pi^+$ branching ratios for $\eta_c(1S)$ and
$\eta_c(2S)$ \cite{Asner:2003wv}.

The third type of reaction uses flavour tagging
\cite{Abe:2002rb,Abe:2004ww}, see Fig. \ref{ddecay} in section
\ref{Flavour tagging}. In $e^+e^-$ annihilation, two $c\bar c$ pairs
can be produced. If a J/$\psi$ is observed, the recoiling system must
be a $c\bar c$ state. The virtual photon enforces negative $C$ parity
of the final state and hence positive $C$ parity for the recoiling
$c\bar c$ system.  The $c\bar c$ mass spectrum recoiling against the
reconstructed J/$\psi$ in inclusive $e^+e^- \to J/\psi\,X$ events
produced in $e^+e^-$ annihilation at $\sqrt s =10.6$\,GeV/c$^2$  is
shown in Fig. \ref{fig:etac2s} (right). Peaks due $\eta_c(1S)$,
$\chi_{cJ}(1P)$ -- mostly $\chi_{c0}(1P)$ -- and $\eta_c(2S)$ can be
identified.

Instead of reporting the results from the individual measurements, we
compare in Table \ref{tab:etac2s} the PDG mean values of the most
important properties of $\eta_c(1S)$ and $\eta_c(2S)$.

\begin{table}[ph]
\caption{\label{tab:etac2s}
Properties of $\eta_c(1S)$ and $\eta_c(2S)$ \cite{Eidelman:2004wy}.
\vspace{2mm}}
\begin{center}
\renewcommand{\arraystretch}{1.5}
\begin{tabular}{cccc}
\hline\hline
$\eta_c$ state & Mass           & Width     & $\Gamma_{\gamma\gamma}$\\
\hline
$\eta_c(1S)$&$2980.4\pm1.2$\,MeV/c$^2$&$25.5\pm3.4$\,MeV/c$^2$&$6.7^{+0.9}_{-0.8}$\,keV/c$^2$\\
$\eta_c(2S)$& $3638\pm4$\,MeV/c$^2$   &   $14\pm7$      & $1.3\pm0.6$\,keV/c$^2$ \\
\hline\hline
\end{tabular}
\renewcommand{\arraystretch}{1.0}
\end{center}
\end{table}

The two-photon width measures the density of the wave function at the
origin which is smaller by $0.19\pm 0.09$ for $\eta_c(2S)$ compared to
$\eta_c(1S)$ (see last column, Table \ref{tab:etac2s}). In positronium,
this factor is 1/8.  The reduction of the $e^+e^-$ widths of $\psi(2S)$
and $\psi(3770)$ charmonium and of bottomonium vector states -- for
which recently very precise results were reported \cite{Duboscq:2006de}
-- is much smaller:

\bc\renewcommand{\arraystretch}{1.6}
\begin{tabular}{lccclccc}
${\Gamma_{ee}(\psi(2S)) \over{ \Gamma_{ee}(J/\psi(1S))}}$       & = &
$0.393 \pm 0.008$\,  &\quad &
${\Gamma_{ee}(\psi(3770)) \over{ \Gamma_{ee}(J/\psi(1S))}}$        & = &
$0.045\pm 0.006$  & \cite{Eidelman:2004wy}\,\nonumber  \\
${\Gamma_{ee}(\Upsilon(2S))  \over{ \Gamma_{ee}(\Upsilon(1S))}}$ & = &
$0.461 \pm 0.008 \pm 0.003$\, &\quad &
${\Gamma_{ee}(\Upsilon(3S))  \over{ \Gamma_{ee}(\Upsilon(1S))}}$ & = &
$0.318 \pm 0.007 \pm 0.002$ & \cite{Duboscq:2006de}\..\nonumber    \\
\end{tabular}
\renewcommand{\arraystretch}{1.0}\ec

The $e^+e^-$ width $\Gamma_{ee}(\psi(3770))$ is much smaller than
expected for a $3S$ level; $\psi(3770)$ is (dominantly) the charmonium
$1^3D_1$ state.

The discovery of the $h_c(1P)$ and $\eta_c(2S)$ resonances marks an
important step: all charmonium states below the $ D\bar D$
threshold predicted by quark models have been found, and no extra state.

\subsubsection{\label{The 12 rule and the rhopi puzzle}
The 12\% rule and the $\rho\pi$ puzzle}

The largest J/$\psi$ decay fraction supposedly proceeds via 3-gluon
intermediate states. Appelquist and Politzer \cite{Appelquist:1974zd}
have shown that this fraction decreases for $\psi(2S)$. Using
perturbative arguments, they find
\be \frac{\mathcal B_{\psi'\to ggg}}{\mathcal B_{\psi\to ggg}} =
\frac{\Gamma_{e^+e^-}(\psi(2S))\cdot\Gamma_{tot}(\psi(1S))}
     {\Gamma_{e^+e^-}(\psi)(1S))\cdot\Gamma_{tot}(\psi(2S))} =
                                              (12.37\pm  0.03)\%\,
\ee
This is the famous 12\% rule: hadronic decays are expected to be
reduced to 12\% when $\psi(2S)$ hadronic decays are compared to
J/$\psi$ (=$\psi(1S)$) decays. A large fraction of $\psi(2S)$ does not
decay hadronically via three-gluon intermediate states; the
largest fraction goes instead into J/$\psi$ plus hadrons. We note in
passing that the same argument can be made for the $\chi_{cJ}$ decays
via $n$ gluons, $\Gamma_{\chi\to ng}$. The branching fractions for
decays into $ \Lambda\bar\Lambda$ are:
\bc\renewcommand{\arraystretch}{1.6}
\begin{tabular}{cccccc}
\hline\hline
J/$\psi$&$\chi_{c0}(1P)$&$\chi_{c1}(1P)$&$\chi_{c2}(1P)$&$\psi(2S)$&$\to
\Lambda\bar\Lambda$\\ $15.4\pm1.9$& $4.4\pm 1.5$ & $2.4\pm 1.0$  &
$2.7\pm 1.3$&$2.5\pm 0.7$ &$\times 10^{-4}$\\ \hline\hline
\end{tabular}
\renewcommand{\arraystretch}{1.0}\ec
The branching fraction for $\Lambda\bar\Lambda$ decays from $\psi(2S)$
is reduced to $(16\pm 5)$\% of the fraction for J/$\psi$; the reduction
is compatible with the 12\% rule. For the $\chi_{cJ}$ states, the mean
reduction is about $(19\pm5)$\% of the J/$\psi$ branching fraction to
$\Lambda\bar\Lambda$. In all cases, the error in the individual
quantities is large enough to cover the 12\% rule.

\begin{figure}[ph]
\bc
\includegraphics[width=0.72\textwidth]{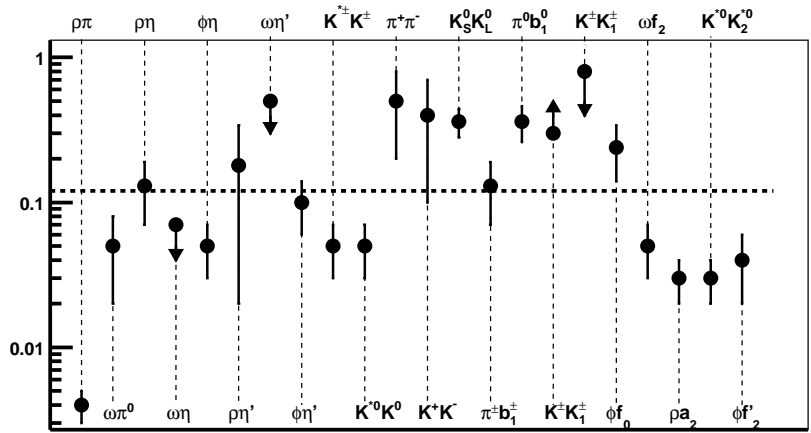}
\ec
\caption{\label{12percent}The ratio of branching fractions of
$\psi(2S)$ and J/$\psi$ decays into hadronic final states. The data are
from the compilation of Brambilla {\it et al.} \cite{Brambilla:2004wf}.
The vertical line indicates the 12\% rule.
}
\end{figure}

In general, the 12\% rule is valid only approximately. Fig.
\ref{12percent} gives a survey of results compiled by Brambilla {\it et
al.} \cite{Brambilla:2004wf}. There seems to be no obvious preference
why some of these ratios are large and some small. The suppression of
$\rho\pi$ from $\psi(2S)$ is most striking, it is known as $\rho\pi$
puzzle. In the report of Brambilla {\it et al.} \cite{Brambilla:2004wf},
several interpretations of the $\rho\pi$ suppression are discussed. For
$\psi(3770)$, $\pi^+\pi^-\pi^0$ is the only hadronic decay mode
observed so far \cite{Ablikim:2005cd}. No $\rho\pi$ contribution was
found.

\subsubsection{\label{Two-pion transitions}
Two-pion transitions}

\begin{figure}[tb]
 \begin{center}
\begin{tabular}{cccc}
\hspace{-2mm}\includegraphics[width=0.23\textwidth,height=3.5cm]{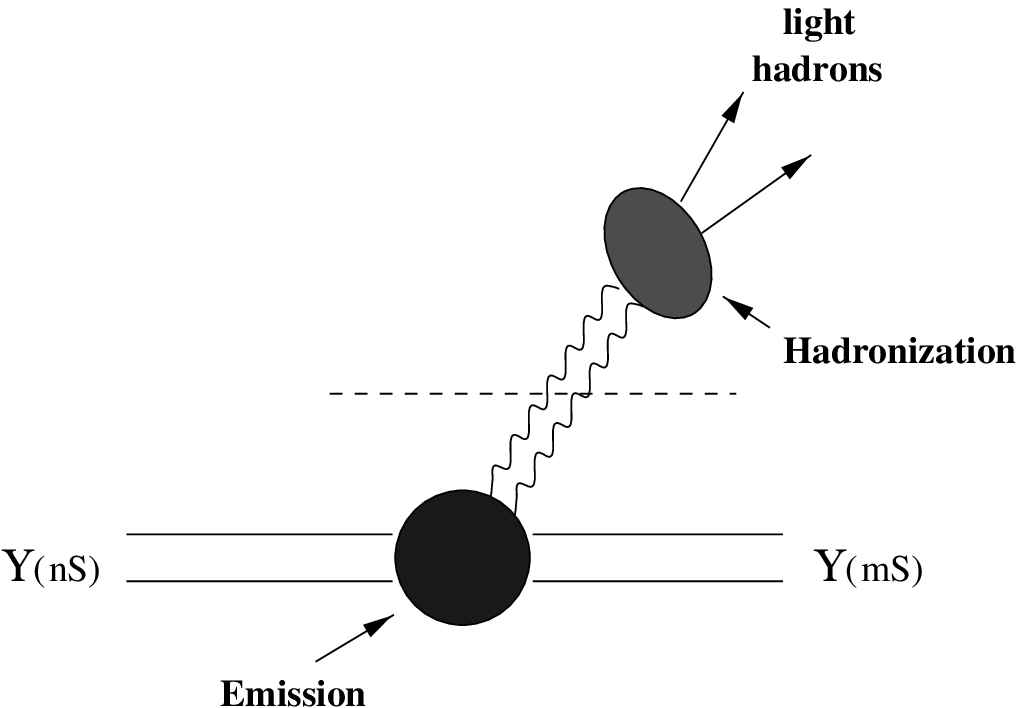}&
 \includegraphics[width=0.23\textwidth,height=4cm]{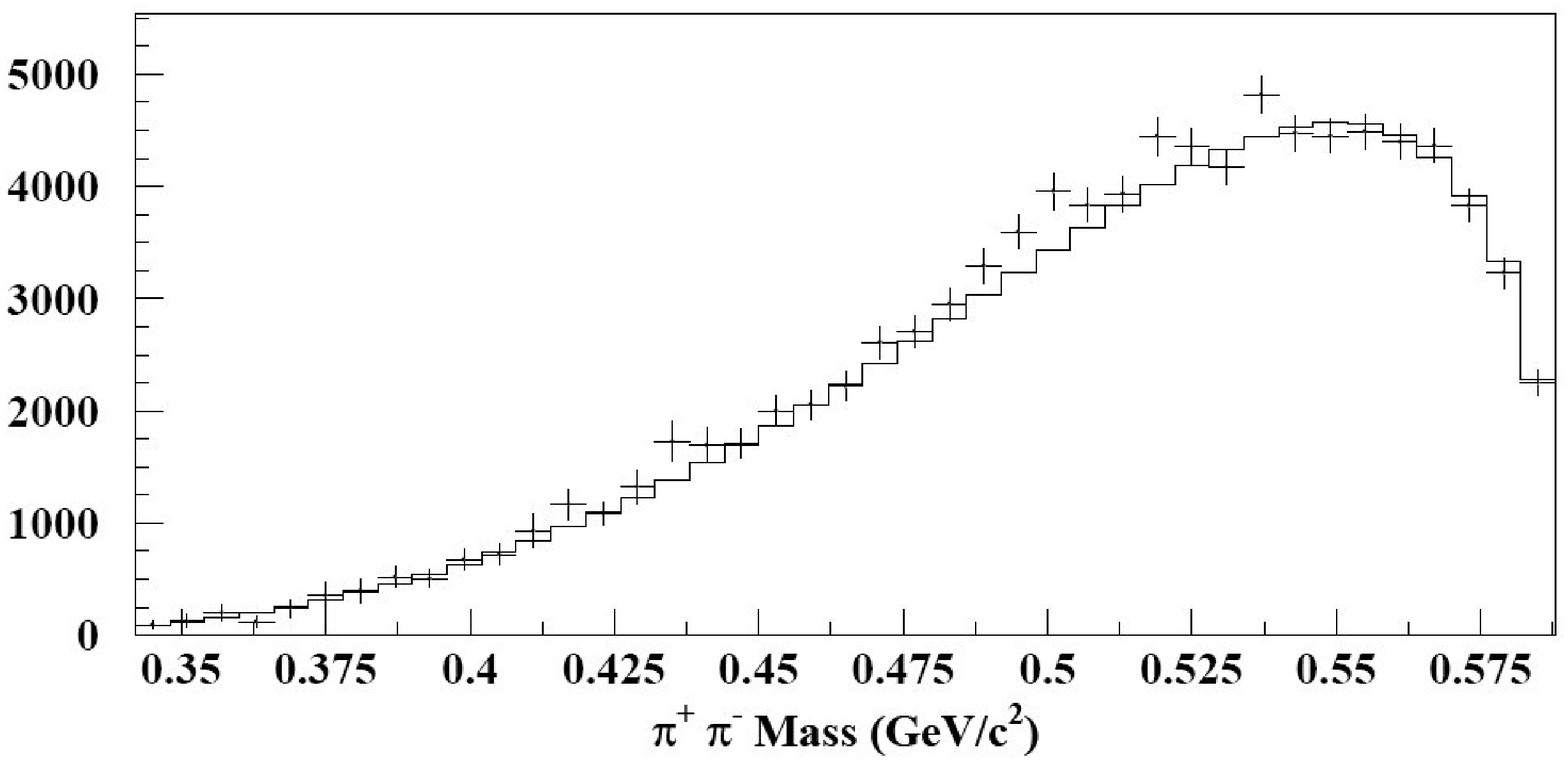}&
\includegraphics[width=0.23\textwidth,height=4cm]{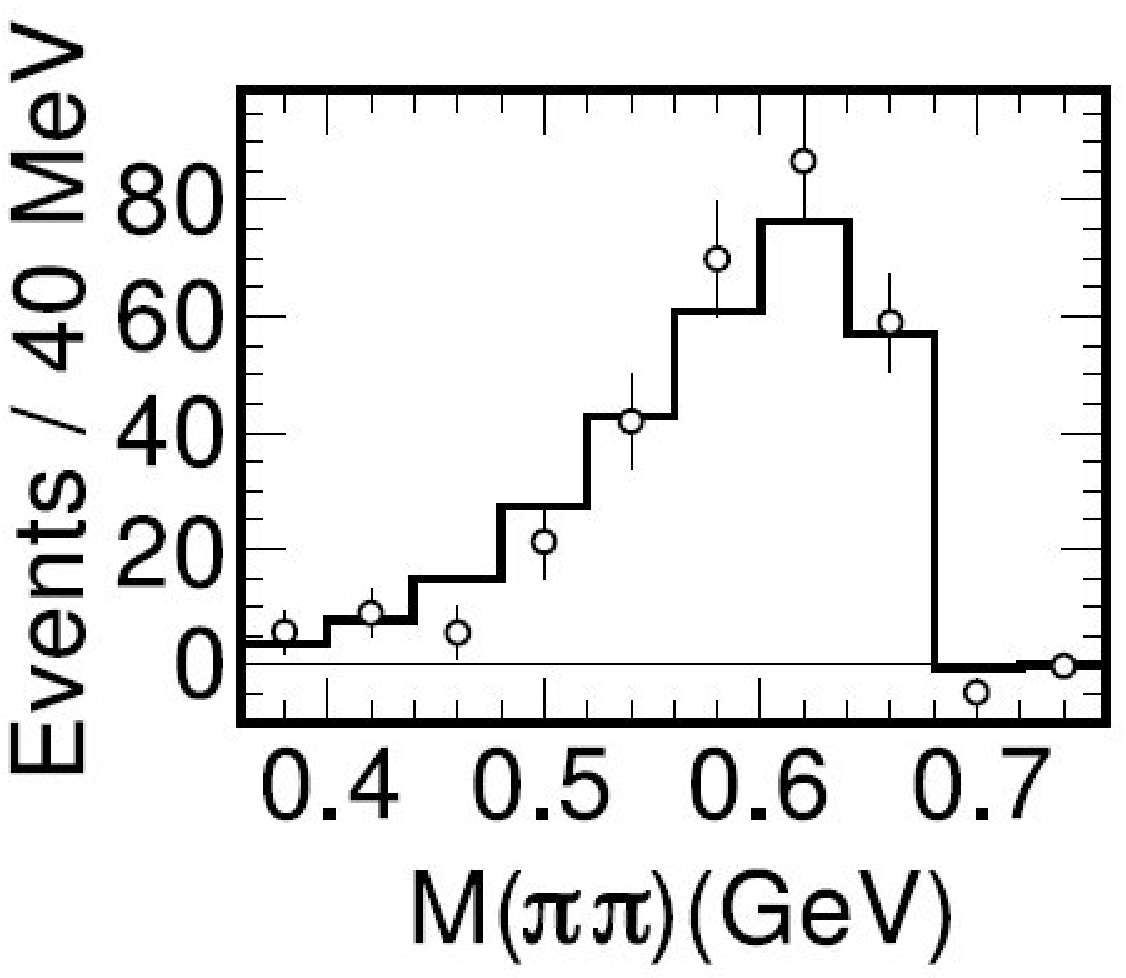}&
 \includegraphics[width=0.23\textwidth,height=4cm]{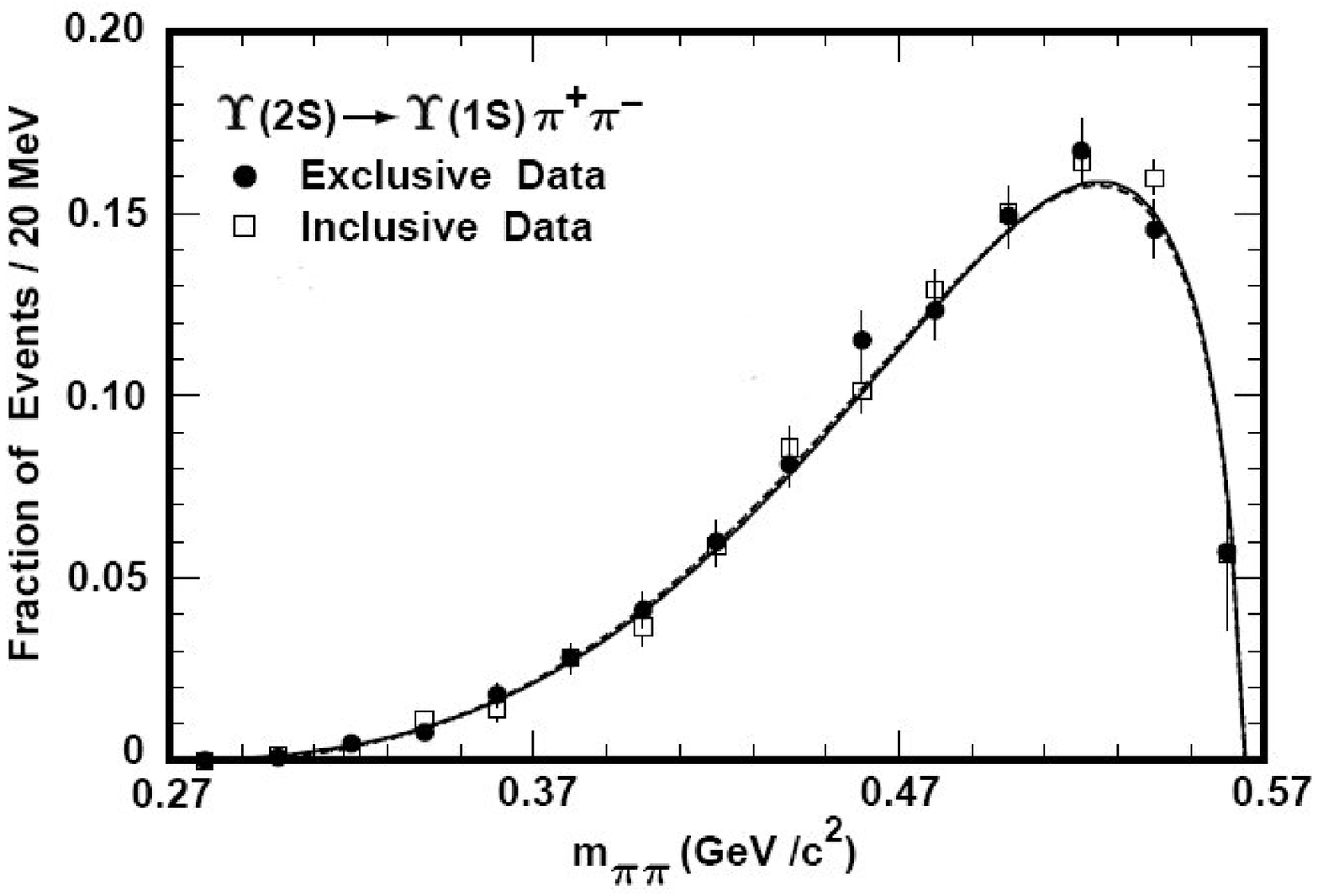} \\
\hspace{-2mm}\includegraphics[width=0.23\textwidth,height=4cm]{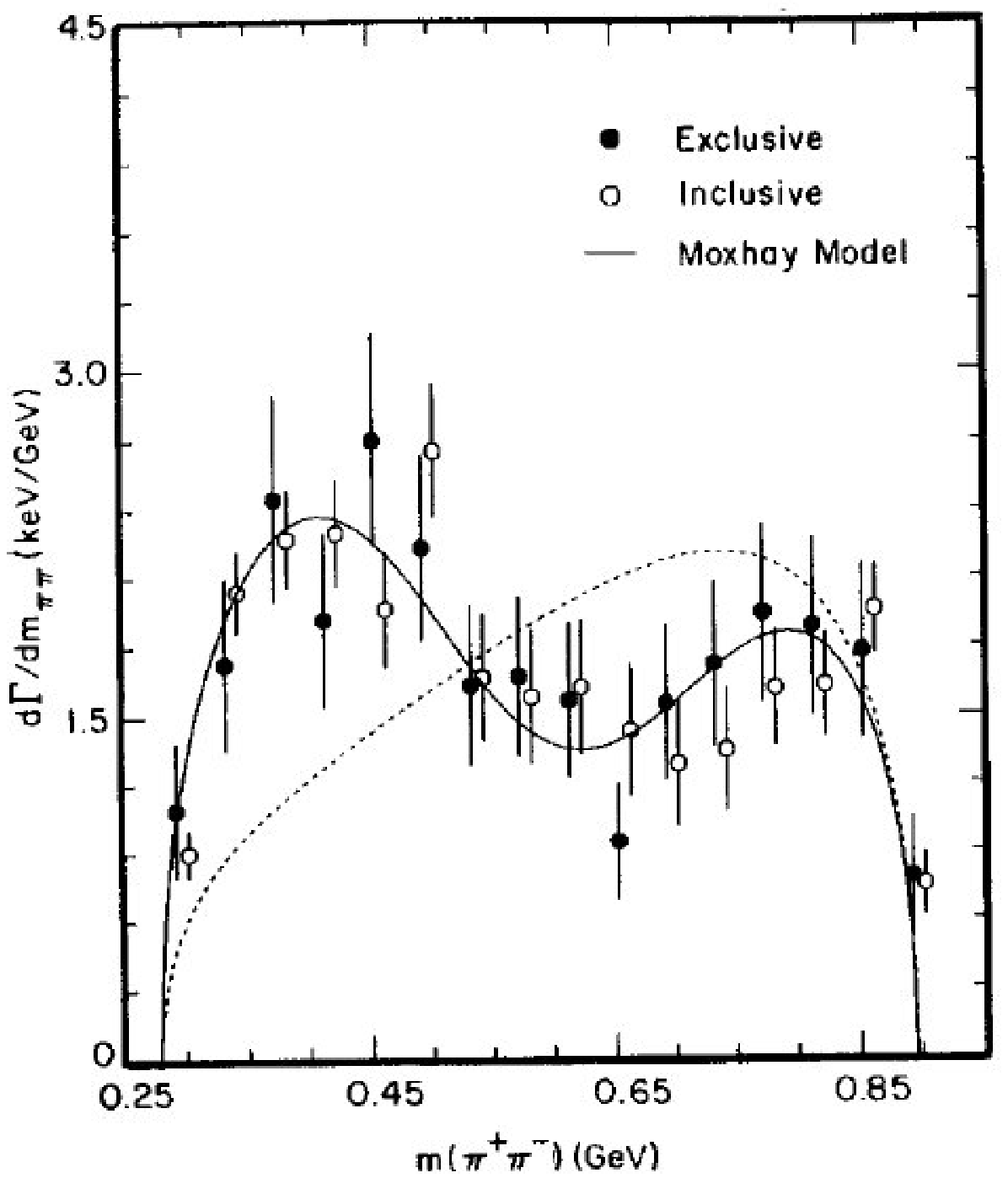}&
\includegraphics[width=0.24\textwidth,height=4cm,angle=-0.5]{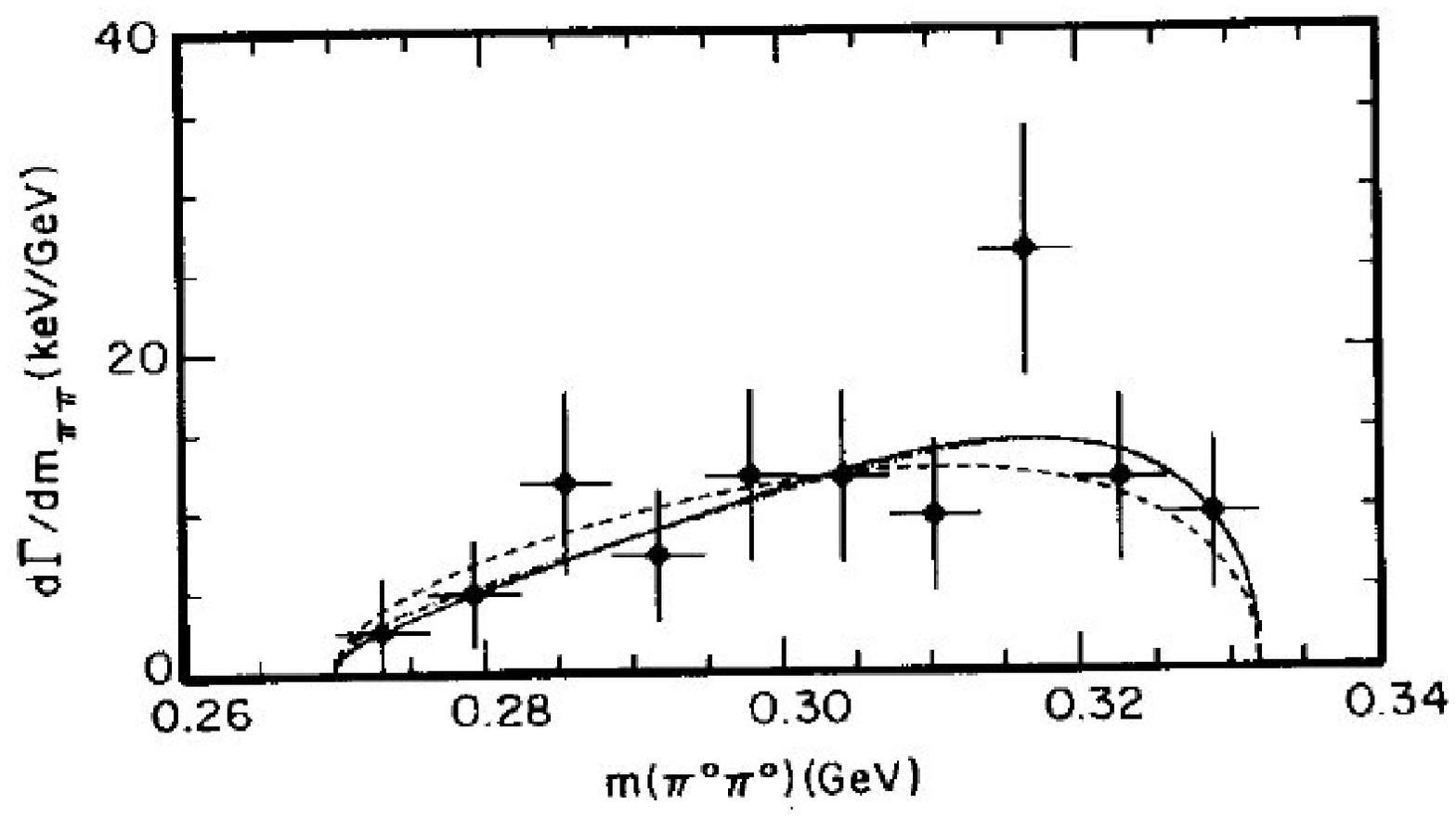}&
 \includegraphics[width=0.23\textwidth,height=4cm]{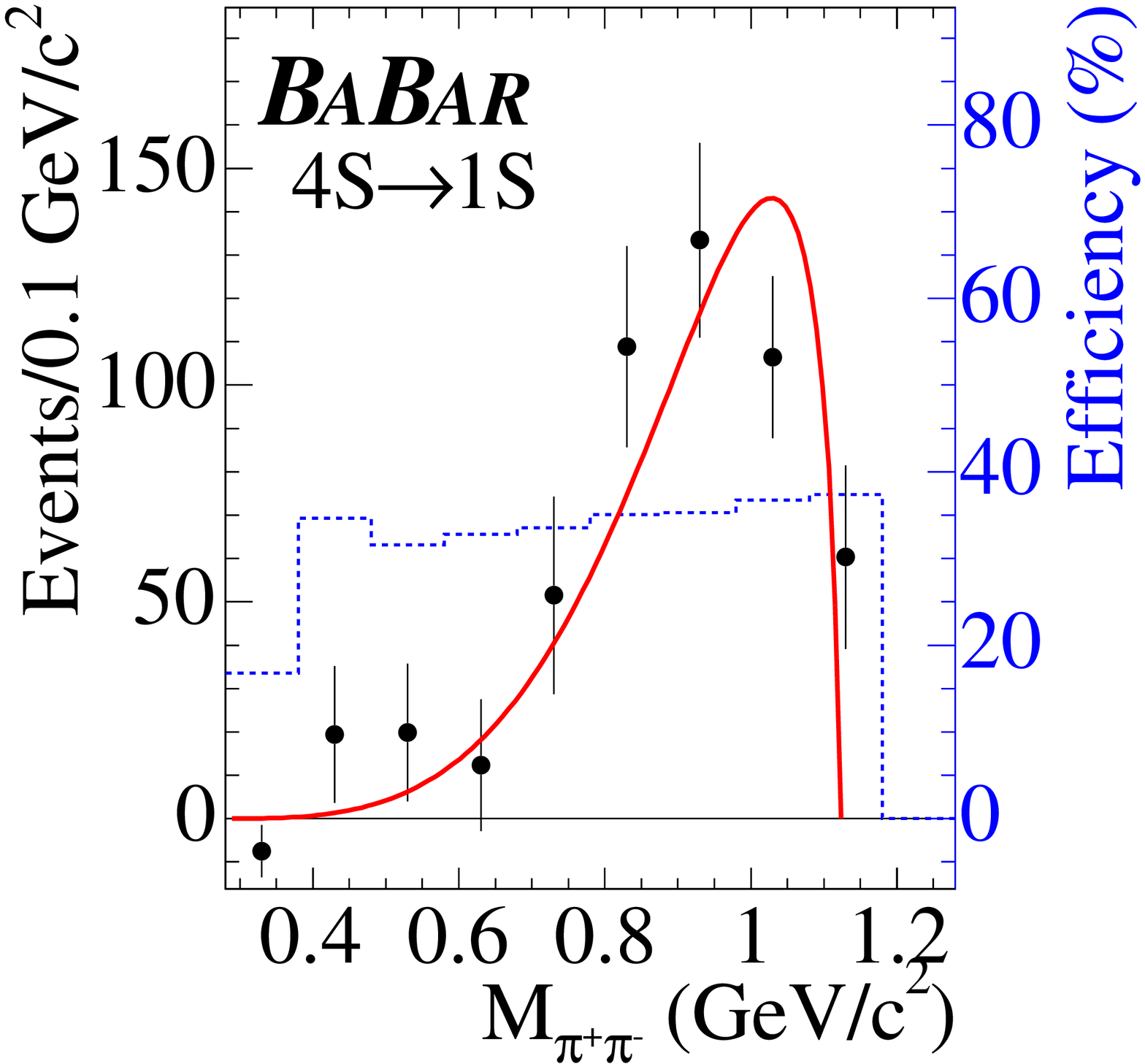}&
 \includegraphics[width=0.23\textwidth,height=4cm]{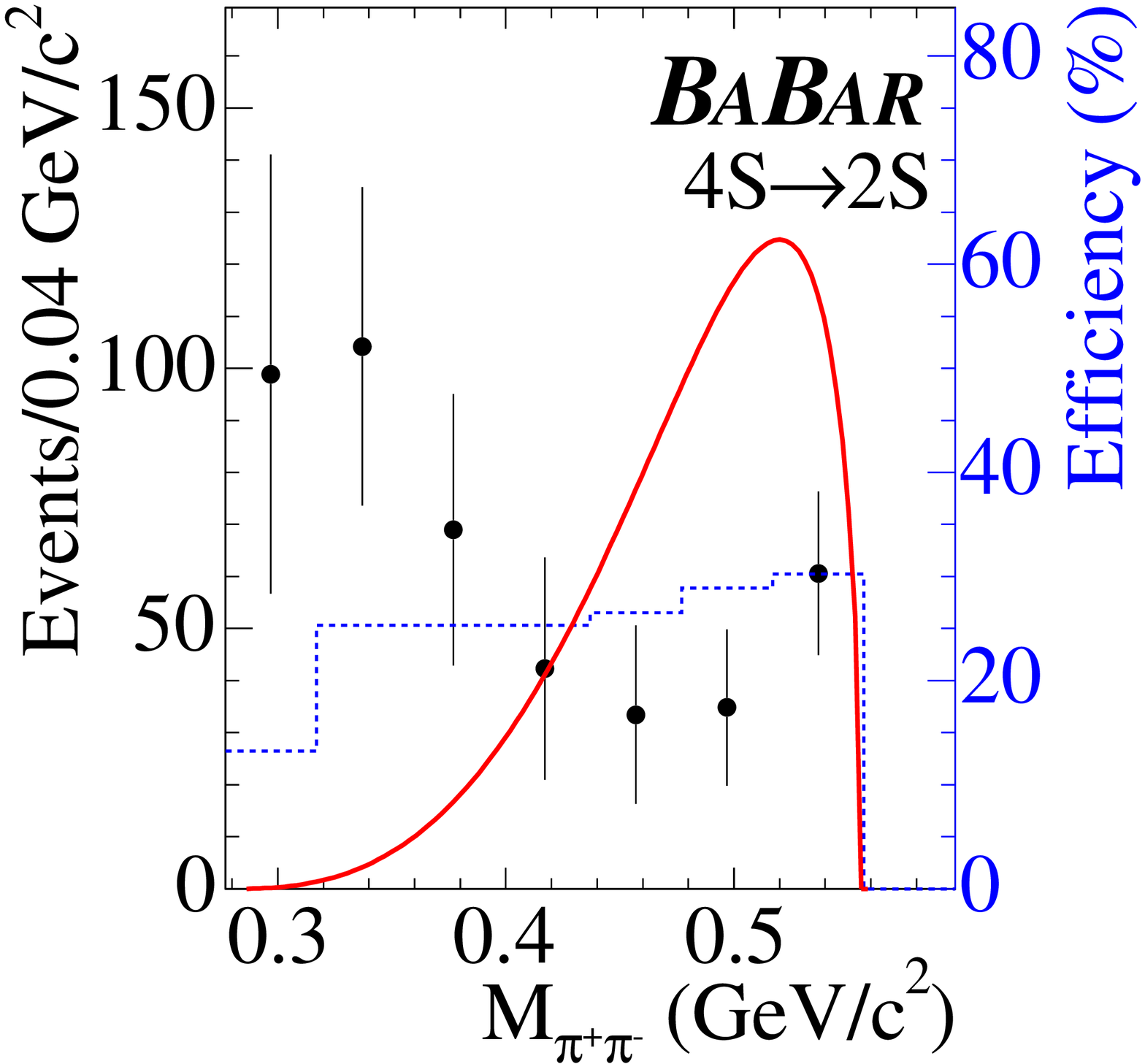} \\
\end{tabular}
\end{center}
\caption{\label{heavy-pipi}
a) $\Upsilon(nS)\to\Upsilon(mS)\pi\pi$ decays as a two-step
process with the emission of two gluons followed by hadronisation to
pion pairs. Other diagrams: $M_{\pi\pi}$ mass distributions from
two-pion emission of heavy quarkonia. The solid lines represent fits or
expected distributions;  (b) $\psi(2S)\to\pi\pi J\psi$
\cite{Bai:1999mj}; (c) for $\psi(3770)\to\pi\pi
J\psi$ \cite{Adam:2005mr}, the latter distribution contains a 1/3
contribution from $\psi(2S)\to\pi\pi J\psi$ decays; (d)
$\Upsilon(2S)\to \pi\pi\Upsilon(1S)$ \cite{Alexander:1998dq}; (e)
$\Upsilon(3S)\to \pi\pi\Upsilon(1S)$; (f) $\Upsilon(3S)\to
\pi\pi\Upsilon(2S)$  \cite{Brock:1990pj,Butler:1993rq}; (g)
$\Upsilon(4S)\to \pi\pi\Upsilon(1S)$; (h) $\Upsilon(4S)\to
\pi\pi\Upsilon(2S)$ \cite{Aubert:2006bm} (see also
\cite{Sokolov:2006sd}). The dotted histograms in (g) and (h) represent
the efficiency. } \end{figure}

Transitions between bound $c \bar{c}$ or $b \bar{b}$ states give access
to their dynamics at short distances. The emission of a $\pi\pi$ pair
is considered to proceed in a two step process by the emission of two
gluons followed by hadronisation to pion pairs, as indicated in
Fig.~\ref{heavy-pipi}a. Due to the small mass differences between
initial and final $Q\bar Q$ system, the gluons are soft and cannot be
handled by perturbative QCD.  In models, the gluon fields are expanded
into a multipole series, the pion hadronisation matrix element is
calculated using current algebra, PCAC, and gauge invariance.

Data are shown in Fig. \ref{heavy-pipi}b-h) for $\psi(2S)\to\pi\pi
J\psi$,  $\psi(3770)\to\pi\pi J\psi$,
$\Upsilon(2S), \Upsilon(3S), \Upsilon(4S)\to \pi\pi\Upsilon(1S)$, and
$\Upsilon(3S)$ and $\Upsilon(4S)\to \pi\pi\Upsilon(2S)$. From the
low-energy $\pi\pi$ interactions, the $\pi\pi$ mass spectra are
expected to exhibit an intensity increasing with the $\pi\pi$ mass, in
agreement with most data in Fig. \ref{heavy-pipi}. Only $\Upsilon(3S)
\to \pi^+ \pi^- \Upsilon(1S)$ and $\Upsilon(4S) \to \pi^+ \pi^-
\Upsilon(2S)$ are significantly different. Obviously, the differences
cannot be assigned to $\pi\pi$ interactions. The anomalous behaviour in
Fig. \ref{heavy-pipi}e,h could signal the existence of an unknown
tetraquark resonance \cite{Voloshin:1982ij,Belanger:1988hs,%
Anisovich:1995zu}. However, the $\Upsilon(3S) \to \pi^+ \pi^-
\Upsilon(1S)$ decay mode is suppressed, and interfering non-leading
diagrams could be the cause for the deviations from the expected shape
\cite{Chakravarty:1992zt}. The $\rm r\hat{o}le$ of nodes in the wave functions of
radially excited states is not fully explored. Fore a more detailed
discussion, we refer the reader to a recent review on this subject
\cite{Voloshin:2006ce}.

\subsection{\label{The J/psi states above the Dbar D threshold}
The  J/$\psi$ states above the $D\bar D$ threshold}

A few years ago, the charmonium region above the $D\bar D$ threshold
was uncharted territory. Now, five states are known, one
reasonably well established and observed in different reactions and by
different collaborations. The remaining four states were observed with
weak statistical evidence only, quantum numbers are often unknown.
Further experimental research is certainly needed. The best studied
resonance is called $X(3872)$.

The new states, in particular the $X(3872)$, are surprisingly narrow
and are often close to important thresholds. These facts motivated
interpretations of some of the new states as hadronic molecules or
tetraquark states. Hadronic molecules and tetraquark states with hidden
charm are in the discussion since very long \cite{Bander:1975fb,%
DeRujula:1976qd,Voloshin:1976ap,Ader:1981db,Tornqvist:1994aa}.

Before discussing evidence for the new meson resonances and their
interpretation, we have to define what we mean when calling a
resonance a molecule, tetraquark state, or a $q\bar q$ meson. Any meson
may have a complicated Fock space expansion, with $q\bar q$, molecular,
and tetraquark components in the wave function. One possibility would
be, e.g., to call a meson a molecule when the largest component of its
wave function can be decomposed into two colourless objects even though
it has a $q\bar q$ `seed'. We call such states $q\bar q$ mesons
having molecular properties or a molecular character when we imply
that the seed is required for the meson to exist, and define molecules
and tetraquark mesons as states falling outside of the $q\bar q$
classification. Molecular states and $q\bar q$ states could mix and
form two mesons with a more complex structure. In this case, one of the
states is still -- from the quark model point of view --
supernumerous. In this case we call one state a $q\bar q$ meson
with a molecular component mixed in, the other state a molecule with a
$q\bar q$ component. Our main concern is, e.g., if $X(3872)$ adopts
the position of the $\chi_{c1}(2P)$ charmonium state or do we expect an
additional state to fill that slot\,?

\subsubsection{\label{X(3872)}
$X(3872)$}

\begin{figure}[pb]
\bc
\begin{tabular}{ccc}
\hspace{-3mm}\includegraphics[width=0.35\textwidth,height=0.23\textwidth,clip=]{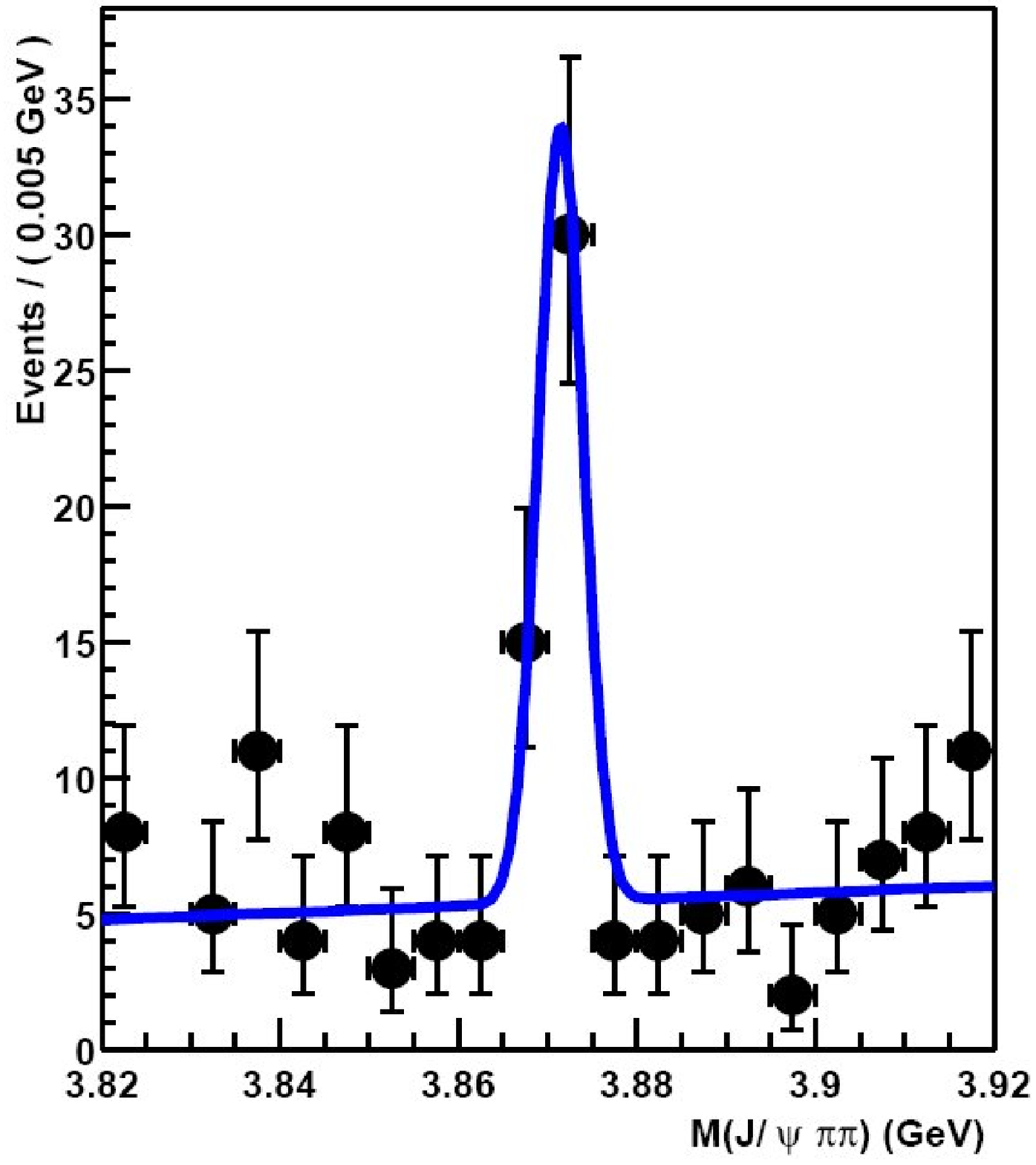}&
\hspace{-5mm}\includegraphics[width=0.35\textwidth,height=0.25\textwidth,clip=]{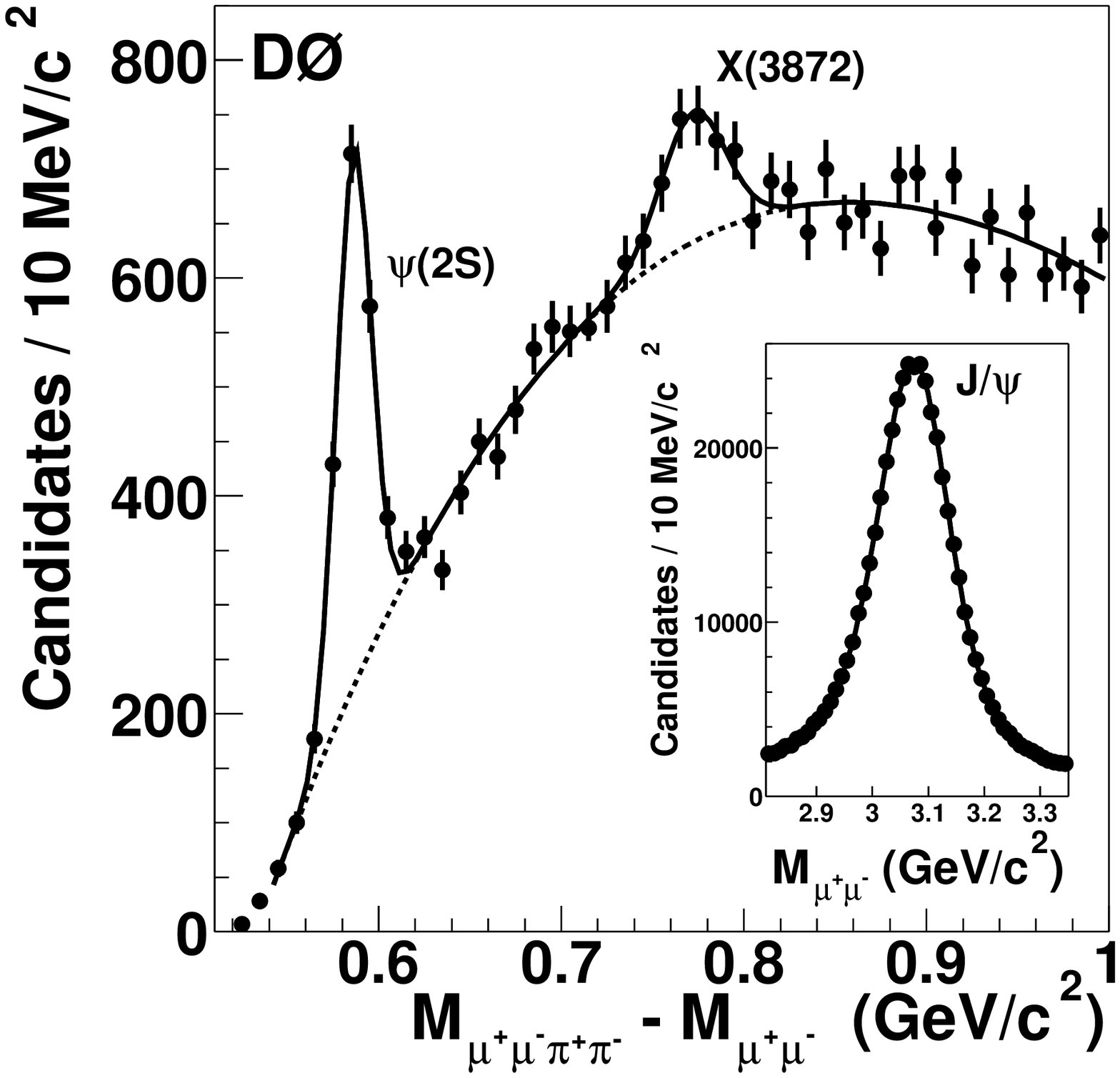}&
\hspace{-7mm}\includegraphics[width=0.32\textwidth,height=0.25\textwidth,clip=]{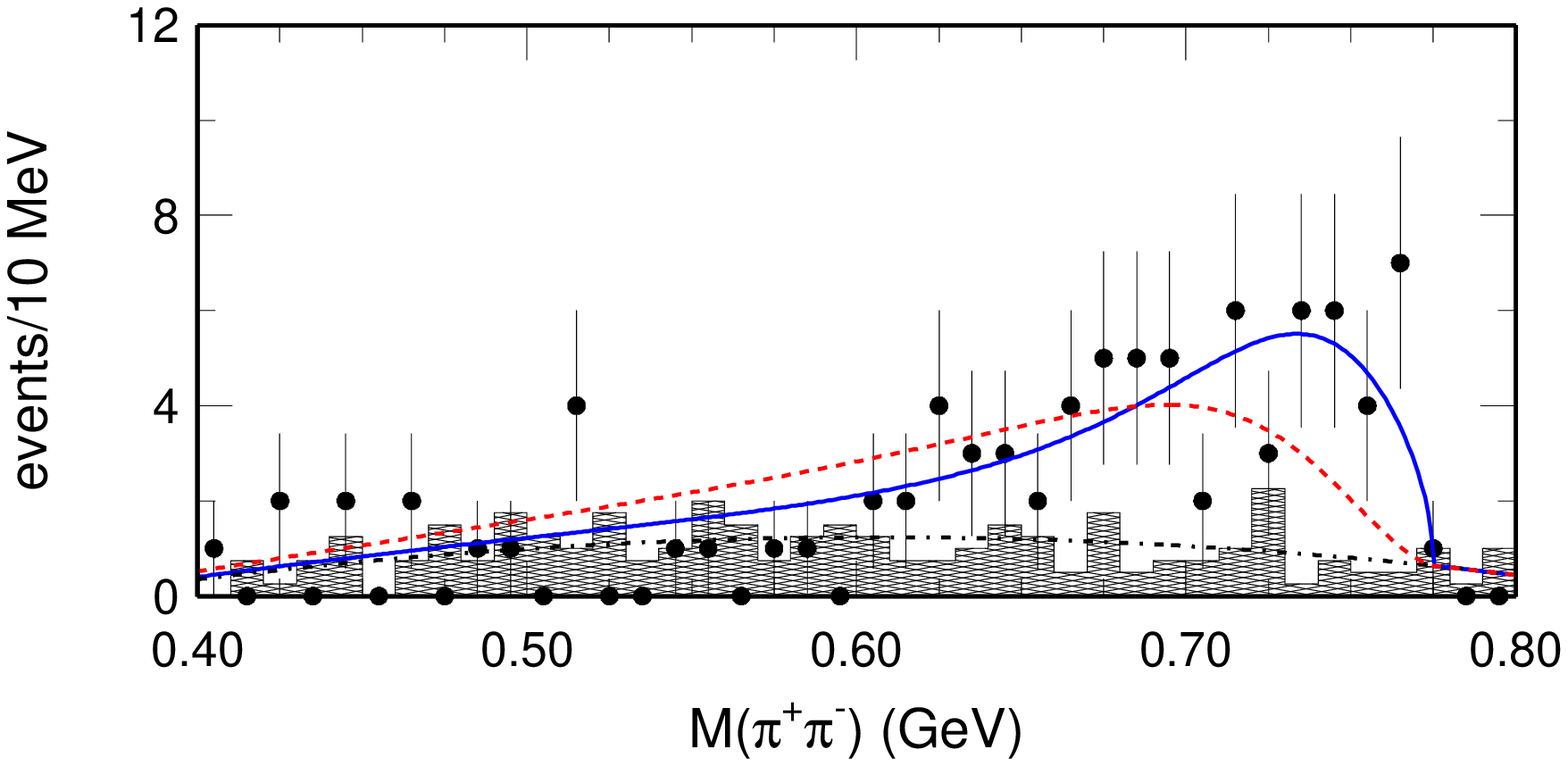} \\
\hspace{-4mm}\includegraphics[width=0.35\textwidth,height=0.25\textwidth,clip=]{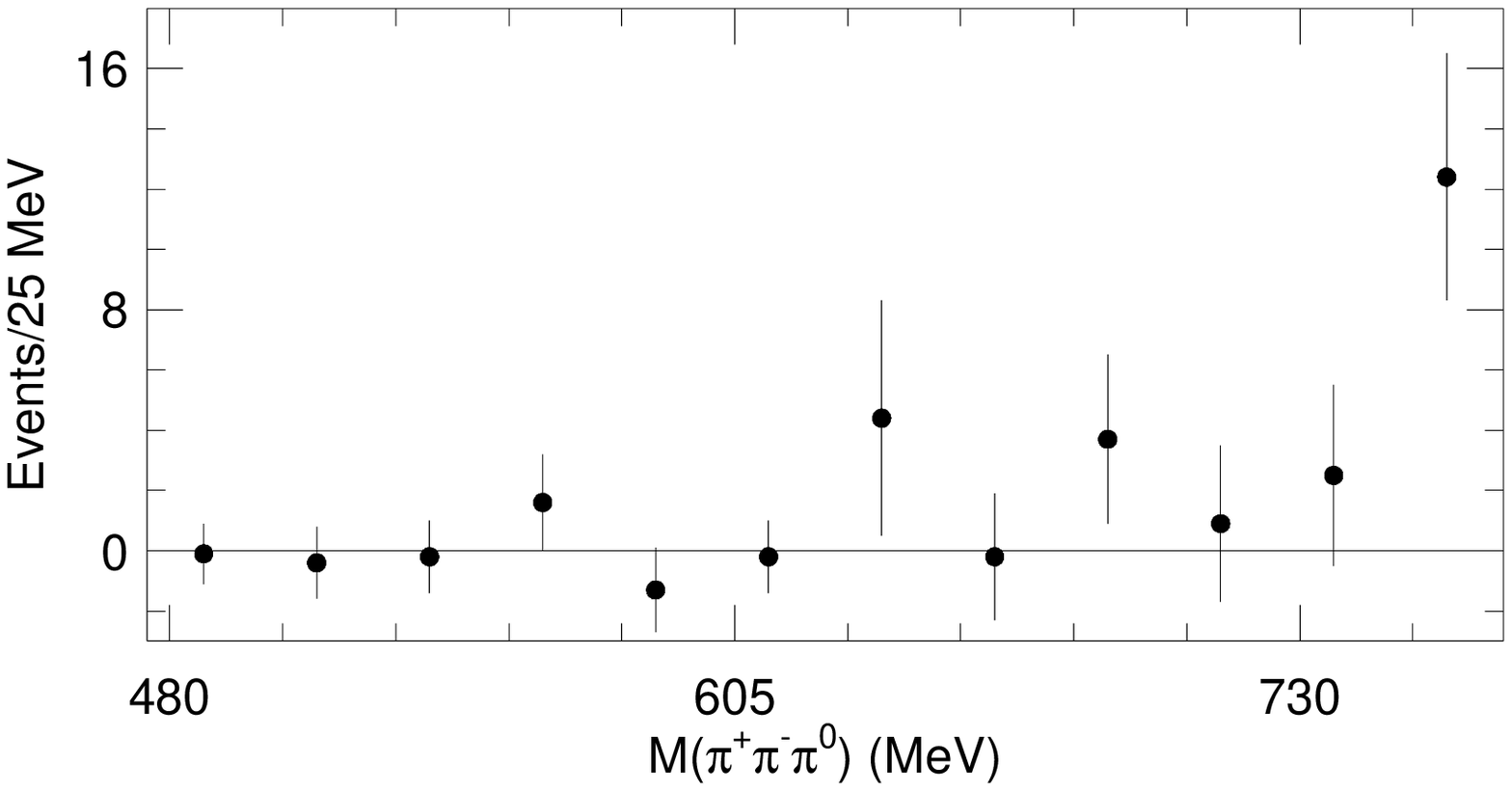}&
\hspace{-8mm}\includegraphics[width=0.35\textwidth,height=0.25\textwidth,clip=]{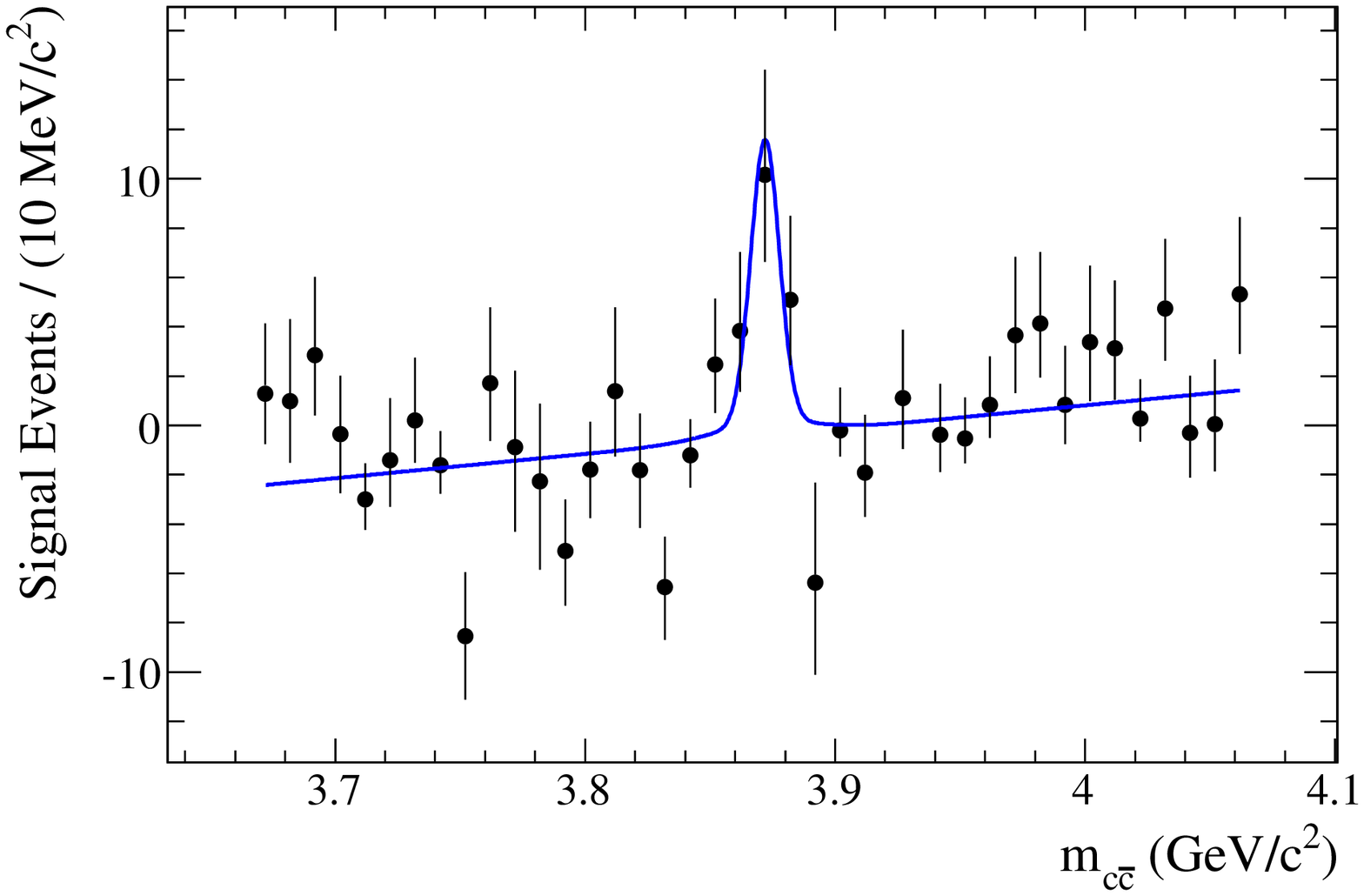}&
\hspace{-6mm}\includegraphics[width=0.35\textwidth,height=0.255\textwidth,clip=]{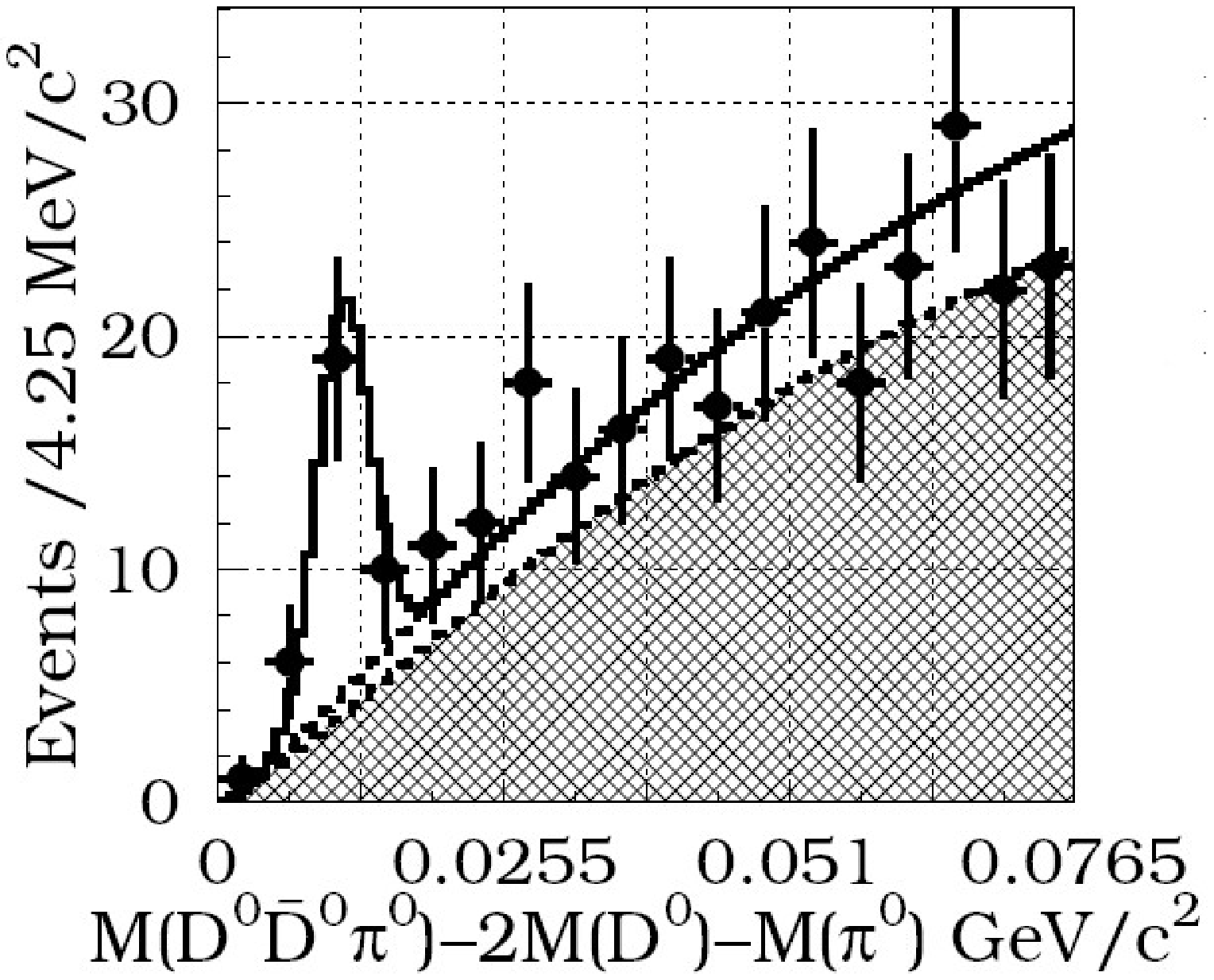} \\
\end{tabular}\vspace{-88mm}\\
\hspace{-32mm}a\hspace{105mm}b\hspace{18mm}c\vspace{41mm}\\
\hspace{-33mm}d\hspace{103mm}e\hspace{20mm}f\vspace{39mm}\\
\ec
\caption{\label{fig:3872}
a:  The $X(3872)$ in  $ B^+$ decays to $ K^+ (\pi^+\pi^-J/\psi)$
\cite{Choi:2003ue}. b: The $\Delta M = M(\mu^{+} \mu^{-} \pi^{+}
\pi^{-}) - M(\mu^{+}\mu^{-})$ mass difference with $\psi(2S)$ and
$X(3872)$. The insert shows the intermediate J/$\psi$
\cite{Abazov:2004kp}. c: The $\pi^+\pi^-$ recoiling against the
J/$\psi$ is compatible with a $\rho$ and vanishing orbital angular
momentum between  J/$\psi$ and the $\rho$ \cite{Abe:2005iy}. d: The
number of $ B^+ \to K^+(\pi^+\pi^-\pi^0 J/\psi)$ events as a function
of the $\pi^+\pi^-\pi^0$  mass \cite{Abe:2005ix}. e: The $\gamma J/\psi$
invariant mass distribution \cite{Abe:2005ix}. f: Evidence for
$X(3872)\to D^0\bar D{}^0\pi^0$ \cite{Gokhroo:2006bt}.}
\end{figure}

\paragraph{Discovery and the $X(3872)$ quantum numbers.}
The  $X(3872)$ was discovered by the BELLE collaboration in the
reaction chain $ B^+ \to K^+ X(3872)$, $X(3872)\to\pi^+\pi^-J/\psi$
\cite{Choi:2003ue}. Fig. \ref{fig:3872}a shows the $\pi^+\pi^-J/\psi$
mass distribution recoiling against a $K^+$ in $B^+$ decays with a
strong peak above little background. It is reassuring that $X(3872)$
is now observed, in the same decay mode, by three other collaborations.
The BaBaR experiment confirmed the existence of the $X(3872)$ in the
same reaction, the $D\O$ \cite{Abazov:2004kp} and the CDF2
\cite{Acosta:2003zx,Abulencia:2005zc} (see Fig. \ref{fig:3872}b)
collaborations reported observation of  $X(3872)$ in $p\bar{p}$
collisions at $\sqrt{s}$= 1.96 TeV. The Particle Data Group gives a
mean value $M=3871.2\pm0.5$\,MeV/c$^2$ and a width
$\Gamma<2.3$\,MeV/c$^2$ at 90\% confidence level. Due to its decay into
of J/$\psi\pi^+\pi^-$ and its narrow width, it must contain a charm and
an anti-charm quark. In $\bar pp$ collisions, $X(3872)$ and $\psi(2S)$
are produced with very similar production patterns
\cite{Abazov:2004kp,Bauer:2004bc}. Obviously, $X(3872)$ is produced like a
regular $c\bar c$ meson, and we anticipate that at short distances,
$X(3872)$ should have a $c\bar c$ structure. Important further
contributions were made to determine its properties.

The $\pi^+\pi^-J/\psi$ invariant mass distribution is presented in Fig.
\ref{fig:3872}c. It was noticed already in \cite{Choi:2003ue} that a
large fraction of $X(3872)\to \pi^+\pi^- J\psi$ decays comes from
high-mass $\pi^+\pi^-$ pairs. The $\pi^+\pi^-$ pair could be in $S$-wave
or form a $\rho$. A charmonium decay into J/$\psi\,\rho$ breaks isospin
and may be unlikely, but it was shown that the $\pi^+\pi^-$ invariant
mass distribution enforces a $\rho$ as intermediate state
\cite{Bugg:2004sh}. The angular momentum between J/$\psi$ and $\rho$
was restricted to $L=0$ or 1 \cite{Abulencia:2005zc}. Possible
interpretations of the new state were given early
\cite{Close:2003sg,Swanson:2003tb,Wong:2003xk,Barnes:2003vb,%
Braaten:2004ai,Braaten:2004fk,Voloshin:2004mh,Tornqvist:2004qy,%
Eichten:2004uh,Suzuki:2005ha,AlFiky:2005jd}.

The BaBaR and BELLE collaborations searched for other $X(3872)$ decay
modes. These are observed with low statistics and further studies are
certainly welcome. The $X(3872)$ was seen to decay into $\omega
J/\psi$ \cite{Abe:2005ix}, $\gamma J/\psi$ \cite{Abe:2005ix,Aubert:2006aj}, and $
D^0D^0\pi^0$ \cite{Gokhroo:2006bt}. The evidence for these decays is
shown in Fig. \ref{fig:3872}d,e,f, respectively. The $X(3872)$ mass is
below the nominal J/$\psi\,\omega$ threshold, hence only virtual
$\omega$ mesons contribute to the $X(3872)\to \pi^+\pi^-\pi^0 J/\psi$
decay mode. In Table \ref{tab:3872}, production and decay properties of
$X(3872)$ are listed. Neglecting further $X(3872)$ decay modes, we find
a total branching ratio $\mathcal{B}\{{ B^+\to K^+}
X(3872)\}=1.30_{-0.34}^{+0.20}\cdot 10^{-4}$.

\begin{table}[pb]
\caption{\label{tab:3872}$X(3872)$ production rates and decay
modes in $B$ decays in comparison to production rates of other
charmonium states.\vspace{2mm}}
\bc \renewcommand{\arraystretch}{1.5}
\begin{tabular}{cccrc}
\hline\hline
${ B^+\to K^+} X(3872)$&$X(3872)$ decay mode&\multicolumn{2}{c}{
Branching ratio}&Ref\\
\hline & $ D^0D^0\pi^0$                                 &
$1.07\pm 0.31_{-0.33}^{+0.19}$&$\cdot 10^{-4}$ & \cite{Gokhroo:2006bt}
\\
       &  $\pi^+\pi^-J/\psi$                                       &
          $(1.14 \pm 0.20 )$&$ \cdot 10^{-5}$ &\cite{Eidelman:2004wy}  \\
      &  $\pi^+\pi^-\pi^0J/\psi$                                 &
           $(1.1 \pm 0.5 \pm 0.4)$&$ \cdot 10^{-5}$ & a\\
       &  $\gamma J/\psi$                                           &
    $(1.8 \pm 0.6 \pm 0.1 )$&$ \cdot 10^{-6}$
&\cite{Abe:2005ix}   \\
       &  $\gamma J/\psi$                                           &
$(3.3 \pm 1.0 \pm 0.3 )$&$ \cdot 10^{-6}$       &\cite{Aubert:2006aj}\\
\hline ${ B^+\to K^+} X(3872)$&&$1.30_{-0.34}^{+0.20}$&$\cdot
10^{-4}$    &\\ ${ B^+\to
K^+}\chi_{c0}(1P)$&&$1.6_{-0.5}^{+0.4}$&$\cdot 10^{-4}$ &
\cite{Eidelman:2004wy}\\
${ B^+\to K^+} \chi_{c1}(1P)$&&$5.3\pm0.7 $&$\cdot 10^{-4}$  &
\cite{Eidelman:2004wy}\\
${ B^+\to K^+} \chi_{c^2}(1P)$&&$<2.9 $&$\cdot 10^{-5}$  &
\cite{Eidelman:2004wy}\\
\hline\hline
\end{tabular}
\vspace{2mm}

a: using $\mathcal B\left\{ X(3872)\to\pi^+\pi^-\pi^0J/\psi\right\} /
\mathcal B\left\{ X(3872)\to\pi^+\pi^-J/\psi\right\}=1.0\pm0.4\pm 0.3$
\cite{Abe:2005ix}.
\ec
\renewcommand{\arraystretch}{1.0}
\end{table}

\begin{figure}[pt] \bc
\includegraphics[width=0.4\textwidth]{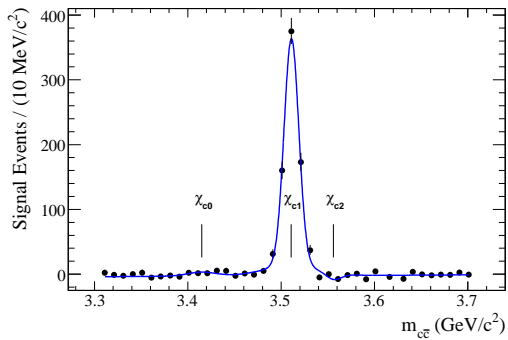}\vspace{-4mm}
 \ec
\caption{\label{chibabar}The $\chi_{cJ}$ states in $B$ decays. The $\chi$
states are identified via their radiative decay into J/$\psi$
\cite{Aubert:2006aj}. }
\end{figure}

The radiative $X(3872)$ decay into J/$\psi$, the J/$\psi\,\rho$ (with
$L=0$) and the J/$\psi\,\omega$ decay restrict the $X(3872)$
quantum numbers to $J^{PC}=0^{++}, 1^{++}$, and $2^{++}$. For a
$J^{PC}=0^{++}$ or $2^{++}$ state, the decay into $ D\bar D$ is
allowed, and these states should be broader. Hence $1^{++}$ quantum
numbers are most likely. This assumption is supported by the pattern of
the $\chi_{cJ}(1P)$ state observed by BaBaR. The $\chi_{c1}$ is the
only $\chi$ state produced with a significant strength, see Fig.
\ref{chibabar}. The branching ratios for decays into $K^+$ and
$\chi_{cJ}(1P)$ states, corrected for their $\gamma J\psi$ decay
fractions, are given in the lower part of Table \ref{tab:3872}. The CDF
collaboration measured angular distributions and correlations of the
$X(3872)\to J/\psi\pi^+\pi^-$ decay mode and constrained the possible
spin, parity, and charge conjugation parity of the $X(3872)$ to $J^{PC}
= 1^{++}$ and $2^{-+}$ \cite{Abulencia:2006ma}. For further
discussions, we assume $J^{PC} = 1^{++}$.

The BaBaR collaboration searched for a charged partner of $X(3872)$ in
the decay mode $X^-\to J/\psi\,\pi^-\pi^0$ in $B$ meson decays $ B^0\to
X^- K^+$ and $ B^-\to X^- K^0_S$ and determined the $X(3872)$ isospin
to $I=0$ \cite{Aubert:2004zr}, hence the $X(3872)$ quantum numbers
correspond to $\chi_{c1}(3872)$. It remains to be seen if $X(3872)$ can
be identified with the $\chi_{c1}(2P)$ charmonium state or if it
requires a different interpretation.

\paragraph{Molecular and tetraquark interpretations of $X(3872)$.}

There is a number of reasons not to identify $\chi_{c1}(3872)$ with
the $c\bar c$ state $\chi_{c1}(2P)$. The most striking feature is the
strong isospin breaking. The branching ratios to J/$\psi\,\rho$ and
J/$\psi\,\omega$ are similar in magnitude; if $\chi_{c1}(3872)$ is a
loosely bound $ D^0D^{0*}$ molecule (decoupled from $ D^+D^{-*}$ due to
the mass gap), then isospin is maximally broken. If $\chi_{c1}(3872)$
is interpreted as a tetraquark state, isospin breaking can be
implemented as well but charged partners of $X(3872)$ are predicted;
they were searched for by BaBaR \cite{Aubert:2004zr}, with negative
result. Matheus, Narison, Nielsen, and Richard \cite{Matheus:2006xi}
argued that $B$ decays may prefer decays via excited $c\bar c$ pairs
recoiling against a cluster of light quarks and that production of
neutral states with hidden charm could be favoured. Hence charged
partners of  $X(3872)$ should be searched for in other reactions, too.

The second argument is based on its low mass, compared to quark-model
calculations. The model of Barnes, Godfrey and Swanson
\cite{Barnes:2005pb}, e.g., reproduces very well the $\chi_{cJ}(1P)$
masses (with deviations of less than 10\,MeV/c$^2$) and predicts
3925\,MeV for $\chi_{c1}(2P)$. Lattice QCD puts the state to 4010
\cite{Okamoto:2001jb} or 4067\,MeV/c$^2$ \cite{Chen:2000ej}. This
50-200\,MeV/c$^2$ mass discrepancy is too large to be bridged by
parameter tuning. On the other hand, the $X(3872)$ mass is just at the
$DD^*$ threshold\footnote{\footnotesize The $D^0D^{0*}$ threshold is at
$(3871.2\pm0.6)$\,MeV/c$^2$, the $ D^+D^{-*}$ threshold at $(3879.3\pm
0.6)$\,MeV/c$^2$.} which makes it a good candidate as molecular bound
state. A recent lattice calculation gave a mass of the $\chi_{c1}(2P)$
state of 3853\,MeV/c$^2$ \cite{Chen:2007vu}. Obviously, there is
some flexibility in the mass calculation.

Arguments based a calculation of $B^+\to X(3872)K^+$ in the pQCD
approach are shown to be in conflict with an interpretation of $X(3872)$
as $2^3P_1$ charmonium state \cite{Liu:2007uj}.

Some other important predictions fail however when compared to
experimental numbers. The ratio \cite{Abe:2005ix,Aubert:2006aj}
\vspace{-5mm}

\begin{equation}
{\mathcal B\left(  X(3872)\to \gamma J/\psi \right)\over \mathcal B \left(
X(3872)\to \pi^+\pi^- J/\psi \right) } = 0.22\pm 0.06
\end{equation}

is much larger than the value 0.007 predicted for a $D^0D^{0*}$
molecule. For a $c\bar c$ state, the ratio can be as large as 0.11 even
though also smaller values have been calculated \cite{Swanson:2004pp}.
Note that the fractional radiative yield for  $\chi_{b1}(2P)\to\gamma
\Upsilon(1S)$ is $(8.5\pm1.3)$\%, that for $\chi_{b1}(2P)\to\gamma
\Upsilon(2S)$ $(21\pm4)$\%. Ref. \cite{Swanson:2004pp} has shown that
this ratio depends critically on the $X(3872)$ internal structure.

For the absolute value, only an upper limit can be given
using the $\chi_{c1}(3872)$ width $\Gamma<2.3$\,MeV/c$^2$ and the
branching ratios of Table \ref{tab:3872}. The radiative rate
$\chi_{c1}(3872)\to\gamma J/\psi$ is bound by
\begin{equation}
\Gamma_{\chi_{c1}(3872)\to\gamma{J}\psi}< 250\,{\rm keV/c^2}
\end{equation}

provided that there are no large unseen decay modes. This upper limit
can be compared to the $\chi_{c1}(1P)\to\gamma{J}\psi$ radiative width
of 288\,keV/c$^2$.

A second weak point of the molecular interpretation is the $X(3872)\to
D\bar D{}^0\pi^0$ decay branching ratio. The $D^0\bar D^0\pi^0$ system
shows a peak at $3875.4\pm 0.7 ^{+1.2}_{-2.0}$\,MeV mass
\cite{Gokhroo:2006bt}. In \cite{:2007rv}, $M=3875.1\pm 1.1\pm
0.5$\,MeV and $\Gamma=3.0^{+4.6}_{-2.3}\pm0.9$\,MeV were reported. If
this state is identified with $X(3872)$, as suggested in
\cite{Hanhart:2007yq}, we have

\begin{equation}
{{\mathcal B}\left( X(3872)\to D^0\bar D^0\pi^0 \right)\over {\mathcal B}\left(
X(3872)\to \pi^+\pi^- J/\psi \right) } = 9.4^{+3.6}_{-4.3}
\end{equation}

while the prediction of the molecular model \cite{Swanson:2004pp} is 0.054.
In the fits by Hanhart {\it et al.} \cite{Hanhart:2007yq}, the shape of
the $X(3872)$ resonance is strongly distorted. The authors conclude
that $X(3872)$ has to be of a dynamical origin; it is interpreted as a
virtual state. Of course, it is difficult to exclude a $c\bar c$
fraction in $X(3872)$ which may serve as a seed to create the observed
state.

The observed rates in $B^+$ and $B^0$ decays do not correspond to the
molecular picture, neither. Experimentally \cite{Gokhroo:2006bt}

\begin{equation}
{B\left( B^0\to
X(3872)K^0 \right)\over B \left( B^+\to X(3872)K^+ \right) } \approx
1.62 \end{equation}

while the prediction is less than 0.1 \cite{Swanson:2004pp}.

In the limit of large quark-mass ratios, tetraquarks consisting of two
heavy and two light quarks should be stable against falling apart into
two heavy-light mesons or into a heavy quarkonium and a light-quark
meson.  The light-quark degrees of freedom cannot resolve the closely
bound system of two heavy quarks. Thus a tightly bound $\bar b\,\bar c$
or  $\bar b\,\bar b$ system in colour $\bf 3$ and two light quarks
$q\,q$ may form a heavy-baryon configuration where the heavy quark is
replaced by two heavy (anti-)quarks \cite{Chow:1994mu}. This exotic
configuration is expected to provide more stability than the molecular
(non-exotic) system consisting of two heavy-light colour-neutral mesons
\cite{Ader:1981db,Brink:1998as,Gelman:2002wf}.

In a model calculation, Janc and Rosina \cite{Janc:2004qn} suggest that
the most stable tetraquark configuration should not be of that of a
heavy baryon but rather a molecular system. Sum rules give tetraquark
masses $M_{T_{bb}}= (10.2\pm 0.3)$\,GeV/c$^2$, which is below the
$\bar{B}\bar{B}^*$ threshold, and $M_{T_{cc}}= (4.0\pm 0.2)$\,GeV/c$^2$
\cite{Navarra:2007yw}.

\begin{figure}[pt]
\bc\includegraphics[width=0.47\textwidth]{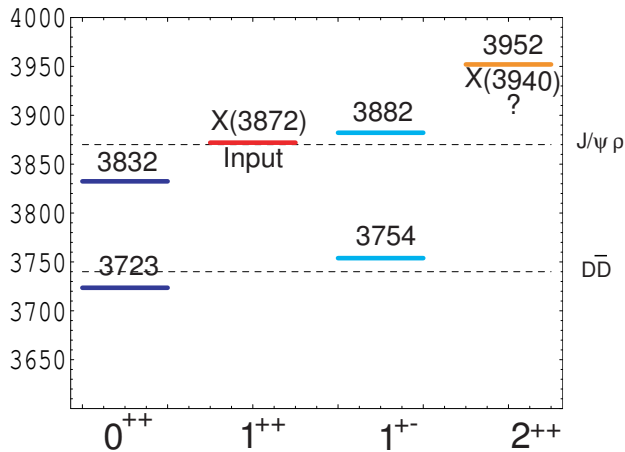}
\vspace{-6mm}\ec
\caption{Spectrum of tetraquark charmonium states \cite{Maiani:2004vq}.}
\label{maianiFig}
\end{figure}

Maiani {\it et al.} see $X(3872)$ as tetraquark charmonium and predict
the spectrum of new tetraquark charmonia including charged isospin
partners of $X(3872)$ \cite{Maiani:2004vq}. Tetraquark states were
constructed from diquarks in colour triplet scalar and vector clusters
interacting by spin-spin interactions. Using the $X(3872)$
as input, Maiani {\it et al.} predict the spectrum of new diquark
charmonia shown in Fig. \ref{maianiFig}.
H{\o}gaasen {\it et al.} \cite{Hogaasen:2005jv} postulate that
$X(3872)$ is a loosely bound tetraquark $cu\bar c \bar u$ and $cd \bar
c \bar d$ state \cite{Hogaasen:2005jv}. In \cite{Buccella:2006fn}, the
interaction of two quarks and two antiquarks in $S$-wave is calculated
including flavour-symmetry breaking.  Karliner and Lipkin
\cite{Karliner:2006hf} suggested that $bn\bar b\bar n$ and $bn\bar
c\bar n$ tetraquarks may fall below the $ B\bar B$ and $ B \bar D$
thresholds. The (four-fold) attractive forces between quarks and
antiquarks may overcome the (twofold) repulsive forces between the
light-quark and the heavy-quark pairs, and attraction prevails.
These new states may have exotic electric charge and their decays might
have striking decay patterns. The calculation might, however, depend on
the coupling scheme chosen: the interactions of two quarks with one
antiquark are individually attractive; the diquark and the antiquark
feel repelling forces. The charmonium spectrum is calculated by
Kalashnikova \cite{Kalashnikova:2005ui} in a coupled channel model with
couplings to $D\bar D$, $D\bar D^*$, $D^*\bar D^*$, $D_s\bar D_s$,
$D_s\bar D_s^*$, and $D_s^*\bar D_s^*$, thus generating a virtual bound
state just above $D\bar D^*$ threshold. Coupled channels in $S$-wave
may lead to cusps at the opening at thresholds; $X(3872)$ may very well
be such a cusp as underlined by Bugg \cite{Bugg:2004rk}.

Tetraquark models postulate a mass difference between $B^+\to
X_{\uub}K^+$  and $ B^0\to X_{\ddb}K^0$. BaBaR determined this mass
difference to $1.7\pm 1.1\pm 0.2$, a value not inconsistent with zero
\cite{Aubert:2004zr}. If molecular forces are responsible for binding the
$X(3872)$, it is difficult to escape the conclusion that a isovector
partner should be observed as well as the isoscalar $X(3872)$.

Summarizing, there are strong arguments that tetraquark states should
exist in the heavy quark limit. Models can be made which predict
tetraquarks in the charmonium system but no proof exists that in the
$cc\bar u\bar d$, $cb\bar u\bar d$, or even $bb\bar u\bar d$ system,
the heavy-quark limit is reached and that these flavour-exotic mesons
are stable against colour rearrangement and do not just fall apart.

\paragraph{Other interpretations of $X(3872)$.}

Other interpretations were attempted as well. Seth proposed
$X(3872)$ to be a vector glueball \cite{Seth:2004zb}, Li
suggested a $c\bar c g$ hybrid interpretation \cite{Li:2004st}.
Both conjectures are at variance with the large radiative yield.

\paragraph{$X(3872)$ as $\chi_{1c}(2P)$ charmonium state.}

The highest obstacles for a $c\bar c$ interpretation of $X(3872)$ are
its mass and the isospin breaking effects in their J/$\psi\,\omega$
\begin{table}[pb]
\caption{\label{tab:chicb}Masses of $\chi$ states
(in MeV/c$^2$). Cursive numbers are obtained by simple scaling. The precision
is truncated at 1\,MeV/c$^2$. \vspace{2mm}} \bc
\renewcommand{\arraystretch}{1.5}\begin{tabular}{ccccccccc}\hline\hline
\multicolumn{4}{c}{$\chi_{bJ}(nP)$ states} &                         &
\multicolumn{4}{c}{$\chi_{cJ}(nP)$ states} \\                   \hline
$\chi_{b0}(1P)$ &9.859 & $\chi_{b0}(2P)$ &10.233 & \quad &
$\chi_{c0}(1P)$ & 3415 & $\chi_{c0}(2P)$ &$\sim${\it 3840}           \\
$\chi_{b1}(1P)$ &9.893 & $\chi_{b1}(2P)$ & 10.255 & \quad &
$\chi_{c1}(1P)$ & 3511 & $\chi_{c1}(2P)$ &$\sim${\it 3900}           \\
$\chi_{b2}(1P)$ &9.912 & $\chi_{b2}(2P)$ & 10.268 & \quad &
$\chi_{c2}(1P)$ & 3556 & $\chi_{c2}(2P)$ & $3929\pm 5\pm 2$          \\
\hline\hline \end{tabular} \renewcommand{\arraystretch}{1.0}       \ec
\end{table}
and  J/$\psi\,\rho$ decays. Concerning the mass, there is of course
some uncertainty in the absolute scale of the predictions.  Below we
will identify the $Z(3930)$ with $\chi_{c2}(2P)$. It has a mass of
$3931\pm 4\pm 2$\,MeV/c$^2$, its predicted mass \cite{Barnes:2005pb} is
3972\,MeV/c$^2$. This is a 40\,MeV/c$^2$ discrepancy, too. We try to
use the bottomonium masses to estimate the mass of the $\chi_{c1}(2P)$
charmonium state see Table \ref{tab:chicb}. Simple scaling leads to
3900\,MeV/c$^2$ as expectation for the $\chi_{c1}(2P)$ mass, about
28\,MeV/c$^2$ above the observed $\chi_{c1}(3872)$ mass. If the
charmonium state $\chi_{c1}(2P)$ were a separate state at about
3900\,MeV/c$^2$, it should have been observed in the $ D^0D^0\pi^0$
mass distribution. Of course, a threshold can attract a pole position.
This is a phenomenon known from the $K\bar K$ threshold and $a_0(980)$
and $f_0(980)$.

The most likely solution seems to be that $\chi_{c1}(3872)$ is the
charmonium $\chi_{c1}(2P)$ state strongly coupled to $D^0D^{0*}$. The
$D^{\pm}D^{\mp *}$ threshold is 8\,MeV higher, the coupling to
$D^{\pm}D^{\mp *}$ plays no dynamical $\rm r\hat{o}le$ in the decay
thus creating large isospin breaking effects. As long as no isospin
partner of $X(3872)$ is discovered, we must conclude that the \ccb\
component plays a decisive $\rm r\hat{o}le$ in binding the molecular
$D^0D^{0*}$ system. In the sense of our introductory discussion,
$X(3872)$ is a $c\bar c$ meson of molecular character. When probed with
high momenta like in $p\bar p$ collisions, the $c\bar c$ component
prevails; at large separations the molecular character is revealed by
its strong affinity to the $D^0D^{0*}$ system.

\subsubsection{\label{The X,Y,Z resonances at 3940 MeV}
The $X,Y,Z$ resonances at 3940\,MeV/c$^2$}

\begin{figure}[pb] \begin{tabular}{ccc}
\hspace{-5mm}\includegraphics[width=0.39\textwidth,height=0.35\textwidth,clip=]{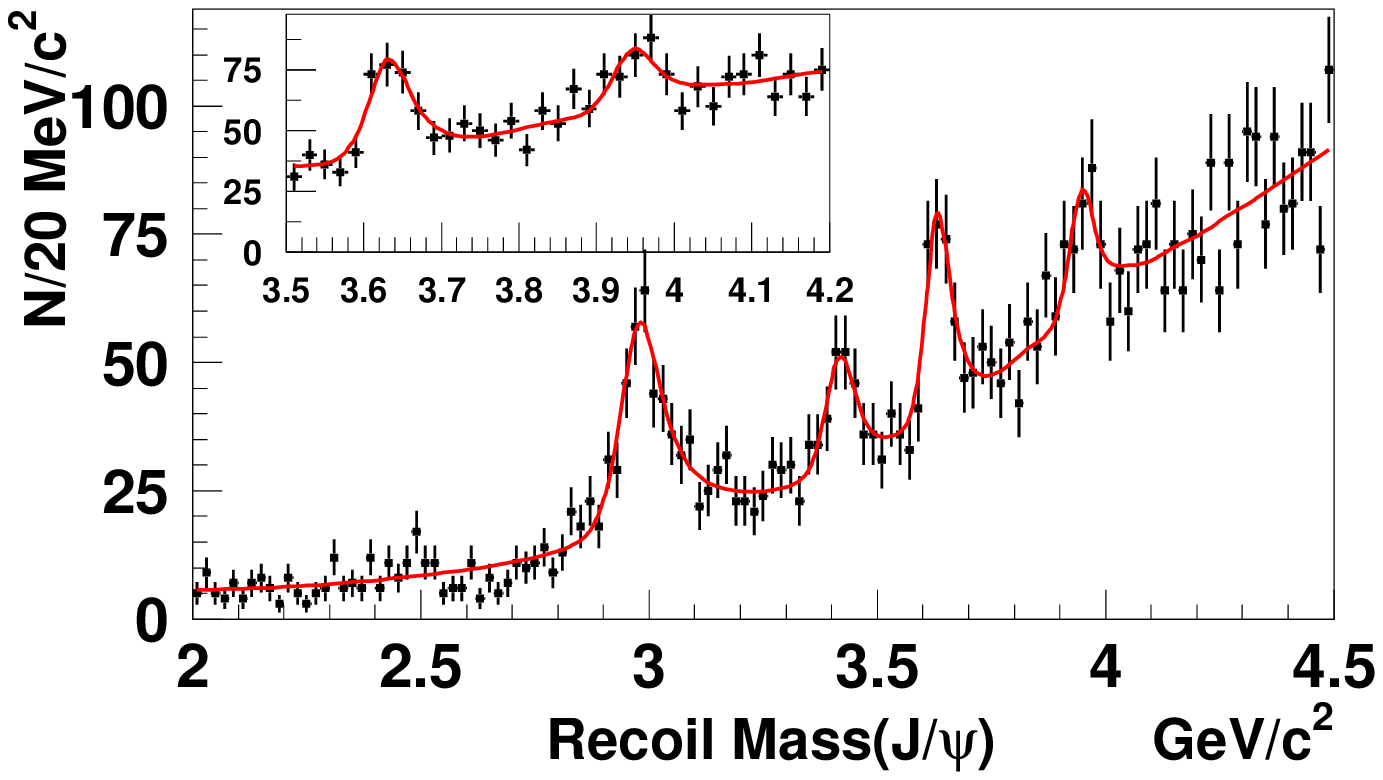}&
\hspace{-3mm}\includegraphics[width=0.3\textwidth,height=0.35\textwidth,clip=]{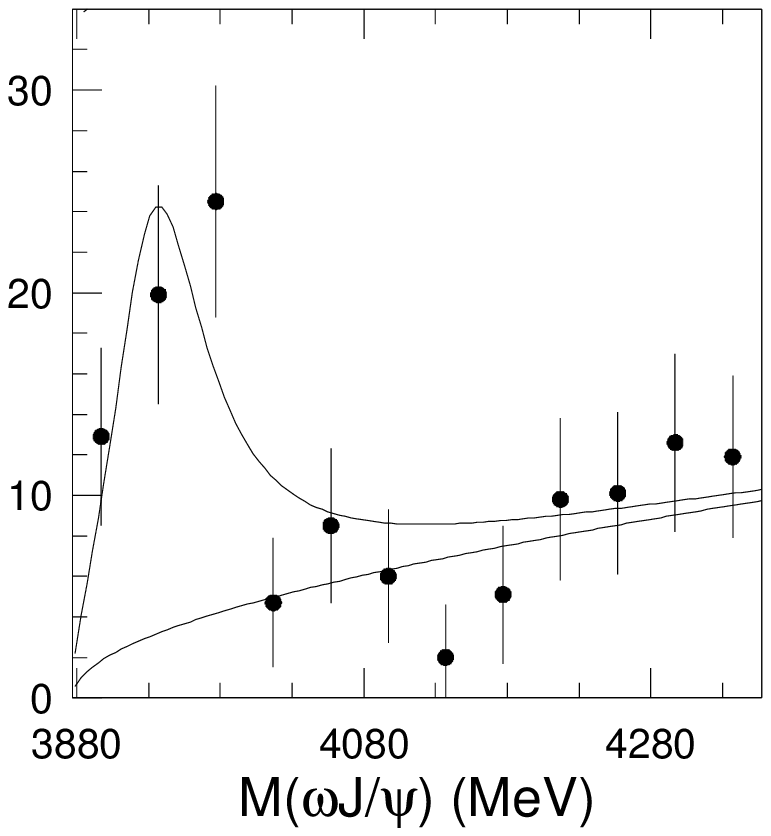}&
\hspace{-3mm}\includegraphics[width=0.3\textwidth,height=0.35\textwidth,clip=]{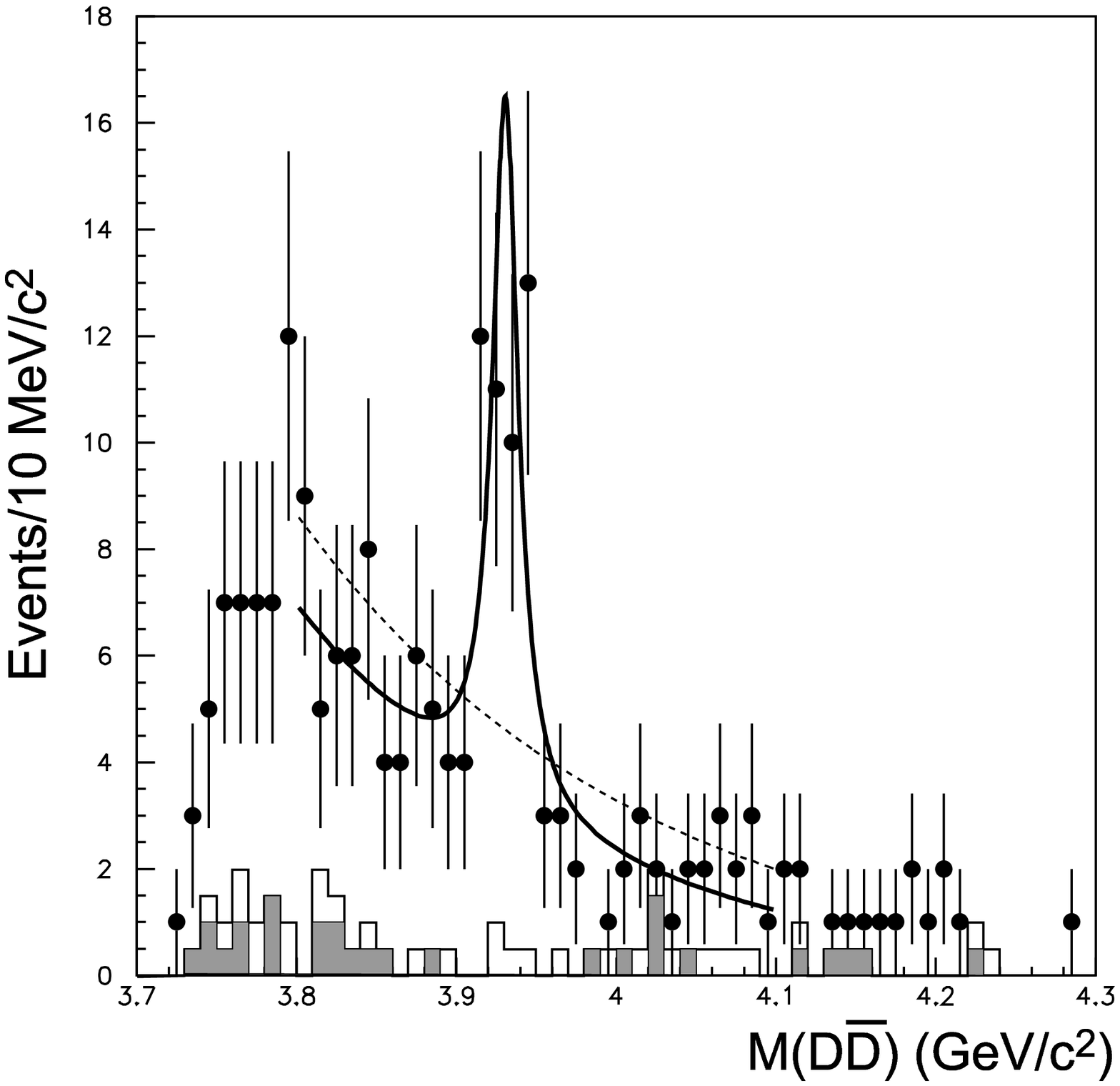}
\end{tabular}\vspace{-60mm}\\
\phantom{a}\hspace{55mm}a\hspace{54mm}b\hspace{50mm}c\vspace{50mm}\\
\caption{\label{fig:X3940}
a: Mass distribution in $e^+e^- \to J/\psi +$ charm
\cite{Abe:2005hd}. The four peaks correspond to $\eta_c(1S)$, a
convolution of $\chi_{cJ}(1P)$ states,  $\eta_c(2S)$ and $X(3940)$.
b: $\omega J/\psi$ mass distribution from $ B \to K\pi\pi\pi
{  J}/\psi$ decays \cite{Abe:2004zs}. c: The $ D\bar D$ mass
distribution from two-photon fusion \cite{Uehara:2005qd}.
 }
\end{figure}

\paragraph{The $X(3940)$} was observed by the BELLE collaboration in
the missing mass distribution recoiling against a J/$\psi$ in $e^+e^-$
annihilation, see  Fig. \ref{fig:X3940}a. Its mass and width were
determined to $(3.943 \pm 0.006 \pm 0.006)$\,GeV/c$^2$ and
$\Gamma<52$\,MeV/c$^2$ at 90\% C.L. \cite{Abe:2005hd,Abe:2007jn}. It is
produced in the same way as $\eta_c(1S)$ and $\eta_c(2S)$. Thus it is a
very good candidate for the $\eta_c(3S)$ resonance. Evidence for
$DD^*$ decays of $X(3940)$ was also reported \cite{Abe:2005hd}. The mass
is unexpectedly low, 4043\,MeV/c$^2$ is expected in the model of
Barnes, Godfrey and Swanson \cite{Barnes:2005pb}, 3900\,MeV/c$^2$ from our simple
scaling.

\paragraph{The $Y(3940)$} was suggested by the BELLE collaboration in
the reaction  $ B\to K\pi\pi\pi {\rm J}/\psi$ with the three pions
forming an $\omega$ meson \cite{Abe:2004zs}, see Fig. \ref{fig:X3940}b.
The $S$-wave Breit Wigner mass and width were determined to be $3943 \pm
11 \pm 13$ MeV/c$^2$ and $87 \pm 22 \pm 26$ MeV/c$^2$, respectively.
The decay into J/$\psi\,\omega$ is very strange: a charmonium state at
3930\,MeV/c$^2$ can decay into $ D\bar D$ or $D^{\ast} D$ while
$\omega J/\psi$ is OZI forbidden. The decay mode resembles the
$f_0(1810)$ which is observed as $\phi\omega$ threshold enhancement in
radiative J/$\psi$ decays \cite{Ablikim:2006dw}. In section \ref{Scalar
mesons in radiative J/psi decays}, $f_0(1810)$ will be interpreted as a
flavour octet state with large $\frac{1}{\sqrt 6}|\uub\ssb + \ddb\ssb
-2\uub\ddb>$ admixture; thus identification of $Y(3940)$ as \ccb\ state
with a $\chi_{c0}(2P)=\frac{1}{\sqrt 3}|\uub\ccb + \ddb\ccb +
\ssb\ccb>$ component does not seem unlikely. The J/$\psi\,\omega$ decay
mode is then rather natural, the mass disagrees however with the
expected 3840\,MeV/c$^2$ using our scaling (see Table \ref{tab:chicb}).
The quark model \cite{Barnes:2005pb} yields a similar value,
3852\,MeV/c$^2$.

The width is compatible with what to expect. The $\chi_{c2}(1P)$ width
is $2.06\pm0.12$\,MeV/c$^2$, the ($\chi_{c0}(1P)$) width is five times
broader, $10.4\pm0.7$\,MeV/c$^2$. With a $\chi_{c2}(2P)$ width of
$29\pm10$\,MeV/c$^2$, we can expect a $\chi_{c0}(2P)$ width of
100--200\,MeV/c$^2$. Its high mass component decays into $\omega
J/\psi$. This picture requires the $\chi_{c0}(2P)$ to decay into $
D\bar D$ with a similar coupling constant as into J/$\psi\,\omega$.

\paragraph{The $Z(3930)$} was observed by the BELLE collaboration
in $\gamma\gamma \to D\bar D$ with a mass of $3931\pm4\pm 2$\,MeV/c$^2$
and a width of $20 \pm 8 \pm 3$\,MeV/c$^2$ \cite{Uehara:2005qd}. The
mass distribution is reproduced in Fig. \ref{fig:X3940}c. The $D\bar D$
helicity distribution was determined to be consistent with $J=2$, hence
it was identified as $\chi_{c2}(2P)$. Its mass and decay mode are
consistent with this identification. The predicted $\chi_{c2}(2P)$ mass
\cite{Barnes:2005pb} is 3972\,MeV/c$^2$.

\subsubsection{\label{A new ccb vector state Y(4260)}
A new \ccb\ vector state $Y(4260)$}

\begin{figure}[pb]
\bc\begin{tabular}{ccc}
\includegraphics[width=0.32\textwidth,height=40mm]{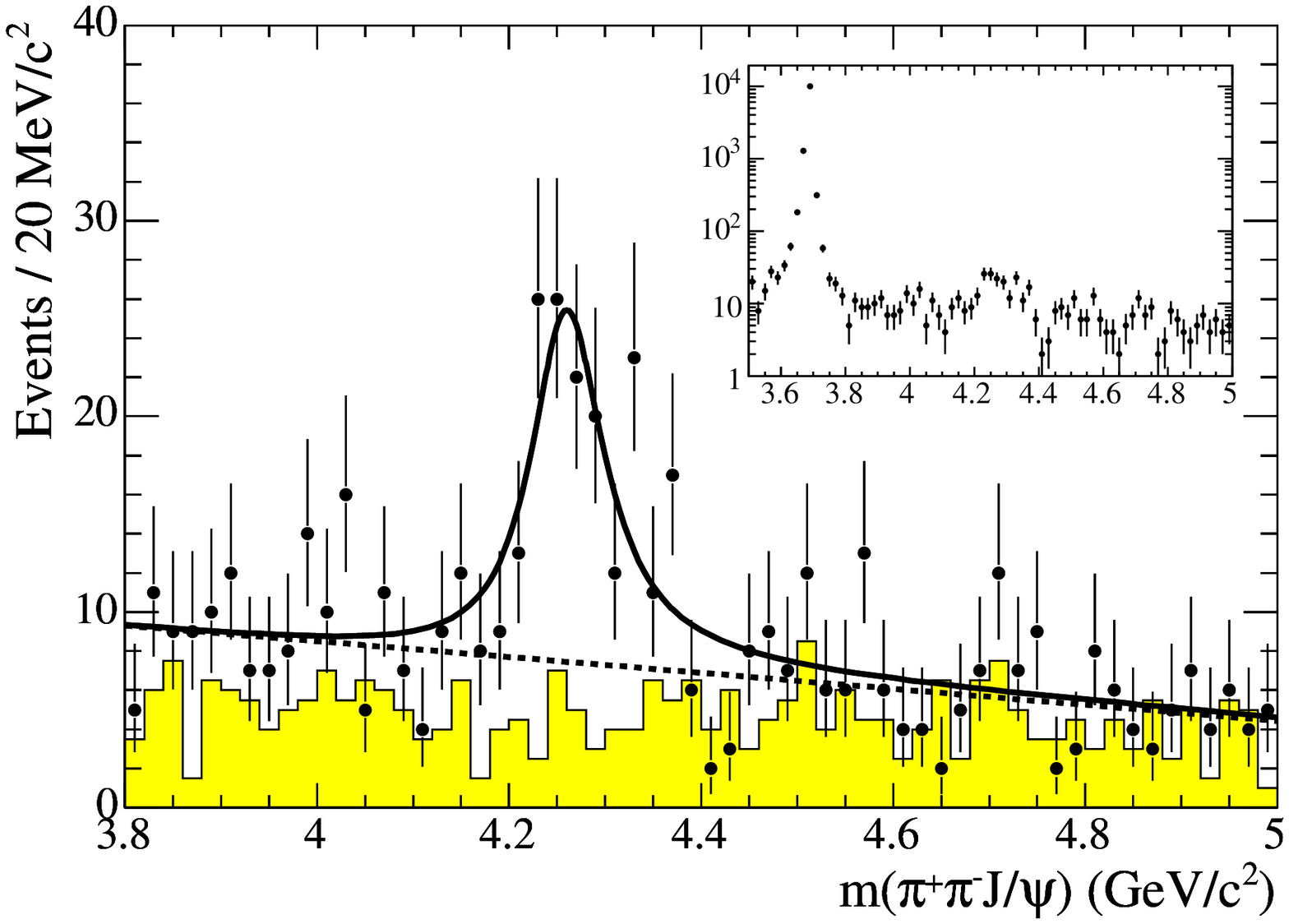}&
\hspace{-3mm}\includegraphics[width=0.32\textwidth,height=40mm]{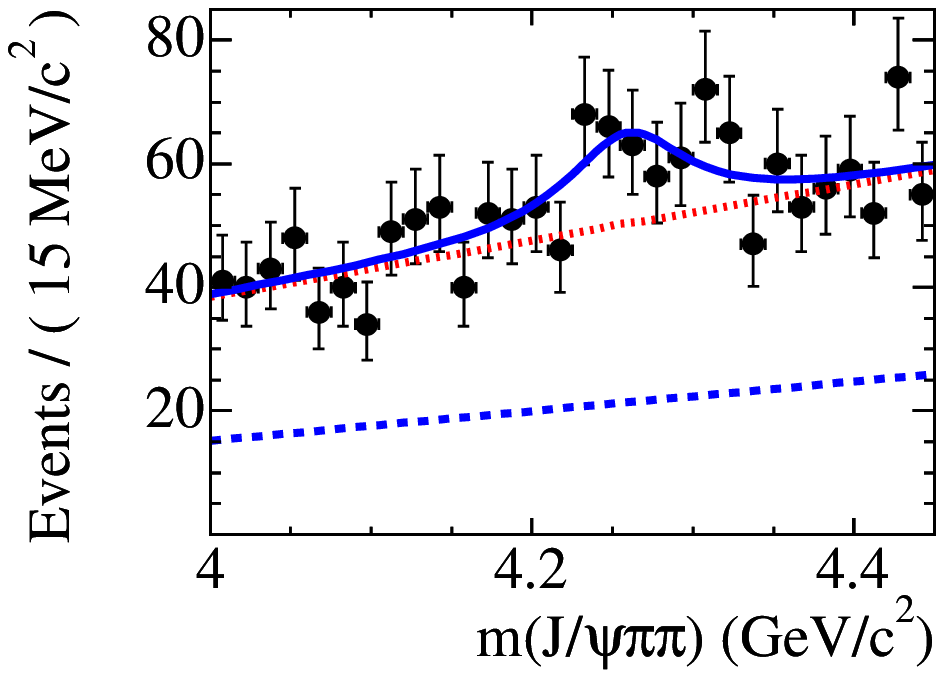}&
\hspace{-3mm}\includegraphics[width=0.32\textwidth,height=40mm]{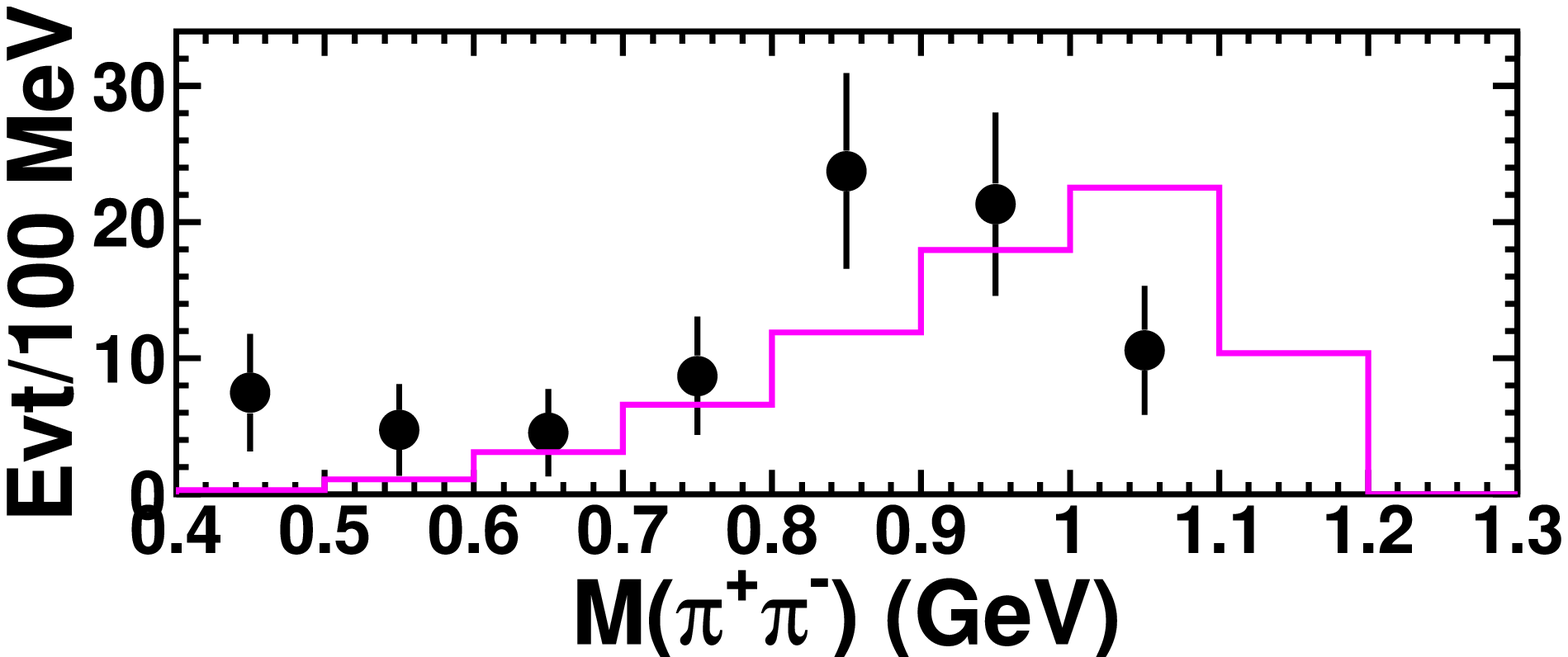}
\end{tabular}\vspace{-42mm}\\
\phantom{a}\hspace{-32mm}a\hspace{60mm}b\hspace{50mm}c\vspace{32mm}\\
\ec
\caption{\label{fig:Y4260}
a: The $Y(4260)$ observed in $e^+e^- \to
\gamma_{ISR}{J}/\psi\,\pi\pi$ \cite{Aubert:2005rm}. The inset shows
the same spectrum over a wider range including the $\psi(2S)$. b:
There is weak evidence for $Y(4260)$ from $ B\to K J/\psi\,\pi\pi$
decays \cite{Aubert:2005zh}. c: The $\pi\pi$ mass distribution
recoiling against the J/$\psi$ and the phase space distribution
\cite{Coan:2006rv}.} \end{figure}

The $Y(4260)$ is a stumbling stone. It was observed by the BaBaR
collaboration as an enhancement in the $\pi\pi J/\psi$ subsystem in the
initial state radiation (ISR), in $e^+e^- \to \gamma_{\rm ISR}
J/\psi\, \pi\pi$ \cite{Aubert:2005rm} (see Fig. \ref{fig:Y4260}, left).
Due to the production mode, $Y(4260)$ must have $J^{PC}=1^{--}$. The
mass was determined to $4259 \pm 8 \pm 4$ MeV/c$^2$ and the width to $88 \pm
23 \pm 5$ MeV/c$^2$.  $Y(4260)$ was confirmed in a
preliminary analysis of the BELLE collaboration \cite{Abe:2006hf}. The
CLEO collaboration observed $Y(4260)$ in a $e^+e^-$ scan
\cite{He:2006kg} and found the ratio of $Y(4260)\to \pi^+\pi ^- J/\psi$
and $Y(4260)\to \pi ^0 \pi ^0 J/\psi$ events to be consistent with
$Y(4260)$ being isoscalar \cite{Coan:2006rv}. Three events due to
$Y(4260)\to K^+K^- J/\psi$ were reported corresponding to a $\pi\pi
/K\bar K$ ratio of about 1/4. A BaBaR search for  $Y(4260)\to p\bar p$
yielded a null result \cite{Aubert:2005cb}. The BaBaR collaboration
searched for $Y(4260)$ in $B$ decays studying the reactions  $B^0 \to
{J}/\psi\, \pi^+\pi^-K^0$ and $B^- \to J/\psi\, \pi^+ \pi^-K^-$
\cite{Aubert:2005zh}. A weak  3-$\sigma$ signal shown in Fig.
\ref{fig:Y4260}b was observed which is compatible with the $Y(4260)$
parameters.

The partial width

\be
\label{Y4260-hybrid}\Gamma(Y(4260)\to e^+e^-)\, Br(Y(4260) \to
J/\psi\, \pi\pi) = 5.5 \pm 1.0\,^{+0.8}_{-0.7} \ {\rm eV/c^2}
\ee

determined in \cite{Aubert:2005rm} can be compared to the rate when
$Y(4260)$ is replaced by $\psi(2S)$:

\be \Gamma(\psi(2S)\to e^+e^-)\,
Br(\psi(2S)\to J/\psi\, \pi\pi) = 672 \pm 45 \ {\rm eV/c^2}. \ee

Either, $Y(4260)$ must have an unusual small $e^+e^-$ width or a very
small coupling to J/$\psi\,\pi\pi$ (or both).

The $Y(4260)$ yield in $B$ decay is anomalously low, too. The product
branching ratio \be \mathcal B(B^- \to K^-Y) \mathcal B(Y \to J/\psi\,
\pi\pi) = (2.0 \pm 0.7 \pm 0.2) \cdot 10^{-5} \ee was interpreted as
90\% C.L. upper limit at $2.9\cdot 10^{-5}$ \cite{Aubert:2005zh}. The
value is two orders of magnitude smaller than the branching fractions
for $K$\,J/$\psi$ or $K$\,$\psi(2S)$. Again, either the coupling $Y \to
J/\psi\, \pi\pi$ is small, or the  $Y(4260)$ production rate is much
smaller than that of  $\psi(nS)$ states.

With $\psi(3686)$,
$\psi(3770)$, $\psi(4040)$, $\psi(4160)$, $\psi(4415)$ as $\psi(2S)$,
$\psi(1D)$, $\psi(3S)$, $\psi(2D)$, and $\psi(4S)$, respectively, and a
$\psi(3D)$ at an expected mass at about 4.5\,GeV/c$^2$, there seems to
be no \ccb\ slot available for the $Y(4260)$; its mass is hence a
further argument in favour of its exotic nature.

Various explanations for $Y(4260)$ have
been proposed. Its, at the first glance, anomalous properties prompted
the suggestion that it might be a $c\bar c$ hybrid
\cite{Zhu:2005hp,Close:2005iz,Kou:2005gt}. Maiani {\it et al.}
\cite{Maiani:2005pe} interpreted the $Y$ as first orbital excitation of
a tetraquark $[cs]_S[\bar c \bar s]_S$ state; additional states were
predicted at somewhat lower masses. Alternatively, the $Y$ could
be a $\chi_{c1}\,\rho$ molecule bound by $\sigma$ exchange
\cite{Liu:2005ay}. A $\Lambda_c\,\bar\Lambda_c$ baryonium state is a
further possibility to understand the $Y(4260)$ \cite{Qiao:2005av}.
The latter suggestion finds support in the claim for a $\phi f_0$
resonance at 2175\,MeV/c$^2$ \cite{Aubert:2006bu} and its possible
interpretation as $\Lambda\bar\Lambda$ bound state.

Further arguments in favour of an exotic nature can be made by
analysing the $\pi\pi$ invariant mass distribution recoiling against
the J/$\psi$ shown in Fig. \ref{fig:Y4260}c. The distribution is
incompatible with the expectation based on normal $\pi\pi$ $S$-wave
interactions produced without any hindrance factor. Bugg
\cite{Bugg:2007kr} fitted this spectrum, and a similar spectrum from
Belle, assuming an intermediate $\pi J/\psi$ resonance. The best fit
required orbital angular momentum $L=1$ between the recoil pion and the
$\pi J/\psi$ resonance, and gave $M = 4080$\,MeV/c$^2$ and $\Gamma =
280$\,MeV/c$^2$. The resonance is suggested to be a tetraquark state.
This is an interesting conjecture but the reader should go back to Fig.
\ref{heavy-pipi}. If $Y(4260)$ is of exotic nature, then it may act as
point-like source for decays, and the $\pi\pi$ invariant mass
distribution can be ascribed to final-state interactions. If it is
a $\psi$ radial excitation, its nodes have a significant impact.

The $Y(4260)$ resonance was searched for in the inclusive $e^+e^-$
annihilation cross section \cite{Mo:2006ss}. The cross section exhibits
a dip-bump-dip structure in the $\sqrt s=4.20-4.35$\,GeV/c$^2$ region
which makes it difficult to extract a reliable estimate for a possible
$Y(4260)$ contribution. Rosner assigned this dip to the opening of the
$DD_1$ channel \cite{Rosner:2006vc}. Llanes-Estrada
\cite{Llanes-Estrada:2005hz} has shown that the inclusive $e^+e^-$
annihilation into hadrons can be explained quantitatively by $S$-D wave
interference and assuming the coupling to two pseudoscalar mesons to
have positive signs for $^3S_1$ and negative signs for $^3D_1$ states.
The $\Upsilon(4260)$ was included in the fit as $\psi(4S)$ state.
In \cite{Mo:2006ss}, different dynamical assumptions  were made to
determine an upper limit for the $Y(4260)\to e^+e^-$ width. The largest
contribution was acceptable in fits in which the $\psi^{\prime}$
resonances were described by interfering Breit-Wigner amplitudes of
constant widths with arbitrary phases. The series of resonances
included $\psi(3770)$, $\psi(4040)$, $\psi(4160)$, $Y(4260)$, and
$\psi(4415)$; a linear non-interfering background was added. The
$Y(4260)$ was given a special treatment: since it is suspected not to
belong to the $\psi^{\prime}$ series, it was considered approppriate to
add it partly incoherently to the cross section. It was hence split into
two contributions, one interfering Breit-Wigner amplitude, a
non-interfering part describing $Y(4260)$ as threshold phenomenon. The
fraction of the coherent contributions was varied between 0.7 to 0.3,
and upper limits for $\Gamma_{Y(4260)\to e^+e^-}$ were derived;
580\,eV/c$^2$ for 70\% coherence, 460\,eV/c$^2$ for 50\%, and
280\,eV/c$^2$ for 30\% coherence. These numbers maybe compared to
$\Gamma_{\psi(4160)\to e^+e^-}=830\pm70$\,eV/c$^2$ and
$\Gamma_{\psi(4415)\to e^+e^-}=580\pm70$\,eV/c$^2$. The $e^+e^-$ width
of $Y(4260)$ is not suspicious. Full coherence was not permitted in
\cite{Mo:2006ss}, extrapolating to full coherence or interpolating
the two values for $\psi(4160)$ and $\psi(4415)$ yields our estimate
of {\it 720}\,eV/c$^2$ which we give in Table \ref{psiprimes}.

From eq. (\ref{Y4260-hybrid}) and the $Y(4260)\to e^+e^-$ width, we
derive
\be
\label{Y4260-hyb}
\Gamma(Y(4260)\to J/\psi\pi^+\pi^-) > 670\pm240  {\rm keV/c^2}.
\ee

The yield is larger than naively expected, in particular when compared
to the corresponding width for the $\psi(3686)(2S)$ resonance. However,
the $\pi\pi$ $S$-wave is open for a wider range in $Y(4260)$ decays.
The BESII collaboration has fitted new data on the $R$ value in
$e^+e^-$ annihilation \cite{Ablikim:2007gd}; fits including $Y(4260)$
were not attempted.

In Table \ref{psiprimes} we compare charmonium vector states assuming
that $Y(4260)$ is $\psi(4S)$. The comparison looks very
favourable. An exotic interpretation of $Y(4260)$ may therefore not
really be enforced by data. Very recently, the BELLE collaboration
reported two further resonant structures in the $\pi^+\pi^-\psi(2S)$
invariant mass distribution, at $4361\pm 9\pm 9$\,MeV/c$^2$ with a
width of $74\pm 15\pm 10$\,MeV/c$^2$, and at $4664\pm 11\pm
5$\,MeV/c$^2$ with a width of $48\pm 15\pm 3$\,MeV/c$^2$
\cite{Wang:2007ea}. We conjecture that these might be the $3D$ and $6S$
states, respectively.

\begin{table}[pb]
\caption{\label{psiprimes}Charmonium states with
$J^{PC}=1^{--}$ in our interpretation.
\vspace{2mm}} \bc \renewcommand{\arraystretch}{1.6}
\begin{tabular}{cccccccc}
\hline\hline
J/$\psi$&$\psi(3686)$&$\psi(3770)$&$\psi(4040)$&$\psi(4160)$&$Y(4260)$&$\psi(4415)$\\
&\quad$2S$\ \quad&\quad$1D$\quad&\quad$3S$\quad&\quad$2D$\quad&\quad$4S$\quad&\quad$5S$ \\
\hline
$\Gamma_{e^+e^-}$&$2.48\pm0.06$&$0.242^{+0.027}_{-0.024}$&$0.86\pm0.07$
&$0.83\pm0.07$& {\it 0.72} &$0.58\pm0.07$& keV/c$^2$\\
$\Gamma_{\rm J\psi\pi^+\pi^-}$&$107\pm5$&$44\pm8$&
$<360$ &$<330$& $670\pm240$ &-& keV/c$^2$\\
$M_{\psi(nS)}-M_{J/\psi}$       &
\quad589\ \quad&\quad674\ \quad&\quad943\quad&\quad1056\
\quad&\quad1163\quad&\quad1318&MeV/c$^2$\\
$M_{\Upsilon(nS)}-M_{\Upsilon}$ &
\quad563\ \quad&&\quad895\quad&&\quad1119\quad&    & MeV/c$^2$ \\
\hline\hline
\renewcommand{\arraystretch}{1.0}
\end{tabular}
\ec
\end{table}

\subsubsection{Conclusions}

There are four positive-parity charmonium states expected in the mass
range from the $\chi_{c2}(1P)$ to 4\,GeV/c$^2$, and four states are
found, $X(3872)$, $X(3940)$, $Y(3940)$, and $Z(3930)$. Based on their
production and decay characteristics, they can be identified with
$\chi_{c1}(2P)$, $\eta_{c}(3S)$, $\chi_{c0}(2P)$, and $\chi_{c2}(2P)$,
respectively. The $Y(4260)$ finds a natural interpretation as
$\psi(4S)$ state. These conclusions were reached independently by
Colangelo and collaborators \cite{Colangelo:2006aa}. Does this imply
the attempts  to understand these particles as exotic states were a
failure? Certainly not; the tetraquark and/or molecular picture focuses
the attention on the dynamics of production processes and decay rates
and augment the \qqb\ picture. Furthermore, the models serve as
stringent guides to search for new phenomena. But from time to time it
is necessary to recall that the quark model provides a solid
foundation; claims for physics beyond the quark model have to survive
critical scrutiny.

\subsubsection{A charged charmonium state ?}

After completion of the review, a narrow $\psi'\pi^{\pm}$ resonance was
reported by the Belle collaboration with a statistical evidence
exceeding 7$\sigma$ \cite{:2007wg}. It was observed in $B$ decays to $K
\pi^+\psi'$, using various $\psi'$ decay modes. The resonance, called
$Z(4430)$, has $4433\pm 4\pm 1$\,MeV/c$^2$ mass and a width of $\Gamma
= 44^{+17}_{-13}$$^{+30}_{-11}$\,MeV/c$^2$ width. It is produced with
a branching fraction
\be
\mathcal B (B\to K Z(4430))\cdot \mathcal B (Z(4430)\to \pi^+\psi') =
(4.1\pm1.0\pm1.3)\cdot 10^{-5}
\ee
It is the first charged resonance with hidden charm and can evidently
not belong to the charmonium family, not even have a $c\bar c$ seed.
Rosner noticed that the $Z(4430)$ mass is not far from $D^*
\bar{D}_1(2420)$ threshold and proposed that the state is formed via the
weak $b\to c\bar cs$ transition, creation of a light-quark pair, and
rescattering of the final-state hadrons \cite{Rosner:2007mu}. The decay
$Z(4430)\to J\psi\pi^{\pm}$ is not observed: the rescattering mechanism
is effective only close to the production threshold.

The $Z(4430)$ state is clearly at variance with ideas presented above.
Its confirmation as resonant state (and not as effect of threshold
dynamic) is very important.

\subsection{\label{Heavy--light quark systems}
Heavy-light quark systems}

\subsubsection{\label{D and Ds mesons and their low-mass excitations}
$D$ and $D_s$ mesons and their low-mass excitations}

\begin{figure}[pb]
\begin{center}
\begin{tabular}{ccc}
\hspace*{-2mm}\includegraphics[width=0.32\textwidth,height=0.30\textwidth,clip=]{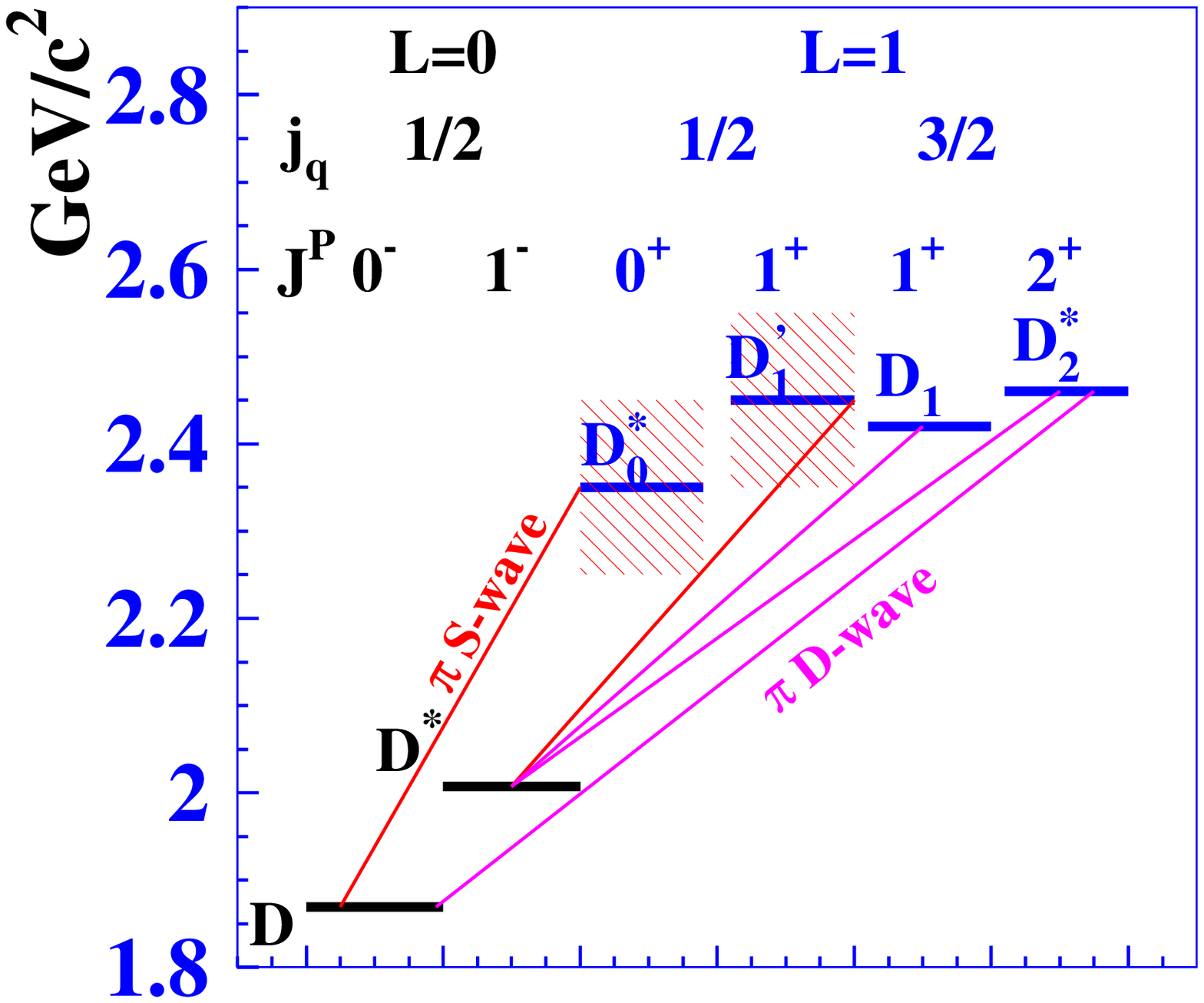}&
\hspace*{-2mm}\includegraphics[width=0.32\textwidth,height=0.30\textwidth]{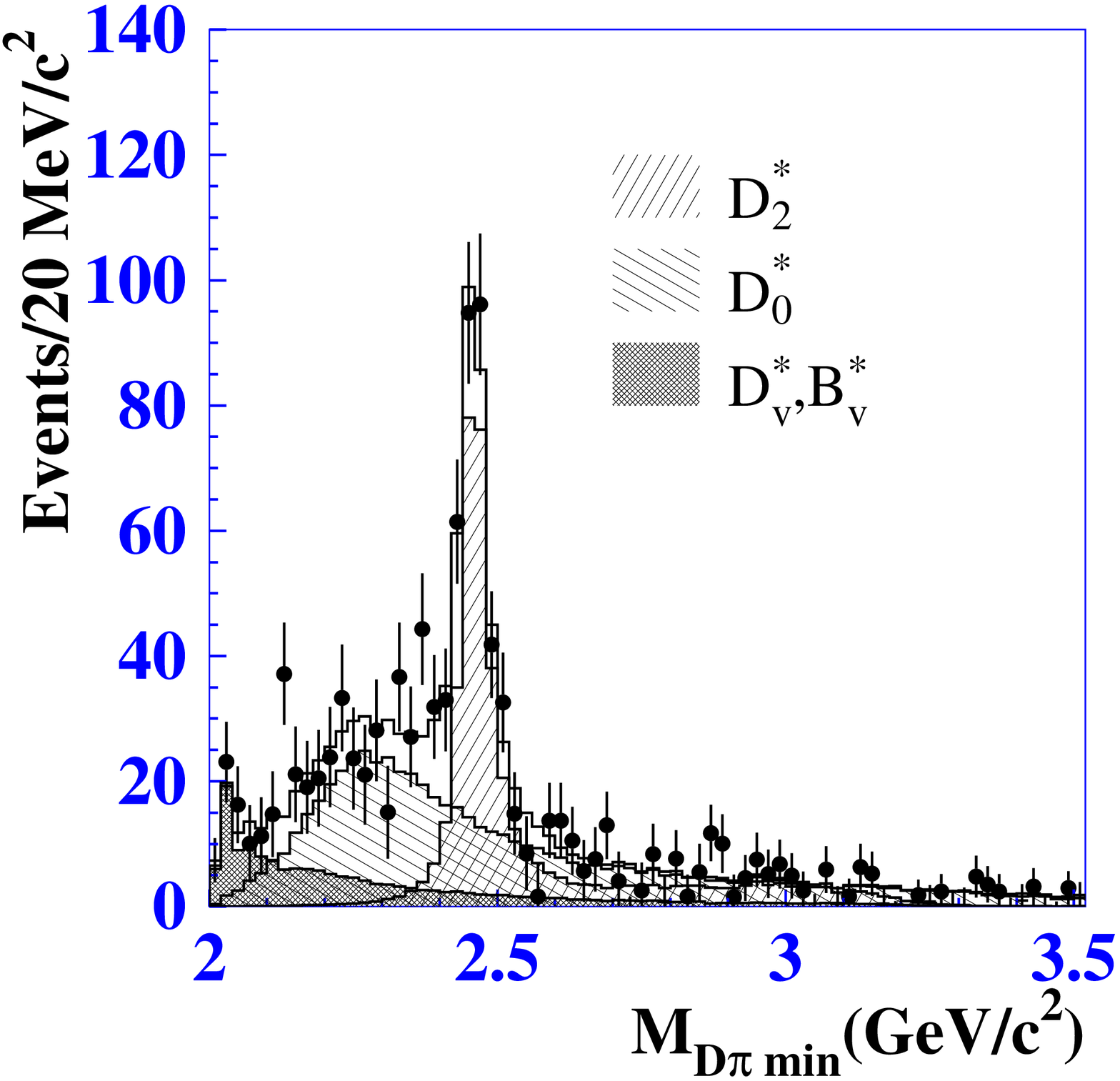}&
\hspace*{-4mm}\includegraphics[width=0.32\textwidth,height=0.31\textwidth]{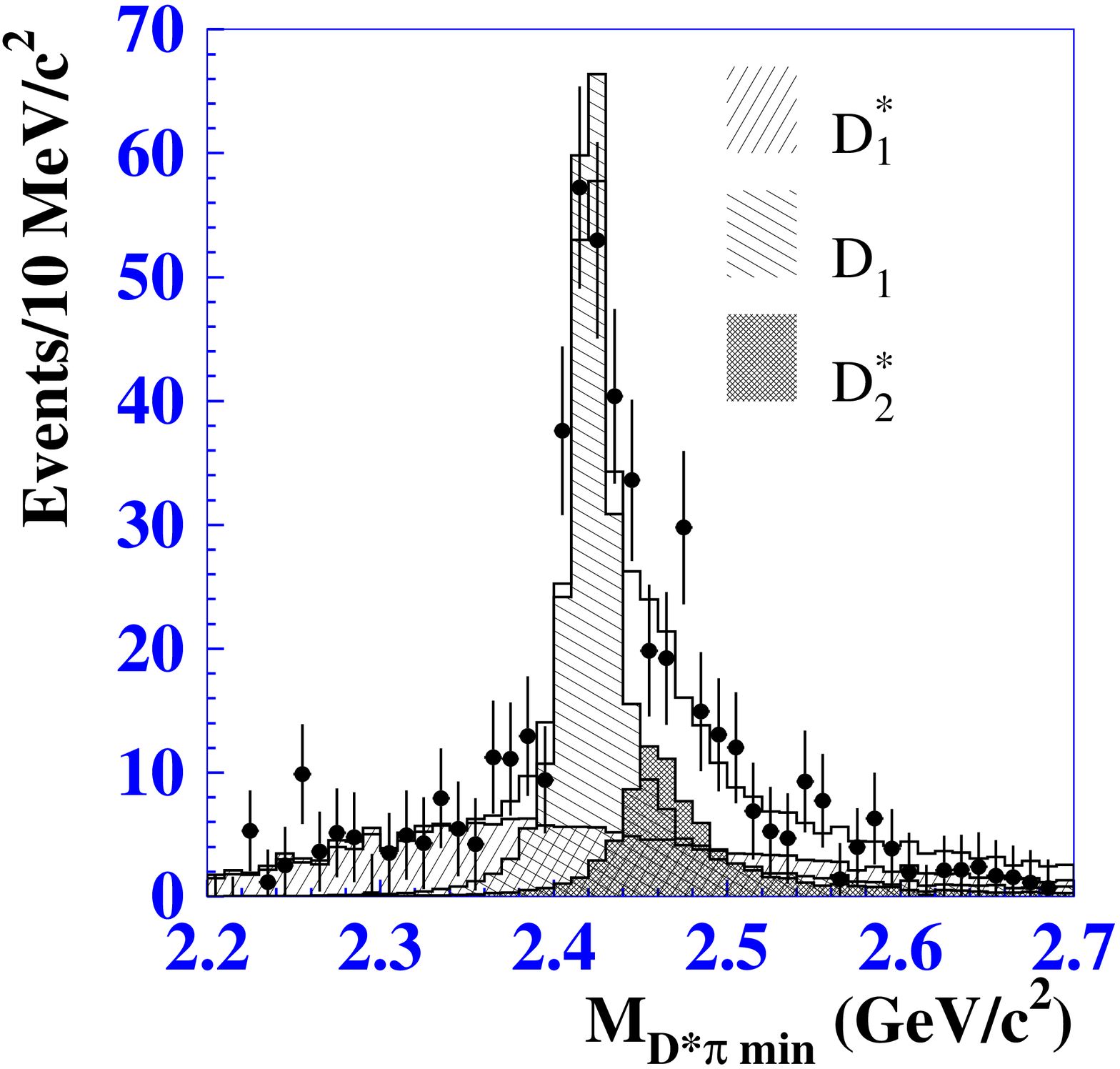}\\
\vspace*{-5.7 cm} & \\
{\hspace*{4.4cm}\bf a)}&{\hspace*{-2.7cm}\bf b)}&{\hspace*{-2.9cm}\bf
c)}\\
\vspace*{3.9 cm} & \\
\end{tabular}
\end{center}
\caption{\label{fig:d2p-d3p}
(a) Spectrum of $D$-meson excitations. The lines indicate allowed
single-pion transitions. In the heavy quark limit, the light-quark spin
and orbital angular momentum couple to a total light-quark angular
momentum $j_q$. In this approximation, one $D_1$ decays into $D\,\pi$
only via $D$-wave.
(b) The background-subtracted minimal $D\,\pi$ mass
distribution in $B^-\to D^+\pi^-\pi^-$ decays. The hatched histograms
show the $D^*_0$ and $D^*_2$ contributions and the additional
contributions from virtual intermediate states. The open histogram
shows the coherent sum of all contributions.
(c) The hatched histograms show the $D_{1}$, $D^{\prime}_{1}$,
and $D^{*}_{2}$ contributions, the open histogram is a coherent sum of
all contributions \cite{Abe:2003zm}.}
\end{figure}

\paragraph{$D$ Mesons:} Figure~\ref{fig:d2p-d3p}a shows the spectrum of
$D$-meson excitations. The states of lowest mass are the $D$ and the
$D^*$ and the four $P$-wave states. In the heavy-quark limit, the
heavy-quark spin ${\vec s}_c$ decouples from the other degrees of
freedom and the total angular momentum of the light quark ${\vec
j}_q=\vec{L}+{\vec s}_q$ is a good quantum number. This yields the
following spin-parity and light-quark angular-momenta decompositions:
$0^+(j_q=1/2),~1^+(j_q=1/2),~1^+(j_q=3/2)$ and $2^+(j_q=3/2)$, which
are usually labelled as $D^*_0,~D'_1,~D_1$ and $D^*_2$, respectively.
The two $j_q=3/2$ states are narrow due to the angular-momentum barrier
and have $\sim$20\,MeV/c$^2$ widths. The $j_q=1/2$ states decay via
$S$-waves and are expected to be quite broad. The BELLE collaboration
studied these states in $B$ decays into $D\,\pi\pi$ and $D\,\pi\pi\pi$
\cite{Abe:2003zm}. Further support for the broad states was reported by
the FOCUS \cite{Link:2003bd} and the CDF \cite{Abulencia:2005ry}
collaborations.

The partial wave analysis of events due to $ B^-\to D^+\pi^-\pi^-$
decays using $ D^*_0$ and $ D^*_2$ as only isobars did not give a fully
satisfactory description of the $ D^+\pi^-\pi^-$ Dalitz plot. Virtual
vector states as additional isobars improved the fit without changing
significantly the $ D^*_0$ and $ D^*_2$ parameters. The fit results are
shown in Fig. \ref{fig:d2p-d3p}b. The narrow $ D^*_2$ is readily
identified but there is a substantial wide `background' which is mostly
assigned to scalar $D\,\pi$ interactions. Closer inspection reveals
a spike at threshold; it is described by a virtual $D^*_v$ (and a
second one, called $B^*_v$). The spin-parity of the virtual states was
not tested. Below (in section \ref{Scalar states from the sigma to
chib0(1P)}) we suggest the existence of a subthreshold $ D^*_0(1980)$.
The spike and/or the black area in Fig. \ref{fig:d2p-d3p}b could be a
trace of this hypothesised state. A low mass $D^*_0$ (at
2030\,MeV/c$^2$) was also suggested by Beveren and Rupp
\cite{vanBeveren:2003kd}.

For the reaction $ B^{-}\to D^{*+}\pi^{-}\pi^{-}$, an unbinned
likelihood fit in the four-dimensional phase space was performed to
extract the amplitudes and phases of the contributing isobars. The fit
results shown in Fig. \ref{fig:d2p-d3p}c exhibit three components, two
narrow ones and a broad one, in line with the expectations based on
Fig. \ref{fig:d2p-d3p}a.

Instead of presenting the results on masses and widths of $D$ resonances
from \cite{Abe:2003zm} from the fits described above, we quote these
quantities in Table \ref{DDecays} from the latest Review of Particle
Properties \cite{Eidelman:2004wy}. Masses and widths will be discussed
in connection with the corresponding states of the $D_s$ family.

\begin{figure}[pt]
\vspace{-12mm}
\bc
\begin{tabular}{ccc}
\hspace{8mm}\includegraphics[width=0.45\textwidth,height=0.32\textwidth]{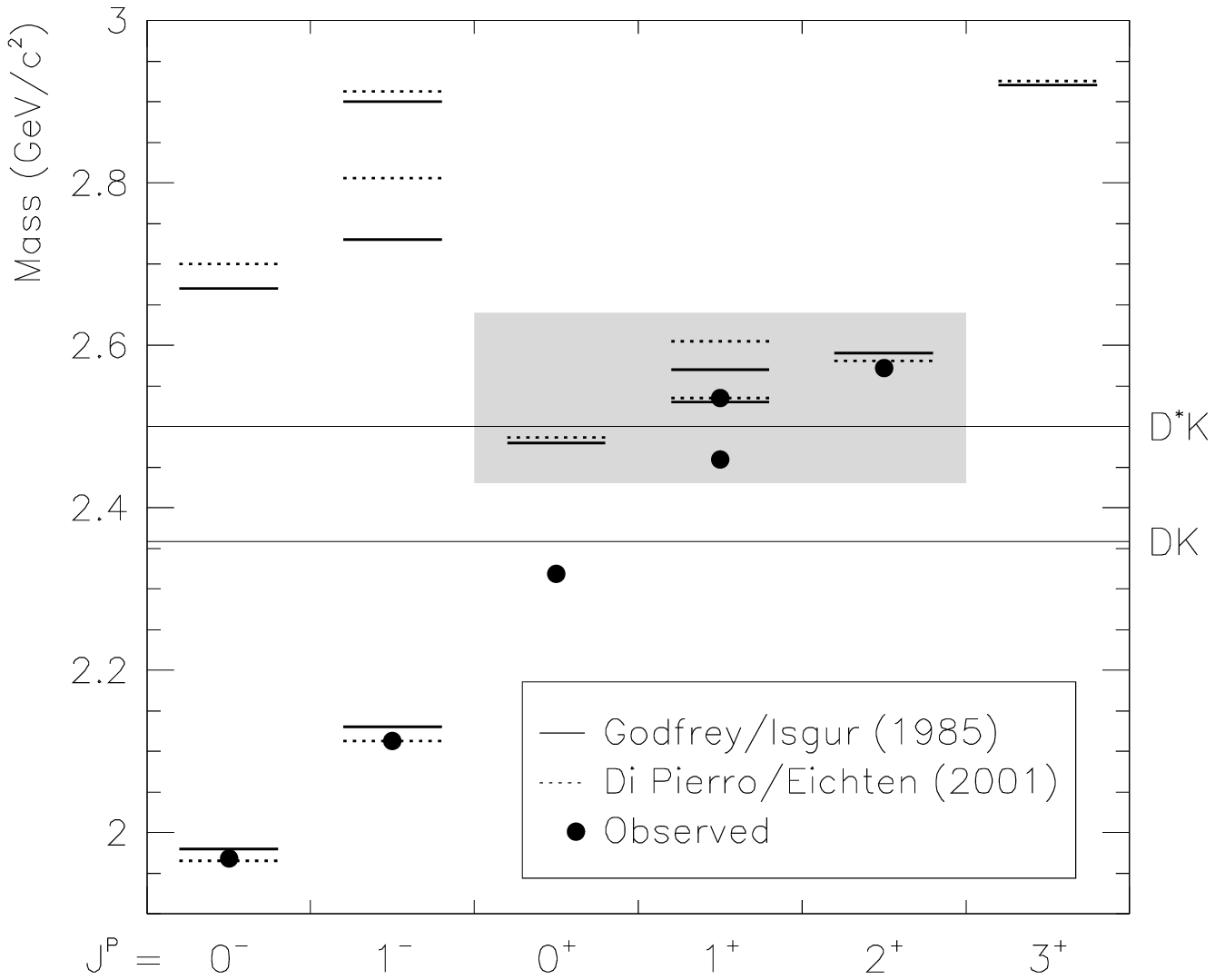}&
\hspace{-8mm}\includegraphics[width=0.45\textwidth,height=0.34\textwidth]{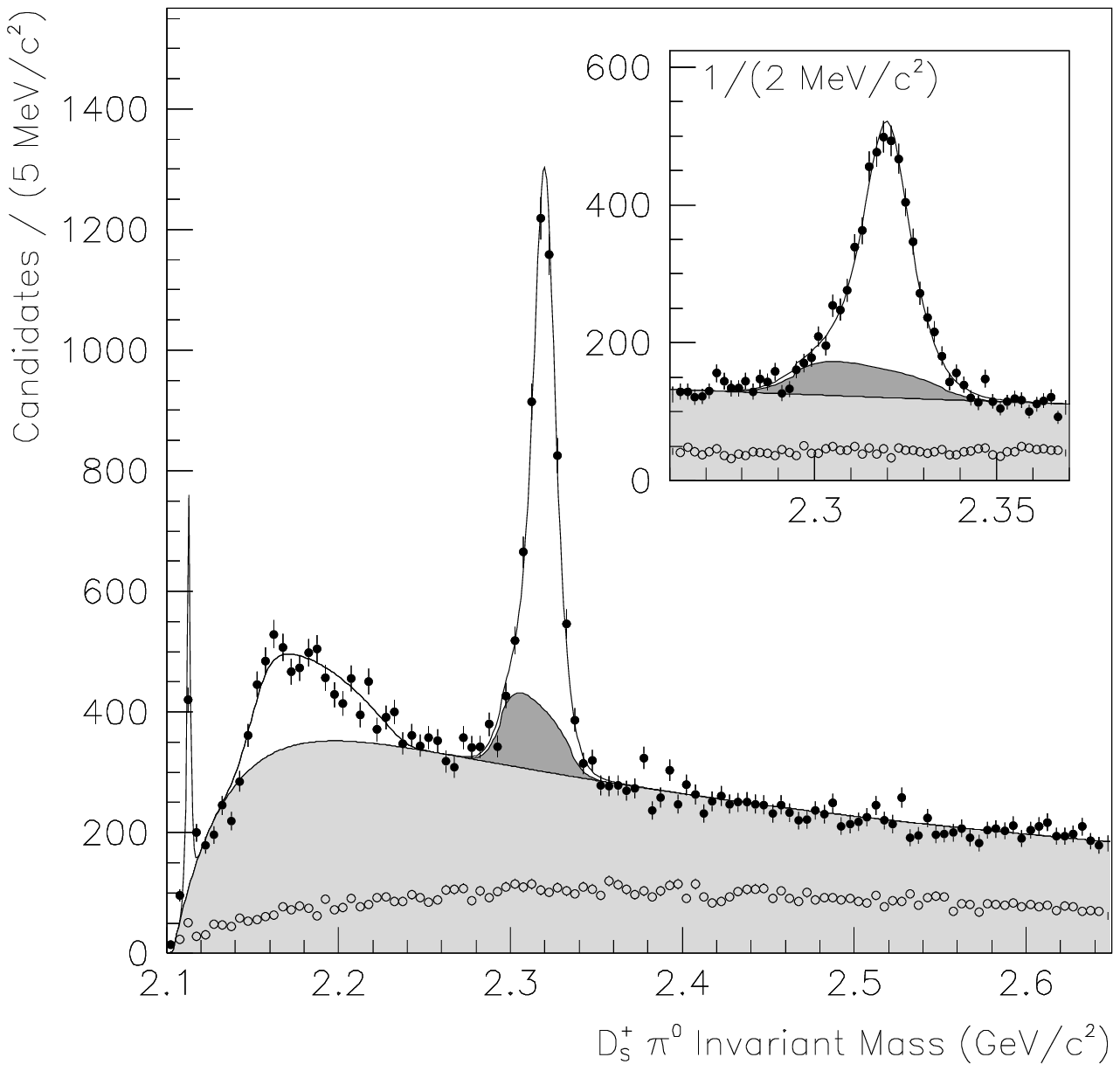}
\end{tabular}
\vspace{-8mm}
\ec\caption{\label{dsspectr}
Left: The spectrum of $D_s$ excitations as predicted by Godfrey and
Isgur~\cite{Godfrey:1985xj} (solid lines) and Di Pierro and
Eichten~\cite{DiPierro:2001uu} (dashed lines) and as observed by
experiment (points). The $D\,K$ and $ D^*\,K$ mass thresholds are
indicated by two horizontal lines. The $J^P=0^+$ state at 2317\,MeV/c$^2$ has
an unexpected low mass.
Right: The $D_{s0}^*(2317)^+$. The invariant mass distribution for
(solid points) $D_s\,\pi^0$ candidates and (open points) the
equivalent using the $D_s$ side bins. The fit includes
combinatory background (light grey) and the reflection from $
D_{sJ}(2460)^+\to D_s^{*}(2112)^+\pi^0$ decay (dark grey). The insert
shows an extended view of the $ D_{s0}^*(2317)^+$ mass region
\cite{Aubert:2006bk}.
}
\end{figure}

\paragraph{$D_s$ mesons:}
Fig. \ref{dsspectr}a shows the expected low-lying $c\bar s$
excitations. The two ground states and two resonances above the
$D^*\,K$ thresholds were known since long
\cite{Kubota:1994gn,Albrecht:1995qx} from CLEO and Argus and were used
to define the quark model parameters. The other two states were thought
to have escaped discovery due to their large (expected) widths. It thus
came as a complete surprise when $D_{s0}^*(2317)$ was discovered by
BaBar \cite{Aubert:2003fg} as narrow and low-mass resonance, in the
inclusive $ D_s^+\pi^0$ invariant mass distribution produced in
$e^+e^-$ annihilation at energies near 10.6\,GeV/c$^2$. Fig. \ref{dsspectr}
(right) shows the $ D_s\,\pi^0$ invariant mass distribution with an
impressive $D_{s0}^*(2317)^+$ signal \cite{Aubert:2006bk}. Among other
(continuous) background contributions, reflections from other decay
sequences are an intriguing source since they can mimic wrong signals.
Particularly dangerous are reflections from $D_s^{*}(2112)^+\to
D_s\gamma$ decays in which an unassociated $\gamma$ particle is added
to form a false $\pi^0$ candidate, and from $ D_{sJ}(2460)^+\to
D_s^{*}(2112)^+\pi^0$ decay in which the $\gamma$ from the $
D_s^{*}(2112)^+$ decay is missing. The reflections also occur in
side bands and can thus be controlled quantitatively. A $D_{s0}^*$
resonance should decay into $D$ plus $K$. However, the $D_{s0}^*(2317)$
mass is below the $D\,K$ threshold. Its mass is even below the
$D_0^*(2350)$ mass, in spite of a $n$ quark being replaced by a $s$
quark. Since its natural decay mode is forbidden, $D_{s0}^*(2317)$ must
decay via an isospin violating mode into $D_s\,\pi$, and is narrow. The
new resonance is now confirmed in several new data sets
\cite{Besson:2003cp,Abe:2003jk,Krokovny:2003zq,Aubert:2004pw,Aubert:2006bk}.
It must have natural spin-parity, $J^{PC}=0^{++}$ were proposed and
found to be consistent with later data having higher statistics. In
particular the absence of radiative transitions to the ground state
supports this assignment.

\begin{figure}[pt]
\begin{center}
\begin{tabular}{ccc}
\includegraphics[width=0.38\textwidth,height=0.40\textwidth]{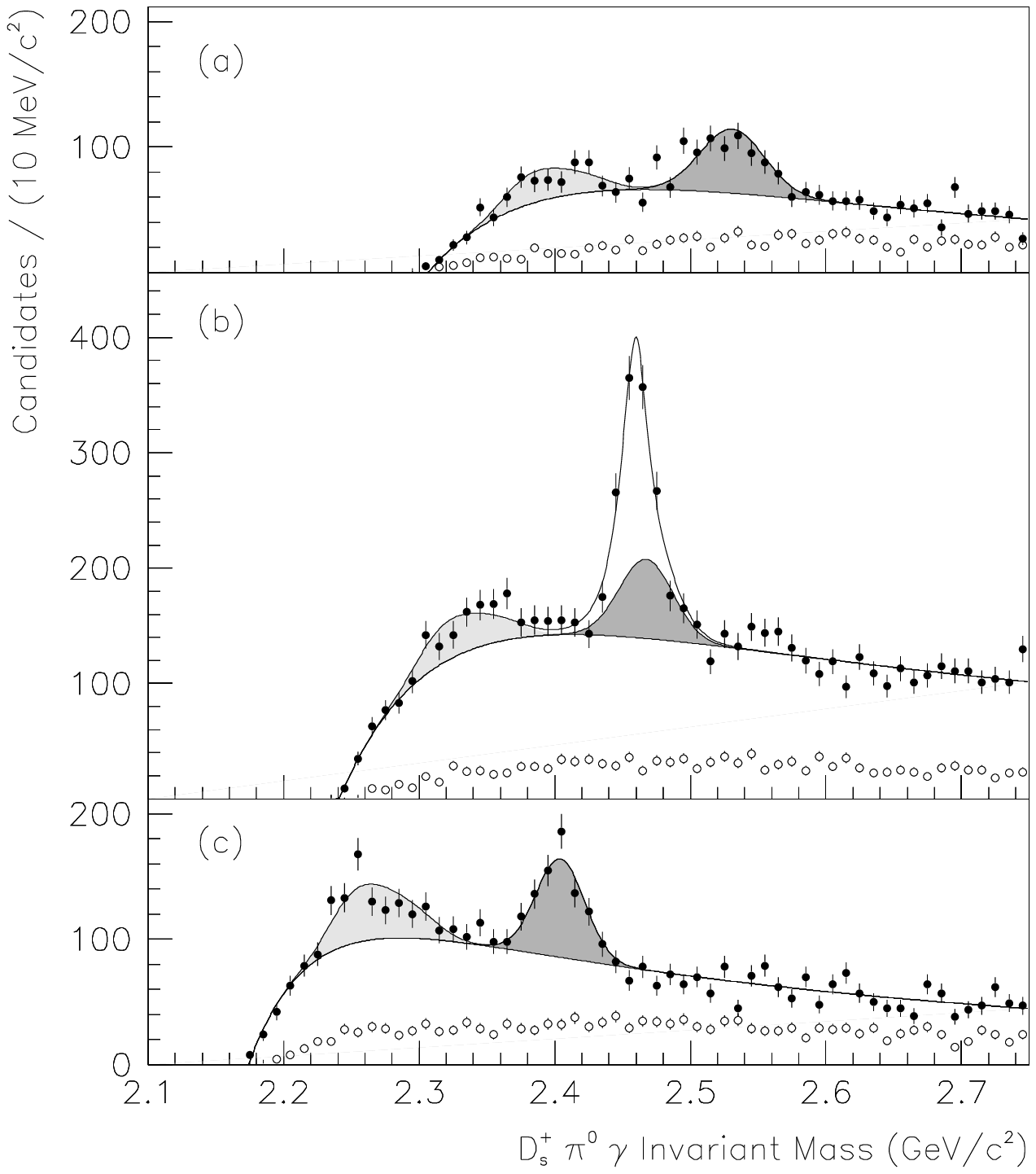}&
\hspace{-12mm}\includegraphics[width=0.38\textwidth,height=0.40\textwidth]{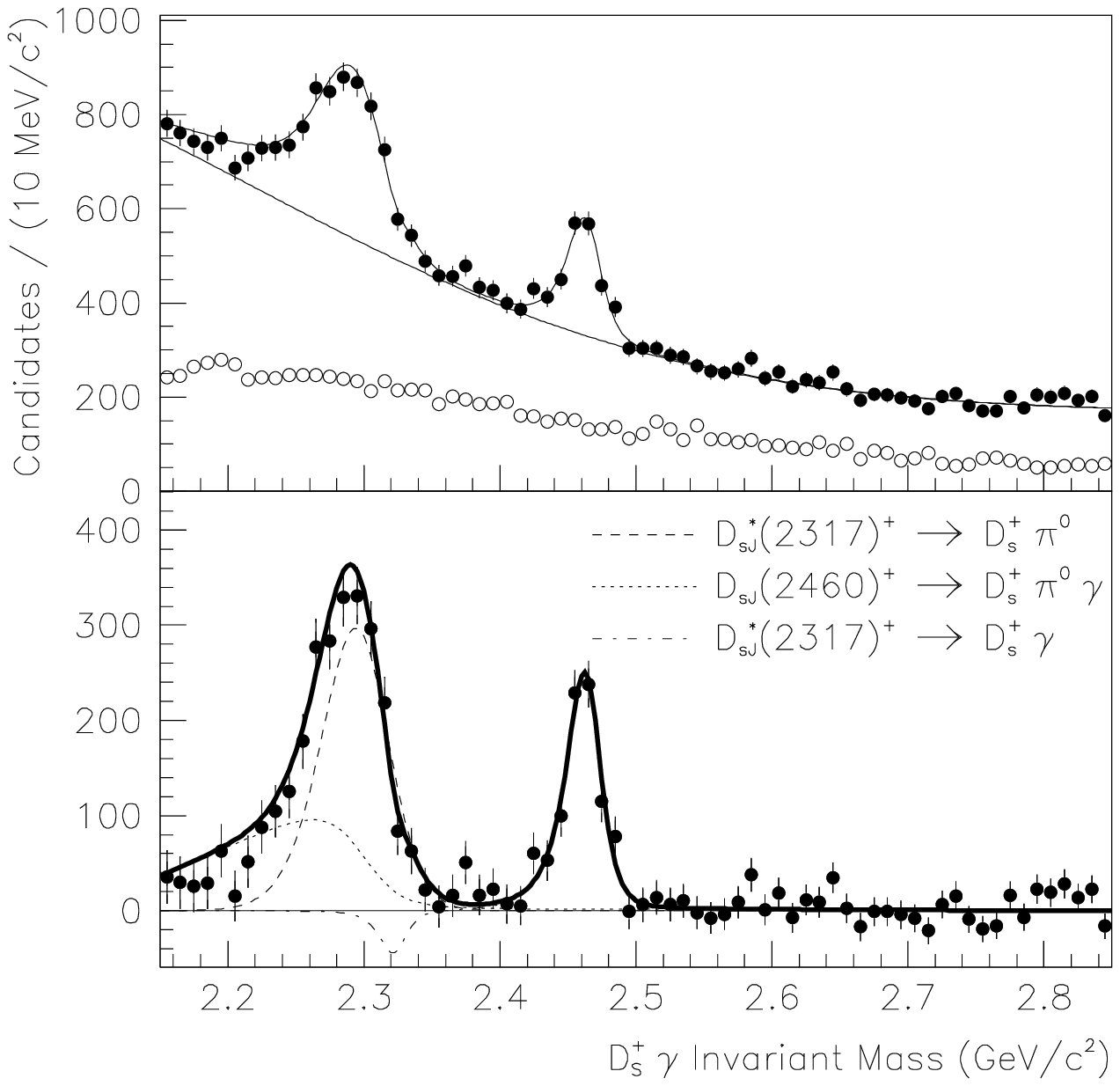}&
\hspace{-12mm}\includegraphics[width=0.30\textwidth,height=0.30\textwidth]{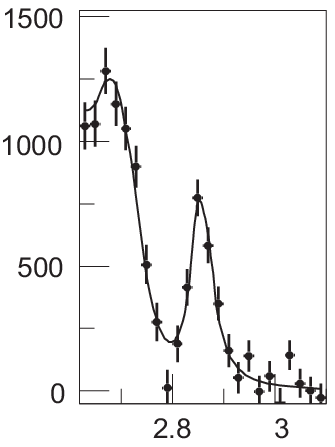}
\end{tabular}
\vspace{-6mm}
\end{center}
\caption{\label{Ds}
Left: The $ D_{sJ}(2460)^+\to\gamma D_s^{*}(2112)^+,
D_s^{*}(2112)^+\to  D_s^+\pi^0$ decay. Shown is the invariant mass
distribution of $ D_s\,\pi^0\gamma$ candidates for the $ D_s$
signal (solid points) and $ D_s$ side bands (open points). The photon
energy may be compatible with a transition to $D_s^{*}(2112)^+$ (b) and
be too soft (a) or too hard (c).  The dark grey (light grey) region
corresponds to the predicted contribution from the $
D_{s0}^*(2317)^+$ ($D_s^{*}(2112)^+$) reflection \cite{Aubert:2006bk}.
centre: The $ D_{sJ}(2460)^+\to\gamma D_s^+$ decay. Contributions
due to reflections are indicated dashed, dashed-dotted and dotted
lines. The contribution $ D_{s0}^*(2317)^+\to D_s\gamma$ is fitted
to make a negative contribution (compatible with zero)
\cite{Aubert:2006bk}. Right: Observation of $ D_{sJ}(2860)^+$.
Background-subtracted D\,K invariant mass distributions for the sum of
$ D^0$ decays into $ K^-\pi^+$ and $ K^-\pi^+\pi^0$, and $
D^+$ decays into $ K^-\pi^+\pi^+$ \cite{Aubert:2006mh}.
}
\end{figure}

The second missing resonance turned out to have a low mass and to be
narrow, too. The $D_{sJ}(2460)^+$ cannot decay into $D\,K$ if $J=1$;
its mass falls below the $D^*\,K$ threshold. It was discovered
shortly after $D_{s0}^*(2317)$ \cite{Krokovny:2003zq} in pion emission
to the $D_s^{*}(2112)^+$ vector state and in direct radiative
transitions to the $D_s^+$ ground state. In Fig. \ref{Ds} evidence for
these two decay modes is displayed. The selection rules in decays of
the $D_{s0}^*(2317)^+$ and $ D_{sJ}(2460)^+$ states are consistent with
$J^P=0^+$ and $1^+$, respectively. Masses and widths of the low-mass
excitations of $D_s$ are summarised in the lower part of Table
\ref{DDecays}.

\begin{table}[pt]
\caption{\label{DDecays}
Masses and width of $D$ an $D_s$ mesons and
their excitations quoted from the Review of Particle Properties
\cite{Eidelman:2004wy}. If statistical and systematic errors are
given, their quadratic sum is listed below. All quantities are given in
MeV/c$^2$, except the $D$-meson life times which are given in
fs (=$10^{-15}$s) units. \vspace{2mm}}
\renewcommand{\arraystretch}{1.5} \bc\begin{tabular}{cccccccc}
\hline\hline
Charge & $D$ & $D^*$ & $D^*_0$ & $D_1'$ & $D_1$ & $D^*_2$  \\
\hline
 $\pm$\quad M& $1869.3\pm 0.4$ & $2010.0\pm 0.4$&$2403\pm38$&
         $2423.4\pm3.1$&                    & $2459\pm4$ \\
 \qquad$\Gamma$  & $1040\pm 7$\,fs & $0.096\pm0.023$&   $238\pm 42$&
$25\pm 6$&                                               & $29\pm5$ \\
\hline
0 \quad M& $1864.5\pm 0.4$  & $2006.7\pm 0.4$&$2352\pm 50$&
$2422.3\pm 1.3$&       $2427\pm 36$& $2461.1\pm1.6$\\
 \qquad$\Gamma$      &$410.1\pm 1.5$\,fs& $<2.1$         &$261\pm 50$
&$20.4\pm 1.7$& $384^{+130}_{-105}$& $43\pm 4$\\ \hline Charge
&$D_s$&$D_s^*$& $D^*_{s0}$&$D_{s1}'$&$D_{s1}$&$D^*_{s2}$&\\\hline
$\pm$\quad M& $1968.2\pm 0.5$ & $2112.0\pm 0.6$&$2317.3\pm0.6$&
         $2458.9\pm0.9$& $2535.35\pm0.61$  &$2573.5\pm1.7$ \\
 \qquad$\Gamma$ & $98.85\pm 0.30$ &$<1.9$&   $<4.6$& $<5.5$& $<2.3$
                                              & $15^{+5}_{-4}$ \\
\hline\hline
\end{tabular}
\ec\renewcommand{\arraystretch}{1.0}
\end{table}

It is illuminating to compare the excitation energies for the three
doublets with $(L=0,j_q=1/2)$, $(L=1,j_q=1/2)$, $(L=1,j_q=3/2)$. We use
the masses of the charged $D$ mesons when available. For the two
multiplets with $(L=0,j_q=1/2)$ and $(L=1,j_q=3/2)$, the mass
differences between $D_s$ and $D$ excitations are 99, 112, 98,
112\,MeV/c$^2$, respectively, consistent with a mass difference between
$s$ and $d$ quark of $\approx 105$\,MeV/c$^2$. The third multiplet
resists such a simple reasoning. The mass difference for the $D_{s1}'$
and $D_{1}'$ is 37\,MeV/c$^2$, for the $D_{s0}^*$ and $D_{0}^*$ it
depends on the experiment. Taking the BELLE value, the difference is
-86\,MeV/c$^2$, and +9\,MeV/c$^2$ for the FOCUS value.

The masses of the $(L=1,j_q=1/2)$ doublet are weird. If we expect all
four $P$ states to have similar masses, the $D$ meson excitations
concur with it, and the $D_{s0}^*(2317)$ has exotic properties. While
$D^*_0(2400)$ is about 350\,MeV/c$^2$ above the $D\,\pi$ threshold,
$D_{s0}^*(2317)$ is 40\,MeV/c$^2$ below the $D\,K$ threshold. Hence the
states do behave anomalous.

In any case, with these states, the magic number of 6 low-mass states
is reached and the spectrum of $S$- and $P$-wave $c\bar n$ and $c\bar s$
states covered. Correspondingly, Bardeen, Eichten and Hill
\cite{Bardeen:2003kt} interpreted the two new resonances
$D_{s0}(2317)$ and $D_{sJ}(2460)$ as $c \bar s$ $(0^+, 1^+)$ spin
parity partners of the $(0^-, 1^-)$ doublet, and demonstrated that
symmetry relation between their partial decay widths derived from
chiral symmetry are fulfilled.

Recent lattice QCD calculations \cite{Lin:2006vc} gave
an excitation energy from the $D_s^{*}(2112)^+$ vector to the
$D_{s0}^{*}(2112)^+$ scalar state of $(328\pm40)$\,MeV instead of the
observed 205\,MeV mass gap. For their nonstrange partners the
corresponding mass difference is 440\,MeV on the lattice and
$(342\pm50)$\,MeV on the experimental side. In agreement with quark
model results \cite{Godfrey:1985xj,DiPierro:2001uu}, the masses of
$D_{s0}(2317)$ and $D_{sJ}(2460)$ are to low to support a
straightforward interpretation as $c\bar s$ states; alternative
interpretations have therefore been suggested.

Barnes, Close and Lipkin argued that $D_{s0}(2317)$ is a $D\,K$
molecule and predicted the scalar $c\bar s$ state at 2.48 GeV
\cite{Barnes:2003dj}. The latter state is predicted to couple strongly
to the $D\,K$ channel and the isoscalar $D\,K$ molecule $D_{s0}(2317)$
could be formed. The molecular forces lead to a small isovector
admixture which explains its narrow width and the isospin violating
decay mode $D_s \pi$. $D_{s0}(2317)$ is proposed to be a dominantly I=0
$D\,K$ state with some I=1 admixture.

In the unitarised meson model of Van Beveren and Rupp
\cite{vanBeveren:2003kd}, $D_{s0}(2317)$ is created by a strong
coupling to the $S$-wave $D\,K$ threshold. The standard $c\bar{s}$
charm strange scalar meson $D_{s0}$ is predicted at about 2.79\,GeV and
with 200\,MeV width. In \cite{vanBeveren:2005ha}, the $D_{s0}(2317)$ is
linked to the $\kappa$ when the quark masses are lowered, and to
$\chi_{c0}$ for larger quark masses.

I.~W.~Lee, T.~Lee, Min, and Park show that the one-loop chiral
corrections could be responsible for the small $D_{s0}(2317)$ mass thus
supporting the picture of the new resonance as composed of a
heavy quark and a light valence quark \cite{Lee:2004gt}. Kim and Oh
used QCD sum rules study to argue that the $D_s (2317)$ should be a
bound state of scalar-diquark and scalar-antidiquark and/or
vector-diquark and vector-antidiquark \cite{Kim:2005gt}.

An interpretation of $D_{sJ0}(2317)$ and $D_{sJ}(2460)$ as ordinary $c
\bar s$ mesons derived from their decay patterns was given by Wei,
Huang, and Zhu \cite{Wei:2005ag} who use light cone QCD sum rules.
Through $\eta-\pi^0$ mixing, their pionic decay widths are calculated
which turn out to be consistent with the experimental values (or upper
limits).

The unusual properties of the $D_{s0}(2317)$ and $D_{sJ}(2460)$
encouraged speculations that they could belong to a family of $c\bar{s}
n \bar{n}$ tetraquark states \cite{Cheng:2003kg,Terasaki:2003qa,%
Maiani:2004vq,Dmitrasinovic:2006uk,Terasaki:2006wm}. Tetraquark states
have however a common problem: many more states are predicted than
found experimentally. The molecular picture is supported by
calculations of strong $D_{s0}^*$ to $D_s\,\pi^0$ and radiative
$D_{s0}^*$ to $D_s^*\gamma$ decays \cite{Faessler:2007gv} and the
leptonic decay constants $f_{D_{s0}^*}$ and $f_{D_{s1}}$
\cite{Faessler:2007cu}.

We come back to the discussion at the beginning of section \ref{The
J/psi states above the Dbar D threshold}. So far, there is no evidence
for supernumerosity of states but some states certainly show an
anomalous behaviour. In our language, we conclude that the states with
$(L=1,j_q=1/2)$ are $c\bar n$ or $c\bar s$ quark model states but their
coupling to the $D\,\pi$ or $D^*\,\pi$ ($D_s\,\,\pi$ or $D_s^*\,\pi$)
thresholds leads to strong molecular and/or tetraquark components. More
detailed information on the wave functions can probably be deduced from
radiative transitions of $D_{s0}^*(2317)$ and $D_{sJ}'(2460)$ mesons
to the $D_s^*$ state \cite{Godfrey:2002rp}. Since the absolute width of
the $D_{s0}^*(2317)$ is expected to be in the order of 100\,keV/c$^2$,
only ratio of radiative to hadronic widths are experimentally
accessible. The results, collected in Table \ref{D-rad} are consistent
with $c\bar q$ descriptions of $D_{s0}^*(2317)$ and $D_{sJ}'(2460)$
\cite{Bardeen:2003kt} but do not enforce this interpretation.

\subsubsection{\label{Further mesons with open charm}
Further mesons with open charm}
\begin{table}[pt]
\caption{\label{D-rad}
Ratio of radiative and strong decay widths of $D_{s0}^*(2317)$
and $D_{sJ}'(2460)$.\vspace{2mm}}
\begin{center}
\renewcommand{\arraystretch}{1.5}
\begin{tabular}{cccc}
\hline\hline & BELLE \cite{Krokovny:2003zq,Abe:2003jk}
& BaBaR \cite{Aubert:2004pw,Aubert:2006bk} & CLEO \cite{Besson:2003cp} \\ \hline
${\Gamma_{ D^*_{sJ}(2317)\rightarrow D_{s}^{\ast }\gamma
 }}/{ \Gamma_{ D^*_{sJ}(2317)\rightarrow D_{s}\pi
^{0} }}$ & $<0.18$ & & $<0.059$ \\
${\Gamma_{D_{sJ}(2460)\rightarrow D_{s}\gamma  }}/{
\Gamma_{ D_{sJ}(2460)\rightarrow D_{s}^{\ast }\pi ^{0} }}$
& $0.55\pm0.13\pm0.08$ & $0.375\pm0.054\pm0,057$ &$<0.49$ \\
$\frac{\Gamma_{ D_{sJ}(2460)\rightarrow D_{s}^{\ast }\gamma
 }}{\Gamma_{ D_{sJ}(2460)\rightarrow D_{s}^{\ast }\pi
^{0} }}$ & $<0.31$ & &$<0.16$  \\
${\Gamma_{ D_{sJ}(2460)\rightarrow D^*_{sJ}(2317)\gamma
 }}/{\Gamma_{ D_{sJ}(2460)\rightarrow
D_{s}^*\,\pi^0  }}$  &  &   & $ < 0.58$ \\
\hline\hline \end{tabular}
\renewcommand{\arraystretch}{1.0}
\end{center} \end{table}

A few higher-mass states were also reported. So far, there is evidence
only for one additional state belonging to the $D$ meson family, with
quantum numbers  $J^{PC}=1^{--}$. The $D^*(2640)$ was observed by
Delphi at LEP in $Z^0$ decays \cite{Abreu:1998vk}. It has a mass
difference to the ground state $D^*(2010)$ of 630\,MeV/c$^2$ which
compares well with those of other vector meson, but its width is
anomalously narrow, less than 15\,MeV/c$^2$. The $D^*(2640)$ was
searched for by the Opal collaboration; an upper limit, incompatible
with the Delphi result, was given \cite{Abbiendi:2001qp}. If a broad
resonance is present, but at the statistical limit of its
observability, then its is sometimes seen due to a statistical
fluctuation as a narrow peak. Without or opposite a statistical
fluctuation, the resonance is ascribed to the background.

In the $D_s$ family, a few states were reported. A  $D_{sJ}^{*+}(2630)$
was observed by the SELEX collaboration at FNAL \cite{Evdokimov:2004iy}
in the final states $D^0K^+$ and $D_s \eta$ produced inclusively in a
hyperon ($\Sigma^-$) beam. The mass is determined to $2632.6 \pm 1.6$
MeV/c$^2$ and the width to be less than 17\,MeV/c$^2$ (at 90\% c.l.).
Due to the decay mode, it must have natural spin-parity. The ratio of
the partial widths

\begin{equation}\label{2630r} {\Gamma(D_{sJ}^{*+}(2630) \to
D^0K^+) \over \Gamma(D_{sJ}^{*+}(2630) \to D_s\,\eta)} = 0.16 \pm 0.06.
\end{equation}

is unexpected since the D\,K mode is favoured by phase
space. The width is unexpectedly narrow as well. These properties
encouraged searches by other collaborations, by FOCUS (quoted from
\cite{Swanson:2006st}), BaBaR \cite{Aubert:2004ku}, and CLEO
\cite{Galik:2004mi}. No evidence for the $ D_s(2630)^{*+}$ was found.
However, SELEX is the only experiment using a hyperon beam, and it may
be that this is a more favourable production mode.

The  $D_{sJ}^{*+}(2630)$ properties may evidence an exotic nature of
the state. Van Beveren and Rupp \cite{vanBeveren:2004ve} interpret the
state in their coupled channel model. The model predicts further,
partly narrow, ${D_s^*}'$ states at 2720, 3030, and 3080\,MeV/c$^2$. Maiani
{\it et al.} suggest \cite{Maiani:2004xg} that the $ D_s(2630)$ is a
$[cd][\bar d \bar s]$ tetraquark. If the state does not mix with $[c
u][\bar u \bar s]$, the decay to $ D^0K^+$ would be isospin violating,
thus making the $ D_s(2630)$ narrow. Among other predictions, a
close-by state with charge $+2$ decaying into $D_s^+\pi^+$ and $D^+K^+$
should exist.

 Y.-R. Liu, Dai, C. Liu, and Zhu \cite{Liu:2004kd} demonstrated that
the ratio (\ref{2630r}) can be reproduced quantitatively when the
$D_{sJ}(2632)$ is interpreted as the $J^P=0^+$ isoscalar member of the
${\bf 15}$ multiplet of tetraquarks. the meson has the quark content
${1\over 2\sqrt{2}}(ds\bar{d}+sd\bar{d}+su\bar{u}+us\bar{u}-2ss\bar{s})
\bar{c}$ which leads to the relative branching ratio

\begin{equation}
{\Gamma(D_{sJ}^+(2632)\rightarrow {D}^0 K^+)\over
{\Gamma(D_{sJ}^+(2632)\rightarrow D_s^+\eta)}} =0.25\; .
\end{equation}

We point out that the wave function can be reduced to $s\bar c$. In our
language, the $D_{sJ}(2632)$ is a quark model state with a tetraquark
component. If, as we believe, the $s\bar c$ component is essential for
the binding, its flavour exotic partners -- predicted in \cite{Liu:2004kd}
-- should not exist. Due to the large decay probability of the
tetraquark component, this part dominates the decay properties of the
state.

\vspace{-5mm}
\begin{figure}[pt]
\bc
\includegraphics[width=0.6\textwidth,height=0.35\textwidth]{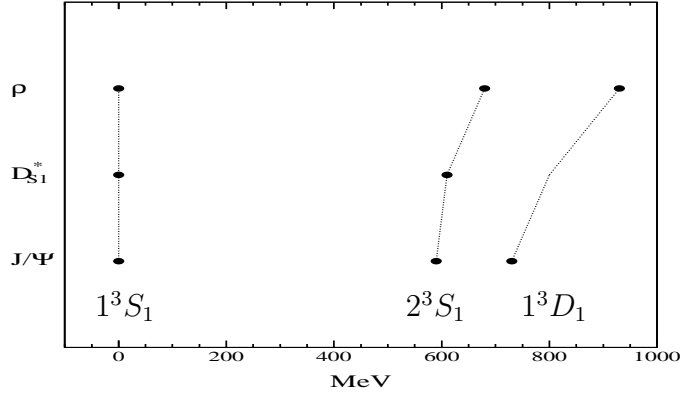} \\
\vspace{-15mm}
$1^3S_1$\hspace{32mm} $2^3S_1$ \hspace{6mm}$1^3D_1$ \\
\vspace{8mm}
\ec
\caption{\label{vector-ds}
Excitation energies of vector states above the $1^3S_1$ level. Dotted
lines connect states of different flavour content but the same
spectroscopic quantum numbers. The charmonium states are (from
left to right) J/$\psi$, $\psi(2S)$ and $\psi(1D)$; the
isovector vector states are $\rho(770)$, $\rho(1450)$ and $\rho(1700)$
which are likely the $1^3S_1$, $2^3S_1$ and $1^3D_1$ quark-model
states. The $ D_{s1}^*(2715)$ is naturally assigned to a $2^3S_1$
configuration.} \end{figure}

A $c\bar s$ vector excitation was observed in a Dalitz plot
analysis of $ B^+ \to \bar D^0 D^0 K^+$ decays \cite{unknown:2006xm,Brodzicka:2007aa}
in the $ D^0 K^+$ mass spectrum. Its mass and width are $(2715 \pm
11^{+11}_{-14})$ and $(115 \pm 20^{+36}_{-32})$ MeV/c$^2$,
respectively. Its spin-parity is $J^P = 1^-$. Fig. \ref{vector-ds}
shows a comparison of $ D_{s1}^*$ states with excitations of
J/$\psi$ and $\rho$. An identification of $ D_{s1}^*(2715)$ as
$D_{s1}^*(2S)$ is very convincing even though an exotic interpretation
as tetraquark state has also been suggested \cite{Wang:2007nf}.

The so-far highest mass  $c\bar s$ state was observed by the BaBaR
Collaboration \cite{Aubert:2006mh}. Fig. \ref{Ds} (right) shows
the sum of the $ D^0 K^+$ and $ D^+ K^0_S$ mass distribution. Due
to the decays, it must have  natural spin-parity $0^+$, $1^-$, $2^+,
\ldots$. Its mass and width are $2856.6 \pm 1.5 \pm 5.0$ and $48 \pm
7 \pm 10$\,MeV/c$^2$. With weak evidence, a further state is reported
with $M=2688\pm 4\pm2,\Gamma=112\pm7\pm36$\,MeV/c$^2$.

Van Beveren and Rupp suggest both states to have scalar
quantum numbers and identify the $ D_{sJ}^*(2860)$ as the radial
excitation of the $D_{s0}^*(2317)$ while the $ D_{sJ}^*(2690)$  is
assigned to a dynamical pole originating from unitarity constraints
\cite{vanBeveren:2006st}. Close, Thomas, Lakhina and Swanson use a
modified quark-quark scalar potential which allows them to treat the
$D_{s0}^*(2317)$ as ordinary $c\bar s$ meson; based on the masses, they
suggest to identify $ D_{sJ}^*(2690)$ with $ D_{s1}^*(2715)$ and
to interpret  $ D_{sJ}^*(2860)$ as  $ D_{s0}^*(2P)$ state
\cite{Close:2006gr}. The $ D_{sJ}^*(2860)$ may also be
 a $3^-(^3D_3)$ state \cite{Colangelo:2006rq}. The centrifugal barrier
then explains its narrow width.

Based on a discussion of widths and decay modes, Bo Zhang et al.
\cite{Zhang:2006yj} interpret $ D_{sJ}(2715)$ as the
$1^{-}(1^{3}D_{1})$ $c\bar{s}$ state (although the $1^{-}(2^{3}S_{1})$
assignment is not excluded). An interpretation of $D_{sJ}(2860)$ as
$1^{-}(2^{3}S_{1})$ or $1^{-}(1^{3}D_{1})$ candidate was considered to
be unlikely, and an assignment as $0^{+}(2^{3}P_{0})$ or
$3^{-}(1^{3}D_{3})$ $c\bar{s}$ state more plausible. Obviously, more
experimental data are needed to assign quantum numbers reliably.

\subsubsection{\label{Mesons with open beauty}
Mesons with open beauty}

Mesons with open beauty, $b\bar n$ mesons, correspond closest to the
classical picture of hydrogen atoms and are best suited to test ideas
on heavy flavour symmetry. Experimentally, the study of $B$ meson
excitations is still in its infancy. One expects, as in the case of $D$
meson excitations, a vector resonance $ B^*$, two narrow states
$ B_1$ and $ B_2^*$, and two broad states  $ B_0^*$ and
$ B_1$.

\begin{figure}[pt]
\bc
\includegraphics[width=0.52\textwidth,height=5cm]{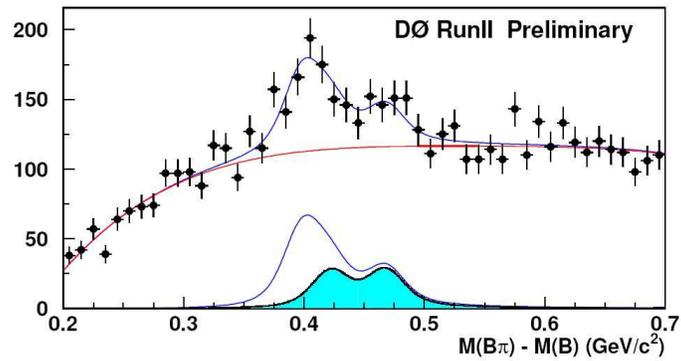}
\vspace{-3mm}
\ec
\caption{\label{fnalB}
The B$\pi-$B mass difference with the results of a fit. The 50\,MeV
photon from B$^*\to\gamma$B was not reconstructed, hence $ B_2^*$
decays to $ B_0^*\,\pi$ are shifted in mass. The background subtracted
spectrum shows contributions assigned to $ B_2^*$ decays to $
B_0^*\,\pi$ and to  $ B\pi$ (shaded area) and from $ B_1$ decays
$ B_0^*\,\pi$ \cite{Cheu:2004zc}.}
\end{figure}

A pilot study was made by D$\O$ at FNAL \cite{Cheu:2004zc}. The $B$
mesons were reconstructed from their J/$\psi\, K$ and J/$\psi\, K^*$
decays. From the inclusive B$\pi$ mass spectrum shown in Fig.
\ref{fnalB}, the mass of the two expected 'narrow' resonances were
deduced. \vspace{-3mm}
\begin{eqnarray}
 B_1 \ \ \qquad& M & = 5724\pm 4\pm 7\ {\rm MeV/c^2} \nonumber
\\ B_2^*\ \qquad\ & M_{ B_2^*} - M_{ B_1}& = 23.6\pm 7.7\pm 3.9\ {\rm
MeV/c^2}. \nonumber \end{eqnarray} The widths of both states were
assumed to be equal and fitted to $\Gamma=23\pm 12\pm 6$\,MeV/c$^2$.
The results do not yet deepen our insight into heavy quark physics but
rather demonstrate the feasibility of such studies.

\markboth{\sl Meson spectroscopy} {\sl High orbital and radial
excitations}  \clearpage\setcounter{equation}{0}\section{\label{High radial and orbital excitations of light mesons}
Radial and orbital excitations of light mesons}

\subsection{\label{Patterns at high l, n}
Patterns at high $l, n$}

The masses of the lightest glueballs are predicted to fall into the 1.5
to 2.5\,GeV/c$^2$ mass range. This is a mass interval in which a rich
spectrum of meson resonances is expected, and indeed observed. This
spectrum may however contain not only $q\bar q$ mesons but hybrid
excitations and multiquark clusters as well. Thus it is essential to
develop a good understanding of the `background' of $q\bar q$ mesons in
order to identify anomalies and to establish properties of states
falling outside of this classification scheme, for their
supernumerosity, for their production modes, or for their unusual
decays. In the last few years, significant progress was achieved in the
study of highly excited hadrons. The comprehensive analysis of various
final states in $p \bar p$ annihilation in flight
\cite{Anisovich:2000ut,Anisovich:2001pn,Anisovich:2002su,Anisovich:2002sv}
revealed a large number of resonances with masses up to 2.4\,GeV/c$^2$.
The total antiproton-proton annihilation cross section is saturated in
all partial waves accessible at a given momentum, and meson resonances
are excited with low and high orbital and radial quantum numbers. The
simultaneous analysis of several final states gives access to all
(non-exotic) quantum numbers. Many of the resonances are observed in
several final states; the internal consistency of the results and the
consistency of some results with earlier observations lends further
credibility to the findings.

\begin{figure}[h!] \bc
\includegraphics[width=0.5\textwidth]{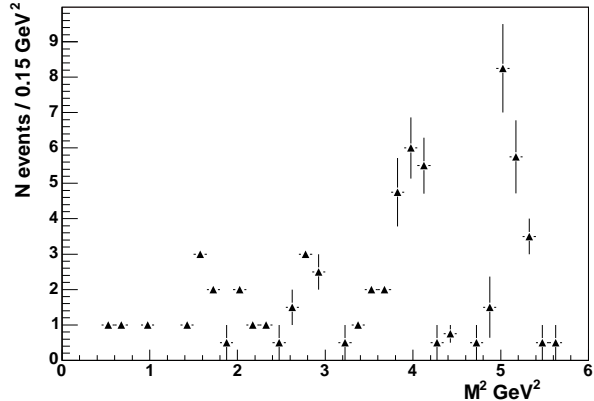} \ec
\caption{\label{fig:allmesons}
The density of states versus their squared masses. }
\end{figure}

The multitude of resonances excludes a discussion of each individual
state. Instead, we treat the data on a statistical basis. In Fig.
\ref{fig:allmesons} the number of states as a function of their squared
mass is plotted. Pseudoscalars and scalars are not included in this
figure for reasons discussed in the introduction (section \ref{The
light meson spectrum and quark model assignments}). Strange particles
are omitted as well. Those states which are not well confirmed are
taken in the histogram with weight $W=0.5$ and error $E=0.5$. In
assigning the significance to the states we follow the judgement of Bugg
\cite{Bugg:2004xu}.
The figure demonstrate clustering of states very clearly
\cite{Bugg:2004xu}. An analogous effect of clustering is known in
baryon spectroscopy \cite{Klempt:2002vp,Afonin:2006ts}. The spacing
between bumps in Fig. \ref{fig:allmesons} is about $1.1$ GeV/c$^2$,
very similar to that in baryons.

\begin{figure}[h!]
\bc
\includegraphics[width=1.0\textwidth]{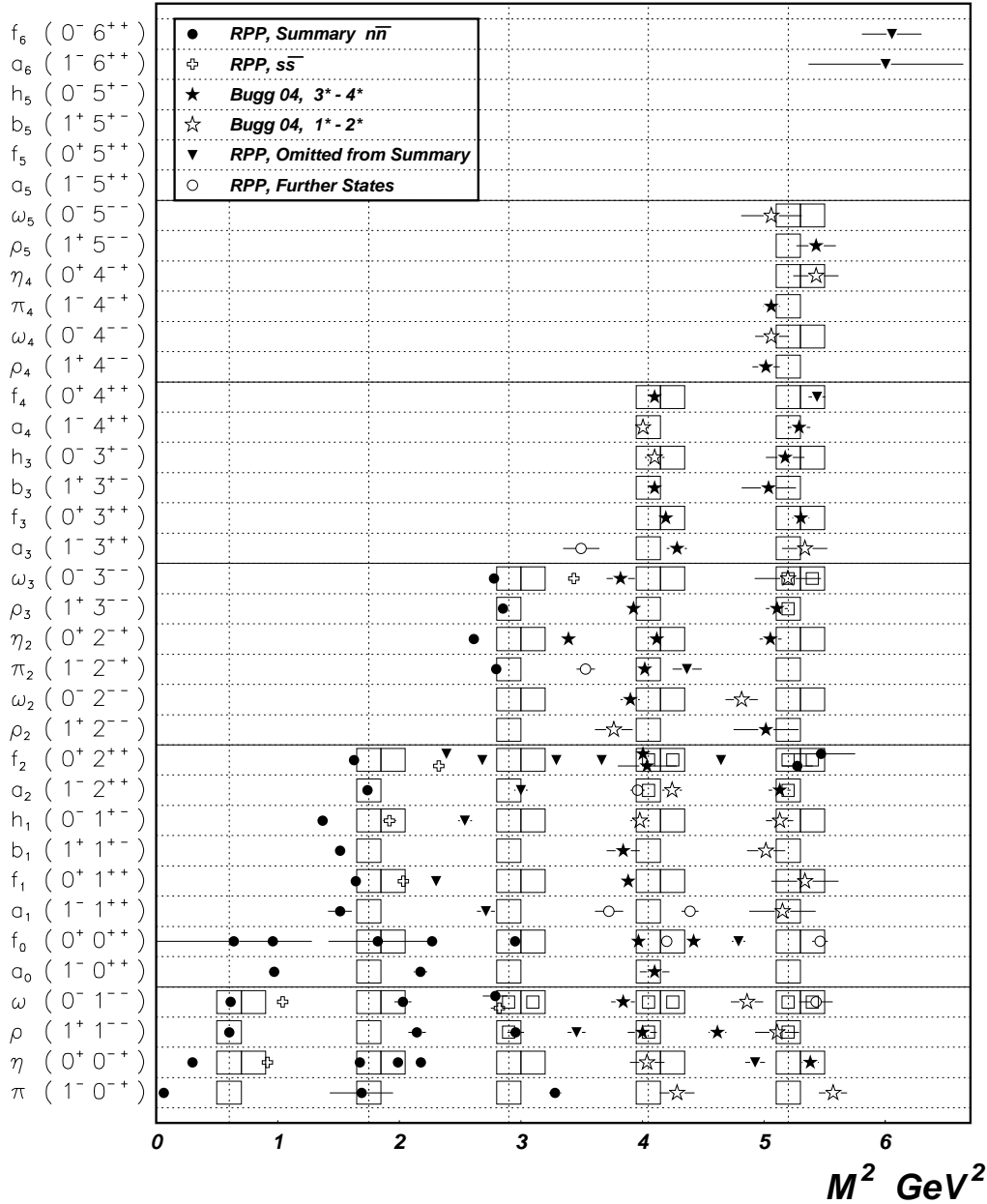}
\ec
\caption{\label{fig:globalview}
The pattern of light quark meson states (adapted from
\cite{Afonin:2007jd}). }
\end{figure}

A more detailed picture of light-quark mesonic states is given in
Fig. \ref{fig:globalview}.
In this figure all light quark mesons with known quantum
numbers $I^G J^{PC}$ are plotted as a function of $M^2$. The ordering of
states follows expectations from potential models for $q \bar q$
mesons. The first four lines are occupied by states with zero orbital
momentum, the next eight lines by states with $L=1$, etc. States
with different creditability are plotted using different symbols.  The
empty boxes indicate the position of states in an oversimplified model
in which masses of mesons are proportional to $l+n$, where $l$ is the
orbital and $n$ the radial quantum number. Mesons of zero isospin have
two nearby boxes for $n \bar n$ and $s \bar s$ states or for $SU(3)$
singlet and octet states. Some boxes are doubled because two different
states ($J=l+1,n$ and $J=l-1,n+2$) can sit there.

From this amusing picture we conclude that at sufficiently high $J$ and
$M$, meson trajectories are nearly linear in both, in the $M^2,J$ and
in the $M^2,n$ plane. The trajectories are parallel and equidistant.
The linearity of the trajectories and the $l,n$ degeneracy were
predicted in a model based on the dual superconductor mechanism of
confinement \cite{Baker:2002km} and a model guided by nonperturbative
QCD \cite{Karch:2006pv}, see section \ref{Regge trajectories} in the
introduction. For sake of convenience, we reproduce eq.
(\ref{string})
$$ M_n^2(l) =
2\pi\sigma\left(l+n+\frac{1}{2}\right)\,.
$$
which shows that the squared masses are linear in $l$ and $n$, and
degenerate in $n+l$.

The $n+l$ degeneracy has been interpreted as evidence for restoration
of chiral symmetry in highly excited mesons
\cite{Glozman:2002cp,Glozman:2002kq}. A new QCD scale
$\Lambda_{CSR}=2.5$\,GeV/c$^2$ is suggested at which chiral symmetry is
restored \cite{Swanson:2003ec,Afonin:2006vi}. The string model and the
conjectured restoration of chiral symmetry thus both lead to a $n+l$
degeneracy of excited states. The two models make however different
predictions for `stretched' states, for states with $J=l+s$. The string
model predicts no parity partners for $a_2(1320)$--$f_2(1270)$,
$\rho_3(1690)$--$\omega_3(1670)$, $a_4(2040)$--$f_4(2050)$,
$\rho_5(2350)$--$\omega_5$, $a_6(2450)$--$f_6(2510)$. On the contrary,
there is no reason why these isospin doublets should not be accompanied
by chiral partners, if chiral symmetry would be restored in highly
excited mesons. Experimentally, there are no chiral partner for any of
these 10 states. Hence data do not support the hypothesis of
chiral symmetry restoration. Similar arguments have been made for
baryons \cite{Klempt:2002tt} for which restoration of chiral symmetry
is suggested as well \cite{Glozman:1999tk,Cohen:2002st}. From a
theoretical side,  Jaffe, Pirjol and Scardicchio \cite{Jaffe:2005sq}
concluded that massless pions preclude chiral symmetry be restored in
the hadron mass spectrum and suggested other reasons for the observed
parity doublets like suppression of the violation of the flavour
singlet axial symmetry $U(1)_{A}$ \cite{Jaffe:2006jy}.

Fig. \ref{fig:globalview} bares another important consequence: the
 bulk of observed mesons are compatible with a $q \bar q$ assignment,
with at most a few exceptions. These striking regularities suggest that
at sufficiently high mass and $J$, exotic states (glueballs, hybrids,
multiquark states, multimeson states) either do not exist or are not
produced in $p \bar p$ annihilation. Only a few states fall outside
of the regular pattern; they deserve special attention and will be
discussed below (in section \ref{Hybrids candidates}).

One final word of caution. The majority of high-mass entries in Fig.
\ref{fig:globalview} stems from one single experiment. Due to the early
closure of LEAR, only a rather limited number of antiproton momenta was
used and there is the need to extend the measurements to lower and to
higher momenta.  Rare channels suffered from low statistics. A
confirmation with improved statistics and improved {methods} (e.g. an
improved detector, a polarised target) is certainly highly desirable.
Such experiments will become feasible again when the antiproton project
at GSI comes into operation.

\subsection{\label{Systematic of qbarq mesons in planes}
Systematic of $q\bar q$ mesons in $(n,M^2)$ and $(J,M^2)$ planes}

In recent years, the Gatchina group has performed fits to a large
number of data sets. The fits have grown in time to increasing
complexity. Earlier variants can be found in \cite{Anisovich:1995jy,%
Anisovich:1996tk,Anisovich:1996qj,Anisovich:1997pe,Anisovich:1997qp,%
Anisovich:1997ye,Anisovich:1997rd}; the results presented here are
taken from \cite{Anisovich:2000kx} and \cite{Anisovich:2002ij}. Recent
fits include the CERN-Munich data on $\pi^- p\to\pi^+\pi^- p$; GAMS
data on $\pi^- p\to \pi^0\pi^0 n$, $\eta\eta n$ and $\eta\eta' n$ at
small transferred momenta, $|t|<0.2$\,GeV$^2$/c$^2$, and on $\pi^- p\to
\pi^0\pi^0 n$ for $0.30<|t|<1.0$\,GeV$^2$/c$^2$; BNL data on $\pi^-p\to
\pi^0\pi^0n$ at (squared) momentum transfers $0<|t|<1.5$\,GeV/c$^2$ and
on $\pi^- p\to K\bar K n$; Crystal Barrel data on $ p\bar p$
annihilation at rest in liquid H$_2$ into $\pi^0\pi^0\pi^0$,
$\pi^0\pi^0\eta$, $\pi^0\eta\eta$, $\pi^+\pi^-\pi^0$, $ K^+K^-\pi^0$, $
K^0_SK^0_S\pi^0$, $ K^+K^0_S\pi^-$ and $ \bar pn$ annihilation  in liquid
D$_2$ into $\pi^0\pi^0\pi^-$, $\pi^-\pi^-\pi^+$, $ K^0_SK^-\pi^0$, and $
K^0_SK^0_S\pi^-$.  Data on $\pi^0\pi^0\pi^0$, $\pi^0\pi^0\eta$ for
antiprotons stopping in gaseous H$_2$ were used to constrain
annihilation contributions from $S$-wave and $P$-wave orbitals of the $
p\bar p$ atom. Such analyses have the distinctive advantage that a
complete meson spectrum emerges which sums the combined knowledge from
very different reactions into one unified picture.

\begin{figure}[pt]
\bc\begin{tabular}{cccc}
\hspace{1mm}\includegraphics[width=0.23\textwidth,height=0.23\textwidth]{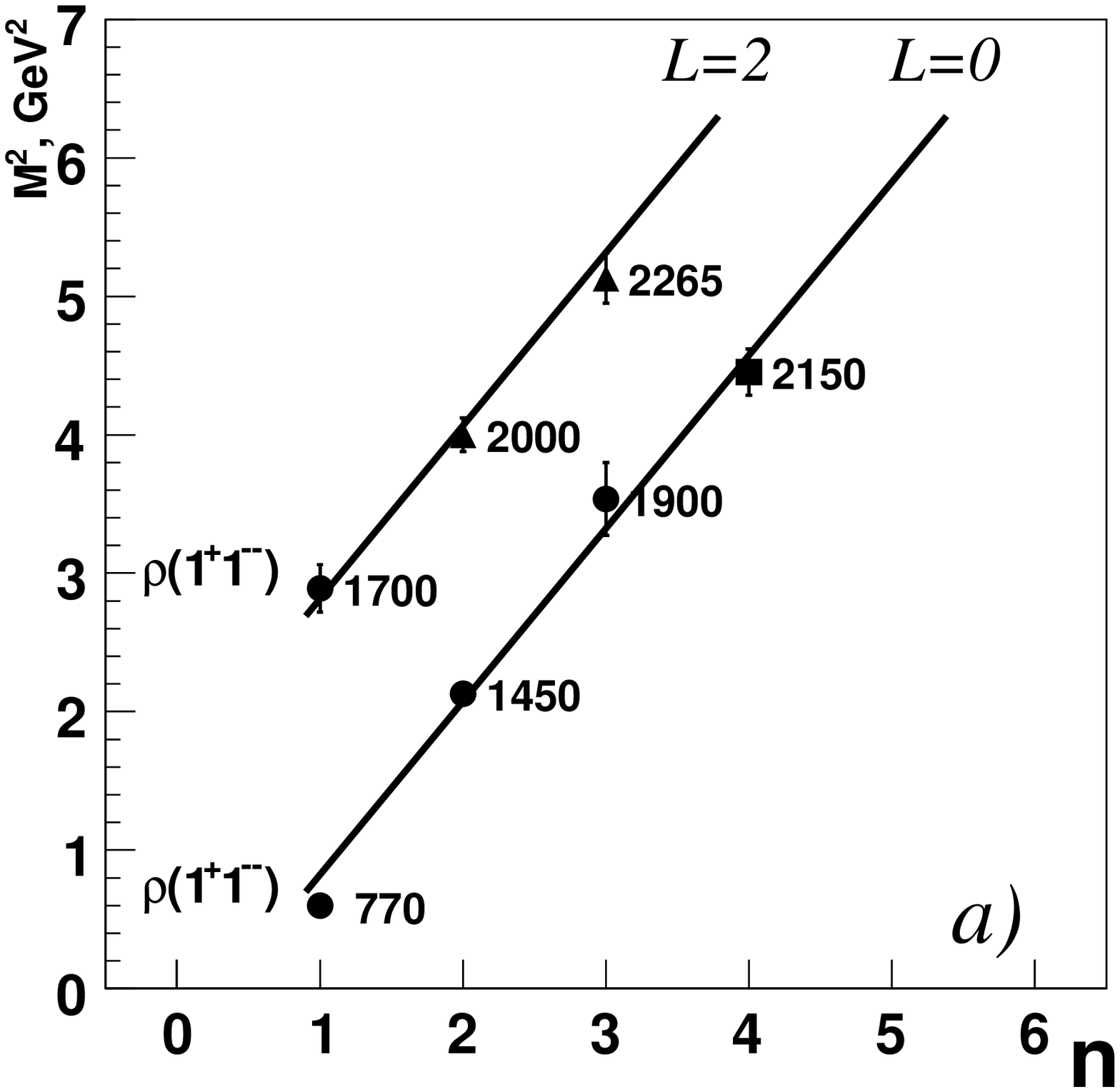}&
\hspace{1mm}\includegraphics[width=0.23\textwidth,height=0.23\textwidth]{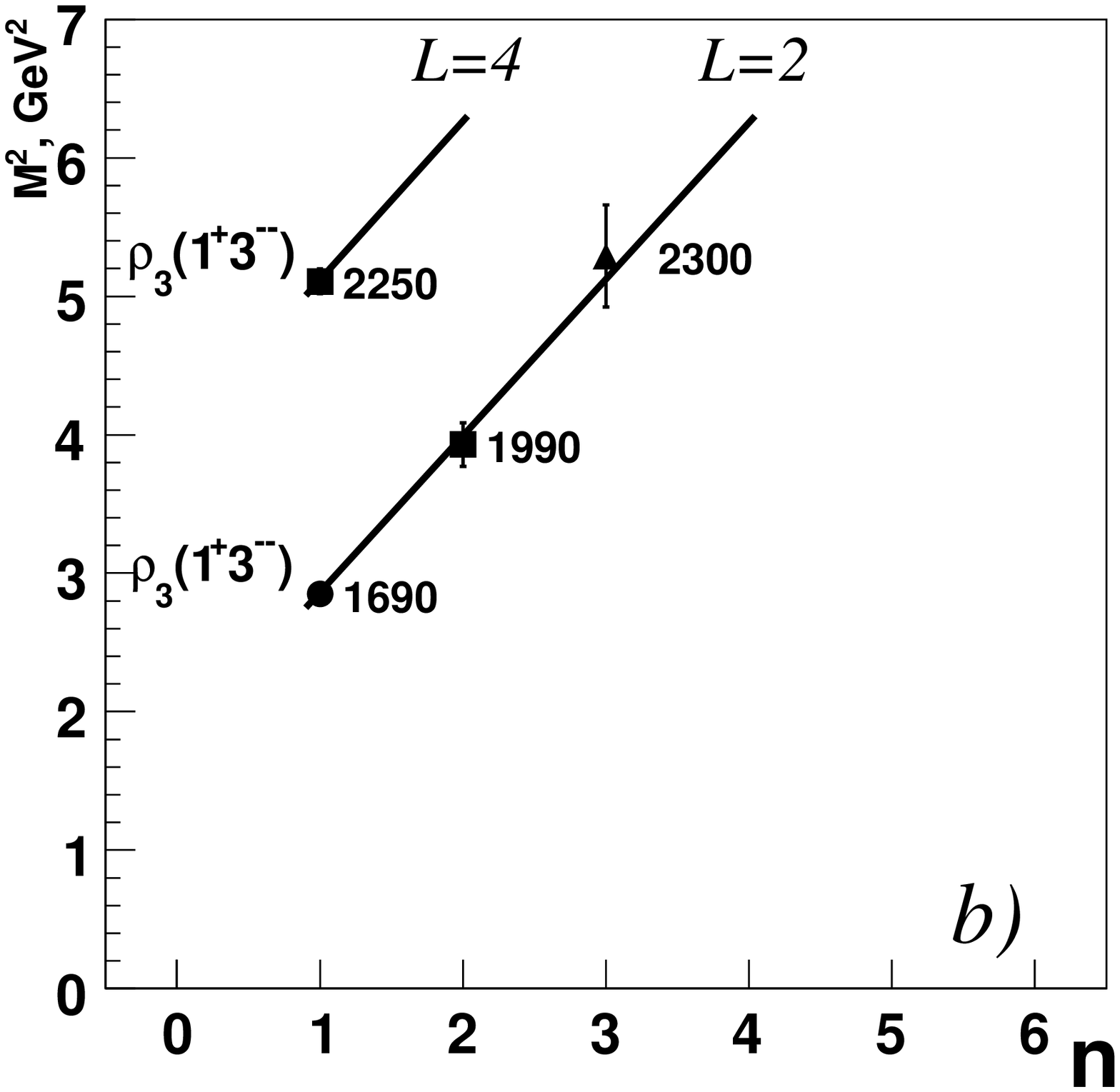}&
\hspace{1mm}\includegraphics[width=0.23\textwidth,height=0.23\textwidth]{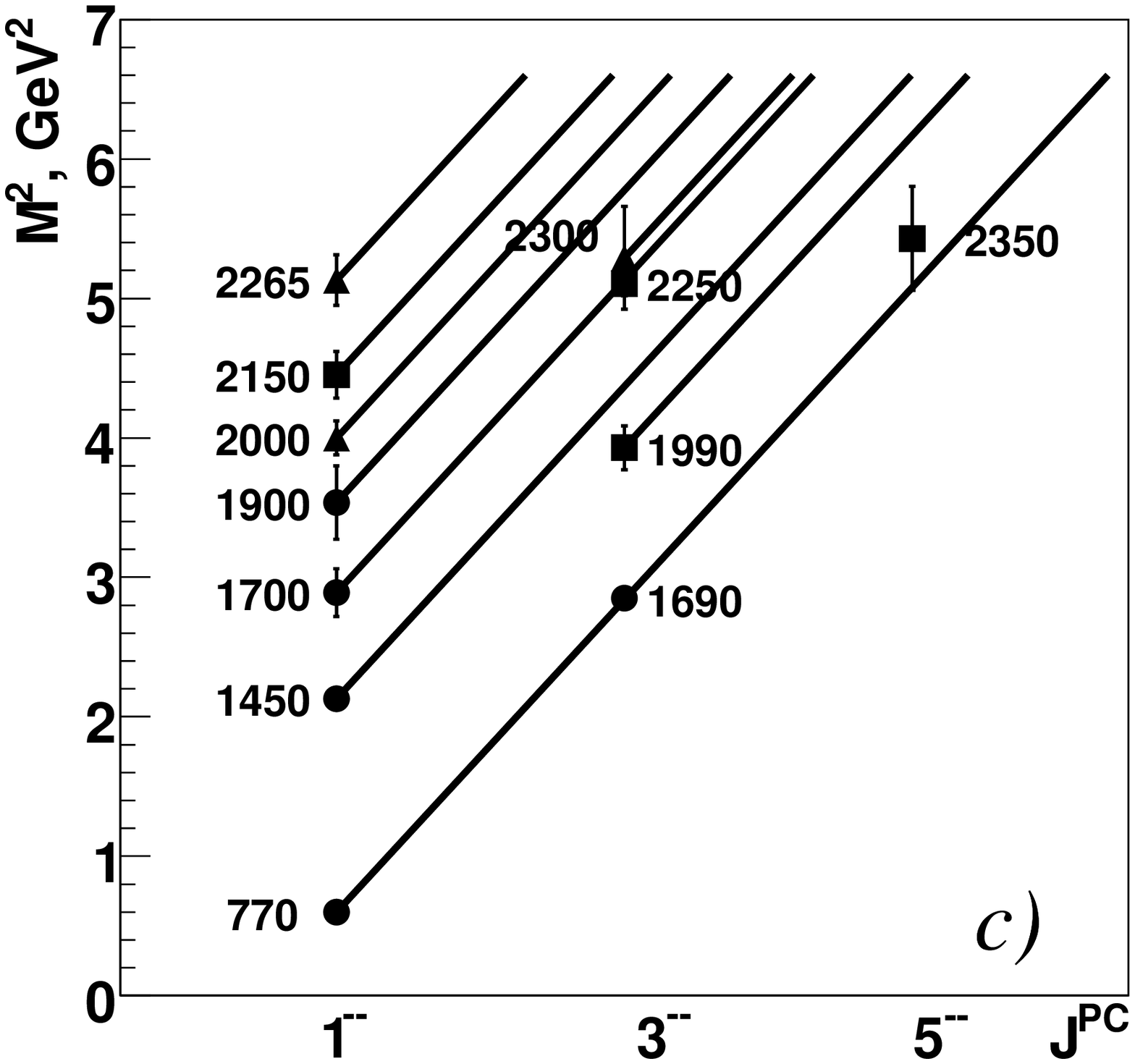}&
\hspace{1mm}\includegraphics[width=0.23\textwidth,height=0.23\textwidth]{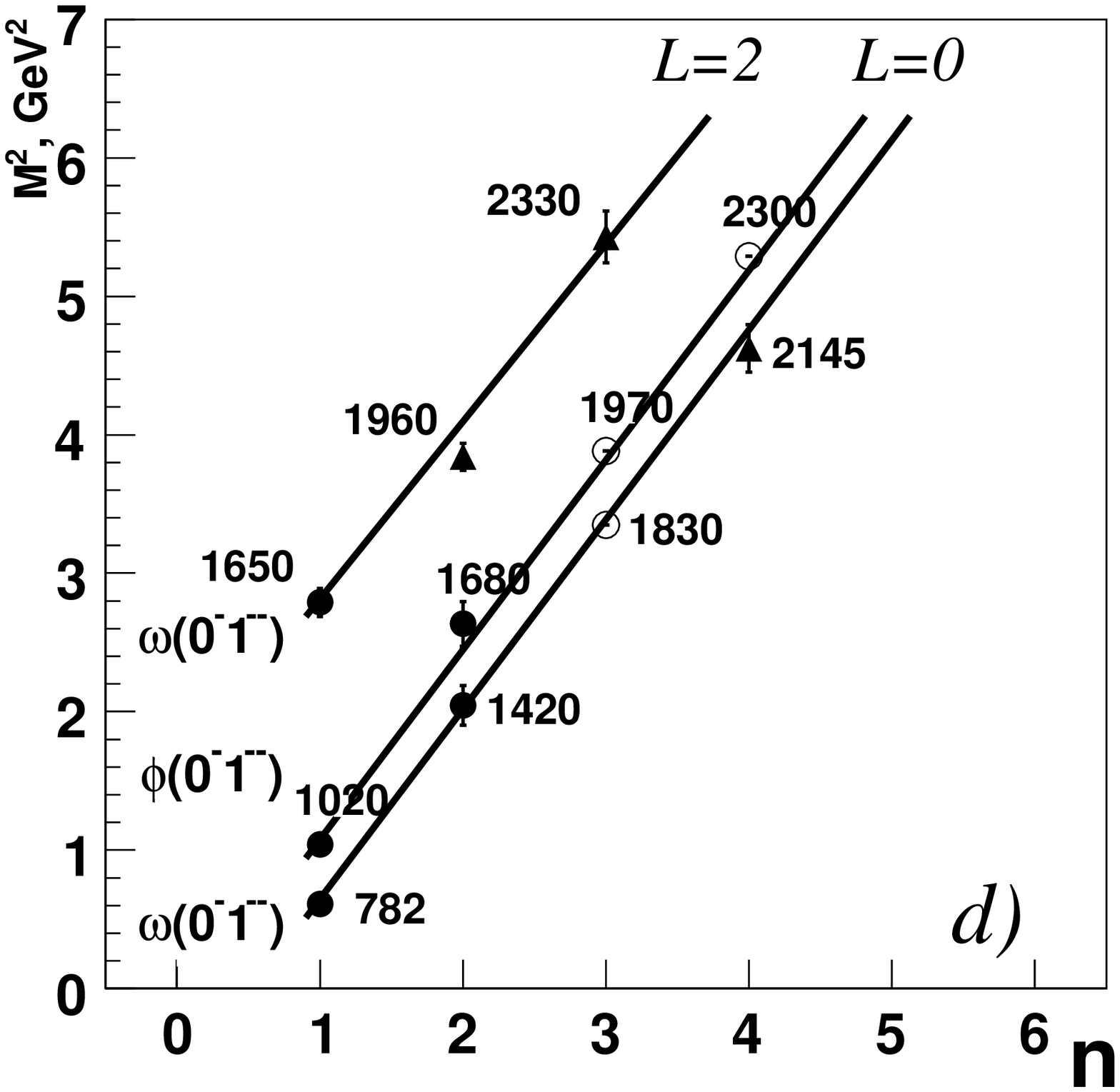}\vspace{5mm}\\
\hspace{1mm}\includegraphics[width=0.23\textwidth,height=0.23\textwidth]{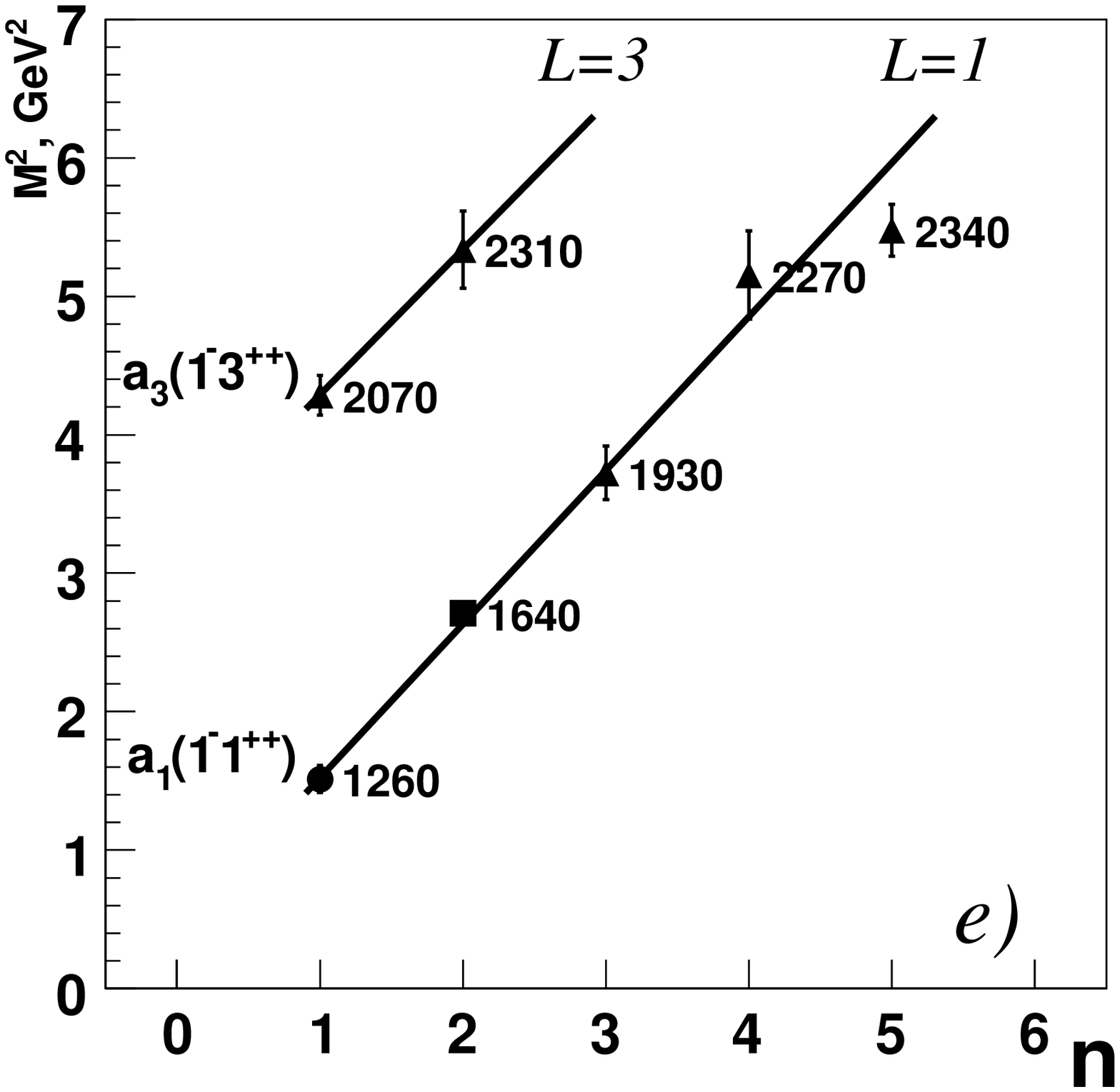}&
\hspace{1mm}\includegraphics[width=0.23\textwidth,height=0.23\textwidth]{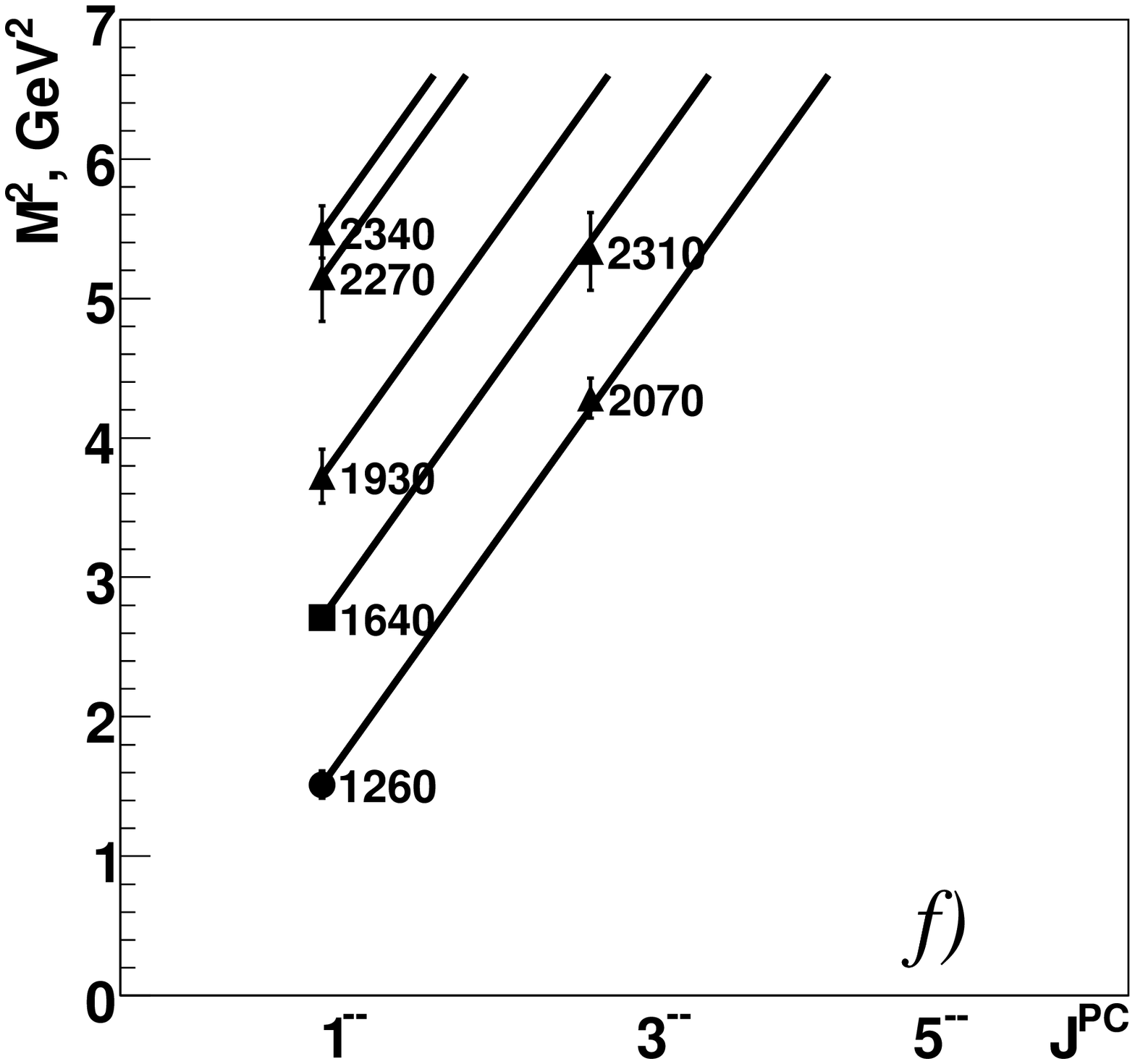}&
\hspace{1mm}\includegraphics[width=0.23\textwidth,height=0.23\textwidth]{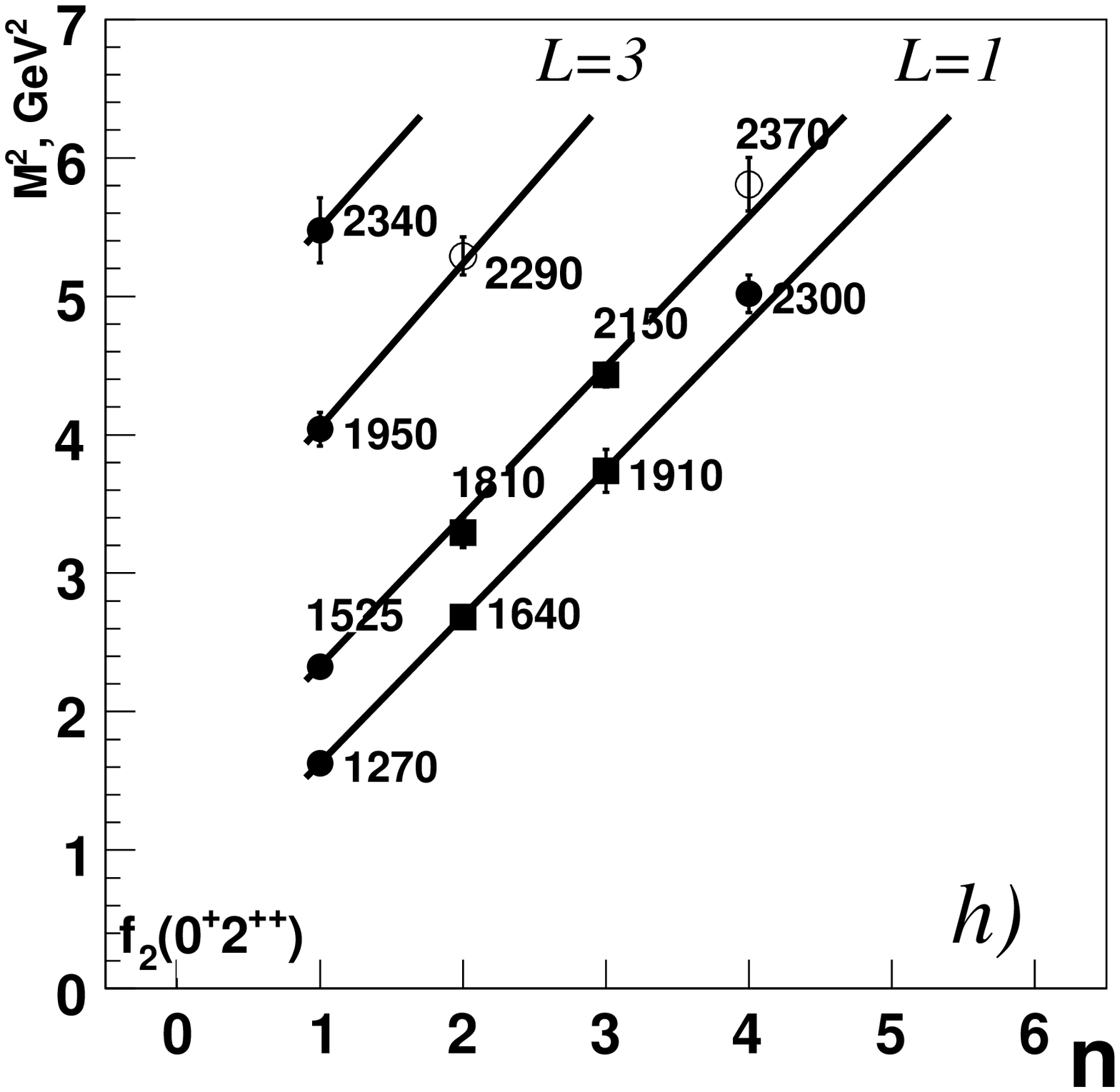}&
\hspace{1mm}\includegraphics[width=0.23\textwidth,height=0.23\textwidth]{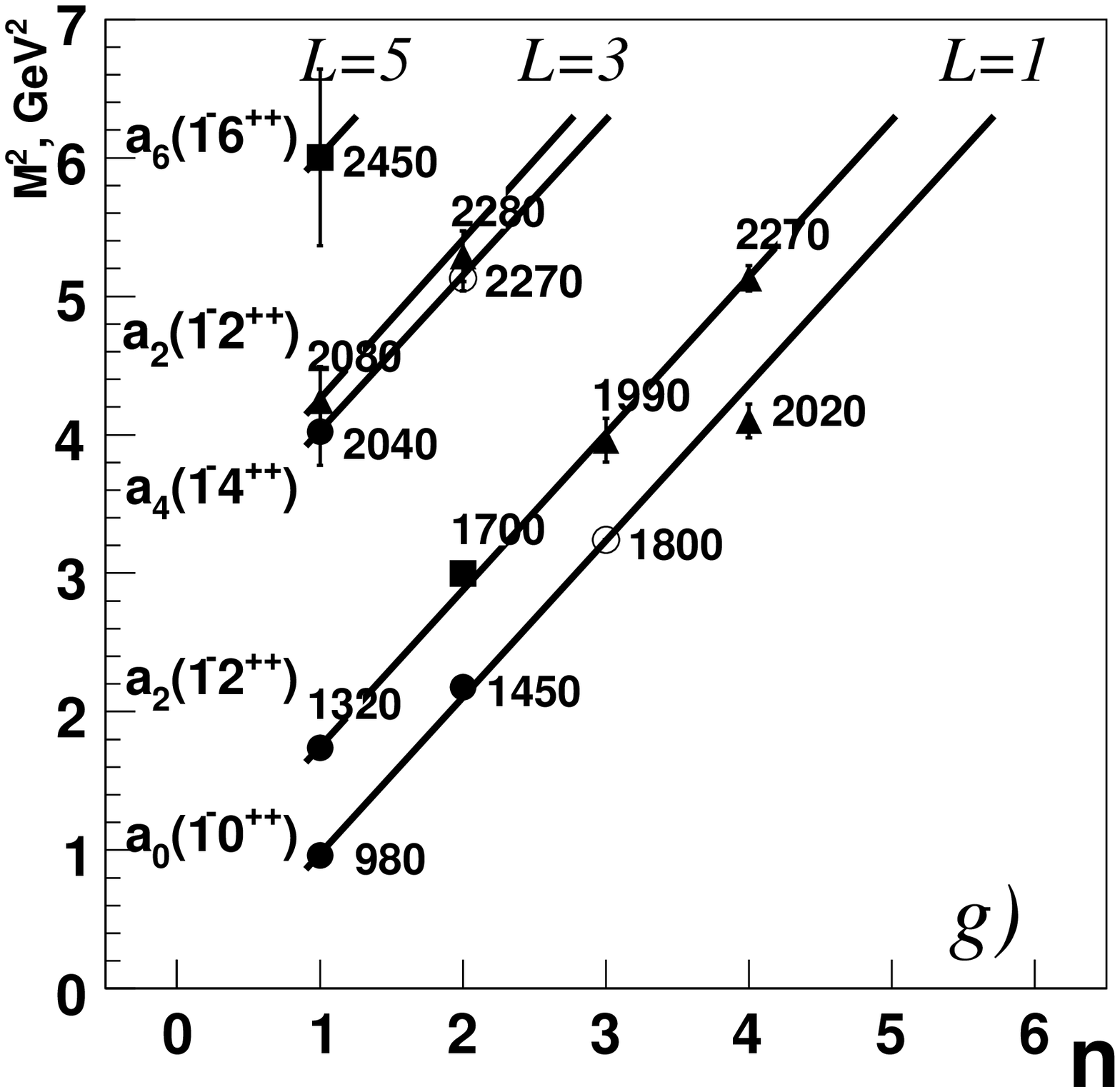}\vspace{5mm}\\
\end{tabular}\ec
\caption{\label{AAS_planes}
Squared meson masses as functions of $n$ and $J$. Particles from the
PDG {\it Summary Table} are represented by a {\Large$\bullet$},
those {\it omitted from the Summary Table} by a $\small\blacksquare$.
$\Large\blacktriangle$ represents data given by Anisovich {\it et al.}
\cite{Anisovich:2000kx} and listed as {\it Further States} by the PDG,
open circles predicted by their classification. a) $(n,
M^2)$-trajectories of $\rho(1^{--})$ $^3S_1$ and $^3D_1$ mesons as
functions of $n$; b) $\rho(3^{--})$ $^3D_3$ and $^3G_3$ trajectories as
functions of $n$; c) Regge trajectories for $\rho(770) (1^3S_1)$,
$\rho(1450) (2^3S_1)$, $\rho(1700) 1(^3D_1)$, $\rho(1900) (3^3S_1)$,
$\rho(2000) 2(^3D_1)$, $\rho(2150) (4^3S_1)$, $\rho(2265) 1(^3D_1)$;
$\omega(1^{--})$ and $\phi(1^{--})$ $^3S_1$ and $\omega(1^{--})$
$^3D_1$ trajectories as functions of $n$. e)
 $(n, M^2)$-trajectories of $a_1(1^{++})$ $^3P_1$ and $a_3(3^{++})$
$^3F_3$ mesons and f) the corresponding Regge trajectories; g,h)  $(n,
M^2)$-trajectories for $a_{2J}$ and $f_{2J}$ mesons with even spin.
}
\end{figure}

The results of the Gatchina group on scalar and tensor mesons deserve a
deeper discussion (see sections  \ref{JPC=2++} and \ref{The
Anisovich-Sarantsev picture}, respectively). Here we concentrate on
general issues from a wider perspective. A.V. Anisovich, V.V.
Anisovich, and Sarantsev \cite{Anisovich:2000kx} have assigned orbital
and radial quantum numbers to mesonic resonances they have found in the
fits described here, in the data on $p\bar p$ annihilation in flight
described in section \ref{Patterns at high l, n}, and to resonances
reported elsewhere \cite{Eidelman:2004wy}, and suggested a vision how
to interpret the spectrum. In the mass region up to $M<2400$ MeV/c$^2$ the
masses of mesons are plotted in $(n,M^2)$ and $(J,M^2)$ planes, thus
suggesting their classification in terms of $n\; ^{2S+1}L_J \; q\bar q$
states. All trajectories are linear, with nearly the same slopes as
expected from the relation $M_n^2(l) = 2\pi\sigma\left(l+n+\frac{1}{2}
\right)$. The Figs. \ref{AAS_planes}a-h characterise meson resonances
by their mass; their production and decay pattern is not used to check
if the interpretation based on their masses is consistent. Nevertheless,
the figures can serve as guides to identify mesons beyond the $q\bar q$
scheme. Further plots are shown in the section on pseudoscalar (Fig.
\ref{psaas}) and scalar mesons (Fig. \ref{scalarplot}).

The strength of such a unified picture is evident and in most cases,
the scheme provides a viable classification. Obviously, there is one
missing state in the series of $\rho$ radial excitations consisting of
$\rho(770)$, $\rho(1450)$, $\rho(2150)$; the missing state is predicted
to have a mass of about 1830\,MeV/c$^2$. This is a welcome hint for further
searches. The $\rho(1700)$ is the leading resonance of the $^3D_1$
series of states in which the intrinsic orbital angular momentum $l=2$
makes a large contribution; a small admixture of $^3S_1$ is of course
not excluded. Based on the mass spectrum depicted in Fig.
\ref{AAS_planes}a, it is unlikely that a hybrid vector meson is part of
the game. A conjecture based on an unexpected decay pattern of
$\rho(1450)$ will be discussed below. It is proposed that a
hybrid vector meson may hide in the spectrum since the observed decay
modes are inconsistent with a pure $q\bar q$ interpretation of
$\rho(1450)$. This is a weak point of the classifications suggested
in Fig. \ref{AAS_planes}: the decay modes which provide important
information on the flavour structure of mesons were not considered.
There is thus a potential danger in this approach: it is unclear if the
scheme survives when one of the resonances disappears or shows decay
modes incompatible with the assignment based on its mass. There are a
few further examples where one has to be cautious:

In section \ref{Pseudoscalar mesons} on pseudoscalar mesons it will
be shown that $\eta(1295)$ cannot be a $q\bar q$ meson. How stable is
then the classification against taken this meson out? The $\eta_2(1870)$
resonance is produced and decays rather as $n\bar n$ and not as $s\bar
s$ state; that is the reason to discuss this state below as hybrid
candidate. In Fig. \ref{psaas} it is nevertheless interpreted as $s\bar
s$ partner of $\eta_2(1645)$. In Fig. \ref{scalarplot} the $a_0(980)$
is needed as lowest mass scalar isovector state. Likewise, the Gatchina
interpretation requires $f_0(980)$ to be a $q\bar q$ state. Both these
mesons are often interpreted as non-$q\bar q$ objects. The
$\sigma(485)$ is tentatively interpreted as additional state beyond
$q\bar q$ spectroscopy generated by an amplitude singularity related to
confinement, its twin brother $\kappa(700)$ is not considered. In the
Gatchina scheme, $a_0(980)$ and $f_0(980)$, $\sigma(485)$, and
$\kappa(700)$ are three unrelated phenomena. The $f_0(1300)$ with
$\rho\rho$ as most prominent decay mode is interpreted in Fig.
\ref{scalarplot} as $s\bar s$ state. The plots make very detailed
predictions, that is of course its strength. But details may be at
variance with new knowledge or interpretations. We will come back to
some of these questions in the sections on pseudoscalar mesons and on
the interpretation of scalar mesons.

\subsection{\label{Threshold dynamics}
Threshold dynamics}

The importance of $S$-wave thresholds has often been underlined. Here we
remind of some recent work
\cite{Tornqvist:2003na,Bugg:2004rk,vanBeveren:2006ua,Rosner:2006vc}.
The best studied example is of course the $K\bar K$ threshold which
gave rise to many interpretational controversies. At BES II and BELLE,
several threshold enhancements in baryon-antibaryon systems were
observed which will be discussed briefly in section
\ref{Baryon-antibaryon threshold enhancements}. At the $p\bar p$
threshold, a dip is observed in $e^+e^-\to 6\pi$ which points to an
interpretation within $N\bar N$ interaction physics. $0^{-+}$ quantum
numbers are suggested for the $p\bar p$ threshold enhancement, the dip
in $e^+e^-\to 6\pi$ is $1^{--}$, so it is unclear if the two
observations are related. A threshold bump at $m=2.175$ GeV/c$^2$ has
been observed at BaBaR \cite{Aubert:2006bu} in the radiative-return
reaction $e^+ e^- \to \phi f_0(980)$. Due to its production and decay,
it might be a $\phi$ radial excitation; $\Upsilon(nS)$ and $\psi(nS)$
have a high chance to decay into their respective ground state plus
$\pi\pi$ in the $S$-wave, too. A $\pi_2(1870)$ was observed at the
threshold of its $a_2(1320)\eta$ decay in diffractive scattering at VES
\cite{Amelin:1995fg}. In section \ref{Heavy--quark spectroscopy} we
have discussed several further states which are found close to their
respective decay channels. We mention $X(3872)$, $Y(3940)$,
$D_{s0}(2317)$, and $D_{s1}(2460)$. The three states $X(1812)$,
$\phi(2175)$, and $Y(3940)$ have unusual decay modes; we will suggest
strong tetraquark components for $X(1812)$ ($=f_0(1760)$) and $Y(3940)$
($=\chi_{c0}(2P)$) to explain their anomalous decay.

In our view, thresholds have a significant impact on the precise mass
and width of resonances. Thresholds can generates cusps, attract poles,
but -- as a rule -- they do not generate new poles.

\subsection{\label{Hybrids candidates}
Hybrids candidates}

There are a few cases where experimental findings disagree with
expectations based on the quark model. These states could be glueballs,
hybrids, or multiquark resonances. The most suspicious cases are found
in the scalar and pseudoscalar mass spectra; these will be discussed in
separate sections. Here we report on evidence for further mesons which
resist a straightforward interpretation as $q\bar q$ mesons.

\subsubsection{\label{JPC=1{--}}
$J^{PC}=1^{--}$}

First evidence for the existence of a higher-mass vector meson was
found in bubble chamber data on $p\bar p \to \omega\pi^+\pi^-$
\cite{Frenkiel:1973un,Ballam:1973rd} at a mass of about 1250\,MeV/c$^2$.
Later, the joint analysis of data from various experiments on ${\rm
e}^+{\rm e}^-$ annihilation and photoproduction of final-states
containing two or four pions proved the existence of $\rho(1450)$
and $\rho(1700)$ \cite{Donnachie:1986pk,Clegg:1989mp}. In the isoscalar sector,
three states are known,  $\omega(1420), \omega(1650)$
\cite{Cordier:1981zs,Donnachie:1988ws}, and $\phi(1680)$
\cite{Bisello:1991kd}. The $\omega$ resonances are readily assigned to
two states expected in the quark model, the first radial excitation
$2^3S_1$ and the first $l=2$ orbital excitation $1^3D_1$, Fig.
\ref{vector-ds} in section \ref{Further mesons with open charm}. Two
$\phi$ states are, of course, expected but only one was reported so far.

In the reaction J/$\psi \to K^+K^-\pi^0$, the BES collaboration found a
broad enhancement in the $K^+K^-$ invariant mass distribution. A
partial wave analysis shows that the structure has $J^{PC}=1^{--}$.
Since it is recoiling against a pion, its quantum numbers are thus like
those of the $\rho$. When fitted with a Breit-Wigner amplitude, a pole
was found at the position $(1576^{+49}_{-55}(stat)^{+98}_{-91}(syst))
{\rm MeV/c^2} - i(409^{+11}_{-12}(stat)^{+32}_{-67}(syst)) {\rm
MeV/c^2}$ \cite{Ablikim:2006hp}. This is an important finding: the
$K^+K^-$ enhancement is not easily assigned to any quark model state.
It shows that there are background amplitudes which could be
wide tetraquark resonances or generated by meson-meson interaction
dynamics via $t$-channel exchanges. Of course, it is very difficult to
prove a resonant character of such a wide background, extending over
more than 1\,GeV/c$^2$. Possibly, a slow background phase motion due to
molecular interactions may as well be compatible with the data.

This classification of the states $\rho(1450)$ and $\rho(1700)$ meets
with some problems \cite{Donnachie:1999re}: \par - The experimental
ratio of the $e^+e^-$ width of the $\rho(1700)$ to that of the
$\rho(1450)$ is too large in comparison with calculations in
non-relativistic limit where this ratio is zero. \\ - The decay widths
of these states do not follow predictions of the $^3P_0$ model which
often gave realistic estimates for hadronic partial widths. According
to model predictions, the $4\pi$ decays of $\rho_{2S}$ are small and
$\rho_{1D}$ should decay into $h_1 \pi$ and $a_1 \pi$ with large and
nearly equal probability. From this we would expect $\sigma(e^+e^- \to
\pi^0 \pi^0 \pi^+ \pi^-) > \sigma(e^+e^- \to \pi^+ \pi^- \pi^+ \pi^-)$,
in contradiction with observations. Yet, the flux tube model --
assuming  $\rho(1450)$ to be a hybrid -- gave no reasonable agreement
with data neither.

The assumption of an additional hybrid $\rho_H$ state helps to solve
these problems \cite{Donnachie:1999re}. This hypothetical state is
expected to decay mainly into $a_1 \pi$ and not into $h_1 \pi$. Thus
the ratio $(\pi^0 \pi^0 \pi^+ \pi^-)$/ $(\pi^+ \pi^- \pi^+ \pi^-)$ moves
into the right direction. Mixing of the two $q\bar q$ states with
$\rho_H$ can reproduce the observed decay pattern.

The hybrid state needed for this scheme has to be relatively light, $M
\approx 1.6$ GeV/c$^2$, otherwise it does not mix sufficiently with the
conventional $\rho$ state. Such a low mass of a vector hybrid imposes
severe restrictions on the mass scale of other light hybrids due to
predicted mass ordering $0^{-+} < 1^{-+} < 1^{--} < 2^{-+}$ which is
valid in different models \cite{Isgur:1985vy}. The missing state of
this trio could be $\rho(1900)$ seen in $6\pi$ final state
\cite{Frabetti:2003pw}, or a low mass state $\rho(1200)$
\cite{Amsler:2004kn}.

Recently, the BES collaboration has reported an extremely broad vector
state with $(1576^{+49}_{-55}(stat)^{+98}_{-91}(syst))$\,MeV/$c^2$ mass
and $(818^{+22}_{-24}(stat)^{+64}_{-134}(syst))$\,MeV/$c^2$ width
\cite{Ablikim:2006hp}. Due to its production in J/$\psi$ decays into
$\pi\,(K\bar K)$, its isospin is likely $I=1$. No attempt has been made
so far to see if this state can cure these problems. In
\cite{Liu:2007qi} it was shown that interference effects due to final
state interactions could produce an enhancement around 1540 MeV in the
$K^+K^-$ spectrum but the expected intensity is smaller than the
observed one. Interpretations are given by a number of authors
\cite{Guo:2006hd,Wang:2006gj,Ding:2006vk,Karliner:2006xc} within
tetraquark scenarios.

\subsubsection{\label{JPC=1{++}}
$J^{PC}=1^{++}$}

The flux tube model predictions
\cite{Close:1994hc,Barnes:1995hc,Barnes:1996ff,Page:1998gz} for the
decay widths of $J^{PC}=1^{++}~~2P$ state and of $1^{++}$ hybrid are
collected in the Table \ref{tab:1++}.
\begin{table}[h] \begin{center}
\caption{\label{tab:1++}
Partial widths (in MeV/c$^2$) of $2P (q \bar q)$ and hybrid
$a_1(1700)$\vspace{2mm}} \renewcommand{\arraystretch}{1.5}
\begin{tabular}{ccccccccc}
\hline\hline
 & $\rho \pi$ & $\rho \omega$ & $\rho(1470) \pi$& $b_1 \pi$ &$f_1 \pi$& $ f_2 \pi$ & $K^* K$ \\ \hline
$2P$ & 57 & 15 & 41 & 0 & 18 & 39 & $18 \div 30$ \\
$H$ & $28 \div 40$ & 0 & $50 \div 70$ & 0 &$4 \div 60$&$2 \div 75$&$18 \div
24$\\
\hline\hline
\end{tabular}
\renewcommand{\arraystretch}{1.5}
\end{center}
\end{table}

The most selective decay mode is $\rho\omega$ which is supposed to be
suppressed for a hybrid and strong for a $2P$ $q\bar q$ state. The two
alternatives differ also by their $D/S$ ratio in $a_1(1700)\to\rho\pi$
decay. For a $a_1$ hybrid, $S$-wave dominance is expected, for a
2D-state, the $D$-wave should exceed the $S$-wave.

\begin{figure}[htb] \begin{center}
\includegraphics[width=.4\textwidth,height=.35\textwidth]{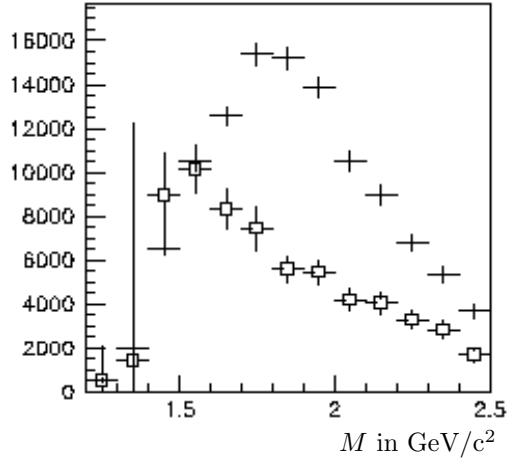}\vspace{-7mm}\\
\phantom{r}\hspace{4cm}{\footnotesize $M$ in GeV/c$^2$}\vspace{-3mm}
\end{center}
\caption{\label{fig:1++}
The intensities of $J^{PC}=1^{++}$ waves in the reaction $\pi^- N \to
\omega \pi^- \pi^0 N$: crosses - sum of all $J^{PC}M^{\eta}=1^{++}0^+$
waves, squares - sum of two waves \cite{Amelin:1999gk}. See text for
further discussion.}
\end{figure}

Experiments exhibit a strong $J^{PC}=1^{++}$ signal at $M \approx
1.7$\,GeV/c$^2$ in $\rho \pi$ \cite{Amelin:1995gu,Chung:2002pu}, $f_1 \pi$
\cite{Ryabchikov:1995js,Kuhn:2004en} and $\omega \rho$
\cite{Amelin:1999gk}. The shapes of these signals are different, the
visible width varies from $\Gamma \approx 300$ MeV/c$^2$ in the $\rho
\pi$ channel to $\Gamma \approx 600$ MeV/c$^2$ in $\omega \rho$. If we
assume that the source of these signals is one resonance, then the
decay pattern is consistent with a conventional $2P$ meson.
The $D/S$ ratio of the $\rho \pi$ channel ($D>S$) supports the $2P$
interpretation. The $b_1 \pi$ channels is predicted to be suppressed
for both, the quark-model $a_1(1700)(2P)$ and the hybrid $a_1(1700)(H)$.
This suppression is observed in data; this may
serve as indication that the models give a realistic estimate of decay
widths. The data are presented in Fig. \ref{fig:1++}. Shown are the sum
of all partial wave contributions,
$S_1(\rho\omega)+S_1(\rho_{1450}\omega), P_1(b_1\pi), D_1(\rho\omega),
D_2(\rho\omega)$, and the sum of the $S_1(\rho\omega)$ and
$P_1(b_1\pi)$ waves.  The latter two waves obviously do not contribute
to the $a_1(1700)$ signal. The letter $S,P,D$ denotes the orbital
angular momentum, the subscript the total spin.

The absence of a hybrid meson signal in $J^{PC}=1^{++}$ channel
poses a real problem. According to \cite{Barnes:1996ff} this meson
should be narrow $\Gamma \approx 100$ MeV/c$^2$ and have $\rho \pi$ as
one of its dominant decays. Thus, the signal is difficult to hide. For
example, it has to be seen in the charge exchange reaction $\pi^- p \to
\pi^+ \pi^- \pi^0 n$ where experiment does not show any trace of a
$1^{++}$ resonance in the $\approx 1.6 \div 2.0$\,GeV/c$^2$ mass region.

\subsubsection{\label{JPC=2++}
$J^{PC}=2^{++}$ and search for the tensor glueball}

The Particle Data Group \cite{Eidelman:2004wy} lists 14 isoscalar
tensor mesons with masses below 2.4\,GeV/c$^2$ and considers 6 of them
as established, 6 states are omitted from the Summary Table, and 2 are
listed as Further States. In this mass range, 12 quark model states are
expected (see Fig. \ref{fig:globalview}). It would certainly be too
naive to claim on this basis that there is abundance of tensor mesons.
The problem is of course that the results come from a variety of
different experiments and analyses. Some measurements could be wrong;
more trivially, some states could be listed twice. The two candidates
$f_2(1565)$ and $f_2(1640)$ are, e.g., very likely one state; the mass
shift could be a measurement error or could be caused by the
$\omega\omega$ threshold. The only realistic chance to bring order into
this mess are analyses which describe all most sensitive data with one
set of amplitudes.

Such analyses have been performed by the Gatchina group; their results
on tensor states were reported in \cite{Anisovich:2005xd,%
Anisovich:2005iv,Anisovich:2006ma}. Data on two-photon fusion into
$\pi^+\pi^-\pi^0$ \cite{Shchegelsky:2006es} and into $K^0_SK^0_S$
\cite{Shchegelsky:2006et} from L3 at LEP (see section \ref{The L3
experiment}) provided information on the two lowest-mass tensor nonets.
The results are reproduced in Table \ref{tab:ksks}. For the
ground-state nonet, the $K^*_2(1430)$ serves as a solid anchor for the
mass scale; there is no $K^*_2$ known which could play this $\rm
r\hat{o}le$ for the second nonet; an additional $K^*_2$ should hence be
expected at about 1650\,MeV/c$^2$. Both nonets are found to be nearly
ideally mixed; at the first glance, this is surprising given the large
mass difference between $a_2(1730)$ and $f_2(1560)$ which are
interpreted as isospin partners. The mixing angle is however determined
from the $f_2(1560)$ and $f_2(1750)$ couplings to $K^0_SK^0_S$ and not
from the GMO formula.

\begin{table}[ph]
\caption{\label{tab:ksks}
Parameters of resonances observed in the reaction $\gamma\gamma\to
K^0_SK^0_S$. The values marked with stars are fixed by using other
data.\vspace{2mm}} \begin{center} \renewcommand{\arraystretch}{1.4}
\begin{tabular}{cccccccc}
  \hline\hline
  ~ & \multicolumn{3}{c}{First nonet} &\quad&
  \multicolumn{3}{c}{Second nonet} \\
  \hline
  ~ & $a_2(1320)$ & $f_2(1270)$ & $f'_2(1525)$ &
         & $a_2(1730)$ & $f_2(1560)$ & $f_2(1750)$ \\
  \hline
 Mass (MeV) & $1304\pm 10$ & $1277\pm 6$ & $1523\pm 5$ &
            & $1730^*$ & $1570^*$ & $1755\pm 10$ \\
\hline Width (MeV) & $120\pm 15$ & $195\pm 15$ & $104\pm 10$ &
            & $340^*$ & $160^*$ & $67\pm 12$ \\
\hline
 $g^L$ (GeV) & $0.8\pm 0.1$ & $0.9\pm 0.1$ & $1.05\pm0.1$ &
              & \multicolumn{3}{c}{$0.38\pm 0.05$} \\
\hline
$\varphi$ (deg) & \multicolumn{3}{c}{$-1\pm 3$} &
              & \multicolumn{3}{c}{$-10^{+5}_{-10}$} \\
\hline\hline
\end{tabular}
\renewcommand{\arraystretch}{1.0}
\end{center}
\end{table}

For the higher-mass region, the authors use the Crystal Barrel data on
$\bar pp$ in flight discussed in section \ref{Patterns at high l, n}
and data on  $\bar p p\to \pi^+\pi^-$ data obtained with a polarised
target \cite{Eisenhandler:1975kx}. Five tensor states were required to
describe the data, $f_2(1920)$, $f_2(2000)$, $f_2(2020)$, $f_2(2240)$,
$f_2(2300) $:

\renewcommand{\arraystretch}{1.4}
\bc
\begin{tabular}{ccc}
 Resonance &  Mass (MeV/c$^2$) & Width (MeV/c$^2$) \\
$f_2(1920) $&$ 1920\pm 30 $&$ 230\pm 40$ \nonumber \\
$f_2(2000) $&$ 2010\pm 30 $&$ 495\pm 35$ \nonumber \\
$f_2(2020) $&$ 2020\pm 30 $&$ 275\pm 35$ \label{tensor-as}\\
$f_2(2240) $&$ 2240\pm 40 $&$ 245\pm 45$ \nonumber \\
$f_2(2300) $&$ 2300\pm 35 $&$ 290\pm 50$\, \nonumber .
\end{tabular}
\ec\renewcommand{\arraystretch}{1.0}

Antiproton-proton annihilation couples primarily to the $n\bar n$
states. Therefore, the three states observed at BNL in the reaction
$\pi^-p\to\phi\phi n$ by Etkin {\it et al.}
\cite{Etkin:1982bw,Etkin:1987rj} were included as additional states in
their discussion. In a recent reanalysis of the BNL data
\cite{Longacre:2004jn}, new mass values were given; due to their decay
into $\phi\phi$, the three states $f_0(2120)$, $f_0(2340)$, and
$f_0(2410)$ should be discussed as $s\bar s$ states or as glueballs.
The group advocated since long that a tensor glueball must be present
in this mass range since the $\phi\phi$ final state can be reached only
via gluonic intermediate states and the  $\phi\phi$ yield was found to
be unexpectedly large, incompatible with the hypothesis that it might
be due to $\omega-\phi$ mixing.

The pair of states $f_2(1920)$ + $f_2(2140)$ are supposed to be
contained in the $3^3P_2$ nonet, $f_2(2240)$ + $f_2(2410)$ in the
$3^3P_2$ nonet, and $f_2(2020)$ + $f_2(2340)$ in the $1^3F_2$ nonet.
The $f_2(2300)$ resonance is interpreted as $2^3F_2$ quark-model state.
Its $s\bar s$ partner remains to be discovered. Thus all states are
mapped onto quark-model expectations, except for one state at
2\,GeV/c$^2$. It has a much larger width and its couplings to final
states are consistent with the assumption that it is flavour blind.
Hence the data are compatible with the presence of a tensor glueball at
2\,GeV/c$^2$. Its mass is unexpectedly low, from lattice calculation
2.4\,GeV/c$^2$ would be expected, or 1.6 times the scalar glueball
mass. Certainly, more work is needed to solidify this interesting
conjecture.

\subsubsection{\label{JPC=2{-+}}
$J^{PC}=2^{-+}$}

These quantum numbers are probably most convenient ones to search of
non-$q \bar q$ states. The quark model $q\bar q$ states with
$J^{PC}=2^{-+}$ are relatively narrow and not distorted by mixing of
$J=L+1$ and $J=L-1$ states. And the flux tube model predicts hybrids
with these quantum numbers to be narrow as well
\cite{Close:1994hc,Barnes:1995hc,Page:1998gz,Barnes:1996ff}. The
dominant (predicted) decay modes, $\pi_2(H) \to \rho \pi$ and $\pi_2(H)
\to f_2 \pi$, are experimentally well accessible.

There are several experimental indications suggesting the existence of
$\pi_2(1880)$. A first signal was observed by the ACCMOR collaboration
\cite{Daum:1979ix} in the reaction $\pi^- p \to \pi^+ \pi^- \pi^- p$.
A resonance-like signal was reported in the $2^-D_0^+(f_2 \pi)$ wave.
The signal itself was confirmed by VES \cite{Amelin:1995gu}. However,
the signal could be described as well by interference of $\pi_2(1680)$
with a broad state $\pi_2(2100)$. Further evidence for the $\pi_2(1880)$
resonance, in its decay to $a_2 \eta$, was reported in $\pi^-  N \to
\eta \eta \pi^- N$ \cite{Amelin:1995fg} and in $p\bar p$ annihilation
in flight into $\eta \eta \pi^0 \pi^0$ \cite{Anisovich:2001hj}.
The reaction $\pi^{-} p\rightarrow\pi^{+}\pi^{-}\pi^{-}\pi^{0}\pi^{0}p$
was analysed at BNL \cite{Lu:2004yn}. Three isovector $2^{-+}$ states
were seen in the $\omega\rho^{-}$ decay channel, the well known
$\pi_{2}(1670)$ and evidence for two further states, $\pi_{2}(1880)$
and $\pi_{2}(1970)$. The two resonances $\pi_{2}(1670)$ and
$\pi_{2}(1970)$ match perfectly to the pattern displayed in Fig.
\ref{fig:globalview}, the $\pi_{2}(1880)$ does not. In the reaction
$\pi^{-} p\rightarrow f_1(1285)\pi^-p$, $\pi_{2}(1670)$ and
$\pi_{2}(1970)$ were found but not $\pi_{2}(1880)$ \cite{Kuhn:2004en}.
Thus evidence for the existence of this state is very suggestive but
not forcing. If its existence is assumed, it is a viable hybrid
candidate.

Similar observations have been made in the $\eta_2$ partial wave. Apart
from the well established $\eta_2(1645)$, a high-mass state is reported
at $M=2030\pm5\pm15$\,MeV/c$^2$, $\Gamma=205\pm10\pm15$\,MeV/c$^2$
\cite{Anisovich:2000mv}. According to the systematic of Fig.
\ref{fig:globalview}, these two mesons well suited to represent the
$1^1D_2$ and $2^1D_2$ quark model states. The third state,
$\eta_2(1870)$, falls in between the  two quark model states and is a
natural isoscalar partner of $\pi_2(1880)$. In the isoscalar sector,
two $1^1D_2$ states are of course expected but the $\eta_2(1870)$
production modes, central production and $p\bar p$ annihilation, makes
an $s\bar s$ interpretation of $\eta_2(1870)$ unlikely.

The  $\eta_2(1870)$ in central production is observed in its
$a_2(1320)\pi$, $a_0(980)\pi$ and $f_2(1270)\eta$ decay mode in the
$\eta\pi\pi$ final state \cite{Barberis:1999be}, and in central
production of four pions in the $a_2(1320)\pi$ isobar
\cite{Barberis:1999wn}. In $p\bar p$ annihilation it was observed just
above threshold in $f_2(1270)\eta$ with $L = 0$ \cite{Adomeit:1996nr},
and in its $a_2(1320)\pi$ decay mode \cite{Anisovich:2000mv}. The
latter analysis determines the $a_2(1320)\pi /f_2(1270)\eta$ decay mode
ratio to $1.27\pm 0.17$, presumably not incompatible with the ratio 4,
predicted by Barnes, Close, Page, and Swanson \cite{Barnes:1996ff}
and by Page, Swanson, and  Szczepaniak \cite{Page:1998gz}.

These two resonances, $\pi_2(1880)$ and $\eta_2(1870)$, are the best
evidence for the existence of hybrids with non-exotic quantum numbers.
If confirmed, they change our view of hadron spectroscopy. On the other
hand, if hybrids really exist as independent identities, many more of
them should be expected and it is a miracle that we have evidence for
some 150 mesons, and all but two are $q\bar q$ compatible. For this
reason, we still maintain our general suspicion that hybrids, like
tetraquark states, do not show up as separate particles but are rather
part of the hadronic wave function. Observable resonances may need a
$q\bar q$ component which provides additional attractive forces. A way
to scrutinise this conjecture is the search for resonances in exotic
partial waves where $q\bar q$ components are forbidden by conservation
laws.


\markboth{\sl Meson spectroscopy} {\sl Exotic mesons}
\clearpage\setcounter{equation}{0}\section{\label{Mesons with exotic quantum numbers}
Mesons with exotic quantum numbers}
\subsection{Model predictions for exotic mesons}
\subsubsection{Hybrid mesons}
Mesons with exotic quantum numbers are non-\qqb\ objects. They may be
hybrid mesons ($q\bar qg$), multiquark states ($q\bar qq\bar q ...$)
or multimeson states ($M_1~M_2 ...$). Exotic quantum numbers are, e.g.
\be J^{PC}_{exotics}= 0^{--},~~ 0^{+-},~~1^{-+},~~2^{+-},~~3^{-+} .....
\ee The existence of hybrid mesons was suggested in 1976 by Jaffe and
Johnson \cite{Jaffe:1975fd} and Vainsthein and Okun
\cite{Vainshtein:1976ke}. Often, the symbolic notation $q \bar q g$ is
used which reminds of the excitation of the gluon field between the
$q\bar q$ pair. The spectrum of hybrid mesons was calculated in
different models. A collection of results is presented in Table
\ref{predict}. \begin{table} [H] \caption{\label{predict} Prediction of
various models for the mass of the lightest hybrid meson.\vspace{2mm}}
\begin{center}
\renewcommand{\arraystretch}{1.4}
\begin{tabular}{ccc}
\hline\hline Model & Mass, Gev/c$^2$ & References \\
\hline Bag model
& $ 1.3\div 1.4$ & \cite{Chanowitz:1982qj}   \\ Flux-tube model & $
1.8\div$ 2.0    &
\cite{Isgur:1985vy,Close:1994hc,Barnes:1995hc,Page:1998gz} \\ Sum rules
& $1.3 \div $ 1.9   &
\cite{Balitsky:1982ps,Latorre:1985tg,Narison:1999hg}   \\ Lattice QCD&
$1.8 \div 2.3$ & \cite{Bernard:2003jd}\\ Effective Hamiltonian & $2.0
\div 2.2$ & \cite{Cotanch:2001mc}\\ \hline\hline
\end{tabular} \renewcommand{\arraystretch}{1.0} \end{center}
\end{table}
From the table we may conclude that the lightest $I^G J^{PC} = 1^-
1^{-+}$ state should have a mass in the $1.7 - 2.2$ GeV/c$^2$ region
even though smaller values are not ruled out. Early estimates assumed
$\eta(1410)$ to be a pseudoscalar hybrid \cite{Chanowitz:1982qj}, and
suggested low hybrid masses. For exotic states with quantum numbers of
$I^G J^{PC} = 1^- 1^{-+}$, the conventional name is now $\pi_1$.
Earlier publications called it $\hat\rho$ or $M$. The mass of
the lightest hybrids with nonexotic quantum numbers $J^{PC}=0^{-+}$
is predicted in the same region \cite{Isgur:1985vy}.\\

In recent lattice calculations \cite{Hedditch:2005zf}, the mass of the
$1^{-+}$ exotic meson -- created with hybrid interpolating fields -- was
explored with light-quark masses approaching 25 MeV/c$^2$ ($m_\pi / m_\rho
\simeq 1/3$). The results indicate that the $1^{-+}$ exotic partial wave
exhibits significant curvature close to the chiral limit, suggesting
that previous linear extrapolations, far from the chiral regime, may
have overestimated the mass of the $1^{-+}$. It was found that the
$1^{-+}$ mass can be as low as  $M_{\pi_1} \approx 1600$ MeV/c$^2$. In line
with this observation, Thomas and Szczepaniak\cite{Thomas:2001gu}
argued that chiral extrapolations may be difficult for exotics and that
small quark masses are required to get reliable results.

The $\pi_{1}$ partial widths for decays to $\rho\pi$, $b_1\pi$,
$f_1\pi$, $\eta'\pi$ and $\eta\pi$ were calculated in different models
with very different results. In the potential quark model with a
constituent gluon \cite{Tanimoto:1982eh,Iddir:1988jd}, as well as in
the flux tube model \cite{Isgur:1985vy,Close:1994hc,Page:1998gz}, the
decays of a hybrid meson into a pair of mesons having identical spatial
wave functions are forbidden; hence decays into one $S$-wave and one
P-wave meson are favoured. This selection rule leads to the
suppression of the decay $\pi_{1} \to \rho \pi$ due to the assumed
similarity of the wave functions for $\rho$ and $\pi$ mesons. The
predicted width is in the region of a few MeV/c$^2$, even a width of a
few tens of MeV/c$^2$ can be accepted in the flux-tube model
\cite{Close:2003af}. Relativistic calculations \cite{Poplawski:2004qj}
confirmed the $S+P$ rule. On the other hand, calculations within QCD
spectral sum rules gave  $10 \div 100$ MeV/c$^2$ \cite{DeViron:1985xn}
or even $0.6$ GeV/c$^2$ \cite{Latorre:1985tg} for the $\rho\pi$ width.
The dominant decays $\pi_1 \to b_1\pi$ and $\pi_1 \to f_1 \pi$ have
quite large widths: $\Gamma_{\pi_1 \to b_1\pi} =100 \div 500$ MeV/c$^2$,
and $\Gamma_7\pi_1 \to f_1 \pi^0 = 50 \div 150$ MeV/c$^2$ were
predicted \cite{Iddir:1988jd,Page:1998gz,Burns:2006wz,McNeile:2006bz}.

Decays of $J^{PC}=1^{-+}$ states into two pseudoscalars can be used to
shed light onto their flavour structure. In \cite{Chung:2002fz} it was
pointed out that an isovector $1^{-+}$ state can belong to either the
$SU(3)$ octet or to the decuplet-antidecuplet. The former can decay only
into $\eta_1\pi$, while the latter can decay only into $\eta_8\pi$, where
$\eta_1$ and $\eta_8$ are singlet and octet $\eta/\eta'$ combinations.
Hence an octet state will decay into $\eta\pi$ and $\eta'\pi$ with a
ratio of the squared matrix elements $|M(\pi_1\to\eta \pi)|^2/|M(\pi_1
\to \eta' \pi)|^2 = \tan^2 \Theta_{PS}\simeq0.1$; a
decuplet-antidecuplet state will decay with a ratio $\tan^{-2}
\Theta_{PS}\simeq10$. In other words, octet $1^{-+}$ states decay
mainly into $\eta'\pi$, and decuplet -antidecuplet states into
$\eta\pi$. This result is valid for any $J^{PC}=1^{-+}$ systems; it
does not use arguments related to a gluon-enrichment of the $\eta'$ or
$\eta' \pi$ state. Hybrids states with $J^{PC}=1^{-+}$ being members of
an $SU(3)$ octet have to demonstrate this unusual ratio of $\eta'
\pi/\eta \pi$ branching fractions. As a guide for this ratio we can use
the branching ratios of J/$\psi$ radiative decays to $\eta'$ and
$\eta$. If radiative decays couple to the flavour singlet component
only, this ratio is given by $\tan^{-2}\Theta_{PS}\simeq10$ as well.
Experimentally, the ratio is $R_{\eta' \pi /\eta \pi} \approx 5$. QCD
spectral sum rules expect a width of $\pi_1 \to \eta' \pi$ decays which
is relatively small, $\Gamma_{\eta' \pi} \approx 3$ MeV/c$^2$
\cite{Narison:1999hg}. However, if the coupling of $\eta'$ to two
gluons through the anomaly is included into the model,  the width could
be as large as $\Gamma_{\eta' \pi} \approx 1$ GeV/c$^2$
\cite{Frere:1988ac}.

\subsubsection{Multiquark states} In the limit of $SU(3)$ symmetry two
quarks in flavour 3 combine to $3\otimes 3 = \bar 3 + 6$ and two
antiquarks to $ 3 + \bar 6$. From these, the irreducible representations
of the $SU(3)$ group can be constructed for the $qq\bar q\bar q$ system:
\begin{eqnarray}
(\bar 3 +6)\otimes (3 +\bar 6)~~~=~~~ \bar 3
\otimes 3~~+~~ \bar 3 \otimes \bar 6~~+~~6 \otimes 3~~+~~6 \otimes \bar
6~~~= \nonumber\\ 1~+~8~+~8~+~10~+~8~+~10~+~1~+~8~+~27   \end{eqnarray}
A large number of tetraquark states should be expected from the four
different octets and the $10+\bar 10$ and $10-\bar 10$ multiplets. The
$10+\bar 10$ and $10-\bar 10$ representations as well as the octets
include exotic states with $J^{PC}=1^{-+}$. The 27-plet has even spin
only and does not contribute to $J^{P}=1^{-}$. The $10+\bar 10$,
$10-\bar 10$ and 27-plet representations include flavour exotic states.

As a rule, the decays of these states to mesons are
superallowed and therefore these bag-model objects do not exist as
$T$-matrix poles \cite{Jaffe:1978bu}. On the other hand, quarks in
multiquark states can cluster to $(qq) (\bar q \bar q)$ systems.
Contrary to naive expectation based on our experience with QED, the
forces between two quarks can be attractive. The strongest attraction
is expected in the system of two different quarks $q$ and $q'$ in a
colour-SU(3)-antitriplet spin-singlet state. The lowest scalar nonet
($a_0(980), f_0(980), \sigma, \kappa$) is the favorite candidate for
$(q q) (\bar q \bar q)$ exotics \cite{Jaffe:1976ig}. Within the bag
model, the mass of the lightest  $J^{PC}=1^{-+}$ tetraquark state is
about $M\approx1.7$ GeV/c$^2$. As this state has a number of
superallowed decay channels it is likely too broad to be identified
unambiguously. In section \ref{Dynamical generation of resonances and
flavour exotics} we will argue that $(qq) (\bar q \bar q)$ systems do
not bind without additional $q\bar q$ forces. This is good news for
hybrids with exotic quantum numbers: If a resonance with exotic quantum
numbers is found, it is unlikely to be a tetraquark state. Hence it
must be a hybrid.

\subsubsection{Molecular states}

A large number of states with different quantum numbers including
exotic ones can be generated dynamically from a meson pair $M_1~M_2$.
The forces between the mesons could be sufficiently attractive to
form bound systems, in particular close to their thresholds. Barnes,
Black and Swanson \cite{Barnes:2000hu} have calculated two meson states
with exotic quantum numbers with a quark-interchange model. In exotic
waves,  the annihilation process $q~q~\bar q~\bar q \to~q~\bar q$ is
forbidden and therefore the interaction of the two mesons $M_1$ and
$M_2$ is driven mainly by the exchange of quarks (antiquarks) from
different mesons. The model describes rather well the low-energy
scattering amplitudes in annihilation-free channels like $I=2~~ \pi
\pi$ and $I=3/2~~ K \pi$ or low-energy nucleon-nucleon scattering. In
particular, the model gives small negative phase shifts for $K^+ \pi^+$
$P$-wave scattering, in agreement with experiment
\cite{Estabrooks:1977xe}. Low-energy  $\rho \pi \to \rho \pi$
scattering in the exotic $J^{PC}=1^{-+}$ wave and $P$-wave $\eta \pi
\to \eta \pi$ scattering were studied as well. In both processes, the
phase motion was found to be small; binding of the $\rho \pi$ or $\eta
\pi$ system in the exotic $1^{-+}$ partial wave does not occur.

The exotic wave $J^{PC}=1^{-+}$ was studied with an effective
Lagrangian approach. Chan and Haymaker \cite{Chan:1974rb} calculated
the scattering amplitudes for scattering of two pseudoscalar mesons in
the framework of a $SU(3) \times SU(3)~~\sigma$ model involving only
scalar and pseudoscalar mesons. Low energy scattering of two
pseudoscalars were reasonably well described; repulsive forces were
predicted for the $J^{PC}=1^{-+}$ wave. Bass and Marco
\cite{Bass:2001zs} included a gluonic potential to generate
contributions to the $\eta$ and $\eta'$ masses. The model does not
exclude exotic resonances in the $\eta'\pi$ $P$-wave. Achasov and
Shestakov \cite{Achasov:2001iq} suggested a model for scattering if a
vector and a pseudoscalar meson into two pseudoscalar mesons,
constructed tree-amplitudes from an `anomalous' effective interaction
of vector and pseudoscalar mesons with subsequent unitarisation, and
calculated exotic-wave amplitudes with $J^{PC}=1^{-+}$ for the
reactions $\rho\pi\to\rho\pi$, $\rho\pi\to\eta'\pi$,
$\eta\pi\to\eta\pi$, $\eta\pi\to (K^*\bar K + \bar K^*)$, and others.
Depending on the choice of free parameters of the model,  various
resonant-like amplitudes could be generated in different channels.
General, Wang, Cotanch and Llanes-Estrada \cite{General:2007bk} use an
Hamiltonian approach to arrive at the conclusion that molecular-like
configurations involving two color singlets are clearly favoured
compared to hybrid (tetraquark) configurations in which a $\bar qq$
pair (or two pairs) carry colour.

\subsection{Experimental results for $J^{PC}=1^{-+}$}
The lightest isovector state with these quantum numbers could decay to
$ \pi \eta ~$, $\pi \eta' ~$, $\pi \rho ~$, $ \pi b_1 ~$, $\pi f_1 ~$,
$a_0 \rho~$, $a_1 \eta ~$, $\pi \rho'$. Most of these channels were
studied at various places using GAMS/NA12 at IHEP-CERN, E179 at KEK,
Crystal Barrel at CERN,  E852 at BNL, Obelix at CERN and VES at IHEP
(see Table \ref{table-1-+}). \subsubsection{The wave $J^{PC} = 1^{-+}$
in the $\eta \pi$ channel.} 
\begin{figure}[pb]
\begin{center} \includegraphics[width=0.9\textwidth]{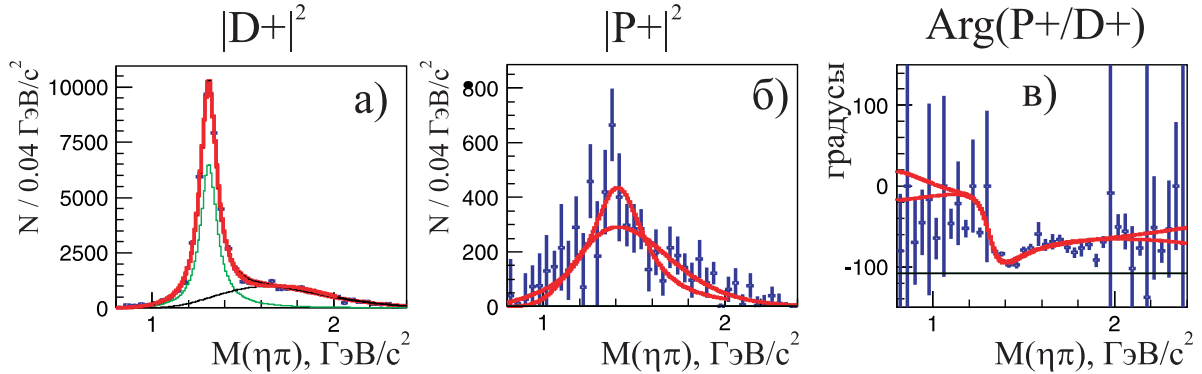}
\caption{
Results of partial-wave analysis of the $\eta \pi^-$ system:
a) intensity of $\rm D_+$ wave, ~ b) intensity of $\rm P_+$ wave,
c) phase difference $\rm P_+/D_+$ \cite{Dorofeev:2001xu}.}
\label{ex:etpi}
\end{center}
\end{figure}

Diffractive reactions could be especially
effective to produce hybrid mesons, as these mesons have some
additional gluon-like component in their wave function and could have a
strong coupling to the Pomeron. After a very early first observation of
an $J^{PC}=1^{-+}$ exotic wave in 1981 \cite{Apel:1981wb}, high
statistics data became available from VES and KEK in 1993
\cite{Beladidze:1993km,Aoyagi:1993kn}\footnote{An observation of
$J^{PC}=1^{-+}$ in $\eta \pi$-channel by GAMS \cite{Alde:1988bv} was
revised in \cite{Prokoshkin:1992yp}.}.  In the VES experiment, events
due to the exclusive reactions $\pi^- Be \to X Be$ (where $X$ means the
aforementioned decay channels) at $p_{\pi} = 37$ GeV/c were selected
and subjected to a partial wave analysis (PWA). Waves with orbital
angular momenta of $L$=0,~1, and 2 ($L = S, P, D$) and orbital angular
momentum projection of $M= 0$ and 1 onto the Gottfried-Jackson axis
with both natural and unnatural parity exchange were included in the
analysis, i.e. the waves $S,P_0, P_+, P_-, D_0, D_+, D_-$ were
tested. It turned out that waves with unnatural parity exchange are not
significant. The results of the VES collaboration on this channel were
published in \cite{Beladidze:1993km,Dorofeev:2001xu}. The dominant wave
$J^{PC}=2^{++}$ is peaking at the mass of the $a_2(1320)$ meson. There
is a clear signal in the exotic wave  $J^{PC} = 1^{-+}$ centered at $M
\approx 1.4 $ GeV/c$^2$ (Fig. \ref{ex:etpi}). The mass-dependent fits
to the PWA results were reported in \cite{Dorofeev:2001xu}.
\begin{table}[pt]
\begin{center}
\caption{\label{table-1-+}Experimental studies of $1^{-+}$ states.}
\vspace*{2mm}
\renewcommand{\arraystretch}{1.5}
\begin{tabular}{llcc} \hline\hline
Experiment & Reaction & final state & Reference \\ \hline
VES  &   Diffraction; charge exchange;  &$ \eta\pi^-, \eta' \pi^-, \rho \pi^-$&\cite{Beladidze:1993km,Amelin:1995gu,Amelin:1999gk}\\
     & 28-,37-GeV/c $\pi^-$ beam                                 &$ f_1 \pi^-, b_1 \pi^-, \eta' \pi^0$&\cite{Khokhlov:2000tk,Amelin:2004ns,Amelin:2005ry}\\
E179 & Diffraction, 6.3-GeV/c~~ $\pi^-$beam   & $ \eta\pi^-$ & \cite{Aoyagi:1993kn,Amelin:2005ry} \\
Crystal Barrel & $p \bar p$ annihilation  &$\eta
\pi^{\pm}, \eta \pi^0 $   &  \cite{Abele:1998gn,Abele:1999tf}   \\
               &to $\eta \pi^+ \pi^-, \eta \pi^- \pi^0, \pi^-3\pi^0, \pi^+
 \pi^- \pi^0 \omega $&$\rho \pi, b_1 \pi$&\cite{Dunnweber:2004vc,Baker:2003jh}\\
Obelix &$p \bar p$ annihilation to $\pi^+ \pi^- \pi^+ \pi^-$&$\rho \pi$&\cite{Salvini:2004gz}\\
GAMS/NA12& Charge exchange; $\pi^-$~~32, 38, 100 GeV/c& $ \eta \pi^0$&
\cite{Alde:1999gh}\\
E852 & Diffraction; charge exchange; 18 GeV/c &$ \eta\pi^-,\eta'\pi^-,\rho\pi^-$&
\cite{Thompson:1997bs,Adams:1998ff,Chung:1999we,Chung:2002pu}\\
     & $\pi^-$ beam  &$ f_1 \pi^-, b_1 \pi^-$&\cite{Ivanov:2001rv,Dzierba:2003fw,Kuhn:2004en,Lu:2004yn}\\
\hline\hline
\end{tabular}
\renewcommand{\arraystretch}{1.0}
\end{center}
\end{table}

These results were beautifully confirmed by the E-852 collaboration
\cite{Thompson:1997bs}. When overlayed, both data sets are
indistinguishable in their distributions demonstrating the reliability
of both experiments with respect to detector performance, data
reconstruction and partial wave analysis. Differences which turn up in
the publications are due to the mass-dependent fits to the results of
the mass-independent partial-wave analyses, and due to different
underlying assumptions in these fits. Are meson-meson interactions
dominated by resonances or do background amplitudes play an important
$\rm r\hat{o}le$\,?  For $P$-wave meson-meson scattering, this question is unsolved
but it may serve as an indication that $\pi N$ scattering even in
the $\Delta(1232)$ region proceeds not only via $\Delta(1232)$
formation but also via a background $P$-wave $\pi N$ scattering
amplitude.

In the KEK experiment \cite{Aoyagi:1993kn} at $p_{\pi}=6.3$ GeV/c, a
very different shape of the $P_+(\eta \pi^-)$ wave was found. The
dominant $D_+$ wave exhibited a clear $a_2(1320)$; in the exotic
$P_+$ wave, amplitude and phase followed exactly the $D_+$ wave.
In a mass-dependent fit, an exotic resonance with parameters of the
$a_2(1320)$ shows up. This coincidence may suggest that the $P_+$
wave signal could be driven by leakage from the dominant $a_2(1320)$
resonance, either due to imperfections of the PWA model or due to
experimental effects like acceptance, resolution or background.
Possibly, the PWA model with seven waves ($S, P_0, P_+, P_-, D_0,
D_+, D_-$) in the $\eta \pi$ system could be oversimplified: At
$p_{\pi^-}$=6.3\,GeV/c contributions from baryon resonances could be
important and lead to distortions in the $\eta \pi$ system. The
compatibility of the E-852 and VES results (taken at 18\,GeV/c and at
28 and 37\,GeV/c, respectively) suggests that at these momenta, baryon
resonances are well enough separated and have no impact on the mesonic
system.

The data in Fig. \ref{ex:etpi} exhibit a dominant $a_2(1320)$ in the
$D_+$ and a clear bump at $M \approx 1.4$\,GeV/c$^2$ in the (exotic)
$P_+$ partial wave. The E-852 collaboration
\cite{Thompson:1997bs,Chung:1999we} finds that the data are consistent
with a simple ansatz, assuming contributions from two resonances, one
in each partial waves. The $\rm D_+$ wave returns the parameters of
$a_2(1320)$, for the $P_+$ partial wave, mass and width are determined
to~~$M=1370\pm 16~^{+50}_{-30}$~~;~~$\Gamma= 385 \pm 40^{+\ 65}_{-105}$.
This observation is referred to as $\pi_1(1400)$ by the Particle Data
Group \cite{Eidelman:2004wy}.

A systematic study was carried out in \cite{Dorofeev:2001xu} to
investigate if the observations can be explained without introducing an
exotic resonance. The data were fitted with $a_2(1320)$ and $P_+$ and
$D_+$ backgrounds in the form
\be
A_B=LIPS_l \times (m-m_t)^{\alpha}\exp{(-\beta (m-m_t))},
\label{bg-ves}
\ee
where $LIPS_l(m)$ is the phase space volume,
$m_t$ the threshold mass, and $\alpha$, $\beta$ shape parameters. An
arbitrary (constant) phase is allowed for both background
amplitudes. The background amplitudes are interpreted as meson-meson
interaction dynamics originating from $t$-channel exchange currents.
The fit without $\pi_1(1400)$ gives a perhaps acceptable $\chi^2$ of
244 for 149 degrees of freedom. If $\pi_1(1400)$ with E-852 parameters
is added, $\chi^2$ improves significantly by 57 units for 2 additional
parameters (amplitude and phase). The two fits are presented and
compared to the data in Fig.~\ref{ex:etpi}.

A judgement if the $\rm P_+$ wave carries a resonance is not only a
statistical question. Obviously, there are spin-flip and
spin-nonflip contributions at the baryon vertex leading to a production
density matrix of rank two.  For a system of two pseudoscalars, a
rank-2 density matrix can not be reconstructed in a model-independent
way due to a continuous ambiguity. Hence some doubts remain as to
the existence of a resonant $\rm P_+$ wave in the $\pi\eta$ channel.

The Indiana group \cite{Szczepaniak:2003vg} tested the idea that $t$
channel exchange forces might give rise to a background amplitude which
could mimic $\pi_1(1400)$. A model respecting chiral symmetry for
meson-meson interactions was constructed in which these interactions
were expanded in terms of the relative momenta\footnote{Even though
related, the analysis is not using an effective field theory approach
based on a chiral Lagrangian with low-energy constants fixed from
other processes.}. The production amplitudes were supposed to follow the
momentum dependence of the interactions. In this way, $\pi\eta$
$P_+$-wave interactions were constructed which are very similar to
$\pi\pi$ $S$-wave interactions. The latter were characterised by
the $\sigma$ pole; as a consequence, $\pi_1(1400)$ is considered as
$\sigma$-type phenomenon in $\pi\eta$ $P_+$-wave interactions. In
the words of the authors of \cite{Szczepaniak:2003vg}, $\pi_1(1400)$ is
`not a QCD bound state' but rather generated dynamically by meson
exchange forces. This difference is subtle; the reader may like to have
a look at the discussion in section~\ref{Dynamical generation of
resonances and flavour exotics}.

The $\eta\pi$ $P_+$-wave was also studied in the charge exchange
reaction $\pi^- p \to \eta \pi^0 n$ \cite{Alde:1999gh,Dzierba:2003fw}.
The $\rm P_+$ wave is clearly seen in both experiments. Mass-dependent
fits of the amplitudes required the introduction of a resonance but
yielded resonance parameters which were different in the three ranges
of $t$ studied in \cite{Dzierba:2003fw}. The authors concluded that
there is no evidence for a resonant signal at $M \approx
1.4$\,GeV/c$^2$. Of course, this does not rule out the existence of
$\pi_1(1400)$. The exotic $\eta\pi^0$ $P_+$-wave is produced by
$t$-channel exchange of isospin-$1$ natural-parity objects like $\rho$.
From the absence of $\pi_1(1400)$ in $\pi^- p \to \eta \pi^0 n$ it can
only be inferred that the $\pi_1(1400)\to\rho\pi$ coupling  is small in
comparison to its coupling to $\pi$-Pomeron or $\pi$-$f_1(1285)$. On
the other hand, these results point to a significant nonresonant
background in the exotic $\pi\eta$ $P_+$-wave channel.

\begin{figure}[pb] \vspace*{3mm}
\begin{minipage}[c]{0.48\textwidth}
\includegraphics[width=0.96\textwidth]{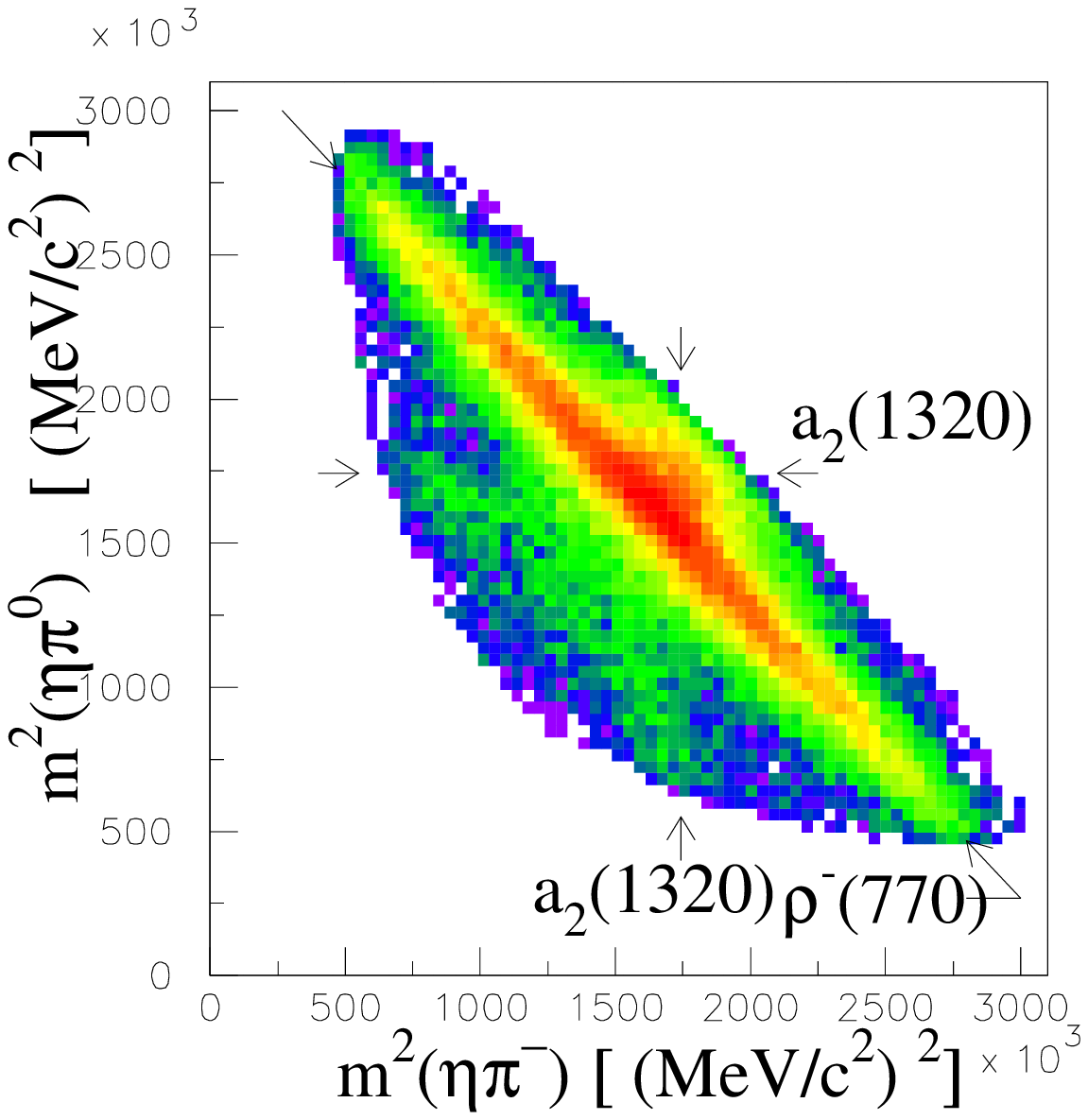}
\end{minipage}
\begin{minipage}[c]{0.48\textwidth}
\begin{tabular}{cc}
\hspace{-5mm}\includegraphics[width=0.48\textwidth]{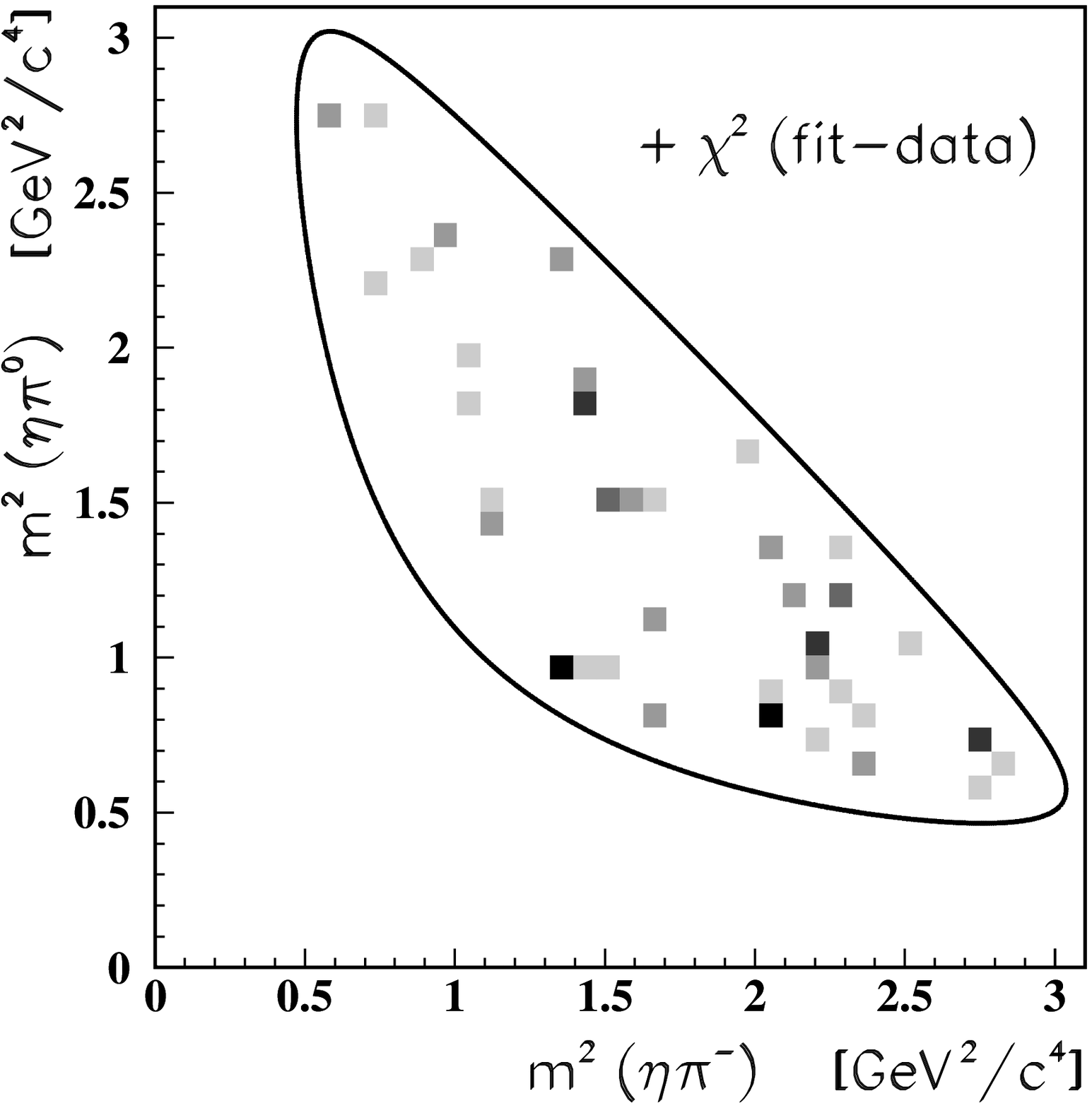}&
\hspace{-5mm}\includegraphics[width=0.48\textwidth]{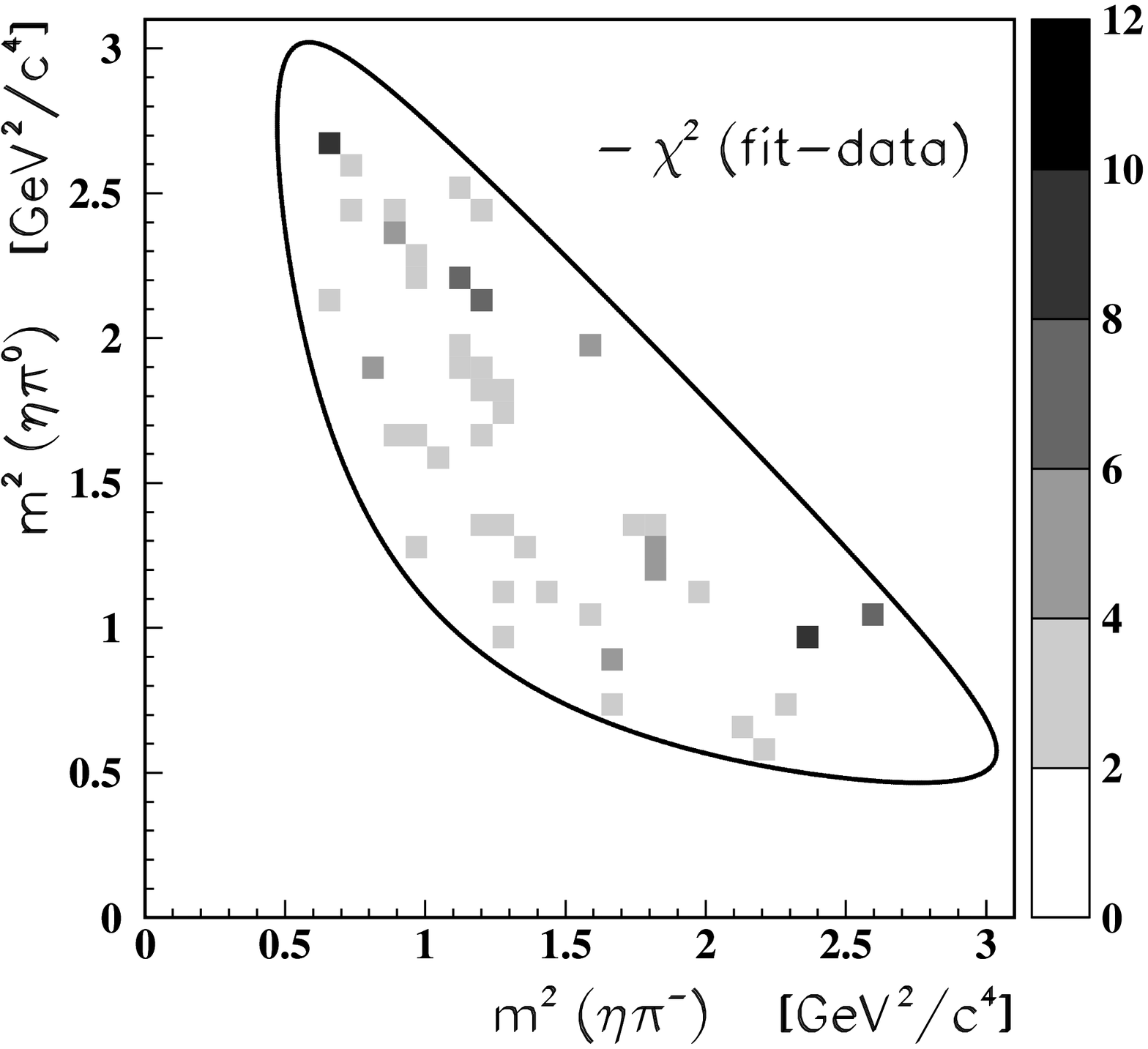}\vspace{-5mm}\\
\hspace{-5mm}\includegraphics[width=0.48\textwidth]{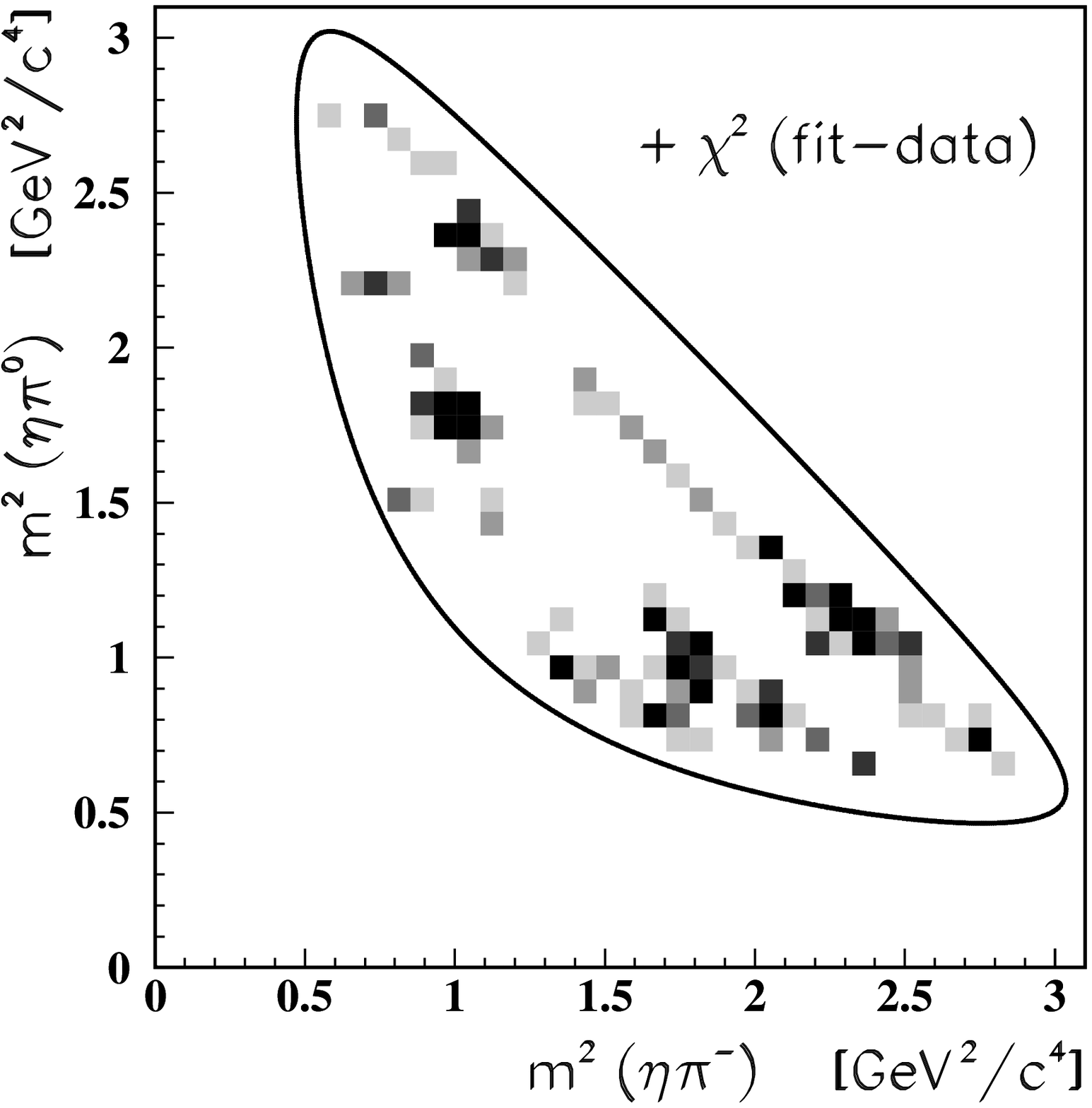}&
\hspace{-5mm}\includegraphics[width=0.48\textwidth]{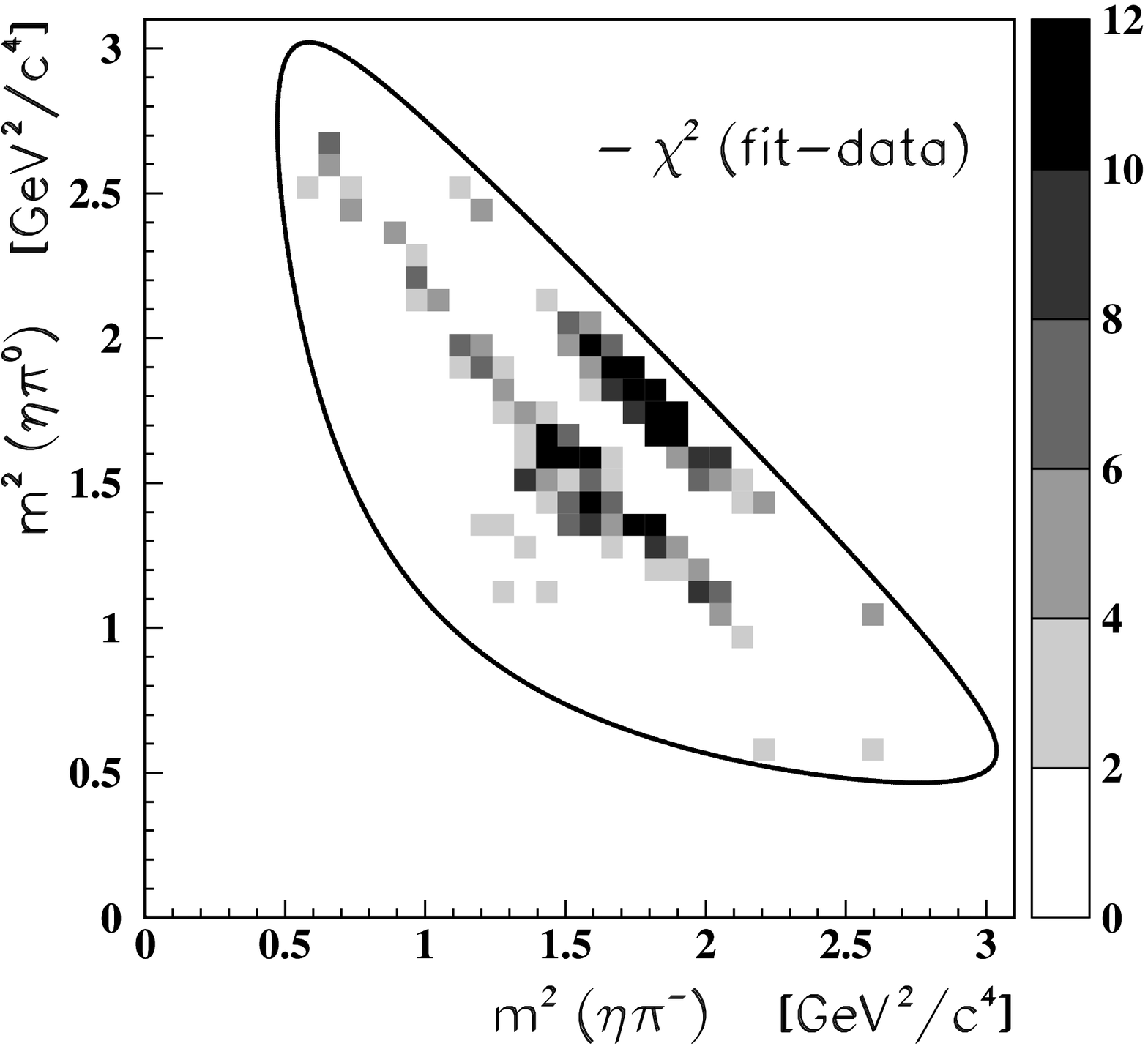}\\
\end{tabular}
\end{minipage}
\caption{Left: Dalitz plot for $\bar pn \to \pi^- \pi^0 \eta$, right:
Differences between data and fit (in $\chi^2$) over the Dalitz plot.
The $\pi\eta$ $P$-wave is included in (a,b) but not in (c,d). The + sign
indicates that the fit exceeds the data, the - sign that data exceed
the fit. All fits include the $\rho(770), a_2(1320), a_0(980)$ and
$a_0(1450)$ \cite{Abele:1998gn}.
}
\label{CB_dp_exot}
\end{figure}

The Crystal Barrel Collaboration confirmed the existence of the exotic
$\pi\eta$ $\rm P_+$-wave in an entirely different reaction: in $ \bar p
n \to \pi^- \pi^0 \eta$ and $\bar p p \to 2\pi^0 \eta$
\cite{Abele:1998gn,Abele:1999tf}. The evidence for $\pi_1(1400)$
comes from the first reaction; in the reaction $\bar p p \to 2\pi^0
\eta$, a small signal was unraveled by comparison of this reaction in
liquid and gaseous hydrogen where the fraction of annihilation
contributions from atomic $S$ and $P$ states is different. Only the
combined fit of these two data sets gave positive evidence for a
small contribution from an exotic wave. A reanalysis
\cite{Sarantsev:2004tn} fitted the $\bar p p \to 2\pi^0 \eta$ data
without the need for a narrow exotic resonance. In the annihilation
reaction  $\bar pn \to \pi^- \pi^0 \eta$, the exotic wave is rather
strong, however, and provided a contribution to the final states which
is about 1/3 of the $a_2(1320)$ contribution. The Dalitz plot is shown
in Fig.~\ref{CB_dp_exot}. If fitted with only conventional mesons, a
poor description of the data is achieved; the addition of the exotic
$\pi_1(1400)$ gave an excellent fit. The fit and the data are compared
in Fig.~\ref{CB_dp_exot}. In the lower part the amplitudes of the
resonances allowed by conservation laws were optimised but the exotic
meson was omitted.

The exotic partial wave was described by a Breit-Wigner resonance. The
best fit was obtained for $M = 1400\pm20\pm20$\,MeV/c$^2$ and
$\Gamma=310\pm 50^{+50}_{-30}$\,MeV/c$^2$ where the first error is a
statistical and the second a systematic error estimated from a variety
of different fits.

The Crystal Barrel and Obelix collaborations found  a $J^{PC}=1^{-+}$
wave in $(\rho \pi)$ in $p \bar p$ annihilation to four pions. Crystal
Barrel reported this observation only in a conference contribution
\cite{Dunnweber:2004vc} and reported $M,\Gamma\sim 1400, \sim
400$\,MeV/c$^2$. The Obelix parameters for this wave are $M,\Gamma
=(1384\pm 28), (378\pm 58)$\,MeV/c$^2$ \cite{Salvini:2004gz}. The
results are not incompatible with the findings reported for the
$\pi\eta$ $P$-wave. A major point of concern is the production
mechanism. The $\pi\eta$ $P$-wave is seen to be produced from spin
triplet states of the $ N\bar N$ system, in particular from the $^3S_1$
state. In contrast, the exotic $\rho\pi$ wave comes from spin singlet
states, dominantly from the $^1S_0$ state. Hence these must be
different objects. It was pointed out by Sarantsev (A. Sarantsev,
private communication, 2005) that in $p \bar p$ annihilation to
$\rho\pi\pi$, rescattering processes in the final state could lead to
logarithmic singularities which could mimic a pole singularity. This
argument holds true for $\eta\pi\pi$ as well but no calculation has
ever been reported.

From a wider view, the claim for an exotic $\pi_1(1400)$ resonance in
$\pi\eta$ interactions is rather problematic. $P$-wave states in
the $\eta_8 \pi$ channel belong to a $SU(3)$ decuplet
\cite{Chung:2002fz}, the suggested resonance cannot be a hybrid.
One can assume that $\pi_1(1400)$ belongs to a decuplet-antidecuplet
family of multiquark or mesomolecular states. The
decuplet-antidecuplet includes also $ K^+ \pi^+$ $P$-wave states.
The amplitude of  $ K^+ \pi^+ \to K^+ \pi^+$ scattering was studied
in \cite{Estabrooks:1977xe}. It was found that the scattering amplitude
is dominated by $S$-wave while the $P$-wave is strongly suppressed.
The $P$-wave phase shift is less than a few degrees from threshold to
$M \approx 1.8$ GeV/c$^2$ and does not leave room for a resonance in
this mass region\footnote{The $ K^+ \pi^+ \to K^+ \pi^+$ data are
also at variance  with the interpretation of
\protect\cite{Szczepaniak:2003vg} which requires a sizable phase motion
in $\pi\eta$ $\rm P_+$-wave}. Based on SU(3) symmetry, very little phase
motion should be expected for the $\pi\eta$ $\rm P_+$-wave. Donnachie and
Page \cite{Donnachie:1998ty} suggest that the 1.4 GeV enhancement in
the VES and E852 $\eta\pi$ data can be understood as interference
between a non-resonant Deck-type background and a hybrid resonance at
1.6\,GeV.  A hybrid should not decay into $\pi\eta_8$ due to
SU(3) arguments \cite{Chung:2002fz}. In \cite{Donnachie:1998ty}, the
suppression of hybrid $\eta\pi$ decays is circumvented by rescattering
of intermediate $b_1(1235)-\pi$ into $\eta\pi$. Of course, this is
difficult to accept when the suppression is due to SU(3). The mass of
$\pi_1(1400)$ is very close to the $f_1(1285)\pi$ threshold. A
$f_1(1285)\pi$ virtual bound state or a cusp may contribute to the
scattering amplitude as well.

In summary, all experiments agree that there is a substantial
contribution of the exotic $P_+$-wave to $\pi\eta$ interactions.
The data are well described once a resonant $\pi_1(1400)$ is
introduced. Due to SU(3) symmetry arguments, the observation cannot be
due to a hybrid, the roots have to be in tetraquark dynamics. However,
in tetraquark systems in decuplet-antidecuplet, no substantial phase
motion is expected.  Fits to the diffractively produced data without
$\pi_1(1400)$ but including $t$-channel exchange dynamics were
successfully performed, too; the dynamical origin of the non-resonant
background amplitudes was however not yet traced back to meson-meson
interactions in a convincing way.

\subsubsection{The wave $J^{PC} = 1^{-+}$ in the $\eta' \pi$ channel.}

Even if interpreted as resonance, the $\pi_1(1400)$ cannot belong to
the family of hybrid mesons due to SU(3) arguments. The $P_+$-wave in
$\eta^{\prime}\pi$ is a much better place where a true hybrid might be
found. Fig.~\ref{ex:et1pi-} shows the $\eta'\pi^-$ system produced in a
diffractive-like reaction at $p_{\pi_-} =18$\,GeV/c. The data are from
the E-852 collaboration \cite{Thompson:1997bs,Chung:1999we}; VES using
a beam at $p_{\pi_-} = 37$\,GeV/c showed similar distributions
\cite{Beladidze:1993km}. The main results of the partial wave analysis
of both experiments can be summarised as follows:

\begin{figure}
\begin{minipage}[c]{0.50\textwidth}
\includegraphics[width=\textwidth]{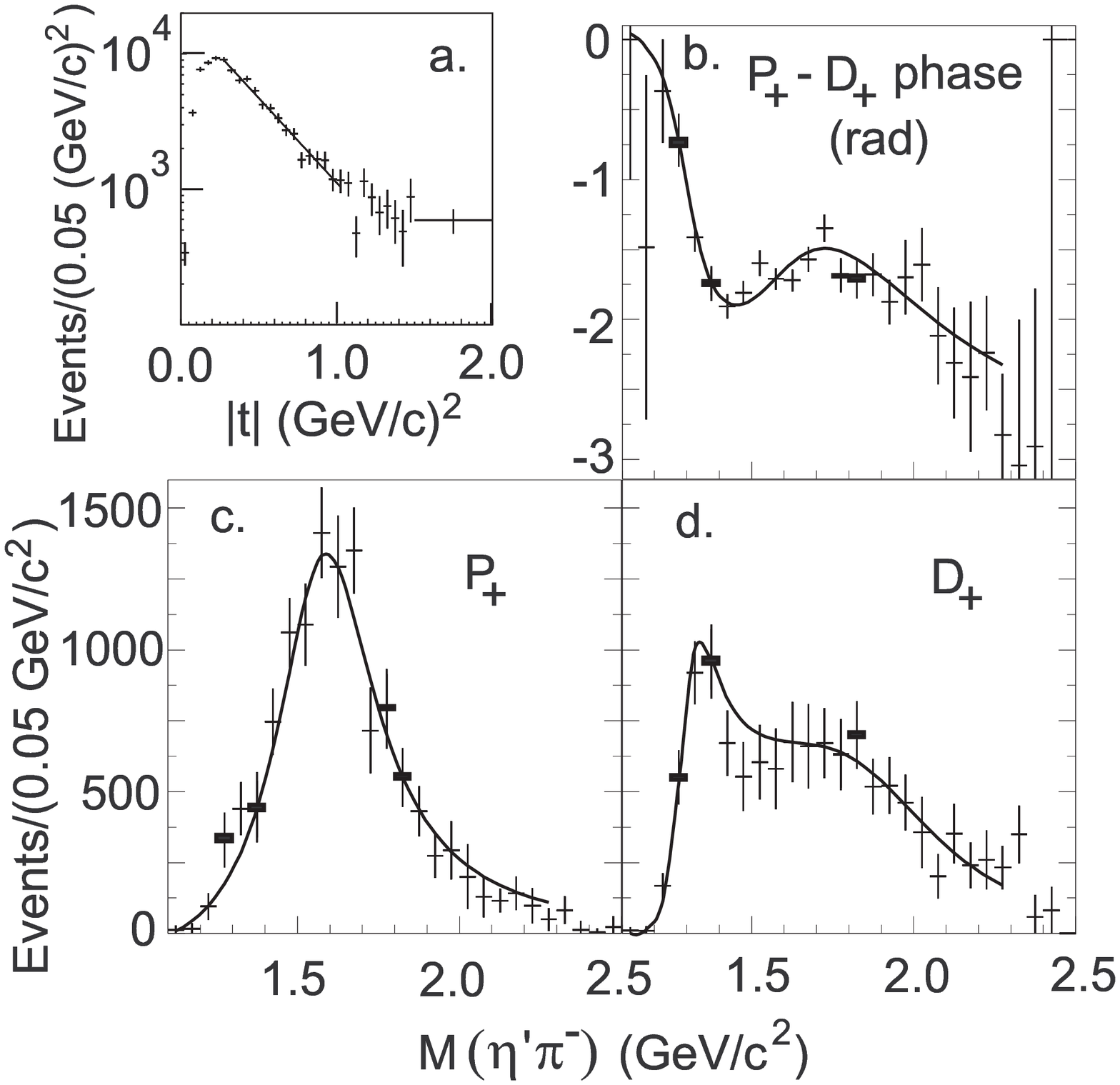}
\end{minipage}
\begin{minipage}[c]{0.50\textwidth}
\includegraphics[width=0.8\textwidth]{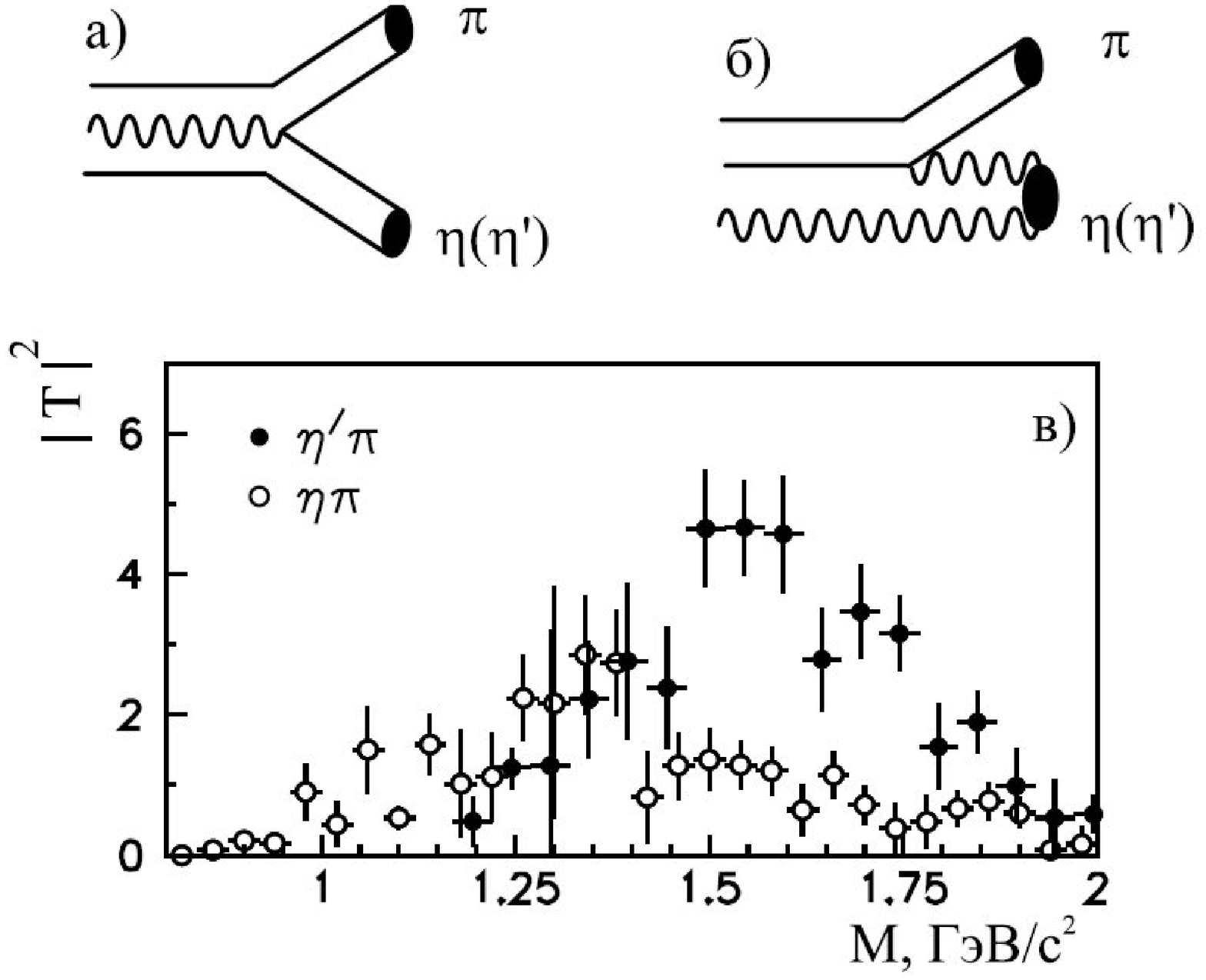}
\end{minipage}
\caption{\label{ex:et1pi-}
Left panel, BNL data \cite{Chung:1999we}: (a) The acceptance-corrected
$|t|$ distribution fitted with the function $f(t)=ae^{b|t|}$ (solid
line).  (b), (c), (d) results of a mass-independent PWA and a
mass-dependent fit (solid curve) for the $\rm P_+$ and $\rm D_+$ partial waves
and their phase difference. (b) The $(P_{+}-D_{+})$ phase difference.
(c) Intensity of the $P_+$ and (d) of the $D_+$ partial wave. Right
panel: (a) OZI-allowed and (b) OZI-forbidden diagrams for the decay of
a hybrid meson and (c) squares of matrix elements for the $1^{-+}$ wave
in the $\eta \pi^-$ and $\eta' \pi^-$ systems. }
\end{figure}

 - waves with unnatural parity exchange are strongly suppressed;

- the wave $1^{-+}$ is the dominant one, it has a maximum at $M \approx
1.6$ GeV/c$^2$;

- in the wave $J^{PC}=2^{++}$ there is a $a_2(1320)$ signal and a broad
bump at $M = 1.6\div 2.0$ GeV/c$^2$.

- a $4^{++}$ wave is also clearly observed (but documented only for
BNL).

The intensity of the $1^{-+}$ wave at $M \approx 1.6$ GeV/c$^2$ is much
higher than that in the $\eta\pi$ channel (Fig. \ref{ex:et1pi-});
therefore this $1^{-+}$ state belongs to $SU(3)$ octet. This octet
dominance looks very natural for diffractive-like reactions which are
mediated by Pomeron exchange at sufficiently high energy. In the
collision of octet pions with (SU(3) singlet)
Pomerons\footnote{\footnotesize The \ssb\ suppression in the soft
Pomeron induces however an octet component.}, only octets can be
produced.

The E-852 collaboration performed a fit to these partial waves with the
following ingredients: the $2^{++}$ wave was described by $a_2(1320)$
and either one broad (550 to 750\,MeV/c$^2$) or two narrower Breit-Wigner
tensor resonances, the  $4^{++}$ wave by the $a_4(2040)$ meson in its
$\eta^{\prime}\pi^-$ decay mode. The dominant, exotic (non-$q\bar{q})$
$1^{-+}$ partial wave was described as a resonance with a mass of
$1.597 \pm 0.010^{+0.045}_{-0.010}$ GeV/c$^2$ and a width of $0.340 \pm
0.040 \pm 0.050$ GeV/c$^2$. The exotic partial wave is produced with a
$t$ dependence which is different from that of the $a_2(1320)$ meson,
indicating different production mechanisms for the two states.

The VES collaboration performed tests on the stringency with which these
claims can be made. They fit the data with $a_2(1320)$ and a
phenomenological background amplitude for the $2^{++}$ and $1^{-+}$
with the shape given in eq.~(\ref{bg-ves}). These assumptions yield
also a perfect fit. Hence the decisive question (for both reactions,
$\rm\pi^-p\to n\pi^-\eta$ and $\rm\pi^-p\to n\pi^-\eta^{\prime}$) is if
$t$ channel exchange forces between $\pi$ and $\eta$
($\eta^{\prime}$) can lead to significant background amplitudes. It is
well known that a large fraction of the $\pi\pi$ $S$-wave phase shift
and the magnitude of the scattering amplitude can be understood in
terms of $t$-channel $\rho$ exchange. An additional phase amplitude
and phase shift is provided by $f_2(1270)$ exchange (see e.g.
\cite{Wu:2003wf}). Since the $\rho$ and $f_2(1270)$ couplings are
known, absolute predictions can be made. Likewise, after projection
onto the $P$-wave, absolute predictions of the background amplitude due
to $a_2(1320)$ exchange are possible and should be made.

The Indiana group \cite{Szczepaniak:2003vg} tried to expand their model
on $\pi\eta$ interactions to $\pi\eta^{\prime}$ but did not succeed to
describe the data without a resonance at 1600\,MeV/c$^2$ and with
300\,MeV/c$^2$ width.

\begin{figure}[ltb]
\begin{center}
\includegraphics[width=.6\textwidth]{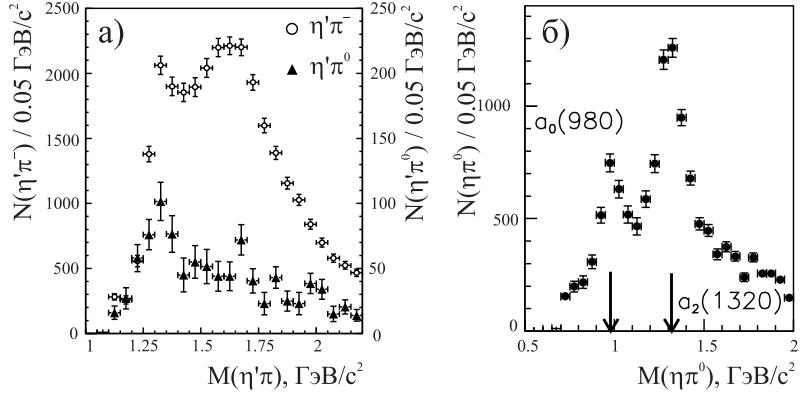}
\end{center}
\caption{\label{ex:neutr}
The mass spectra of (a) $\eta'\pi^-,~\eta'\pi^0$, and (b) $\eta\pi^0$.}
\end{figure}

The  $\eta' \pi^0$ system was studied GAMS, E-852 and VES by charge
exchange \cite{Alde:1999gh,Dzierba:2003fw,Amelin:2004ns}. The results
are again compatible; Fig.~\ref{ex:neutr} shows $\eta\pi^0$ and
$\eta'\pi^0$ mass spectra from VES.
The $\eta\pi^0$ spectrum is dominated by the $a_2^0(1320)$ meson. The
$\eta'\pi^0$ spectrum is also dominated by $a_2^0(1320)$ production, in
spite of a strong threshold suppression ($\sim p^5$). The signal at $M
\approx 1.6$ GeV/c$^2$ which dominates the $\eta'\pi^-$ spectrum has
disappeared in $\eta'\pi^0$. The ratio of $P$-wave intensities in the
$\eta' \pi^0$ and $\eta \pi^0$ channels is $$ R=|T^{\eta'
\pi^0}_{P+}|^2/|T^{\eta\pi^0}_{P+}|^2 \approx 0.1 \pm 0.1. $$ The
suppression of the $\eta' \pi^0$ signal in charge exchange can be used
to derive an upper limit on the $\pi_1(1600)$ branching ratios.

\be
Br(\pi_1(1600) \to \rho \pi)\times Br(\pi_1(1600) \to \eta' \pi) \le 3
\times 10^{-3}
\label{1600rhopi}
\ee

Obviously, the $\rho\pi$ coupling is small. The contribution from
$\rho\pi\to\eta^{\prime}\pi$ $P$-wave scattering is of the order of the
contribution expected from $\rho\pi\to\eta_8\pi$ $P$-wave scattering via
SU(3)$_{(10,\bar 10)}$ amplitudes and standard $\eta_1$-$\eta_8$
mixing. Hence the octet $\rho\pi\to\eta^{\prime}\pi$ $P$-wave scattering
amplitude must be very small and is compatible with zero.

\subsubsection{Partial wave analyses of the diffractively produced
$\pi^+\pi^-\pi^-$ system}
The wave $J^{PC}=1^{-+}$ in the $\pi^+\pi^-\pi^-$ system is highly
controversial. It was studied by two experiments in diffractive-like
reactions: by the VES collaboration \cite{Amelin:1995gu,Amelin:2005ry}
at $p_{\pi} =37$ GeV/c and by the E-852 collaboration at $p_{\pi} =18$
GeV/c \cite{Adams:1998ff,Chung:2002pu}. A new BNL data sample with
10-fold increased statistics was reported in \cite{Dzierba:2005jg}.

Most of the basic parameters of the PWA models are similar in all three
analyses. Up to 45 waves with total angular momenta ranging from $J=0$
to $J=4$ were used in
\cite{Amelin:1995gu,Kachaev:2001jj,Amelin:2005ry}, 21 to 27 waves in
\cite{Adams:1998ff,Chung:2002pu} and 36 waves in \cite{Dzierba:2005jg}.
The $J^{PC}=1^{-+} \rho^0 \pi^-$ wave was included in PWA with three
different states: $M^{\eta} = 0^-,~1^-$ and $1^+$. In the first BNL
analysis and in \cite{Dzierba:2005jg}, a spin-density matrix of rank 1
or 2 was used to account for the possibility of two incoherent
contributions to the reaction. In the analysis of the Protvino data, the
spin-density matrix was taken in a general form without any restrictions
of its rank. The reason for this approach is that any PWA model is
based on some assumptions and approximations. These model
imperfections could lead to a leakage from the most intensive
to less intensive waves and generate spurious artificial signals. This
effect seems unavoidable in a PWA model with density matrix of
rank 1. PWA models based on density matrices of arbitrary rank give
more freedom to waves how to interfere and leakage from intensive waves
is not so dangerous. Hence it is less likely to produce spurious
results; the risk is that small signals could be washed out.

\begin{figure}[htb]
\begin{center}
\includegraphics[width=.6\textwidth]{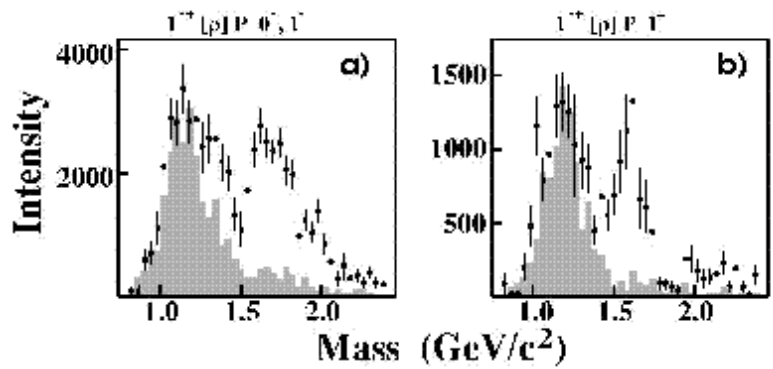}
\end{center}
\caption{\label{3pi}
Wave intensities of the $1^{-+}(\rho \pi)P$ exotic waves: (a)
the $M^{\eta}=0^-$ and $1^-$ waves combined; (b) the $M^{\eta}=1^+$
wave. The 21-wave rank-1 PWA fit to the data is shown as points with
error bars; the shaded histograms show estimated contributions from all
non-exotic waves due to leakage. }
\end{figure}

The results of BNL experiment on $J^{PC}=1^{-+}$ waves are presented in
Fig. \ref{3pi}.
Clear peaks
of comparable intensities are seen at $M \approx 1.6 $ GeV/c$^2$, both
in natural parity exchange ($J^{PC}M^{\eta}=1^{-+}1^+$) and in
unnatural parity exchange ($J^{PC}M^{\eta}=1^{-+}0^-$ and
$J^{PC}M^{\eta}=1^{-+}1^-$). The distributions were fitted with a
Breit-Wigner amplitude, the fit gave mass and width
\be
\label{mw1600}
M = 1593 \pm 8^{+ 29}_{-47} {\rm MeV/c}^2, ~~~\Gamma = 168 \pm
20^{+150}_{-12} {\rm MeV/c}^2.
\ee
The flatness of phase difference $\phi(1^{-+}1^+(\rho \pi)) -
\phi(2^{-+}0^+(f_2 \pi))$ in presence of the well established resonance
$\pi_2(1670)$ in the wave $J^{PC}M^{\eta}=2^{-+}0^+(f_2 \pi)$ supports
the resonance interpretation of the signal. The total intensity of the
exotic wave $J^{PC}=1^{-+}$ at $M \approx 1.6$ GeV/c$^2$ is about
$20\%$ of the $a_2(1320)$ signal at its maximum.

\begin{figure}[pt]
\includegraphics[width=.49\textwidth]{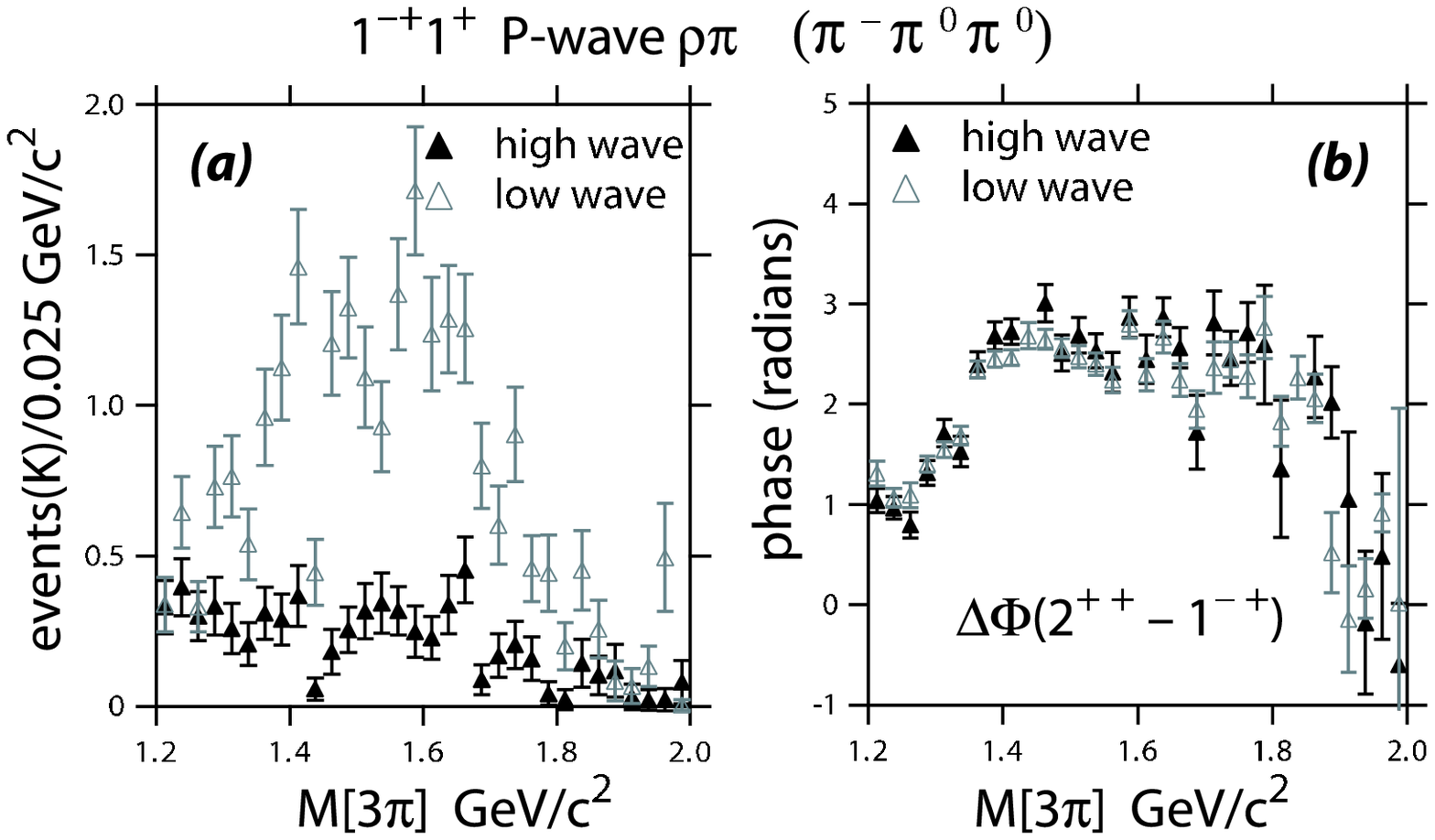}
\includegraphics[width=.49\textwidth]{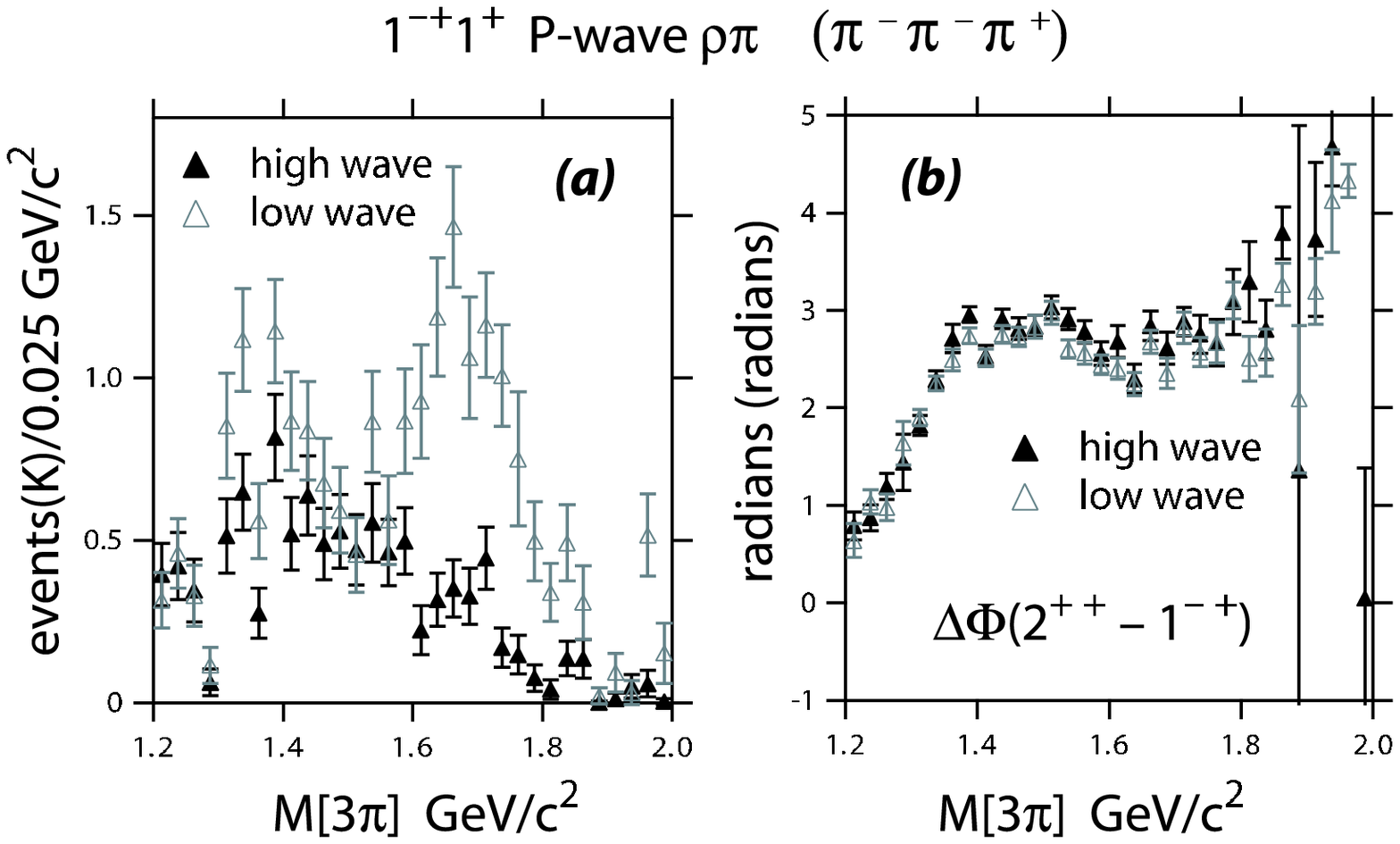}
\caption{\label{exotic_indiana}
Left panel: (a)  The $1^{-+}1^+$ \ $P$-wave $\rho\pi$ partial wave in the
neutral mode ($\pi^-\pi^0\pi^0$) for the high-wave set PWA and the
low-wave set PWA and (b)~the phase difference $\Delta\Phi$ between the
$2^{++}$ and $1^{-+}$ for the two wave sets.
Right panel: (a)  The $1^{-+}1^+$ \ $P$-wave $\rho\pi$ partial wave in
the charged mode ($\pi^-\pi^-\pi^+$) for the high-wave set PWA and the
low-wave set PWA and (b)~the phase difference $\Delta\Phi$ between the
$2^{++}$ and $1^{-+}$ for the two wave sets \cite{Dzierba:2005jg}.
}
\end{figure}
In the second analysis of BNL data (with increased statistics), 20
partial waves were used (called low-wave set) and, alternatively,  35
partial waves (called high-wave set). Fig.~\ref{exotic_indiana} shows
the results for the exotic $\rho\pi$ $P$-wave. While the peak is
clearly visible in the low-wave set, it has disappeared in the
high-wave set.

\begin{figure}[pb]
\begin{center}
\includegraphics[width=.8\textwidth]{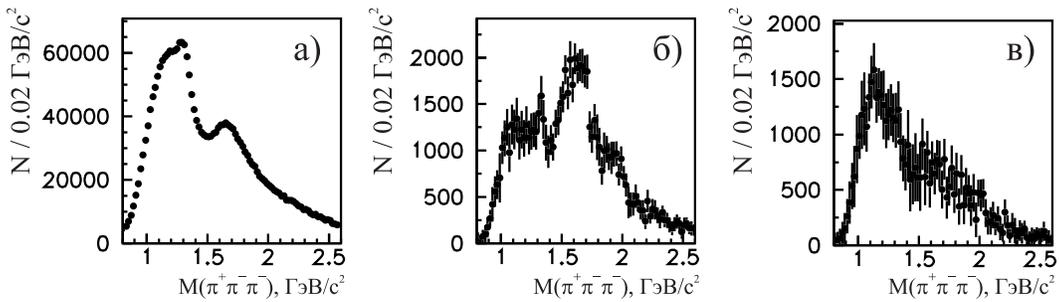}
\end{center}
\caption{\label{3pi_VES}
Results of PWA of the $\pi^+ \pi^- \pi^-$ system: (a) total intensity;
intensity of the $J^{PC}M^{\eta}=1^{-+}1^+$ wave in the $\rho \pi$
channel(b) under the assumption of full coherence of the $2^{-+}$ waves
and (c) in the case where this assumption is not used \cite{Amelin:2005ry}.}
\end{figure}

A similar observation was made by the VES collaboration analysing an
even higher-statistics data set
\cite{Amelin:1995gu,Kachaev:2001jj,Amelin:2005ry}. The $3\pi$ mass
distribution (a) shows a large enhancement due to $a_1(1260)/a_2(1320)$
production, followed by a peak which is dominantly due to
$\pi_2(1670)$. This wave is rather strong in the data; when full
coherence is required in the PWA mode, a signal in $J^{PC}=1^{-+}$
waves of both parity-exchange naturalities is observed, likely due to
leakage from the much stronger $2^{-+}$ wave \cite{Amelin:2005ry}. The
1600\,MeV/c$^2$ bump in Fig. \ref{3pi_VES}(b) disappears in the full model
allowing for an arbitrary incoherence in the $2^{-+}$ wave, see
Fig. \ref{3pi_VES}(c). In the unnatural parity exchange sector
($M^{\eta} =0^-, 1^-$), the intensity of the exotic signal at $M
\approx 1.6$ GeV/c$^2$ is compatible with zero.

A broad signal is seen in natural parity exchange ($M^{\eta} = 1^+$)
with maximum at $M \approx 1.2$ GeV/c$^2$. This low mass peak can be
assigned to leakage from the very intensive wave to $J^{PC}=1^{-+}$.
The intensity is about $2\%$ of the $a_2(1320)$ signal. Because of its
smallness and broadness, the signal is not believed to represent a true
exotic signal.

The three analyses of VES and E-852 data arrive at very different
results and it is hard to decide objectively, which analysis is right.
The approach in \cite{Amelin:2005ry} and \cite{Dzierba:2005jg}
allowing for more flexibility in the fit is certainly more
conservative. However fitting is an art, and it is not inconceivable
that too many fit parameters lead to an overparametrisation which
whashes out a signal which is observed only when just the right waves
are introduced and no additional spurious waves. At Hadron05, Adams
showed results of the partial wave analysis of Dzierba et al. (claiming
absence of $\pi_1(1600)\to\rho\pi$) on the $2^{-+}$ and $1^{-+}$ waves.
Adams fitted the data of Dzierba et al. in a new mass-dependent fit
which returned -- within errors -- the $\pi_1(1600)$ Breit-Wigner
parameters.

The BNL-E852 collaboration studied the stability of the $1^{-+}$ signal
and its phase motion with respect to the $2^{-+}$ wave.  The amount of
the $1^{-+}$ wave and the width proved to be unstable; it varied
from a clear, unmistakable signal with 'narrow' width to a vanishingly
small number of $1^{-+}$ events and a very wide width, depending on the
number of partial waves introduced and, in particular, if rank 1 or 2
of the density matrix was assumed in the analysis. It is clear that, if
one attempts to claim a new state or its decay mode, its proof must
remain stable against the number waves introduced or the rank of the
density matrix. For this reason they relied on the phase motion
$\Delta\phi=\phi(1^{-+})-\phi(2^{-+})$. In all of their numerous fits,
the phase motion remained stable. So the authors were confident
that their choice of the phase motion made sense. The instability in
the amount and the width of $\pi_1(1600)\to\rho\pi$ led to the highly
asymmetrical error in the width given in eq. (\ref{mw1600})
\cite{Adams:1998ff,Chung:2002pu}.

The Breit-Wigner parameters of the observed $\rho\pi$ exotic signal
\cite{Adams:1998ff,Chung:2002pu} coincide within errors with those of
the signal seen in the $\eta'\pi$ channel. However, the $\pi_1(1600)$ in
$\eta'\pi$ is seen in natural parity exchange, and not in unnatuarl
parity exchange. From this observation, eq. (\ref{1600rhopi}) was
derived.  The signal in $\rho\pi$ is observed with comparable
intensities in both, natural and unnatural parity exchange. Hence there
must be two different mass degenerate objects, one decaying into
$\rho\pi$, the other one into $\eta'\pi$. The need for two resonances
degenerate in mass is difficult to accept and, given the differences in
the published approaches and results, the $\pi_1(1600)\to\rho\pi$
decay remains a controversial issue.

\subsubsection{The wave $J^{PC}=1^{-+}$ in the $\omega \pi \pi$ channel.}
The wave $J^{PC}=1^{-+}$ in the $\omega(\pi^+\pi^-\pi^0) \pi^-\pi^0$
channel was studied by VES \cite{Amelin:1999gk,Amelin:2005ry} and E-852
\cite{Lu:2004yn}. Three isobars $\omega \rho,~ b_1 \pi$ and
$\rho_3 \pi$ were included in the PWA model, with waves in the range $J
\le 4,~L \le 3$, and $m \le 1$, with $J$ as total angular momentum, $L$
as decay orbital-angular momentum, and $m$ as projection of $J$
onto the beam axis.

\begin{figure}[ph]
\begin{center}
\includegraphics[width=0.6\textwidth]{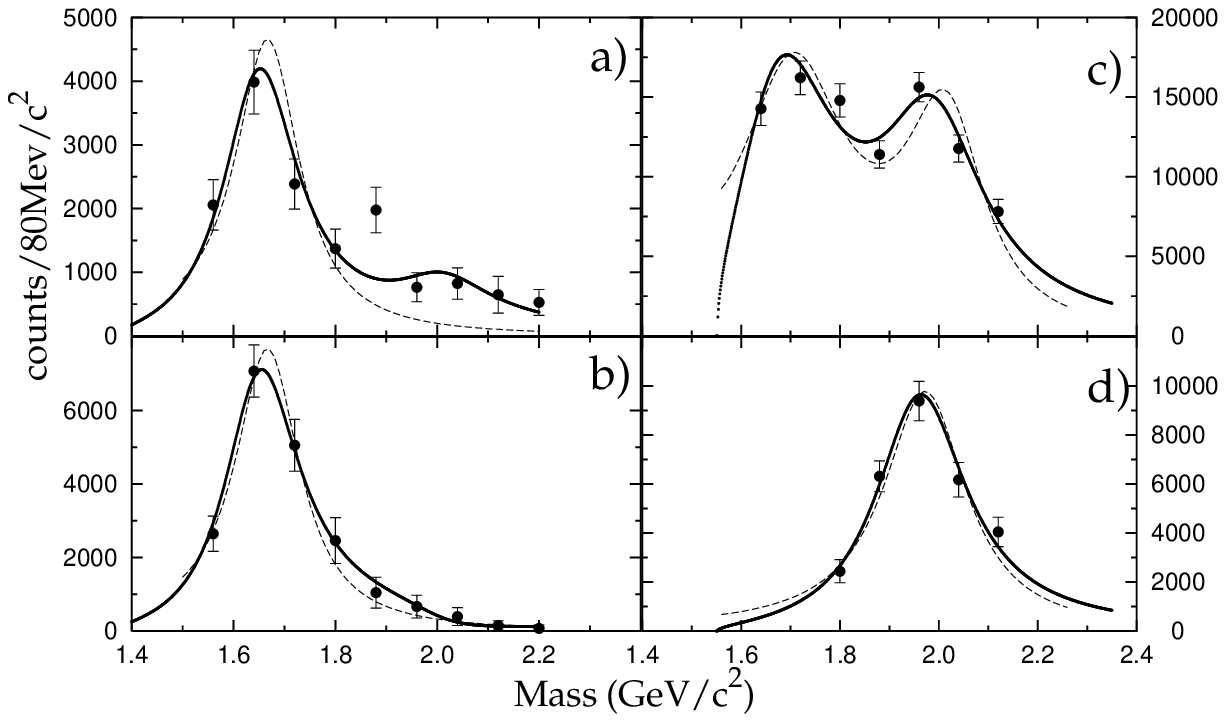}
\end{center}
\caption{Wave intensity for (a) $1^{-+}(b_1\pi)^S_1 1^+$, (b)$1^{-+}(b_1\pi)^S_1
 0^-$, (c) $2^{++} (\omega \rho)^S_2 1^+$, and (d) $4^{++}(\omega \rho)^D_2
 1^+$. The solid line is the result of Breit-Wigner fit for two poles and
 dashed line is for one.}
\label{ompipi}
\end{figure}

The results of E852 are summarised in Fig.~\ref{ompipi}. Two bumps at
$M \approx 1.7$ GeV/c$^2$ and at $M \approx 2.0$ GeV/c$^2$ are seen in
the dominant $2^{++} (\omega \rho)^S_2 1^+$ wave, a further peak in the
$a_2(1320)$ region is not shown. There are two intensive exotic waves
$1^{-+}(b_1\pi)^S_1 1^+$ and $1^{-+}(b_1\pi)^S_1 0^-$ peaking at $M
\approx 1.6$ GeV/c$^2$. These three waves can be fitted with four
Breit-Wigner resonances $a_2(1700),~a_2(2000),~\pi_1(1600)$ and
$\pi_1(2000)$. Mass and width of $\pi_1(1600)$ were determined to

$~~~M = 1664\pm 8\pm 10$\,GeV/c$^2$, $\Gamma = 184\pm 25\pm
28$\,GeV/c$^2$.

The production characteristics resembles that of the $\pi_1(1600)$ in
$\rho\pi$; both have nearly equal intensities in natural and unnatural
parity exchange. One more exotic resonance is suggested at $M \approx
2.0$ GeV/c$^2$ to fit the tail in the wave $1^{-+}(b_1\pi)^S_1 1^+$.

In the VES experiment, the dominant $2^{++} (\omega \rho)^S_2 1^+$ wave
has a broad bump at $M \approx 1.7$ GeV/c$^2$, the signal at $M
\approx 2.0$ GeV/c$^2$ is not seen. The exotic wave $1^{-+}(b_1\pi)$ is
significant only in natural parity exchange and gives a broad
contribution of low intensity, about 15\% of the leading $2^{++}$ wave.
A fit with a Breit-Wigner amplitude shows that the data are consistent
with resonant behaviour of the amplitude with $M\sim 1.6,
\Gamma=0.33$\,MeV/c$^2$ \cite{Dorofeev:2001xu}.

The three signals in $b_1(1235)\pi$, $\eta^{\prime}\pi$ and $\rho\pi$
have the relative strength $1:1\pm0.3:1.6\pm0.4$
\cite{Khokhlov:1999aq}. But the coherence study \cite{Kachaev:2001jj}
suggested to remove $\rho\pi$ from this comparison.

\subsubsection{The wave $J^{PC}=1^{-+}$ in the $f_1 \pi$ channel.}
\begin{figure}[pb]
\begin{center}
\includegraphics[angle=-90,width=.6\textwidth]{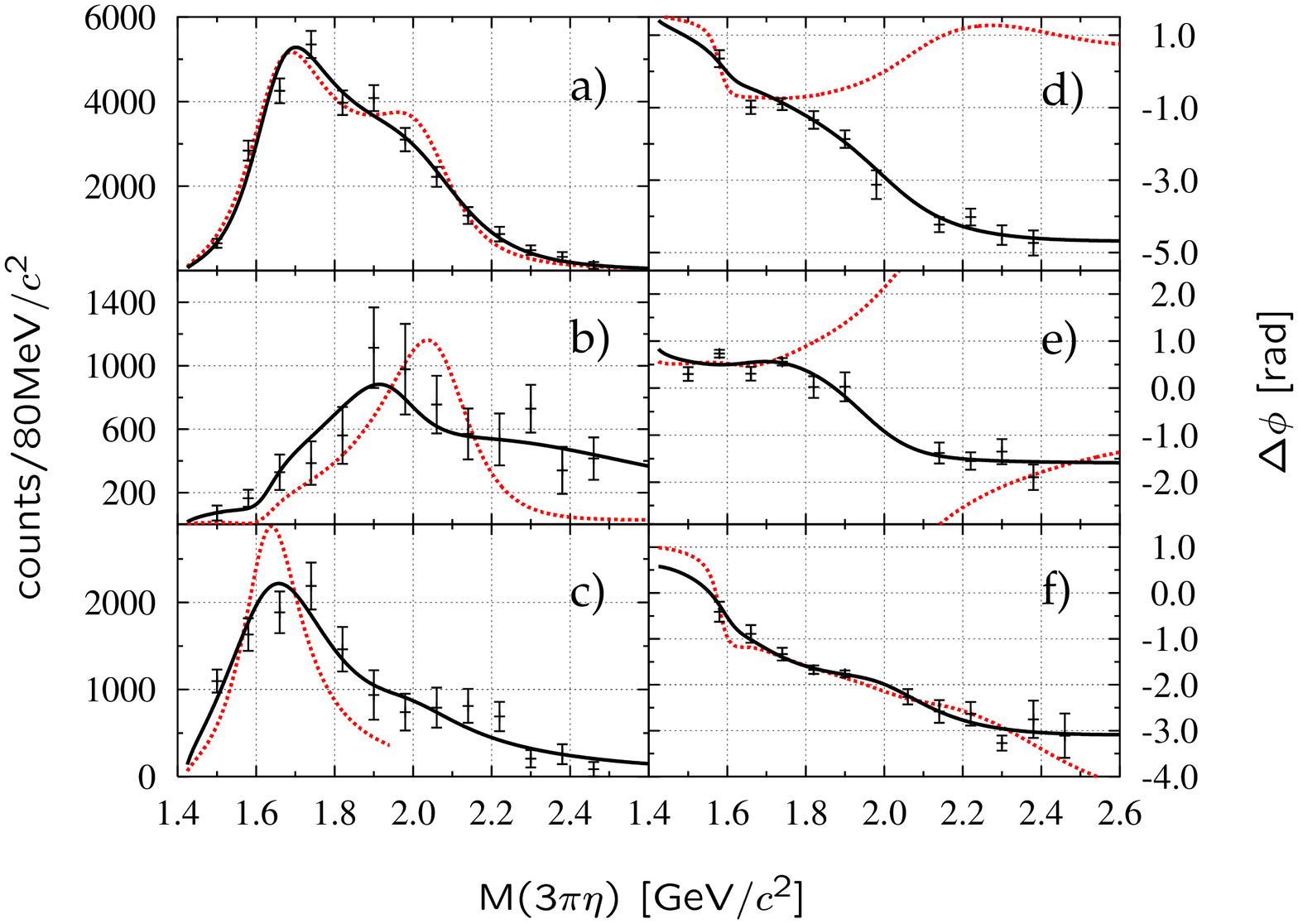}
\end{center}
\caption{PWA
results: $f_1(1285) \pi^-$ intensity distributions (a) $1^{++}0^+f_1
\pi^- P$, (b) $2^{-+}0^+ f_1 \pi^- D$, (c) $1^{-+} 1^+ f_1 \pi^- S$ and
phase difference distributions (d) $\phi(1^{-+})-\phi(2^{-+})$, (e)
$\phi(1^{-+} - \phi(1^{++})$, (f) $\phi(1^{++}) - \phi(2^{-+})$. The
results from a last squares fit are overlaid as the solid line (two
poles in the $1^{-+} f_\pi$ wave) and the dashed line (one pole in the
$1^{-+}$ wave) \cite{Kuhn:2004en}.}
\label{ex:f1pi}
\end{figure}

The E-852 results on the reaction $\pi^- A \to \eta \pi^+ \pi^- \pi^- A$
are published in \cite{Kuhn:2004en} and those of VES in
\cite{Amelin:2005ry}. The results of the partial wave
analyses of both experiments are again very similar, and it is
sufficient to show, in Fig. \ref{ex:f1pi}, one data set only.

The dominant wave is $J^{PC}=1^{-+}(f_1 \pi)$. It is broad and
structureless with a maximum at $M \approx 1.7$ GeV/c$^2$ (Fig.
\ref{ex:f1pi}(a)). The signal in the wave $J^{PC}M^{\eta}=1^{-+}1^+$ is
clearly seen  (Fig. \ref{ex:f1pi} (c)). It is
produced via natural parity exchange; it resembles in production
characteristics the $\eta^{\prime}\pi$ exotic wave.

The shape of the signal is very similar to the shape of the
$1^{++}0^+(f_1\pi)$ wave. The phase differences $\phi(1^{-+})-
\phi(1^{++})$ and $\phi(1^{-+})-\phi(2^{-+})$ both fall by $\approx
2$rad as $M_{\eta 3\pi}$ increases from $M=1.5$ GeV/c$^2$ to $M=2.4$
GeV/c$^2$. Hence the $1^{++}$ and $2^{-+}$ phases rise faster than
$\phi(1^{-+})$. The number of $1^{++}$ and $2^{-+}$ resonances which
the data can accommodate thus determines the number of  $1^{-+}$
resonances.

The E-852 collaboration fits the PWA intensity distributions and phase
differences with a superposition of Breit-Wigner resonances in all
channels. Any type of smooth background is not admitted. In this
resonance-dominated approach, several resonances are needed to
get a good fit to the distributions. These are listed, with the results
on masses and width, in Table~\ref{f1pibw}.
\begin{table}[H]
\caption{\label{f1pibw} Results of the mass-dependent fit.
\vspace*{2mm}}
\begin{center}
\renewcommand{\arraystretch}{1.4}
\begin{tabular}{c c c}
\hline\hline
Wave & Mass [MeV/c$^2$] & $\Gamma$ [MeV/c$^2$] \\ \hline
  &  $1714$ (fixed)  &  $308$ (fixed) \\
\raisebox{2.5mm}[2mm][0mm]{$1^{++}0^{+}f_{1}\pi P$}  &  $2096 \pm 17
\pm 121$  &  $451 \pm 41 \pm 81$ \\ \hline
  &  $1676$ (fixed)  &  $254$ (fixed) \\
$2^{-+}0^{+}f_{1}\pi D$  &$2003\pm 88\pm 148$&$306\pm 132\pm 121$ \\
  &  $2460 \pm 328 \pm 263$  &  $1540 \pm 1214\pm 718$\\ \hline
  &  $1709 \pm 24 \pm 41$  &  $403 \pm 80 \pm 115$\\
\raisebox{2.5mm}[2mm][0mm]{$1^{-+}1^{+}f_{1}\pi S$}  &  $2001 \pm 30
\pm 92$  &  $333 \pm 52 \pm 49$ \\ \hline\hline \end{tabular}
\renewcommand{\arraystretch}{1.4}
\end{center}
\end{table}

\subsubsection{Conclusion on $J^{PC}=1^{-+}$ exotics}
Before coming to conclusions on $J^{PC}=1^{-+}$ exotics, the
assumptions will be discussed which are made in the data analysis.  As
a rule, a partial wave analysis is performed in two steps. The first
step is the mass-independent PWA which serves to construct the
parameters of a density matrix $\rho_{ij}$ which represents the final
state $X$ with mass $M$. In a second step, these parameters are fitted
in some model in a mass-dependent fit.

The isobar model relies on the basic assumption that all processes
involved can be decomposed in two-body  subprocesses. For example, the
reaction $\pi^- p \to \pi^+ \pi^- \pi^- p$ is considered as the sum of

$~~~\pi^- p\to A_i p$ with subsequent decays of $A_i\to B_j\pi$ and $B_j\to\pi\pi$

$~~~\pi^- p \to C_i D_j $ with subsequent decays $C_i\to \pi\pi$ and $D_j\to\pi p$

$~~~\pi^- p \to \pi E_j $ with subsequent decays $E_i\to\pi F_j$ and $F_j\to\pi p$

In general, each partial wave may have resonant and non-resonant
contributions. But often it is assumed that all decay channels are
saturated with reasonably narrow resonances. Even with these
approximations, the large quantity of free parameters may lead to the
necessity of various truncations like a reduction in the list of
isobars due to limitations in $J_{max}$ of contributing isobars.

A first assumption which is generally made supposes that the three
reactions listed above live in different corners of the phase space
and can be separated by kinematical cuts. At sufficiently high energy,
this is the case, but at  moderate energies the assumption does not
need to be absolutely true. But it is essential for the extraction of
weak exotic $J^{PC}=1^{-+}$ wave from some reactions like $\pi^- p \to
\eta \pi^- p$. The problem is that the $\eta p$ and $\pi^- p$ systems
are very different in nature and have different isospin decompositions.
Therefore the angular distribution of the $\eta \pi$ system on
$\theta_{GJ}$ will be asymmetric as soon as nucleon excitations are not
completely ruled out. Within a standard PWA model, this asymmetry is
then assigned to the interference of the dominant $D_+$ wave and a
wave of opposite parity, $P_+$. Thus, the effect can generate an
effective resonant-like exotic wave $J^{PC}=1^{-+}$ in the $a_2$ region.

Another potential sources of false signals are parametrisations of
isobars and integrations over $t$. The different partial waves depend
on $t$ in different ways; integration over $t$ decreases therefore the
coherence between partial waves. Quite generally, any model
imperfections decrease the coherence and may generate false signals in
low intensity waves by leakage from most intensive waves.
The importance of leakages depends on the PWA model. In case of a
density matrix of rank one with $n$ waves, there are $n$ different
functions to absorb these imperfections. As a result, leakage should be
expected at the level of $\alpha/n$, where $\alpha \approx 0.1$ is the
scale of imperfection of the model. A density matrix of arbitrary
rank has $n(n+1)/2$ different functions resulting to a significant
reduction of the leakage problem. Therefore a density matrix of highest
possible rank is recommended for the study of low intensity waves. Of
course, higher-rank fits require significantly higher statistics since
more parameters need to be determined.

Summarizing the progress in studies of the $J^{PC}=1^{-+}$ hybrid
mesons achieved since the first observation of the exotic wave we
arrive at the following conclusions:

\begin{itemize}

\item Exotic waves are observed in numerous final states with
comparable intensities in diffraction-like reactions. Only one
of them is confirmed in a non-diffractive process, in $ \bar pp$
annihilation.

\vspace{2mm}\item Even at low energies, several different intermediate
states are separately identified due to their different production
characteristics:

 \begin{enumerate}

\vspace{2mm}\item $\pi\eta$ in SU(3) $(10,\bar 10)$ with strong
intensity at 1400\,MeV/c$^2$.

\vspace{2mm}\item Possibly $\rho\pi$ at  1400\,MeV/c$^2$ in $\bar pp$
annihilation, but from a different initial $ \bar pp$ state.

\vspace{2mm}\item $\pi\eta^{\prime}$ and $f_1(1285)\pi$ in natural
parity exchange, with strong intensity at 1600\,MeV/c$^2$.

\vspace{2mm}\item $b_1(1230)\pi$ and $\rho\pi$ in natural and unnatural
exchanges, with strong intensity at 1600\,MeV/c$^2$.

\end{enumerate}

\vspace{2mm}\item Results requiring unnatural parity exchange are
controversial.

\vspace{2mm}\item The data are consistent with both, resonant and
non-resonant interpretations; a decision which interpretation is
correct requires calculations of meson-meson interaction amplitudes.

\vspace{2mm}\item There is a chance that the lowest mass exotic hybrid
has been discovered. It would have  1600\,MeV/c$^2$ mass and a width of
300\,MeV/c$^2$. It couples to $\pi\eta^{\prime}$ and $f_1(1285)\pi$.
However, more work is required to establish or to reject this
possibility.

\end{itemize}

\subsection{Further $J^{PC}$ exotic waves}

\subsubsection{$J^{PC} = 0^{+-}, 2^{+-}$}

Predictions for quantum numbers, masses, widths and branching ratios of
hybrid mesons were calculated in the flux-tube model \cite{Isgur:1985vy}.
The lowest hybrids are the states with total orbital angular momentum
$l=1$ and one phonon of transverse flux-tube vibration which carries
one unit of angular momentum around the $q \bar q$ axis; positive and
negative parity eigenstates are found. Combined with the quark spin
$S=0,~1$, there are eight low-mass hybrid states with
$$J^{PC}= ~1^{\pm \pm};~~~0^{\pm \mp},~1^{\pm \mp},~2^{\pm \mp}.$$
Three out of these eight states have exotic quantum numbers, $J^{PC}=
~0^{+-} ,~1^{-+},~2^{+-}$. Isovectors  and isoscalars are expected at
$M \approx 1.8 \div 2.0$ GeV/c$^2$. The isovector exotic state with
$J^{PC}=1^{-+}$ appears to be the most convenient one for experimental
studies. It is predicted to be relatively narrow ($\Gamma \approx 0.2$
GeV/c$^2$), it can be produced by a pion beam in diffractive-like
reaction, and its dominant decay modes $f_1 \pi,~b_1 \pi$ are suitable
for detection and for partial-wave analysis .

Other predicted isovector mesons with exotic quantum numbers
($J^{PC}=0^{+-},~2^{+-}$) are much less appealing. These states are
found to be wide $\Gamma > 0.5$ GeV/c$^2$, their dominant decay modes
end up in a tetrapion final state which is difficult for the
partial wave analysis; the cross sections for production in a pion beam
must be relatively small due to the positive $G$-parity of the states.

For searches of isoscalar exotic states, the quantum numbers
$J^{PC}= 0^{+-}, 2^{+-}$ look most promising. These states are not very
wide $\Gamma < 0.5$ GeV/c$^2$ and have convenient decay modes like $b_1
\pi$. The reaction $\pi^- p \to b_1 \pi n$ requires $b_1$ exchange and
should have a very specific broad distribution as a function of squared
transfer momentum $t$ .

Up to now all these states with exotic quantum numbers except of
$J^{PC}=1^{-+}$ remain unchartered territory.

\subsubsection{Isospin exotics}
Non-$q \bar q$ resonances could manifest themselves in isospin exotic
channels, like $I=2$ for nonstrange mesons and $I=3/2$ for mesons with
strangeness. Detailed experimental studies of $\pi^+ \pi^+$ and $K^+
\pi^+$ scattering \cite{Hoogland:1977kt,Estabrooks:1977xe} clearly show
negative (repulsive) phases from thresholds to $M \approx
1.8$\,GeV/c$^2$, in $\pi^+ \pi^+$ $S$- and $D$-waves and in $K^+
\pi^+$ $S$- and $P$-waves. These results agree with calculations based
on an Effective Lagrangian \cite{Chan:1974rb} and on a Quark Exchange
model \cite{Barnes:2000hu} (but are at variance with
the narrow $\pi^+\pi^+$ resonance at 1420\,MeV
reported by the Obelix collaboration \cite{Filippi:2002rd}).

\begin{figure}[pb]
\begin{center}
\includegraphics[width=.7\textwidth]{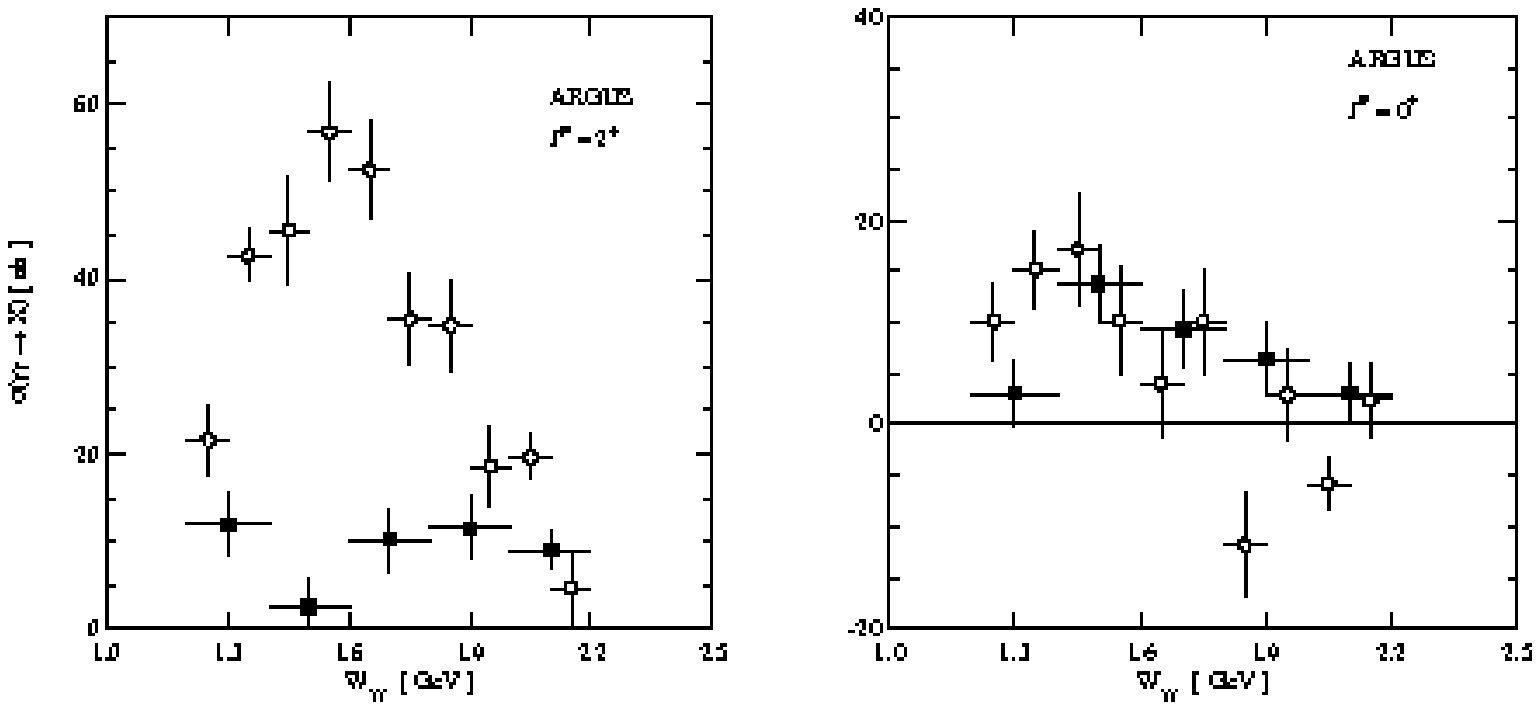} \end{center}
\caption{Cross sections for the dominant amplitudes $\gamma \gamma \to
\rho^0 \rho^0$ (open circles) and $\gamma \gamma \to \rho^+ \rho^-$ (full
squares) \cite{Albrecht:1996gr}.}
\label{fig:rhorho}
\end{figure}

The situation is different in the isotensor $I=2~~~\rho \rho $ channel.
In two-photon fusion into two $\rho$ mesons, in $\gamma \gamma \to
\rho^0 \rho^0$, a resonance-like enhancement at $M \approx 1.6$
GeV/c$^2$  was observed in several experiments. The quantum numbers
of this signal are \cite{Albrecht:1990cr}: $(J^P,~J_z,~ S) = (2^+,~\pm
2,~ 2)$ (where $J^P$ are spin and parity, $J_z$ spin component in the
beam direction and $S$ the total spin of two $\rho$'s). The $\rho^0
\rho^0$ signal is much stronger than one in $\gamma\gamma\to\rho^+
\rho^-$ (Fig. \ref{fig:rhorho}).

This result was confirmed by L3
\cite{Achard:2003qa,Achard:2004ux,Achard:2004us,Achard:2005pb} with
much higher statistics. If the $\rho^0\rho^0$ bump has a defined
isospin $I=0$ or $I=2$, the ratio of $\rho^+ \rho^-$ to $\rho^0 \rho^0$
should be $2$ (for $I=0$) or $1/2$ (for $I=2$); $I=1$ is forbidden for
$\rho^0 \rho^0$. The experimental ratio is about $1/5$. This can
be achieved assuming interference between a isotensor resonance ($I=2$)
with some isoscalar contribution. The existence of tetraquark states
with $I=0$ and $I=2$ and their destructive interference in $\gamma
\gamma \to \rho^+ \rho^-$ and constructive interference in $\gamma
\gamma \to \rho^0 \rho^0$ was predicted a long time ago by Achasov,
Devyanin and Shestakov  \cite{Achasov:1982bt} and by Li and Liu
\cite{Li:1981ez}. The data were thus be claimed to prove the existence
of tetraquark isotensor resonances at $M\approx 1600$ MeV/c$^2$, $\Gamma
\approx 500$ MeV/c$^2$ \cite{Achasov:2000pj,Achasov:1999ha,Achasov:1999cu}.
Calculations within QCD sum rules \cite{Wei:2004tc} do no exclude
such resonances. However, an interpretation of the data as generated by
Pomeron exchange between the two $\rho^0$ -- present due to vector
meson dominance -- does not seem unlikely.

Several other final states with two vector mesons were studied in
$\gamma \gamma$ fusion, $\omega \omega,~ \omega \rho^0$, $K^{0*} \bar
K^{0*}$, $K^{+*} \bar K^{-*}$, $\rho^0 \phi$ and $\omega \phi$
\cite{Albrecht:1994hg,Albrecht:1999zu}. Very wide threshold
enhancements are seen in most of these channels. The  cross sections
vary from $\approx 60$ nb for $\gamma\gamma\to\rho^0\rho^0$ to $\approx
2$ nb for $\gamma \gamma \to \rho \phi$. A clear $q^2 \bar
q^2$-resonance-dominated view did not emerge from the studies. The
ratio of cross section for $~K^{0*} \bar K^{0*}$ and $~K^{+*} \bar
K^{-*}$ is about $1/8$. This number does not fit into any model with
$S$-channel resonances. If a resonance would contribute, it must
decay to $\rho \phi$; experimentally, this channel is strongly
suppressed compared to $~K^{+*} \bar K^{-*}$. These features
indicate that nonresonant dynamics play a decisive $\rm r\hat{o}le$. We have to
underline that broad threshold enhancements can appear without
$s$-channel resonance contributions; this is the case, for
example, in the reaction $K^+ p \to K^0 \Delta^{++}$.

A description of the data without $s$-channel resonances can be achieved
within a phenomenological model based on the assumption of $t$-channel
factorisation \cite{Alexander:1990bn}. In this model the cross section
for $\gamma \gamma \to \rho^0 \rho^0$ can be calculated absolutely from
experimental data on the reaction $\gamma p \to \rho p$. Even though
data are successfully described by this model, it does not give a
detailed microscopic picture of the processes (and does not aim for
this). Quark-model calculations of these reactions
are reviewed in \cite{Rosner:2004nh}.

In conclusion we have to say that the dynamics of two mesons in
isotensor configuration is not yet fully understood. A detailed partial
wave analysis of channels like $\pi^+ \pi^+$, $\rho^+ \rho^+$, $\pi^+
\pi^+ \pi^0$, $\pi^+ \pi^+ \eta$, $\pi^+ \pi^+ \eta'$, $\pi^+ \pi^+
\omega$ in different reactions could bring a better understanding
of multiquark and multimeson dynamics, and possibly lead to new
insights.


\markboth{\sl Meson spectroscopy} {\sl Pseudoscalar mesons}
\clearpage\setcounter{equation}{0}\section{\label{Pseudoscalar mesons}
Pseudoscalar mesons}

The radial excitations of the pseudoscalar ground--state nonet provide
answers to important questions: how do radial excitation energies of
light mesons compare to the mass gap between $\eta_c(1S)$ and
$\eta_{c}(2S)$? Due to their Goldstone nature, pseudoscalar
ground-state mesons (except $\eta^{\prime}$) have a much smaller mass
than vector mesons have; is this pattern reproduced for $n=2$ (and
higher) excitations? The mixing angle of the ground-state nonet is far
from the ideal mixing angle; what is its value for pseudoscalar radial
excitations? The $\eta^{\prime}$ mass is large compared to the $\eta$
mass due to the $U_A(1)$ anomaly; how large is the corresponding mass
difference for radial excitations? At the end of this section these
questions will be resumed to see to what extend data can provide
answers to them.

\subsection{\label{The pi radial excitations}
The $\pi$ radial excitations}

\subsubsection{\label{The pi(1300)}
The $\pi(1300)$}

The $\pi(1300)$ was first reported by  Daum {\it et al.}
\cite{Daum:1979iv} in 1981 in a (successful) attempt to identify the
$a_1(1260)$ in $ \pi^- p$ scattering. The Deck effect (see section
\ref{The Deck effect}) provides a substantial background in the
$\rho\pi$ $S$-wave and carries a phase motion in itself. Hence it was
difficult to isolate the $a_1$. In addition to the $a_1(1260)$ in the
$\rho\pi$ $S$-wave, a resonance in the $\rho\pi$ $P$-wave was found, with
quantum numbers $ (I^G)J^{PC} = (1^-)0^{-+}$ and a mass of $\sim
1400$\,MeV/c$^2$. Its mass and quantum numbers suggested its
interpretation as first radial excitation of the pion. The existence of
$\pi(1300)$ was confirmed in several similar experiments, even though
the measured masses scattered over a wide range
\cite{Bonesini:1981sx,Aaron:1980zk,Bellini:1982ec,Zielinski:1984mt}.

The VES collaboration observed a clear low-mass enhancement in the
$\pi(\pi\pi)_{S-wave}$ \cite{Amelin:1995gu}, but demonstrated that
there is a strong influence of the Deck effect for this partial wave
which may fake a $\pi(1300)$ at a mass of about 1200\,MeV/c$^2$. The
authors of \cite{Amelin:1995gu} therefore refused to give any mass or
width for the $\pi(1300)$. The importance of the Deck effect for
$\pi(1300)\to\pi(\pi\pi)_{S-wave}$  decays was not discussed in the
early papers, and some of those claims may have been premature.

The E852 collaboration \cite{Chung:2002pu} saw the same enhancement and
observed a similar phase motion in the $\pi(\pi\pi)_{S-wave}$ as VES did.
The fit to the $\pi(\pi\pi)_{S-wave}$ proved to be very unstable and,
in view of the unexplored influence of the Deck effect, observation of
$\pi(1300)\to\pi(\pi\pi)_{S-wave}$ was not claimed. In the
$(\pi\rho)_{P-wave}$, the partial wave analysis did reveal a clear
resonant behaviour which was interpreted as $\pi(1300)$. Mass and width
are listed in Table~\ref{tab:1300}.

\begin{table}[pb]
\caption{\label{tab:1300} The $\pi(1300)$ mass and width from
$\pi(1300)\to\rho\pi$ decays \vspace{2mm}}
\bc
\renewcommand{\arraystretch}{1.4}
\begin{tabular}{ccc}
\hline\hline
M                  & $\Gamma$          & Ref. \\
$\sim 1400$        &                   &  \cite{Daum:1979iv} \\
$1343\pm 15\pm 24$ & $449\pm 39\pm 47$ & \cite{Chung:2002pu} \\
$1375\pm 40$       & $268\pm 50$       & \cite{Abele:2001pv}\\
\hline $1373\pm 25$& $358\pm 70$       & mean \\
 \hline\hline
\end{tabular}
\renewcommand{\arraystretch}{1.0}
\ec
\vspace{-3mm}
\end{table}

The Obelix collaboration reported an analysis of $\bar pp\to
\pi^+\pi^+\pi^-\pi^-$~\cite{Bertin:1997vf}. The solution comprised a
$\pi(1300)$ with mass and width $(M, \Gamma )=(1275\pm 15, 218\pm
100)$\,MeV/c$^2$ decaying dominantly to $\pi(\pi\pi)_{S-wave}$. In
the Crystal Barrel experiment, data on $\bar pp\to 5\pi^0$ were
analysed \cite{Abele:1996fr}. The best solution had a very low
$\pi^{\prime}$ mass at 1122\,MeV/c$^2$. However, this low mass was
incompatible with the $\pi(1300)$ masses obtained when data on $\bar
pn\to \pi^-4\pi^0$ and $\bar pn\to 2\pi^-2\pi^0\pi^+$ were studied in a
combined analysis. In the latter analysis, the optimum $\pi(1300)$
parameters were $(M, \Gamma )=(1375\pm 40, 286\pm 50)$\,MeV/c$^2$. The
$\pi(1300)$ was seen only in its $\rho\pi$ decay, the coupling to
$\pi(\pi\pi)_{S-wave}$ converged to zero. The collaboration
reported an upper limit of 15\% for the $\pi(\pi\pi)_{S-wave}$
decay partial width.  Reanalysing the $\bar pp\to 5\pi^0$ data,  a
nearly equally  good fit was achieved without introduction of a
$\pi(1300)$~\cite{Abele:2001pv} while other resonance parameters were
not affected (in particular the scalar mesons did not change their
masses and widths). The results from \cite{Abele:1996fr} are therefore
superseded by the more recent values, given in~\cite{Abele:2001pv} and
reproduced in Table~\ref{tab:1300}.

In view of the importance of the Deck effect, we retain only results in
which the $\rho\pi$ decay mode surpasses in magnitude the
$\pi(\pi\pi)_{S-wave}$ decay mode. The existence of the latter decay
mode is of course not excluded, but we do not consider it as
experimentally established.

In the $^3P_0$ model, the $\rho\pi$ is expected to be the dominant mode
of the $\pi(1300)$ if interpreted as 2S $q\bar q$ state.
In~\cite{Barnes:1996ff}, a partial width of

\vspace{-5mm}
\begin{equation}
\label{g1300}
\Gamma(\pi(1300) \rightarrow \pi \rho) = 209 \; {\rm MeV/c^2} \ .
\end{equation}\vspace{-7mm}

is predicted while the $\pi(\pi\pi)_{S-wave}$ decay width is expected
to be small. With the latter decay mode being small, there is
reasonable consistency between the value (\ref{g1300}) and the average
of Table~\ref{tab:1300}. Thus the observed $\pi(1300)$ is consistent
with expectations for a 2$^1$S$_0$ $q\bar q$ state.

\subsubsection{ \label{The pi(1800)}
The $\pi(1800)$ }

The $\pi(1800)$ was first reported by Bellini~\cite{Bellini:1982ec} and
studied extensively by the VES collaboration. Its decay into
$\pi(\pi\pi)_{S-wave}$ and $\pi(\eta\eta)_{S-wave}$ are the most
prominent modes~\cite{Amelin:1995gu}. Further decay modes are
$\pi(\eta\eta^{\prime})$ \cite{Beladidze:1993km}, and $K^*_0(1430)\pi$
\cite{Berdnikov:1994kc}. The VES collaboration also reported
$\pi(1800)\to\omega\pi\pi$ decays~\cite{Amelin:1999gk} even though the
$\omega\pi\pi$ resonance mass was found at a much lower value.
Surprisingly, the $\pi(1800)$ is not seen in $\rho\pi$ or $
K^*\bar{K}$, both of these decay modes are suppressed
\cite{Amelin:1995gu,Chung:2002pu,Berdnikov:1994kc}.

\begin{figure}[pt]
\bc
\includegraphics[width=0.65\textwidth,height=0.38\textheight]{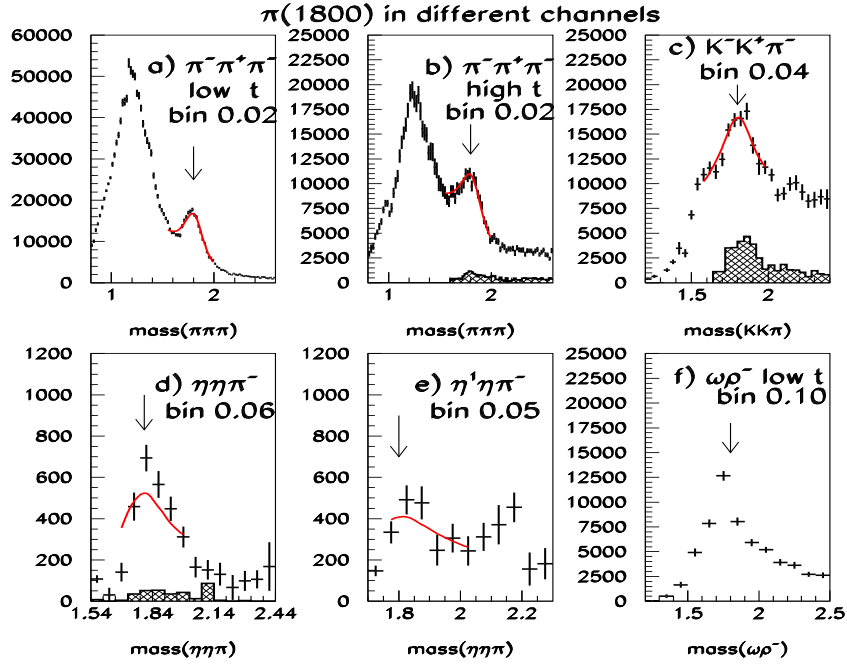}
\vspace{-3mm}
\ec
\caption{\label{pic:ves1800}
 Mass spectra for $(I^G)J^{PC}=(1^-)0^{-+}$ wave. The data points
 represent the results of a mass-independent partial wave analysis,
 the solid curve a fit using a $\pi(1800)$ with $\Gamma=270$\,MeV/c$^2$ and
 a polynomial background amplitude with  varying phases (constrained by
 a 2$^{nd}$-order polynomial).
}
\end{figure}

\begin{table}[pb]
\caption{\label{tab:pi1800}
Ratios of $\pi(1800)$ partial widths. The relative widths are not
corrected for further decay modes; the correction factors are
given in the $4^{\rm th}$ column. \vspace{2mm}}
\bc
\renewcommand{\arraystretch}{1.4}
\begin{tabular}{lccc}
\hline\hline Final state      & subchannel     & relative width  &decay
fraction\\ \hline $\pi^+\pi^-\pi^-$&                 &   1 &   \\
                 & $(\pi\pi)_{S-wave}\pi^-$  &   $1.1\pm 0.1$    & 2/3 \\
                 &$f_0(980)\pi^-$ &  $0.44\pm 0.15$   & 2/3$\cdot$0.84\\
                 &$f_0(1500)\pi^-$&  $0.11\pm 0.05$   & 2/3$\cdot$0.35\\
                 & $\rho^0\pi^-$  &$<0.02$ at 90\% c.l.& 1/2\\
$K^+K^-\pi^-$&                    &   $0.29\pm 0.10$  & \\
                 & $K^*(892)K^-$&$<0.03$ at 90\% c.l.& 1/4\\
$\eta\eta\pi^-$  &                &   $0.15\pm 0.06$  &  \\
                 & $a_0(980)\eta$&   $0.13\pm 0.06$   & 0.8\\
                 &$f_0(1500)\pi^-$&  $0.012\pm 0.005$ & 0.05\\
$\eta\eta^{\prime}\pi^-$&$f_0(1500)\pi^-$&$0.026\pm 0.010$ & 0.019\\
\hline\hline
\end{tabular}
\renewcommand{\arraystretch}{1.0}
\ec
\end{table}
Recently, the VES collaboration performed a combined analysis of all
reactions \cite{Nikolaenko:2004tz,Amelin:2005ry}, with only the
$\rho\omega$ channel excluded. Mass and width of the global fit are
compatible with $M=1800$\,MeV/c$^2$ and $\Gamma=270$\,MeV/c$^2$. From
the fit, ratios of partial width were determined which are reproduced
in Table~\ref{tab:pi1800}. Data and fit are shown in
Fig.~\ref{pic:ves1800}.

Some notations and values in the Table need explanation. For the
$(\pi\pi)_{S-wave}$ with mass up to $ \approx 1.4$\,GeV/c$^2$, the
parametrisation of Au, Morgan and Pennington \cite{Au:1986vs} was
chosen. The $f_0(980)$ is listed as a separate resonance; its
contribution was subtracted from the $S$-wave. The sum of decay
probabilities into $(\pi^+\pi^-)_{S-wave}\pi^-$ and into $f_0
(980)_{\pi^+\pi^-}\pi^-$ exceeds the total width of the
$(\pi^-\pi^+\pi^-)$ channel due to strong interference of these two
channels. The $ K^-K^+\pi^-$ channel is not divided into the $
(K^+K^-)_{S-wave} \pi^-$ and $K^-(K^+\pi^-)_{S-wave}$ isobars as their
separation proved to be very model dependent.

Table~\ref{tab:pi1800} also contains the fraction with which isobars
are observed in the given final state. To give partial decay widths is
problematic since the results from E782 and VES are often incompatible.
Furthermore, the $\pi(1800)\to\rho\omega$ contribution and high
multiplicity decays are unknown. Based on the graphs in
Fig.~\ref{pic:ves1800} we estimate the $\rho\omega$ contribution and
assume that no other decay modes exist. The partial decay widths
of Table \ref{1800partialwidth} thus present an educated guess.

\begin{table}[pt]
\caption{\label{1800partialwidth}
Estimated decay partial widths of $\pi(1800)$. The estimate assumes
absence of high multiplicity decay modes. There could be two different
mesons in the mass region (see text for a discussion). \vspace{2mm}}
 \bc
\renewcommand{\arraystretch}{1.2}
\begin{tabular}{lr}
\hline\hline
$\Gamma_{\pi(1800)\rightarrow\pi^- (\pi\pi)_{S-wave}}$&$ 100\pm
25$\,{\rm MeV/c$^2$}\\ $\Gamma_{\pi(1800)\rightarrow\pi^- f_0(980)}$&$
50\pm 20$\,{\rm  MeV/c$^2$}\\ $\Gamma_{\pi(1800)\rightarrow\pi^-
f_0(1500)}$&$ 25\pm 15$\,{\rm MeV/c$^2$}\\
$\Gamma_{\pi(1800)\rightarrow\rho\omega}$&$ 60\pm 30$\,{\rm
MeV/c$^2$}\\ $\Gamma_{\pi(1800)\rightarrow a_0(980) \eta}$&$ 16\pm
8$\,{\rm MeV/c$^2$}\\ $\Gamma_{\pi(1800)\rightarrow\rho\pi}$&$ 3\pm
3$\,{\rm MeV/c$^2$}\\ $\Gamma_{\pi(1800)\rightarrow K^*K}$&$ 6\pm
6$\,{\rm MeV/c$^2$}.\\ \hline\hline \end{tabular}
\vspace{2mm}\renewcommand{\arraystretch}{1.0} \ec
\begin{center}
\caption{\label{tab:pi1800dec}
Model predictions for $\pi_H(1800)$ and for $\pi_{3S}$
\cite{Close:1994hc,Barnes:1995hc,Page:1998gz,Barnes:1996ff}.
\vspace{2mm}}
\renewcommand{\arraystretch}{1.4}
\begin{tabular}{ccccccc} \hline\hline
 & $\rho \pi$ & $\rho \omega$ & $\rho(1470) \pi$ & $f_0(1370) \pi$ & $
f_2 \pi$ & $ K^* K$ \\ \hline $3S$ & 30 & 74 & 56 & 6 & 29 & 36 \\
$\pi_H$ & $30 \div 50$ & 0 & $10 \div 50$ & $40 \div 170$ & $3 \div 8$ & $ 5 \div 15$ \\
Expt. & $<6$ & $60\pm 30$ & small & $100\pm 25$ & small & $ <12$ \\
\hline\hline
\end{tabular}
\renewcommand{\arraystretch}{1.0}
\end{center}
\end{table}

The decay pattern is not easily understood. Two directions were
tried: the assumption that the 1800\,MeV/c$^2$ region houses two states
with pion quantum numbers \cite{Barnes:1996ff}, and the hypothesis that
special selection rules govern $\pi(1800)$ decays \cite{Amelin:2005ry}.
In the mass range of the $\pi_1(1800)$, two pseudoscalar resonances are
predicted, the second radial excitation of the pion and a hybrid meson.
To distinguish these two possibilities we have to rely on model
predictions. In Table \ref{tab:pi1800dec}, predictions for the two
alternatives are collected
\cite{Close:1994hc,Barnes:1995hc,Page:1998gz,Barnes:1996ff}.

\begin{itemize}

\item A hybrid $\pi_H(1800)$ and a pseudoscalar radial excitation
$\pi_{3S}$ are both predicted to have a $\rho\pi$ partial width of
30\,MeV/c$^2$, incompatible with the experimental finding. This may
serve as a warning that the identification of the nature of $\pi(1800)$
is model dependent. Other well established pseudoscalar resonances like
$\pi(1300)$ and $ K(1460)$ decay dominantly into a vector and a
pseudoscalar meson. The pattern resembles the $\rho\pi$ puzzle in
charmonium decays: the decay fraction J/$\psi\to\rho\pi$ is large while
$\psi^{\prime}\to\rho\pi$ is suppressed.

\vspace{2mm}\item A distinctive feature of $\pi(1800)$ is the
$f_0(1370)\pi$ decay mode\footnote{\footnotesize In section \ref{Scalar
mesons above 1 GeV} it is argued that $f_0(1370)$ is not a resonance
with a proper phase motion.}. A hybrid $\pi_H(1800)$ is predicted to
have a large $f_0(1370)\pi$ partial width (up to 170\,MeV/c$^2$) while
$\pi_{3S}$ is predicted to have a partial decay width for this decay of
6\,MeV/c$^2$ only.

\vspace{2mm}\item A hybrid $\pi_H(1800)$ should decouple from
$\rho\omega$ while a large partial decay width (74\,MeV/c$^2$) is
predicted for $\pi_{ 3S}$.

\vspace{2mm}\item The isobars $\rho(1470)\pi$ and $f_2(1270)\pi$ were
not needed in the partial wave analysis; presumably they are small. The
$\pi(3S)$ radial excitation is however expected to decay into these
modes.

\vspace{2mm}\item The $ K^* K$ contribution is compatible with zero; a
$\pi_{ 3S}$ state is predicted to couple significantly to $ K^*
K$.

\end{itemize}

The data seem to favour the hybrid interpretation except for the large
$\rho\omega$ decay width which should vanish for a hybrid. This partial
width is predicted to be strong for $\pi(3S)$, in agreement
with the experimental result. Hence there is the possibility that two
states have been observed, the $\pi_{ 3S}$ and a hybrid $\pi_H$, as
pointed out by Barnes~\cite{Barnes:1996ff}. The $\rho\omega$ signal
peaks at a lower mass than the other mass distributions in
Fig.~\ref{pic:ves1800}. The authors in~\cite{Amelin:1999gk} quote
$(M,\Gamma)=(1737\pm5\pm15, 259\pm19\pm6)$\,MeV/c$^2$ which seems
inconsistent with the nominal mass value. We also note that
$\pi(1800)$ peaks at a higher mass, at $\sim 1830$\,MeV/c$^2$ in its
$\pi(\eta\eta)_{S-wave}$ and $\pi(\eta\eta^{\prime})$ decay modes which
go largely via $f_0(1500)\pi$.

The decays into $\pi (\pi\pi)_{S-wave}$ pose a difficulty. In the
VES analysis the $(\pi\pi)_{S-wave}$ includes the 1300\,MeV/c$^2$
region. In this mass region, there is the $f_0(1370)$. In  Table
\ref{tab:pi1800dec}, the full $\pi (\pi\pi)_{S-wave}$ is ascribed
to the $f_0(1370)$ even though it is known that it contributes only
little to $\pi (\pi\pi)_{S-wave}$. The $\pi (\pi\pi)_{S-wave}$
background needs to be subtracted, unseen $f_0(1370)$ decay modes would
need to be accounted for. Both is impossible at the moment. Hence the
$\pi(1800)\rightarrow\pi^-f_0(1370)$ decay mode is essentially unknown
and cannot be used to assign $q\bar q$ or hybrid wave functions to
$\pi(1800)$.

The VES collaboration argued that the observed decay pattern indicates
strong OZI violation in $\pi(1800)$ decays \cite{Amelin:2005ry}. The
ratio of decays $\pi(1800)\to K^+K^-\pi^-$/$\pi(1800)\to
\pi^+\pi^-\pi^-$ is rather large, and also the decay rate $\pi(1800)\to
f_0(980)\pi^-$ is unexpectedly high since $f_0(980)$ couples mainly to
$s\bar s$ and not to $n\bar n$. Note however, that other radial
excitations have significant $(\pi\pi)_{S-wave}$ decay modes as well,
like $\Upsilon(2s)\to \Upsilon(1s)(\pi\pi)_{S-wave}$, $\omega(1470)\to
\omega (\pi\pi)_{S-wave}$ or $\eta(1405)\to \eta (\pi\pi)_{S-wave}$.

Summarising, we do not  consider the large branching to $f_0\pi$ as
convincing evidence for an exotic nature of the $\pi1800$, and it
will be discussed as regular second radial excitation of the pion."

\subsection{\label{The K radial excitations}
The $K$ radial excitations}

\subsubsection{\label{The K(1460)}
The $K(1460)$}

The $K(1460)$ was observed in two experiments, at SLAC by Brandenburg
\emph{et al.}~\cite{Brandenburg:1976pg} and at CERN by
Daum \emph{et al.}~\cite{Daum:1981hb}.
The SLAC experiment reported the resonance in its $
K(\pi\pi)_{S-wave}$ decay. In view of the difficulties in
identifying the $\pi(1300)$ in its $\pi(\pi\pi)_{S-wave}$ decay
mode, we disregard this observation. The CERN analysis identified
three decay modes of the $ K(1460)$ with \be \Gamma_{
K(1460)\to K^*(892)\pi}\sim 109 \qquad \Gamma_{ K(1460)\to
K\rho}\sim 34 \qquad \Gamma_{ K(1460)\to K^*_0(1430)\pi}\sim
117 \qquad \ee
The state was not observed in the LASS
experiment.

\subsubsection{\label{The K(1830)}
The $K(1830)$}

There is one indication for the second radial excitation of the
Kaon from a study of the $\phi K$ final state in the Omega spectrometer
at CERN. Mass and width $(M, \Gamma \sim 1830, 250)$\,MeV/c$^2$ were
reported~\cite{Armstrong:1982tw}.

\subsection{\label{Isoscalar resonances}
Isoscalar resonances}

\subsubsection{\label{The eta(1295)}
The $\eta(1295)$}

In 1979, there was a claim for a new meson resonance with quantum
numbers $(I^G)J^{PC}=(0^+)0^{-+}$ at a mass of 1275\,MeV/c$^2$ and
70\,MeV/c$^2$ width from a phase-shift analysis of the $\eta\pi^+\pi^-$
system produced in a 8.45\,GeV/c pion beam \cite{Stanton:1979ya}. The
resonance is now called $\eta (1295)$; it was later confirmed in other
analyses \cite{Ando:1986bn,Birman:1988gu,Alde:1997vq,Manak:2000px,%
Adams:2001sk} of experiments studying pion-induced reactions.

This state has a decisive influence on the interpretation of the nonet
of pseudoscalar radial excitations. It is observed at a mass of
$1294\pm 4$\,MeV/c$^2$, below the mass of the $\pi(1300)$, and a width
of $\Gamma = 55\pm 5$\,MeV/c$^2$ \cite{Eidelman:2004wy}, much narrower
than its isovector 2$^1$S$_0$ partner. It has been reported to decay
into $a_0(980)\pi$ and $\eta\pi\pi$ and this may explain why it is so
narrow: the most prominent decay mode of the $\pi(1300)$ is,
supposedly, $\rho\pi$, and there is no corresponding $\eta(1295)$ decay
mode. The $\eta(1295)$ is the lowest-mass pseudoscalar meson with $I=0$
above the $\eta^{\prime}$ and is thus naturally interpreted as radial
excitation of the $\eta$ as proposed by Cohen and Lipkin
\cite{Cohen:1978ge}. It is nearly degenerate in mass with the
$\pi(1300)$, hence the pseudoscalar radial excitations are expected to
be ideally mixed. The isoscalar partner of $\eta(1295)$ should then
have a mass of 1500\,MeV/c$^2$~\cite{Cohen:1978ge} and should, as
$\bar ss$ state, decay dominantly into $ K\bar{K}^* + K^*\bar K$.

\subsubsection{\label{The eta(1440)}
The $\eta(1440)$}

The $\eta(1440)$ was discovered in 1965 in $p\bar p$ annihilation at
rest into $(K\bar K\pi)\pi^+\pi^-$~\cite{Armenteros:1963}. Mass, width
and quantum numbers were determined to be $M = 1425 \pm 7, \Gamma =
80\pm 10$\,MeV/c$^2$, and $J^{PC} = 0^{-+}$~\cite{Baillon:1967}. In
parallel, a further resonance was found in the charge exchange reaction
$ \pi^- p \to n K\bar K\pi$, using a 1.5 to 4.2\,GeV/c pion
beam~\cite{Dahl:1967ad}. It had mass and width $M = 1420 \pm 20, \Gamma
= 60\pm 20$\,MeV/c$^2$; quantum numbers $J^{PC} = 1^{++}$ were suggested
on the basis of the production characteristics. Even though the quantum
numbers were different, both particles were called E-meson.

In 1980, the MARKII collaboration observed a strong signal in radiative
J/$\psi$ decays into $ (K^{\pm}K^0_S\pi^{\mp})$ \cite{Scharre:1980zh}.
Mass and width, $M = 1440 \pm 20, \Gamma = 50\pm 30$\,MeV/c$^2$ were
compatible with those of the E-meson with which the signal was
tentatively identified. In \cite{Edwards:1982nc}, pseudoscalar quantum
numbers were found and the E-meson was renamed to $\iota (1440)$ to
underline the claim that a new meson was discovered and that this new
meson was the $\iota^{\rm st}$ observed glueball. These results
prompted a reanalysis of the bubble chamber data~\cite{Baillon:1967}.
The $J^{PC} = 0^{-+}$ quantum numbers were confirmed and $J^{PC} =
1^{++}$ quantum numbers were shown to be incompatible with the data
\cite{Baillon:1982ap}. The Crystal Ball collaboration studied
J/$\psi\to\gamma (K^+K^-\pi^0)$ and $\eta\pi\pi$
\cite{Edwards:1982nc,Edwards:1983pv}; the partial wave analysis
confirmed $J^{PC} = 0^{-+}$ quantum numbers. The Asterix experiment at
LEAR deduced pseudoscalar quantum numbers for the state using arguments
based on $\bar pp$ annihilation dynamics \cite{Duch:1989sx}, in
agreement with the partial wave analysis of Crystal Barrel data on
$\bar pp\to \eta\pi^+\pi^-\pi^0\pi^0$ \cite{Amsler:1995wz}.

\begin{figure}[pt]
\bc
\includegraphics[width=0.5\textwidth,height=0.40\textwidth]{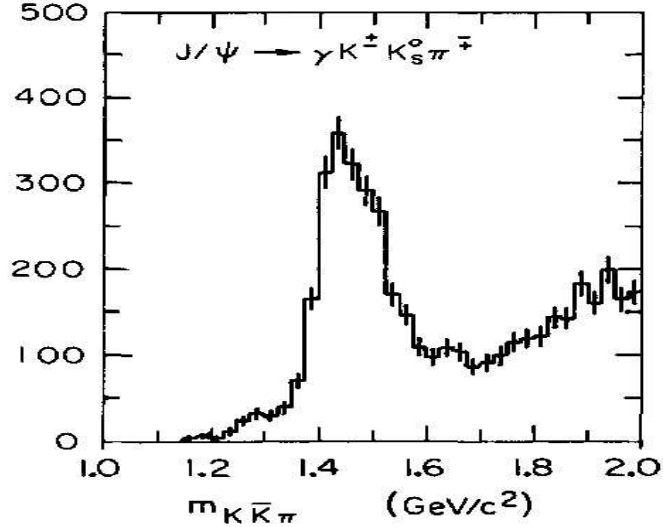}
\ec
\caption{\label{pic:jpsi2giota}
The $K\bar K\pi$ invariant mass distribution produced in radiative
J/$\psi$ decays by the MarkIII collaboration \cite{Kopke:1988cs}. The
structure is resolved into two pseudoscalar states and a $J^{PC} =
1^{++}$ resonance. }  \end{figure}

Higher statistics revealed that the peak at 1440\,MeV/c$^2$ has a more
complex structure \cite{Bai:1990hs,Augustin:1990ki}, see
Fig.~\ref{pic:jpsi2giota}. The MarkIII collaboration suggested a
pattern of three states, two pseudoscalar states at
$M=1416\pm8^{+7}_{-5}; \Gamma= 91^{+67}_{-31}{^{+15}_{-38}}$\,MeV/c$^2$
(mainly $a_0(980)\pi$) and $M=1490{^{+14}_{-8}}{^{+3}_{-6}};
\Gamma= 54{^{+37}_{-21}}{^{+13}_{-24}}$\,MeV/c$^2$ (mainly $ K^*K$),
and a $J^{PC} = 1^{++}$ resonance at 1440\,MeV/c$^2$~\cite{Bai:1990hs}.
DM2 found two pseudoscalar states, a low-mass (1420\,MeV/c$^2$)
component coupling to $ K^*K$ and a high-mass (1460\,MeV/c$^2$)
component decaying into $a_0(980)\pi$. Furthermore, DM2 studied
radiative decays into $\eta\pi\pi$ (with  $a_0(980)\pi$ as isobar)
which peaks at the lower mass. They conclude that the 1440 mass region
was not yet fully understood.

In a partial wave analysis of BES data on J/$\psi\to\gamma (K^{\pm}
K^{0}_{S}\pi^{\mp })$ \cite{Bai:1990hs,Bai:2000ss}
and to $ \gamma (K^+K^-\pi^{0})$ \cite{Bai:1998eg}, the $\eta
(1440)/\iota$ region was fitted with a Breit-Wigner amplitude with
$s$-dependent widths.  Decays into $ K^*K$, $ \kappa K$, $\eta \pi
\pi$ and $\rho \rho$ were included ($ \kappa$ refers to the $ K\pi$
S-wave). At a $ K\bar{K}\pi$ mass of $\sim 2040$ MeV/c$^2$, a second peak
with width $\sim 400$ MeV/c$^2$ was seen; $J^P = 0^-$ was preferred over
$1^+$ and $2^-$ respectively.  The authors suggested the state as
candidate for a $0^-$ $s\bar sg$ hybrid partner of $\pi (1800)$.

\par
The $\eta(1440)$ region was also studied in hadronic reactions.
At BNL, it was reported in $ \pi^- p$ charge exchange and in \pbp\
annihilation in flight. With an 8\,GeV/c pion beam \cite{Chung:1979fb}
and in \pbp\ annihilation \cite{Reeves:1986jg}, the $\eta(1440)$ was
reported at 1420\,MeV/c$^2$ and with $a_0(980)\pi$ as dominant decay mode
while in a 21\,GeV/c pion beam, the most prominent feature was a
pseudoscalar meson at 1460\,MeV/c$^2$ mass decaying into $
K^*\bar{K}+c.c.$ \cite{Rath:1989rt}.

The $\eta (1440)$ is a very strong signal, one of the strongest, in
radiative J/$\psi$ decays. The radial excitation $\eta (1295)$ is not
seen in this reaction; hence the $\eta (1440)$ must have a different
nature. At that time it was proposed (and often still is
\cite{Eidelman:2004wy,Minkowski:2002nf,Faddeev:2003aw}) to be a
glueball.

\begin{table}[ph]
\caption{\label{eisglue}
Popular interpretation of the spectrum of pseudoscalar radial
excitations \cite{Eidelman:2004wy,Minkowski:2002nf,Faddeev:2003aw}.}
\renewcommand{\arraystretch}{1.4}
\vspace*{2mm}
\begin{center}
\begin{tabular}{ccccc} \hline\hline
$\pi$      & $\eta$     &            &$\eta^{\prime}$&  $K$ \\
$\pi(1300)$&$\eta(1295)$&$\eta(1405)$&$\eta(1475)$   &$K(1460)$ \\
$n\bar n$  &$n\bar n$   &  glueball  &$s\bar s$   &$n\bar s,\bar n s$\\
\hline\hline
\end{tabular}
\renewcommand{\arraystretch}{1.0}
\end{center}
\end{table}

\subsubsection{\label{The split eta(1440)}
The split $\eta(1440)$}

The Obelix collaboration at LEAR~\cite{Nichitiu:2002cj} studied the
reaction $ p\bar p\to \pi^+\pi^- K^{\pm}K^0_S\pi^{\mp}$ at 3 different
densities. The $\pi^+\pi^-$ system recoiling against
$K^{\pm}K^0_S\pi^{\mp}$ has very little energy and is hence likely in
S-wave. Likewise, there is no angular momentum to be expected between
$(\pi^+\pi^-)$ and  $(K^{\pm}K^0_S\pi^{\mp})$. Then, the most likely
initial state of the $ p\bar p$ atom from which annihilation occurs
has the quantum numbers of the $(K^{\pm}K^0_S\pi^{\mp})$ system
\cite{Duch:1989sx}. This selection rule helps in the partial wave
analysis.

\begin{figure}[pb]
\bc
\includegraphics[width=0.98\textwidth,height=0.22\textheight]{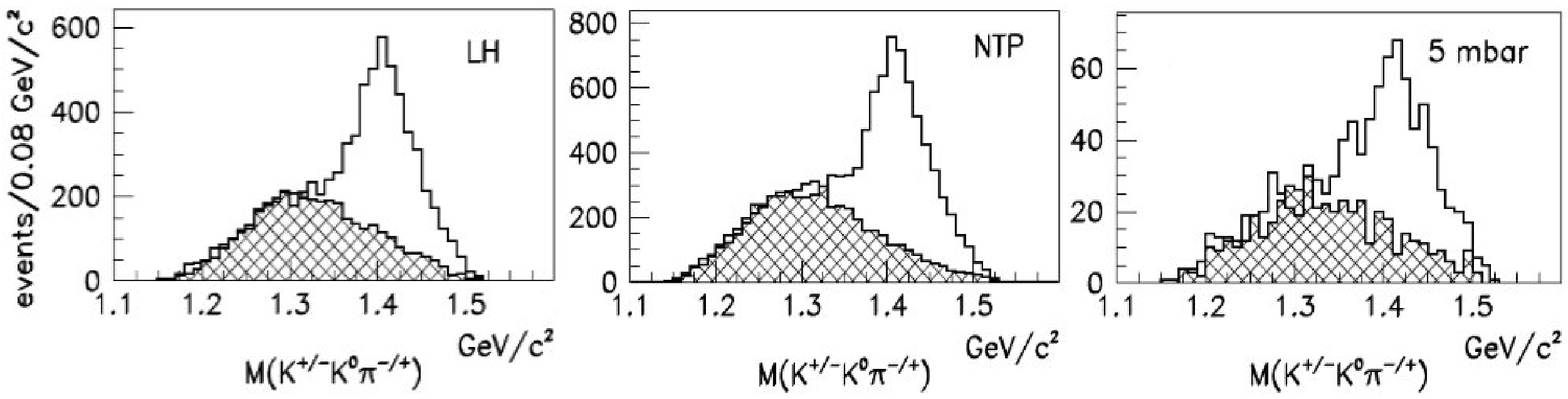}
\vspace{-3mm}
\ec
\caption{\label{obelixeta}
 The $K^{\pm}K^0_S\pi^{\mp}$ invariant mass distribution from $ p\bar
p$ annihilation at rest at 3 different target densities
\cite{Nichitiu:2002cj}. There are two entries per event. The hatched
area is the expected background from the 'wrong'
$(K^{\pm}K^0_S\pi^{\mp})$ combination.} \end{figure}

The Obelix partial wave analysis  confirmed the conjecture of MARKIII
that the $\eta (1440)$ is split into two components, an $\eta(1405)\to
a_0(980)\pi$ with $M = 1405 \pm 5, \Gamma = 56\pm 6$\,MeV/c$^2$ and an
$\eta(1475)\to  K^*\bar K +\bar K^*K$ with $M = 1475 \pm 5, \Gamma =
81\pm 11$\,MeV/c$^2$. Even though a bit low in mass, the $\eta(1475)$ fits
well the properties expected for a $\bar ss$ pseudoscalar radial
excitation. The $\eta(1405)$ finds no 'slot' in the spectrum of $\bar
qq$ resonances, it is an intruder, possibly a glueball. Since 2002, the
Particle Data Group supports this interpretation of the pseudoscalar
mesons with the caveat that lattice gauge calculations predict the
pseudoscalar glueball mass well above 2000\,MeV/c$^2$.

Two quantitative tests have been proposed to test if a particular
meson is glueball-like: the stickiness \cite{Chanowitz:1984cb}
and the gluiness \cite{Close:1996yc} which are expected to be close to
unity for normal $q\bar q$ mesons. They were introduced in section
\ref{Two--photon fusion}, eqs. (\ref{sticky}) and (\ref{gluish}). The
L3 collaboration determined~\cite{Acciarri:2000ev} the stickiness to
$S_{\eta(1440)}=79\pm 26$ and the gluiness ($G$) to
$G_{\eta(1440)}=41\pm 14$. No distinction was made in
\cite{Acciarri:2000ev} between $\eta(1405)$ and $\eta(1475)$. These
numbers can be compared to those for the $\eta '$ for which $S_{\eta '}
= 3.6 \pm 0.3$ and $G_{\eta '} = 5.2 \pm 0.8$ was determined, with
$\alpha_s(\rm 958\,MeV/c^2)=0.56\pm0.07$. Also $\eta'$ is `gluish', but
much more the $\eta(1405)$. The $\eta(1405)$ has properties as expected
from a glueball.

The L3 data were challenged by the CLEO collaboration
\begin{figure}[pb]
\bc
\includegraphics[width=0.6\textwidth,height=0.38\textheight]{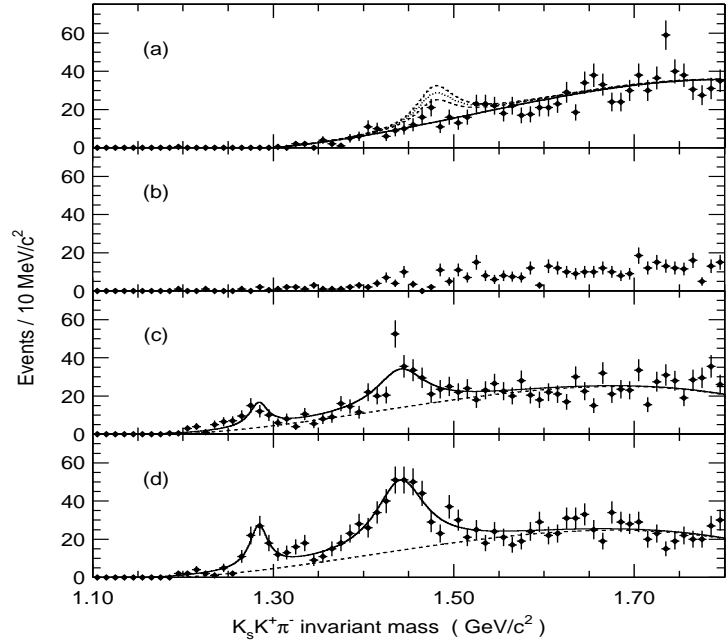}
\ec
\caption{\label{fig_data_mass_kskp_7}
The distributions of the $ K^0_S K^\pm \pi^\mp$ invariant mass from
the CLEO experiment \cite{Ahohe:2005ug}. The distributions are shown
for (a) $p_\perp \le 100 {\rm ~MeV/c}$, (b) $100 {\rm ~MeV/c} \le
p_\perp \le 200 {\rm ~MeV/c}$ and (c) $200 {\rm ~MeV/c} \le p_\perp \le
600 {\rm ~MeV/c}$ in the untagged mode, and (d) for all $p_\perp$ in
the tagged mode. The dashed curves in (a) show the strength of the
expected $\eta(1440)$ signal according to the L3 results
\protect\cite{Acciarri:2000ev}. The solid curves represent fits with
resonances and polynomial non-interfering combinatory backgrounds.
Figure (a) shows dotted and dashed distributions expected from L3,
using their $\gamma\gamma\to\eta(1440)$ yield and error. }
\end{figure}
\cite{Ahohe:2005ug}. Fig.~\ref{fig_data_mass_kskp_7} shows their $
K^0_S K^\pm \pi^\mp$ mass distributions for different bins of the
transverse momentum transfer $p_\perp$ from $e^+e^-$ to the $ K^0_S
K^\pm \pi^\mp$ system. For two nearly real photons, $p_\perp$ is small;
two on-shell photons do not couple to $J^P=1^+$ mesons (Young's
theorem) while pseudoscalar meson decays to $2\gamma$ are allowed. The
absence of signals at 1285\,MeV/c$^2$ and 1440\,MeV/c$^2$ in
Fig.~\ref{fig_data_mass_kskp_7} proves that the signal observed for
large $p_\perp$ is due to axial vector mesons. With no pseudoscalar
state observed, it is difficult to arrive at any definite conclusion
concerning their nature. Both, pseudoscalar radial excitations and a
pseudoscalar glueball (if it exists in this mass range) may have small
couplings to $\gamma\gamma$.

The $\rho\gamma$ decay mode is disturbing for the glueball
interpretation. At BES, $\eta (1295)$ and $\eta (1440)$ were studied in
J/$\psi\to(\rho\gamma)\gamma$ and $\to(\phi\gamma)\gamma$
\cite{Bai:2004qj}. A peak below 1300\,MeV/c$^2$ was assigned to
$f_1(1285)$ even though a small contribution from $\eta(1295)$ was not
excluded. The $\eta(1440)$ (observed at 1424\,MeV/c$^2$) was seen to
decay strongly into $\rho\gamma$ and not into $\phi\gamma$. Neither a
glueball nor a $\bar ss$ state should decay radiatively into a $\rho$
meson. A small $\bar uu+\bar dd$ component in the $\eta(1475)$ wave
function could be responsible for this decay mode but then, a large
$\phi\gamma$ decay rate should be expected. This however is not
observed. The only escape is the assumption that the $\rho\gamma$
events stem from the $f_1(1420)$. The spin-parity analysis prefers
pseudoscalar quantum numbers for the peak at 1424\,MeV/c$^2$ but
$1^{++}$ is not completely ruled out.

\subsubsection{\label{Isoscalar resonances revisited}
Isoscalar resonances revisited}

There are severe inconsistencies in the scenario presented above
(in section~\ref{The eta(1295)}-\ref{The split eta(1440)}). The
$\eta(1295)$ is seen only in pion induced reactions.  Radiative
J/$\psi$ decays show an asymmetric peak in the $\eta(1440)$ region;
therefore $\eta(1405)$ and the $\eta(1475)$ are both produced in
radiative J/$\psi$ decays. The $\eta(1295)$ as isoscalar partner must
then also be produced, but it is not - at least not with the expected
yield.

There are a few reasons not to accept $\eta(1295)$ as established
resonance. It is approximately degenerate in mass and width
 with $f_1(1285)$. This is potentially dangerous. The matrix element
for a pseudoscalar meson decaying into three pseudoscalar mesons does
not have any peculiarities, all distributions are isotropic and flat.
This decay looks like a `garbage can'. Any imperfection in the
description of the dominant isoscalar $1^{++}$ wave leads inevitably
to the appearance of a spurious signal in the $0^{-+}$ wave. A number
of imperfections could lead to such a feedthrough: not
perfectly understood acceptance or resolution, not fully justified
assumptions in the PWA model like coherence of the $1^{++}$ and/or
the $0^{-+}$ wave in production, a non-perfect parametrisation of
$f_1(1285)$ decays or any other imperfection. The feedthrough signal
has a Breit-Wigner amplitude including a phase motion, but the
parameters of the two resonances are similar.

The L3 observation of $\gamma\gamma$ fusion into $\eta (1440)$ but not
into $\eta (1295)$ is problematic. The $n\bar n$ state should have
larger $\gamma\gamma$ couplings than a glueball or $s\bar s$ state.
However, CLEO finds neither $\eta (1295)$ nor $\eta (1440)$, and there
is no conflict (but also no evidence for $\eta(1295)$, neither).

\begin{figure}[pt]
\begin{center}
\begin{tabular}{cc}
\hspace*{-2mm}\includegraphics[width=0.4\textwidth,height=0.3\textwidth]
{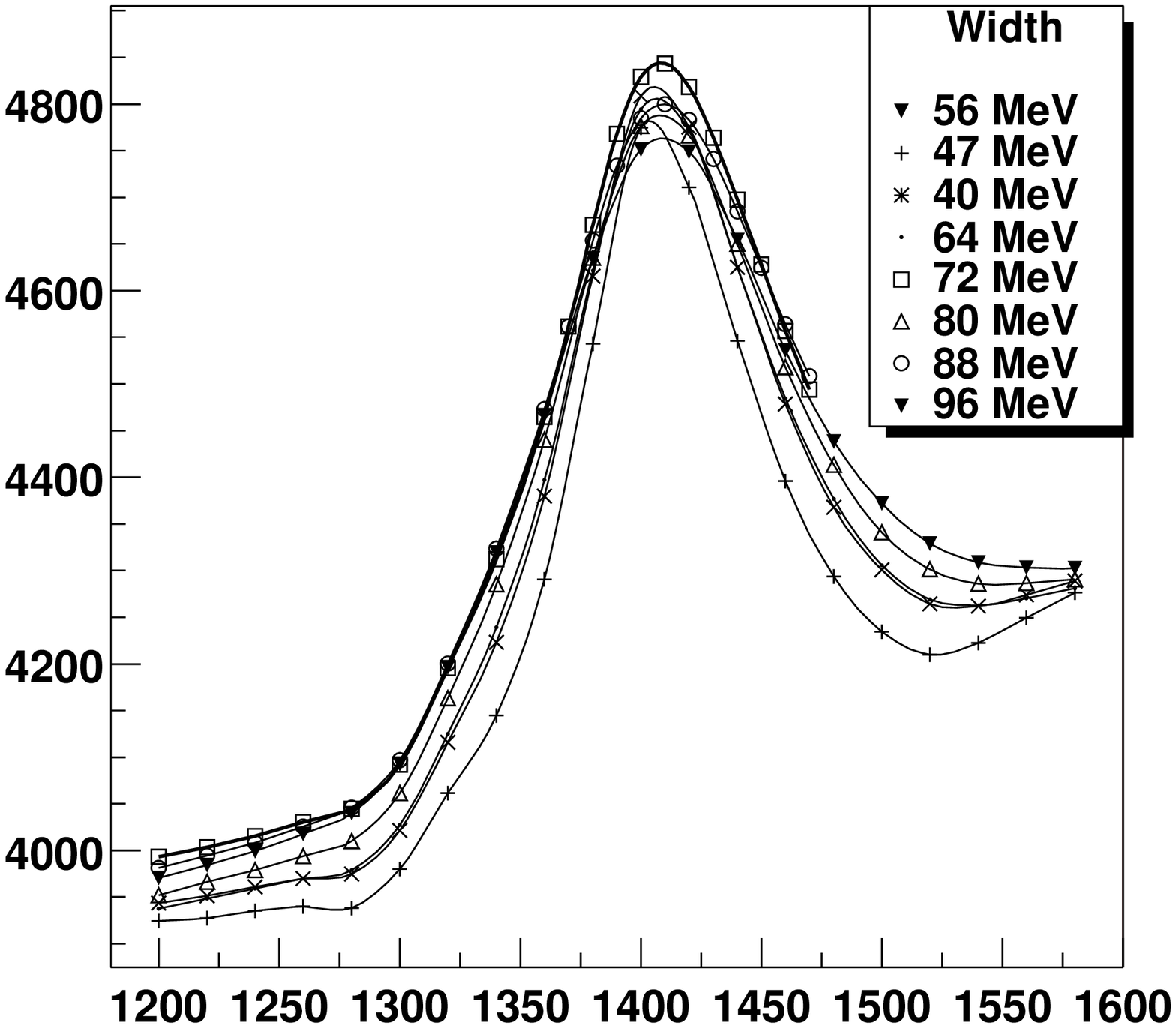}
&
\hspace*{-4mm}\includegraphics[width=0.4\textwidth,height=0.3\textwidth]
{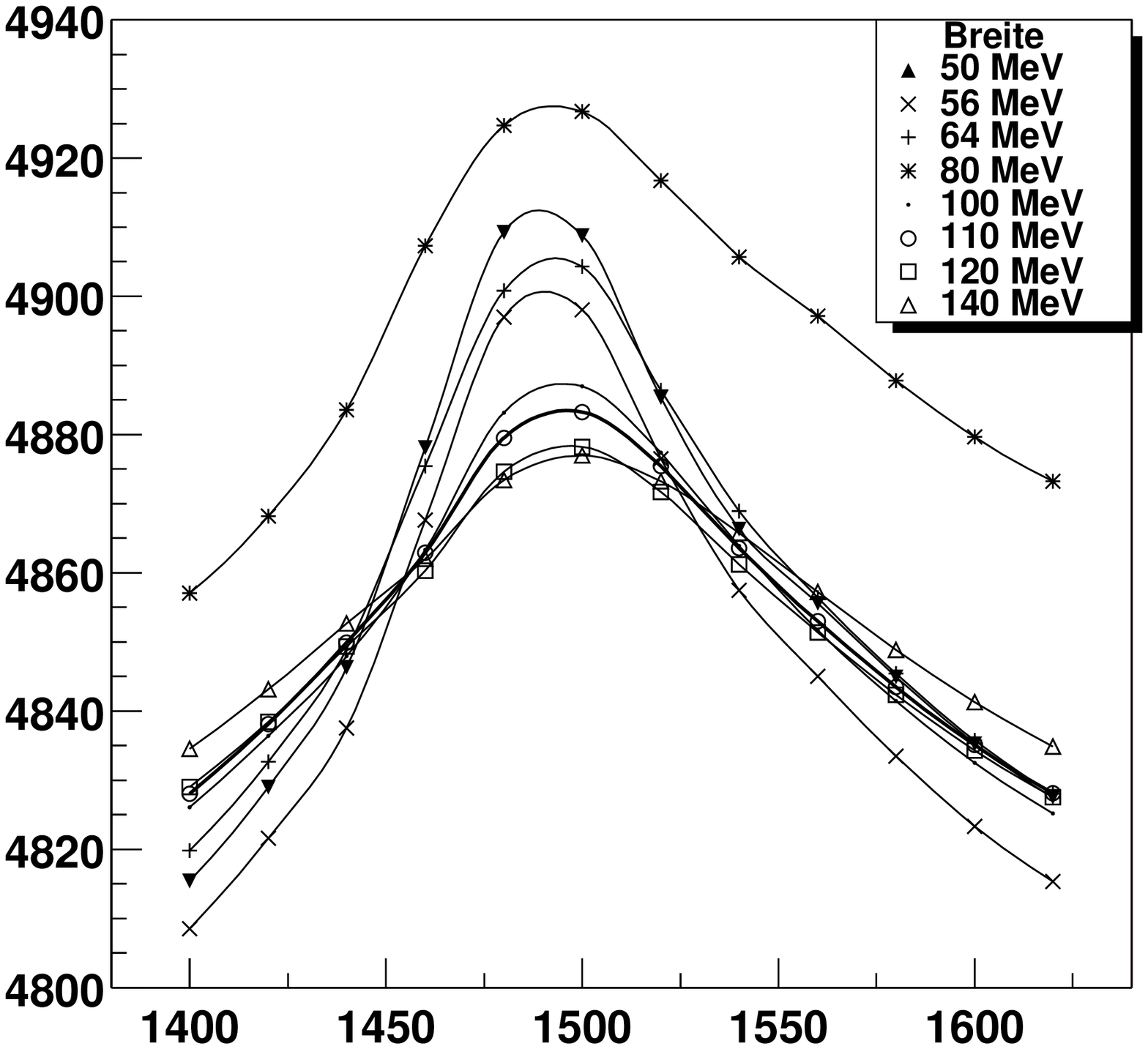} \\
\vspace{-50mm}
\end{tabular} \\
\hspace{-45mm}a \hspace{65mm} b\hspace{40mm}\\
\vspace{40mm}
\end{center}
\caption{\label{escan}Scan for a $0^+ 0^{- +}$ resonance with different
widths\protect\cite{Reinnarth}. a: The likelihood optimises for
$M=1407\pm 5, \Gamma = 57\pm 9$\,MeV/c$^2$. The resonance is identified
with the $\eta(1405)$. b: A search for a second pseudoscalar resonance
(right panel) gives evidence for the $\eta(1475)$ with $M=1490\pm 15,
\Gamma = 74\pm 10$\,MeV/c$^2$.} \end{figure}

The Crystal Barrel collaboration searched for the $\eta (1295)$ and
$\eta (1440)$ in the reaction $p\bar p\to\pi^+\pi^-\eta (xxx)$, $\eta
(xxx)\to\eta\pi^+\pi^-$. The search was done by assuming the presence
of a pseudoscalar state of given mass and width; mass and width were
varied and the likelihood of the fit plotted. Fig.~\ref{escan}a shows
such a plot \cite{Reinnarth}; a clear pseudoscalar resonance signal is
seen at 1405\,MeV/c$^2$. Two decay modes are observed, $a_0(980)\pi$
and $\eta(\pi\pi)_{S-wave}$  with a ratio $0.6\pm0.1$.

A scan for an additional $0^+ 0^{- +}$ resonance provides no evidence
for the $\eta (1295)$ but for a second resonance at 1490\,MeV/c$^2$,
see Fig.~\ref{escan}b, with $M=1490\pm 15 ,\Gamma=74\pm 10$MeV/c$^2$.
It decays to $a_0(980)\pi$ and $\eta(\pi\pi)_{S-wave}$  with a
ratio $0.16\pm0.10$. This data could be interpreted as first evidence
for $\eta(1475)\to\eta\pi\pi$ decays.

Amsler and Masoni reinforced the existence of $\eta(1295)$ by claiming
additional evidence from four further observations, unrelated to pion
induced reactions (see \cite{Eidelman:2004wy}, page 591).

\begin{itemize}

\item In an analysis of the reaction $p\bar
p\to\eta\pi^+\pi^-\pi^0\pi^0$ \cite{Amsler:1995wz}, the $\eta(1405)$
was studied and the decay mode $\sigma\eta$ was observed for the first
time. The likelihood of the fit improved when a $\eta(1295)$ was added.
No further study was reported; in particular $\eta(1295)$ was not
replaced by $f_1(1285)$ so that the fit likelihoods cannot be compared.

\vspace{2mm}\item In \cite{Anisovich:2001jb}, the partial wave analysis
of the reaction $p\bar p\to\eta\pi^+\pi^-\pi^+\pi^-$  was directed to
determine properties of $f_0(1370)$. A small $\eta(1295)$ signal was
found but at 1255\,MeV/c$^2$. Studies at Bonn demonstrated that the
peak was faked by the Monte Carlo simulation in which the trigger
condition was not sufficiently well reproduced (the same Monte Carlo
data sets were used in both analyses). The small enhancement at
1260\,MeV/c$^2$ in Fig. \ref{escan} is the effect which remains once
the trigger condition was described more precisely in the Monte Carlo
simulation.

\vspace{2mm}\item In another paper on $p\bar
p\to\eta\pi^+\pi^-\pi^+\pi^-$ by Amsler {\it et al.}
\cite{Amsler:2004rd} a low-mass peak $X(1285)$ was assumed to be the
$f_1(1285)$. A fit with $f_1(1285)$ mass and width gave the best
description. However, the possibility that the peak is due to the $\eta
(1295)$ was not excluded.

\vspace{2mm}\item The DM2 data on J/$\psi\to\gamma\eta\pi\pi$ showed a
small peak at 1265\,MeV/c$^2$ and a larger one at 1400\,MeV/c$^2$
\cite{Augustin:1990ki}. The partial wave analysis preferred pseudoscalar
quantum numbers for the peak at 1265\,MeV/c$^2$ even though there is no
comment on this aspect in the paper. The second peak at 1400\,MeV/c$^2$
is interpreted as $f_1(1420)$/c$^2$. The $K\bar K\pi$ pseudoscalar wave
was fitted with two resonances in the pseudoscalar partial wave at
1420\,MeV/c$^2$ in $K^*K$ and at 1460\,MeV/c$^2$ in $a_0(980)$, and one
$f_1(1460)$. These findings contradict in most aspects more recent
results based on larger statistical samples. We refuse to use these
data as argument in favour of the existence of $\eta(1295)$.
\end{itemize}

We conclude that even a large number of cases which are individually
not convincing do not make up a convincing argument. A major further
point against the existence of $\eta(1295)$ are internal
inconsistencies: the interpretation of $\eta(1295)$ and $\pi(1300)$ as
radial excitations would imply that the nonet of pseudoscalar radial
excitation were ideally mixed. It is then not possible that the $s\bar
s$ state $\eta(1475)$ is produced strongly in $\bar pp$, the $n\bar n$
state $\eta(1295)$ weakly. This behaviour would contradict violently
what we have learned about the $\bar pp$ annihilation process
\cite{Klempt:2005pp}. There is no way to understand why the yield
of the $s\bar s$ state $\eta(1475)$ in radiative J/$\psi$ decays is
large and the $\eta(1295)$ hardly visible. Hence $\eta(1295)$ cannot be
a $q\bar q$ meson. On the other hand, we do not expect glueballs,
hybrids or multiquark states so low in mass. An interpretation of $\eta
(1295)$ as exotic particle is hence unlikely even though not excluded.
A possible origin of $\eta(1295)$ might be interference of $f_1(1285)$
with the $\eta(\pi\pi)_{S-wave}$ Deck amplitude, faking
$\eta(1295)$.  Here, it is excluded from the further
discussion.

The next puzzling state is the $\eta (1440)$. It is not produced as
$\bar ss$ state but decays with a large fraction into $ K\bar K\pi$
and it is split into two components. It was suggested in
\cite{Klempt:2004xg} that the origin of these anomalies is due to a
node in the wave function of the $\eta(1440)$. This node has an
impact on the decay matrix elements as calculated by Barnes {\it et
al.}\cite{Barnes:1996ff} within the $^3P_0$ model.

\begin{figure}[pb]
\vspace*{3mm}
\begin{center}
\begin{tabular}{ccc}
\includegraphics[width=0.25\textwidth,height=50mm,clip=on]{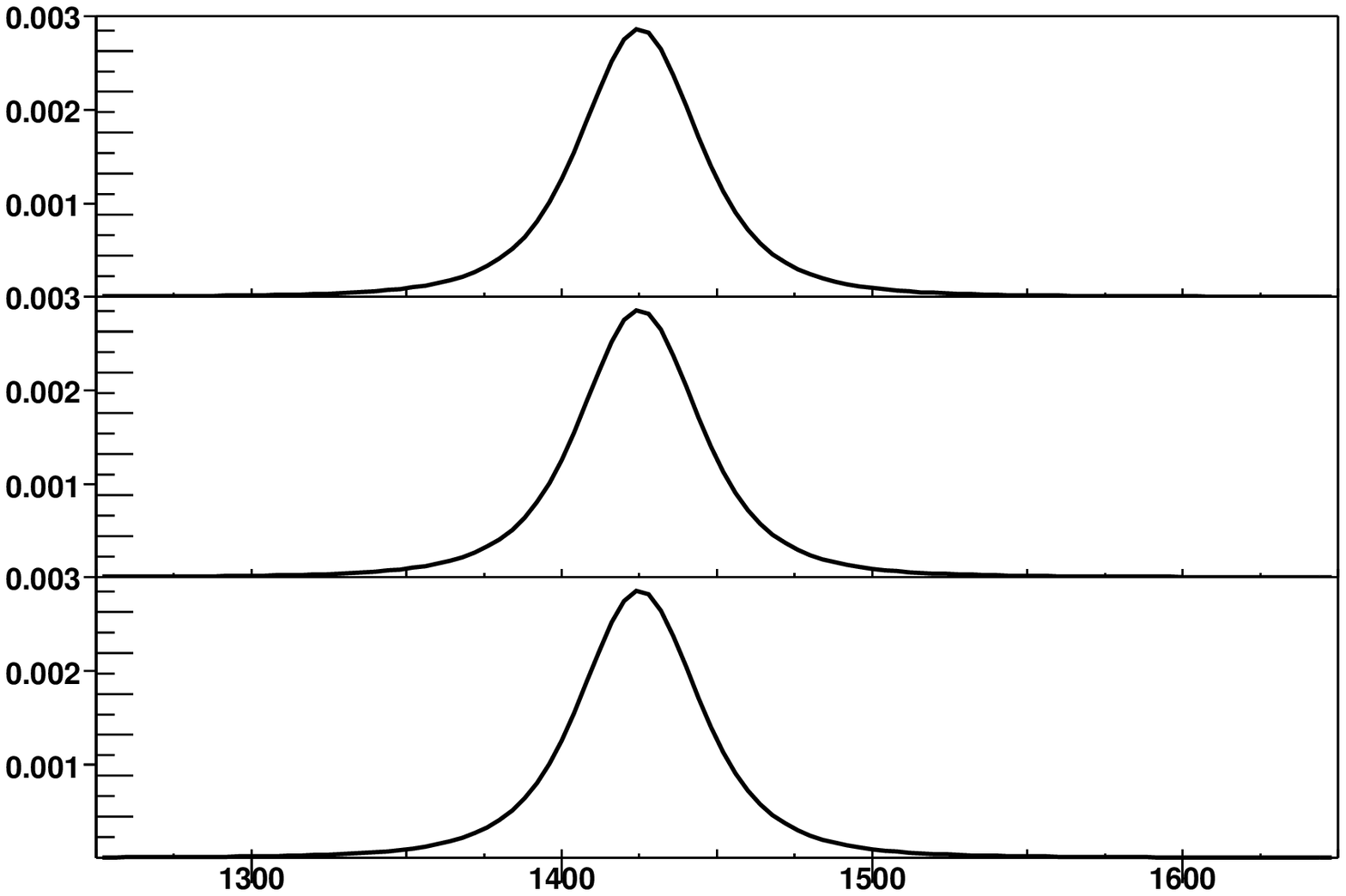}&
\hspace*{-2mm}\includegraphics[width=0.25\textwidth,height=50mm,clip=on]{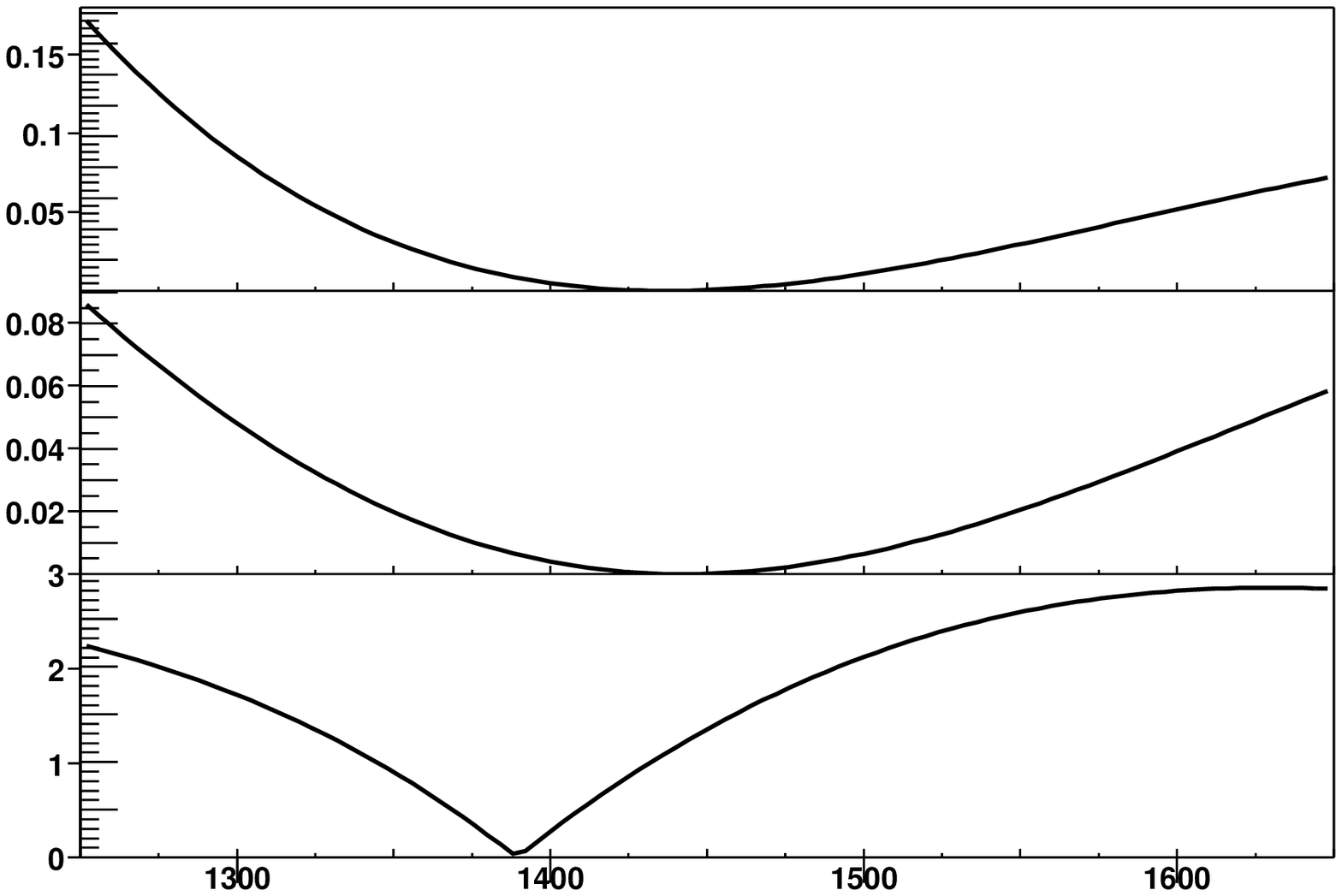}&
\hspace*{-2mm}\includegraphics[width=0.25\textwidth,height=50mm,clip=on]{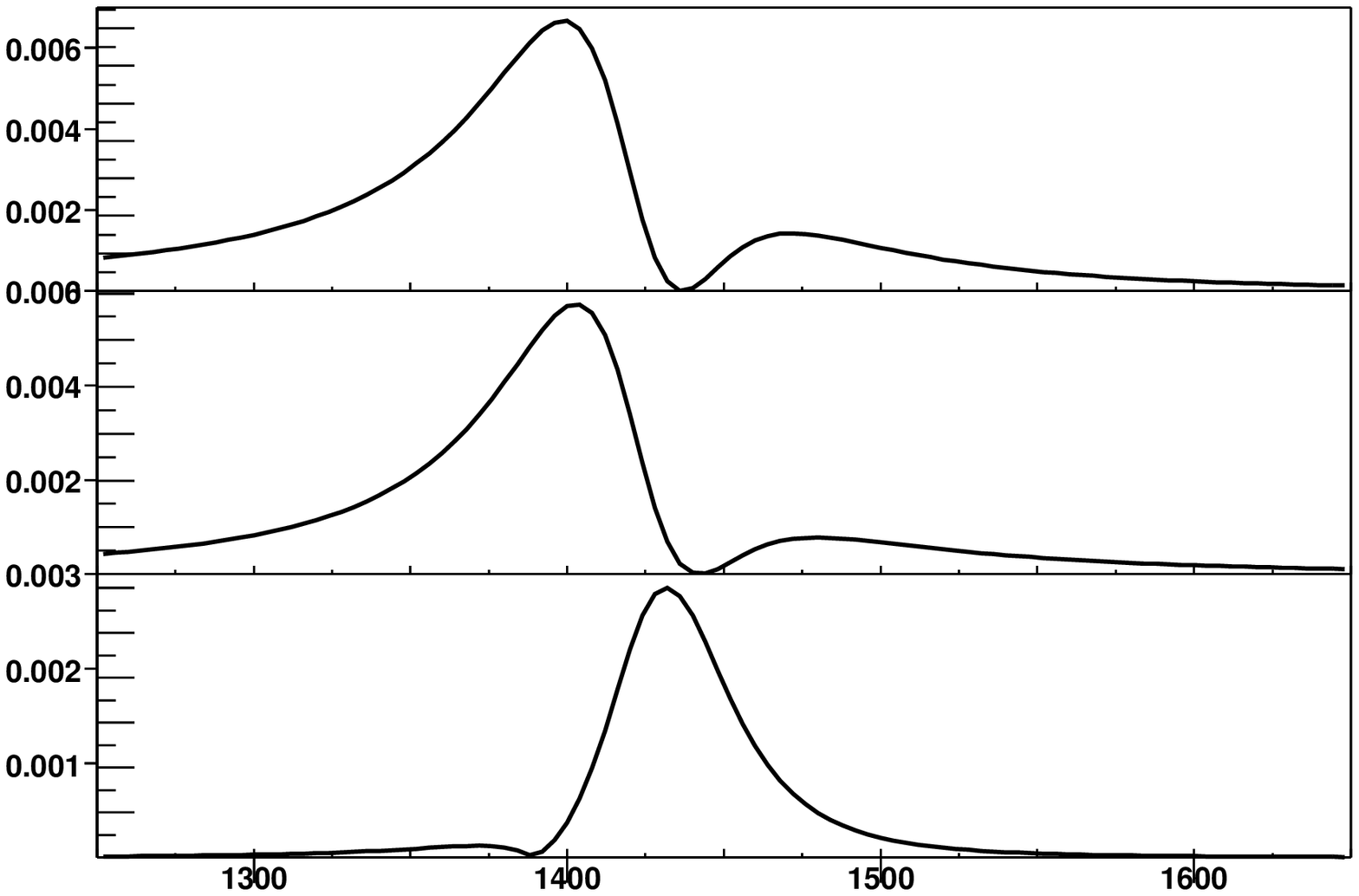}
\end{tabular}\vspace{-48mm}\\
\footnotesize\hspace{35mm}a1\hspace{42mm}a2\hspace{42mm}a3\vspace{11mm}\\
\footnotesize\hspace{35mm}b1\hspace{42mm}b2\hspace{42mm}b3\vspace{11mm}\\
\footnotesize\hspace{35mm}c1\hspace{42mm}c2\hspace{42mm}c3\\
\vspace{11mm}
\end{center}
\caption{\label{node}
Amplitudes for $\eta(1440)$ decays to $a_0\pi$ (a),
$(\pi\pi)_{S-wave}\eta$ (b), and $ K^*\bar K$ (c); the
Breit-Wigner functions are shown (1), then (2) the squared decay
amplitudes using eqs.~(\protect\ref{rad_1}),(\ref{rad_2}),  and (3)
the resulting squared transition matrix element. } \end{figure}

The matrix elements for decays of the $\eta (1440)$ as a radial
excitation (=$\eta_R$) depend on spins, parities and decay momenta
of the final state mesons. For $\eta_R$ decays to $ K^*K$, the
matrix element is given by
\be
\label{rad_1}
f_P = \frac{2^{9/2}\cdot 5}{3^{9/2}}\cdot x\left(1-\frac{2}{15}x^2
\right).
\vspace{2mm}
\ee
In this expression, $x$ is the decay momentum in units of
400\,MeV/c; the scale is determined from comparisons of measured
partial widths to model predictions. The matrix element vanishes
for $x=0$ and $x^2 = 15/2$, or $p=1$\,GeV/c. These zeros have
little effect on the shape of the resonance.
\par
The matrix element for $\eta_R$ decays to $a_0(980)\pi$ or
$(\pi\pi)_{S-wave}\eta$ has the form
\be
\label{rad_2}
f_S = \frac{2^{4}}{3^{4}}\cdot \left(1-\frac{7}{9}x^2 +
\frac{2}{27}x^4 \right)
\ee
and vanishes for  $p=0.45$\,GeV/c. The amplitude for $a_0(980)\pi$
decays vanishes at 1440\,MeV/c$^2$. This has a decisive impact
on the shape, as seen in Figure~\ref{node}. Shown are an undistorted
Breit-Wigner function, the transition matrix elements for three
$\eta(1440)$ decay modes as given by Barnes {\it et al.}
\cite{Barnes:1996ff}, and the product of the squared matrix elements
and a Breit-Wigner distribution with mass 1420\,MeV/c$^2$ and width
60\,MeV/c$^2$.

The $\eta(1440)\to a_0(980)\pi$ and $\to K^*K$
mass distributions have different peak positions;  at approximately the
$\eta(1405)$ and $\eta(1475)$ masses. Hence there is no need to
introduce $\eta(1405)$ and $\eta(1475)$ as two independent states.
One $\eta(1420)$ and the assumption that it is a radial excitation
describe the data. Of course, the $ ^3P_0$ model for meson decays is
a model. Model-independent is however the observation that zeros in the
wave functions can lead to distortions in the final states observed. If
two resonances are found having identical quantum numbers and very
close in mass, extreme care must be taken before far-reaching claims on
the abundances of meson resonances in that partial wave can be made.

\begin{figure}[pt]
\vspace*{-4mm}
\begin{center}
\includegraphics[width=0.35\textwidth,height=0.28\textwidth]{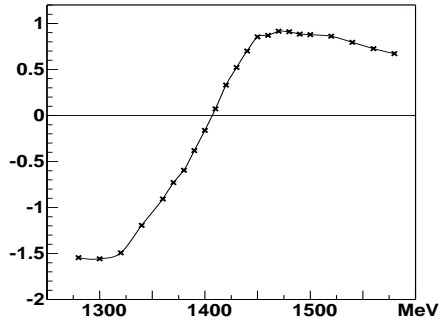}
\end{center}
\vspace*{-2mm}
\caption{\label{phase}
Phase motion of the $a_0(980)\pi$ isobars in $p\bar p$ annihilation
into $4\pi\eta$. In the mass range from 1300 to 1500\,MeV/c$^2$ the
phase varies by $\pi$ indicating that there is only one resonance in
the mass interval. The $(\pi\pi)_{S-wave}\eta$ (not shown) exhibits the
same behaviour~\protect\cite{Reinnarth}. } \end{figure}

The conclusion that $\eta(1405)$ and $\eta(1475)$ is one single
resonance can be tested further by following the phase motion of the
$a_0(980)\pi$ or $(\pi\pi)_{S-wave}\eta$ isobar\cite{Reinnarth}.
The phase changes by $\pi$ and not by 2$\pi$, see Fig.~\ref{phase}.

Summarising, the results for the radial excitations of pseudoscalar
mesons are as follows:

\begin{itemize}

\vspace*{2mm}\item The $\eta(1295)$ is not a $q\bar q$ meson. The
source of the signal might be the Deck effect and feedthrough from the
$f_1(1285)$ wave.

\vspace*{2mm}\item The $\eta (1440)$ wave function has a node leading
to two apparently different states $\eta(1405)$ and $\eta(1475)$.

\vspace*{2mm}\item  There is only one $\eta$ state, the $\eta(1440)$,
in the mass range from 1200 to 1500 MeV/c$^2$ and not 3\,!

\vspace*{2mm}\item The $\eta(1440)$ is the radial excitation of the
$\eta$. The radial excitation of the $\eta'$ is expected at about
1800\,MeV/c$^2$.

\end{itemize}

\subsection{\label{Higher--mass eta excitations}
Higher-mass $\eta$ excitations}

\subsubsection{\label{Selection rules}
Selection rules}

\begin{table}[pb]
\caption{\label{psselect}SU(3) selection rules for excited $\eta$
states. \vspace{2mm}} \begin{center} \renewcommand{\arraystretch}{1.4}
\begin{tabular}{ccccc}
\hline\hline
$\eta_1\to PS + V$ &    &\qquad &$\eta_8\to PS + V$ & $ K^*K$  \\
$\eta_1\to PS + S$ & $f_{0(1)}\eta_1$, $f_{0(8)}\eta_8$, $a_0\pi$, $ K^*_0K$ &
&$\eta_8\to PS + S$ &$f_{0(1)}\eta_8$, $f_{0(8)}\eta_1$, $a_0\pi$, $
K^*_0K$\\ $\eta_1\to V + V$& $\omega\omega$,
$\rho\rho$, $ K^*_0K^*_0$&&$\eta_8\to V + V$ &$\omega\omega$,
$\rho\rho$, $ K^*_0K^*_0$\\ \hline\hline \end{tabular}
\renewcommand{\arraystretch}{1.0}
\end{center}
\end{table}
We begin this section with a Table (\ref{psselect}) which
summarises selection rules for two-body decays of $\eta$ excitations
into pseudoscalar, scalar, and vector mesons. Decays into
two pseudoscalar or two scalar mesons are forbidden due to parity and
charge conjugation conservation. The selection rules are expressed by
the SU(3) singlet and octet components of the (pseudo-)scalar mesons.
The $ K^*K$ decays are particularly interesting since they identify
octet states. For that reason, $\eta(1440)$ must be a flavour octet
state or at least have a large octet component.

\subsubsection{\label{eta excitations from radiative J/psi decays}
$\eta$ excitations from radiative J/$\psi$ decays}

There are several candidates for pseudoscalar isoscalar excitations
above 1.5\,GeV/c$^2$. MarkIII reported evidence for dominance of the
$I^G(J^{PC})=0^+(0^{-+})$ partial wave below 2\,GeV/c$^2$ in radiative
J/$\psi$ decays into $\rho\rho$ \cite{Baltrusaitis:1985nd} and
$\omega\omega$ \cite{Baltrusaitis:1985zi}. Summarizing MarkIII
results, Eigen reported strong pseudoscalar activity below 2 GeV/c$^2$ for
$\rho\rho$, $\omega\omega$ and, above their respective thresholds, for
$ K^*K^*$ and $\phi\phi$ \cite{Eigen:1990cm}. The analysis of DM2
data suggested 3 states, $\eta(1500)$, $\eta(1800)$, and $\eta(2100)$,
decaying into $\rho\rho$ \cite{Bisello:1988as}, and possible evidence
for a $\phi\phi$ pseudoscalar meson at 2.24\,GeV/c$^2$
\cite{Bisello:1990qn}. Bugg noticed~\cite{Bugg:1995jq} that the three
DM2 peaks coincide in mass and width with three peaks in the $\eta\eta$
mass spectrum produced in $\bar pp$ annihilation in flight into
$\pi^0\eta\eta$. Any $\eta\eta$ resonance must have natural spin-parity
quantum numbers; a reanalysis of Mark3 data on J/$\psi\to 4\pi$ (the
DM2 data were no longer available) suggested that the decay mode should
not be in the $\rho\rho$ pseudoscalar wave but in the $(\pi\pi)_{\rm
S-wave}(\pi\pi)_{S-wave}$ $0^{++}$ scalar wave \cite{Bugg:1995jq}.
The analysis was repeated with BES data; the scalar quantum numbers of
the three states were confirmed \cite{Bai:1999mm} suggesting the three
$\rho\rho$ resonances at 1500, 1760, and 2100\,MeV/c$^2$  should be
interpreted as $f_0$ mesons in their $\sigma\sigma$ decays.

\begin{figure}[pb]
\begin{center}
\begin{tabular}{ccc}
\hspace*{-2mm}\includegraphics[width=0.48\textwidth,height=40mm,clip=on]{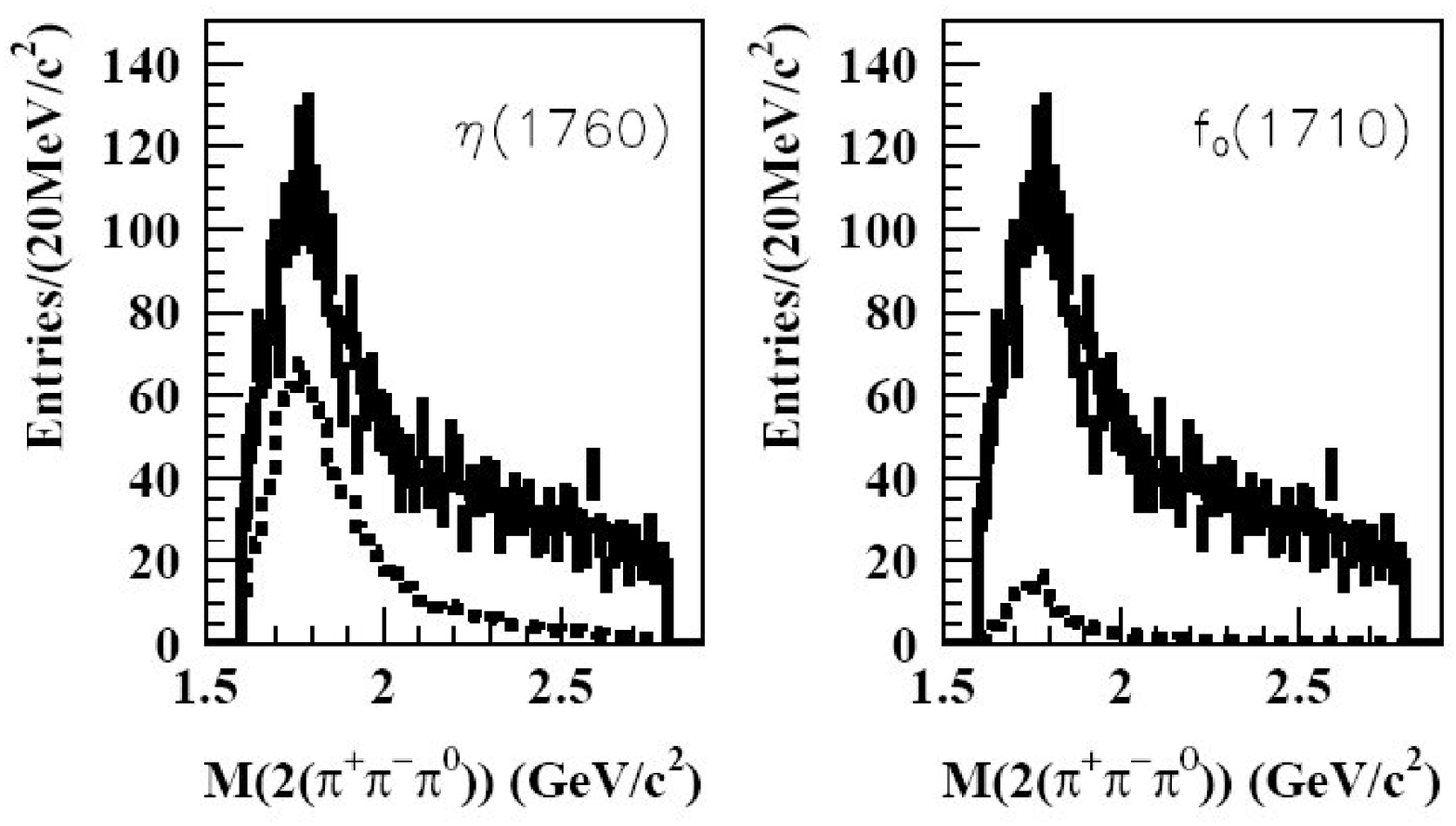}&
\hspace*{-2mm}\includegraphics[width=0.23\textwidth,height=40mm,clip=on]{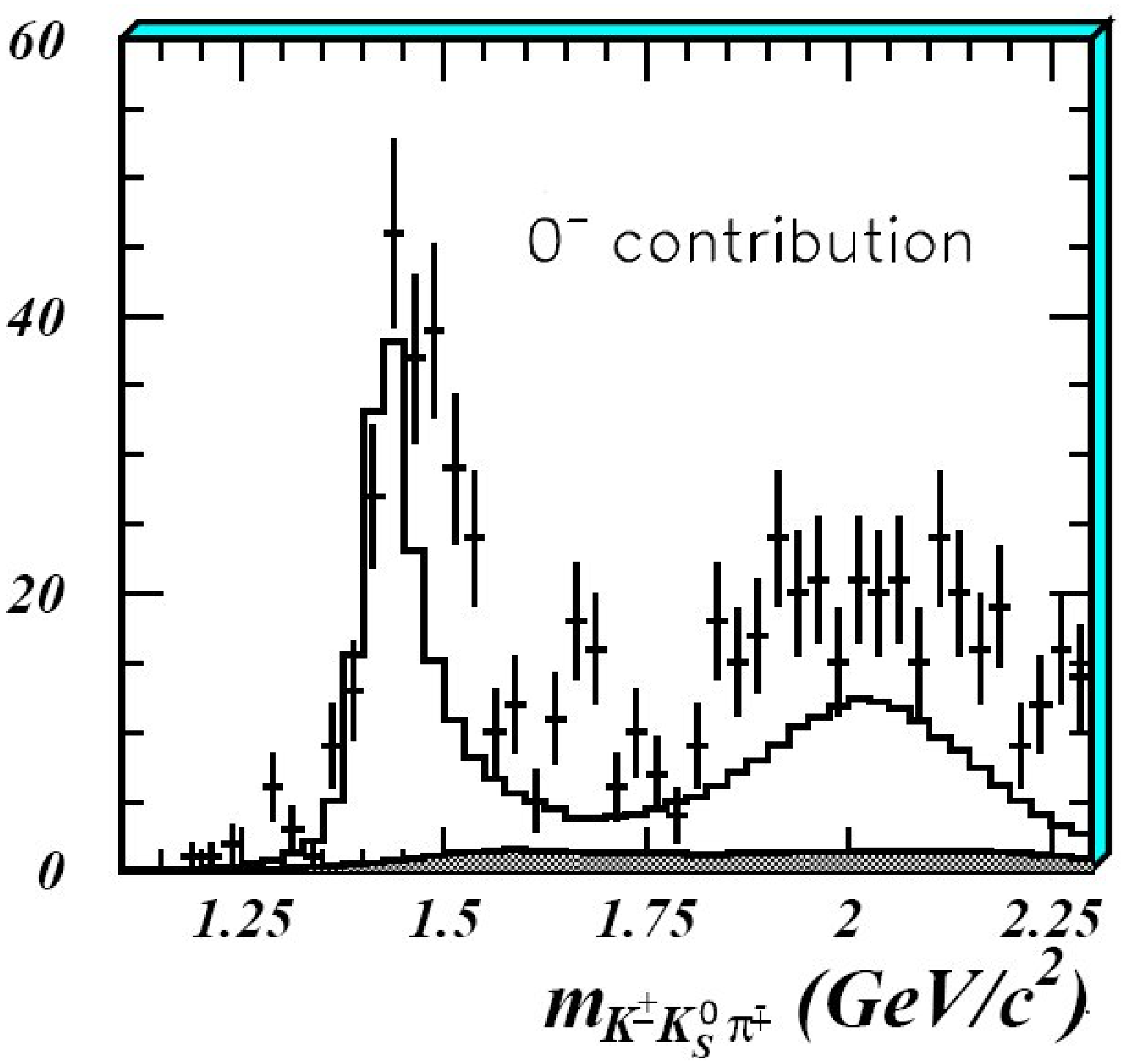}&
\hspace*{-2mm}\includegraphics[width=0.24\textwidth,height=41mm,clip=on]{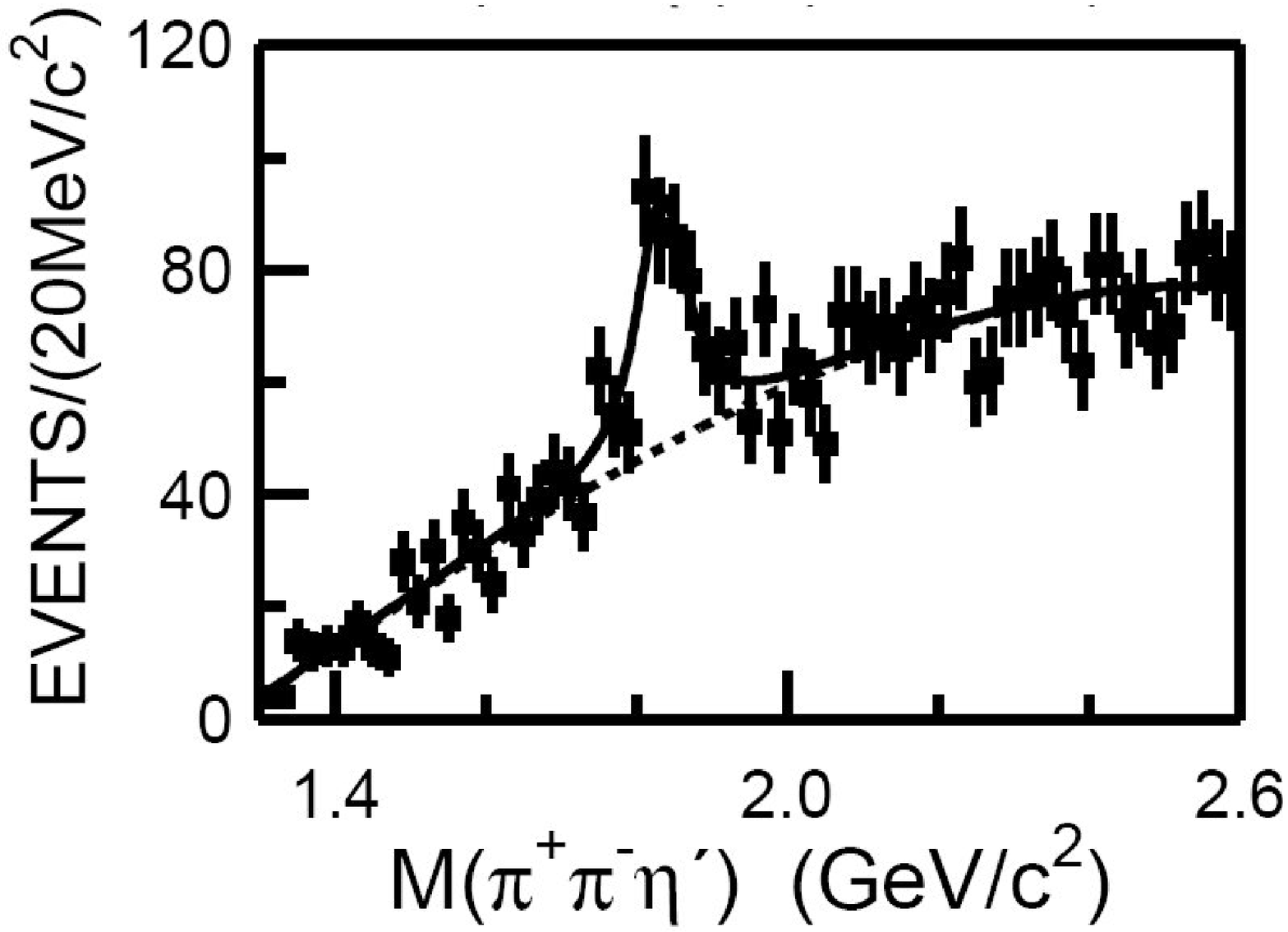}
\end{tabular}\vspace{-40mm}\\
\footnotesize\hspace{-18mm}a\hspace{41mm}b\hspace{35mm}c\hspace{44mm}d\hspace{18mm}\\
\vspace{32mm}\end{center}
\caption{\label{bes-3}Radiative decays of J/$\psi$ to $\omega\omega$
with pseudoscalar and the scalar wave as determined from PWA
\cite{Ablikim:2006ca}.   J/$\psi$ decays into $\gamma K\bar K\pi$
\cite{Bai:2000ss} show $\eta(1440)$ and a broad $\eta(2040)$. In
radiative J/$\psi$ decays into $\pi^+\pi^-\eta^{\prime}$ a narrow peak
at 1835\,MeV/c$^2$ is observed \cite{Ablikim:2005um}. } \end{figure}

Recently, BES reported results from a partial wave analysis of
radiative J/$\psi$ decays into $\omega\omega$ \cite{Ablikim:2006ca}.
Fig. \ref{bes-3} shows a strong enhancement in the
$\omega\omega$ invariant mass distribution at 1.76 GeV/c$^2$.  A
partial wave analysis found pseudoscalar quantum numbers to be dominant
with small scalar and tensor contributions. The large yield stimulated
the collaboration to discuss if the $\eta(1760)$ has a significant
glueball content.

A pseudoscalar $\omega\omega$ resonance at 1760\,MeV/c$^2$, glueball or
$q\bar q$ state, should couple to $\rho\rho$. Isospin invariance would
require equal radiative rates for $\rho^0\rho^0$ and $\omega\omega$.
This is incompatible with data. For the moment, we assume that the
$\rho\rho$ signal at 1760\,MeV/c$^2$ is completely of pseudoscalar
nature (in agreement with \cite{Baltrusaitis:1985nd,Bisello:1988as} and
in contrast to \cite{Bugg:1995jq,Bai:1999mm}. The measured yields are
$(1.44\pm 0.12\pm 0.21)\cdot 10^{-3}$ for $\rho\rho$
\cite{Bisello:1988as} and $(1.98\pm 0.08\pm 0.32)\cdot 10^{-3}$ for
$\omega\omega$ \cite{Ablikim:2006ca}, respectively. The $\rho\rho$
yield should be 3 times larger than that for $\omega\omega$. Thus
isospin is badly broken, by about a factor 4. When the $\rho\rho$
signal has scalar quantum numbers \cite{Bugg:1995jq,Bai:1999mm}, this
factor is even larger. There must be a dynamical reason for the large
pseudoscalar $\omega\omega$ contribution. Possibly, the $c\bar c$
converts into a photon plus two gluons which hadronise into two
coloured $\omega$ mesons. They then neutralise in colour in the final
state and undergo final-state interactions. This process is impossible
for two $\rho$ mesons.

Further pseudoscalar signals were reported by the BES collaboration.
In J/$ \psi\to\gamma K\bar K\pi$ \cite{Bai:2000ss}, the $\eta(1440)$
is seen; the PWA gave evidence for a further resonance, above
2000\,MeV/c$^2$ and called $\eta(2050)$ here.  Mass and width are given
in Table \ref{bes1750}. Data on radiative production of $K^*K^*$ show
(as the MarkIII data \cite{Eigen:1990cm}) a wide pseudoscalar
distribution \cite{Bai:1999mk} which can be fitted with a broad
pseudoscalar resonance at $\sim 1800$\,MeV/c$^2$. Data on J/$
\psi\to\gamma\eta\pi\pi$ \cite{Bai:1999tg} confirmed previous findings
on the $\eta(1440)$ (but did not require a splitting into two
components). At higher mass, two pseudoscalar states were observed with
masses and widths collected in Table \ref{bes1750}.

 \begin{table}[pb] \caption{\label{bes1750} BES results on
radiative J/$\psi$ decay into pseudoscalar mesons $X$ and tentative
SU(3) assignments. The $\eta(2000)$ is proposed as flavour-blind
background, possibly the ground-state pseudoscalar glueball.  The
yields are in unit of $10^{-3}$. The singlet/octet assignment is
discussed in the text. } \renewcommand{\arraystretch}{1.4}
\vspace*{2mm} \begin{center} \begin{tabular}{cccllcc} \hline\hline
$X$&SU(3)& Decay & Mass (MeV/c$^2$) & Width (MeV/c$^2$) & Yield
J/$\psi\to\gamma X$& Ref. \\ \hline $\eta$ &8&&&&$0.90\pm 0.10$ &
\cite{Eidelman:2004wy}\\ $\eta^{\prime}$ &1 &&&&$4.71\pm 0.27$ &
\cite{Eidelman:2004wy}\\ $\eta(1440)$&8 &$ K^*K$&&&$2.8\pm 0.6$  &
\cite{Eidelman:2004wy}\\
 &&$a_0\pi+f_0\eta$&&&$0.4\pm0.1$&\cite{Eidelman:2004wy}\\
 &&$\rho\rho$&&&$1.7\pm 0.4$  & \cite{Eidelman:2004wy}\\
\hline
$X(1835)$&&$ \eta^{\prime}\pi^+\pi^-$
& $1833.7\pm 6.1\pm2.7$ & $67.7\pm20.3\pm7.7$&
         $0.22\pm0.04\pm0.04$   & \cite{Ablikim:2005um}     \\
\hline
$\eta(1760)$&8(?)&$\omega\omega$ & $1744\pm10\pm 15$ &
$244^{+24}_{-21}\pm 25$ & $1.98\pm 0.08\pm 0.32$
&\cite{Ablikim:2006ca}\\ &&$\rho\rho$ & $1775\pm20$ & $115\pm50$
& $1.44\pm 0.12\pm 0.21$         &\cite{Bisello:1988as}$^*$\\ &&$
\eta\pi^+\pi^-$&$1760\pm35$&$\sim250$&$1.2\pm0.5$&\cite{Bai:1999tg}\\
\hline
$\eta(2070)$&1(?)&$\rho\rho$ & $2080\pm40$ & $210\pm40$    &
$1.32\pm 0.15\pm 0.21$          &\cite{Bisello:1988as}$^*$\\
&&$ K^*_0K$ & $2040\pm 50$ & $400\pm 90$  &
    $2.1\pm 0.1\pm 0.7$           & \cite{Bai:2000ss} \\
\hline
$\eta(2000)$&1&$
\eta\pi^+\pi^-$&$\sim1800$&&$0.72\pm0.03$&\cite{Bai:1999tg}\\ &&$
K^*K^*$&$1800\pm100$&$500\pm200$ &$2.3\pm 0.2\pm 0.7$
                                              &\cite{Bai:1999mk} \\
&&$ K^*_0K$ & $\sim1800$ & $\sim 1000$                             &
    $(0.58\pm 0.03\pm 0.20)$    &    \cite{Bai:2000ss}\\
&&$ coupled$ &$2190\pm50$  & $850\pm100$                             &
        & \cite{Bugg:1999jc}   \\

\hline\hline \end{tabular} \end{center}
$^*$ The pseudoscalar nature of the signal was rejected in
\cite{Bugg:1995jq,Bai:1999mm}.
\renewcommand{\arraystretch}{1.0}
\end{table}

Bugg, Dong and Zou \cite{Bugg:1999jc} fitted data on radiative decays
of J/$\psi$ to various final state  with a single Breit-Wigner
resonance having s-dependent widths proportional to the available phase
space in each channel. This resonance has a K-matrix mass
$2190\pm50$\,MeV/c$^2$ and a width $\Gamma= 850\pm100$\,MeV/c$^2$.
Within errors, decays are flavour-blind. The resonance was suggested to
be a pseudoscalar glueball.

In the reaction  J/$\psi\to\gamma\pi^+\pi^-\eta^{\prime}$, a narrow
peak called $X(1835)$ was observed with a statistical significance of
7.7$\sigma$. The data are shown in Fig. \ref{bes-3}. A fit with a
Breit-Wigner function gave mass and width $M=1833.7\pm6.1(stat)\pm
2.7(syst)$\,MeV/c$^2$ and
$\Gamma=67.7\pm20.3(stat)\pm7.7(syst)$\,MeV/c$^2$, respectively, and a
yield $\mathcal B (J\psi\to \gamma {\it X})\cdot \mathcal B(X\to
\pi^+\pi^-\eta^{\prime})= (2.2 \pm 0.4_{\rm stat} \pm0.4_{\rm syst})
\cdot 10^{-4}$. No partial wave analysis has been made but based on
production and decay, an interpretation as $\eta(1835)$ is most likely.
In \cite{Huang:2005bc}, $X(1835)$ is suggested to be the 2$^{\rm nd}$
radial excitation of the $\eta^{\prime}$. We rather believe it to be
the 1$^{\rm st}$ $\eta^{\prime}$ radial excitation. In this
interpretation, $X(1835)$ decays into $\eta^{\prime}$ and the singlet
part of  $\sigma(485)$. Decays into an octet $\eta$ and the
$\sigma(485)$ octet fraction might be small leading to the observed
narrow width. The BES collaboration suggests that $X(1835)$ is related
to $p\bar{p}$ mass threshold enhancement, also observed in radiative
J/$\psi$ decays.

Table \ref{bes1750} collects the BES results  on pseudoscalar meson
production in radiative J/$\psi$ decays. There is no clear separation
into strong and weak radiative yields. The commonly used picture of
radiative J/$\psi$ decays in which the $c\bar c$ system radiates off a
photon and two gluons forming a flavour singlet expects large radiative
yields for singlet production and small octet yields. Of course, unseen
decay modes could change the pattern significantly.

The $X(1835)$ is particularly interesting. As it may be related to the
$p\bar{p}$ mass threshold enhancement, we present also
baryon-antibaryon threshold enhancements observed in other reactions
even though there are mostly no quantum number determinations.

\subsubsection{\label{Baryon-antibaryon threshold enhancements}
Baryon-antibaryon threshold enhancements}

\begin{figure}[pb]
\begin{minipage}[c]{0.35\textwidth}
\includegraphics[width=\textwidth,height=1.06\textwidth]{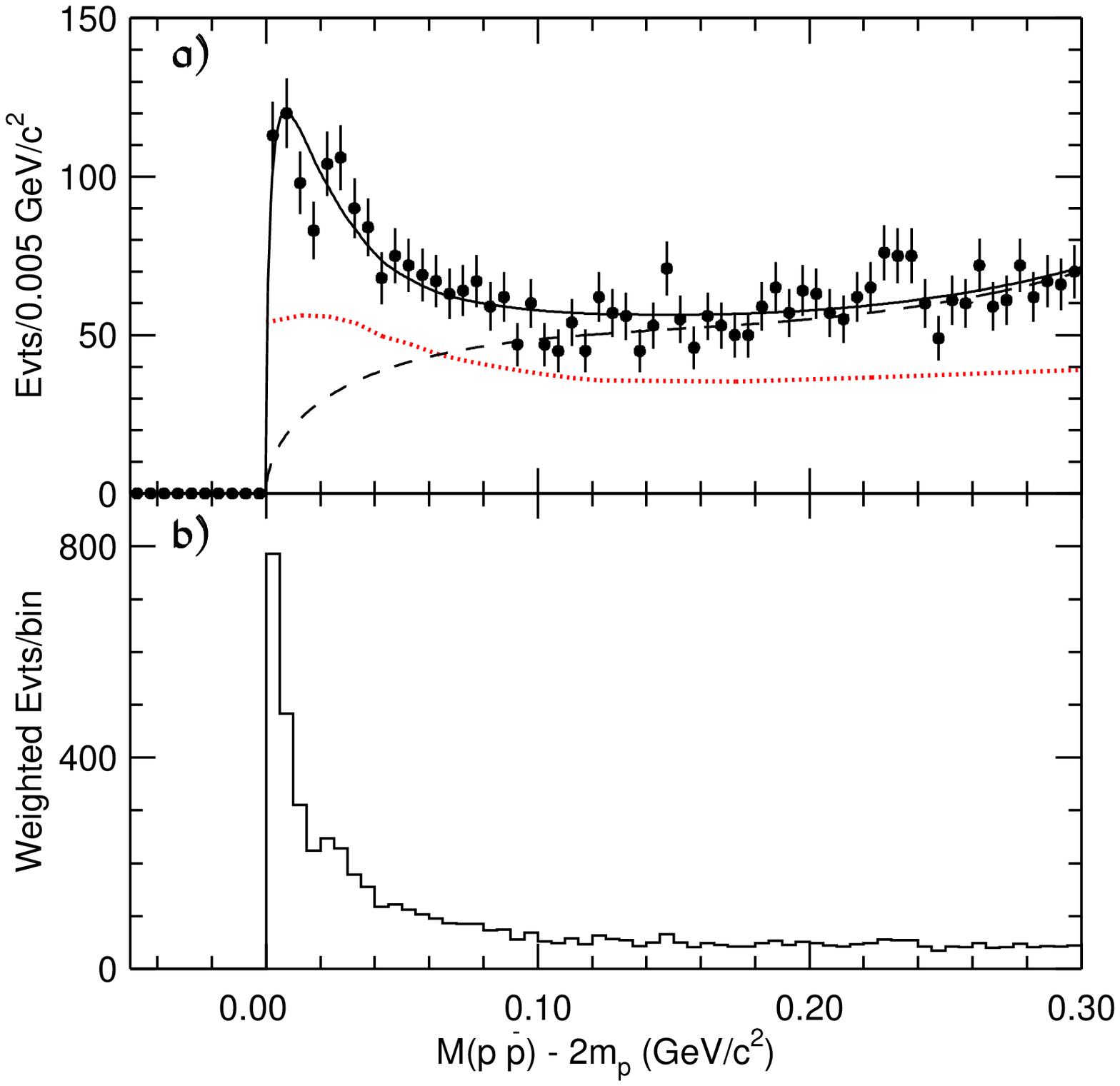}
\end{minipage}
\begin{minipage}[c]{0.64\textwidth}
\begin{tabular}{cc}
\hspace{-3mm}\includegraphics[width=0.45\textwidth,height=0.28\textwidth]{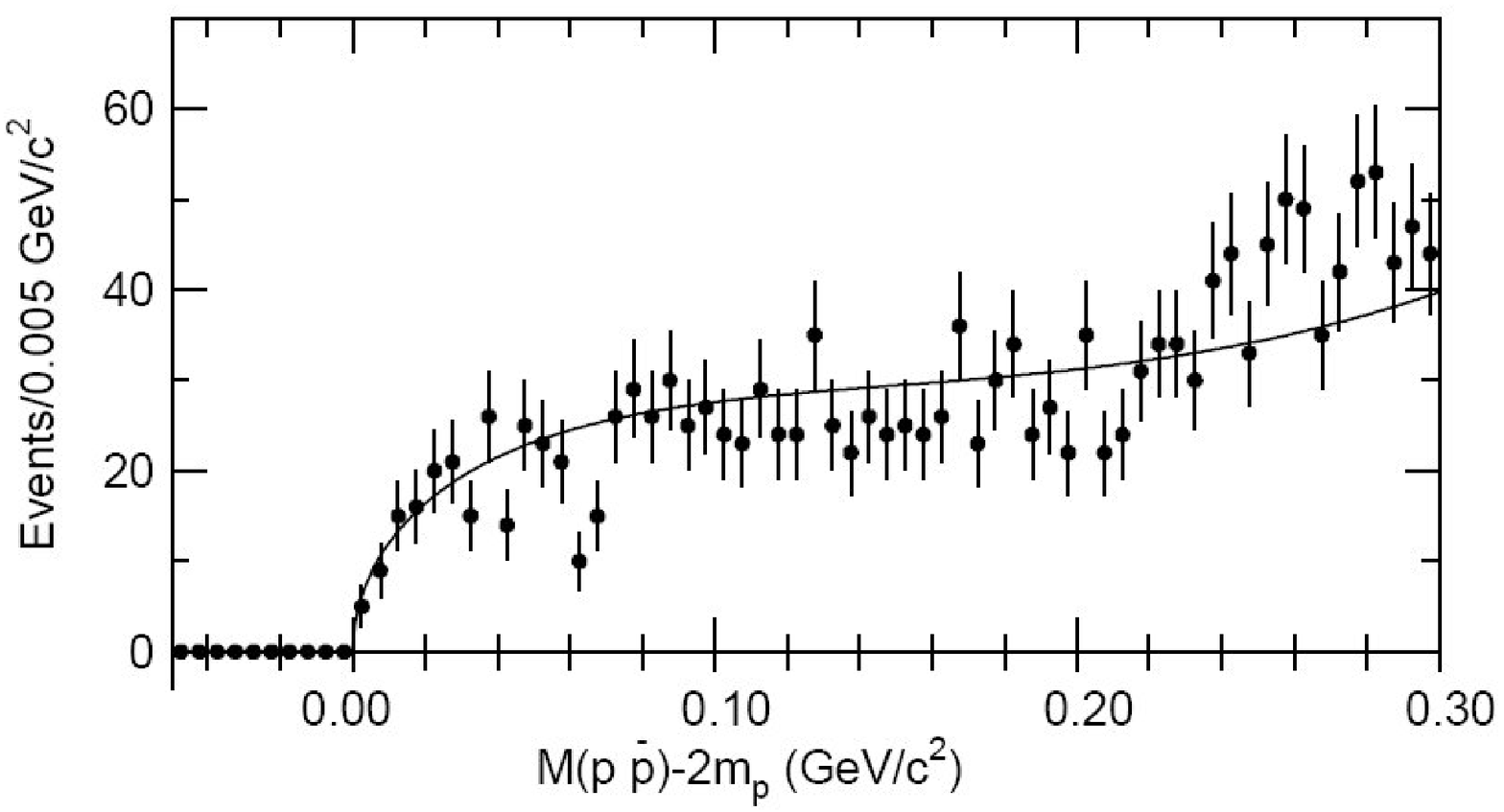}&
\hspace{-3mm}\includegraphics[width=0.45\textwidth,height=0.28\textwidth]{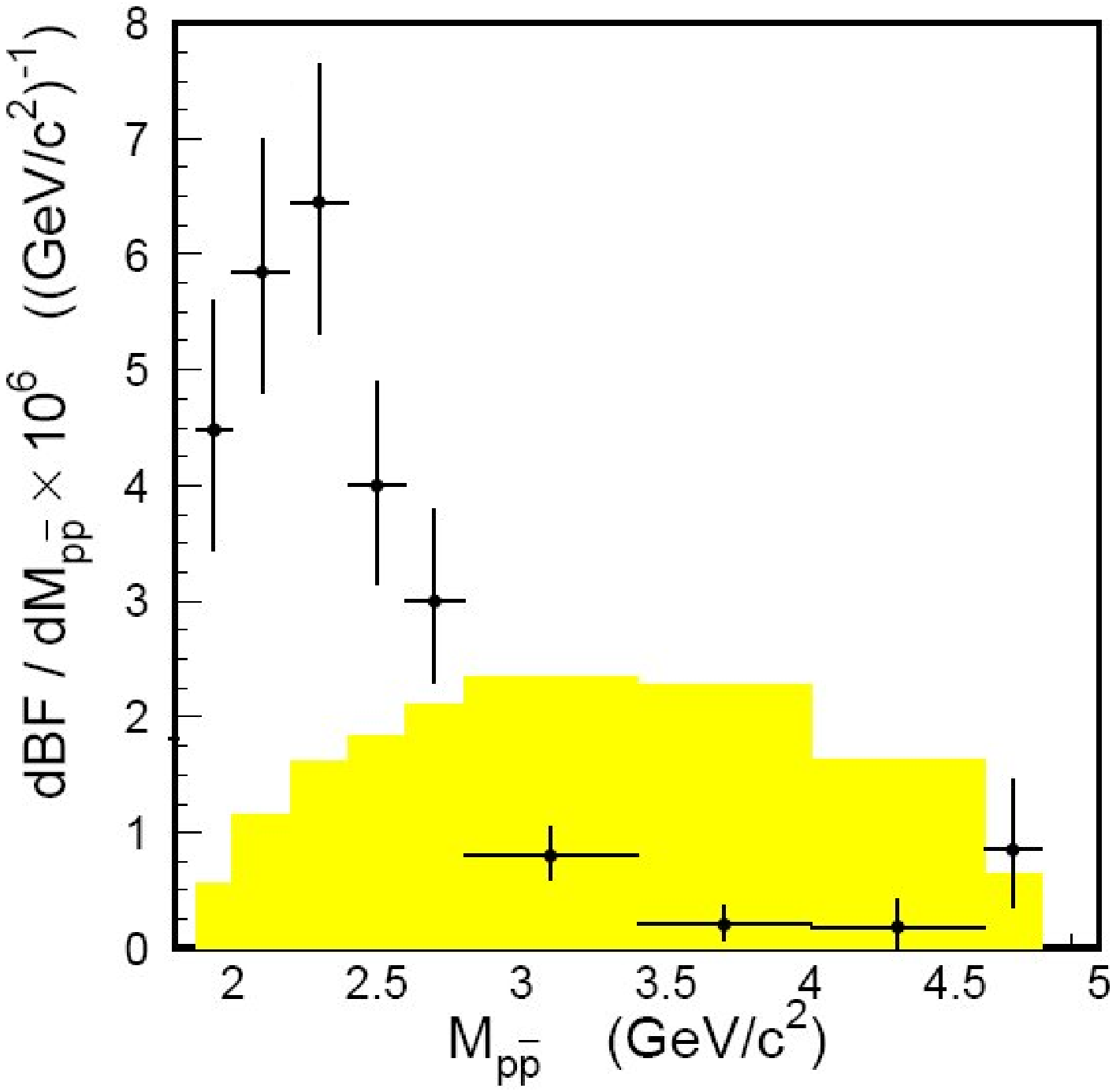}\\
\hspace{-3mm}\includegraphics[width=0.45\textwidth,height=0.28\textwidth]{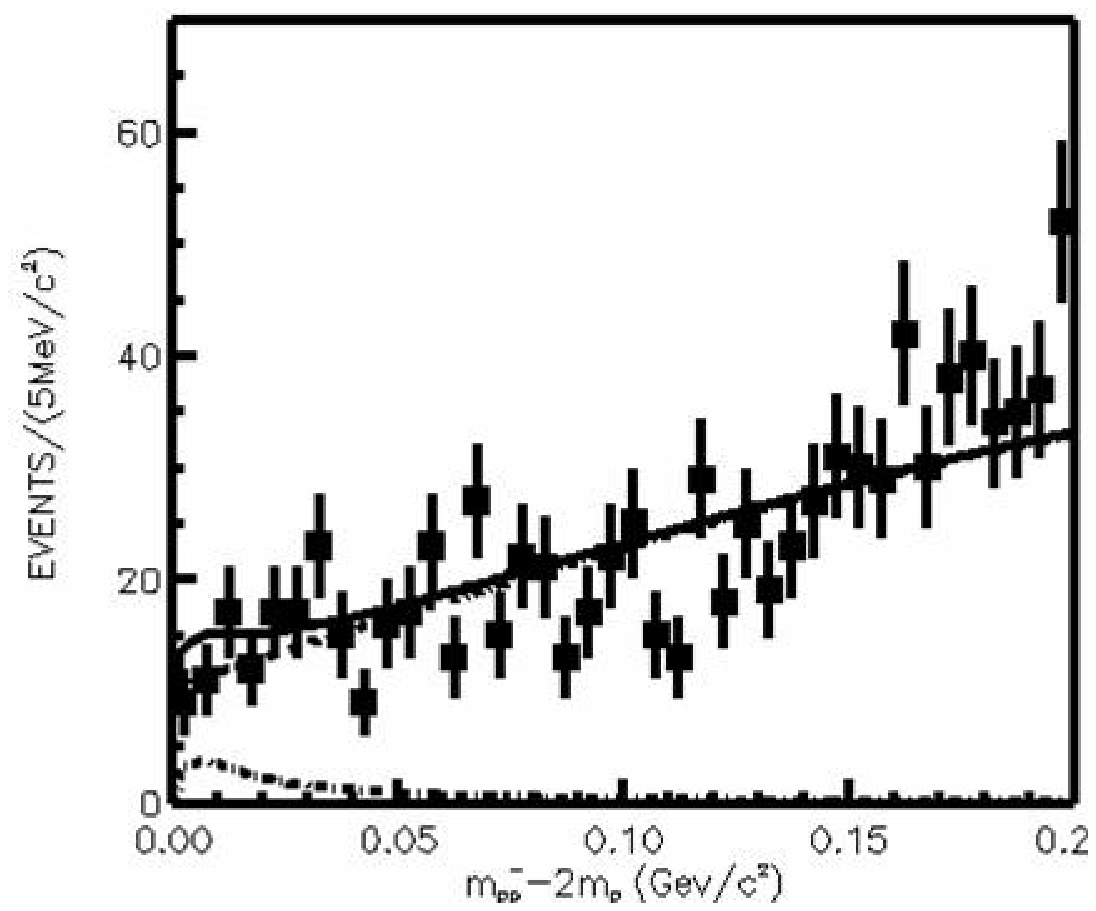}&
\hspace{-3mm}\includegraphics[width=0.46\textwidth,height=0.28\textwidth]{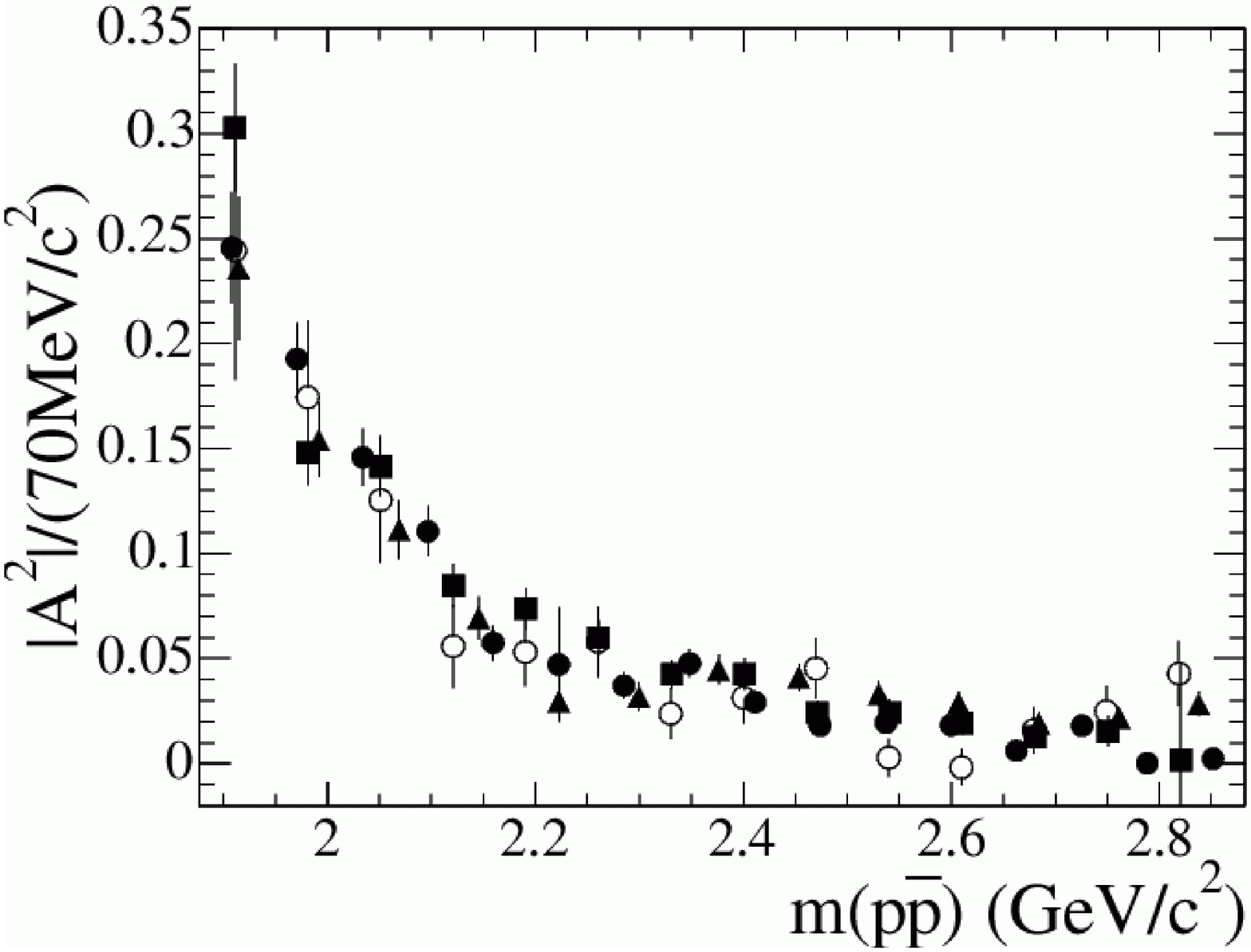}\\
 \end{tabular}\\
 \end{minipage}
\bc
\vspace{-64mm}
\footnotesize\hspace{37mm}c)\hspace{83mm}e)\hspace{-20mm}\vspace{30mm}\\
\footnotesize\hspace{37mm}d)\hspace{83mm}f)\hspace{-20mm}\vspace{25mm}\\
\ec
\caption{\label{ps:BES}
a: The  $M_{ p\bar p}-2m_p$ distribution for J/$ \psi\to\gamma
p\bar p$. The data is fitted with a Breit-Wigner function plus
background function represented by the dashed curve.  The dotted curve
gives the acceptance. b: The  $M_{\ppbar}-2m_p$
distribution with events weighted by $q_0/q$
\cite{Bai:2003sw}. The $M_{\ppbar}-2m_p$ distributions for c)
J/$ \psi\to\pi^0 p\bar p$ (\cite{Bai:2003sw}) and for d)
J/$\psi\to\omega p\bar p$ (\cite{Jin:2006ro}) show no
significant threshold enhancements.  Invariant $ p\bar p$
mass distributions from: e) $B^0\to p \bar p K^0$, BELLE
experiment \cite{Wang:2003yi}; f) phase space-corrected $p\bar{p}$
invariant mass distributions for four decay modes: $ B^0\to \bar D^0
p\bar p$ (triangles), $B^0\to \bar D^{*0} p \bar p$ (open circles),
$B^0 \to D^- p \bar p \pi^+$ (squares), and $ B^0 \to D^{*-} p\bar
p\pi^+$ (closed circles); BaBaR experiment \cite{Aubert:2006qx}.
 } \end{figure}

The BES collaboration \cite{Bai:2003sw} observed a narrow $ p\bar p$
threshold enhancement in radiative J/$\psi$ decay but not in  J/$
\psi\to\pi^0 p\bar p$ nor in J/$ \psi\to\omega p\bar p$ (Fig.
\ref{ps:BES}). It can be fitted with an $S$- or $P$-wave Breit-Wigner
function giving masses of $1859\pm 6$ and $1876\pm0.9$\,MeV/c$^2$,
respectively. The absence of the enhancement in J/$\psi\to\pi^0 p\bar p$
suggests positive $C$-parity; parity conservation and assuming low
angular momenta $\ell=0$ or 1 restrict quantum numbers to $0^{-+}$,
$0^{++}$, $1^{++}$, and $2^{-+}$. The absence of the signal in recoil
to an $\omega$ must be due to a special dynamical selection.  A first
sight of an anomalous behaviour at the $ p\bar p$ threshold had already
been reported by the BELLE collaboration in $B$ decays $B^{\pm}\to p
\bar{p} K^{\pm}$ \cite{Abe:2002ds}. Closer inspection shows, however,
that the Belle and BES observations are very different in nature. The
BELLE structure seen in $B$ decays is much broader (with a width,
estimated from the graph, of about 400\,MeV/c$^2$) and resembles
perhaps the $ p\bar p$ signal observed by BaBaR in different decay
modes of $B$ mesons into  $D\pi p\bar p$ final states
\cite{Aubert:2006qx}. The BES signal has a visible width of about
60\,MeV/c$^2$ and fits give values which are compatible with zero,
hence the two phenomena observed at Belle and BES are likely different.

There have been several attempts to explain the BES result.
The J\"ulich group analysed the data within their $N\bar N$ model and
were able to reproduce the $p\bar p$ data with just one normalization
parameter. The isovector part of $N\bar N$ interactions plays the
dominant $\rm r\hat{o}le$; fine-tuning is possible by adding final-state
interactions with $\pi(1800)$ as intermediate state
\cite{Sibirtsev:2004id}. An extension to $B$ decays is found in
\cite{Haidenbauer:2006au}. Of course, production of isovector mesons
like $N\bar N$ with $I=1$ or $\pi(1800)$ is unexpected in radiative
J/$\psi$ decays; indeed, no visible signal is observed in
$J\psi\to\gamma\pi^+\pi^-\pi^0$ \cite{Jin:2006ro}.
Entem and Fernandez, describing scattering data and mass shifts and
broadening of $p\bar p$ atomic levels in a constituent quark model,
assign the threshold enhancement to final-state interactions
\cite{Entem:2006pc,Entem:2007bb}.
Bugg interpreted the threshold $p\bar p$ peak as a cusp  arising from
the well known threshold peak in $p\bar p$ elastic scattering due to
annihilation \cite{Bugg:2004rk}. Zou and Chiang find that
final-state-interaction makes an important contribution to the
$p\bar{p}$ near-threshold enhancement \cite{Zou:2003zn}. We just
mention (without imposing a relation between the two phenomena) that
the $p\bar p$ threshold is not far from the dip in the six-pion
photoproduction cross section observed by DM2 \cite{Bisello:1981sh},
Focus \cite{Frabetti:2003pw}, and BaBaR \cite{Aubert:2006jq}. Its
interpretation of the dip goes a la mode; DM2 -- with low statistics --
assign the peak below the dip to $\rho(1600)$, the Focus dip is argued
to be a hybrid, BaBaR suggests an interpretation as amalgamation of
several broad resonances.

The BES collaboration associated the $p\bar p$ threshold enhancement
with the $X(1835)$ suggesting a large affinity of both, $ p\bar p$ and
$\eta^{\prime}$, to gluons. From the $ p\bar p$ data, $M = 1831 \pm
7$~MeV/c$^2$ and a width $\Gamma < 153$~MeV/c$^2$ (at the 90$\%$ C.L.)
was obtained, compatible with mass and width  of the
$\eta^{\prime}\pi^+\pi^-$ signal. Hence both signals might be one
comparatively narrow pseudoscalar resonance. The  product branching
ratio for the $X(1835)$ radiative yield $$\mathcal B(J/\psi\to\
X_{\pi\pi\eta'})\cdot B(X_{\pi\pi\eta'}\to \pi^+\pi^-\eta') = (2.2 \pm
0.4(stat) \pm0.4(syst)) \times 10^{-4} $$ can be compared to the
branching ratio for the $p\bar p$ threshold enhancement $$\mathcal
B(J/\psi\to\ X_{p\bar p})\cdot B(X_{p\bar p}\to p\bar p) =  (7.0 \pm
0.4 {\rm (stat)} ^{+1.9}_{-0.8}{\rm (syst)})\times 10^{-5} $$ Hence the
$(p\bar p)/(\pi^+\pi^-\eta)$ ratio is 1/3; from the absence of
$X(1835)$ in the inclusive photon spectrum measured by the Crystal Ball
collaboration \cite{Gaiser:1985ix}, the BES collaboration concluded
that $X$ must have a branching ratio into $p\bar p$ exceeding 4\%
\cite{Jin:2006ro}.

A clue to decide on the different interpretations is possibly provided
by the non-observation of the $p\bar p$ threshold enhancement in
J/$\psi\to\omega p\bar p$. The $\omega$ has the same quantum numbers as
the photon; hence the absence of the signal in the latter data cannot
be due to any kind of selection rule based on conservation of angular
momentum, charge conjugation or parity. On the other hand, the J/$\psi$
radiative yield for $\eta^{\prime}$ exceeds the $\eta$ yield by a
factor 4.8 while $\omega\eta^{\prime}$ has a branching ratio (from
J/$\psi$) which is 10\% only of the $\omega\eta$ yield. The observation
of $X(1835)$ in radiative J/$\psi$ and its non-observation in
$\omega$\,J/$\psi$ may thus hint at its flavour singlet structure.
This conjecture is supported by its observation in
$\eta^{\prime}\pi\pi$, its non-observation in $\eta\pi\pi$ and
assigning a large flavour singlet component to the $\sigma(485)$.

The $X(1835)$ is one of the many examples where the vicinity of a
threshold, in this case $\bar pp$, attracts the mass of a
close by resonance. `Dressing' of the $q\bar q$ singlet meson with
two $q\bar q$ pairs can create $N\bar N$. Final state interactions
enhance the probability of this transition. In this way, the $q\bar q$
meson mixes with the $\bar pp$ final state and its wave function
develops a sizable $\bar pp$ component.

\begin{figure}[pt]
\begin{center}
\begin{tabular}{ccc}
\hspace{-3mm}\includegraphics[width=0.35\textwidth,height=0.28\textwidth]{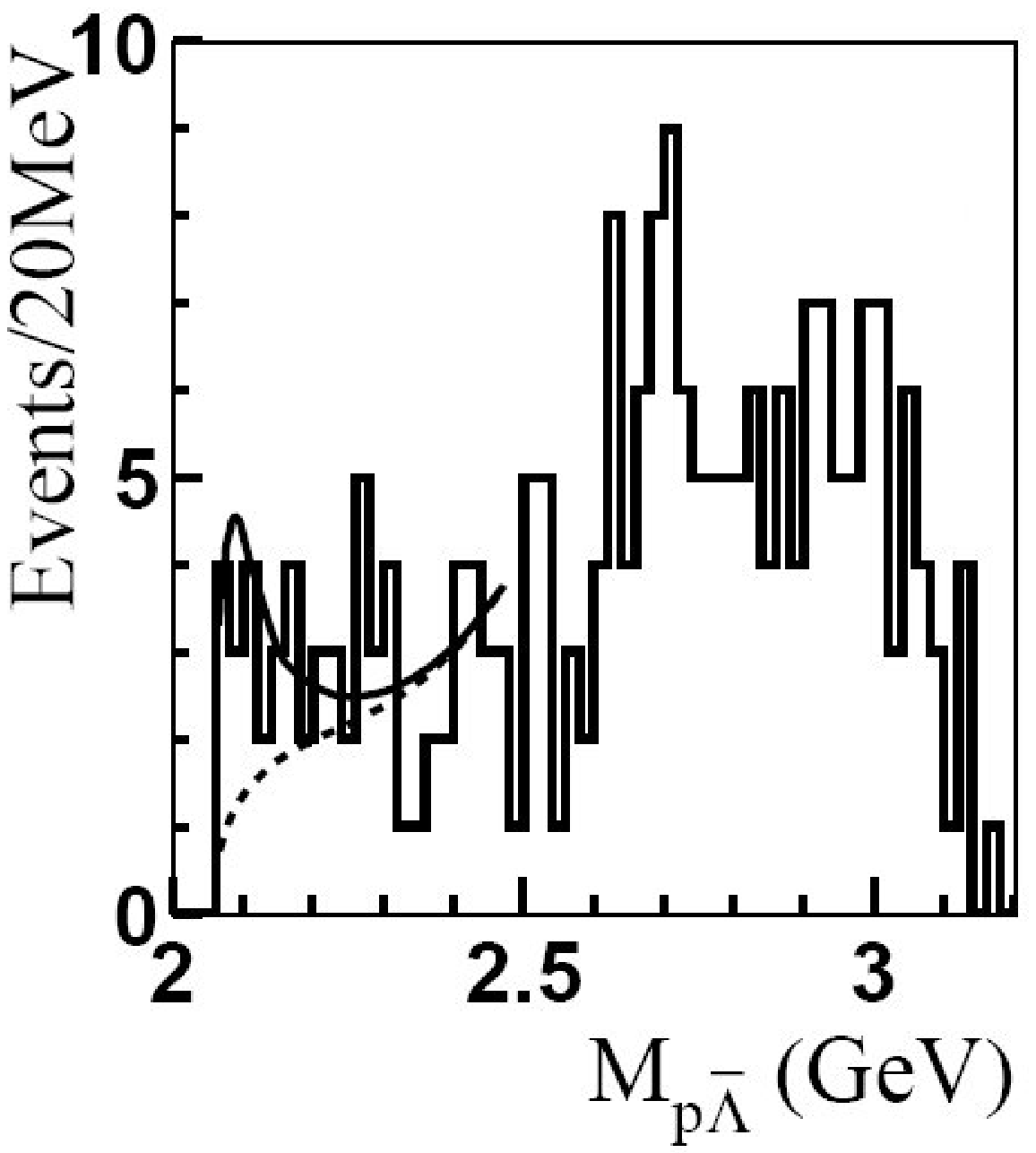}&
\epsfig{file=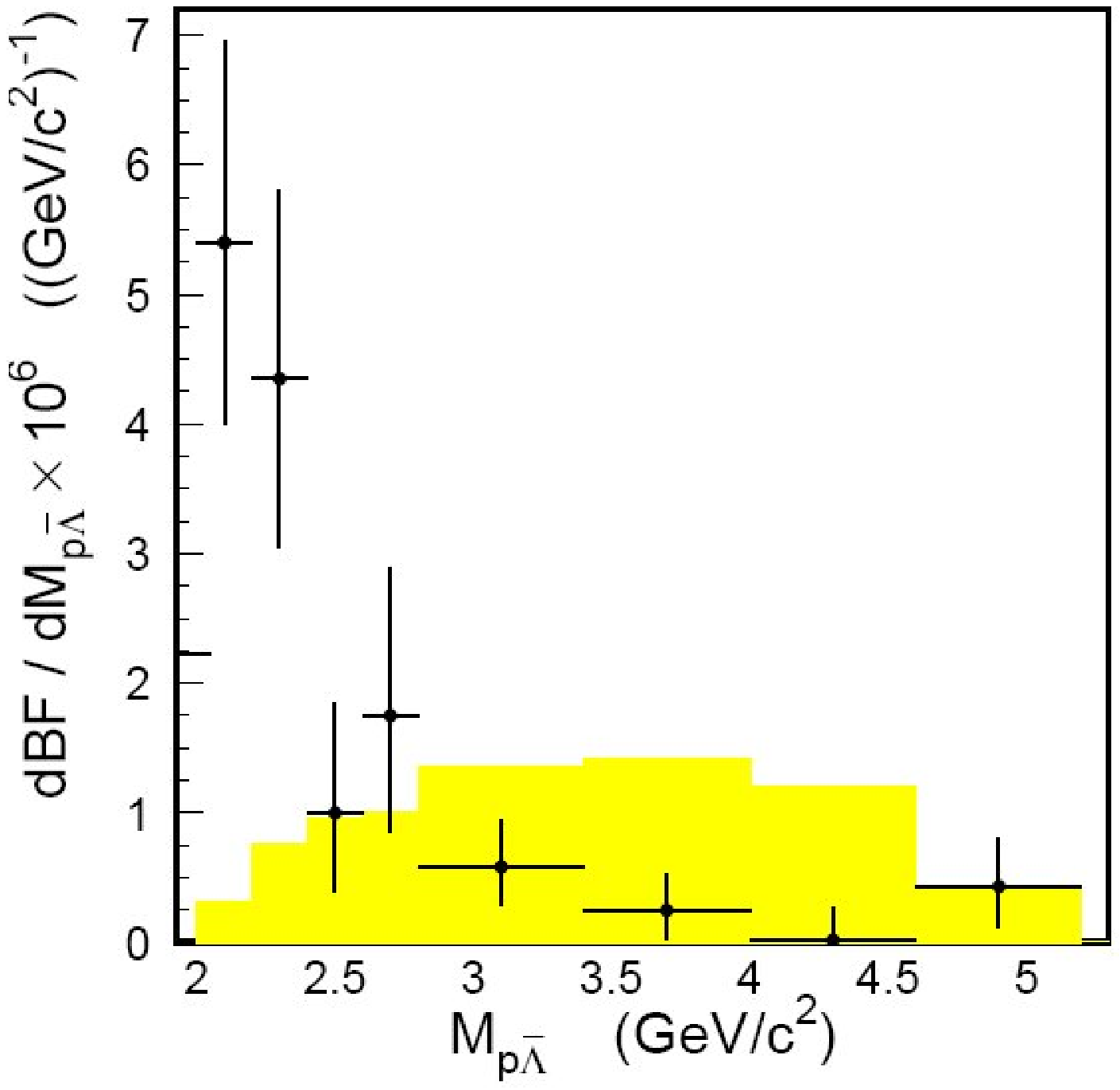,width=0.3\textwidth,height=0.25\textwidth,clip=}&
\epsfig{file=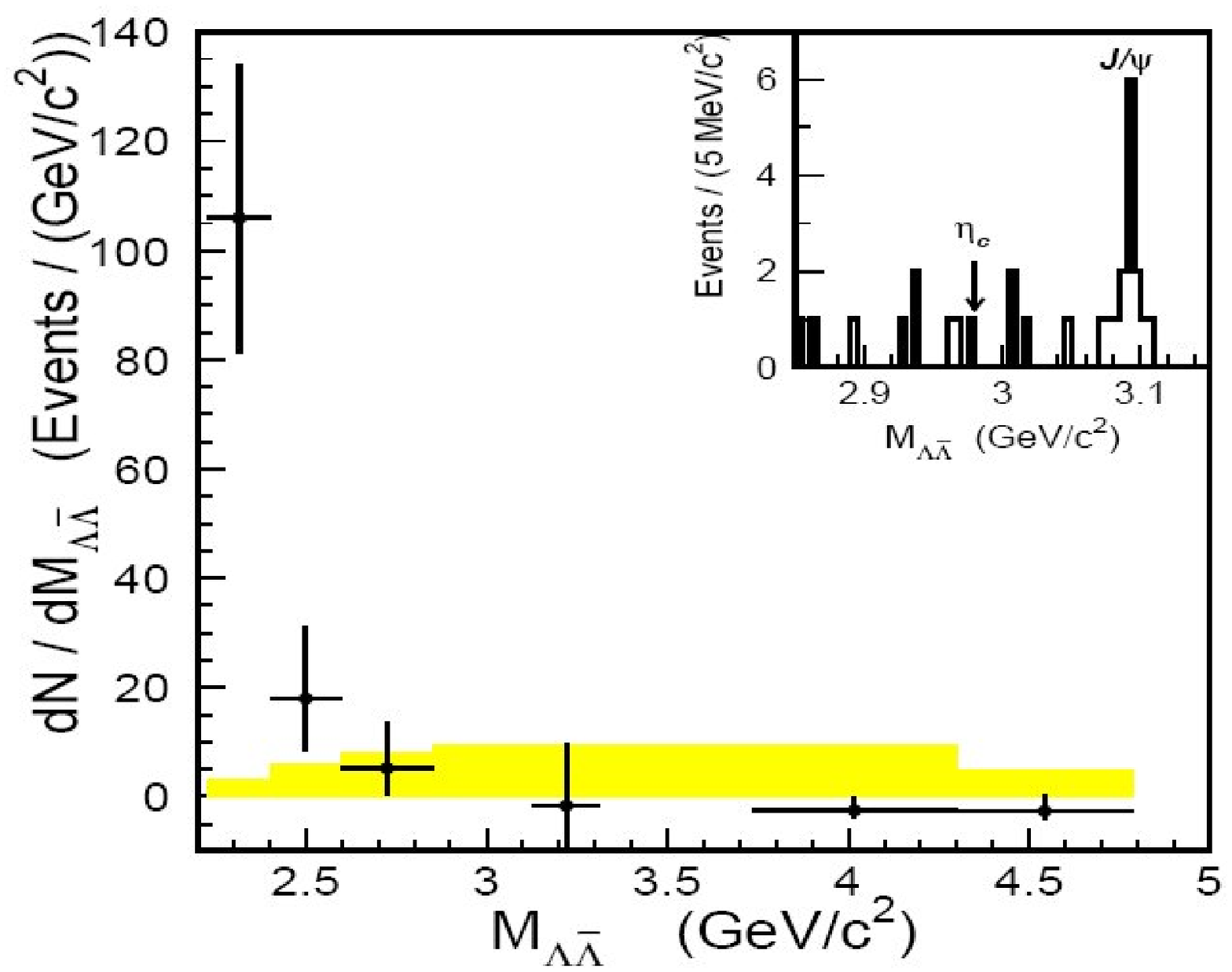,width=0.3\textwidth,height=0.25\textwidth,clip=}\\
\end{tabular}
\end{center}
\caption{\label{ps:Belle}Left: The $p\bar \Lambda$ mass distribution
from J/$\psi\to\gamma p\bar\Lambda$ \cite{Ablikim:2004dj};
 centre: $p\bar \Lambda$ mass distribution from  $ B^0 \to p
\bar\Lambda\pi^-$, right: $\Lambda \bar\Lambda$ mass distribution from $
B^+ \to \Lambda \bar\Lambda K^+$
\cite{Wang:2003yi,Wang:2003iz,Lee:2004mg}.} \end{figure}

Further reactions with a baryon and an antibaryon in the final state
were studied by the BELLE and BES collaborations. The reactions include
J/$\psi\to\gamma p\bar\Lambda$ \cite{Ablikim:2004dj}, $\bar B^0 \to
D^{(*)0} p\bar p$,  $ B^0 \to p \bar\Lambda\pi^-$ $B^+ \to p\bar p
\pi^+, B^0\to p \bar p K^0$, $ B^+ \to p\bar p  K^{*+}$, $B^+\to\Lambda
\bar \Lambda K^+$, $ B^- \to J/\psi \Lambda \bar p$, $B^+\to\bar
\Xi_{c}^0\Lambda_c^+$, and $ B^0\to\bar\Xi_{c}^-\Lambda_c^+$
\cite{Abe:2002tw,Wang:2003yi,Wang:2003iz,Lee:2004mg,Xie:2005tf,%
Chistov:2005zb,Wang:2007as}. In Fig. \ref{ps:Belle} we show the $
p\bar\Lambda$ invariant mass distribution from J/$\psi\to\gamma
p\bar\Lambda$ (BES) and from  $ B^0 \to p \bar\Lambda\pi^-$ and $ B^+
\to \Lambda \bar\Lambda K^+$ (Belle) decays. It is not known at present
if particular spin-parities can be assigned to these threshold
enhancements, and if resonances are involved. Recently, a wide $
p\bar\Lambda$ threshold enhancement was reported for $B^+\to p
\bar\Lambda\pi^0$ \cite{Wang:2007as}. We notice that the
$p\bar \Lambda$ mass distribution observed in radiative J/$\psi$
decays appears to be much narrower than their counterparts from $B$
decays. The BES data can be fitted with both, $S$-wave and $P$-wave
Breit-Wigner amplitudes.  If it is fitted with an $S$-wave Breit-Wigner
resonance function, the mass is $m=2075\pm 12 \pm 5$\,MeV/c$^2$, the
width  $\Gamma =90 \pm 35 \pm 9$\,MeV/c$^2$, and the branching ratio
$BR(J/\psi \rightarrow K^-X) BR (X\rightarrow p\bar\Lambda)$ = $ (5.9
\pm 1.4 \pm 2.0 ) \times 10^{-5} $. It is tempting to interpret this
observation as pseudoscalar $K(2075)$. It would certainly fit very well
into the scheme suggested in Table \ref{radexcps} and Fig.
\ref{pic:m2}. To substantiate this conjecture, a signal has to be seen
in radiative J/$\psi$ production of
 $K\pi$ and $K\pi\pi$ systems.

\subsection{\label{The nonet of pseudoscalar radial excitations}
The nonet of  pseudoscalar radial excitations}

\begin{table}[pb]
\caption{\label{radexcps}
Pseudoscalar radial excitations. The mixing angle is calculated from
the linear GMO formula \ref{GMO1}, the sign from \ref{GMO2}. The
mass differences to the respective ground states are listed as small
numbers. See text for discussion of the states.\vspace{2mm}}
\begin{center} \renewcommand{\arraystretch}{1.4}
\begin{tabular}{cccccrrl}
 \hline\hline
 $1^1S_0$      &$\pi(135)$&K(498)&$\eta(548)$&$\eta(958)$&
$\rm\Theta_{PS}=-24.6^{\circ}$& $\eta_c(2980)$ \\
\hspace{5mm}{\tiny $\delta
M=$}&{\tiny1.24}&{\tiny0.97}&{\tiny0.89}&{\tiny0.88}&&{\tiny0.66 }&
\hspace{-5mm} {\tiny GeV/c$^2$}\\
$2^1S_0$&$\pi(1375)$&K(1460)&$\eta(1440)$&$X(1835)$&$-20.8^{\circ}$&$\eta_c(3638)$\\
\hspace{5mm}{\tiny $\delta M=$ }
 &{\tiny1.66}&{\tiny1.34}&{\tiny1.21}&{\tiny1.11} &&{\tiny0.96}&
\hspace{-5mm} {\tiny GeV/c$^2$}\\
 $3^1 S_0$
&$\pi(1800)$&K(1830)&$\eta(1760)$&$\eta(2070)$&$-30.5^{\circ}$&X(3940)\\
\hline\hline
\end{tabular}
\renewcommand{\arraystretch}{1.0}
\end{center}
\end{table}

With $\eta(1760)$ and $X(1835)$, there are two close-by resonances,
likely with identical quantum numbers but with widths which differ by
a factor 4. Expected are in this mass region the $2^{\rm
nd}$ radial excitation of the $\eta$ and the  $1^{\rm st}$ radial
excitation of the $\eta^{\prime}$. In the SU(3) limit of $\sigma(485)$
being a flavour singlet resonance, singlet $\eta$ excitations decay
into $\eta^{\prime}\sigma(465)$, octets into $\eta\sigma(485)$. The
narrow $X(1835)$ width may thus indicate that it is a flavour singlet
state. Thus we interpret $X(1835)$ as $1^{\rm st}$ $\eta^{\prime}$
radial excitation.

In the region above 2\,GeV/c$^2$, a few observations of isoscalar
pseudoscalar resonances have been reported. Bisello {\it et al.} find
2104\,MeV/c$^2$ from  J/$\psi\to\gamma\rho\rho$ , BES $M=2040\pm
50,\Gamma=400\pm90$\,MeV/c$^2$ from  J/$\psi\to\gamma K\bar K\pi$,
Bugg {\it et al.} report $2190\pm50$\,MeV/c$^2$ and a width $\Gamma=
850\pm100$\,MeV/c$^2$ from an analysis of several final states in radiative
J/$\psi$ decays. The Queen-Mary-St. Petersburg group analysed Crystal
Barrel data on $ p\bar p$ annihilation in flight and reported an
$\eta(2010)$ with $M=2010^{+35}_{-60}, \Gamma=270\pm60$\,MeV/c$^2$,
consistent with $\eta(2040)$. Table \ref{radexcps} collects
pseudoscalar ground states and radial excitations. We combine these
results by introducing an $\eta(2070)$ into the discussion, with a mass
given by the mean of the two results quoting a narrow width. This state
is the most uncertain one in the table. In this mass region, the
$2^{\rm nd}$ $\eta^{\prime}$ and $3^{\rm rd}$ $\eta$ radial excitation
are expected.

\begin{figure}[pt]
\bc
\includegraphics[width=0.6\textwidth]{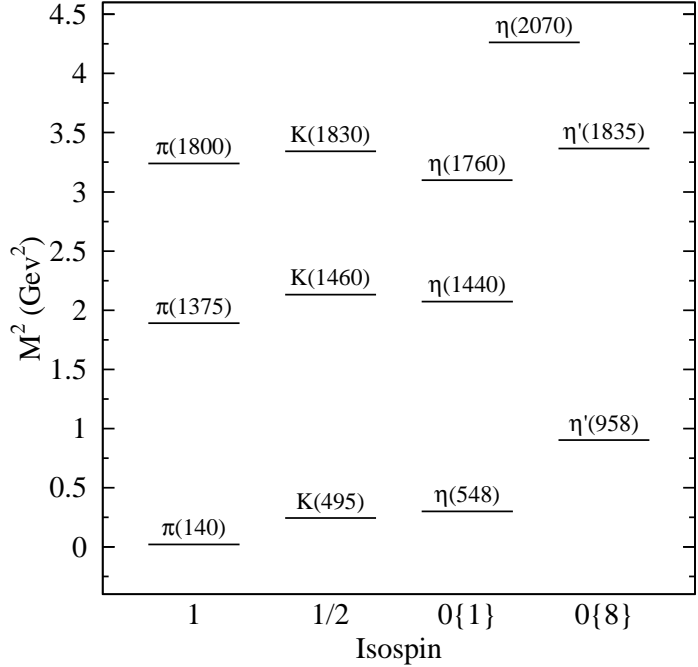}
\ec
\caption{\label{pic:m2}
The pseudoscalar radial excitations. Most of the resonances are well
established. The $\eta(1295)$ has been omitted. $ K(1830)$ and
$\eta(1760)$ are {\it omitted from the summary table} in PDG; for
$\eta(1760)$ there is new evidence from BES. For $X(1835)$, spin and
parity are not measured. $\eta(2070)$ relies on a weak evidence from
two analyses. It might be a flavour singlet or octet state. See text
for discussion and references.
  }
\end{figure}

The X(3940) resonance is seen as a narrow peak in $e^+ e^-$
annihilation in the spectrum of masses recoiling against reconstructed
J/$\psi$ \cite{Abe:2005hd}. The spectrum contains $\eta_c(1S)$ and
$\eta_c(2S)$ and $\chi_{c0}$. Due to its production and decay (assuming
$\ell=0$ or 1) into $ D^*D$ its quantum numbers could  be $J^{PC}=
0^{-+},1^{++}$ or $2^{-+}$. If we reserve $1^{++}$ to X(3872), it seems
likely that X(3940) should be identified as $\eta_c(3S)$.

The systematic behaviour apparent in Table \ref{radexcps} covers the
range from the pion and its excitations to the charmonium states. It is
obvious that the inclusion of an $\eta(1295)$ would be problematic.
The first radial excitations as proposed here still show a singlet-octet
splitting rather than ideal mixing. For the second radial excitations
the uncertainties become large. The singlet-octet mixing angles are not
very different for the three nonets suggesting that the radial
excitations of pseudoscalar mesons are organised into singlets and
octets rather than into $n\bar n$ and $s\bar s$ states. The mixing
angles given in Table \ref{radexcps} are calculated from the Gell-Mann
Okubo mass formula
\be
\label{GMO1}
\tan^2\Theta = \frac{4m_K-m_{\pi}-3m_{\eta}}{-4m_K+m_{\pi}+3m_{\eta'}}.
\ee
This equation does not yield the sign of the mixing angle. One can use
instead
\be
\label{GMO2}
\tan\Theta = \frac{4m_K-m_{\pi}-3m_{\eta}}{m_{\pi}-m_K}
\ee
The latter formula yields mixing angles of $-11.5^{\circ}$,
$-31.1^{\circ}$, and $-70.5^{\circ}$, respectively. These values depend
on small and not precisely known mass differences like
$M_{K(1830)}-M_{\pi(1800)}$ and are considered as less reliable.

In addition to the 'narrow' resonances collected in Fig. \ref{pic:m2},
a wide isoscalar $0^{-+}$ background amplitude is required in some
analyses. It is compatible with being flavour singlet. The most
straightforward interpretation assigns the background amplitude to a
very wide pseudoscalar glueball, with a width of more than 800\,MeV. We
refrain from giving a precise pole position of this elusive object.

One final (trivial) remark. The masses of pseudoscalar and vector
radial excitations are similar; radial excitations have no Goldstone
character.

\subsubsection{\label{Other interpretations}
Other interpretations}

\begin{figure}[ph]
\vspace{2mm}
\bc
\begin{tabular}{cccc}
\hspace{-1mm}\includegraphics[width=0.23\textwidth,height=0.23\textwidth]{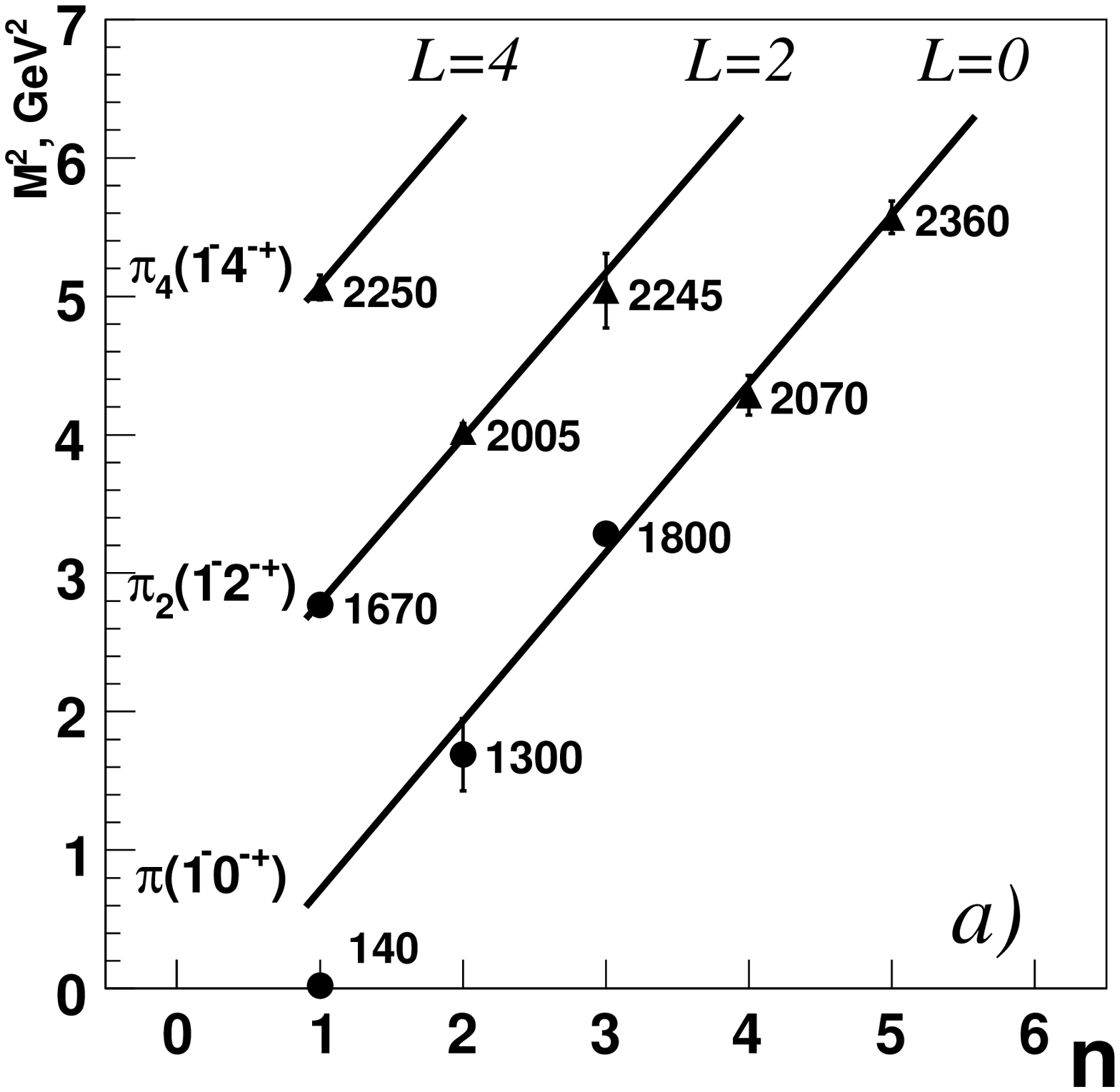}&
\hspace{-3mm}\includegraphics[width=0.23\textwidth,height=0.23\textwidth]{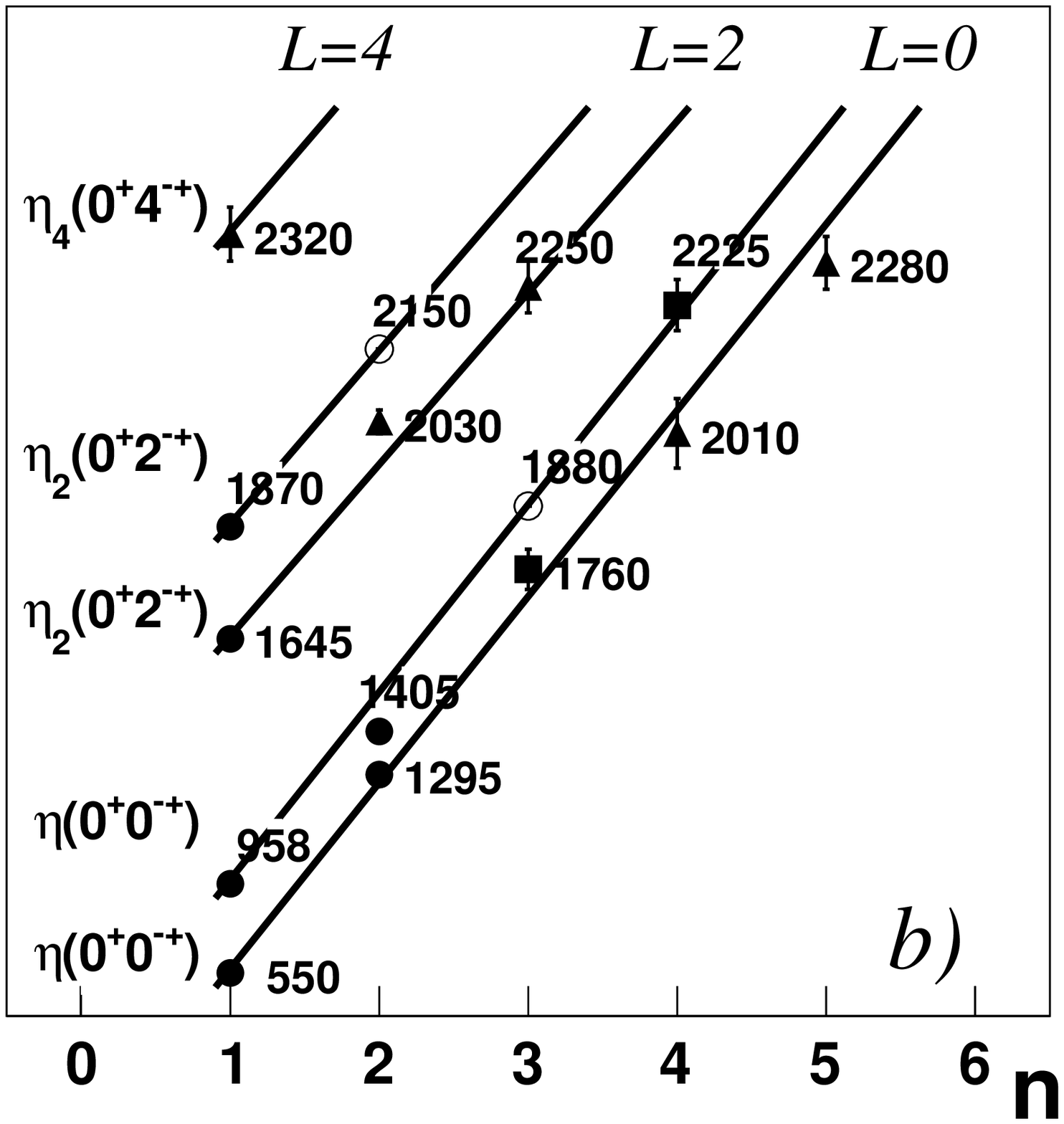}&
\hspace{-3mm}\includegraphics[width=0.23\textwidth,height=0.23\textwidth]{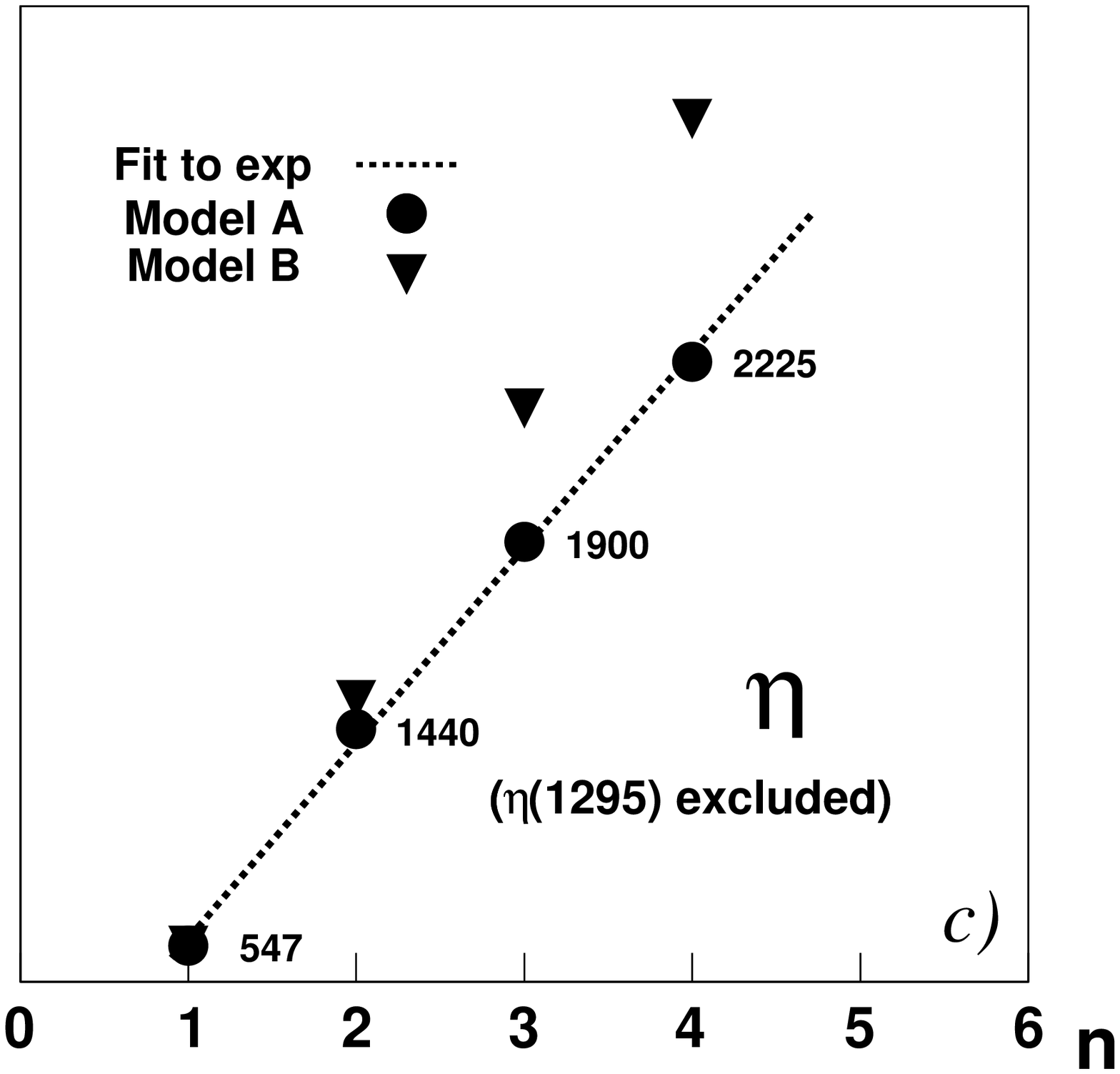}&
\hspace{1mm}\includegraphics[width=0.23\textwidth,height=0.23\textwidth]{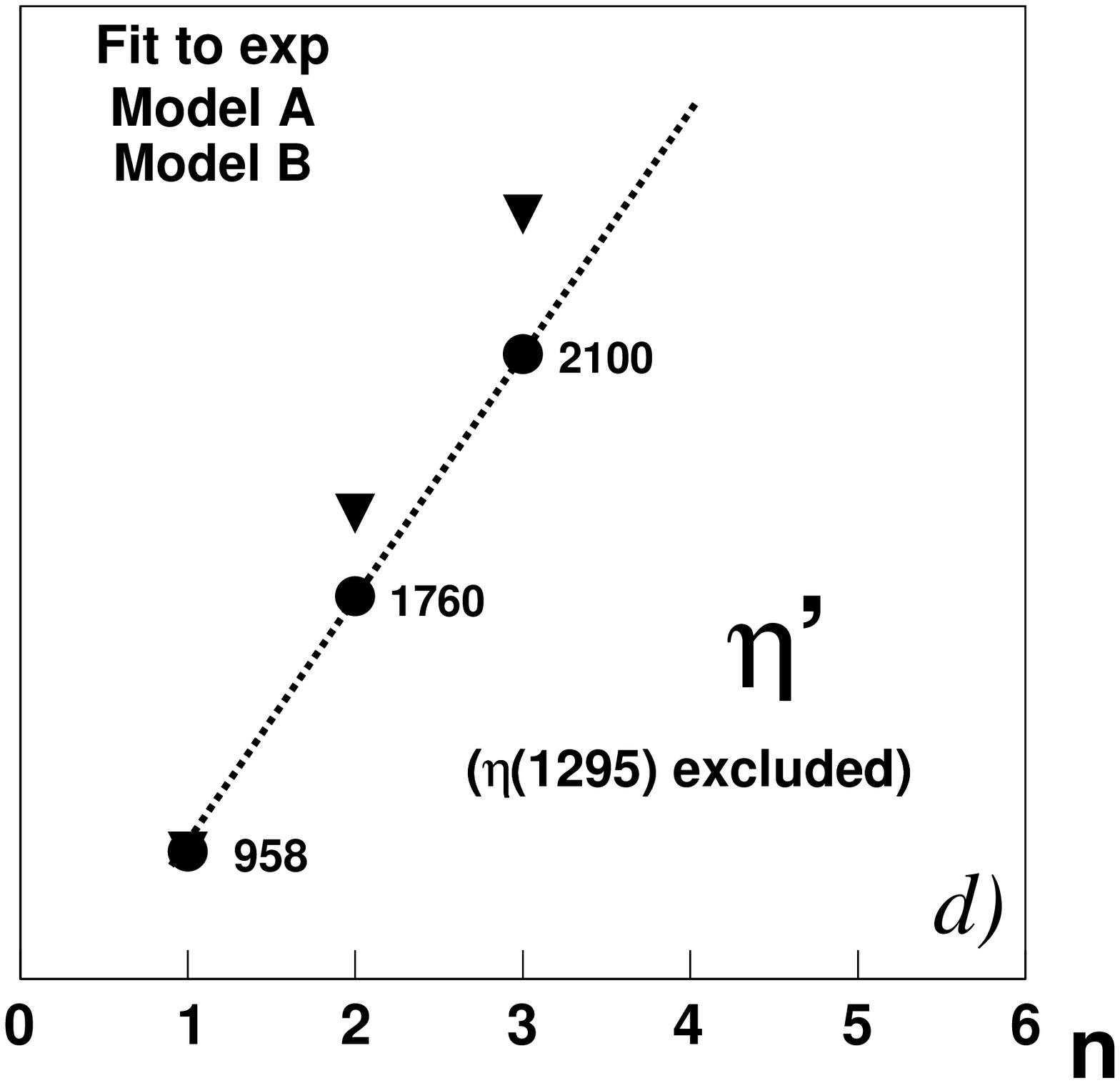}\vspace{5mm}\\
\end{tabular}
\ec
\caption{\label{psaas}
$(n, M^2)$-trajectories. In a) and b), particles from the
PDG {\it Summary Table} are represented by a {\Large$\bullet$}, those
{\it omitted from the Summary Table} by a $\small\blacksquare$.
$\Large\blacktriangle$ represents data given by Anisovich {\it et al.}
\cite{Anisovich:2000kx} and listed as {\it Further States} by the PDG, open
circles predicted by their classification.  The number of
isoscalar mesons is twice larger due to hidden strangeness; $u\bar
u+d\bar d$ and $s\bar s$ components mix to form the observed states.
 a) $(n, M^2)$-trajectories of $\pi$, $\pi_2$, $\pi_4$, and b) $\eta$,
$\eta_2$, and $\eta_4$ excitations. c) and d) give a comparison of
experimental results with two model calculations of the Bonn group
\cite{Ricken:2000kf}. The lines gives a fit though the data, the points
represent results of the calculations. Model B reproduces the data
points when $\eta(1295)$ is omitted. } \end{figure}

The linear mass gaps between the pseudoscalar ground states
$\pi$, $K$,  $\eta$ and $\eta'$ to their first radial excitation
vary slowly and can even be reasonably extrapolated to the $\eta_c$.
The mass square gaps are however about 1.8\,GeV$^2$/c$^4$ for the first
three mesons; for the $\eta'$ it is even larger. This is inconsistent
with the general observation outlined in section \ref{Systematic of
qbarq mesons in planes}. Hence other identifications have been made in
the literature. In Fig. \ref{psaas}a,b the interpretation of $\pi$ and
$\pi_2$, $\eta$, $\eta'$, and $\eta_2$ by A.V. Anisovich, V.V.
Anisovich, and Sarantsev is shown. The mass spectra can also be
organised in linear trajectories when $\eta(1295)$ is omitted (c,d).

\vfill


\markboth{\sl Meson spectroscopy} {\sl Scalar mesons}
\clearpage\setcounter{equation}{0}\section{\label{Scalar mesons}
Scalar mesons: \boldmath$\sigma, \kappa, \delta$\unboldmath, S$^*$}
Scalar mesons, their mass spectrum, their decays and their
interpretation, are one of the hottest subjects in meson spectroscopy;
scalar mesons have been in the centre of common interest for several
decades. The flavour singlet component of scalar mesons carries the
quantum numbers of the vacuum and is thus intimately tied up with the
$\rm r\hat{o}le$ of quark and gluon condensates. The nine scalar mesons
of lowest mass, $a_0(980)$, $f_0(980)$, $f_0(485)$, $ K^*_0(900)$ are
often called with their traditional names $\delta, S^*, \sigma,
\kappa$. Special conferences have been devoted to the question if the
$\sigma(485)$, the low-mass $\pi\pi$ enhancement, is a genuine
resonance, if it is a tetraquark state or generated by molecular
forces, by $t$-channel exchange dynamics \cite{Tornqvist:2000jt}. A
controversy has arisen on the correct formalism to determine the
properties of the $\sigma(485)$ meson and of its twin brother
$\kappa(700)$, on achievements in the past and on the fair use of data
taken by a collaboration \cite{Takamatsu:2006td,Bugg:2007ud}.

The $\sigma(485)$ meson has a long history, and its meaning changed in the
course of time. It was originally introduced to improve the description
of nucleon-nucleon interactions in one-boson-exchange potentials. An
isoscalar and an isovector $\sigma_0$ and $\sigma_1$ were invented.
Modern theories of nuclear forces understand the interaction by
correlated two-pion exchanges \cite{Meissner:2004yy} with no need to
introduce $\sigma_0$ and $\sigma_1$. Nambu \cite{Nambu:1960xd} and
Nambu and Jona-Lasinio \cite{Nambu:1961tp,Nambu:1961fr} described mass
generation in analogy to superconductivity and argued that spontaneous
mass generation is linked to a massless pion field. Following these
ideas,  Delbourgo and Scadron \cite{Delbourgo:1982tv} predicted a
scalar companion of the pion, a $1^3P_0$ $q\bar q$ state, at a mass
corresponding to twice the constituent quark mass, i.e. at 600 to
700\,MeV/c$^2$. While pions get their (small) mass from the distortion
of chiral symmetry by finite light-quark masses, the lowest-mass scalar
mesons acquire their mass by spontaneous symmetry breaking. Thus the
$\sigma(485)$ is sometimes called the Higgs of strong interactions
\cite{Pennington:2005am}.

Experimentally, the existence of the $\sigma(485)$
was, e.g., deduced from $\pi\pi\to\pi\pi$ phase shifts between the
$\pi\pi$ and the $\bar KK$ thresholds. In \cite{Ishida:1995xx}, a
negative background phase was introduced, reflecting a ``repulsive core"
in $\pi\pi$ interactions; the $\sigma(485)$ resonance adds a full
$180^{\circ}$ phase shift. A repulsive background interaction could
stem from left-hand singularities \cite{Markushin:2000kx} which
are usually omitted in fits to the data. Ishida san and collaborators
interpret the $\sigma(485)$ as a relativistic $q\bar q$ state in an
S-wave, making the lowest mass scalar mesons to a nonet of `chiralons'
\cite{Ishida:1997ze,Ishida:2002vr}. Weinberg \cite{Weinberg:1966kf}
showed that the isoscalar $\pi\pi$ $S$-wave scattering amplitude vanishes
when the momenta of the pions go to zero, growing linearly with
increasing $s$. The conflict between this consequence of chiral
symmetry and unitarity requires the existence of a low-mass pole (see
however \cite{Boglione:1996uz}). These qualitative arguments can be
sharpened within Chiral Perturbation Theory (see below). Related to the
$\sigma(485)$ are discussions of the existence of the $\kappa(700)$, a
low-mass $ K\pi$ enhancement. If the $\sigma(485)$ exists but not the
$\kappa(700)$, the $\sigma(485)$ must be related to the spectrum of
glueballs; indeed, QCD spectral sum rules require a scalar glueball
below 1\,GeV \cite{Dosch:2002hc,Narison:2002gv}.

\begin{figure}[pt]
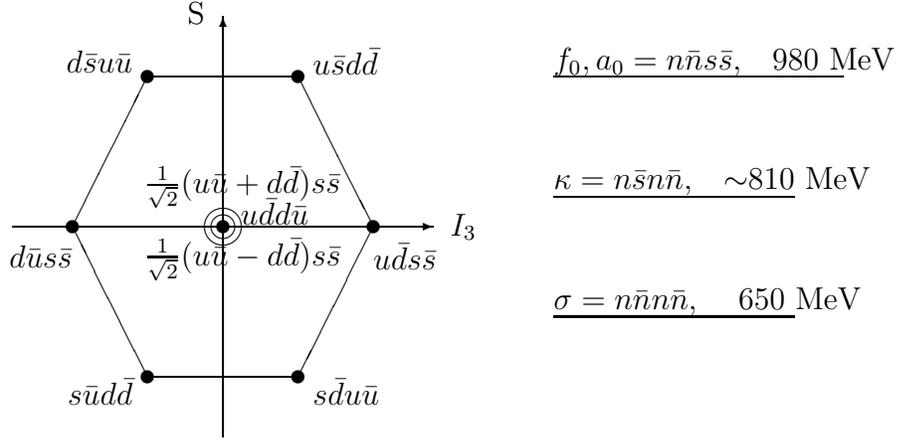
 \bc
\nonet{$d\bar su\bar u$}{$u\bar sd\bar d$}{$s\bar ud\bar d$}{$s\bar
du\bar u$} {$d\bar us\bar s$}{$u\bar ds\bar s$}{$\frac{1}{\sqrt
2}(u\bar u-d\bar d)s\bar s$} {$\frac{1}{\sqrt 2} (u\bar u+d\bar d)s\bar
s$}{$u\bar d d\bar u$} \ec
\vspace*{-8mm}\caption{\label{mes:fig:four-q}
The nonet of lowest-mass scalar mesons and mass
ordering \cite{Jaffe:1976ig}.
 }
\end{figure}

Uncounted is the number of papers on the $f_0(980)$ and $a_0(980)$,
which are fiercely defended as genuine $q\bar q$ states, as tetraquark
states or as $ K\bar K$ molecules, forming -- together with
$f_0(470)$ and $ K^*_0(900)$  -- a full nonet of dynamically
generated resonances. The two twin brothers $f_0(980)$ and $a_0(980)$
have very unusual properties: both have a mass very close to the $
K\bar K$ threshold and a large coupling to $ K\bar K$; Weinstein and
Isgur argued \cite{Weinstein:1982gc,Weinstein:1983gd,Weinstein:1990gu}
that the properties can be explained assuming that they are $ K\bar
K$ molecules. Jaffe \cite{Jaffe:1976ig,Jaffe:1976ih} had combined the
$a_0(980)$, $f_0(980)$, $f_0(470)$, and $ K^*_0(700)$ (with somewhat
different masses) to form a nonet of tetraquark states depicted in
Fig.~\ref{mes:fig:four-q}. The unusual properties were assigned to
their intrinsic structure as tetraquark 
states where the mass scales with the number of $s$ quarks. Their low
mass -- in comparison to 1.3\,GeV expected for $1^3P_0$ $q\bar q$
states -- is due to the absence of an orbital angular momentum barrier,
and due to a tighter binding of $qq$ pairs in a colour $\bf\bar 3$
state (and $\bar q\bar q$ pairs in colour $\bf 3$).

Last not least, a scalar glueball is expected. Its mass is likely to
fall into the 1 to 2\,GeV region; it possibly mixes with $q\bar
q$ states, thus creating a complex pattern of states. An orthodox view
has been developed in which the 3 states $f_0(1370)$, $f_0(1500)$, and
$f_0(1760)$\footnote{\footnotesize The Particle Data Group quotes the
$f_0(1760)$ as $f_0(1710)$. The region may, however, house up to 3
isoscalar scalar resonances, called $f_0(1710)$, $f_0(1790)$, and
$f_0(1810)$. In this report, the unresolved structure is denoted
$f_0(1760)$.} originate from two $q\bar q$ states and a glueball even
though the existence of $f_0(1370)$ is sometimes challenged.

\begin{figure}[pb]
\hspace{5mm}\includegraphics[width=0.6\textwidth,height=0.3\textheight]{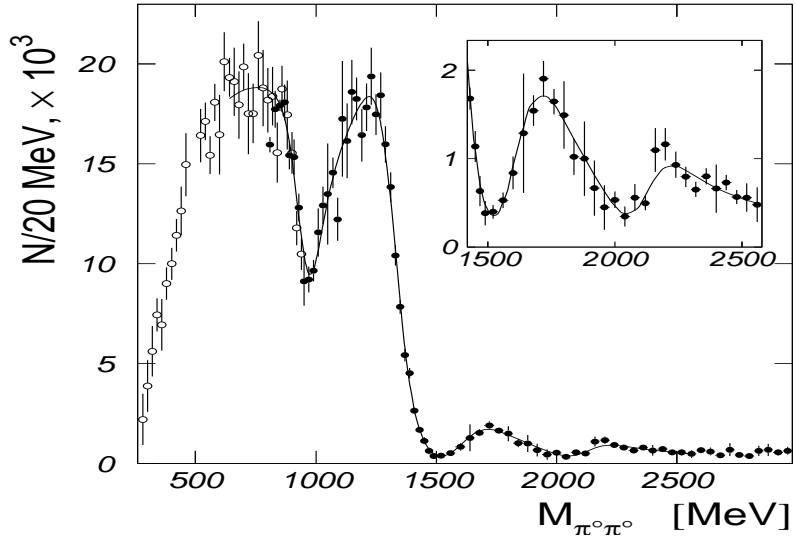}
\caption{\label{fig:gams-scalar}The reaction $\pi^- p
\to p \pi^0\pi^0$ at $100$\,GeV/c for $\mid t\mid < 0.2$\,GeV$^2$
\protect\cite{Alde:1998mc}. Open circles represent data taken at
38\,GeV. The $S$-wave exhibits peaks at 600\,MeV/c$^2$, 1300\,MeV/c$^2$, 1700\,MeV/c$^2$
and at 2200\,MeV/c$^2$ and dips at 980\,MeV/c$^2$, 1500\,MeV/c$^2$ and at 2100\,MeV/c$^2$.}
\end{figure}

A survey of the scalar  intensity can be seen in GAMS data
on $ \pi^- p \to n \pi^0\pi^0$. The process offers the advantage that
odd partial waves cannot contribute. The GAMS collaboration has studied
this reaction and measured angular distribution from 0.8 to 3\,GeV.
Fig.~\ref{fig:gams-scalar} shows the modulus of the $S$-wave amplitude
\cite{Alde:1998mc}. A series of peaks and dips is observed and it is
not straightforward to identify the scalar resonances. Some of the
peaks have traditional names: $\sigma$ (now $f_0(485)$), $\epsilon$,
and the G(1590). The latter peak was first observed by GAMS in the
reaction $\pi^-p\to\eta\eta\,n$ as a peak in the $\eta\eta$ invariant
mass distribution \cite{Alde:1985kp}, the two peaks at 1300 and
1590\,MeV/c$^2$ are now understood as consequence of the $f_0(1500)$
dip \cite{Li:2000jq}. The dips correspond to well known meson
resonances, $f_0(980)$, $f_0(1500)$ and to the less well established
$f_0(2100)$. In other experiments the dips are observed as peaks. For
$f_0(980)$ and $f_0(1500)$ phase motions have been observed; there is
no doubt that these are resonances. It is thus tempting to try to
understand the $\pi\pi$  $S$-wave as a global phenomenon generated by a
very broad object $f_0(1000)$ \cite{Au:1986vs,Minkowski:1998mf},
possibly generated by $t$-channel exchanges, interspersed with $q\bar
q$ resonances represented by dips. The peaks may house additional
resonances, like $f_0(1370)$, $f_0(1710)$, and $f_0(1760)$, and deserve
a deeper discussion.

\subsection{\label{pipi scattering}
$\pi\pi$ scattering}

Scalar resonances coupling to two pions can be studied by scattering
negatively charged pions off protons to produce $\pi^+\pi^-$ or
$\pi^0\pi^0$ pairs. A proton can dissociate into neutron and $\pi^+$,
and scattering may take place off the virtual pion. If the
four-momentum transfer $q$ (with $t=-q^2$) to the proton is small, the
pion is nearly on-shell and the process can be considered as a
scattering experiment in which a beam scatters off a pion
target~\cite{Chew:1958wd}.  A simultaneous analysis
of $\pi^+\pi^-$ and $\pi^0\pi^0$ provides for a decomposition of the
scattering amplitude into isoscalar and isotensor contributions.

A classic experiment to determine the $\pi\pi$ scattering amplitude was
carried out at CERN in the sixties by the CERN-Munich collaboration
(see section \ref{The CERN-Munich experiment}). A 17.2\,GeV/c $\pi^-$
beam  and an unpolarised target were used. The data were expanded into
Legendre polynomials from which $\pi\pi$ amplitudes can be extracted.
In the general case, the $S, P_0, P_+, P_-$ amplitudes and their
relative phases depend on the nucleon helicities. Without using a
polarised target, these partial waves cannot be deduced from the
moments without further assumptions \cite{Ochs:1972mc,Estabrooks:1972tw}.
Without polarisation data, there are no unambiguous solutions and
additional assumptions are necessary. At low $q^2$, the one pion
exchange mechanism dominates which prefers spin flip at the nucleon
vertex. Under these conditions, the rank of the spin-density matrix is
reduced to one and all density matrix elements can be fixed. In
\cite{Hyams:1973zf,Grayer:1974cr} it was thus assumed that for small
values of $t$, the proton spin-flip amplitude is dominant. More
generally speaking, it was assumed that the $\pi\pi$ phase does not
depend on the projection of the orbital angular momentum onto the
flight direction with which it is produced (phase coherence) and that
the ratio of flip and non-flip amplitudes is universal (`spin
coherence').  Later, the experiment (performed by a CERN-Krakow-Munich
collaboration) included a polarised target \cite{Becker:1978ks}, and
the assumptions on spin and phase coherence were verified to a good
approximation in a model-independent analysis. For large $t$, neither
phase nor spin coherence is granted, and feedthrough of higher partial
waves into the scalar wave cannot be excluded without exploiting a
polarised target.

The threshold region is of particular importance. The most precise data
stem from $ K^+\to\pi^+\pi^-e^+\nu_e$ decays \cite{Rosselet:1976pu}
(see \cite{Pislak:2003sv,Masetti:2006kj} for recent data). This type of
data is often included in partial wave analyses in order to constrain
the low energy behaviour of the scattering amplitude. A recent
reanalysis of the CERN-Munich data yielded the $\pi\pi$ $S$-wave
amplitude and phase (and other partial wave amplitudes) in a $\pi\pi$
mass range from 600\,MeV/c$^2$ to 1600\,MeV/c$^2$
\cite{Kaminski:1998ns}. However, it was still not possible to
reconstruct all partial waves unambiguously. There is a long history of
`up' or `down' solutions (and `flat' or `steep') which gave equally
good descriptions of the moments. Steep solutions violate unitarity and
can be discarded. The remaining ambiguity can be resolved by invoking
crossing symmetry. Crossing symmetry relates a partial wave amplitude
to the imaginary part of the amplitude, integrated from threshold to
infinity (Roy's equations \cite{Roy:1971tc}). The integral is dominated
by the threshold region and was evaluated using some approximations. It
turns out that only the `down-flat' is compatible with Roy's equations
and satisfies crossing symmetry \cite{Kaminski:2002pe}. The `down-flat'
solution is also compatible \cite{Kaminski:2001hv} with the BNL
\cite{Gunter:2000am} and GAMS results \cite{Alde:1998mc} which we
discuss further down.

\begin{figure}[pt]
\vspace{6mm} \bc
\begin{minipage}[c]{0.35\textwidth}
\hspace{-5mm}\mbox{\epsfig{file=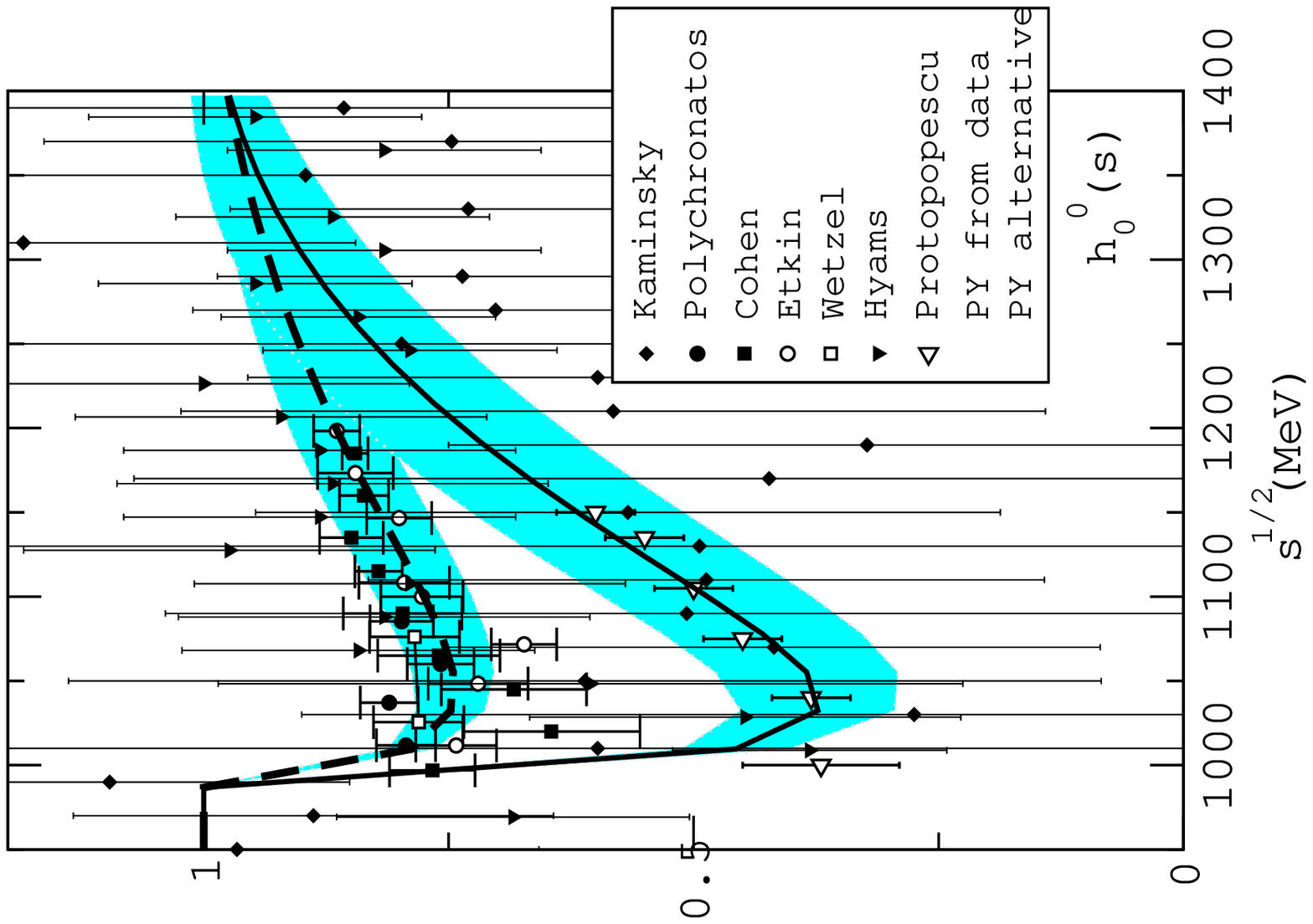,width=6cm,height=60mm,angle=-90}}
\end{minipage}
\begin{minipage}[c]{0.45\textwidth}
\mbox{\epsfig{file=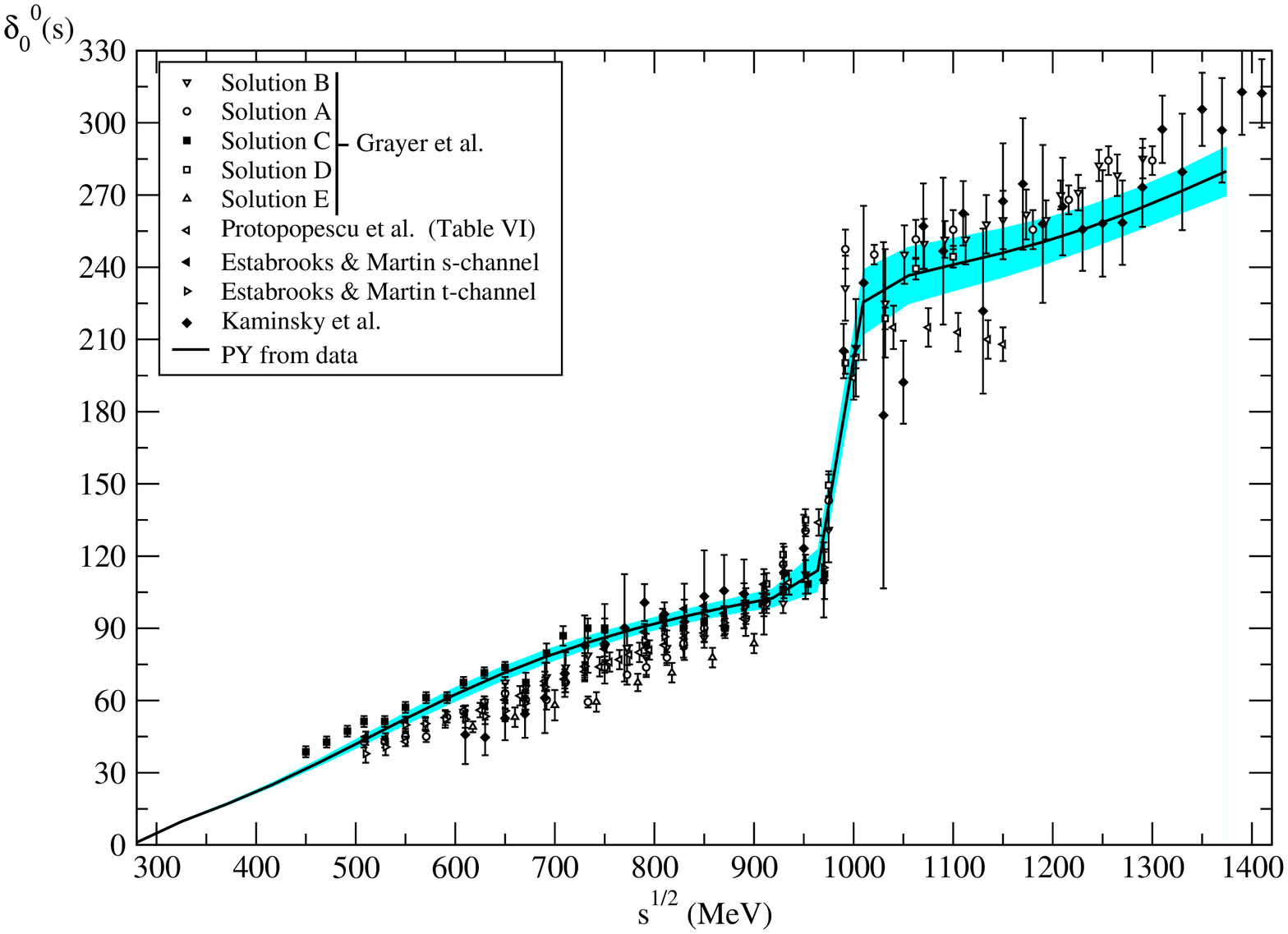,width=6cm,height=80mm,angle=-90}}
\end{minipage}
\ec
\caption{\label{fig:S-wave-pelaez}The $I=0$, $S$-wave
inelasticity and phase shift (with error band) with a fit by Pelaez and
Yndur\'ain \protect\cite{Pelaez:2004vs}. Data points from K decays are
not shown. The inelasticity derived from $\pi\pi$ elastic scattering
(marked PY from data) is much larger than that from $\pi\pi\to K\bar
K$ (PY alternative). }
\end{figure}

Many more experiments and analyses have been carried out
\cite{Hyams:1973zf,Grayer:1974cr,Hyams:1975mc,Au:1986vs,%
Estabrooks:1974vu,Kaminski:1996da,Kaminski:2001hv,Protopopescu:1973sh};
Fig.~\ref{fig:S-wave-pelaez} collects isoscalar scalar phase shifts
deduced from these experiments. The solid line is a fit by Pelaez and
Yndur\'ain \cite{Pelaez:2004vs}, a refined analysis by Kaminski, Pelaez
and Yndur\'ain is presented in \cite{Kaminski:2006yv}.

\begin{figure}[pt]
\begin{tabular}{cc}
\includegraphics[width=0.45\textwidth]{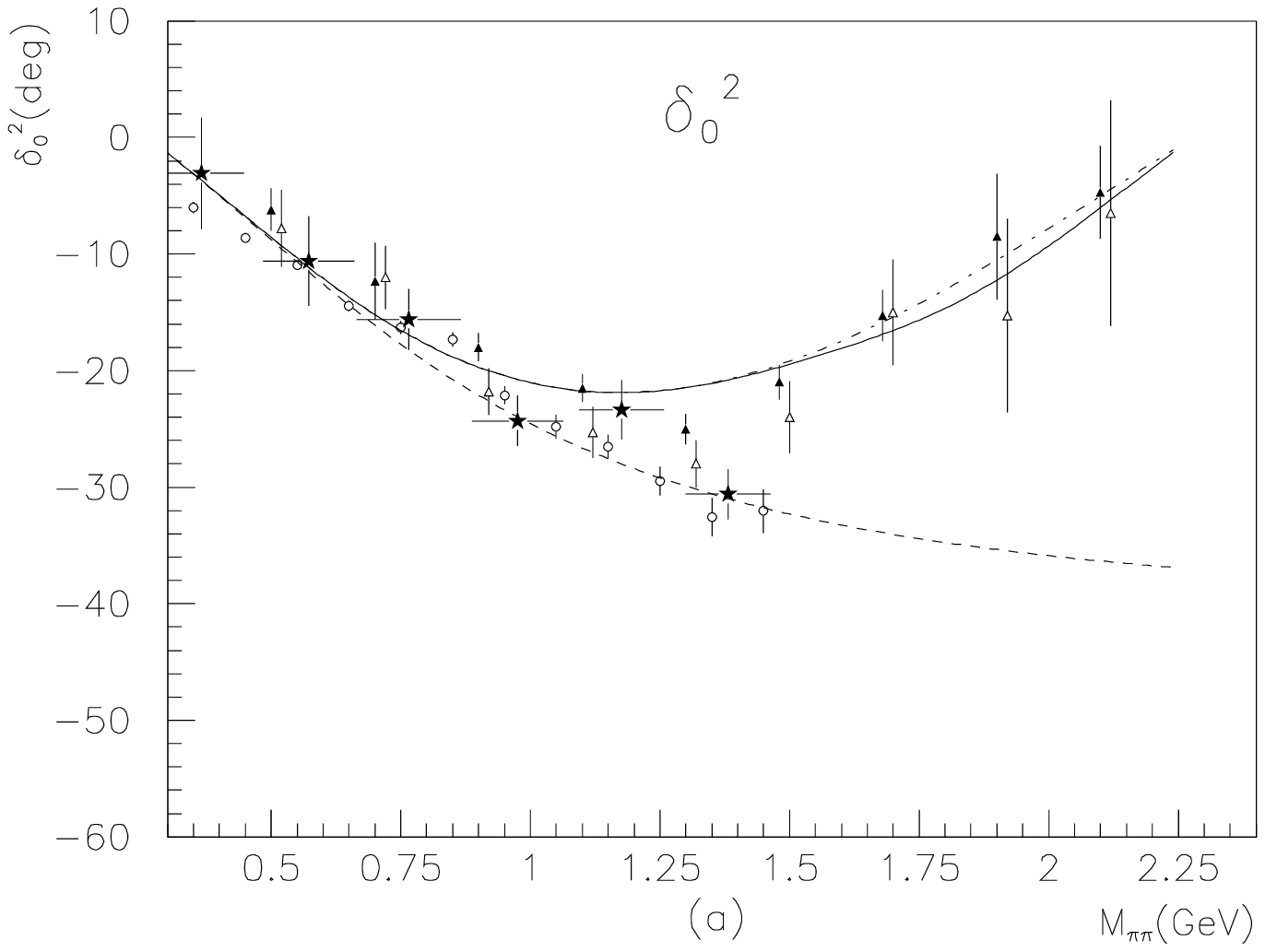}&
\includegraphics[width=0.45\textwidth]{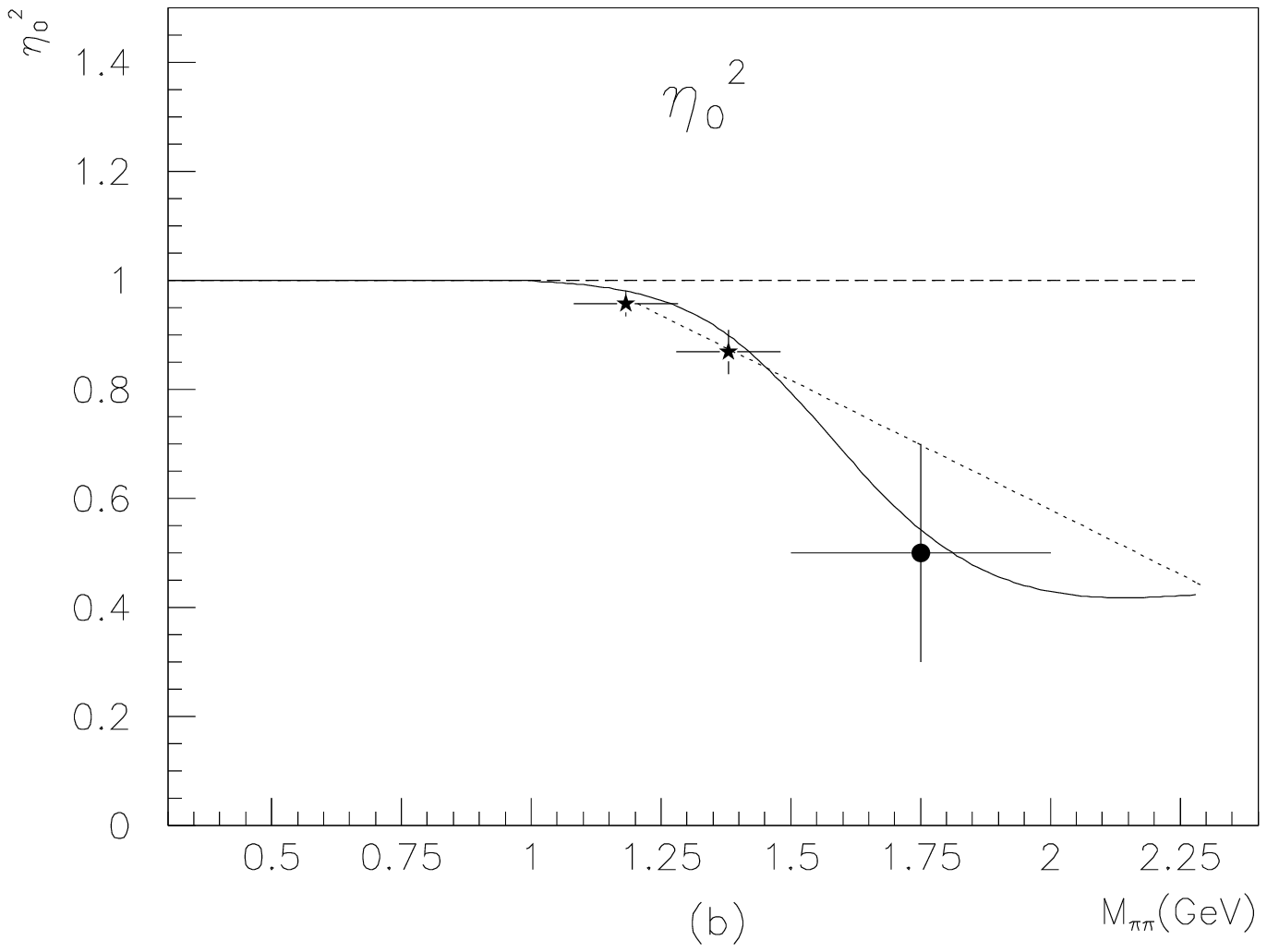}
\end{tabular}
\vspace{-3mm}
\caption{\label{swaveitensor} The $I=2$ $\pi\pi$ $S$-wave phase
shift $\delta_0^2$ (a) and inelastic coefficient $\eta_0^2$ (b). The
experimental data for $\delta_0^2$ are from \cite{Hoogland:1977kt}
(circles), \cite{Cohen:1973yx} (stars), and \cite{Durusoy:1973aj}
(triangles); the data for $\eta_0^2$ are from \cite{Cohen:1973yx}
(stars), and \cite{Durusoy:1973aj} (solid circle). The solid curves
represent a fit which includes $\rho$ and $f_2(1270)$ exchange and a
box diagram with two $\rho$ mesons in the intermediate state. The dashed
curves include only t-channel $\rho$ exchange, the dot-dashed curves
include $\rho$ and $f_2(1270)$ exchange. The dotted line in (b)
is $\eta_0^2=1.53-0.475m_{\pi\pi} (GeV/c^2)$ used in
\cite{Durusoy:1973aj}. Data compilation and fit are from
\cite{Wu:2003wf}. }
\end{figure}

The following observations can be
made. The $\pi\pi$ $S$-wave scattering amplitude vanishes at some small
value of $s$. This is the Adler or Adler-Weinberg zero
\cite{Adler:1964um,Weinberg:1966kf}. With increasing mass $m_{\pi\pi}$,
amplitude and phase rise slowly until they reach the 1\,GeV region
where the phase increases rapidly (from $\sim 90^{\circ}$ to $\sim
240^{\circ}$) and where the amplitude exhibits a dip. These effects are
due to $f_0(980)$; the phase advance signals a resonance, the dip is a
unitarity effect. The coupling of $f_0(980)$ to $ K\bar K$ can be
determined from the intensity missing in $\pi\pi$, and as observed
intensity in the $ K\bar K$ final state. These two quantities disagree
as shown in Fig.~\ref{fig:S-wave-pelaez}a. The dashed and dotted lines
in Fig.~\ref{fig:S-wave-pelaez}a represent fits using either
$\pi\pi\to\pi\pi$ or  $\pi\pi\to K\bar K$ data, respectively. The phase
shifts are fitted directly, the pole structure of the amplitude was not
studied in \cite{Pelaez:2004vs} and we refrain from a discussion of
resonance parameters. At higher masses, the phase continues to increase
slowly and the amplitude approaches the unitarity limit again. There is
a hint that something new may occur in the 1400 to 1500\,MeV/c$^2$
region. The isotensor $S$-wave phase shift and inelasticity are shown
in Fig. \ref{swaveitensor}.

\begin{figure}[pb]
\centerline{
\begin{tabular}{cc}
\includegraphics[width=0.5\textwidth]{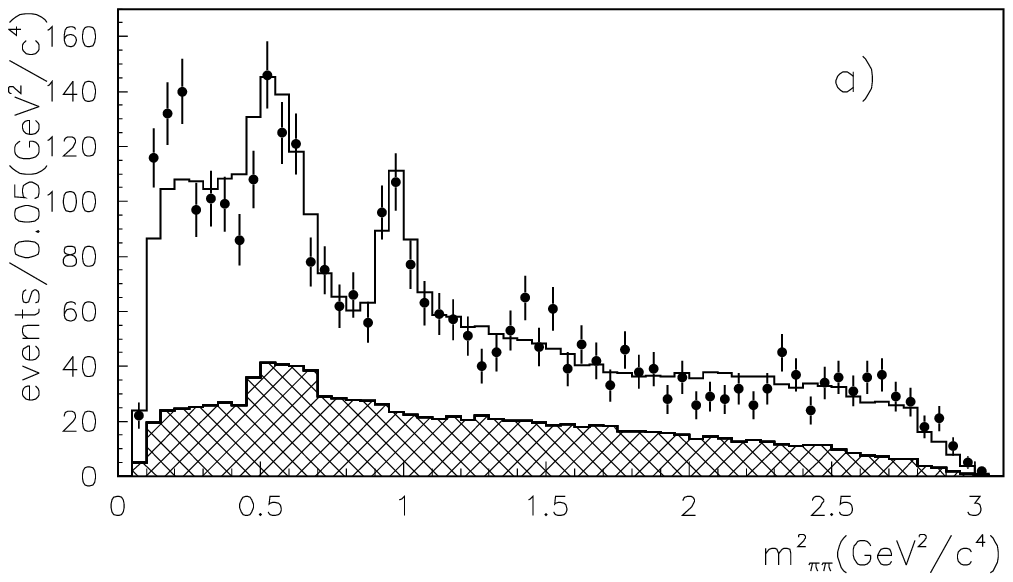}&
\includegraphics[width=0.5\textwidth]{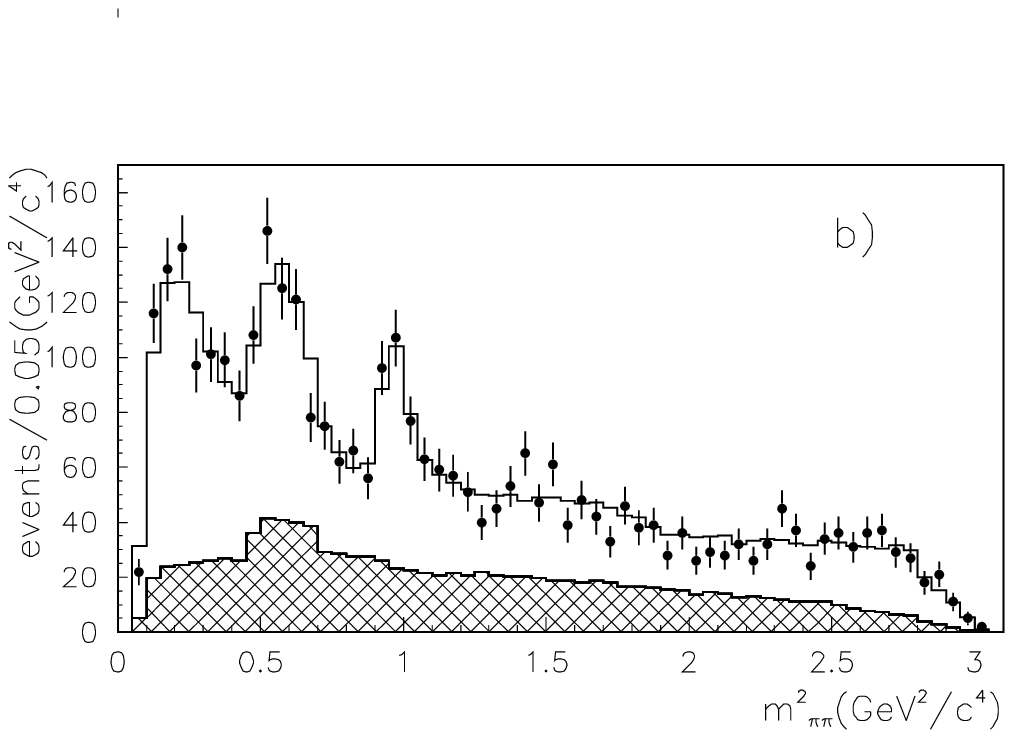}
\end{tabular}
}
\vspace{-3mm}\caption{\label{e791-sigma}  $M_{\pi^+\pi^-}^2$
distribution from $ D^+\to\pi^+\pi^+\pi^-$ decays. Data are shown with
error bars, the histogram represent a fit without (a) and with a
Breit-Wigner resonance (the $\sigma(485)$) \protect\cite{Aitala:2000xu}.}
\end{figure}

The Adler-Weinberg zero is a consequence of chiral symmetry, it forces
$\pi\pi$ interactions to vanish close to the $\pi\pi$ threshold. The
zero is not necessarily present in a production experiment where the
$\sigma(485)$ may appear as a peak. The E791 collaboration analyzed the
Dalitz plot of the reaction $ D^+\to\pi^+\pi^+\pi^-$
\cite{Aitala:2000xu} and observed an enhancement at low masses. It is
clearly seen (Fig.~\ref{e791-sigma}) as a broad bump in the
$M_{\pi^+\pi^-}^2$ projection of the Dalitz plot. In
\cite{Aitala:2000xu} it is treated as Breit-Wigner amplitude. This is
a crude approximation but the data certainly provide evidence for a
substantial low-energy effect in the $\pi\pi$ isoscalar $S$-wave. The
Breit-Wigner mass and width were fitted to $478^{+24}_{-23} \pm 17 $
MeV/c$^2$ and $324^{+42}_{-40} \pm 21 $ MeV/c$^2$, respectively.
Very similar observations have been made by the BES collaboration in a
study of J/$\psi$ decays to $\omega\pi^+\pi^-$. The data will be
discussed in a different context and are shown in
Fig.~\ref{scalars}. A peak at low $\pi\pi$
masses is observed \cite{Ablikim:2004qn} to which a clear phase motion
can be ascribed \cite{Bugg:2004rb}. The phase motion is identical to
the one observed in elastic $\pi\pi$ scattering. Mass and width of the
$\sigma(485)$ were determined to be $(541 \pm 39)$ - $i$ $(252 \pm
42)$\,MeV/c$^2$. The pole was confirmed in a study of $\psi(2S)\to
\pi^+\pi^-J/\psi$ \cite{Ablikim:2006bz}.

Unitarity suggests the phase of the $\pi\pi$ interaction below the
first inelastic threshold in weak or electromagnetic production
processes to be the same as in elastic $\pi\pi$ scattering; this is the
famous final state interaction theorem by Watson
\cite{Watson:1952ji,Watson:1954uc}. In the case of strong production
modes, rescattering between final-state particles may lead to (likely
small) additional phase shifts. It is very suspicious when a new method
to identify the scalar isoscalar phase motion leads to a $\sigma(485)$
having a phase shift from threshold to 800\,MeV/c$^2$ which is compatible
with a full Breit-Wigner resonance \cite{Bediaga:2002au} and thus
incompatible with scattering data. The $\sigma(485)$ is represented by a
unique pole position in the complex energy plane; the pole must
not change when going from scattering to production experiments; both
types of experiments must be fitted with one amplitude. In a scattering
situation, the scalar isoscalar amplitude is conveniently parametrised
by multiplication with a function enforcing the Adler zero at
$s=m_{\pi}^2/2$, e.g. by writing an amplitude which is $\propto
(s-m_{\pi}^2/2)$ in the low-energy region. Individually, isoscalar and
isotensor $\pi\pi$ phase shifts in final-state interactions and in
scattering are identical; their fractional contributions might however
be different in different experimental situations, and the observed
total $\pi\pi$ phase shifts (which has contributions from $I=0$ and
$I=2$) can differ.

Current algebra or the leading order chiral Lagrangian requires
$\pi\pi$ interactions at low energies to be given by
$A=(s-m_{\pi}^2)/F_{\pi}^2$ plus corresponding terms where $s$ is
replaced by $t$ and $u$. However, the amplitude cannot grow forever
$\propto s$ due to unitarity.  The $\sigma(485)$ pole resolves this
conflict between chiral symmetry and unitarity. Mei\ss ner
\cite{Meissner:1990kz} has sketched how final-state interactions
between the two pions create a pole which he calls an `elusive' or
`illusory' particle to underline its non-$\bar qq$ nature. Its pole
position is approximately determined by the pion coupling constant
$F_{\pi}$ and thus not directly related to the physics of $\bar qq$
mesons. Colangelo, Gasser and Leutwyler \cite{Colangelo:2001df} and
Caprini, Colangelo and Leutwyler \cite{Caprini:2005zr} use further
constraints from the Roy equation leading to a twice subtracted
dispersion relation. The subtraction constants were expressed, using
crossing symmetry, in terms of the $S$-wave scattering lengths which are
known with high precision. The approach provides access to the
scattering amplitude in the complex plane and thus to magnitude and
phase below and above the pole. A pole is mandatory, the authors of
\cite{Colangelo:2001df} quote $M=(470\pm 30) -i(295\pm20)$\,MeV/c$^2$
while in \cite{Caprini:2005zr}, $M=441^{+16}_{-8}\,,\;\;
\Gamma_\sigma=544^{+25} _{-18}$\,MeV/c$^2$ is given. The mass is
lower than those from the other papers; the value was criticised by
Kleefeld in \cite{Kleefeld:2007dt} for having neglected pole terms in
the double subtracted dispersion relation. Bugg \cite{Bugg:2006gc}
assigns the low value to the neglect of virtual loops due to $ K\bar
K \to \pi\pi$ and $\eta\eta\to\pi\pi$ below threshold. Including these
channels into a fit to all data gave a pole at $472\pm30 -
i(271\pm30)$\,MeV/c$^2$. Leutwyler \cite{Leutwyler:2006gz} addresses
this question explicitly, suggesting that in Bugg's method, the
inelastic channels pose severe problems resulting in barely
controllable errors (but not in their own method). The pole structure
of the low energy $\pi\pi$ scattering amplitudes was also studied in
\cite{Zhou:2004ms}. Low energy phase shift data were fitted by imposing
chiral unitarisation and crossing symmetry. A $\sigma(485)$ pole
position at $M_\sigma=470\pm 50$, $\Gamma_\sigma=570\pm 50$\,MeV/c$^2$
was found. The threshold parameters were found to be in good agreement
with results using the Roy equations.

The pole structure of the low energy $\pi\pi$ scattering amplitudes was
studied in \cite{Zhou:2004ms} using a proper chiral unitarisation
method combined with crossing symmetry and the low energy phase shift
data. The $\sigma(485)$ pole position was found at $M_\sigma = 470\pm
50\,{\rm MeV/c^2}$, $\Gamma_\sigma=570\pm 50\,{\rm MeV/c^2}$. Kaon
decays, $K\to3\pi$ and $K\to\pi\pi e\nu$ decays (see, e.g.
\cite{Masetti:2006kj} for recent high-precision data from NA48/2)  give
the most precise data on low energy $\pi\pi$ scattering in the
S-wave. In a systematic evaluation of different data sets, Yndurain
(in collaboration with Pelaez) \cite{Yndurain:2007qm} determined a
precise value for the pole position of the $\sigma(485)$ resonance to
be $M_\sigma=(485\pm18) {\rm MeV}$, $\Gamma_\sigma/2=(255\pm12) {\rm
MeV}$. The `Breit-Wigner' mass of the $\sigma(485)$ was determined,
too, and values $M_0=790\pm25 {\rm MeV/c^2}$ and half width of
$\Gamma_0/2=560\pm60 {\rm MeV/c^2}$ were obtained.

We believe the $\sigma(485)$ to be established and quote
\be
M_\sigma=(485\pm40) {\rm MeV}, \Gamma_\sigma=(565\pm60) {\rm MeV}
\ee
as final value which we use for further discussion.

The mass of the  $\sigma(485)$ is driven by the value of $f_{\pi}$. Does
this require the $\sigma(485)$ to be unrelated to $q\bar q$
spectroscopy? The situation remotely resembles the need to introduce
weakly interacting bosons into Fermi's four-point theory of weak
interactions. The linear rise of the $\bar\nu_e e^-$ cross section and
unitarity clash at 300\,GeV; there must be a $W^-$. But of course, this
is not a $\bar\nu_e e^-$ bound state.

\subsection{\label{Kpi scattering}
$K\pi$\ scattering}

\begin{figure}[pt]
\begin{center}
\begin{tabular}{cc}
\includegraphics[width=0.39\textwidth,height=80mm]{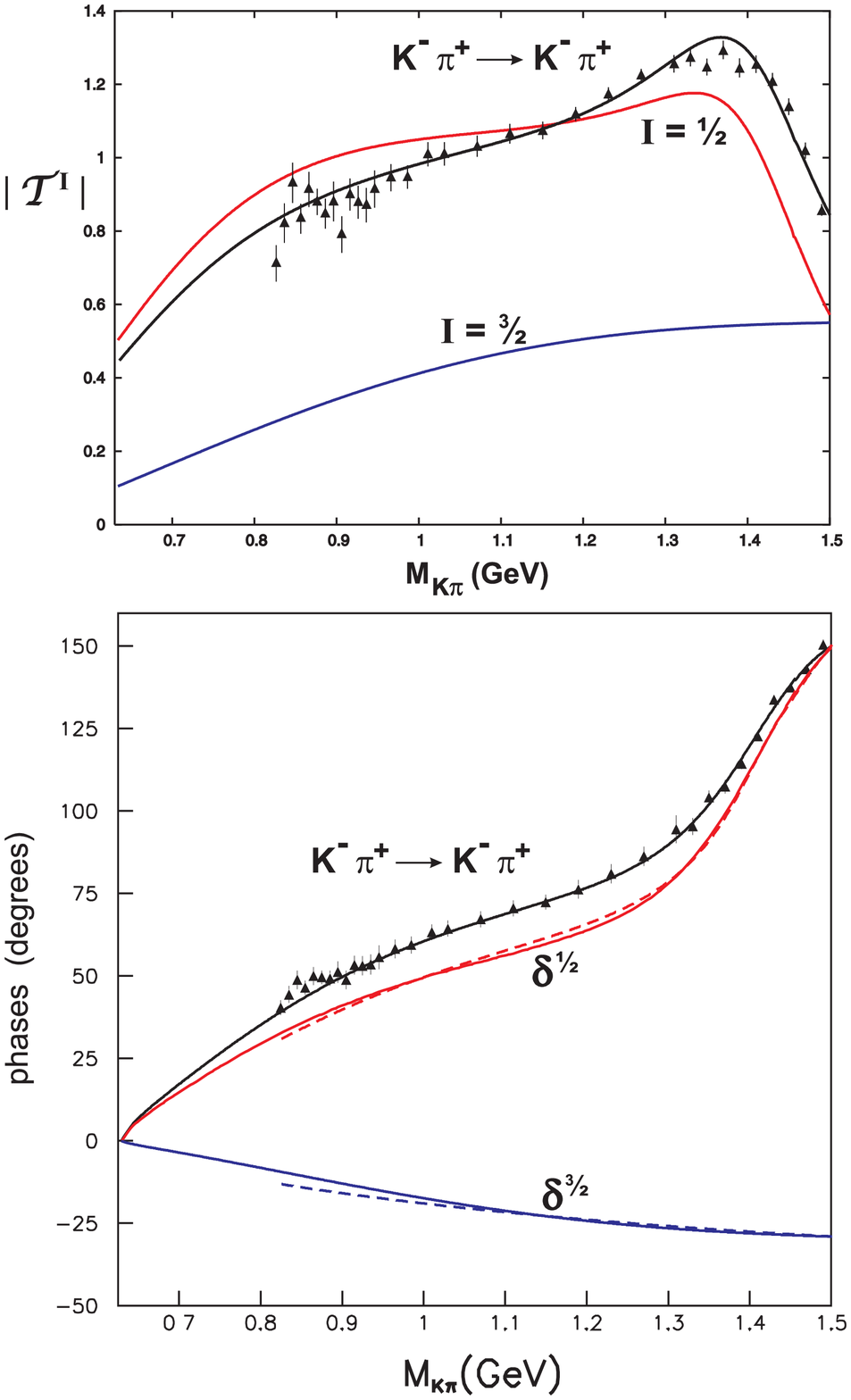} &
\includegraphics[width=0.39\textwidth,height=80mm]{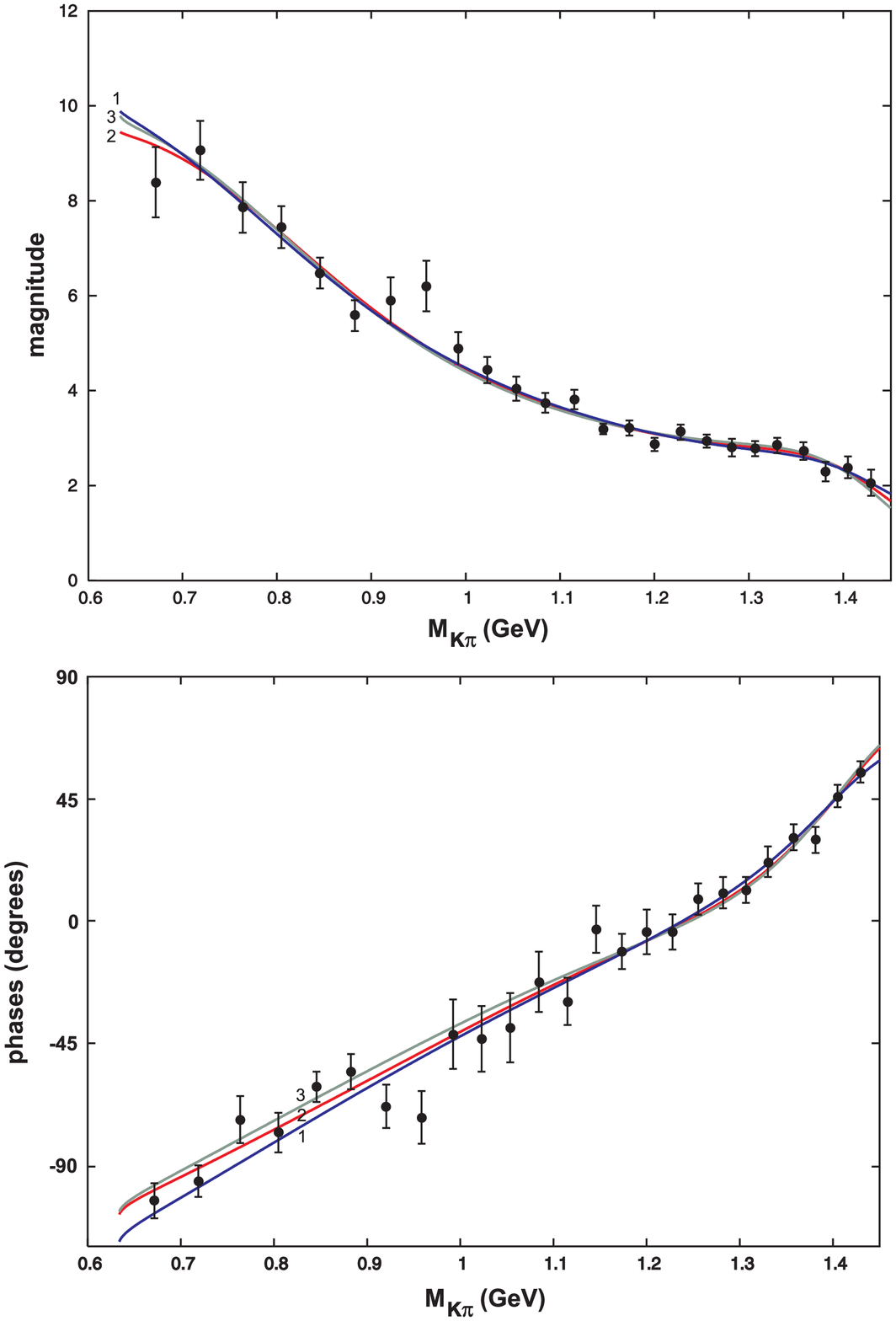}
\end{tabular}\vspace{-57mm} \\
\hspace{7mm}a \hspace{20mm}c \vspace{35mm} \\
\hspace{7mm}b \hspace{20mm}d \vspace{12mm} \\
\end{center}
 \caption{\label{Kpiscatt}Magnitudes (a) and phases (b) of the $I=1/2$
and $3/2$ amplitudes. The data points are from LASS \cite{Aston:1987ir}.
The solid line represent a fit by Boglione and Pennington
\cite{Edera:2005dk} compatible with constraints from current algebra.
Magnitude (c) and phase (d) of the $S$-wave amplitude from E791 results
on D decays. The fit curve (labelled 2) to these data allows for an
$s$-dependent fraction of $I=1/2$ and $3/2$ amplitudes
\cite{Edera:2005dk}. Curves 1 and 3 mark one standard deviation away
from the optimal fit. }
\end{figure}

The LASS experiment at SLAC, carried out in the 70ties of the last
century and  described in session \ref{The LASS experiment}, was the
last one to scatter low-momentum (11\,GeV/c) kaons off protons. For
small momentum transfer to the target proton, the scattering process $
K^-p\to K^-\pi^+n$ is dominated by $K^-\pi^+$ scattering. The two
isospin contributions, $I=1/2$ and $I=3/2$, can be separated by
measuring also $ K^+p\to K^+\pi^+n$. From the angular distributions,
both $S$-wave scattering amplitudes can be determined. The LASS
collaboration described the data (see Fig.~\ref{Kpiscatt}, left panel)
by a $ K^{*}_0(1430)$ and a range parameter which gives an amplitude
rising linearly with $\sqrt s$. Current algebra or Chiral Perturbation
Theory demands an Adler zero of the amplitude at about $(s-m_{
K}^2+\frac12m_{\pi}^2$) and an amplitude rising with $s$
\cite{Bernard:1990kx,Bernard:1990kw}. The range parametrisation is
hence incompatible with chiral symmetry. In production experiments,
there is no Adler-zero suppression, and this is the reason that the
existence of the $\kappa(700)$ became transparent in the analysis of
the E791 Collaboration of data on $ D^{+}\to K^{-}\pi^+\pi^+$
\cite{Aitala:2002kr}. Fig.~\ref{D-decays}d shows the Dalitz plot of the
reaction. The vertical line shows the large contribution of the $
K^*\pi$ isobar. The intensity vanishes in the centre of the $ K^*\to
K\pi$ decay angular distribution due to interference with the $ K^*\pi$
S-wave amplitude: The $P$-wave amplitude has an angular distribution
$\propto\sin\Theta$, the $S$-wave amplitude is constant. Thus the
interference changes from a destructive to a constructive one for low $
K\pi$ masses; the opposite is true for high $ K\pi$ masses. Thus the
scalar phase can be deduced from the interference pattern
\cite{Aitala:2005yh}. The same feature is observed in
Fig.~\ref{D-decays}d with the $\phi$ in $K\bar K$ $P$-wave and
$f_0(980)\to K\bar K$ in $S$-wave. The fit to the data from
\cite{Aitala:2002kr} revealed the existence of a $\kappa(700)$ having
mass and width, respectively, of  $M - i\Gamma/2 =
(797\pm19\pm43)-i(205\pm 22\pm44)$\,MeV/c$^2$ \cite{Aitala:2002kr}, in
agreement with a further analysis by Bugg, finding $(750 \pm30\pm55) -
i(342\pm 60)$\, MeV/c$^2$ \cite{Bugg:2005xx}.

The BES collaboration found further evidence for $\kappa(700)$ from the
reaction J/$\psi\to\bar{K}^*(892)^0K^+\pi^-$ and deduced a pole at
$(841\pm 30^{+81}_{-73}) - i(309\pm 45^{+48}_{-72})$\,MeV/c$^2$ by averaging
values deduced from two different methods. Bugg reanalysed also these
data and found a pole position at $(760\pm 20_{stat}\pm
40_{syst}) -i (420\pm 45_{stat}\pm 60_{syst})$\,MeV/c$^2$
\cite{Bugg:2005ni}. As shown in
\cite{Jamin:2000wn,Zheng:2003rw,Zhou:2006wm}, the LASS data alone yield
a $\kappa(700)$ with (nearly) consistent parameters when analyticity
and chiral symmetry are respected.

How can we reconcile these results with the strict statement of Cherry
and Pennington \cite{Cherry:2000ut}, `there is no $\kappa(900)$'\,?
The authors of \cite{Cherry:2000ut} performed a `model-independent
analytic continuation of the LASS data', determined the number and
position of resonance poles, and concluded that there {\bf is} a
$ K^*_0(1430)$, and {\bf no} $\kappa(900)$. These conclusions were
confirmed in an analysis presented in \cite{Li:2002we}. However, both
groups admitted that the LASS data cannot rule out a $\kappa(700)$ well
below 825 MeV/c$^2$. Both, new data between the $ K\pi$ threshold and
825\,MeV/c$^2$ (where LASS data set in) and the proper inclusion of chiral
symmetry, have finally revealed the existence of a broad $\kappa(700)$.
Very recently, Descotes-Genon and Moussallam used constraints from the
Roy equations for $\pi K$ scattering \cite{Descotes-Genon:2006uk}, the
existence of the $\kappa(700)$ was confirmed and mass and width of $
M_{\kappa} = 658 \pm 13$ and $\Gamma_{\kappa} = 557 \pm 24\,{\rm
MeV/c^2}$ were given.

In our judgement, a low mass $ K^*_0$ exists, which we will call it
$ K^*_0(700)$ or  $\kappa(700)$ with
\be
M_{\kappa}=700\pm 80\,{\rm MeV/c^2}, \qquad \Gamma_{\sigma}=650\pm
120\,{\rm MeV/c^2}\,.
\ee
The central value and the errors are chosen to include the (most
reliable) result of \cite{Descotes-Genon:2006uk}, not rejecting
completely other evaluations giving higher mass values.

\subsection{\label{Isovector pieta interactions}
Isovector $\pi\eta$ interactions}

\subsubsection{\label{The reaction pi-p rightarrow eta pi0 n}
The reaction $\pi^-p\rightarrow\eta\pi^0 n$}

The reaction was studied by the GAMS  collaboration
using the 38 GeV/c $\pi^-$-beam of the IHEP U-70 accelerator
\cite{Alde:1999gh}. The $\eta\pi$ mass spectrum shows two peaks
corresponding to $a_0(980)$ and $a_2(1320)$ formation, no shoulder is
observed which could house the $a_0(1450)$. For  $-t < 1 ~(GeV/c)^2$,
$a_2(1320)$ dominates in the spectrum, at small momentum transfer, for
$-t < 0.05 ~(GeV/c)^2$, the intensities of both peaks are actually
similar. Production of $a_2(1320)$ is dominated by natural spin-parity
exchanges (mainly $\rho$-exchange) in the $t$-channel.  For $a_0(980)$
production, only unnatural exchanges are allowed in the $t$-channel,
and the data are consistent with $\rho_2$ as most significant exchange.
No $a_0(1450)$ was seen in the data. No data on $a_0$ decay into $K\bar
K$ were reported.

\subsubsection{\label{The a_0(980) and f_0(980) from bar pp
annihilation }
The $a_0(980)$ and $f_0(980)$ from $\bar pp$ annihilation }

\begin{figure}[pt] \bc \begin{tabular}{ccc}
\includegraphics[width=0.3\textwidth,height=3.5cm]{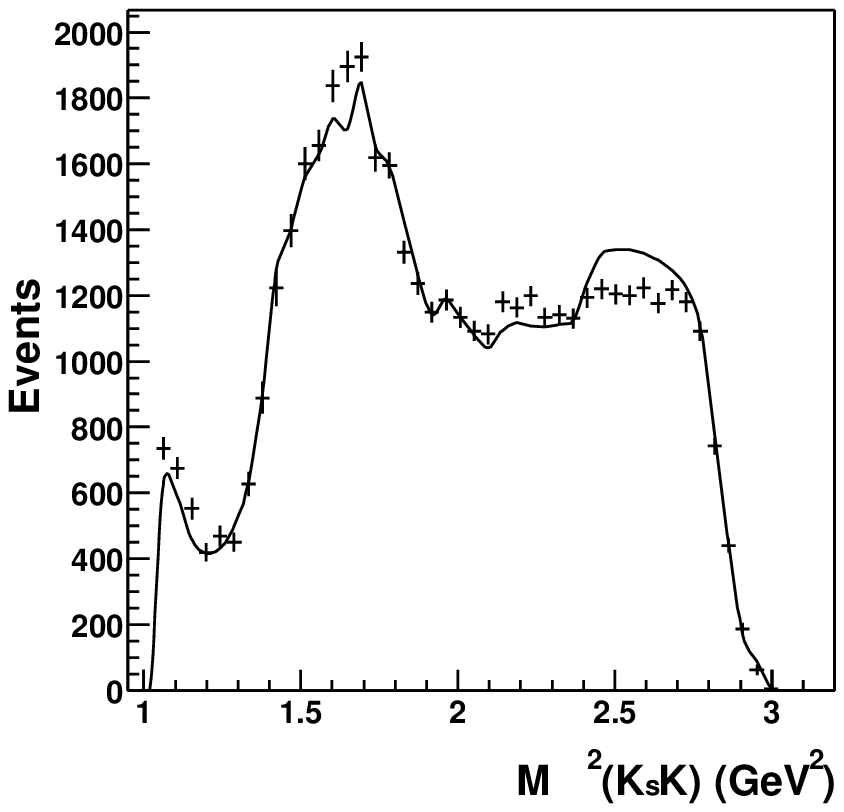} &
\hspace{2mm}
\includegraphics[width=0.3\textwidth,height=3.5cm]{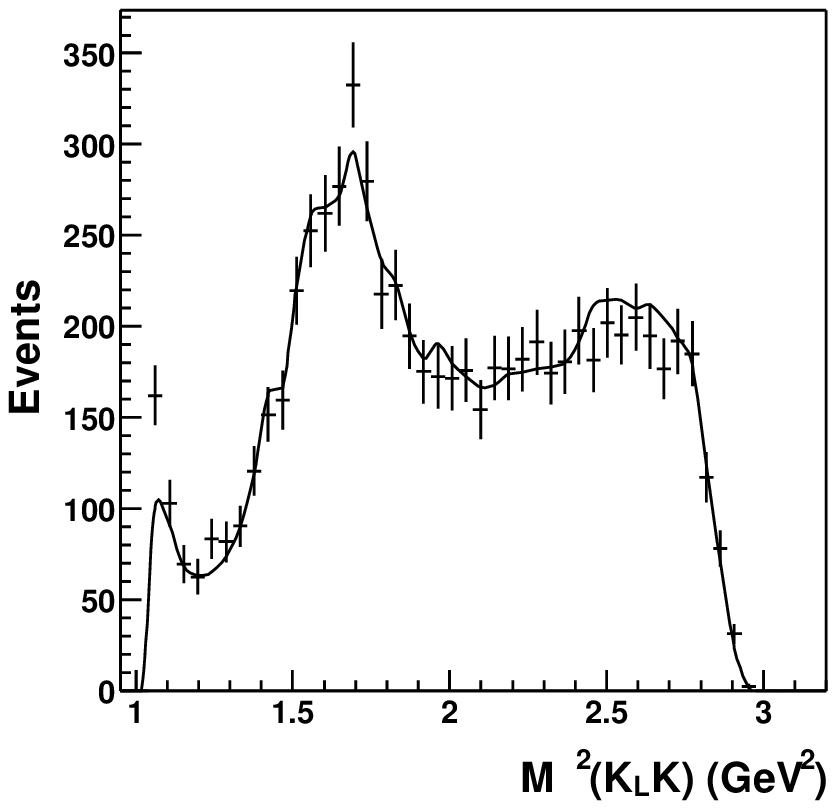}  &
\includegraphics[width=0.3\textwidth,height=3.5cm]{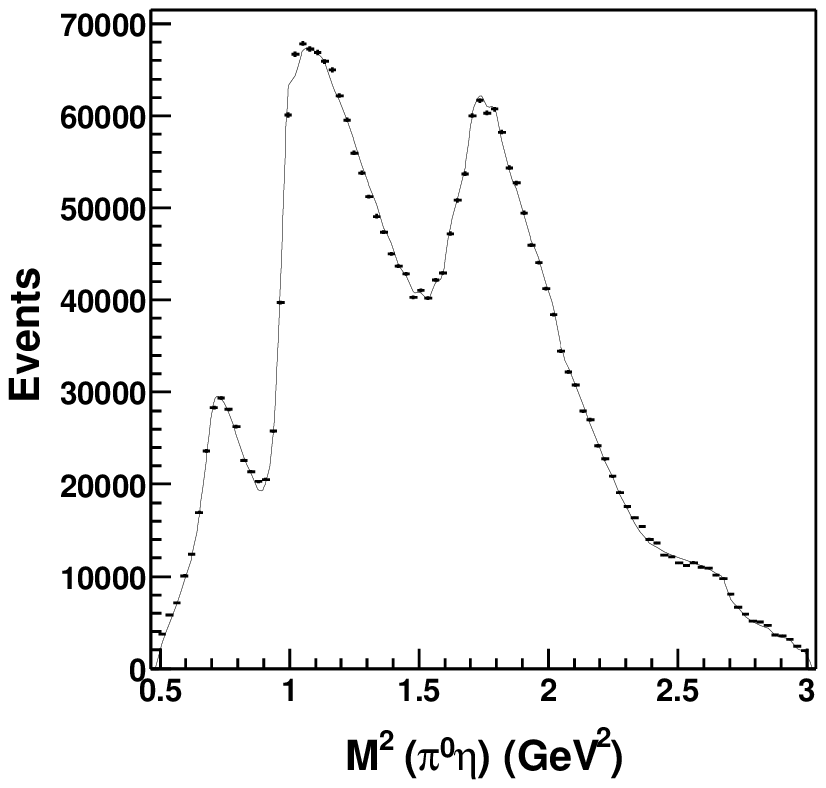}
\vspace{3mm}  \\
\includegraphics[width=0.3\textwidth,height=3.5cm]{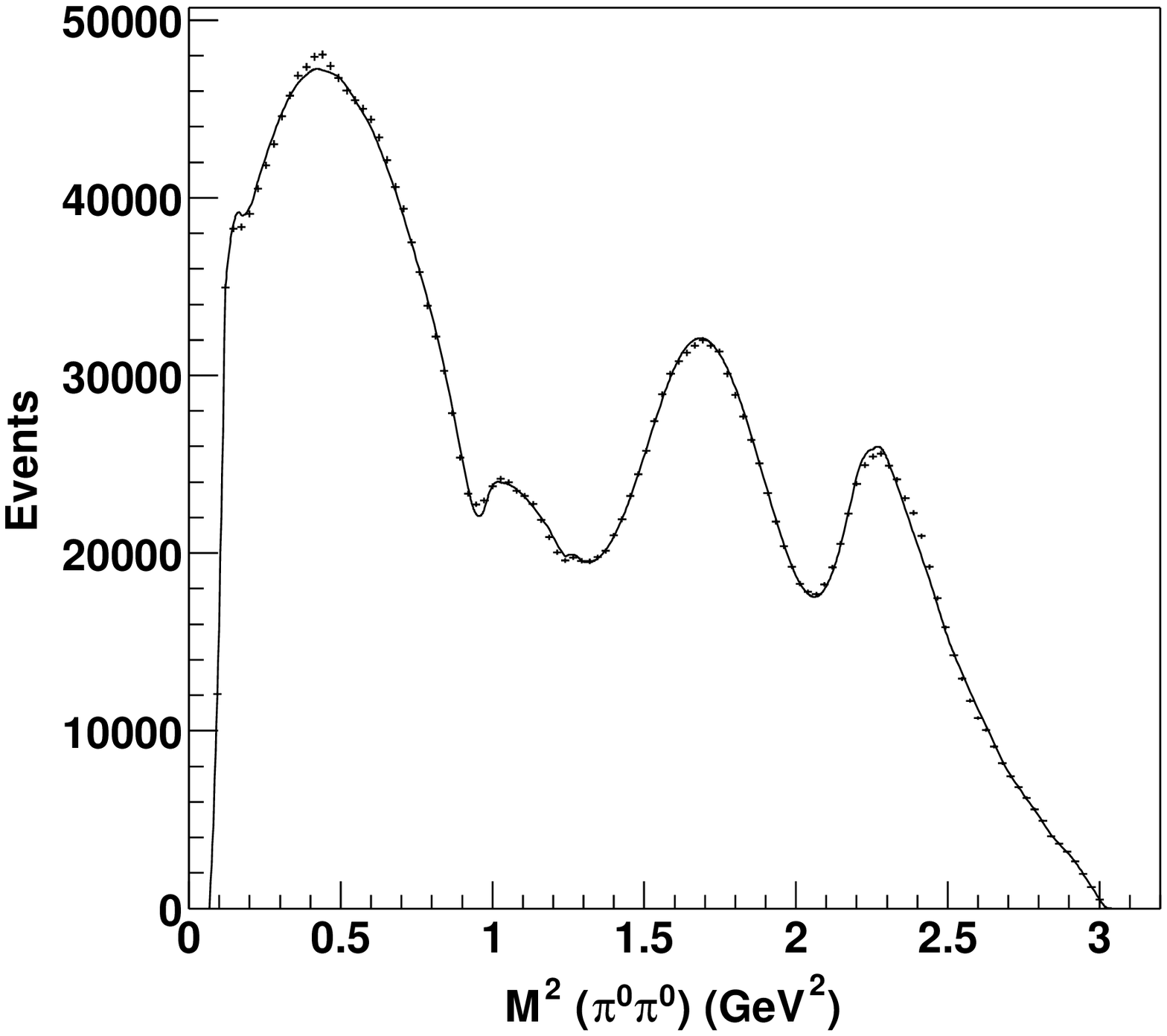}  &
\hspace{2mm}
\includegraphics[width=0.3\textwidth,height=3.5cm]{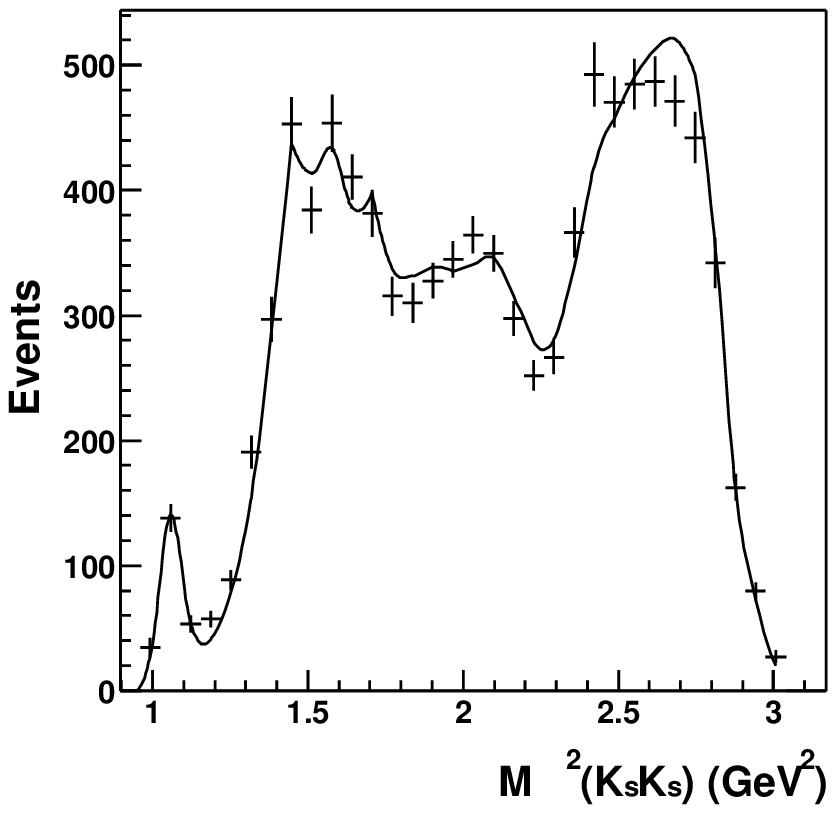} &
\includegraphics[width=0.31\textwidth,height=3.6cm,angle=-1]{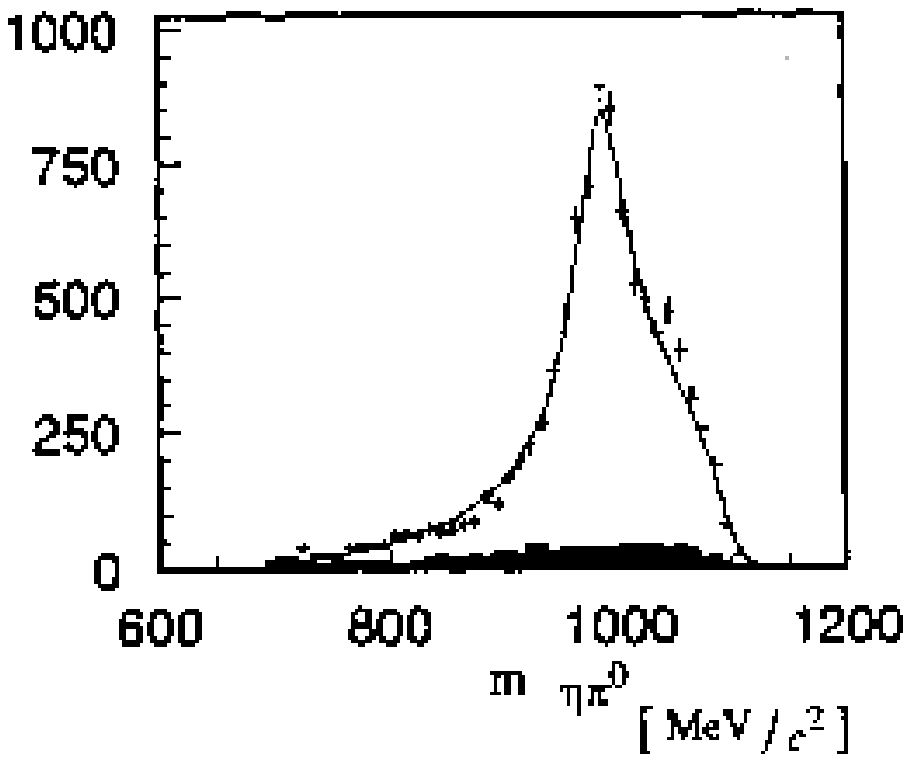}
\end{tabular}\vspace{-75mm}\\
\footnotesize\hspace{20mm}a\hspace{58mm}b\hspace{55mm}c\hspace{-20mm}\vspace{36mm}\\
\footnotesize\hspace{18mm}d\hspace{58mm}e\hspace{54mm}f\hspace{-20mm}\vspace{27mm}\\
\ec
\caption{\label{cbar-a0}Invariant mass distributions from
$ p\bar p$ annihilation at rest. The data are from the Crystal
Barrel experiment, the solid line represents a fit from
\cite{Anisovich:2002ij}. a: $ K^0_LK^{\mp}$ \cite{Abele:1998qd}, b:
$ K^0_SK^{\mp}$,  c: $\pi\eta$ \cite{Amsler:1994pz}, d: $\pi^0\pi^0$
\cite{Amsler:1995gf}, e: $ K^0_SK^0_S$ \cite{Abele:1996nn} and e:
$\pi^0\pi^0$ \cite{Amsler:1995gf}  invariant mass distributions
recoiling against a pion in $ p\bar p$ annihilation at rest.  f:
$\pi\eta$ mass distribution from $ p\bar p\to \omega\pi^0\eta$ with
a fit using a Breit-Wigner amplitude \cite{Amsler:1994ey}.
 }
\end{figure}

The Crystal Barrel detector measured both $\pi\eta$ and
$ K\bar K$ final states and was thus one of the few detectors capable
to determine $a_0(980)$ decays into both final states. Fig.
\ref{cbar-a0}a,b shows $a_0(980)$ in $p\bar p\to \pi^{\pm} K^0K^{\mp}$,
with $ K^0=K^0_L$ (a), $K^0_S$ (b), respectively. In the $\eta\pi$ mass
distribution (c), $a_0(980)$ manifests itself as a fast rise
at 1\,GeV.
The $f_0(980)$ is seen as a dip in $ p\bar p$ annihilation into
3$\pi^0$ (Fig. \ref{cbar-a0}d). Its decay into $ K\bar K$ is difficult
to disentangle from decays of neutral $a_0(980)$; a coupled channel
analysis is needed invoking isospin relations to determine the
$f_0(980)$ and $a_0(980)$ contributions to the $ K^0_SK^0_S$ mass spectrum
shown in Fig. \ref{cbar-a0}e.

Partial widths of resonances are usually defined at their peak
position. This is of course impossible for resonances below an
important threshold. One remedy is to determine ratios of partial
decay widths as ratio of events in different final states. With this
warning, we list the partial widths
\bc
\begin{tabular}{lrrr}
$a_0(980)$ & ${\rm M=984.4\pm 1.3\,MeV/c^2}$ \cite{Amsler:1994ey} &
             ${\rm\Gamma=92\pm 20\,MeV/c^2}$ \cite{Anisovich:1997zw}&
$\frac{\Gamma_{ K\bar K}}{\Gamma_{\pi\pi}}=0.23\pm0.05$\cite{Abele:1998qd}\\
$f_0(980)$& ${\rm M \ = \ 994\pm 5\  MeV/c^2}$ \cite{Anisovich:2002ij}&
             ${\rm\Gamma=64\pm 16\,MeV/c^2}$ \cite{Anisovich:2002ij} &
$\frac{\Gamma_{ K\bar K}}{\Gamma_{\pi\pi}}=0.84\pm 0.02$\cite{Anisovich:2002ij}
\end{tabular}
\ec
which were deduced from the data. The widths are much more uncertain
than the errors suggest. The peak widths are narrow, the decay width
can be as large as 200\,MeV/c$^2$. Closer to the analysis are the
parameters of the Flatt\'e formula (\ref{Flatte}) giving the couplings
instead of the partial widths. For $\pi\eta$ elastic scattering or
scattering into $K\bar K$ via $a_0(980)$ formation, the Flatt\'e
formula reads
\be \frac{d\sigma_i}{dm} \propto
\left|\frac{m_R\sqrt{\Gamma_{\pi\eta}\Gamma_i}}
{m_R^2-m^2-im_R(\Gamma_{\pi\eta}+\Gamma_{K\bar K})}\right|^2.
\label{Flatte}
\ee
The first channel is defined by elastic scattering, with
$\Gamma_{\pi\eta}=\bar g_{\eta}q_{\eta}$, and with $g_{\eta}$ assumed
to be constant in the vicinity of the resonance; $q_\eta$ is
centre-of-mass momentum in the ${\pi\eta}$ system. The partial decay
width to the second ($K\bar K$) channel has a strong energy dependence.
Above the $K\bar K$ threshold, it grows rapidly and the denominator in
eq. (\ref{Flatte}) increases leading to a reduced apparent width in
the $\pi\eta$ mass distribution.
$$ \Gamma_{K\bar K}=\qquad
 \begin{cases}
\qquad \bar g_K \sqrt{m^2/4-m^2_K}   &\qquad {\rm above \ threshold} \\
\qquad i\bar g_K\sqrt{m_K^2-m^2/4}   & \qquad{\rm below \ threshold}
 \end{cases}
$$
The formulae are easily adapted to describe $f_0(980)$.

A compilation of Flatt\'e parameters for $a_0(980)$ and $f_0(980)$ is
reproduced from \cite{Baru:2004xg} in Table \ref{tab:baru}. In the
Table, $m_R$ is the nominal mass of the resonance, $m$ is the actual
invariant mass and $\bar g_\eta$ and $\bar g_K$ are dimensionless
coupling constants that are related to the dimensional coupling
constants $g_{\pi\eta}$ and $g_{K\bar K}$ commonly used in the
literature by $\bar g_\eta=g^2_{\pi\eta}/(8\pi m^2_R)$ and $\bar
g_K=g^2_{K\bar K}/(8\pi m^2_R)$, respectively.

\begin{table}[t]
\caption{\label{tab:baru}
Flatt\'e parameters for the $a_0(980)$ (left) and $f_0(980)$
(right) mesons. The values of $m_R$, $\Gamma_{\pi\eta}$
$\Gamma_{\pi\pi}$ and $E_{B}$ are given in MeV. Values for references
labelled  with the superscript $^a$ are based on a parametrisation
given by Achasov \protect\cite{Achasov:2002ir}. The table is adapted
from  \cite{Baru:2004xg}.
 \vspace{2mm}} \renewcommand{\arraystretch}{1.6} \begin{center}
\vskip 0.2cm \begin{tabular}{ccccccc|ccccccc}
\hline\hline
Ref.&$m_R$&$\Gamma_{\pi\eta}$&$\bar g_{\eta}$&$\bar g_{K}$&$R$&$E_{R}$
&Ref.&$m_R$&$\Gamma_{\pi\pi}$&$\bar g_{\pi}$&$\bar g_{K}$&$R$&$E_{R}$\\
\hline
\cite{Bugg:1994mg}&999&143&0.445&0.516&1.16&7.6&
\cite{Ackerstaff:1998ue} &957  &42.3&0.09&0.97&10.78&-34.3\\
\cite{Teige:1996fi}&1001&70&0.218&0.224&1.03&9.6&
\cite{Akhmetshin:1999di}$^a$&975&149&0.317&1.51&4.76&-16.3\\
\cite{Abele:1998qd}&999&69&0.215&0.222&1.03&7.6&
\cite{Barberis:1999cq}& -- &90&0.19   &0.40&2.11&--\\
\cite{Achasov:2000ku}$^a$&995&125&0.389&1.414&3.63&3.7&
\cite{Aitala:2000xt}&977&42.3&0.09&0.02&0.22&-14.3\\
\cite{Aloisio:2002bs}$^a$&984.8&121&0.376&0.412&1.1&-6.5&
\cite{Achasov:2000ym}$^a$&969.8&196&0.417&2.51&6.02&-21.5\\
\cite{Achasov:2002ir}$^a$&1003&153&0.476&0.834&1.75&11.6&
\cite{Aloisio:2002bt}$^a$&973&256&0.538&2.84&5.28&-18.3\\
\cite{Achasov:2002ir}$^a$&992&145.3&0.453&0.56&1.24&0.6\\
\hline\hline
\end{tabular}
\end{center}
\renewcommand{\arraystretch}{1.0}
\end{table}

It is remarkable how large the spread in the binding energy $E_B$ and
in the ratio $R$ of the coupling constants is. More work is certainly
needed to understand the process-dependence of these quantities. The
large value of $R$ underlines the strong affinity of $f_0(980)$ to the
$K\bar K$ system and can be used to argue in favour of a $K\bar K$
molecular character of the two mesons; see however the discussion in
section \ref{Dynamical generation of resonances and flavour exotics}.

The masses of $f_0(980)$ and $a_0(980)$ are right at the $ K\bar K$
threshold; the impact of the threshold on the $\pi\eta$ mass spectrum
can be seen in Fig. \ref{cbar-a0} (bottom, right) which displays the
$\eta\pi$ spectrum recoiling against an $\omega$ in $ p\bar p$
annihilation at rest. Little kinetic energy is available in this
reaction, a kinematical situation which is not suppressed in $ p\bar
p$ annihilation \cite{Klempt:2005pp}. The small kinetic energies give
the optimum detector resolution, and the non-Breit-Wigner mass
distribution becomes immediately visible. The Crystal Barrel
collaboration did not fit the data with the Flatt\'e formula; also a
study of $ p\bar p\to\omega K\bar K$ was not made.

\subsection{\label{Scalar mesons in radiative phi decays}
Scalar mesons in radiative $\phi$ decays}

\begin{figure}[pt]
\bc
\begin{tabular}{ccc}
\includegraphics[width=0.40\textwidth,height=6cm]{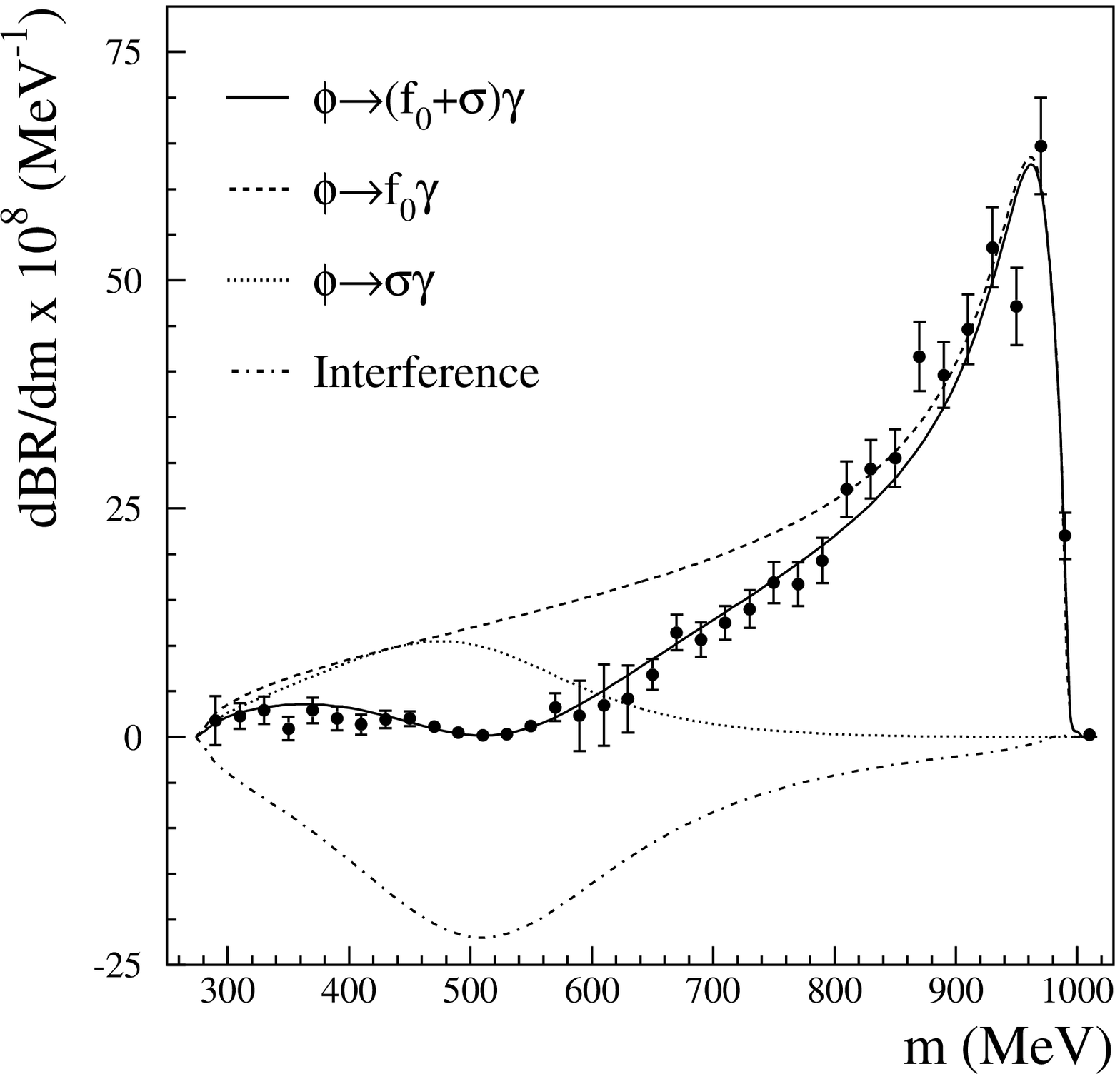}  &
\hspace{-9mm}
\includegraphics[width=0.3\textwidth,height=5cm]{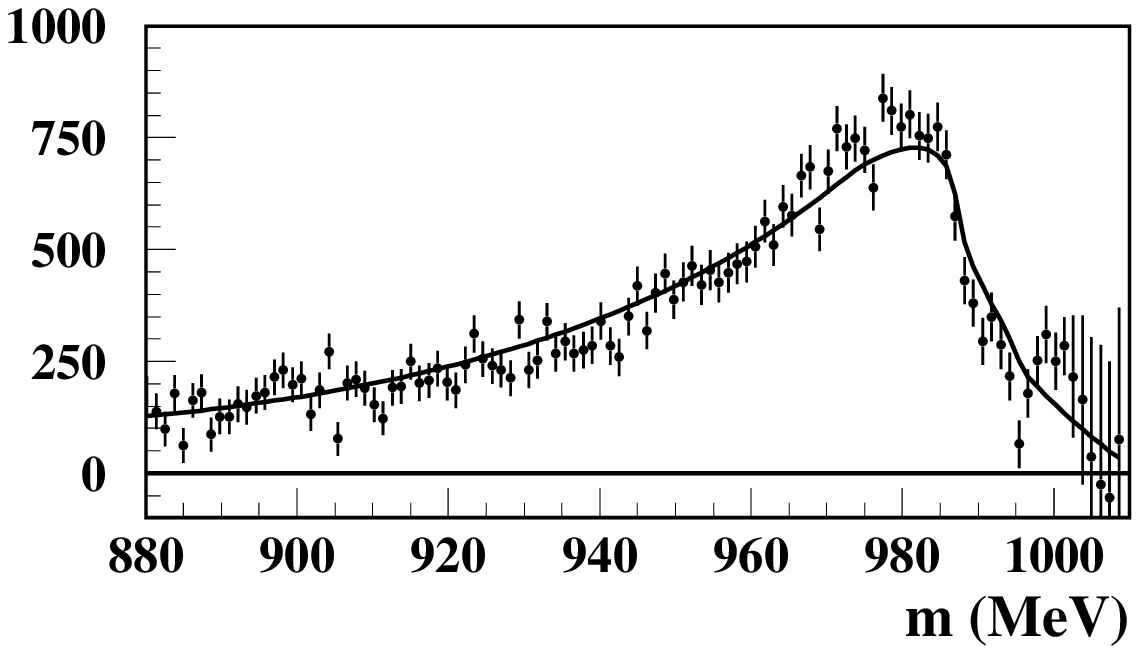} &
\hspace{-2mm}\includegraphics[width=0.3\textwidth,height=5cm]{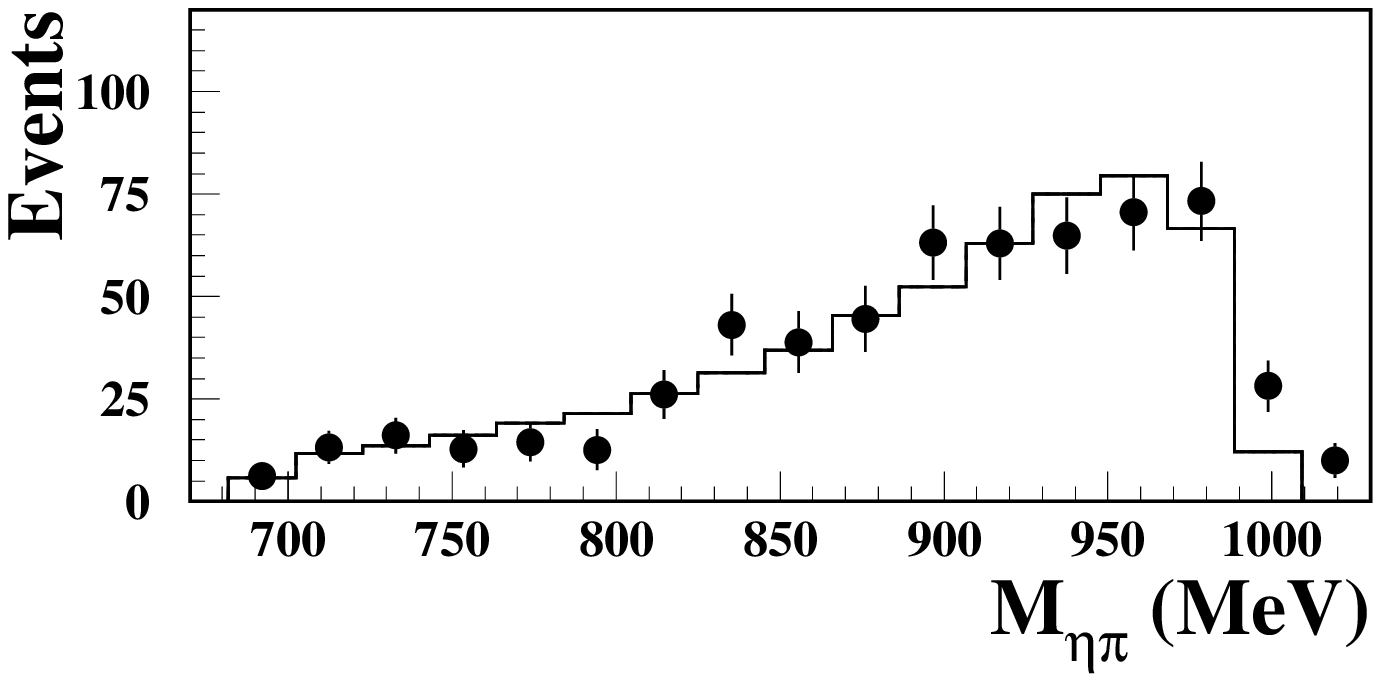}
\end{tabular} \vspace{-55mm}\\
\hspace{-80mm}a \vspace{1mm}\\
\hspace{40mm}b\hspace{53mm}c \vspace{40mm}\\ \ec
\caption{\label{phif0a0}Invariant mass distributions in radiative
$\phi$ decays. a: $\pi^0\pi^0$ \cite{Aloisio:2002bt}, b:
$\pi^+\pi^-$ \cite{Aloisio:2002bs}, c: $\pi^0\eta$
\cite{Ambrosino:2005wk}. } \end{figure}

The open interpretation of the $f_0(980)$ and $a_0(980)$ mesons
initiated experiments aiming at exploring their nature. In the
tetraquark interpretation, the 2-photon width was calculated to be
suppressed \cite{Achasov:1982bt} while large branching ratios were
predicted for radiative $\phi$ decays into $f_0(980)$ and $a_0(980)$
\cite{Achasov:1987ts}. Thus there was hope that their nature can be
identified by a study of radiative $\phi$ decays. The reactions
$\phi(1020)\to\pi^0\eta\gamma$, $\phi(1020)\to\pi^0\pi^0\gamma$, and
$\phi(1020)\to\pi^+\pi^-\gamma$ were first observed in the SND and CMD
experiment at Novosibirsk (see section \ref{VEPP}), and the $a_0(980)$
and $f_0(980)$ contributions were derived from fits to the data
\cite{Achasov:1998cc,Aulchenko:1998xy,Akhmetshin:1999dh,%
Akhmetshin:1999di,Achasov:2000ym,Achasov:2000ku}. The three reactions
were also studied \cite{Aloisio:2002bs,Aloisio:2002bt,Ambrosino:2005wk}
with the Daphne detector at Frascati , described briefly in section
\ref{Daphne}. The latter experiment has the largest statistics; hence
the Daphne results are shown in Fig.~\ref{phif0a0}. The $\pi^0\pi^0$
invariant mass distribution exhibits a dip at about 500\,MeV/c$^2$
which is interpreted in \cite{Aloisio:2002bt} as destructive
interference between two Breit-Wigner amplitudes for $\sigma(485)$ and
$f_0(980)$ production. The results of all three experiments are
summarised in Table~\ref{phidecay}.

\begin{table}[pb]
\caption{\label{phidecay}Branching ratios for radiative $\phi$ decays
(in units of $\cdot 10^{-4}$). The branching ratios for
 $\phi(1020)\to a_0(980)\gamma$ are corrected by using the ratio
$ K\bar K/\eta\pi=0.23\pm0.05$ and $ K\bar K/\pi\pi=0.23\pm0.05$
for $\phi(1020)\to f_0(980)\gamma$. First results from SND were obtained
using $8\cdot 10^{6}\phi$'s, later SMD-2 and SND disposed of  $19\cdot
10^{6}\phi$ events. The first studies of KLOE were based on $53\cdot
10^{6}\phi$'s. In the study of $\phi(1020)\to\pi^+\pi^-\gamma$, an
equivalent of $3\cdot 10^9$ events were recorded. See text for
references. \vspace{2mm}} \bc \renewcommand{\arraystretch}{1.3}
\begin{tabular}{lcccc} \hline \hline
\multicolumn{2}{c}{Reaction} & SND  & CMD-2  & KLOE  \\ \hline
$\phi(1020)\to\pi^0\eta\gamma$ && $0.88\pm0.14$
&$0.90\pm0.24\pm0.10$&\\ $\phi(1020)\to
a_0(980)\gamma$&$a_0(980)\to\eta\pi^0 $ & $0.83\pm 0.23$ &$0.90\pm
0.24\pm 0.10$& $0.74\pm0.07$ \\
$\phi(1020)\to\pi^0\pi^0\gamma$&&$1.14\pm0.10\pm0.12$&$0.92\pm0.08\pm0.06$&$1.09\pm0.03\pm0.05$\\
$3\times\phi(1020)\to
f_0(980)\gamma$&$f_0(980)\to\pi^0\pi^0$&$3.42\pm0.30\pm0.36$&$2.90\pm0.21\pm1.54$&$4.47\pm0.21$\\
$\frac32\times\phi(1020)\to f_0(980)\gamma$&$ f_0(980)\to\pi^+\pi^-$  &&
$1.93\pm 0.46\pm 0.50$ &3.1 -- 3.6\\
\hline \hline \end{tabular}
\renewcommand{\arraystretch}{1.0}
\ec
\end{table}

Looking at Table~\ref{phidecay}
it seems safe to assume that the process  $\phi(1020)\to
f_0(980)\gamma$ is observed with a branching fraction of about $4\cdot
10^{-4}$ of all decay modes. However, Boglione and Pennington
\cite{Boglione:2003xh} criticised the KLOE analysis, reanalysed the
data and found that the result depends crucially on the parametrisation
of the $\pi\pi$ $S$-wave. The $\pi\pi$ $S$-wave from Au, Morgan and
Pennington \cite{Au:1986vs}, of Morgan and Pennington
\cite{Morgan:1993td}, or Anisovich and Sarantsev
\cite{Anisovich:2002ij} gave grossly different results, (0.31 or 0.34 or
$1.92)\cdot 10^{-4}$, respectively. The main reason for the differences
between the $S$-waves from \cite{Au:1986vs,Morgan:1993td} and
\cite{Anisovich:2002ij} are the different $f_0(980)$ pole positions. A
further problem arises from the dip in the $\pi^0\pi^0$ mass
distribution. In \cite{Boglione:2003xh} the dip is created by a zero in
the production amplitude. Even when compatibility with the scattering
amplitudes is maintained, the production amplitude can be shaped using
different parametrisations. As a consequence, very different
results on the branching fraction for  $\phi(1020)\to f_0(980)\gamma$
follow.  It remains to be seen if the dip in Fig.~\ref{phif0a0} is a
real effect or due to a (possibly wrong) background subtraction.
Clearly, the status is unsatisfactory; appropriate methods to analyse
the data need to be developed. A general parametrisation of the
amplitude is proposed in \cite{Isidori:2006we}.

\begin{figure}[pt] \bc
\includegraphics[width=0.6\textwidth]{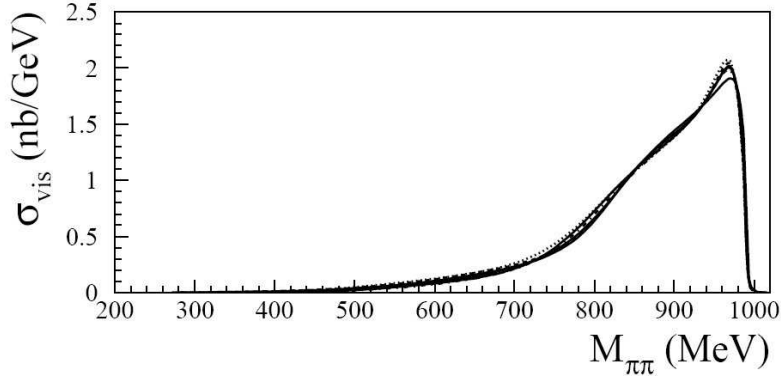}
\vspace{-2mm}
\ec
\caption{\label{pic:daphne_n}
The $\pi^0\pi^0$ invariant mass distribution from Daphne
\cite{unknown:2006hb}. Data points are represented by dots, the thick
  solid line is the result of a fit based on model in which $\phi$
  mesons dissociate in kaon loops radiating off photons.
\vspace{2mm}
  }
\end{figure}

Very recently, the Daphne collaboration reported on a Dalitz plot
analysis of $e^+e^-\to\pi^0\pi^0\gamma$ events collected at $\sqrt{s}
\simeq M_{\phi}$ with the KLOE detector \cite{unknown:2006hb}. The
statistics was increased very significantly; detailed studies were made
to verify the dynamical assumptions of the fit. Data and fit are
reproduced in Fig. \ref{pic:daphne_n}.

In the Kaon Loop model, the two low mass scalars $f_0(980)$ and
$\sigma(485)$ are required to adequately fit the data and a stable
branching ratio of the $\phi\to\pi^0\pi^0\gamma$ process was
obtained:
\begin{eqnarray}
\mathcal B(\phi\to S\gamma\to\pi^0\pi^0\gamma) &=&
   ( 1.07\,^{+0.01}_{-0.03}\,_{\rm fit}\,
  ^{+0.04}_{-0.02}\,_{\rm syst}
  \;^{+0.05}_{-0.06}\;_{\rm mod}) \times 10^{-4} \nonumber \\
    M\!_{f_0} &=& ( 976.8 \pm 0.3_{\rm\,fit}\,^{+0.9}_{-0.6}\,_{\rm
  syst} + 10.1_{\rm\,mod} )\ {\rm MeV/c^2} \nonumber \\
  g_{\!f_{0}K^{+}K^{-}} &=& ( 3.76 \pm
  0.04_{\rm\,fit}\,^{+0.15}_{-0.08}\,_{\rm syst}\,
  ^{+1.16}_{-0.48}\,_{\rm mod} )\ {\rm GeV/c^2}  \\
   g_{\!f_{0}\pi^{+}\pi^{-}} &=&
  ( -1.43 \pm 0.01_{\rm\,fit}\,^{+0.01}_{-0.06}\,_{\rm\,syst}\,
  ^{+0.03}_{-0.60}\,_{\rm mod} )\ {\rm GeV/c^2}\,. \nonumber
 \end{eqnarray}

Like the experimental uncertainties in the determination of the
radiative yield, the theoretical interpretation which one may find in
the literature is far from being unique neither, with the $q\bar q$,
$q\bar qq\bar q$, and the $K\bar K$ molecular picture all being pursued
and compatible with data. In \cite{Anisovich:2004qr} it is shown that
the processes $\phi (1020)\to \gamma\pi\pi$ and $\phi (1020)\to \gamma
f_0(980)$ can well be described assuming $f_0(980)$ to be dominantly
$q\bar q$. Achasov \cite{Achasov:2003cn} and Achasov and Kiselev argue
\cite{Achasov:2005hm} that $a_0(980)$ and   $f_0(980)$ production in
radiative $\phi$ decays gives new evidence in favour of their
tetraquark nature. But also the molecular picture is not excluded
\cite{Kalashnikova:2004ta}. Radiative decays of $\phi$ mesons into
scalars are shown to require a sizable $ K\bar K$ component of the
scalar mesons, but do not allow to discriminate between compact
molecules and $q\bar q$ states when they are allowed to disintegrate
into a kaon loop. Close and Tornqvist \cite{Close:2002zu} propose a
nonet of complex objects: near the centre they are 4-quark states, with
some ($q\bar q$) in $P$-wave; further out they rearrange in colour to two
colourless ($q\bar q$) pairs and finally to meson-meson states.
Beveren and colleagues \cite{vanBeveren:2005ha} show how light and
heavy scalar mesons can be linked to one another by a continuous
variation  of the masses involved. They conclude that all scalar mesons
can be understood as ordinary $q\bar{q}$ states strongly distorted due
to coupled channels, and suggest that labelling scalar mesons as
$q\bar{q}$ states as opposed to dynamical meson-meson resonances makes
no sense, since one kind of pole can be turned into another by tiny
parameter variations. For the moment, we leave the question of the
nature of these states open, and discuss further data. The issue will
be resumed in section \ref{Dynamical generation of resonances and
flavour exotics}.

\subsection{\label{Scalar mesons in 2-photon fusion}
Scalar mesons in 2-photon fusion}

\begin{figure}[pt]
\bc
\begin{tabular}{cc}
\includegraphics[width=0.48\textwidth,height=0.25\textheight]{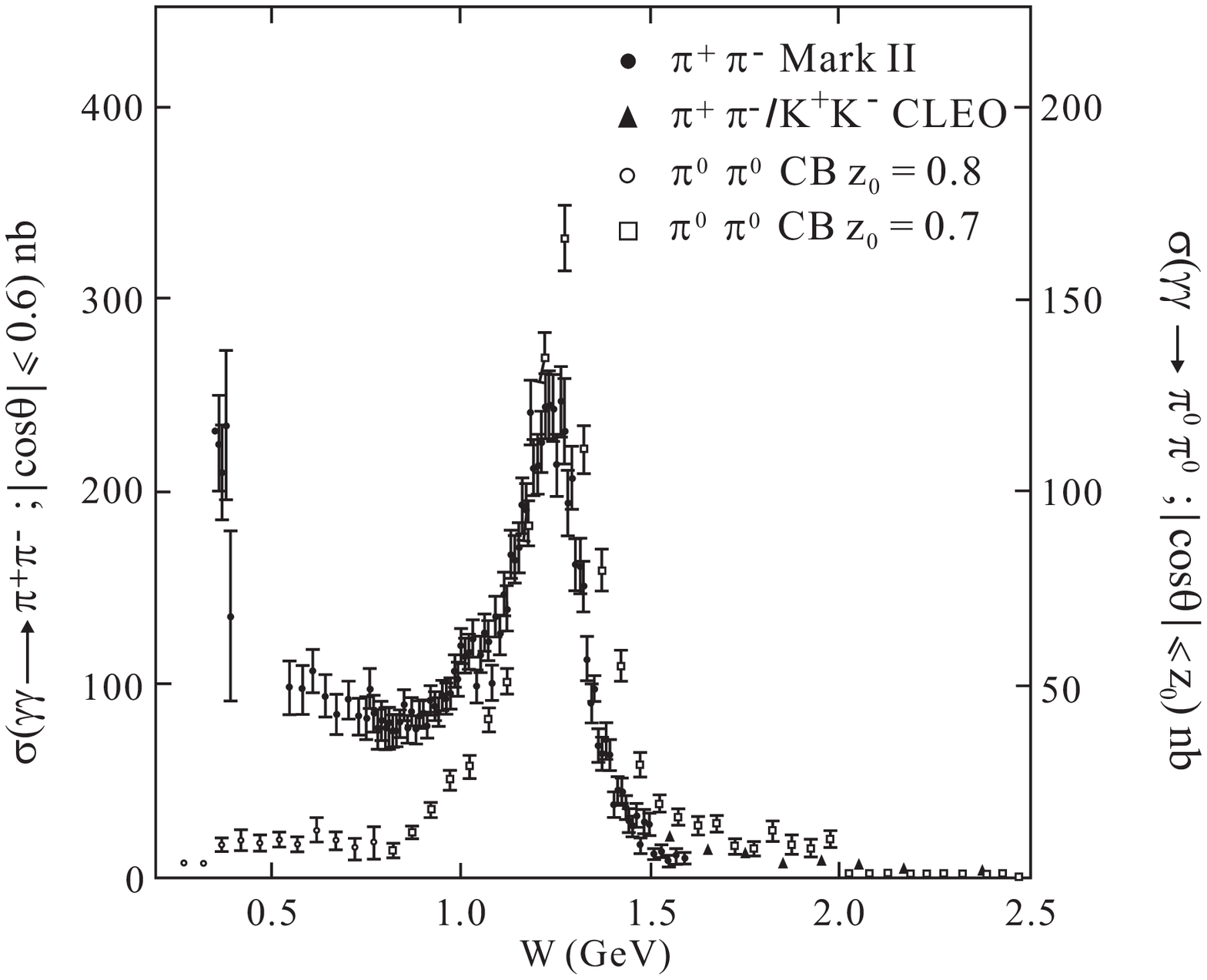}&
\includegraphics[width=0.48\textwidth,height=0.28\textheight]{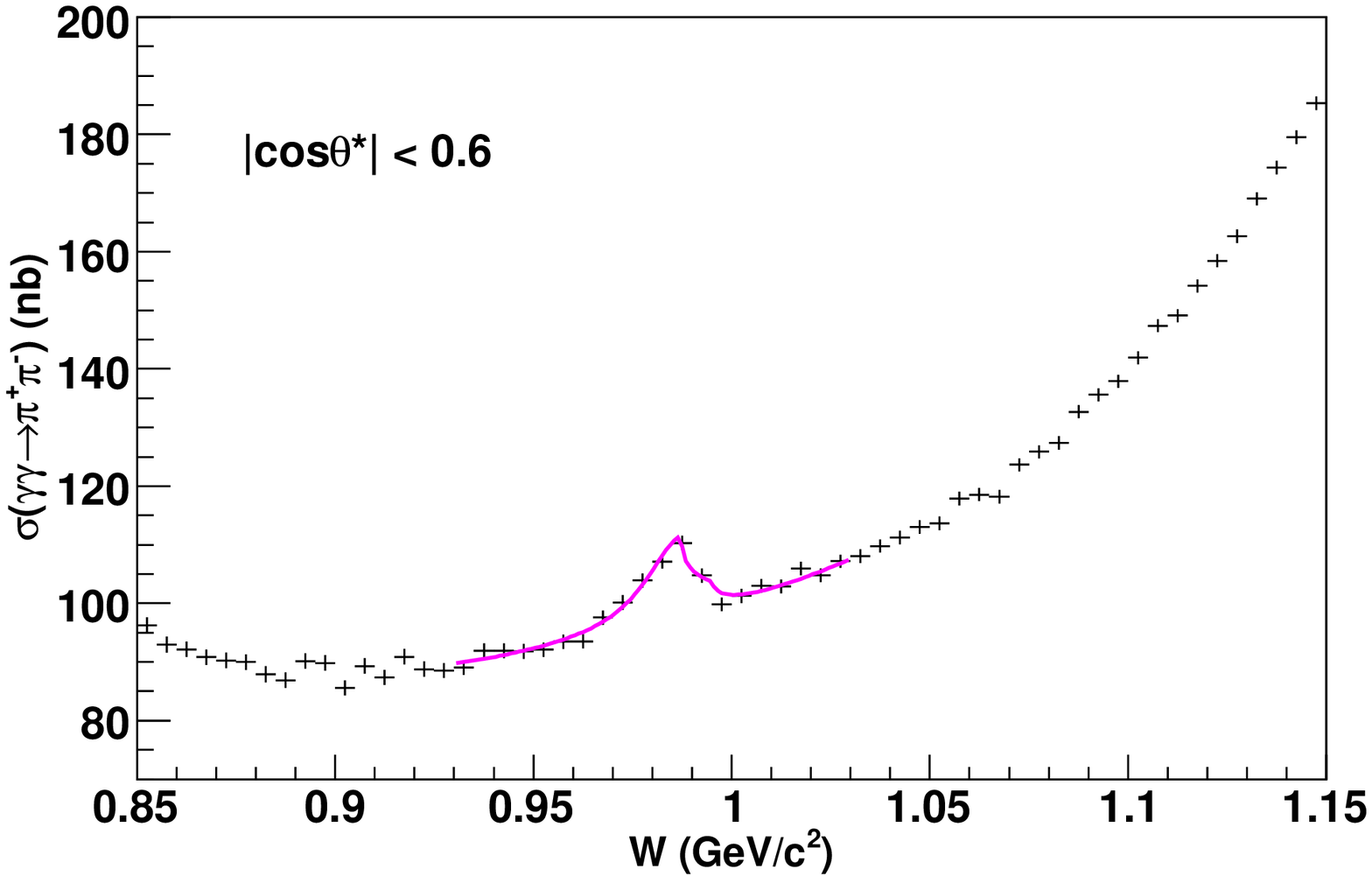}
\end{tabular} \vspace{-60mm}\\
\hspace{7mm}a\hspace{140mm}b\vspace{53mm}\\
\ec
\caption{\label{2gamma}a: Integrated cross-section (within
$|\cos\theta |<0.6$) for $\gamma\gamma\to\pi\pi$ as a function of c.m.
energy $W$ from Mark II \protect\cite{Boyer:1990vu}, Crystal Ball (CB)
\protect\cite{Marsiske:1990hx} and CLEO \protect\cite{Dominick:1994bw}.
b: Expanded view of the $f_0(980)$ region from BELLE
\cite{Mori:2006jj} exhibiting $f_0(980)$ as a clear peak.}
\end{figure}

The radiative width of resonances, their
coupling to two photons, provides another handle on their internal
structure. Fig. \ref{2gamma} shows data on two-photon fusion into a
pion pair from \cite{Boyer:1990vu,Marsiske:1990hx,Dominick:1994bw}.
Identification of scalar mesons in two-photon fusion requires data with
polarised photons and complete angular coverage of the hadron final
state. Such data do not exist; other constraints are needed to make a
partial wave separation possible. Boglione and Pennington
\protect\cite{Boglione:1998rw,Pennington:1999nq} have used several
constraints to deduce the scalar part of the interaction. At low
energies $\gamma\gamma\to\pi\pi$ is dominated by a Born term. Unitarity
requires that the $\gamma\gamma\to\pi\pi$ process is related to
$\pi\pi$ scattering. At sufficiently small masses, constraints from
$\pi\pi$ and $ K{\bar K}$ intermediate states can thus be imposed.
They found two solutions, in which the $f_0(980)$ appears either as a
peak or as a dip. The radiative widths of the $f_2(1270)$, $f_0(980)$
and $f_0(485)$ for the two solutions are listed in
Table~\ref{radwidth}.

\begin{table}[pb]
\caption{\label{radwidth}Radiative widths of states contributing to
$\gamma\gamma\to\pi\pi$ for the two solutions ({\it dip} and {\it peak})
from \cite{Boglione:1998rw} and predicted widths of scalar mesons
for different compositions \cite{Barnes:1985cy} and recent
calculations.} \bc
\renewcommand{\arraystretch}{1.5}
\vspace{2mm}
\begin{tabular}{@{}llllll@{}llllll} \hline\hline & & &&\hspace{-8mm}
$\Gamma(\gamma\gamma)$ keV \\ \hline solution  & $f_2(1270)$
& $\; f_0(980)$ & $f_0(400/1200)$  &$\mid$ \ prediction &$\qquad
n{\overline n}\;$ & $\qquad s{\overline s}\;$ & $\qquad K{\overline
K}\;$ \\ \hline dip                & $\quad 2.64$ & $\quad 0.32$ &
$\qquad 4.7$ &$\mid$ \  $\Gamma(0^{++}\to\gamma\gamma)$   & $\qquad 4.5$ &
$\qquad 0.4$ & $\qquad 0.6$ &\cite{Barnes:1985cy}\\ peak & $\quad 3.04$
& $0.13-0.36$ & $\qquad3.0$ &&&&\qquad  $0.2$& \cite{Oller:1997yg}\\
&&&&&&&$0.22\pm 0.07$ &\cite{Hanhart:2007wa}\\
\hline\hline
\end{tabular} \renewcommand{\arraystretch}{1.0} \ec \end{table}

New data on $\gamma\gamma\to\pi^+\pi^-$ became available from the BELLE
experiment, which supersede all previous work in statistics. A clear
signal for the $f_0(980)$ resonance is observed
\cite{Mori:2006jj,Mori:2007bu}. The mass spectrum was fitted with a
two-channel Breit-Wigner amplitude and interfering and non-interfering
background terms. Mass and partial widths to $\pi^+\pi^-$ and to
$2\gamma$ were determined to

\begin{eqnarray} M^{f_0} = &985.6 ~^{+1.2}_{-1.5}({\rm
stat})
                  ~^{+1.1}_{-1.6}({\rm syst})&~{\rm MeV/c^2},\;
\nonumber\\ \Gamma_{\pi^+\pi^-} ^{f_0}=& 34.2~^{+13.9}_{-11.8}({\rm stat})
              ~^{+8.8}_{-2.5}({\rm syst})~&   {\rm MeV/c^2},\;\\
\Gamma_{\gamma\gamma} ^{f_0}=& 205 ~^{+95}_{-83}({\rm stat})
                    ~_{-117}^{+147}({\rm syst})~&{\rm eV/c^2}, \nonumber
\end{eqnarray}

respectively. Early predictions by Barnes \cite{Barnes:1985cy} gave the
two-photon width for an internal $n\bar n$ or $s\bar s$ structure or a
$f_0(980)$ with molecular origin. Achasov and Shestakov interpreted the
data in a model in which $f_0(980)\to K\bar K$ loops play a decisive
$\rm r\hat{o}le$ \cite{Achasov:2005ub,Achasov:2006rf}; the direct
$f_0(980)\to\gamma\gamma$ coupling is shown to be negligibly small.
Oller and Oset, incorporating chiral symmetry and exchange of vector
and axial resonances in the crossed channels, used kaon loop diagrams
to arrive at their result \cite{Oller:1997yg}. Hanhart, Kalashnikova,
Kudryavtsev, and Nefediev \cite{Hanhart:2007wa} argued for a molecular
picture of $f_0(980)$. Their results are listed in Table
\ref{radwidth}. From an experimental point we find it hard to
differentiate between a $q\bar q$ or $qq\bar q\bar q$ state (the
`seed') decaying into the final state via a $K\bar K$ loop and, on the
other hand, a truly molecular $K\bar K$ state, having no seed bound by
chromo-electromagnetic forces. In a recent paper, Pennington asks ``Can
Experiment Distinguish Tetraquark Scalars, Molecules And $q\bar q$
Mesons?" \cite{Pennington:2007eg}. His answer is yes, with some
conditional ifs, in particular only when the wave function does not
have a complicated Fock space expansion. Applied to the $\sigma(485)$,
the direct answer from its 2-photon coupling assigns a $q\bar q$ nature
\cite{Pennington:2007yt}. Obviously, such predictions have to be taken
with some precaution.

\subsection{\label{Production of f0(980) as a function of t}
Production of $f_0(980)$ as a function of $t$}

The GAMS collaboration reported a peculiar behaviour of $f_0(980)$
\begin{figure}[pb]
\begin{center}
\includegraphics[width=0.7\textwidth,height=11cm]{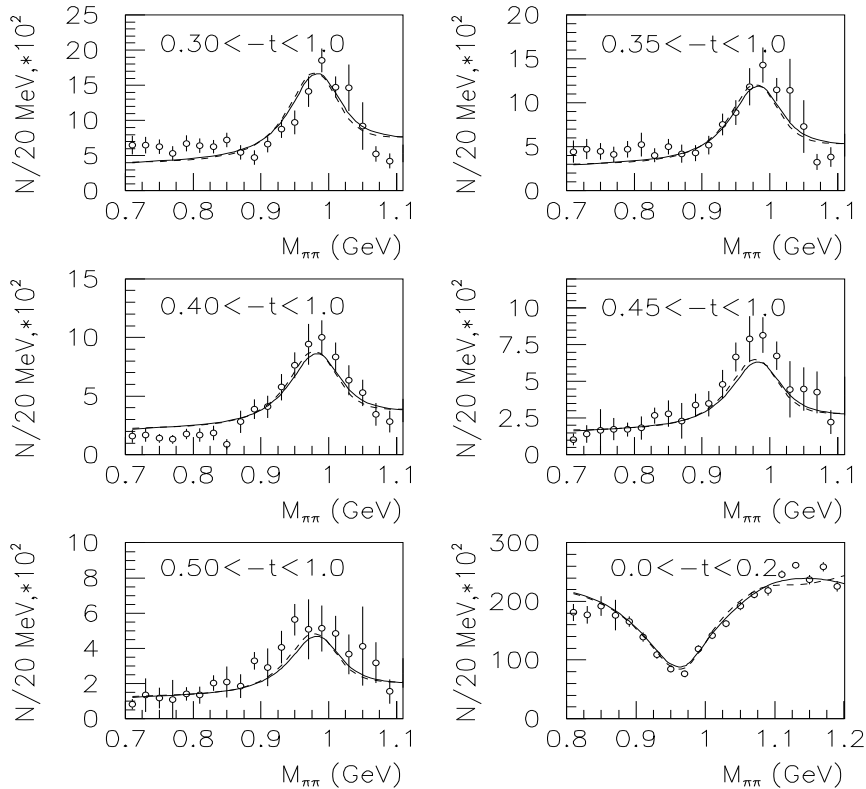}
\end{center}
\vspace{-2mm}
\caption{\footnotesize{The reaction  $\pi^- p\to\pi^0\pi^0n$ as a function
of the momentum transfer to the $\pi\pi$ system \cite{Alde:1994jj}.}}
\end{figure}
production in the reaction $\pi^- p\to n\pi^0\pi^0$. For small
momentum transfers between the proton and the neutron $( -t <
0.1\;\textrm{GeV}^2)$, the scalar amplitudes show a dip around 1 GeV,
while $f_0(980)$ is observed as a peak for large momentum transfers $(
-t > 0.4\;\textrm{GeV}^2 )$. This observation has been interpreted as
evidence for a hard component in the $f_0(980)$
\cite{Anisovich:1995jy,Kondashov:1998uh,Klempt:2000ud,Anisovich:2002us}.
However, the strong dependence of $f_0(980)$-production on the
momentum transfer between proton and neutron can also well be
reproduced by a change of the interference between the resonance
structure and the non-resonant background when $\pi\pi$ and
$a_1(1230)\pi$ scattering are considered, and the data are fully
compatible with the molecular picture of $f_0(980)$
\protect\cite{Achasov:2001fr,Sassen:2002qv,Sassen:2003iz,Achasov:2003xn}.

\subsection{\label{Scalar mesons in fragmentation}
Scalar mesons in fragmentation}

The $f_0(980)$ and $a_0(980)$ wave functions were tested
at LEP by a study of $Z^0$ fragmentation into quark- and gluon jets.
Some total inclusive rates for meson production from $Z^0$
fragmentation are listed in Table \ref{Z0decay}.

\begin{figure}[pb]
\begin{center}
\begin{tabular}{cc}
\mbox{\epsfig{file=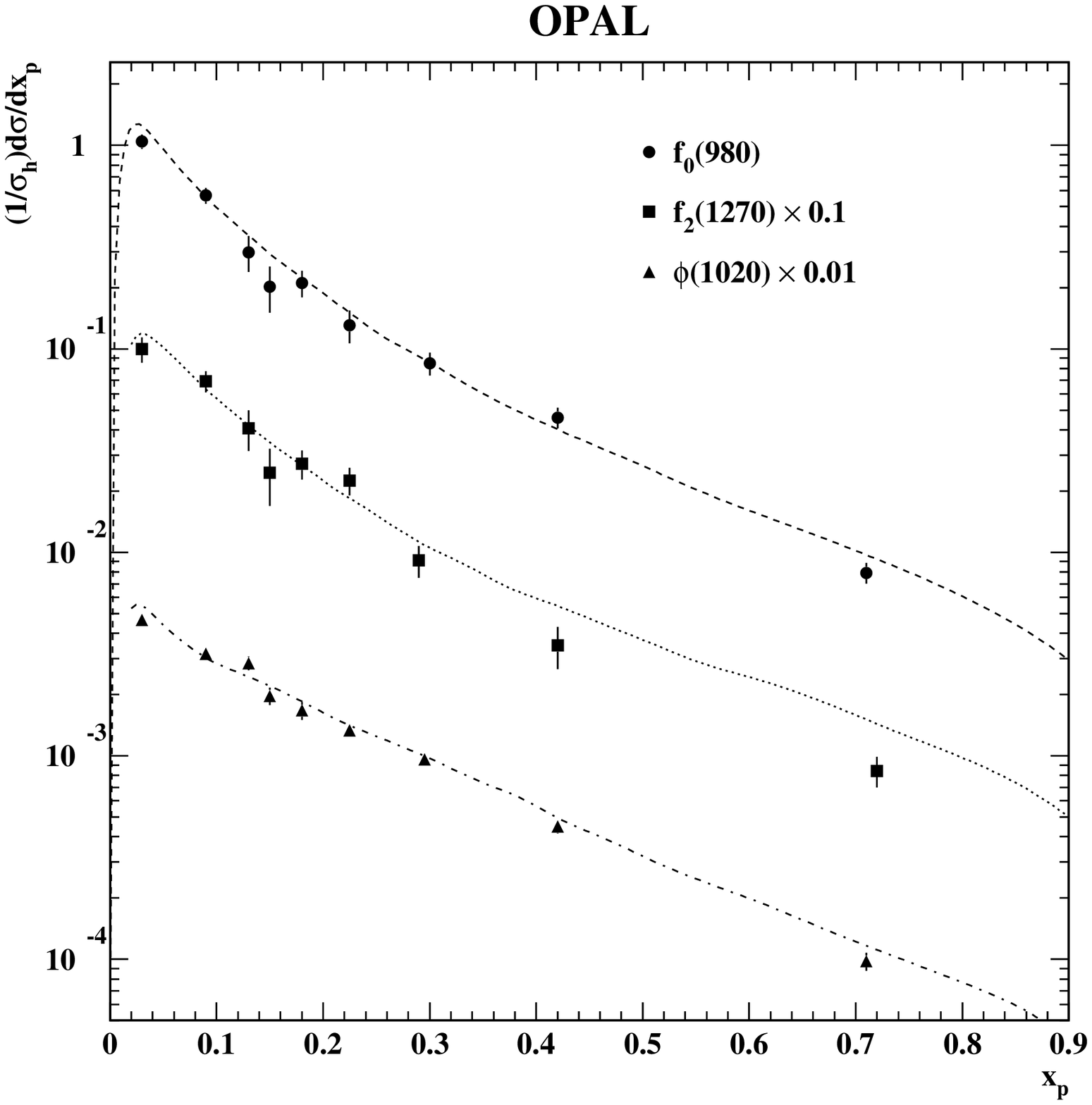,height=0.45\textwidth,width=0.48\textwidth}}&
\hspace{-8mm}\mbox{\epsfig{file=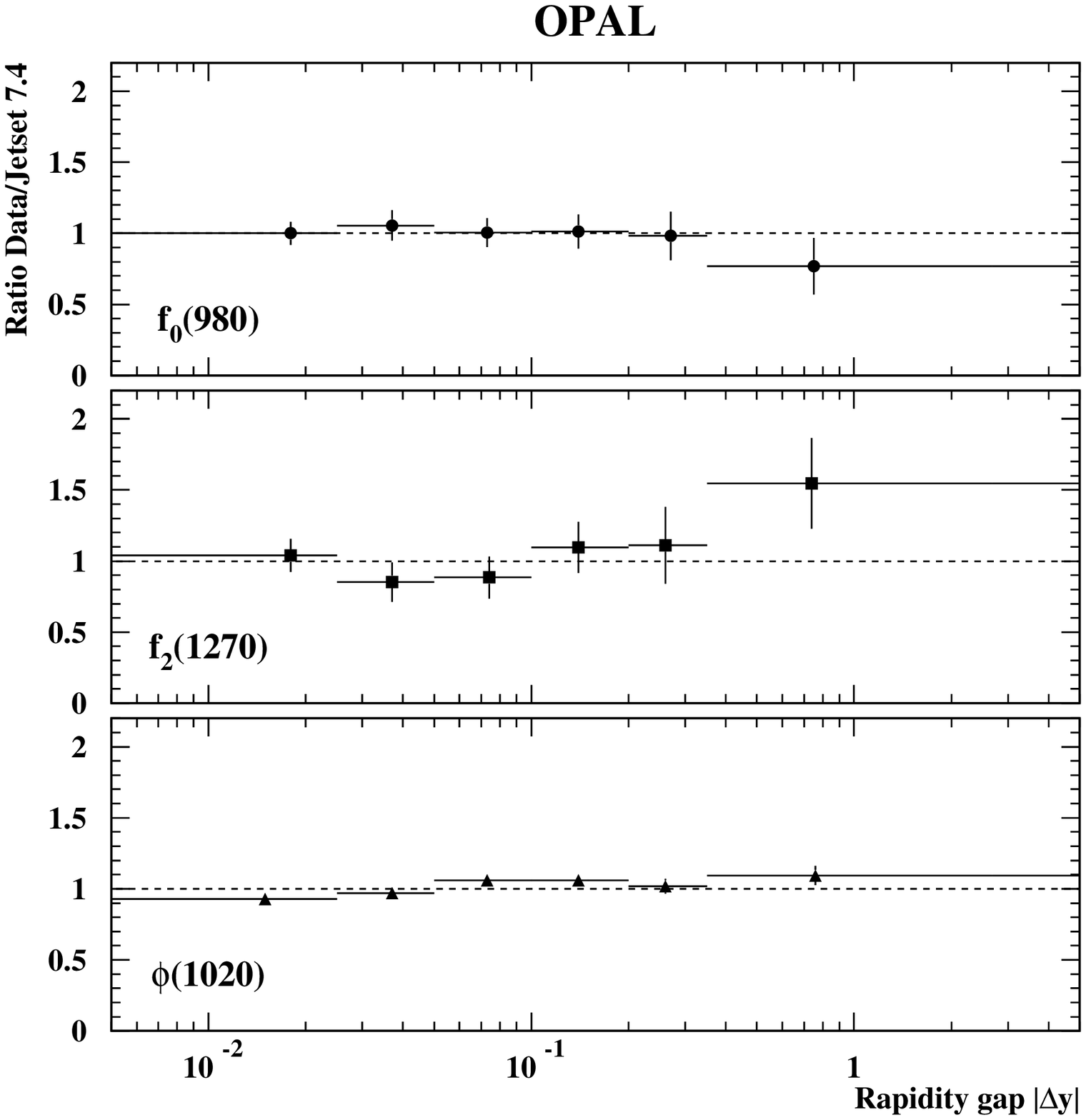,height=0.45\textwidth,width=0.48\textwidth}}
\end{tabular}\vspace{-70mm}\\
\hspace{60mm}a\hspace{78mm}b\vspace{63mm}\\
\end{center}
\caption{\label{multratios}
Production rates for $f_0(980)$, $f_2(1270)$ and $\phi$
compared to JETSET 7.4 simulations. a:  Fragmentation functions
(the $f_2(1270)$ and $\phi$ data are scaled by $\times 0.1$ and
$\times 0.01$ respectively, the Monte Carlo curves are normalised
to the data.  b: Rates for bins of the absolute value
of the rapidity difference between the meson and the nearest charged
particle \protect\cite{Ackerstaff:1998ue}.}
\end{figure}

There is a strong mass
dependence of the production rates. The three mesons \etp , $f_0(980)$
and $a_0(980)$ - which have very similar masses - have production rates
which are nearly identical (the two charge modes of the
$a_0(980)^{\pm}$ need to be taken into account). This fact was
interpreted in \cite{Ackerstaff:1998ue} as evidence that at short
distances, the three mesons have the same internal structure.
Fig.~\ref{multratios} shows the production characteristics of the
$f_0(980)$ compared to those of $f_2(1270)$ and $\phi(1020)$ mesons,
and with the Lund string model of hadronisation in which $f_0(980)$  is
treated as a conventional meson. No difference is observed in any of
these comparisons between $f_0(980)$ and f$_2(1270)$ or $\Phi(1020)$.
The production characteristics of $f_0(980)$ are fully consistent with
the hypothesis that it is a $q\bar q$ state. Further tests were made
\cite{Ackerstaff:1998ue} confirming this conclusion: the total number
of charged particles per event or selection criteria to enhance or
suppress gluon jets or quark jets led to no significant deviations
between data and Monte Carlo simulation assigning a $n\bar n$ nature to
$f_0(980)$ and $a_0(980)$ (at the moment of their creation).
\begin{table}[pt]
\caption{\label{Z0decay}
Yield of light mesons per hadronic $Z^0$ decays as a function of the
mesonic mass \cite{Ackerstaff:1998ue,Boehrer:1996pr}.
\vspace{2mm}
}
\renewcommand{\arraystretch}{1.4}
\bc
\begin{tabular}{lclclc}
\hline \hline
&&\piz    & $ 9.55\pm 0.06\pm 0.75$ & \\
&&\etg & $ 0.97\pm 0.03\pm 0.11$ & \\
\etp    & $ 0.14\pm 0.01\pm 0.02$ & $a_{0}^{\pm}(980)$& $ 0.27\pm
0.04\pm 0.10$& $f_0(980)$    & $ 0.141 \pm 0.007 \pm 0.011$ \\
&& $\phi(1020)$ & $ 0.091\pm0.002\pm0.003$\\
&&$f_2(1270)$   & $ 0.155\pm0.011\pm0.018$ \\
\hline \hline \end{tabular} \ec
\renewcommand{\arraystretch}{1.0}
\end{table}

\subsection{\label{Scalar mesons from chiral symmetry}
Scalar mesons from chiral symmetry}

Oller, Oset, and Pelaez used Chiral Perturbation Theory ($\chi$PT) to
study low-mass scalar meson-meson interactions
\cite{Oller:1997ng,Oller:1998hw,GomezNicola:2001as}. The authors
included order $p^4$ terms in the chiral expansion and  coupled
channels effects. Resonances were constructed using the Inverse
Amplitude Method \cite{Truong:1991gv,Dobado:1996ps} which avoids
conflicts with unitarity. The phase shifts of $\pi\pi$ scattering, the
$\pi\pi\rightarrow K \bar K$ transition amplitude, the phase shifts of
$\pi K$ scattering and the $\pi\eta$ mass distribution for the $a_0$
resonance were well reproduced. The authors suggest that the resonances
$a_0(980),f_0(980),\sigma(485)$, and $\kappa(700)$ form a nonet of
dynamically generated resonances, unrelated to $q\bar q$ spectroscopy
since they have put no ab-initio resonance into the formalism. Of
course, the low energy constants may not be really independent of any
resonance physics. Their conclusions are confirmed by a recent analysis
of van Beveren, Bugg, Kleefeld, and Rupp \cite{vanBeveren:2006ua} in
which generation of $\sigma(485)$, $\kappa(700)$, $a_0(980)$ and
$f_0(980)$ is a balance between attraction due to $q\bar q$ loops and
suppression of the amplitudes in the chiral limit. The immediate
question arises: are there more dynamically generated resonances and
how can they differentiated\,?

In $\chi$PT, the dependence of low energy constants on the number of
colours $N_c$ can be derived. Then masses and widths are known as
functions of $N_c$. The mass of a $q\bar q$ meson should not depend on
$N_c$: it does not matter if the $q\bar q$ pair is red, green or blue
or has additional colour options. The widths however scales with
1/$N_c$ since the newly created $q\bar q$ pair must have colour and
anticolour to match the colour of the mesonic $q\bar q$ pair at the
instant of pair creation.

\begin{figure}[pb]
\bc
\begin{minipage}[c]{0.47\textwidth}
\hspace{-4mm}\mbox{\epsfig{file=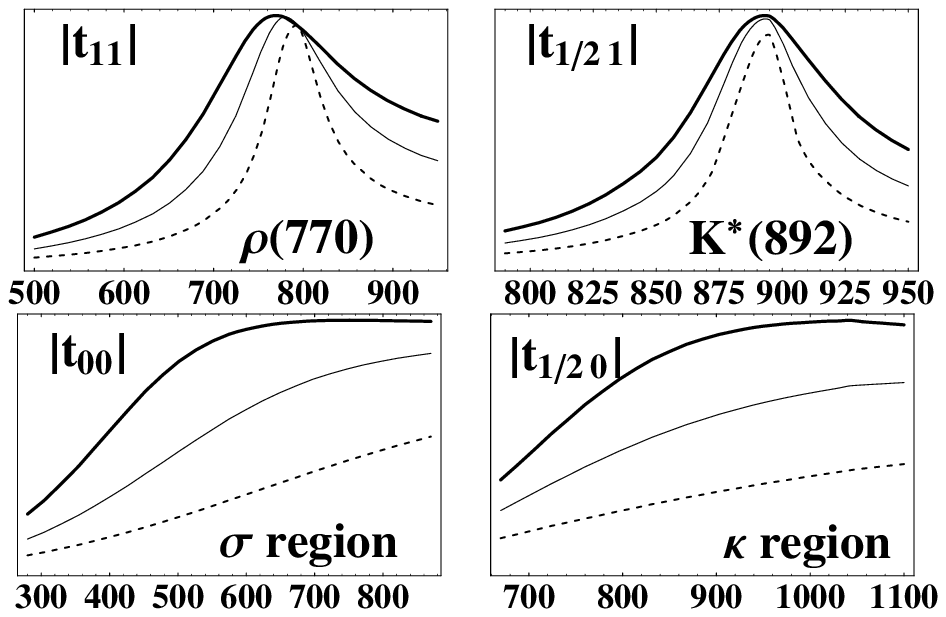,width=\textwidth}}
\end{minipage}
\begin{minipage}[c]{0.47\textwidth}
\mbox{\epsfig{file=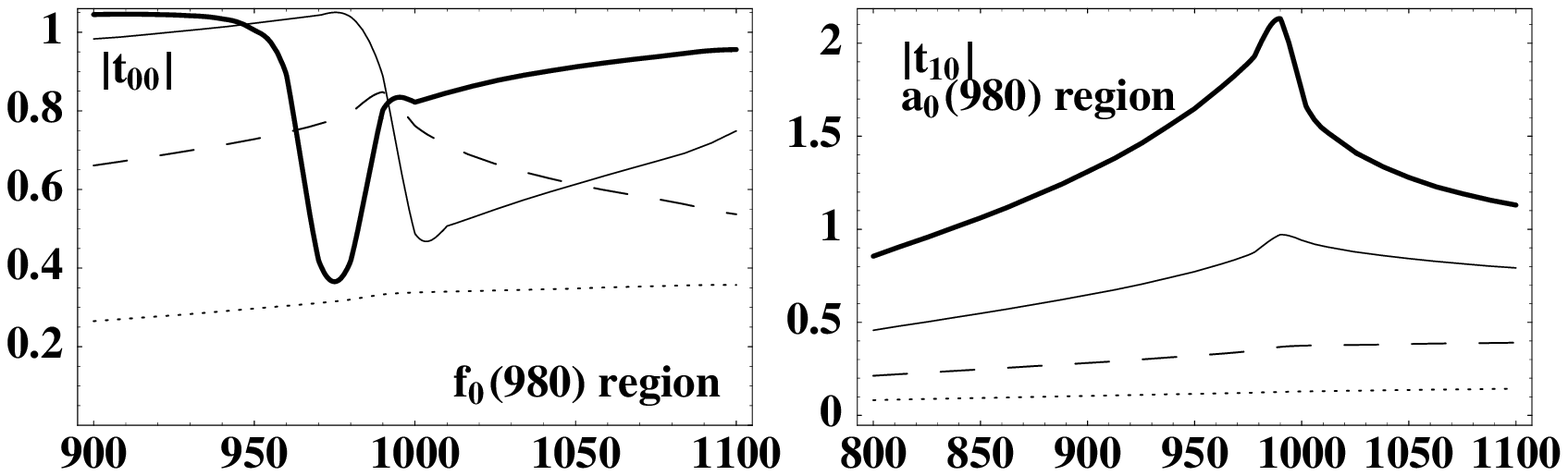,width=1.07\textwidth}}\\
\hspace{-2mm}\mbox{\epsfig{file=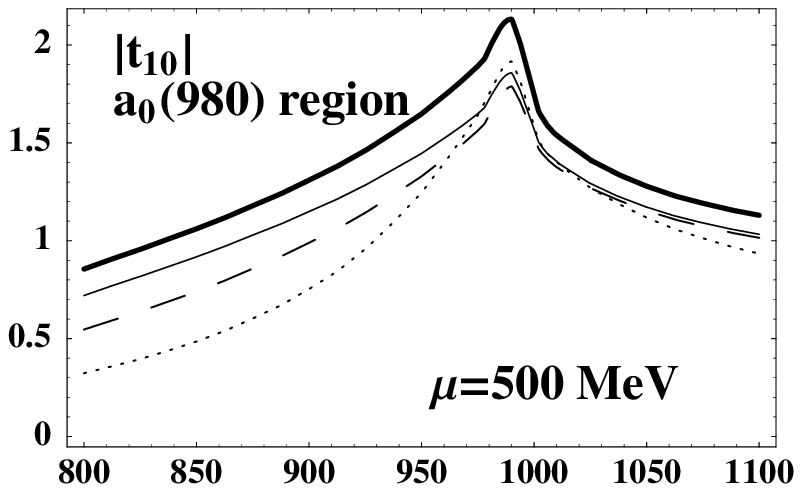,width=0.51\textwidth}}
\end{minipage}
\ec
\caption{\label{fig:nc} $N_c$ dependence of the pole positions
of light vector and scalar mesons \cite{Pelaez:2003dy}. }
\end{figure}
Fig. \ref{fig:nc} shows, for increasing $N_c$, the evolution of the
pole positions of light vector and scalar mesons
\cite{Pelaez:2003dy,Pelaez:2004xp}. The $\rho$ and $ K^*$ masses remain
constant for increasing $N_c$ while the widths become narrower. This is
the behaviour expected for `normal' $q\bar q$ mesons. In the
$\sigma(485)$ and $\kappa(700)$ region, the signal dilutes as it does
in the $f_0(980)$ and $a_0(980)$ region. In the large $N_c$ limit, the
low-mass scalar mesons are lost. This behaviour does not depend
critically on the renormalisation scale chosen in the range
$\mu=0.5-1\,$\,GeV. The behaviour is incompatible with these particles
being $q\bar q$ mesons. Only $a_0(980)$ survives for very small values
of $\mu$. Hence a  $q\bar q$ nature of $a_0(980)$ is not fully
excluded. But in this interpretation, all four states,  $\sigma(485)$,
$\kappa(700)$, $f_0(980)$, and $a_0(980)$, are dynamically generated
poles probably unrelated to the physics of $q\bar q$ states. The $N_c$
dependence of light mesons was also studied by Uehara
\cite{Uehara:2004es}; in agreement with Pelaez it was found that $\rho$
and $ K^*$ survive for large $N_c$ as narrow width resonances while the
scalar mesons below 1\,GeV fade away already when $N_c$ exceeds 6.

\subsection{\label{A low-mass nonet of scalar mesons}
A low-mass nonet of scalar mesons}

There is now solid evidence that there are nine low-mass scalar
mesons. These are $\delta, S^*$, $\sigma(485)$, $\kappa(700)$ or
$a_0(980)$, $f_0(980)$, $f_0(485)$, $ K^*_0(700)$. Likely, they form a
SU(3) nonet. There is, however, still ample room for different
interpretations.

Hanhart, Kalashnikova, Kudryavtsev, and Nefediev \cite{Hanhart:2007wa}
and C. Hanhart, private communication, argue that the scalar nonet,
in particular $f_0(980)$, must be interpreted as $K\bar K$ molecule. If
the pole position of a state is much closer to a threshold than any
intrinsic scale of the problem, one can disentangle the molecular
component from the compact one. The idea is based on analyticity: if
the pole is predominantly generated from the non-analytic pieces that
emerge from the loops of the constituents, then it deserves the name
molecule. If the pole is predominantly generated from quark dynamics,
it can only lead to contributions analytic in $s$, and it should be
called a compact state.

Achasov \cite{Achasov:2003cn} has a different view: he sees the
flow-lying scalar mesons as compact tetraquark states as they result
for instance from the MIT bag model \cite{Chodos:1974je,DeGrand:1975cf}.
The compact object decays virtually into a $K\bar K$ loop which is
responsible for the decay dynamic.

Beveren and collaborators
argue that the strong coupling of $q\bar q$ states to the
meson-meson continuum gives rise to a doubling of states but that it
makes little sense to talk about intrinsic resonances and dynamically
generated poles \cite{vanBeveren:2002mc}. The $\sigma(485)$ and
$\kappa(700)$ still have $q\bar q$ seeds and the $\sigma(485)$ still is
the chiral partner of the pion. The $D_{s0}^*(2317)$ is linked to the
$\kappa(700)$ while the next $c\bar s$ quark model states are predicted
to have a pole position at $M -i\frac12\Gamma = 2800-i200$\,MeV
\cite{vanBeveren:2004bz}, very far off the Godfrey-Isgur prediction
\cite{Godfrey:1985xj}. Obviously it is wrong to consider the quark
model states as independent of the interaction of their decay products;
coupled channel effects and unitarity have a significant impact on the
properties of quark model states even though the seed of these states
may still be of $q\bar q$ nature. It is suggested that the coupling
between $q\bar q$ and meson-meson systems leads to a doubling of states.

Tornqvist \cite{Tornqvist:1995kr}, imposing the Adler zero and
unitarity, identified low-mass scalars as $q\bar q$ objects. He found
`that in particular for the $a_0(980)$ and $f_0(980)$ the $ K\bar K$
component in the wave function is large, i.e., for a large fraction of
the time the $q\bar q$ state is transformed into a virtual $ K\bar
K$ pair. This  $ K\bar K$ component, together with a similar
component of $\eta'\pi$ for the $a_0(980)$, and $\eta\eta$, $\eta\eta'$
and $\eta'\eta'$ components for the $f_0(980)$, causes the substantial
shift to a lower mass than what is naively expected from the $q\bar q$
component alone'. His view finds support in an analysis of $D_s\to
3\pi$ decay via $f_0(980)$ \cite{Deandrea:2000yc}. The $f_0(980)$
resonance is interpreted as $s\bar s$ state with significant $K\bar K$
component. $D_s\to 3\pi$ decays via $f_0(980)$ couple to its $s\bar s$
component.  The methods are tested using the more frequent
process $D_s \to\phi\pi$. In a recent paper by van Beveren, Bugg,
Kleefeld, and Rupp the $\sigma(485)$, $\kappa(700)$, $a_0(980)$ and
$f_0(980)$ nonet is interpreted as balance between attraction due to
$q\bar q$ loops and the suppression of the amplitudes at the Adler
zeros \cite{vanBeveren:2006ua}.

Boglione and Pennington used a toy model to study this phenomenon
\cite{Boglione:2002vv}. Starting with just one bare seed for each
member of a scalar nonet, they generated two isovector states which
they identified with the $a_0(980)$ and the $a_0(1450)$. In the I=1/2
sector, they generated states with masses above 1\, GeV but not the
$\kappa(700)$. The isoscalar sector proved to be the most complicated
one and the outcome was very strongly model dependent but the
occurrence of numerous states was achieved.
Beveren, Pennington and Tornqvist, interpret the scalar mesons as being
generated by both $q\bar q$ mesons and molecular forces. The interplay
of direct QCD forces and molecular forces is suggested to generate two
nonets but all mesons are supposed to have a similar structure: they
are $q\bar q$ mesons with properties which are strongly influenced by
their coupling to the continuum. If this view holds true, we should
expect that scalar mesons do not behave like $1^3P_2$ $q\bar q$ mesons
do; the scalar mesons may be more complicated than their tensor
brothers. Both scenarios agree in the number of states: when the number
of scalar states is counted -- to answer the question if a scalar
glueball hides in the spectrum -- the low-lying scalar meson nonet is
supernumerous. However, there is certainly a strong model dependence in
such a statement.

QCD sum rules were exploited by several authors to shed light onto the
scalar mesons. Sum rules are a powerful technique to study hadrons.
However, there is no direct correspondence between the constituent
quark picture and the operator (current) quark picture. Diquarks
are thus not a meaningful concept is in sum rule approaches
\cite{Zhang:2006xp}, and care must be taken not to overinterpret the
results. Brito, Navarra, Nielsen, and Bracco found that the mesons
decay constants and hadronic couplings are consistent with a tetraquark
structure for the light scalar mesons \cite{Brito:2004tv}. Chen, Hosaka
and Zhu \cite{Chen:2006zh} determine the mass spectra for tetraquark
and $q\bar q$ scalar mesons, and find the tetraquark states below, the
$q\bar q$ mesons above 1\,GeV.  Matheus, Navarra, Nielsen and Rodrigues
da Silva use QCD sum rules to argue against the tetraquark of the light
scalar nonet \cite{Matheus:2007ta}. Obviously, this question is not yet
settled.

The low masses of $\sigma(485)$, $\kappa(700)$, $a_0(980)$ and
$f_0(980)$ -- compared to any quark model prediction -- find a
well-founded explanation by Jaffe's tetraquark picture
\cite{Jaffe:1976ig}. These tetraquark states are not compact objects
like those of Achasov. The difference can be understood discussing a
toy model to illustrate the relation between a `molecular' state and a
state bound by chromodynamic forces, see \cite{Jaffe:2007id} and Jaffe,
private communication. In the toy model, the four quarks are bound very
weakly inside of a strong interaction volume; outside, the $K\bar K$
system is supposed not to interact. If the colour forces are strong
enough to bind the system, its small binding energy forces the system
to become large. Thus $f_0(980)$ does not need to be a compact object
but four quark correlations are still a decisive element. The
$f_0(980)$ is bound by mesonic forces. But the forces are chromodynamic
in origin and do not have a simple representation in terms of
meson-exchange forces between the nucleons.

The 1/$N_c$ expansion identifies the low-mass scalar mesons as
non-$q\bar q$ objects. In the 1/$N_c$ expansion, scalar mesons
certainly behave differently than vector mesons do. This does not imply
that the low-mass scalar mesons have no $q\bar q$ component. Scalar
mesons have a strong coupling to their $S$-wave decays, and
this may already be sufficient to dilute their existence in the large
$N_c$ limit. At present, it is not clear if any scalar meson survives
the 1/$N_c$ test as a $q\bar q$ state.

It would be very illustrative to try to differentiate the 1/$N_c$
behaviour of the $f_1(1420)$ and $f_1(1510)$. Both mesons decay into
$ K^*K$ in $S$-wave. Based on the Gell-Mann-Okubo mass formula and SU(3)
symmetry of meson decays, Gavillet {\it et al.} \cite{Gavillet:1982tv}
concluded that $a_1(1260), f_1(1510), f_1(1285), K_{1A}$ and
$b_1(1260), h_1(1170), h_1(1380),  K_{1B}$ form two nearly ideally
mixed nonets, with $ K_{1A}, K_{1B}$ mixing to form the observed $
K_1(1280)$ and $ K_1(1400)$. The $f_1(1420)$ was interpreted by
Longacre \cite{Longacre:1990uc} as molecular fluctuation between a
pion, orbiting in $P$-wave around a scalar $s\bar s$ core, and a $ K^*K$
in $S$-wave. Based on their  glueball-$q\bar q$ filter (discussed in
section \ref{Scalar resonances from central production}), Close and
Kirk \cite{Close:1997nm} interpreted the $f_1(1420)$ as SU(3) partner
to the $f_1(1285)$ and determined a flavour singlet-octet mixing angle
of approximately 50 degrees. The latter interpretation is adopted by
the Particle Data Group \cite{Eidelman:2004wy}.

There are thus still many dissenting views of the low lying scalar
mesons. Different interpretations of the low-mass region often entails
different interpretations of the mass region above 1\,GeV. One has to
accept that at present, some questions do not find a unique answer: the
$q\bar q$, the tetraquark, and the molecular picture all give e.g.
reasonable descriptions of the process $\phi\to\gamma\pi\pi$ and
$\to\gamma\pi\eta$. Even though we assign $\sigma(485), \kappa(700),
a_0(980)$ and $f_0(980)$ to one nonet, we will, in section \ref{Scalar
mesons and their interpretation} on `Scalar mesons and their
interpretation', present different views as well. We will discuss the
mesons below 1\,GeV as originating from meson-meson dynamics, with or
without $q\bar q$ and $q\bar qq\bar q$ fractions being mixed in. On the
other hand, 1\,GeV exceeds perhaps the range of applicability of chiral
symmetry, and we will also discuss views of scalar mesons in which
$a_0(980)$ and $f_0(980)$ are interpreted as members of a regular nonet
of $q\bar q$ mesons and in which only $\sigma(485)$ and $\kappa(700)$
are generated dynamically (even though we do not share this view). The
$\sigma(485)$ is certainly better established than the $\kappa(700)$;
thus interpretations accepting the $\sigma(485)$ and ignoring the
$\kappa(700)$ will also not be discarded in the discussion. Last not
least, the full nonet of light scalar mesons may have a $q\bar q$ seed
with properties strongly influenced by their couplings to two
pseudoscalar mesons as proposed in section \ref{Dynamical generation of
resonances and flavour exotics}.

\markboth{\sl Meson spectroscopy} {\sl Scalar mesons above 1 GeV}
\clearpage\setcounter{equation}{0}\section{\label{Scalar mesons above 1 GeV}
Scalar mesons above 1\,GeV}

In this section data will be presented and discussed which contribute
to the spectrum of scalar mesons above 1\,GeV. The main focus will be
on the existence and the properties of isoscalar scalar candidates
which have been suggested, $f_0(1370)$, $f_0(1200-1600)$, $f_0(1500)$,
$f_0(1670)$, $f_0(1710)$, $f_0(1790)$, $f_0(1810)$, $f_0(2000)$,
$f_0(2100)$. This is an unexpected large number; the quark model
expects 4 (possibly 6) states, QCD predicts one scalar glueball in this
mass range. Of course, there might be $qq\bar q\bar q$ and hybrid
states in addition. On the other hand, it is also not unlikely that
some of these states do not survive critical discussions of their
viability. In a first scan, we present results from individual
experiments; at the end of this section results from combined fits to
several data sets will be presented. This allows us to define the
impact of individual experiments and to emphasize the virtue of getting
a consistent picture of a large number of meson resonances from fits to
many reactions.

\subsection{\label{Charge exchange}
Charge exchange}

\begin{figure}[pb]
\bc
\includegraphics[width=0.4\textwidth]{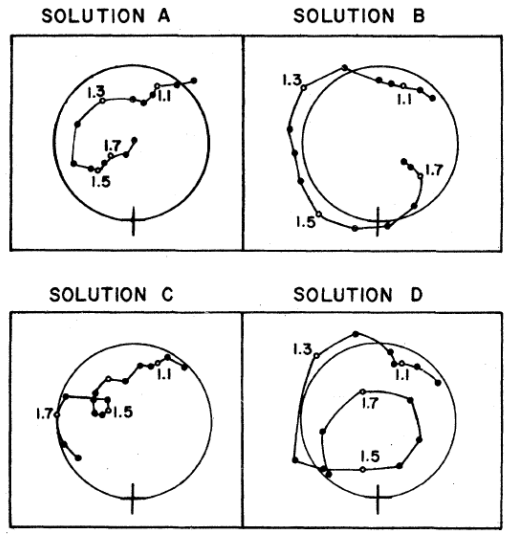}
\ec
\caption{\label{phases-4}
The scalar isoscalar phase from CERN-Munich data on $\pi^- p\to
n(\pi^+\pi^-)_{S-{\rm wave}}$ for four different solutions
\cite{Estabrooks:1974vu,Estabrooks:1975cy}. A $K$-matrix fit
to the yields consistent parameters for $f_0(980)$ and one second
resonance at a mass between 1.5 and 1.6\,GeV
 \protect\cite{Estabrooks:1978de}.  }
 \end{figure}

The CERN-Munich experiment (see section \ref{The
CERN-Munich experiment}) on the charge exchange reaction $ \pi^-p\to
n\pi^+\pi^-$ covered the meson mass range from 600 to 1900\,MeV. The
data are still used to constrain modern partial wave analyses. General
features of apparatus and of the data were presented in
\cite{Grayer:1974cr}, production mechanism and t dependence studied in
\cite{Hyams:1974wr}. The results of energy-dependent and
energy-independent partial wave analyses on the $\pi^+\pi^-$ system
can be found in \cite{Hyams:1973zf,Hyams:1975mc}. Isotensor $\pi\pi$
interactions were studied in \cite{Hoogland:1974cv,Hoogland:1977kt}.
Later, the H$_2$ target was replaced by a polarised target providing
for `model-independent analyses' of the $\pi^+\pi^-$
\cite{Becker:1978ks} and $ K^+K^-$ systems \cite{Gorlich:1979fn}.
Partial wave analyses of the CERN-Munich data were also performed by
Estabrooks and Martin
\cite{Estabrooks:1972tw,Estabrooks:1974vu,Estabrooks:1975cy} and Martin
and Pennington \cite{Martin:1977ff}.  The scalar isoscalar phase motion
above 1\,GeV of the CERN-Munich data is presented in
Fig.~\ref{phases-4} for the four different CERN-Munich solutions.
In all four cases, the phase motion exhibits a resonance-like behaviour
in the 1500\,MeV region, and a hint for a further phase advance above
1.7\,GeV. Estabrooks fitted the data (plus the data below 1\,GeV and
the $ \pi\pi\to  K\bar K$ data) using a two-channel $K$-matrix
\cite{Estabrooks:1978de}. Two resonances and a slowly varying
background amplitude were sufficient to get a good fit. The
$f_0(980)$ mass and width were consistent in all four fits, the
$f_0(1500)$ -- as it is called now -- came out with mass and width
between 1.5 and 1.6\,GeV and 140 and 320\,MeV, respectively.

A recent energy dependent fit to the CERN-Munich data was performed by
L.~Li, B.~S.~Zou, and G.~l.~Li  \cite{Li:2000jq}. Ambiguities were not
discussed, but constraints from knowledge of other experiments was
built into the analysis. The authors used $t$-channel $\rho$ exchange to
fit the $\pi\pi$ isotensor interaction and fixed the $\rho$ exchange
contribution to the isoscalar interaction by the relation
$T^{Born}(I=2) = -\frac12 T^{Born}(I=0)$. This part of the amplitude
gave rise to a $\sigma(485)$ pole at $(0.36-0.53i)$ GeV. Two
relatively narrow resonances were found, $f_0(980)$ and $f_0(1500)$, and
a broad one with a pole at $(1.67-0.26i)$\,GeV. The $f_0(1370)$
and $f_0(1710)$ mesons were not observed.

\begin{figure}[ph]
\bc
\epsfig{file=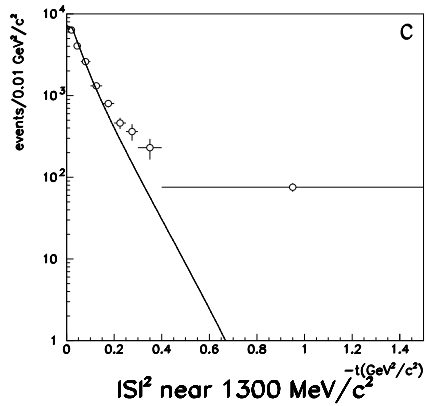,width=0.35\textwidth,height=6cm}
\ec
\caption{\label{fig:piex1}The $S$-wave intensity as a function of $|t|$ at
$1.30\,GeV/c^2$ compared with one-pion exchange. At small $|t|$ data
are well described. At higher $|t|$, additional exchange currents like
$a_1(1260)$ exchange start to contribute~\protect\cite{Gunter:2000am}.
A turnover expected at low $t$ was not observed.}
\end{figure}

The exchange character of the interaction can be
seen in Fig.~\ref{fig:piex1} where the $S$-wave intensity is shown as a
function of the squared momentum transfer $|t|$. The data stem from a
charged exchange experiment (E852) at BNL using a $18.3$\,GeV/c pion
beam~\cite{Gunter:2000am}. A short description of the detector was
given in section \ref{Experiment E852 at BNL}. For large momentum
transfers to the proton, not only pions but also heavier mesons, in
particular $a_1(1260)$, are exchanged.  The quantum numbers of the
exchange particles are restricted by conservation laws. Of course, when
a pion is exchanged, the $\pi(1300)$ can also be exchanged.
Generalising the concept, the exchange of $q\bar q$ pairs is referred to
as Reggeon exchange.

\begin{figure}[t!]
\begin{center}
\begin{tabular}{c}
\vspace*{-2mm}
\epsfig{file=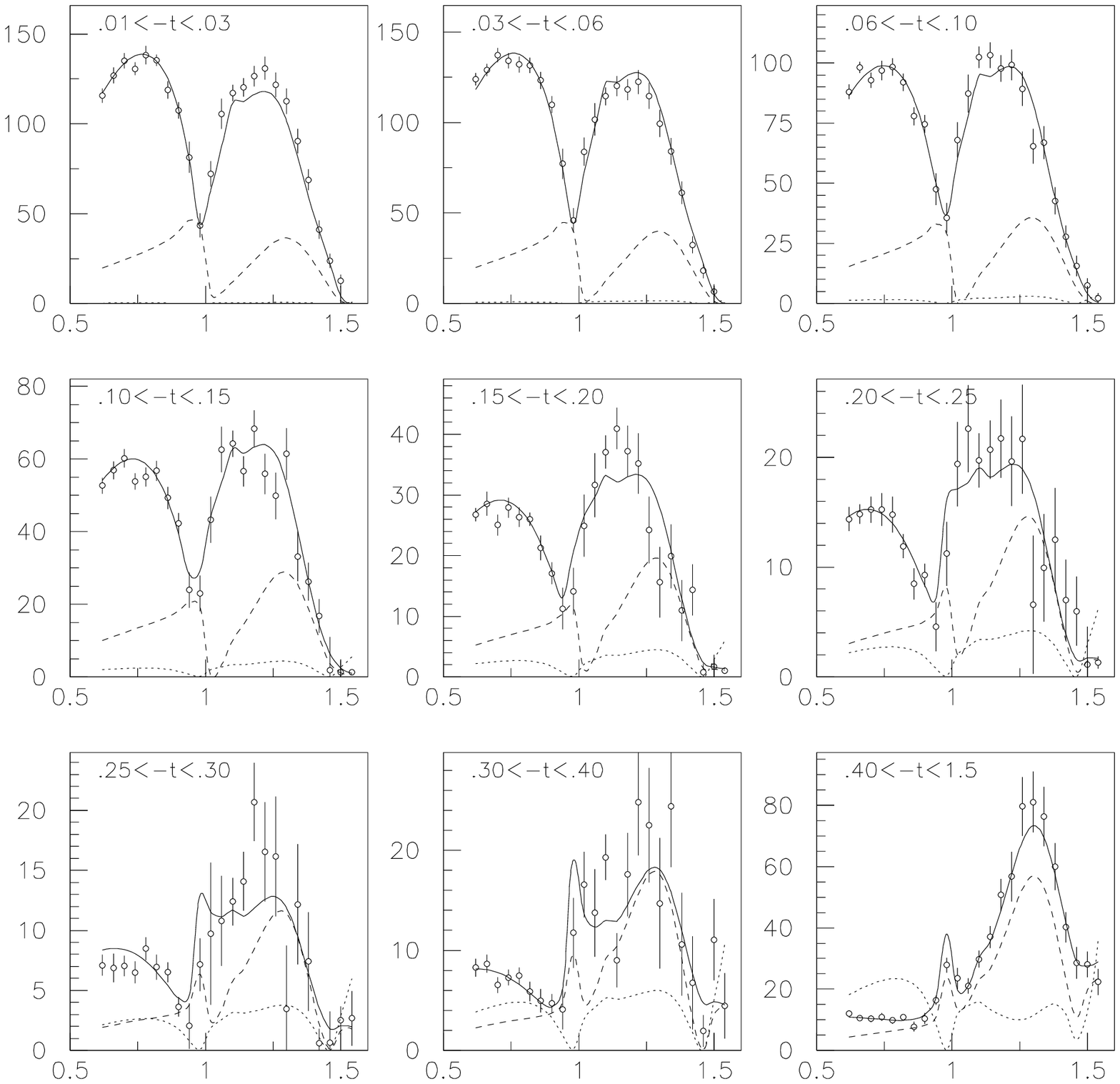,width=0.9\textwidth,height=10cm}\vspace{2mm}\\
\phantom{z}\hfill{\small M [GeV/c$^2$]} \hspace{6cm}\\
\vspace*{-8mm}
\end{tabular}
\caption{\label{fig:piex}The reaction $ \pi^- p \to p
\pi^0\pi^0$ $18.3$\,GeV/c for different $t$-intervals
\protect\cite{Gunter:2000am}. The dashed curve shows the contribution
of the $a_1$-trajectory, the dotted curve of the pion
trajectory~\protect\cite{Anisovich:2002ij}. }
\end{center}
\end{figure}

The $S$-wave mass distributions for different intervals of $t$ are
shown in Fig. \ref{fig:piex} together with a fit by Anisovich and
Sarantsev \cite{Anisovich:2002ij}. The contribution of $a_1(1260)$
deserves some comments. Apart from $\pi$ and $a_1(1260)$ exchanges,
there is the possibility of significant $\pi(1370)$ or $a_2(1320)$
exchange contributions. The $S, P_0, P_+, P_-$ amplitudes and their
relative phases depend on the nucleon helicities \cite{Ochs:1972mc}.
Without using a polarised target, the partial waves $P_0$ and $P_+$
cannot be deduced from the moments without further assumptions. At low
$q^2$ and for pion exchange, the phase between $P_0$ and $P_+$ can be
shown to vanish; the $\pi\pi$ phase is then the same for spin-flip and
spin-non-flip at the $\pi  N$ vertex. This is expected in the
chiral limit and holds true for the reconstructed amplitudes. Under
these conditions, the rank of the spin-density matrix is one, and all
amplitudes can be fixed (apart from the $2^n$ fold ambiguities when the
real part of the amplitudes has $n$ zeros). For large $t$ and complex
exchange currents the hypothesis does not need to be fulfilled, and
the bump in the $\pi\pi$ mass distribution could receive large
contributions from a feedthrough of $f_2(1270)$ production. The warning
is important since the data are used to argue that the existence of
$f_0(1370)$ follows unambiguously from this data (and similar data on
$ \pi\pi\to K\bar K$).

\begin{figure}[t!]
\begin{center}
\begin{tabular}{ccccc}
\epsfig{file=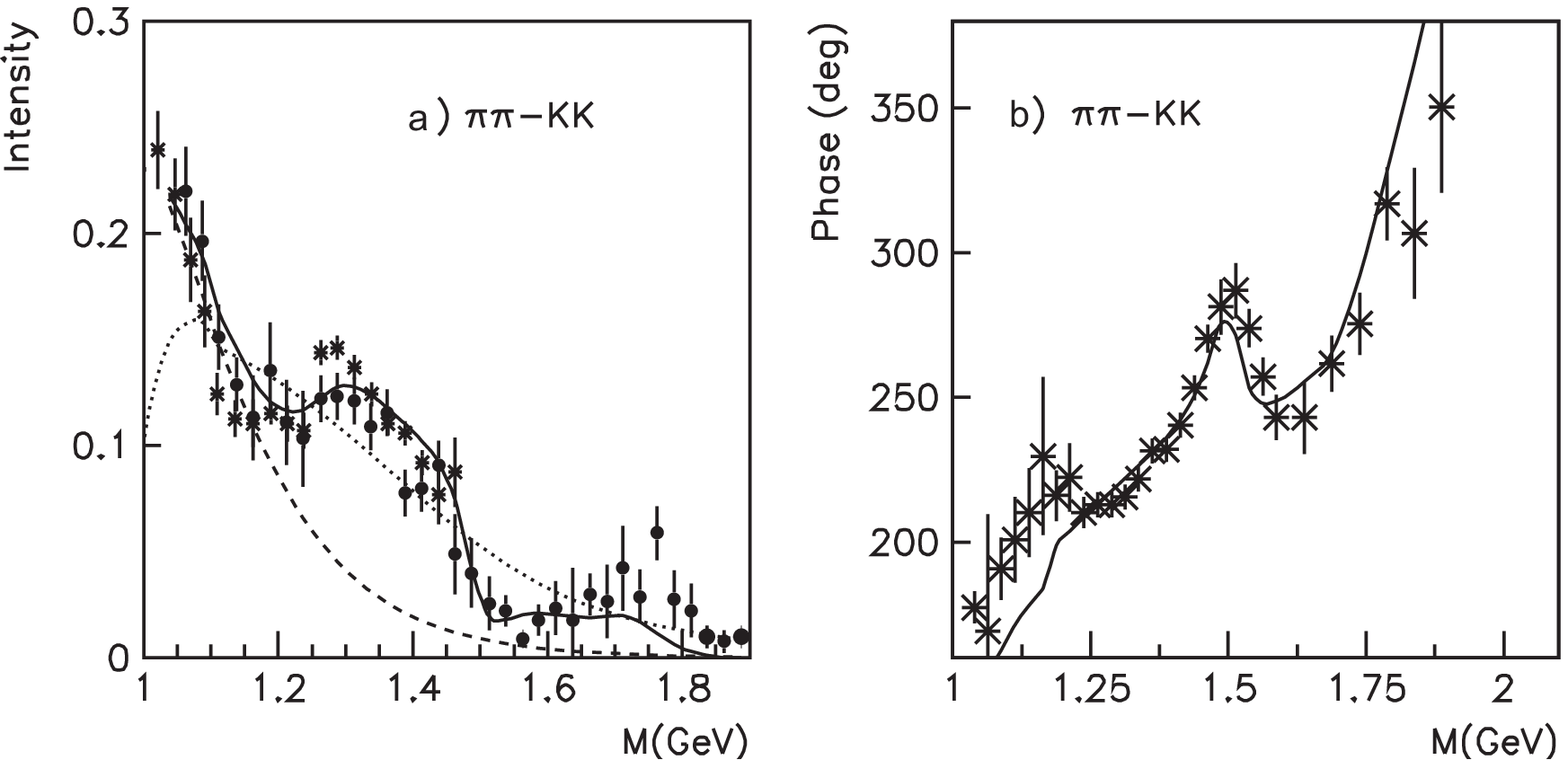,width=0.38\textwidth,height=5cm}&
\epsfig{file=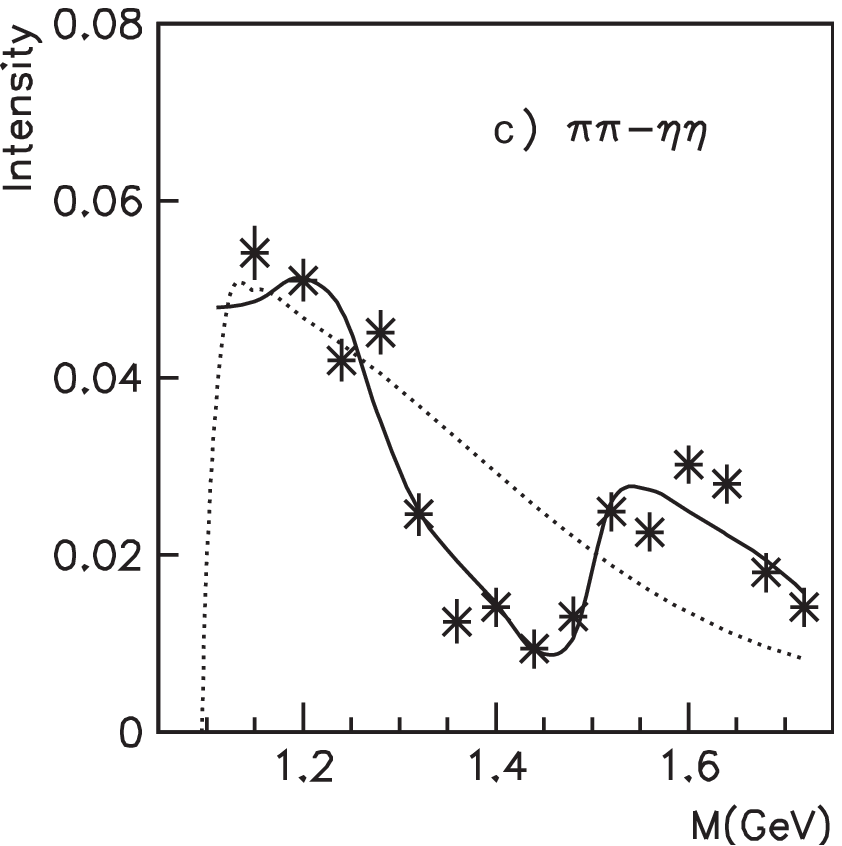,width=0.18\textwidth,height=5cm}&
\epsfig{file=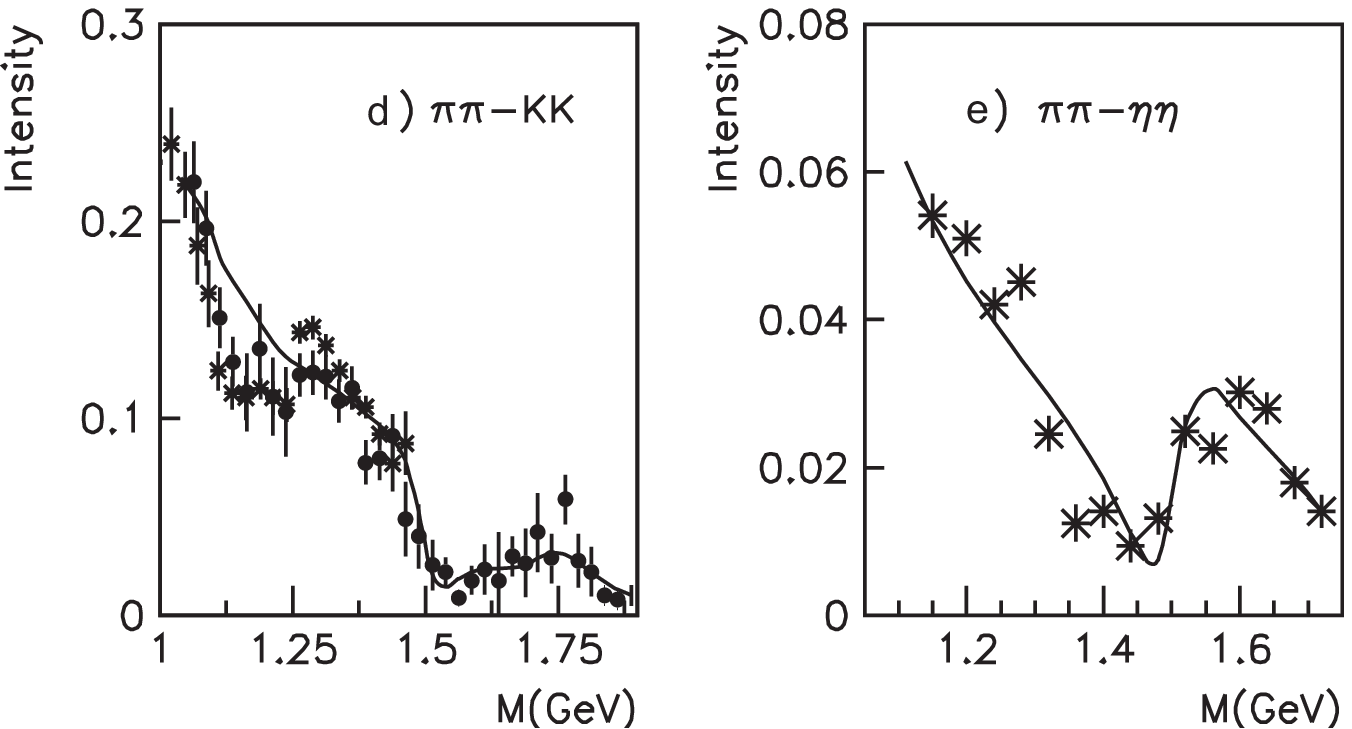,width=0.36\textwidth,height=5cm}
\end{tabular}
\end{center}
\caption{\label{fig:f1370-ev}.
Scalar partial wave produced in $\pi p$-collisions at small
momentum transfers $t$. a + d)  $ \pi \pi \to K\bar K$, crosses show
data from Martin and Ozmutlu, circles those of Lindenbaum and Longacre
\protect\cite{Martin:1979gm,Lindenbaum:1991tq}. b) phases of Etkin
{\textit et al.} \protect\cite{Etkin:1981sg}. c + e) $\pi \pi \to
\eta\eta$, data from Binon et al. \protect\cite{Binon:1983ny}. The
solid lines in a,b,c) show the best fit by Bugg \cite{Bugg:2006sr}
which includes $f_0(1370)$, in d,e) the $f_0(1370)$ resonance is
omitted from the fit. The dashed curve shows the intensity of
$f_0(980)$ and the dotted curves that from $\sigma(485)$. } \end{figure}

Above the $ K\bar K$ threshold, inelastic reactions set in. As seen
in Fig. \ref{fig:gams-scalar}, the $\pi\pi$
S-wave intensity exhibits dips at the $f_0(980)$ and $f_0(1500)$
masses. In the mass range up to 1500\,MeV, open channels are
$\pi\pi\to  K\bar K, \eta\eta$, and $4\pi$. Fig. \ref{fig:f1370-ev}
shows data on the former two channels. The $ K^0_SK^0_S$ mass
distribution falls off from a maximum value due to $f_0(980)$
formation, show a dip-bump structure at and decreases sharply at
1500\,MeV. The $\eta\eta$ mass distribution exhibits a striking dip at
this mass. The data were recently fitted by Bugg \cite{Bugg:2006sr},
with and without introduction of an amplitude for $f_0(1370)$. The $
K^0_SK^0_S$ mass distribution is better described when $f_0(1370)$ is
introduced even though the phase motion is still problematic for
masses below 1.25\,GeV. The discussion if a resonance $f_0(1370)$
exists or not will be resumed in section \ref{The f0(1370) and
f0(1500) resonances}.

\subsection{\label{Scalar resonances from central production}
Scalar resonances from central production}

Double Pomeron Exchange  is one of the reactions which are supposed to
reveal the existence of glueballs. Double Pomeron Exchange ($P P \to
M_X$) is expected to contribute a large fraction to central production
even though the systematic of meson production provides only limited
support for this conjecture. The technique of central production
experiments and a survey of results was presented in section
\ref{Central production}. Pseudoscalar, axial vector and
$J^{PC}=2^{-+}$ mesons seem to be produced mainly via Reggeon-Reggeon
fusion ($R R \to M_X$). For tensor mesons, the available data are not
conclusive while production of scalar mesons follows closely the
expectation from Double Pomeron Exchange: in central production at $\sqrt
s=29.1$\,GeV, the $a_0(980)/f_0(980)$ ratio is $\sim 1/10$, $f_0(1500)$
production is observed, $a_0(1470)$ production not, and the frequency of
scalar mesons as a function of the four-momentum transfer is compatible
with the expected $e^{bt}$ distribution. In this section we restrict
the discussion to scalar mesons and assume that the dominant production
process is Pomeron-Pomeron fusion.

\begin{figure}[b!]
\begin{minipage}[c]{0.49\textwidth}
\epsfig{figure=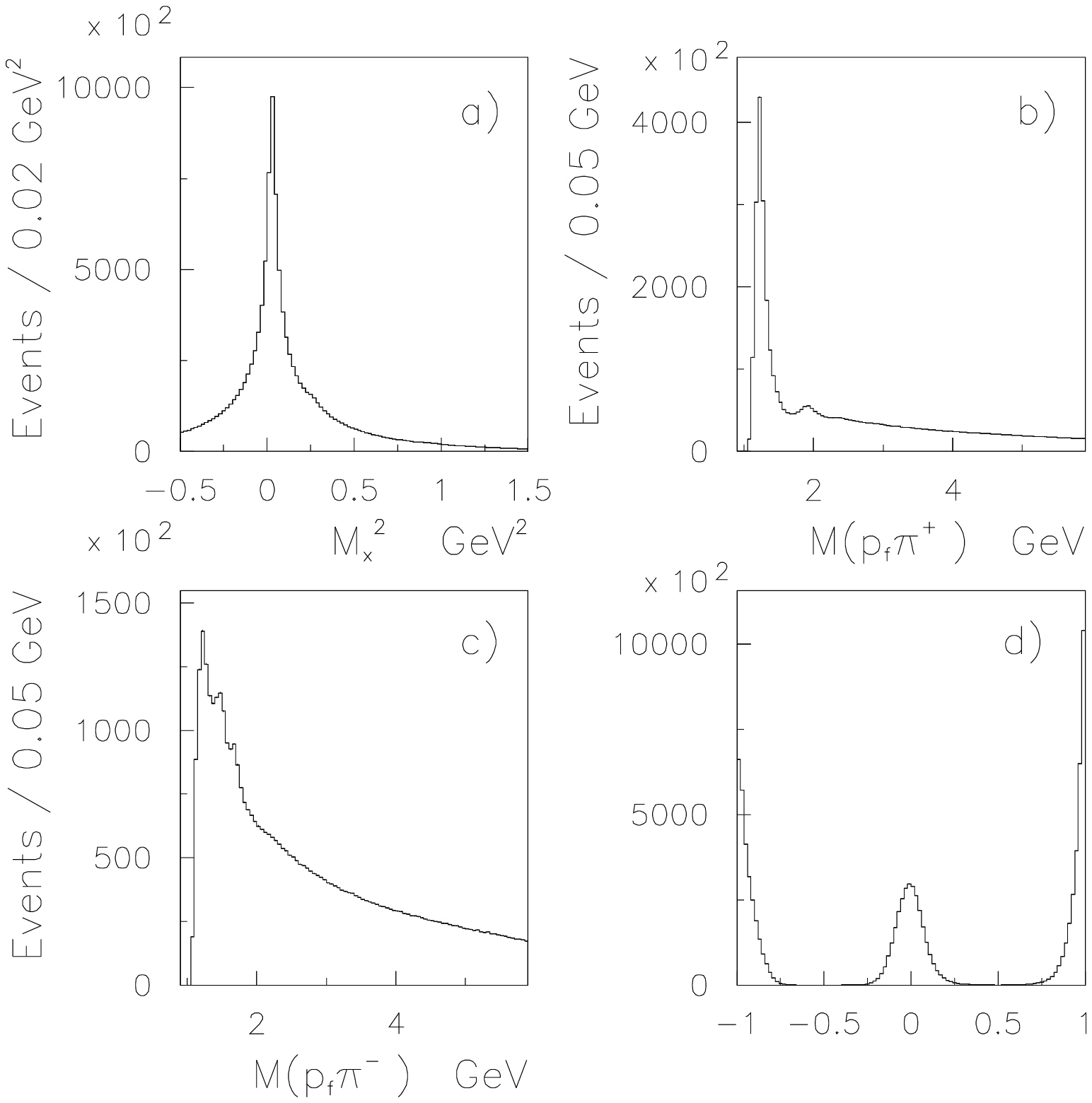,width=7.5cm}
\end{minipage}
\begin{minipage}[c]{0.49\textwidth}
\hspace{-6mm}\epsfig{figure=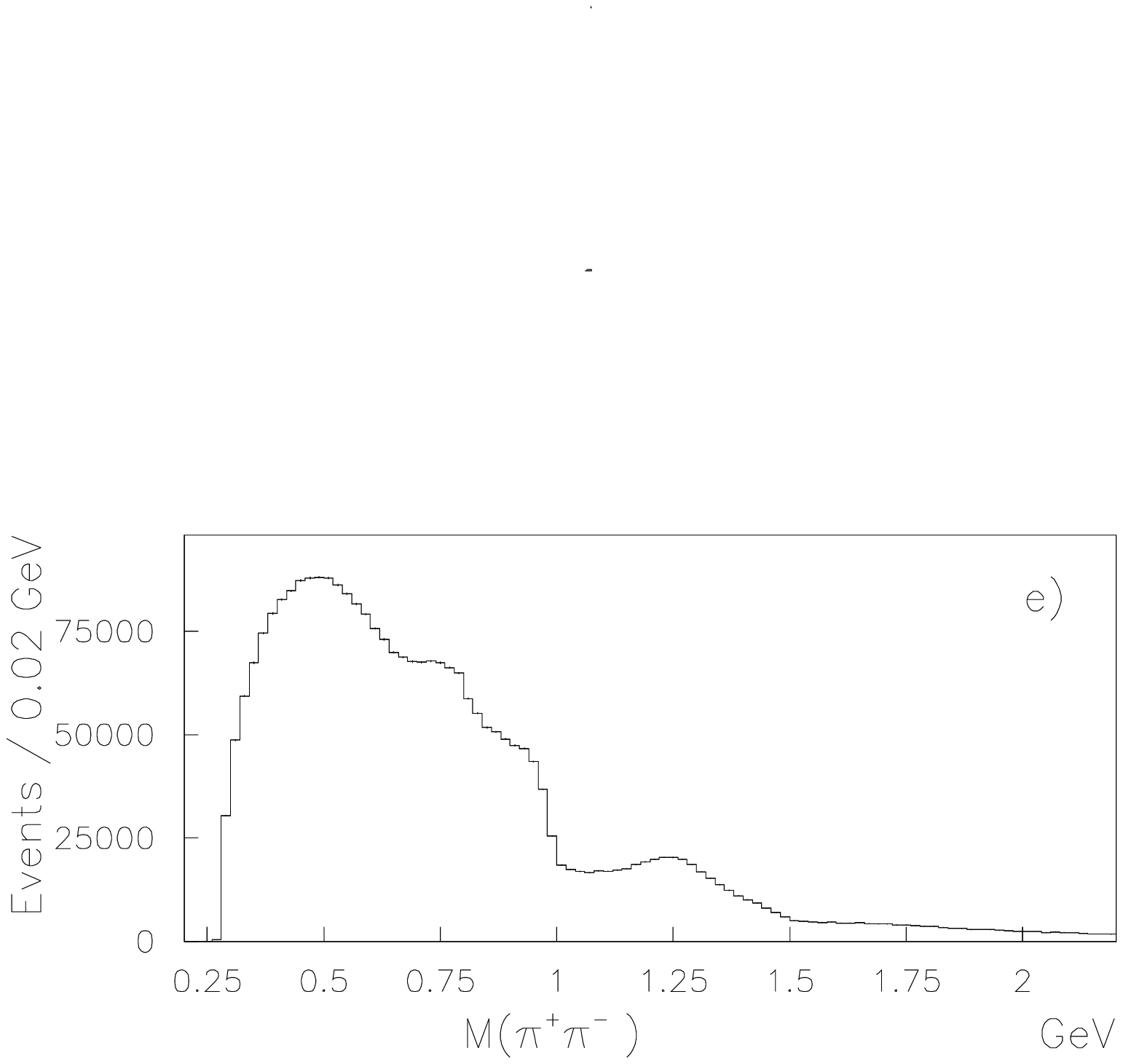,width=8.5cm,height=3.4cm}
\vspace*{2mm} \\
\includegraphics[width=8.5cm,height=3.5cm]{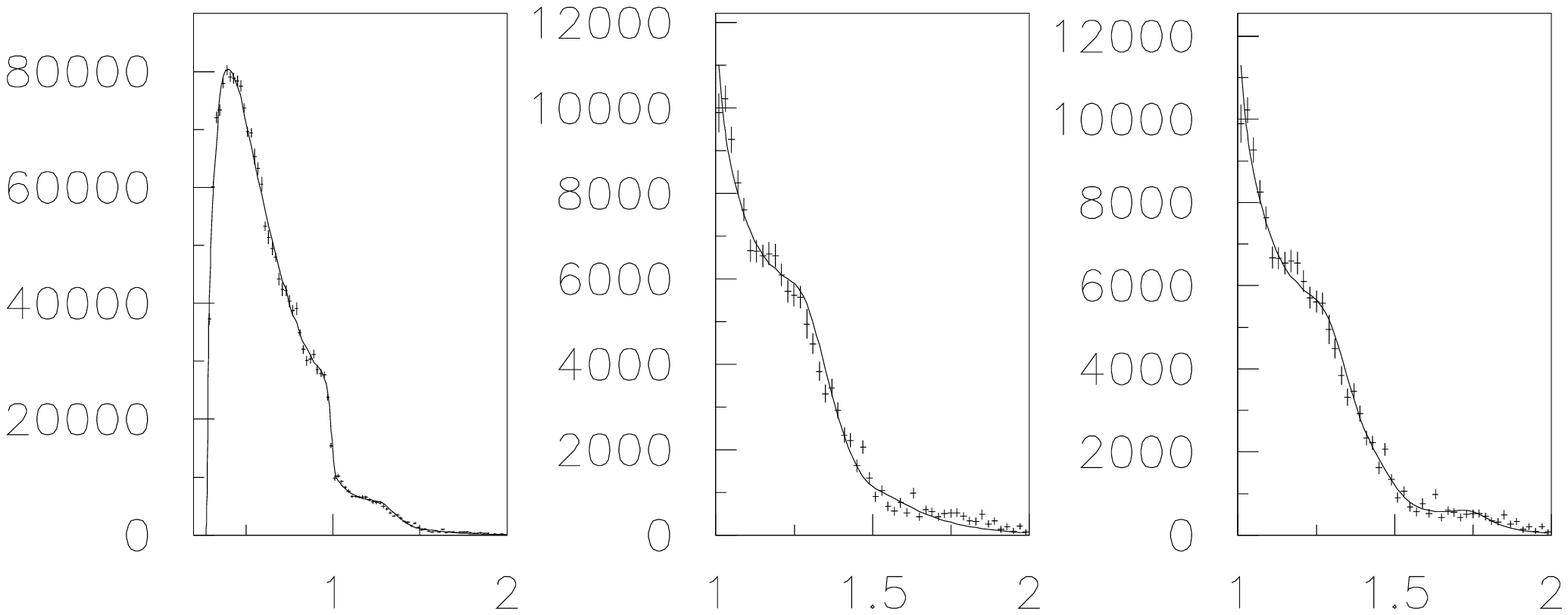}

\vspace{-38mm}
\phantom{rrr}\hspace{24mm}\scriptsize{f)}\hspace{25mm}\scriptsize{g)}\hspace{25mm}\scriptsize{h)} \\
\vspace{25mm}
\end{minipage}
\vspace{-2mm}

\caption{\label{fig:wapm} Extraction of centrally
produced $\pi^+\pi^-$ pairs. a) The Ehrlich mass squared distribution,
b) the $M(p_f\pi^+)$ and c) the $M(p_f\pi^-)$ mass spectra.
d) The $x_F$ distribution for the slow and the fast particle,
and for the $\pi\pi$ system.
e) The centrally produced $\pi\pi$ invariant mass distribution.
The physical solution from the PWA of the $\pi\pi$ final state.
Full mass range (f); the 1 to 2\,GeV region without (f) and with
(h) inclusion of $f_0(1710)$ in the fit \cite{Barberis:2001bs}.
}
\end{figure}

The WA102 results on central production of $\pi^+\pi^-$
\cite{Barberis:2001bs} are summarised in Fig.~\ref{fig:wapm}. The
figure shows (a) the so-called Ehrlich distribution (square of the
transverse missing mass) suggested in \cite{Ehrlich:1969xp} with a peak
at the squared pion mass.
A cut $-0.3 \leq M^2_X \leq 0.2$~$GeV^2$ selects events for further
analysis. The $  p_{fast} \pi^{\pm}$ effective mass spectra are
plotted in Fig.~\ref{fig:wapm}b,c, respectively. In  the $
p_{fast}\pi^+$ mass distribution (Fig.~\ref{fig:wapm}b),
$\Delta^{++}(1232)$ is observed, while in $  p_{fast}\pi^-$
(Fig.~\ref{fig:wapm}c) there is obviously also significant production of
$\Delta^{*}$'s or $ N^*$'s at higher mass. Both $\Delta(1232)$
signals were removed by a cut $M(p_{fast}\pi)$~$>$~1.5~GeV; there is a
similar cut on  $slow$-proton excitations. It was verified by a cut
$M(p\pi)$~$>$~2.0 GeV that production of high-mass baryon
resonances does not lead to significant changes in the final results.
The Feynman $x_F$ distributions for the `slow' and `fast' particles and
the $\pi\pi$ system are shown in Fig.~\ref{fig:wapm}d. The central
region is clearly seen to lie within $|x_F| \leq 0.25$. The resulting
centrally produced $\pi\pi$  mass distribution is shown in
Fig.~\ref{fig:wapm}e. A small $\rho^0(770)$ signal and some $f_2(1270)$
can be seen and a sharp drop at 1\,GeV which is due to interference of
$f_0(980)$ with the $S$-wave background. A second dip develops at
1.5\,GeV due to $f_0(1500)$. The partial wave analysis of this data was
performed assuming the $\pi\pi$ system to be produced by the collision
of two objects (e.g. two Pomerons or Reggeons) and full coherence of
the final state. As discussed above, the latter assumption is a strong
but untested constraint. The data was divided into mass bins and
expanded into multipoles from which the partial wave amplitudes were
derived. A system with $S, P$ and $D$ waves has eight solutions for
each mass bin. In each mass bin, one of these solutions was found from
the fit to the experimental angular distributions while the other seven
can then be calculated by a method described by
Chung~\cite{Chung:1997qd}. The solutions in adjacent mass bins were
required to be analytical function of mass~\cite{Alde:1998mc}. Under
favourable circumstances, the linking procedure leads to eight global
solutions. They all give identical moments but differ in the physical
content.  Differentiation between the solutions requires additional
input such as the requirement that, at threshold, the $S$-wave is the
dominant wave. Apart from the $S$-wave, the data show clear evidence for
$\rho(770)$ in the $P^-$ wave and for $f_2(1270)$ in the $D^-$ wave,
produced dominantly with $m=0$.

The $S$-wave of the surviving solution is shown in Fig.~\ref{fig:wapm}f-h.
The data exhibit a strong threshold enhancement followed by drops in
intensity at 1\,GeV and at 1.5\,GeV as seen already in the global
spectrum.  The $S$-wave was fitted with a background of the form
$a(m-m_{th})^{b}e^{(-cm-dm^{2})}$ -- with $m$ as $\pi\pi$ mass,
$m_{th}$ as $\pi\pi$ threshold mass and $a, b, c, d$ as fit parameters
-- and three interfering Breit-Wigner amplitudes to describe $f_0(980)$,
$f_0(1370)$ and $f_0(1500)$. The fit is shown in Fig.~\ref{fig:wapm}f
for the entire mass range and in Fig.~\ref{fig:wapm}g for masses above
1 GeV. A forth state, $f_0(1710)$, leads to a significant improvement
of the fit, see Fig.~\ref{fig:wapm}h. We quote here parameters (pole
positions on sheet II) which were determined from a coupled channel
analysis on centrally produced $\pi\pi$ and $ K\bar K$ pairs
\cite{Barberis:1999cq}

\begin{center} \vspace*{-2mm}
\begin{tabular}{cccccrc} \renewcommand{\arraystretch}{1.0}
 $f_0(980)$ && $\rm M $&$=$&$\ 987\pm\ 6\pm\ 6$&$-i\ (48\pm 12\pm\ 8)$& MeV\\
$f_0(1370)$ && $\rm M $&$=$&$ 1312\pm 25\pm10$&$ -i(109\pm 22\pm 15)$& MeV\\
$f_0(1500)$ && $\rm M $&$=$&$ 1502\pm 12\pm10$&$ -i\ (49\pm\ 9\pm\ 8)$& MeV\\
$f_0(1710)$ && $\rm M $&$=$&$ 1727\pm 12\pm11$&$ -i\ (63\pm\ 8\pm\ 9)$&
MeV\\ \renewcommand{\arraystretch}{1.0} \vspace*{-2mm} \end{tabular}
 \vspace*{-2mm}\end{center}

which are consistent with the PDG~\cite{Eidelman:2004wy} values for
these resonances.

\begin{figure}[pt]
\begin{minipage}[c]{0.65\textwidth}
\bc
\begin{tabular}{cc}
\hspace{3mm}\includegraphics[width=5.cm,height=7.5cm]{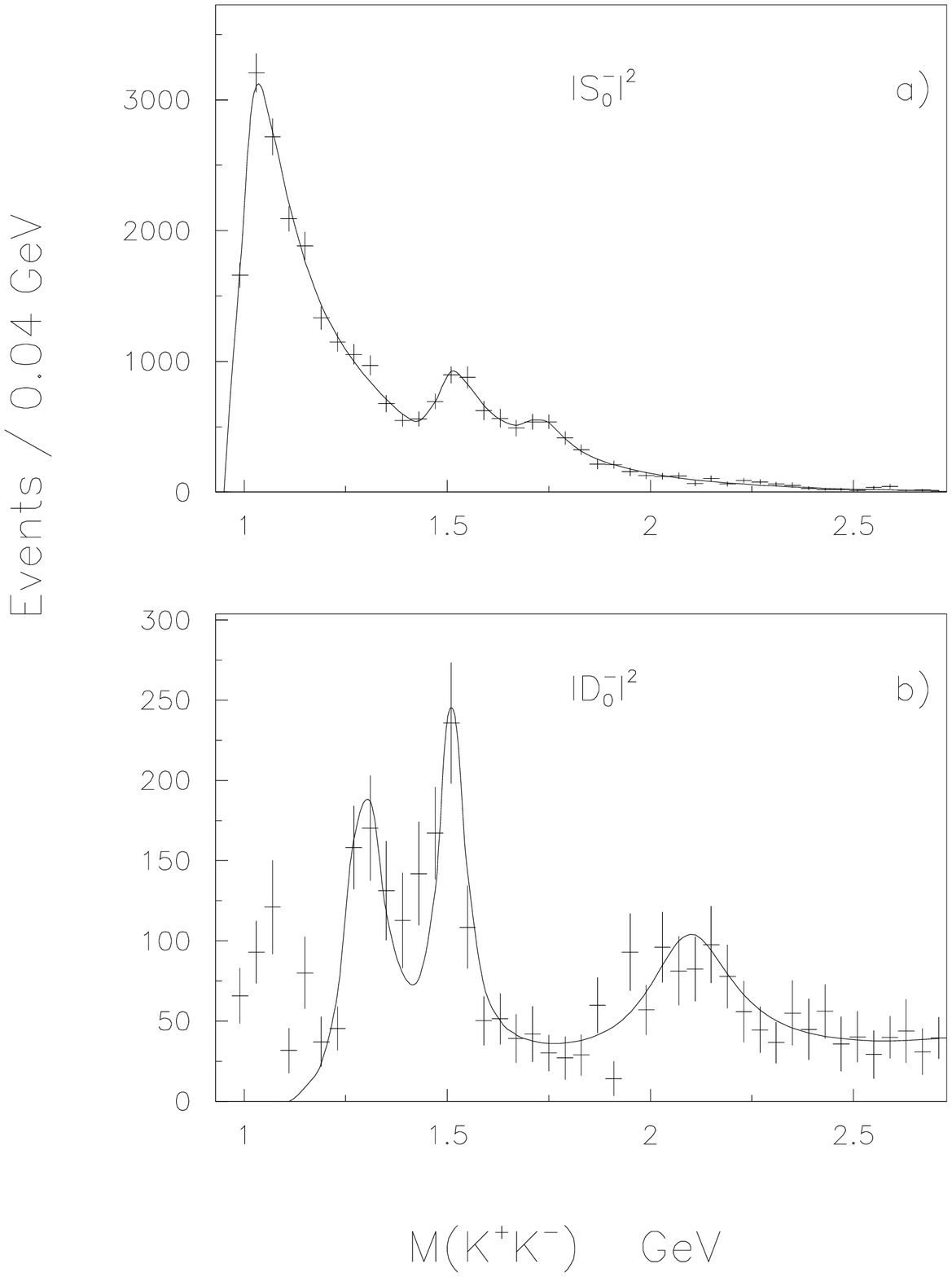}&
\hspace{3mm}\includegraphics[width=5.cm,height=7.5cm]{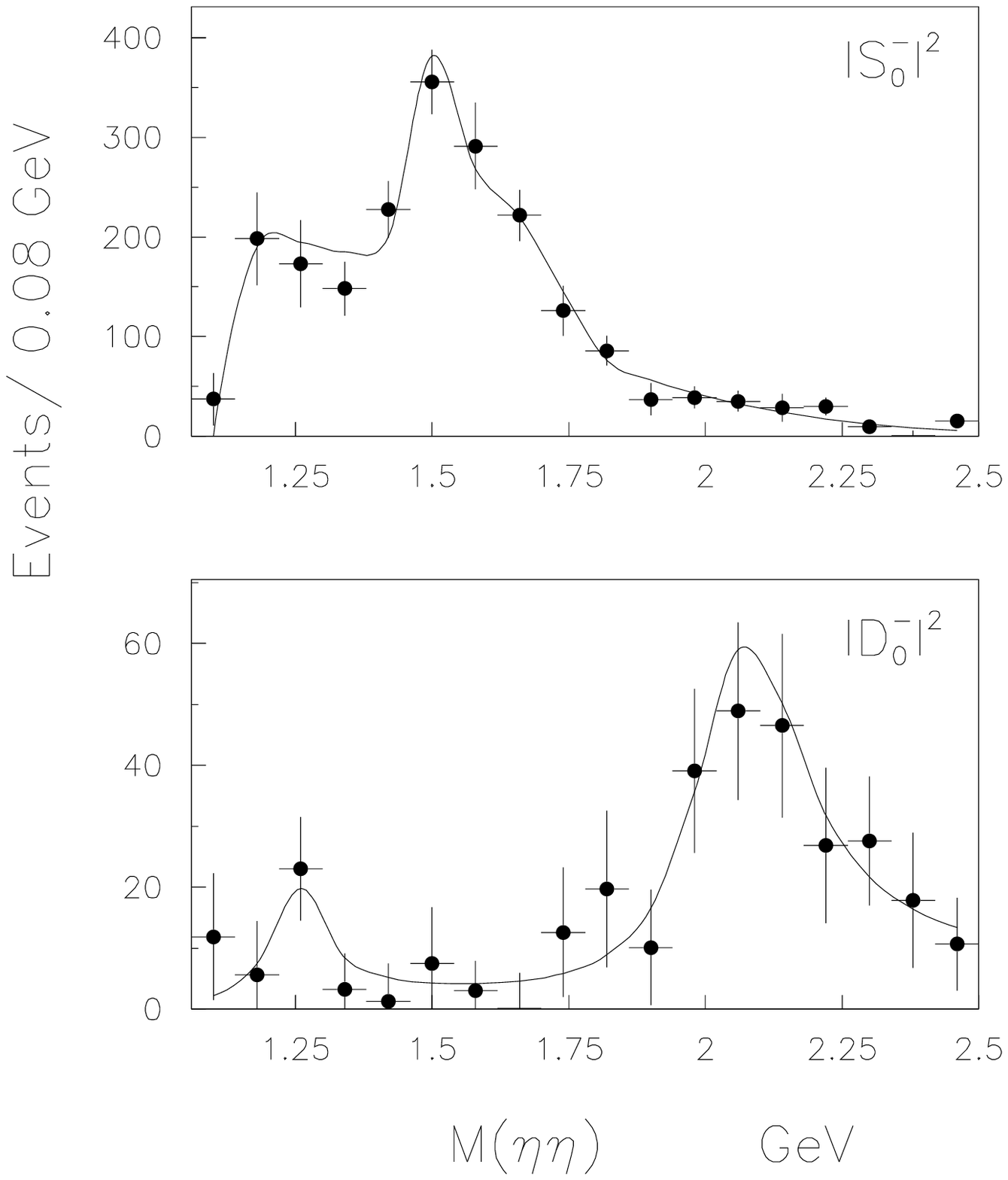} \\
\end{tabular}
\ec

\vspace{-77mm}

\phantom{rrr}\hspace{70mm}\scriptsize{c)} \\

\vspace{27mm}

\phantom{rrr}\hspace{70mm}\scriptsize{d)} \\

\vspace{30mm}

\end{minipage}
\begin{minipage}[c]{0.33\textwidth}
\includegraphics[width=5.cm,height=3.2cm]{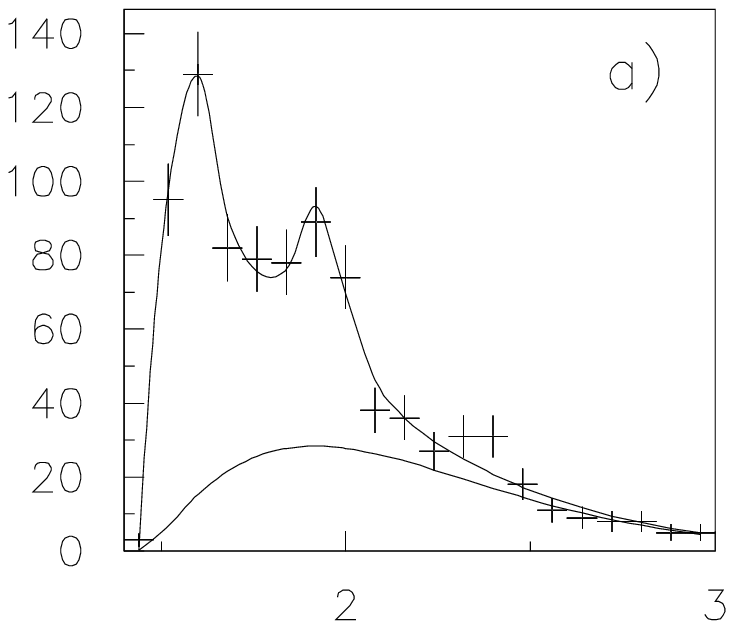}\vspace{3mm}\\
\includegraphics[width=5.cm,height=3.5cm]{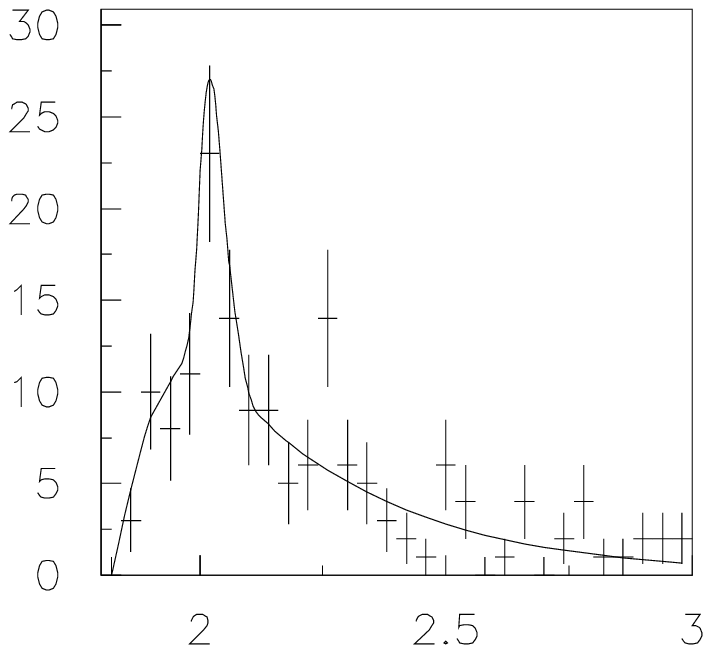}\\
\phantom{rrrrrrrrrrrrrrrrrrrrr}M$(\eta^{\prime}\eta^{\prime})$

\vspace{-78mm}

\phantom{rrr}\hspace{40.5mm}\scriptsize{e} \\

\vspace{27mm}

\phantom{rrr}\hspace{40mm}\scriptsize{f)} \\

\vspace{30mm}

\end{minipage}
\vspace*{-2mm}
\caption{\label{wamult}
The S (a,c) and D (b,d) wave for $ K\bar K$
\cite{Barberis:1999am} and $\eta\eta$ \cite{Barberis:2000cd} in central
production. e) the $\eta\eta^{\prime}$ and f)
$\eta^{\prime}\eta^{\prime}$ mass distributions \cite{Barberis:1999id}.
 }
\end{figure}

In the $ K\bar K$ $S$-wave (Fig. \ref{wamult}a), a strong threshold
enhancement is observed due to $f_0(980)$ with a long tail which may
comprise  $f_0(1370)$, $f_0(1500)$ and $f_0(1710)$. The latter two
resonances are seen as peaks.  Partial decay widths of $f_0(1370)$ and
$f_0(1500)$ will be discussed below (see Table \ref{wabr}), here we
quote
\be
\label{1710ssbar}
\frac{\mathcal B_{f_0(1710) \rightarrow K \overline K}}{\mathcal
B_{f_0(1710) \rightarrow \pi \pi}} = 5.0  \pm 0.6 \pm 0.9 \ee The
$f_0(1710)$ must have a large $s\bar s$ component. The $D$-wave
(Fig. \ref{wamult}b) resonates at the $f_2(1270)$ and $f_2(1525)$
masses, and at 2.15\,GeV. The latter resonance is also seen in the
D-wave $\eta\eta$ mass distribution (Fig. \ref{wamult}d) to which also
$f_2(1270)$ contributes. The scalar wave (Fig. \ref{wamult}c) evidences
$f_0(1500)\to\eta\eta$ decays. There is a shoulder at low $\eta\eta$
masses which is assigned to $f_0(1370)$.

The $\eta\eta^{\prime}$ mass distribution (Fig. \ref{wamult}e) is
interpreted by two Breit-Wigner amplitudes with $(M;\Gamma ) =
(1515\pm12\,;\,55\pm12)$\,MeV and $(1934\pm16\,;\,141\pm41)$\,MeV,
respectively, and a phenomenological background in the form
$a(m-m_{th})^{b}e^{-cm-dm^{2}}$, where $m$ is the $\eta\eta^{\prime}$
mass. The higher-mass resonance was observed before by GAMS
\cite{Alde:1986nx}, also in $\eta\eta^{\prime}$.  It should be
remembered that GAMS had observed an $\eta\eta$ enhancement at 1590,
the so-called G(1590) \cite{Alde:1985kp}. The  $\eta\eta$ distribution
showed a pronounced peak just below 1.6\,GeV which was interpreted as
a resonance. The data is now explained as bump following a dip at the
$f_0(1500)$ mass \cite{Anisovich:2002ij}. Possibly, (Fig. \ref{wamult}e)
should be interpreted by dips at 1500 and 1710\,MeV (and possibly at
2100\,MeV) in an $\eta\eta^{\prime}$ background spectrum. In the
$\eta^{\prime}\eta^{\prime}$ mass distribution (Fig. \ref{wamult}e), a
prominent peak at about 2\,GeV is observed. Its decay angular
distribution suggests tensor quantum numbers. It could be related to
the $D$-wave resonance in $ K\bar K$ and $\eta\eta$ even though the
masses are not fully consistent.

It is interesting to compare the $2\pi$ distributions from central
production (Fig.~\ref{fig:wapm}f) with those on the
$\pi\pi_{S-{\rm wave}}$ from charge exchange in a 100\,GeV pion beam
(Fig.~\ref{fig:gams-scalar}). In charge exchange, there is a background
amplitude with a dip due to the $f_0(980)$. The $\pi\pi$ elastic
scattering amplitude is at the unitarity limit; above the $K\bar K$
threshold the amplitude is still close to it. In central production, a
very large low-mass enhancement is observed. Minkowski and Ochs argue
that this large $S$-wave contribution is a new effect specific for
central production and originating from Pomeron-Pomeron fusion
\cite{Minkowski:2002nf}. The large intensity is however as well expected
from the $1/M^2$ dependence of the central production yield.

\begin{figure}[!ht]
\begin{minipage}[b]{0.65\textwidth}
\includegraphics[width=0.95\textwidth]{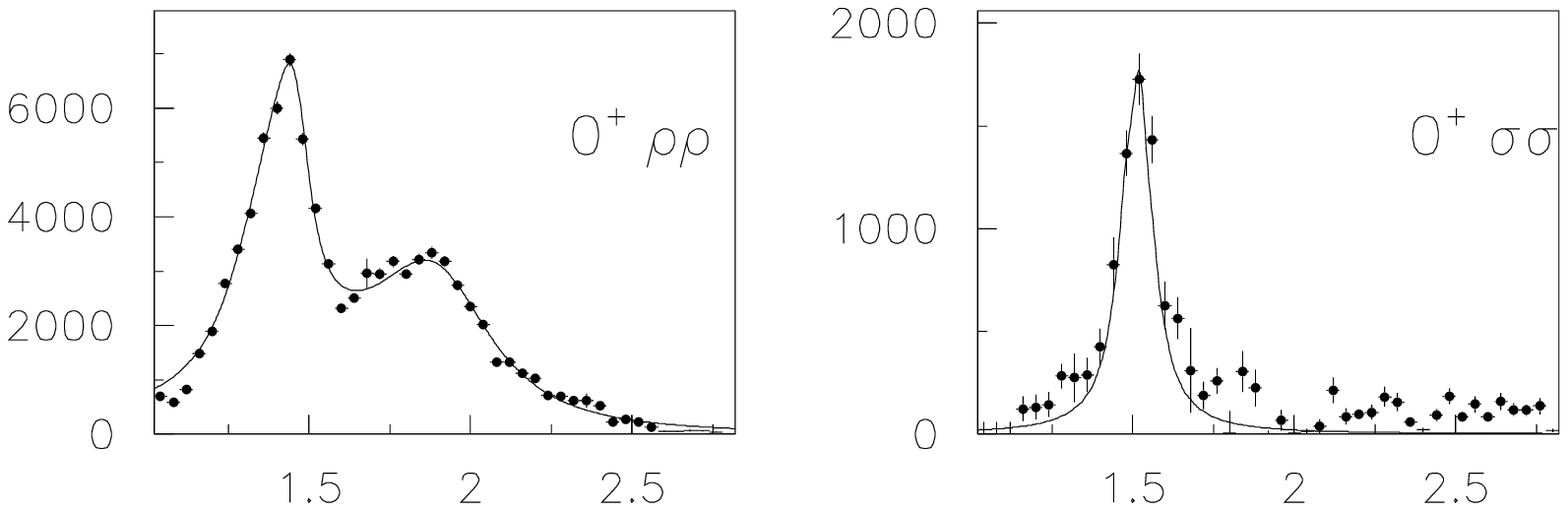}\\
\includegraphics[width=0.95\textwidth]{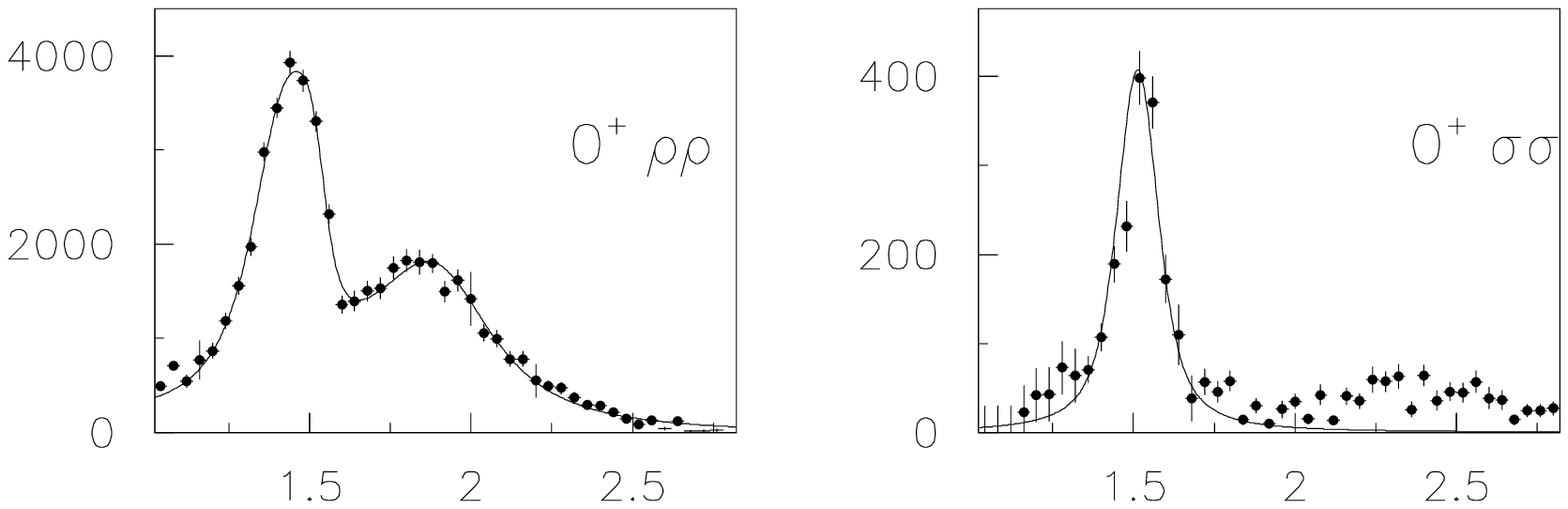}

\vspace{-83mm}

\phantom{rrr}\hspace{45mm}\scriptsize{a)}\hspace{52mm}\scriptsize{c)} \\

\vspace{33mm}

\phantom{rrr}\hspace{45mm}\scriptsize{b)}\hspace{52mm}\scriptsize{d)} \\

\vspace{23mm}

\end{minipage}
\begin{minipage}[b]{0.35\textwidth}
\hspace{-10mm}\includegraphics[width=1.2\textwidth]{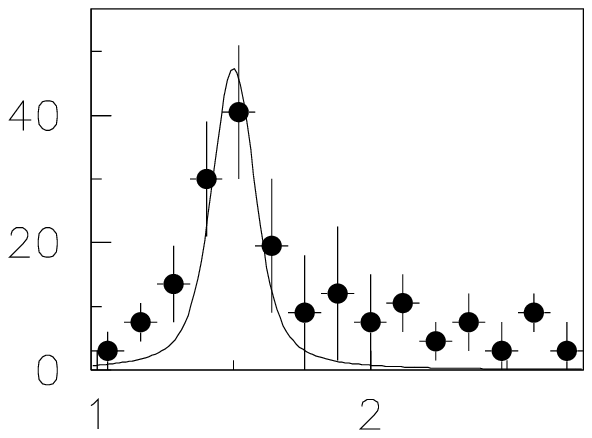}\hspace{10mm}

\vspace{23mm}

\end{minipage}

\vspace{-77mm}

\phantom{rrr}\hspace{162mm}\scriptsize{e)} \\

\vspace{65mm}

\caption{\label{wa4pi}
4$\pi$ invariant mass (in GeV) spectra from central
production.
a) $\rho\rho$ $S$-wave from 2\pip 2\pim ;
b) $\rho\rho$ $S$-wave from from \pip\pim 2\piz ;
c) $\sigma\sigma$ $S$-wave 2\pip 2\pim ;
d) $\sigma\sigma$ $S$-wave from \pip\pim 2\piz ;
e) $\sigma\sigma$ $S$-wave 4\piz .
}
\end{figure}

A bump-dip-bump structure is observed in central production of four
pions, too. Fig.~\ref{wa4pi} shows 4$\pi$ invariant mass spectra from
the WA102 experiment~\cite{Barberis:2000em}. In the $\rho\rho$ scalar
intensity distribution, a large peak at 1370 MeV is seen followed by a
dip in the 1500 MeV region and a further (asymmetric) bump. This
behaviour is in sharp contrast to the $\sigma\sigma$ invariant mass
distribution showing an isolated peak at 1500\,MeV. This is confirmed
by the $4\pi^0$ invariant mass distribution which can receive
contributions from $\sigma\sigma$ but not from $\rho\rho$.

The $f_0(1710)$ is observed neither in $\sigma\sigma$ nor in
$\rho\rho$. Instead, a new scalar isoscalar $\rho\rho$ resonance,
$f_0(2000)$, is suggested. Masses and widths of $f_0(1370)$ and
$f_0(1500)$
\begin{center}
\vspace*{-2mm}
\begin{tabular}{cccccrc}
\renewcommand{\arraystretch}{1.0}
$f_0(1370)$ && $\rm M $&$=$&$ 1309\pm 24$&$ -i(163\pm 26)$&MeV\\
$f_0(1500)$ && $\rm M $&$=$&$ 1513\pm 12$&$ -i( 58\pm 12)$&MeV\\
$f_0(2000)$ && $\rm M $&$=$&$ 1989\pm 22$&$ -i(224\pm 42)$&MeV
\renewcommand{\arraystretch}{1.0}
\vspace*{-2mm}
\end{tabular}
\end{center}
are compatible with the findings from central production of
$\pi\pi$ and $ K\bar K$. The state at 1989\,MeV could be related to the
$\eta\eta^{\prime}$ enhancement in Fig. \ref{wamult}e and to similar
observations in $p\bar p$ annihilation in flight
\cite{Uman:2006xb,Anisovich:2000ae}, see the end of section
\ref{Scalar resonances from pbarp annihilation}. From the data and the
fits, ratios of partial decay rates were derived given in
Table~\ref{wabr}. They will be discussed jointly with results from
\pbp\ annihilation experiments. The $f_0(980)$ couplings were observed
with $g_{ \pi}=0.19\pm0.03\pm0.04$, $g_{ K}=0.40\pm0.04\pm0.04$.

\begin{figure}[!ht] \bc
\begin{tabular}{cc}
\includegraphics[width=6.25cm,height=5.0cm]{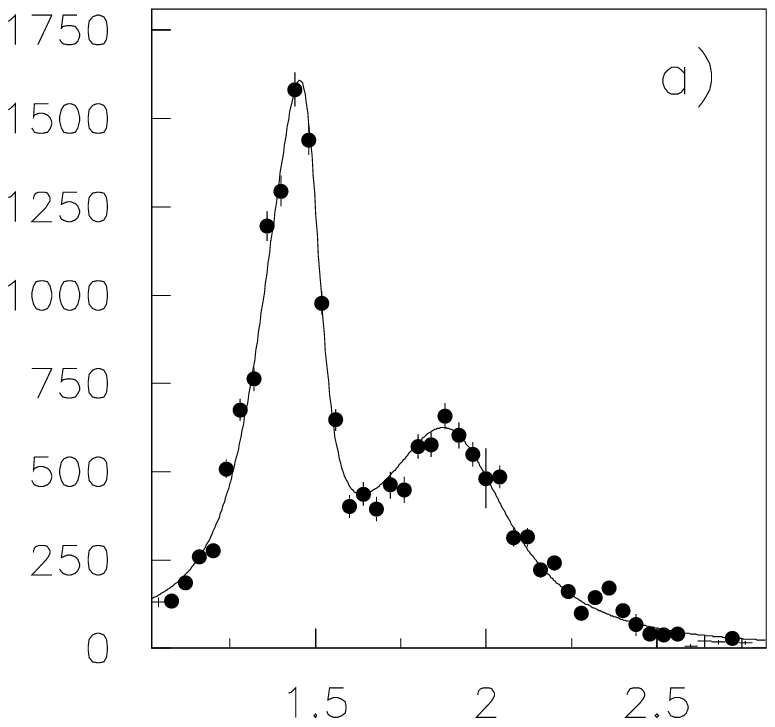}\hspace{6mm}&
\includegraphics[width=6.25cm,height=5.0cm]{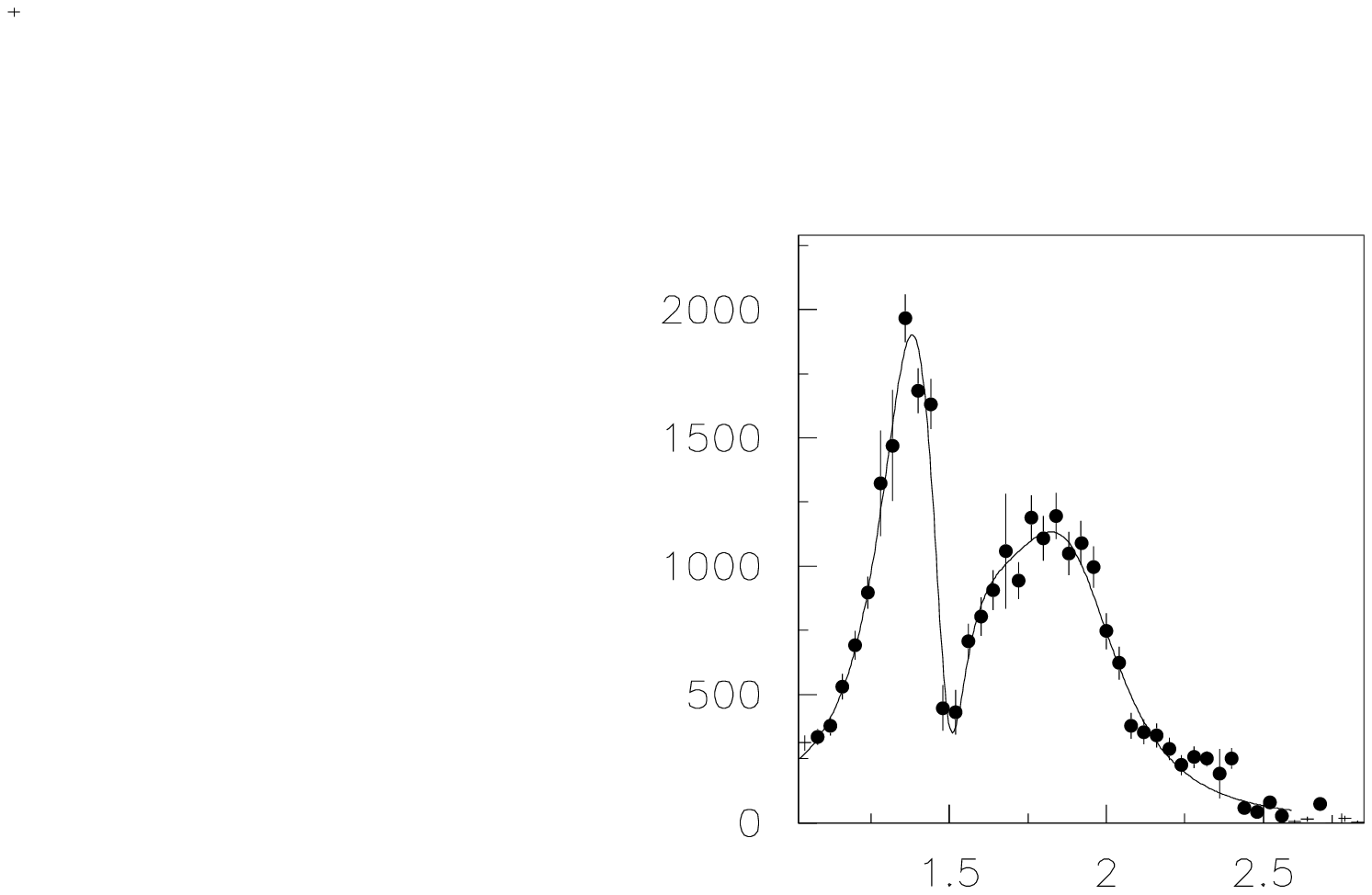}
 \end{tabular}\vspace*{-48mm}\\
\hspace{110mm}b)\vspace*{41mm}\\
\ec
\caption{\label{wadip}
The $J^{PC}=0^{++}$~$\rho\rho$ wave from centrally produced
$\pi^+\pi^-\pi^+\pi^-$ as a function of $\phi$. a) $\mid\phi\mid\leq
45^{\circ}$, b) $\mid\phi-180^{\circ}\mid\leq 45^{\circ}$. }
\end{figure}

The data on central production of four pions
\cite{Barberis:1999wn,Barberis:2000em} are well suited to highlight the
$\rm r\hat{o}le$ of the Close-Kirk glueball filter
\cite{Close:1997pj,Close:1997nm}. Fig.~\ref{wadip} shows the scalar
$4\pi$ mass distribution for events in which the fast and the slow
protons are scattered (a) into the same direction ($\phi\sim
0^{\circ}$) or (b) into opposite directions ($\phi\sim 180^{\circ}$).

A rotation in angle $\phi$ leads to a significant change of the $R R
\to M$ kinematics and to a dramatic change of production mechanism. As
example we consider the symmetrical case where the two scattered
protons reduce their longitudinal momenta by the same value (in the
cms), i.e. $|P_{1L}|=|P_{2L}|$, and obtain equal transverse momenta
$|P_{1T}|=|P_{2T}|$. The transverse momenta $\vec P_{1T}$ and $\vec
P_{2T}$ may point into the same direction, then $\phi$ vanishes; or
they may point into opposite directions with $\phi=180^{\circ}$. For a
specified mass of the centrally produced system, \be \label{kinphi}
M_X^2= 4 (t+P_{L}^2+ P_{T}^2 sin^2(\phi/2))
 )
\ee
holds where $t$ the momentum transfer to system $X$ by each proton.
The $\phi$ dependence observed in Fig. \ref{wadip} is thus related to a
considerable variation in $t$.

Such effects are observed in various reactions. We give a few examples:

- in diffractive production of $\rho \pi$ by pions, the signal in the
$a_1$ region is dominated by the Deck effect at small $-t$ and by a
``true" $a_1$ signal at large $-t$;

- in the reaction $\pi^- p \to \pi \pi n$ , the interference of
$f_0(980)$ with the wide $\pi \pi$ $S$-wave is purely destructive at
small $-t$ due to unitarity and mainly constructive at large $-t$,
where unitarity constraints are relaxed.

Three interpretations were offered to explain the change in the mass
distribution when different $\phi$ ranges are selected.

\begin{figure}[!ht] \bc
\includegraphics[width=0.6\textwidth]{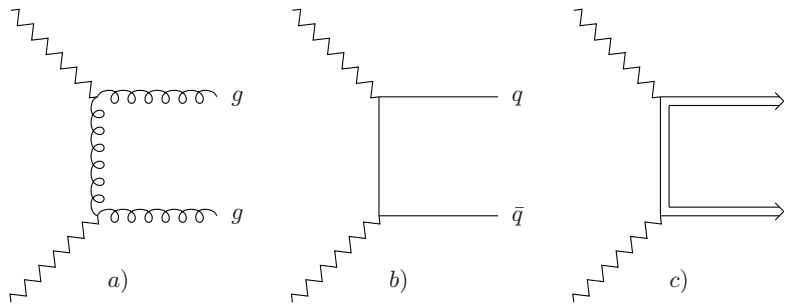}
\ec
\caption{\label{cloki}
Processes contributing to double Pomeron exchange. a) Glueballs may be
produced by couplings of two Pomerons to glue. b) Exchange of a quark
leads to production of $q\bar q$. a) and b) are short range processes.
c) At large distances, mesons can be exchanged between two Pomerons.
}
\end{figure}

\begin{itemize}

\vspace{2mm}\item Close and Kirk compared the $\phi$ dependence of
different mesons and noticed that unconventional states have a large
chance to be produced with $\mid\phi\mid\leq 45^{\circ}$ while
established $q\bar q$ mesons have a better chance to be seen with
$\mid\phi-180^{\circ}\mid\leq 45^{\circ}$
\cite{Close:1997pj,Close:1997nm}. It was suggested that at small
$dk_T$, the production of glueball candidates is enhanced and the
production of mesons is suppressed. At large $dk_T$, there is ample
production of mesons and little production of glueballs. In the fits
\cite{Barberis:1999wn,Barberis:2000em}, the strong enhancement at
$\sim 1.4$\,GeV was assigned to $f_0(1370)$, the $f_0(1500)$ produces a
dip. In this view, three processes contribute to central
production, gluon-gluon fusion (Fig.\ref{cloki}a), $q\bar q$ production
(Fig.\ref{cloki}b) and a general unspecified background.

\vspace{2mm}\item Ochs and Minkowski \cite{Minkowski:1998mf} assigned
the dips to production of $q\bar q$ mesons. The wide background is
produced by diagram Fig.\ref{cloki}a and is called {\it red dragon}. It
is argued that the $f_0(1370)$ does not exist; instead, the slow
background phase motion seen in $\pi\pi$ scattering is due to a very
broad resonance $f_0(1000)$, and this is believed to represent the
scalar glueball.

\vspace{2mm}\item A third explanation of these effects was offered in
\cite{Klempt:2000ud}. The starting point of the interpretation is the
similarity of the dips observed in $\pi\pi$ scattering (Fig.
\ref{fig:gams-scalar}) and in $4\pi$ central production (Fig.
\ref{wadip}b). The peak at 980\,MeV in Fig. \ref{fig:gams-scalar}
disappears gradually when the kinematical regime is changed (Fig.
\ref{fig:piex1}) and may even turn into a peak. The gradual
disappearance of the $4\pi$ dip in Fig. \ref{wadip}a compared to Fig.
\ref{wadip}b suggests that the same mechanism is at work. The dip is
pronounced for small values of $t$. For opposite side scattering, the
object $M$ is produced by nearly real Pomerons ($t' \approx 0$).
Unitarisation of the sum of elastic background amplitude and the
resonant $f_0(1500)$ amplitude causes the observed dip. In case of one
side scattering just the object $M$ is balancing the transverse
momentum of both protons and the object $M$ is produced by highly
virtual Pomerons $t' \approx -P_{T}^2$, and unitarisation has a reduced
influence only. Since a background amplitude and unitarity effects were
not taken into account in the analysis
\cite{Barberis:1999wn,Barberis:2000em}, the results could be
misleading.
\end{itemize}

The spectacular effects (Fig. \ref{wadip} shows one example) are well
explained without requiring introduction of new states. We believe that
two mechanisms compete, producing mesons and a general background. The
latter process is probably more peripheral (smaller $-t$). In
\cite{Klempt:2000ud}, the two processes depicted in Fig.~\ref{cloki}b
and c to are supposed to contribute to the $4\pi$ system: quark
exchange at short distances and a long-range exchange creating a
background amplitude. To produce a $\rho\rho$ background amplitude, a
$\rho$ trajectory needs to be exchanged between the two Pomerons;
to produce $\pi\pi$, a pion trajectory is needed. Thus there is not one
red dragon but a whole family of red dragons. For peripheral
interactions, they exchange e.g. a $\rho$ meson and two $\rho$'s are
seen in the final state.  This process yields a background amplitude.
For large $-t$, the interaction is short range, a quark can be
exchanged and Pomeron-Pomeron fusion leads to production of scalar
$q\bar q$ mesons which interfere destructively with the background
amplitude. Experimental data show that there is no $\sigma\sigma$
background; the $\sigma\sigma$ final state can only be reached via
formation of a $q\bar q$ resonance, see Fig.~\ref{wadip}.


\subsection{\label{Scalar resonances from pbarp annihilation}
Scalar resonances from $ p\bar p$  annihilation}
\begin{figure}[t!]
\bc
\hspace*{-2mm}\begin{tabular}{cccc}
\includegraphics[width=40mm,height=40mm]{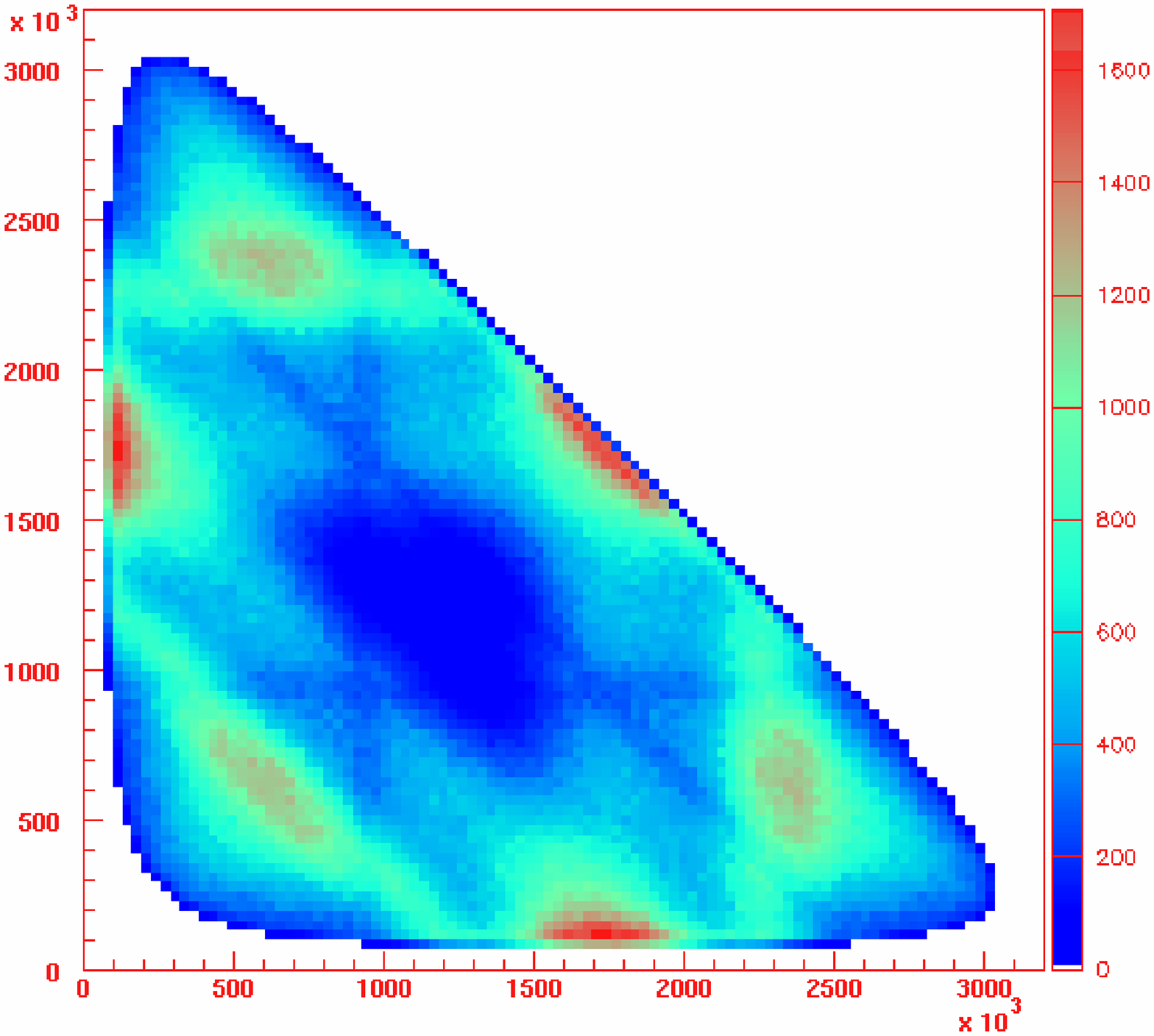}&
\hspace{-2mm}\includegraphics[width=40mm,height=40mm]{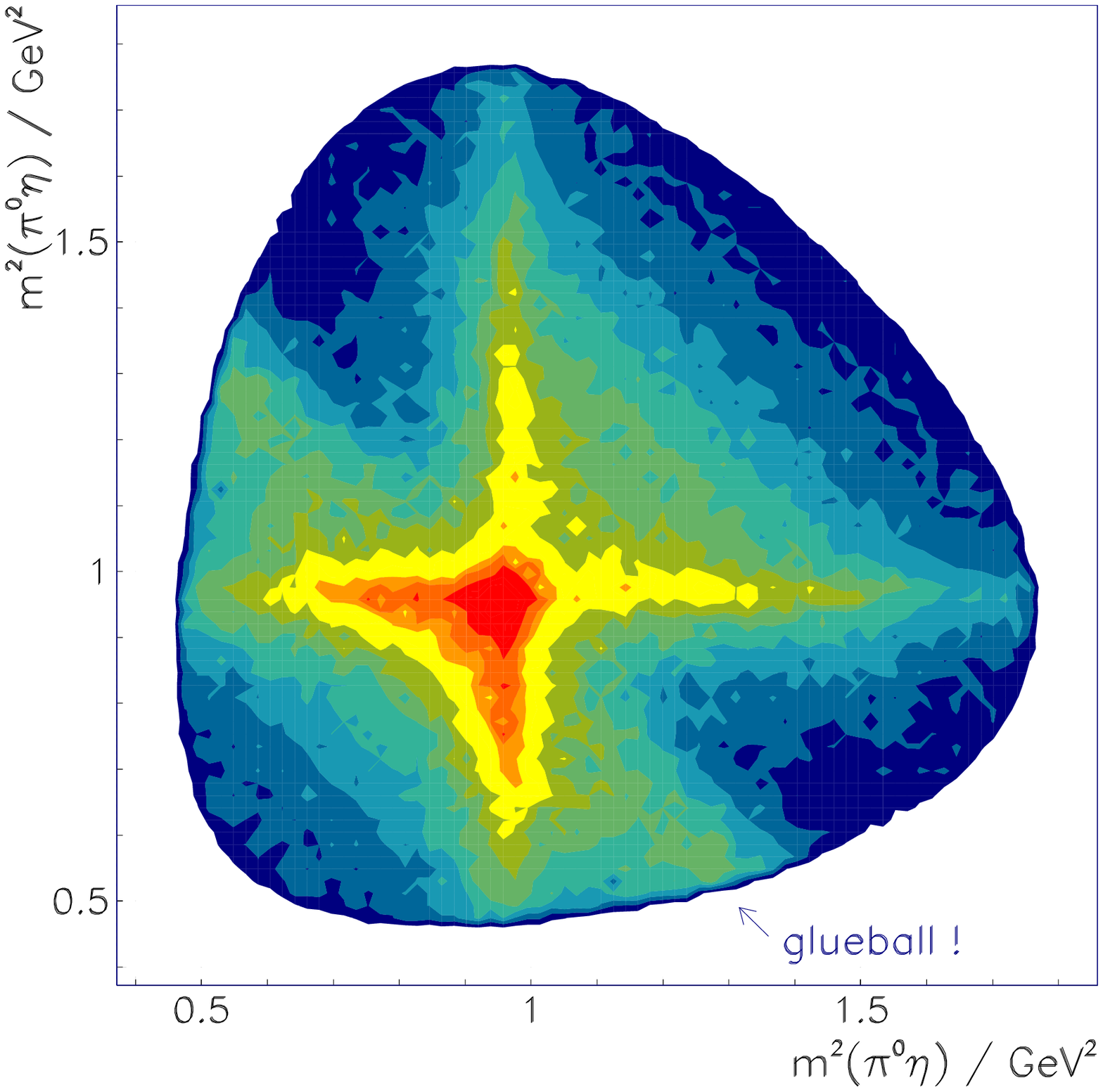}&
\hspace{-4mm}\includegraphics[width=41mm,height=41mm]{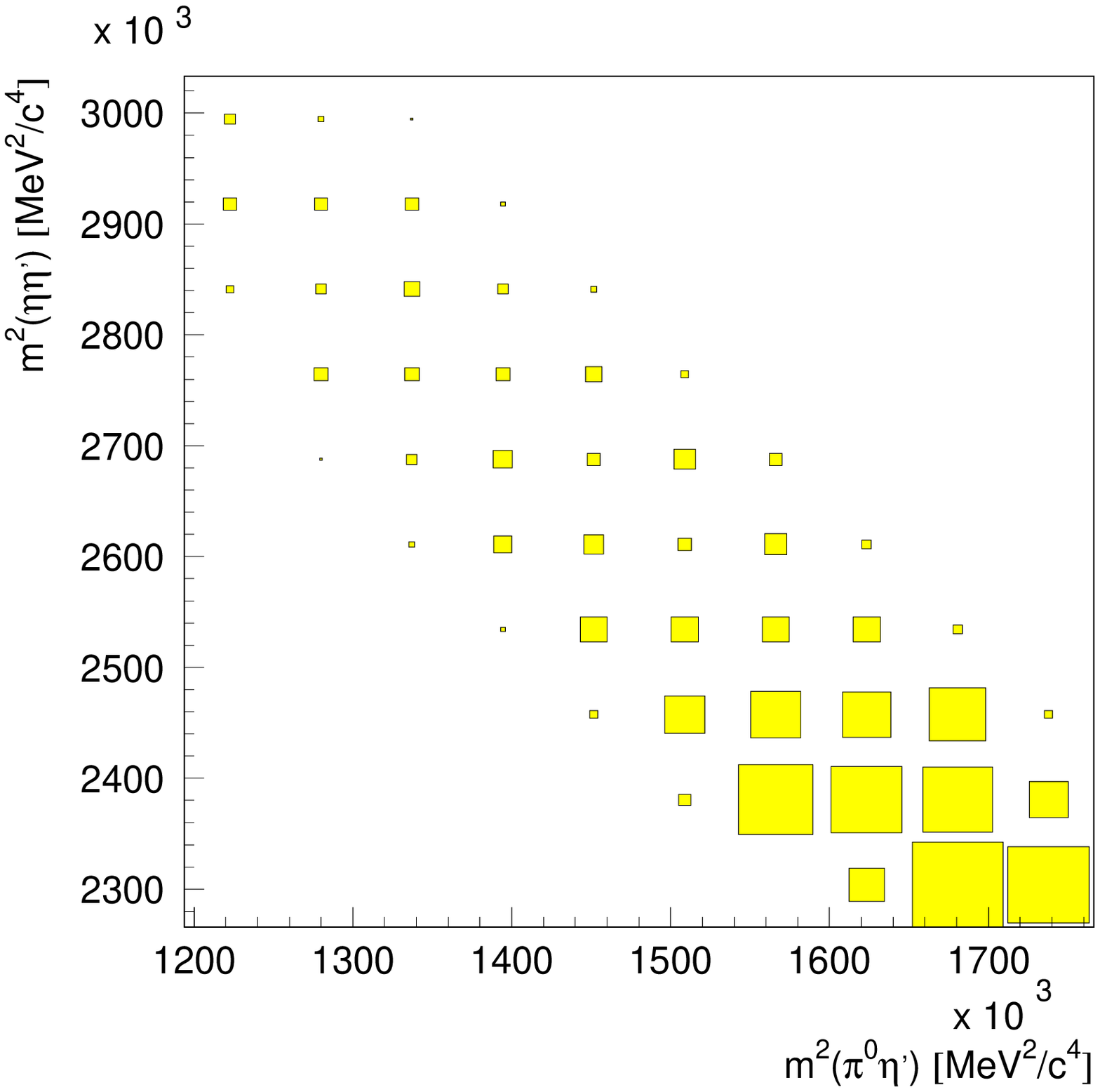}&
\hspace{-6mm}\includegraphics[width=42mm,height=42mm]{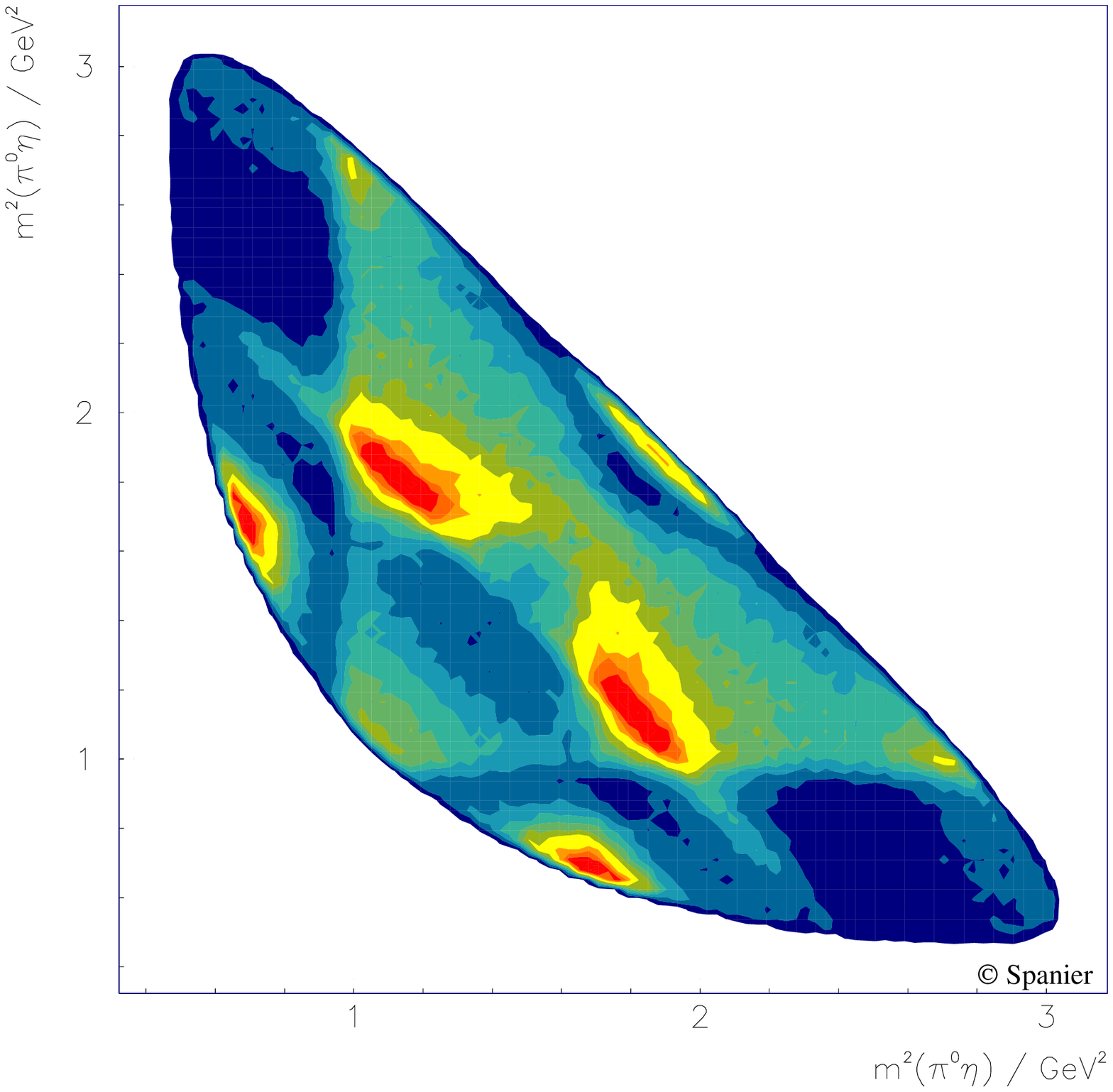}
\vspace*{-40mm}\\
\hspace*{20mm}a&\hspace*{20mm}b&c\hspace*{-20mm}&d\hspace*{-20mm}\vspace*{33mm}\\
\hspace{3mm}\includegraphics[width=40mm,height=40mm]{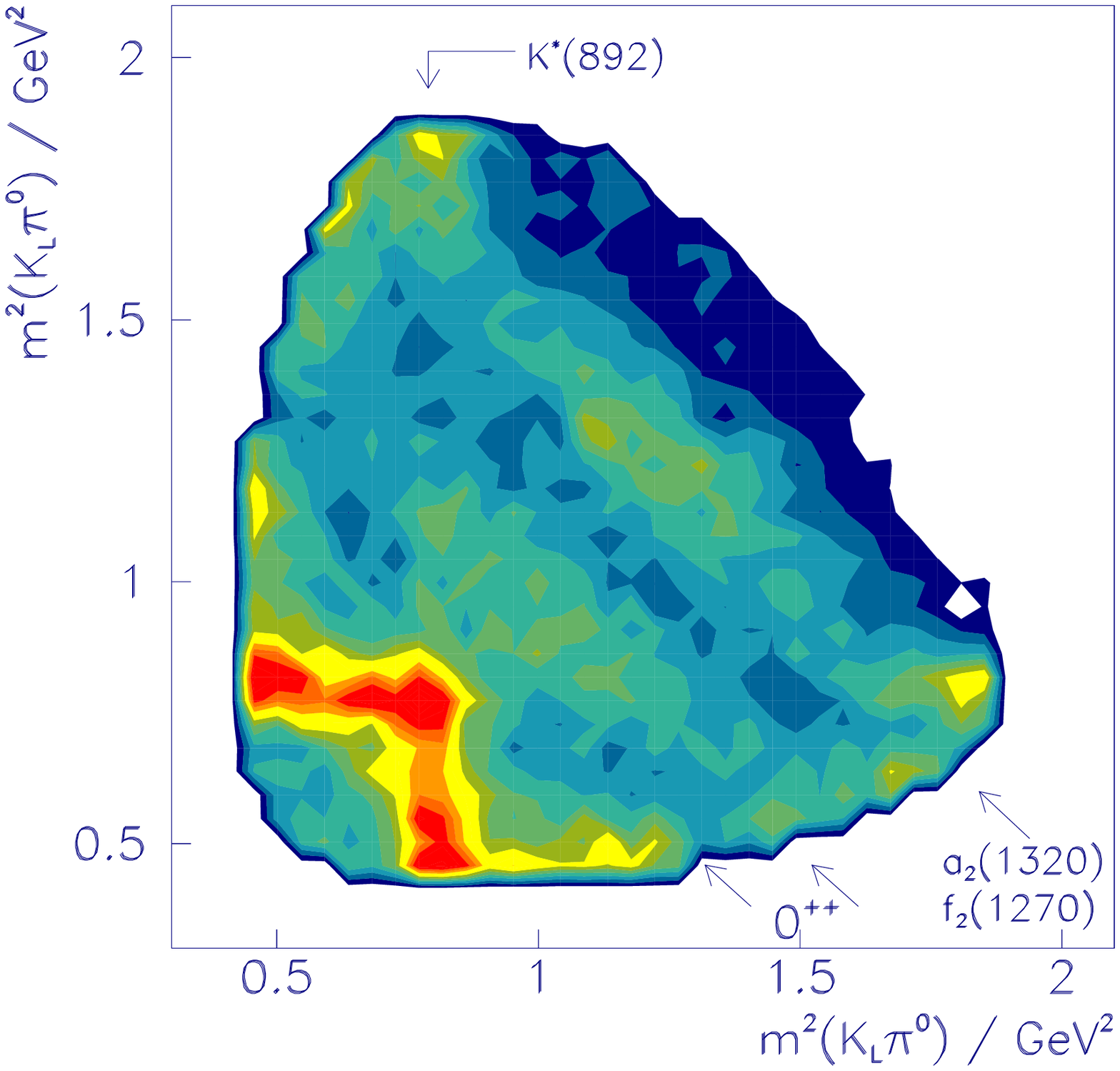}&
\hspace{-2mm}\includegraphics[width=44mm,height=44mm]{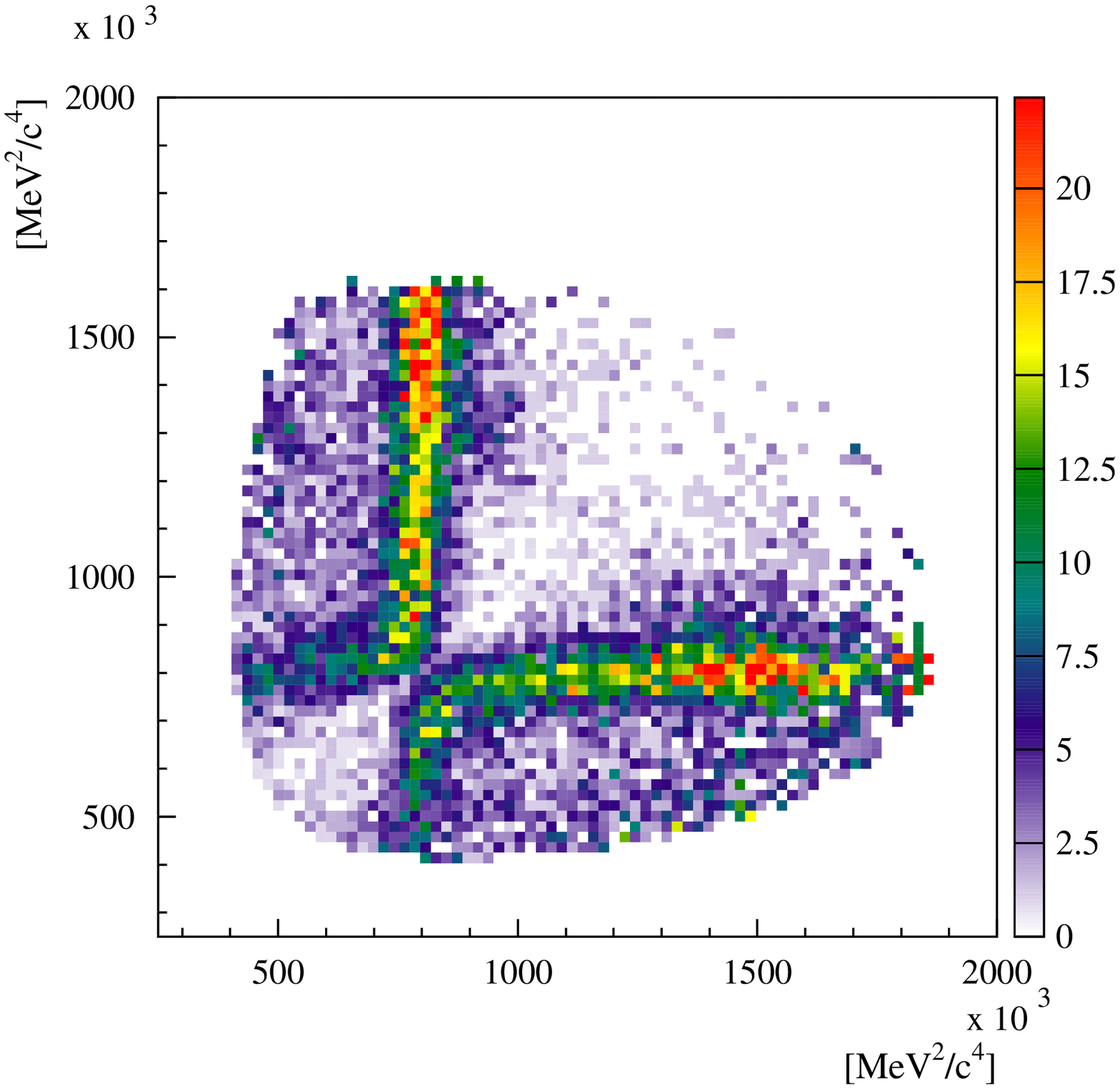}&
\hspace{-4mm}\includegraphics[width=44mm,height=44mm]{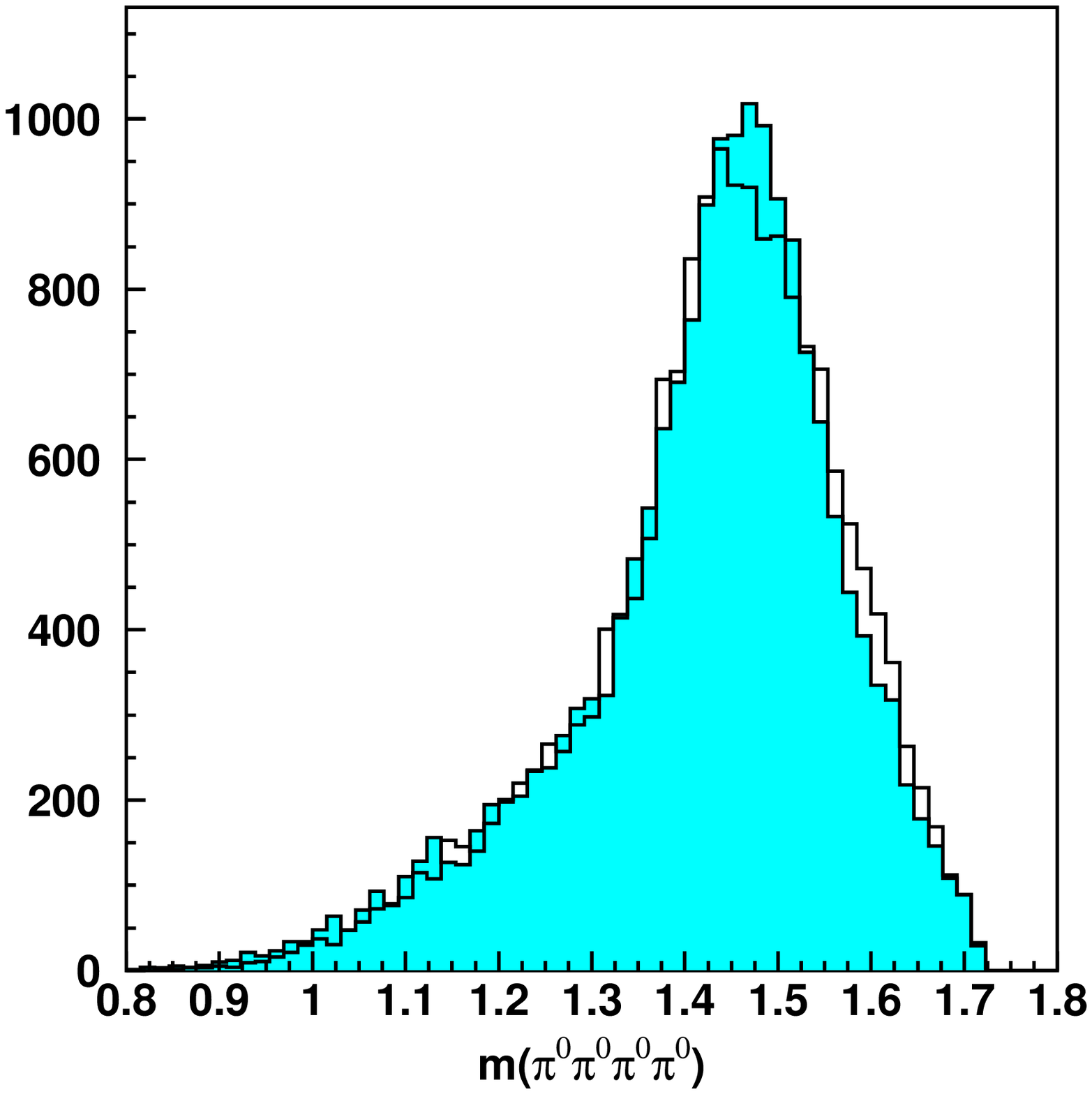}&
\hspace{-4mm}\includegraphics[width=44mm,height=44mm]{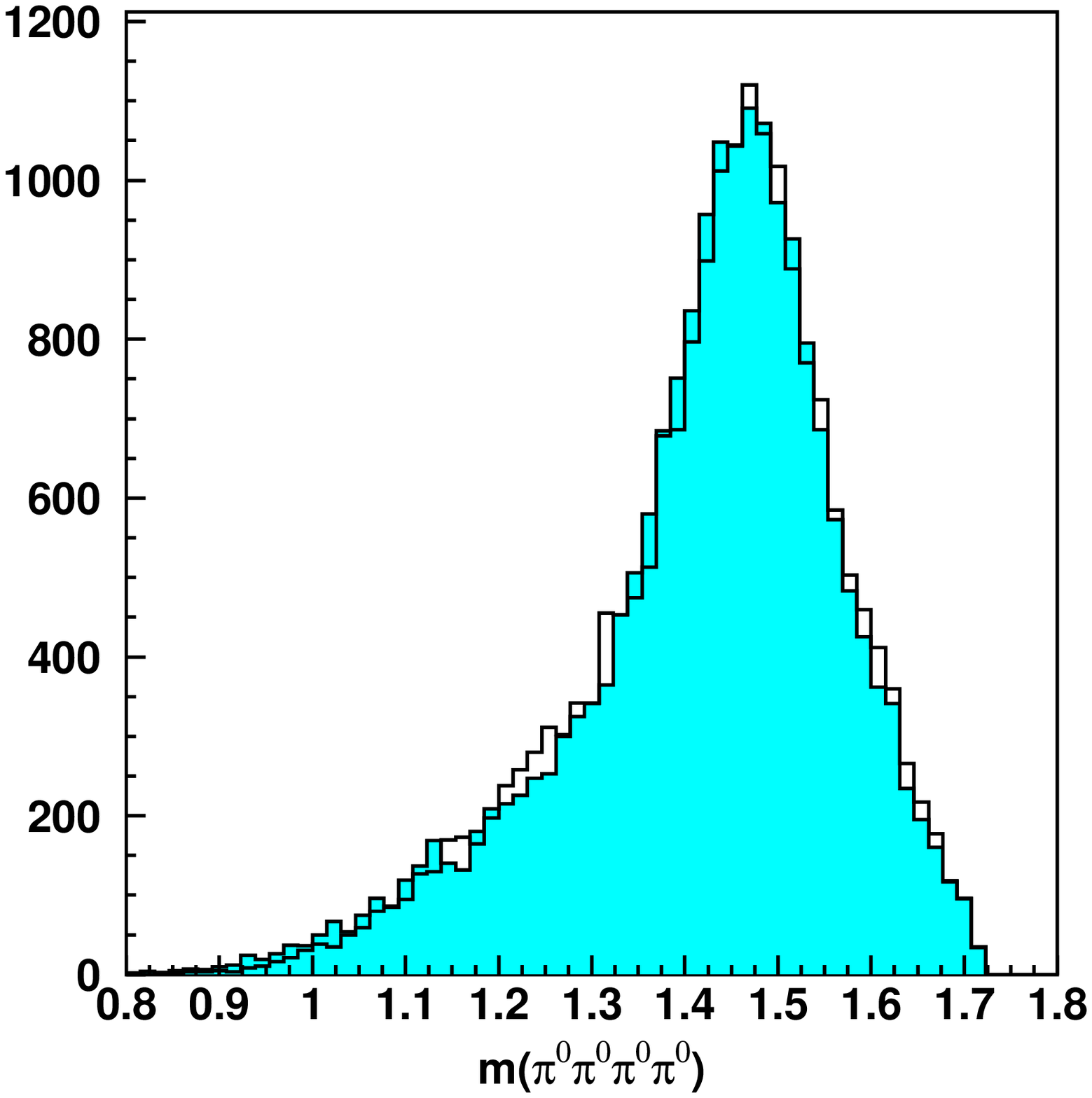}
\vspace*{-40mm}\\
e\hspace*{-20mm}&f\hspace*{-20mm}&g\hspace*{20mm}&h\hspace*{20mm}\vspace*{35mm}\\
\end{tabular}
\ec
\vspace*{-4mm}
\caption{\label{four-dp}
Dalitz plots for $p\bar p$ annihilation at rest into 3$\pi^0$ (a),
$\pi^0 2\eta$ (b), $\pi^0\eta\eta^{\prime}$ (c), $2\pi^0\eta$ (d),
$ K^{\pm}K^0_L\pi^{\mp}$ (e), $ K^0_LK^0_L\pi^0$ (f). Plots (g,h)
show the 4$\pi^0$ invariant mass in the reaction $ \bar pn\to\pi^-
4\pi^0$. A fit (including other amplitudes) with one scalar state
fails; two scalar resonances at 1370 and 1500 MeV give a good fit. Note
that a event-based likelihood fit was performed and not a fit to the
mass projection shown here. The data are from
\protect\cite{Amsler:1995gf,Amsler:1995bz,Amsler:1994ah,Amsler:1994pz,%
Amsler:1994pz,Abele:1996nn,Abele:1998qd,Abele:2001js}.
The $f_0(1500)$ resonances contributes to all reactions except to
(d,f); the $f_0(1370)$ is used to all fits except (c,f). From the
Dalitz plots (a,d,e,f) the properties of $a_0(980), f_0(980)$ and
$a_0(1450)$ are determined. } \end{figure}

The Crystal Barrel collaboration studied scalar mesons in $\bar pp$
annihilation at rest. The detector is briefly described in section
\ref{The Crystal Barrel experiment}, experimental methods are
discussed in \ref{meth:Antiproton-proton annihilation}.
Annihilation in liquid H$_2$ takes place predominantly from $S$-wave
orbitals of the $\bar pp$ atom which is formed when antiprotons come to
rest. The angular momenta are restricted, and scalar mesons are
produced abundantly in this process. Two scalar isoscalar mesons were
discovered, the $f_0(1370)$, $f_0(1500)$, and one isovector $a_0(1450)$.
Their main properties were derived from six Dalitz plots
\cite{Amsler:1995gf,Amsler:1995bz,Amsler:1994ah,Amsler:1994pz,%
Abele:1996nn}, shown in
figure~\ref{four-dp}, and from the analysis of different five-pion
final states~\cite{Amsler:1994rv,Abele:1996fr,Abele:2001js,Abele:2001pv}.
The Dalitz plot for $\bar pp$ annihilation into $3\pi^0$ is shown in
Fig.~\ref{four-dp}a. It is based on 700.000 events. The population along
the $\pi\pi$ mass band marked $f_2(1270)$ increases at the edges of
the Dalitz plot indicating that one $\pi^0$ is preferentially emitted
along the flight direction of the resonance. This is typical for
high-spin resonances. In $\bar pp$ annihilation at rest, resonances
with $J=4$ are hardly produced; the band is produced by
the $f_2(1270)$ decaying with the angular distribution $(3\cos^2\theta
-1)^2$ from $^1S_0$. The $f_0(980)$ appears as a narrow dip in the
$\pi\pi$ $S$-wave.  In addition, a narrow band of about constant
intensity is observed. It is marked $f_0(1500)$ in Fig.~\ref{four-dp}a.
The partial wave analysis revealed contributions from a further scalar
state at 1370\,MeV mass. The enhancement labelled $f_2(1520)$
originates from the constructive interference of the two low-mass
$\pi\pi$ $S$-waves. In addition, some contribution from the $f_2(1565)$
is required in the fit.

The reaction $\bar pp\to \pi^0\eta\eta$ leads to a Dalitz plot
with $2\cdot 10^5$ events; it is shown in Fig.~\ref{four-dp}b.
The $a_0(980)$ decaying to $\eta\pi^0$ is seen as a horizontal and a
vertical band. In the second diagonal, two homogeneously populated bands
can be observed which correspond to two states decaying to $\eta\eta$.
These are marked  $f_0(1370)$ and $f_0(1500)$ in Fig.~\ref{four-dp}b.

In the \piz\etg\etp\ Dalitz plot (Fig.~\ref{four-dp}c) a strong
threshold enhancement in the \etg\etp\ invariant mass is seen; it can
be traced to the observation of $f_0(1500)$ in its $\eta\eta^{\prime}$
decay mode. The $\eta$ and $\eta^{\prime}$ are orthogonal states in
SU(3). The strength of this decay mode indicates a substantial SU(3)
octet component in the $f_0(1500)$ flavour wave function.

The Dalitz plot for $\bar pp\to\pi^0\pi^0\eta$ with $2.8\times 10^5$
events is shown in Fig.~\ref{four-dp}d. It is characterised by a sharp
increase of intensity at a $\pi^0\eta$ mass of $\sim 1$\,GeV/c$^2$ due to
$a_0(980)$ production and a diagonal band due to $f_0(980)$
decaying to $\pi\pi$. The most prominent irregular pattern with
sharp maxima and minima is due to $a_2(1320)\to\eta\pi$. Two corners
(top and right) show additional intensity which requires to introduce a
scalar isovector resonance, the $a_0(1450)$. The strength of the
interference pattern suggests that only one single $\bar pp$ atomic
state ($^1S_0$) makes a significant contribution to this annihilation
mode ($\sim 92$\% \cite{Abele:1999tf}). Similar $P$-wave contributions
are found in \pbp$\to 3\pi$ and $\pi^0\eta\eta$.

The $K^0_LK^0_L\pi^0$ Dalitz plot shown in Fig.~\ref{four-dp}d
is based on events in which one $K^0_L$ is detected in the Crystal Barrel
detector while the second one is missing. Due to the large $K^0_L$ mass,
the final state can be identified in a kinematical fit with a
surprisingly low background of a few \%. It has prominent $K^*$ bands;
it is the interference with the $f_0(1500)$ which makes the intensity
so large in the lower left corner. Finally, the Dalitz plot for
reaction $\bar pp\to K^{\pm}K^0\pi^{\mp}$ with a missing $ K^0_{miss}$
(likely $ K^0_L$) is given in Fig.~\ref{four-dp}f.

The reactions $p\bar p\to \pi^+\pi^-
3\pi^0$~\cite{Amsler:1994rv}, $p\bar p\to 5\pi^0$~\cite{Abele:1996fr},
$p\bar n\to\pi^- 4\pi^0$ ~\cite{Abele:2001js} and  $p\bar n\to 2\pi^-
2\pi^0\pi^+$ ~\cite{Abele:2001pv} were studied to determine meson
decays into 4 pions. The data required two scalar states, $f_0(1370)$
and $f_0(1500)$ with masses and widths given in Table
\ref{tab:f0(1370)}. It was found that the 4$\pi$-decay width of
$f_0(1370)$ is about 6 times larger than the sum of all observed
partial decay widths to two pseudoscalar mesons; the four-pion decays
have important contributions from two pion pairs in $S$-wave and in
$P$-wave (from $\sigma\sigma$ and $\rho\rho$).  The 4$\pi$-decays of the
$f_0(1500)$ represent about half of its total width.

\begin{figure}[pb]
\bc
\includegraphics[width=80mm,height=45mm]{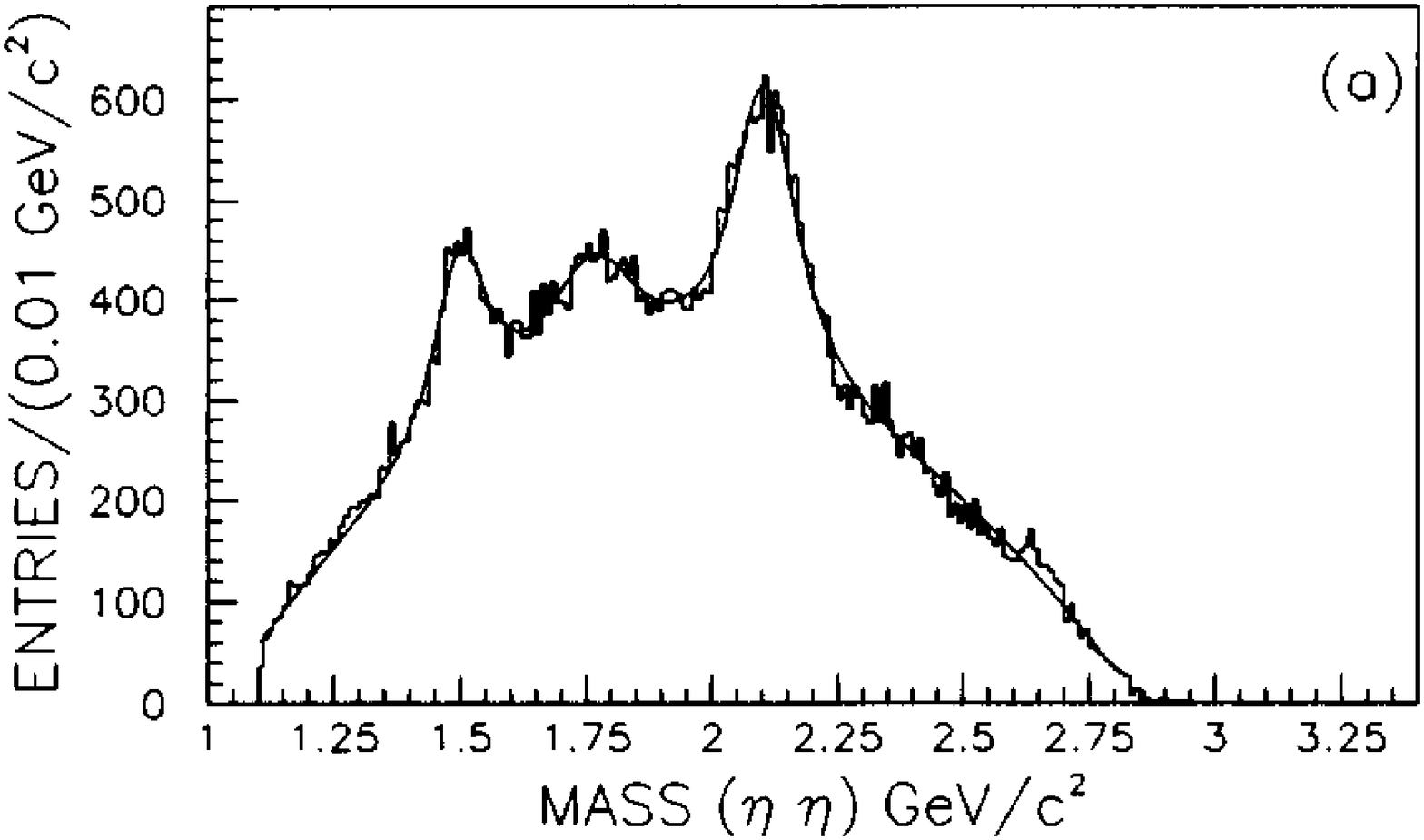}
\ec
\caption{\label{e760_etaeta}
The $\eta\eta$ mass distribution from $ p\bar p\to \pi^02\eta$ at
3\,GeV/c$^2$ cms energy \protect\cite{Armstrong:1993ey}.
}
\end{figure}

\begin{table}[pt]
\caption{\label{tab:f0(1370)}Scalar mesons $f_0(1370)$ and $f_0(1500)$
as observed in $ p\bar p$ annihilation at rest. \vspace{2mm}
}
\bc
\begin{tabular}{ccccccccc}
\hline\hline
\renewcommand{\arraystretch}{1.8}
\hspace{-2mm}$f_0(1370)$\hspace{-15mm}&M
&$\Gamma$&$f_0(1500)$\hspace{-10mm}&M
&$\Gamma$&f.s.&Ref.&Note\\
 \hline
& $1386\pm30$ &$310\pm50$ &&           &          &$ \bar n p\to5\pi$
&\cite{Gaspero:1992gu}&BW \\ & $1330\pm50$ &$300\pm80$ &&
$1500\pm15$&$120\pm25$&$ \bar
pp\to3\pi^0$&\cite{Amsler:1995gf}&Pole\\ & $1360\pm70$ &$450\pm150$&&
$1505\pm15$&$120\pm30$&$ \bar
pp\to\pi^02\eta$&\cite{Amsler:1995bz}&Pole\\ &             &
&& $1545\pm25$&$100\pm40$&$ \bar
pp\to\pi^0\eta\eta'$&\cite{Amsler:1994ah}&BW\\ & $1330\pm50$
&$300\pm80$ && $1500\pm15$&$120\pm25$&$ \bar pp\to\pi^0K\bar
K$&\cite{Abele:1996nn}& Pole \\ & $1395\pm40$ &$275\pm55$ &&
$1500\pm15$&$120\pm25$&$ \bar
pn\to\pi^-4\pi^0$&\cite{Abele:2001js}&BW\\ & $1449\pm20$ &$108\pm33$ &&
$1507\pm15$ &$130\pm20$&$ \bar pp\to3\pi$&\cite{Bertin:1997kh} &BW \\
 \hline\hline
 \renewcommand{\arraystretch}{1.0}
\end{tabular}
\ec
\end{table}
Like CERN, Fermilab developed an intense antiproton source primarily
for high-energy proton antiproton collisions. The E760/E835 experiment
used a fraction of the antiproton beam at low momenta and
a H$_2$ gas jet target, see section \ref{Experiment E835 at FNAL}.
The main motivation were studies of the charmonium states formed in
$ p\bar p$ annihilation. However, some background events were
analysed, $ p\bar p\to 3\pi^0, 2\pi^0\eta$  \cite{Armstrong:1993fh},
$\pi^02\eta$, and $3\eta$ \cite{Armstrong:1993ey}. In particular the
data on $ p\bar p\to \pi^02\eta$ show very exciting structures. The
data are shown in Fig.~\ref{e760_etaeta}. A fit yields 3 resonances
with masses and widths, respectively, of
\bc
\begin{tabular}{rrr}
   $M = 1488\pm 10$\,MeV/c$^2$  &   $M   = 1748\pm 10$\,MeV/c$^2$ & $M=2104\pm 20$\,MeV/c$^2$\\
$\Gamma = 148\pm 17$\,MeV/c$^2$ & $\Gamma = 264\pm 25$\,MeV/c$^2$ &$\Gamma = 203\pm 10$\,MeV/c$^2$.
\end{tabular}
\ec
The first peak is also seen in their $\pi^0\pi^0$ mass distribution
\cite{Armstrong:1993fh}, the other two not. The most natural
interpretation is that all three states are scalar resonances. For
scalar resonances, $\eta\eta$ decays are less suppressed by the
centrifugal barrier than tensor resonances, and thus scalars survive
more visibly (compared to the background) in their $\eta\eta$ decay
mode.

Recently, the data on $p\bar p\to\eta\eta\pi^0$ in flight were subjected
to a partial wave analysis \cite{Uman:2006xb}. Four $f_0$-states
decaying into $\eta\eta$ were reported, with masses and widths,
respectively, of ($1473\pm5$, $108\pm9$); ($1747\pm5$, $188\pm13$);
($2037\pm8$, $296\pm17$); ($2105\pm8$, $236\pm14$). The splitting of
the 2100\,MeV/c$^2$ peak into two scalar states is also suggested by
the QMC-Gatchina analysis of Crystal Barrel data on $p\bar p$
annihilation in flight for 900-1900 MeV/c momenta, reporting masses and
widths of $(2005\pm30), (305\pm50)$\,MeV/c$^2$ and $(2105\pm15),
(200\pm25)$\,MeV/c$^2$ \cite{Anisovich:2000ae}. The agreement between
these two analyses is impressive, not to say seductive; from general
arguments we would expect that two overlapping resonances decouple in
their respective decay modes.

\subsection{\label{The f0(1370) and f0(1500) resonances}
The $f_0(1370)$ and $f_0(1500)$ resonances}

\subsubsection{\label{The f0(1370)/f0(1500) partial decay widths}
The $f_0(1370)/f_0(1500)$ partial decay widths}

Table~\ref{wabr} lists ratios of partial widths of scalar
mesons as derived from $\bar pp$ annihilation at rest and from central
production. The intention is not to replace the PDG average values but
rather to test if the same particles are observed in the two reactions.
For the $f_0(1500)$, this is obviously the case. Most ratios are
reasonably consistent except $\Gamma_{\rho\rho}/\Gamma_{4\pi}$. The
Crystal Barrel value from \pbp\ annihilation is much smaller than the
one from central production (WA102). One source of the discrepancy
could be the inclusion of $\pi(1300)\pi$ and $a_1(1260)\pi$ decays which
were large for $f_0(1500)$ decays in \cite{Abele:2001js} and not
considered in \cite{Barberis:2000em}. Large inconsistencies are
observed for $f_0(1370)$. The three values of $\Gamma_{ K\bar
K}/\Gamma_{tot}$ derived from Crystal Barrel data
\cite{Bugg:1996ki,Anisovich:2002ij,Bargiotti:2003ev} scatter wildly and
are inconsistent with the WA102 result \cite{Barberis:1999cq}.
Suspicious are the small $f_0(1370)$ $\Gamma_{\eta\eta}$ and
$\Gamma_{\sigma\sigma}$ partial widths in central production.
In radiative J/$\psi$ decays into four pions (see section \ref{Scalar
mesons in radiative J/psi decays}) and in $\bar pp$ annihilation, three
scalar mesons are observed, $f_0(1500), f_0(1750), f_0(2100)$, which
all have large $\sigma\sigma$ and $\eta\eta$ couplings. It is strange
that just the $f_0(1370)$ resonance should decouple from
$\sigma\sigma$ and from $\eta\eta$ as found in central production. This
observation and the dip in Fig. \ref{wadip} seem to suggest that
$f_0(1370)$ is generated by $\rho\rho$ interaction dynamics. Background
amplitudes like the one shown in Fig.~\ref{cloki} representing
Pomeron-Pomeron scattering via exchange of isovector mesons ($\pi$ and
$\rho$ trajectories) cannot lead to $\eta\eta$ or ${\sigma\sigma}$;
these are the decay modes missing in central production of the
$f_0(1370)$.

 \begin{table}[pb] \footnotesize{
\caption{\label{wabr}Relative branching ratios of isoscalar
scalar mesons from central production (CP) and \pbp\ annihilation. \vspace{2mm}
}
\renewcommand{\arraystretch}{1.4}
\bc
\begin{tabular}{lcccc} \hline\hline
                     &\multicolumn{2}{c}{$f_0(1370)$}
                     &\multicolumn{2}{c}{$f_0(1500)$}
                      \\
                     & \pbp & CP & \pbp & CP  \\
\hline
$\Gamma_{\pi\pi}/\Gamma_{tot}$      &\multicolumn{2}{c}{$0.15\pm
0.05^1$}
                            &\multicolumn{2}{c}{$0.349\pm 0.023^2$}
                             \\
$\Gamma_{ K\bar K}/\Gamma_{tot}$  &$0.17\pm0.17^3$  & $0.07\pm0.02^4$
                            &$0.070\pm0.014^5$  & $0.09\pm0.02^6$
                      \\
$\Gamma_{\eta\eta}/\Gamma_{tot}$ & $0.022\pm0.013^7$  &
$0.004\pm0.002^8$
                        &$0.058\pm0.012^9$     & $0.063\pm0.011^{10}$
                         \\
$\Gamma_{\eta\eta^\prime}/\Gamma_{tot}$& &
                            & $0.015\pm0.004^{10}$&$0.033\pm0.009^{11}$
                             \\
$\Gamma_{4\pi}/\Gamma_{tot}$         &$0.80\pm0.05^{12}$ &
                            &$0.76\pm0.08^{13}$ & $0.48\pm0.05^{14}$
                             \\
$\Gamma_{\rho\rho}/\Gamma_{4\pi}$   &$0.26\pm0.07^{13}$&$\sim0.9^{14}$
                            &$0.13\pm0.08^{13}$ &  $0.74\pm0.03^{14}$
                             \\
$\Gamma_{\sigma\sigma}/\Gamma_{4\pi}$& $0.51\pm0.09^{13}$& $\sim 0^{14}$
                            &$0.26\pm0.07^{13}$ & $0.26\pm0.03^{14}$
                              \\

\hline\hline
\end{tabular}\vspace{2mm}
\ec
{\scriptsize
 $(1)$: educated guess,  $(2)$: from PDG, otherwise,
mean and spread of results are listed. $(3)$:
\cite{Bugg:1996ki,Anisovich:2002ij,Bargiotti:2003ev}; $(4)$:
\cite{Barberis:1999cq}; $(5)$:
\cite{Bugg:1996ki,Abele:1996nn,Abele:1998qd,Anisovich:2002ij,Bargiotti:2003ev}
$(6)$: \cite{Barberis:1999cq};
$(7)$: \cite{Anisovich:2002ij}; $(8)$: \cite{Barberis:2000cd};
$(9)$: \cite{Amsler:1995bz,Amsler:1995bf,Anisovich:2002ij};
$(10)$: \cite{Amsler:1994ah}; $(11)$: \cite{Barberis:1999id};
$(12)$: \cite{Gaspero:1992gu}; $(13)$: \cite{Abele:2001js};
$(14)$: \cite{Barberis:2000em}.
}
\renewcommand{\arraystretch}{1.0}
}
\end{table}

\subsubsection{\label{Is f0(1370) a resonance ?}
Is $f_0(1370)$ a resonance\,?}

A further point of concern is the absence of any measured $f_0(1370)$
phase motion. The scalar isoscalar amplitude of the CERN-Munich data
was already shown in Fig.~\ref{phases-4}. A rapid change of the phase
is seen at 1000 and 1500\,MeV/c$^2$; a fit with $f_0(980)$ and $f_0(1500)$
but without $f_0(1370)$ describes the data very well. The final
conclusions of the CERN-Munich analysis were phrased by Estabrooks
\cite{Estabrooks:1978de}: `we found no evidence for more than three
resonances' beyond $\sigma(485)$, $f_0(980)$ and $f_0(1500)$. The same
conclusion were obtained by Long Li, Bing-song Zou, and Guang-lie Li
\cite{Li:2002we} who decomposed the scalar amplitudes of the
CERN-Munich data into $t$-channel $\rho$ meson exchange plus
$s$-channel resonances $f_0(X)$. The $\rho$ exchange amplitude was
fixed by the isotensor $\pi\pi\to\pi\pi$ $S$-wave scattering. The fit
required $f_0(980)$, $f_0(1500)$, and a broad $f_0(1670)$ background
amplitude -- tentatively assigned to the scalar glueball -- while
$f_0(1370)$ was not needed.

The phases of the $4\pi$ scalar isoscalar partial wave was studied in
data on \pbp\ annihilation at rest into five pions, $2\pi^-2\pi^0\pi^+$
and $\pi^-4\pi^0$, respectively \cite{Abele:2001js} using a method
described in \cite{Klempt:2006xx}. In that analysis, scans were made in
which the likelihood of fits were determined when the mass of a scalar
resonance was changed in steps while all other variables were refitted.
The scalar isoscalar $4\pi$ intensity was described by one or two
Breit-Wigner resonances. The Breit-Wigner magnitude is multiplied with
a free phase factor $e^{i\phi}$ where the phase is defined relative to
the $\rho(1450)\pi$ isobar. The optimum choice of the phase is made
when the phase of the Breit-Wigner amplitude, shifted by $\phi$, agrees
with the scalar isoscalar phase of the data. Using a single
Breit-Wigner amplitude, the phase $\phi$ (represented by x in
Fig.~\ref{phases-2}a,b) follows closely the expected Breit-Wigner
shape of a $100-120$\,MeV/c$^2$ wide resonance at 1500\,MeV/c$^2$ (solid lines).
When two amplitudes are offered in the fit, the complex sum of these two
amplitudes has a phase which is reproduced in Fig.~\ref{phases-2}a as
small dots. The resulting phase adjusts itself as to mimic the
phase motion of a single resonance. Attempts were made to fit the data
with the $f_0(1370)$ amplitude being replaced by an amplitude with no
phase motion. Then, the sum of the two amplitudes for $f_0(1370)$ and
$f_0(1500)$ cannot mimic a Breit-Wigner resonance and the fit quality
is worse.

Recently, Bugg has reported a `study in depth of $f_0(1370)$'
\cite{Bugg:2007ja}. Fits to four data sets show a highly significant
improvement (in $\chi^2$) when the $f_0(1370)$ is included as
resonance. Obviously, the fit hypothesis `there is no $f_0(1370)$' at
all is worse than the hypothesis that $f_0(1370)$ exists. A study of
the width dependence shows however that the $f_0(1370)$ expected line
shape depends only marginally on its width. When the partial width
$\Gamma_{2\pi}$ of the $f_0(1370)$ was increased to 800\,MeV/c$^2$ (and
$\Gamma_{\rm tot}$ to 1\,GeV/c$^2$), $\chi^2$ of the fit changed very
little.  A very wide $f_0(1370)$ is however fully compatible with the
red dragon of Minkowski and Ochs, and with our view.

\begin{figure}[ph]
\begin{center}
\includegraphics[width=0.4\textwidth]{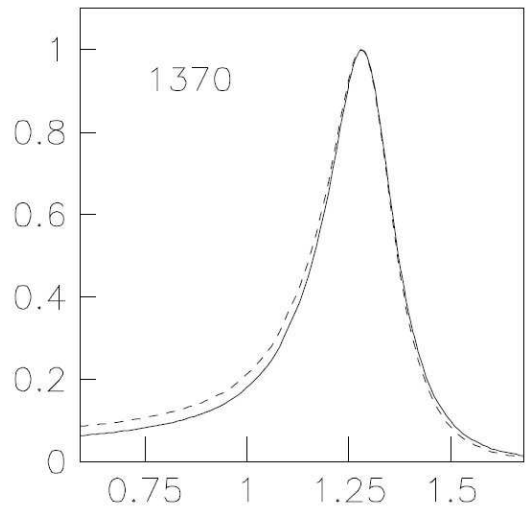}
\end{center}
\caption{\label{w1370}
Line-shapes of $f_0(1370)$ within the phase space of $p\bar p$
annihilation at rest into $3\pi^0$, normalised to 1 at its peak height,
for $\Gamma_{2\pi}=325$\,MeV/c$^2$ (solid line) and 800\,MeV/c$^2$
(dashed line) \cite{Bugg:2007ja}.} \end{figure}

As a conclusion, we do not consider the $f_0(1370)$ as established
resonance. In the discussion (section \ref{Scalar
mesons and their interpretation}) we present interpretations of the
scalar meson spectrum with and without assuming its existence.

\begin{figure}[pt]
\bc
\begin{tabular}{cc}
\includegraphics[width=50mm,height=45mm]{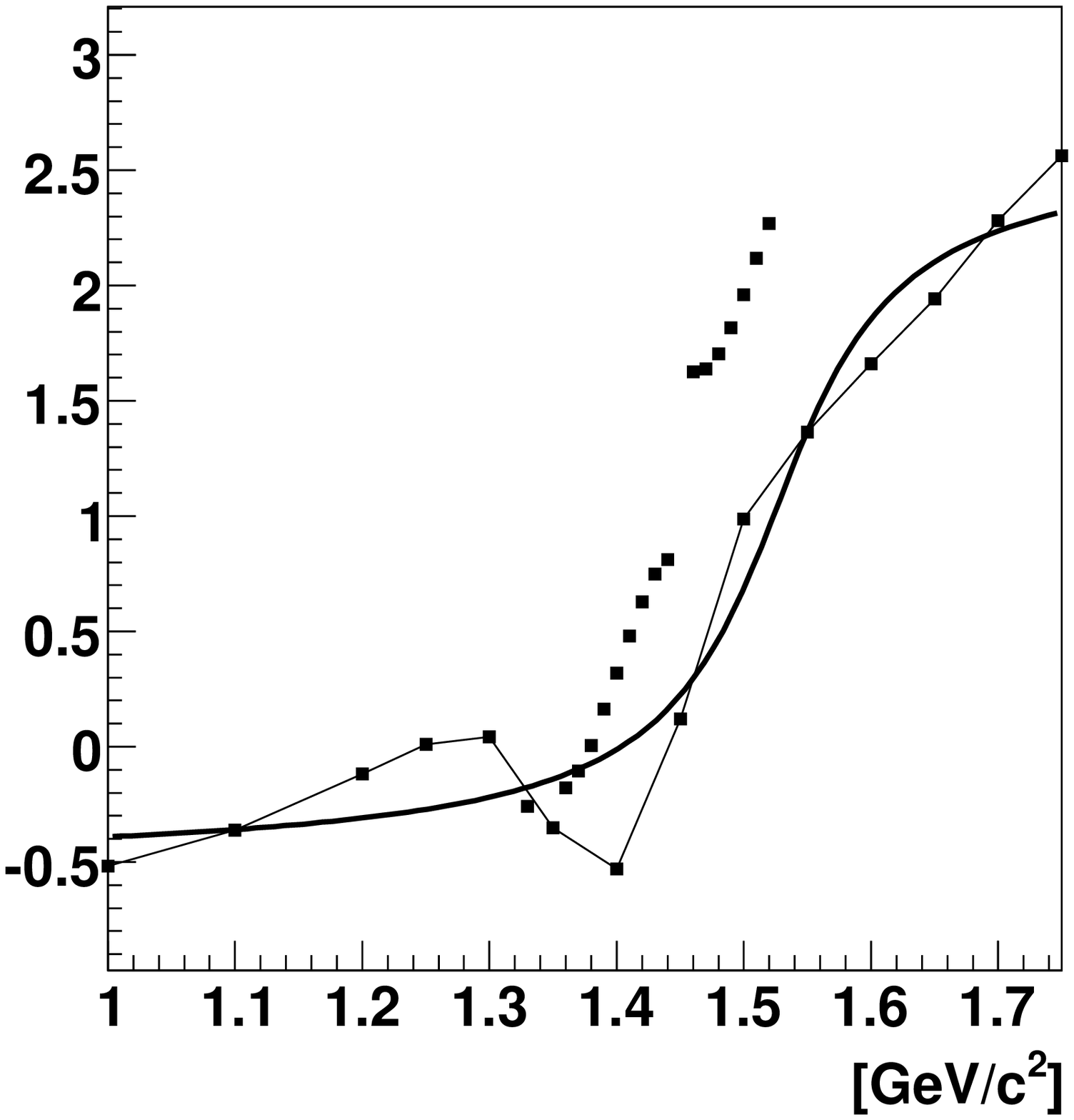}&
\hspace{-3mm}\includegraphics[width=50mm,height=45mm]{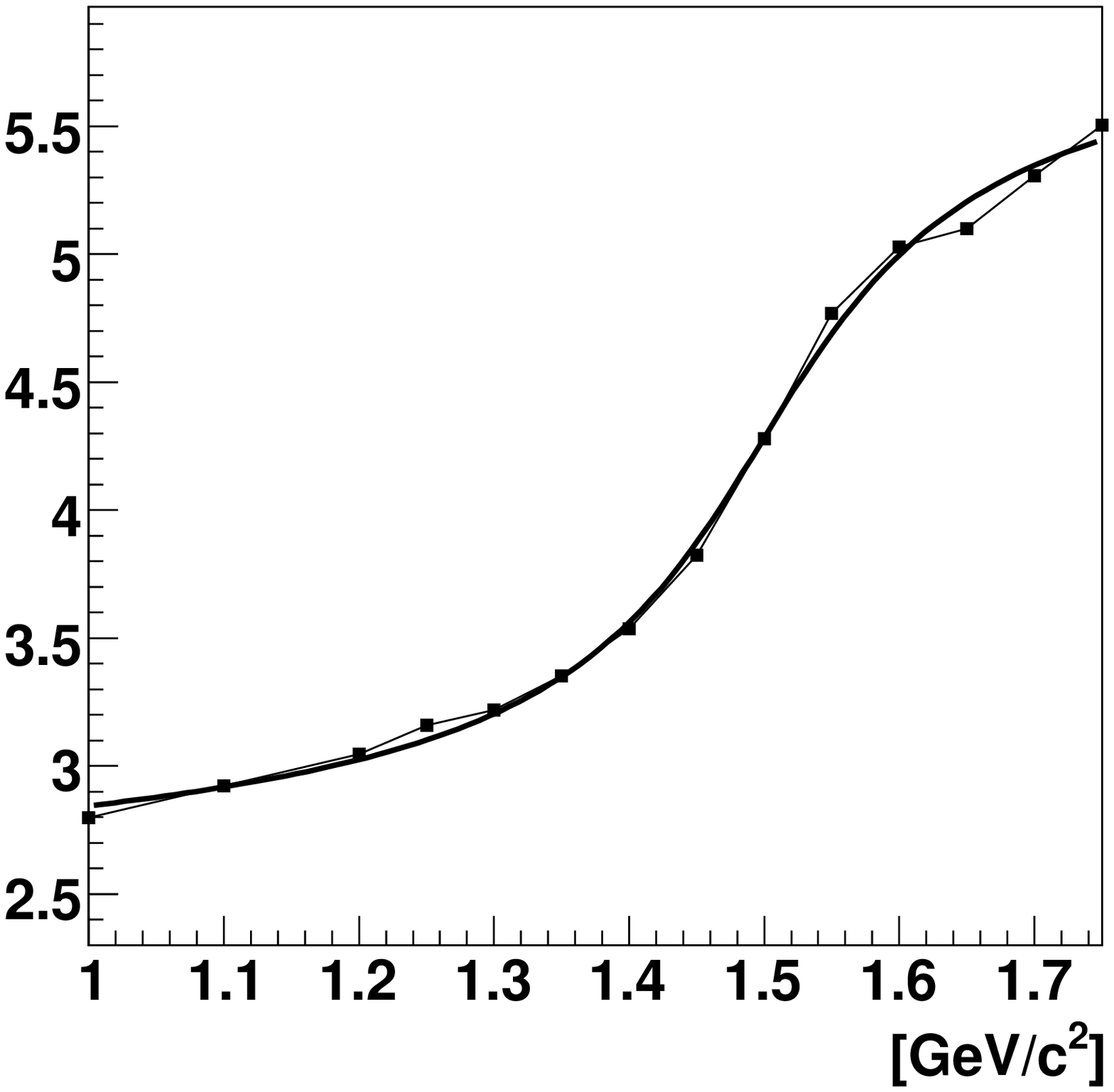}
\end{tabular}  \vspace{-40mm}\\
\hspace{-33mm}a\hspace{50mm}b\\
\vspace{31mm}
\ec
\caption{\label{phases-2}
The scalar isoscalar phase from  $ \bar pn\to
\pi^-(2\pi^0\pi^+\pi^-)_{S-wave}$ (a) and from $ \bar pn\to
\pi^-(4\pi^0)_{S-wave}$ (b). The phases are represented by x
connected with a thin line. The thick line gives the phase motion
expected for $f_0(1500)$. The small squares in (a) give the
effective phase when the scalar intensity is fitted with two
resonances.  }
\end{figure}

\subsection{\label{Scalar mesons in D, Ds and B decays}
Scalar mesons in $ D$, $ D_s$ and $ B$ decays}

\begin{figure}[pb]
\bc
\begin{tabular}{cccc}
\hspace*{-6mm}\includegraphics[width=4.5cm,height=4.2cm]{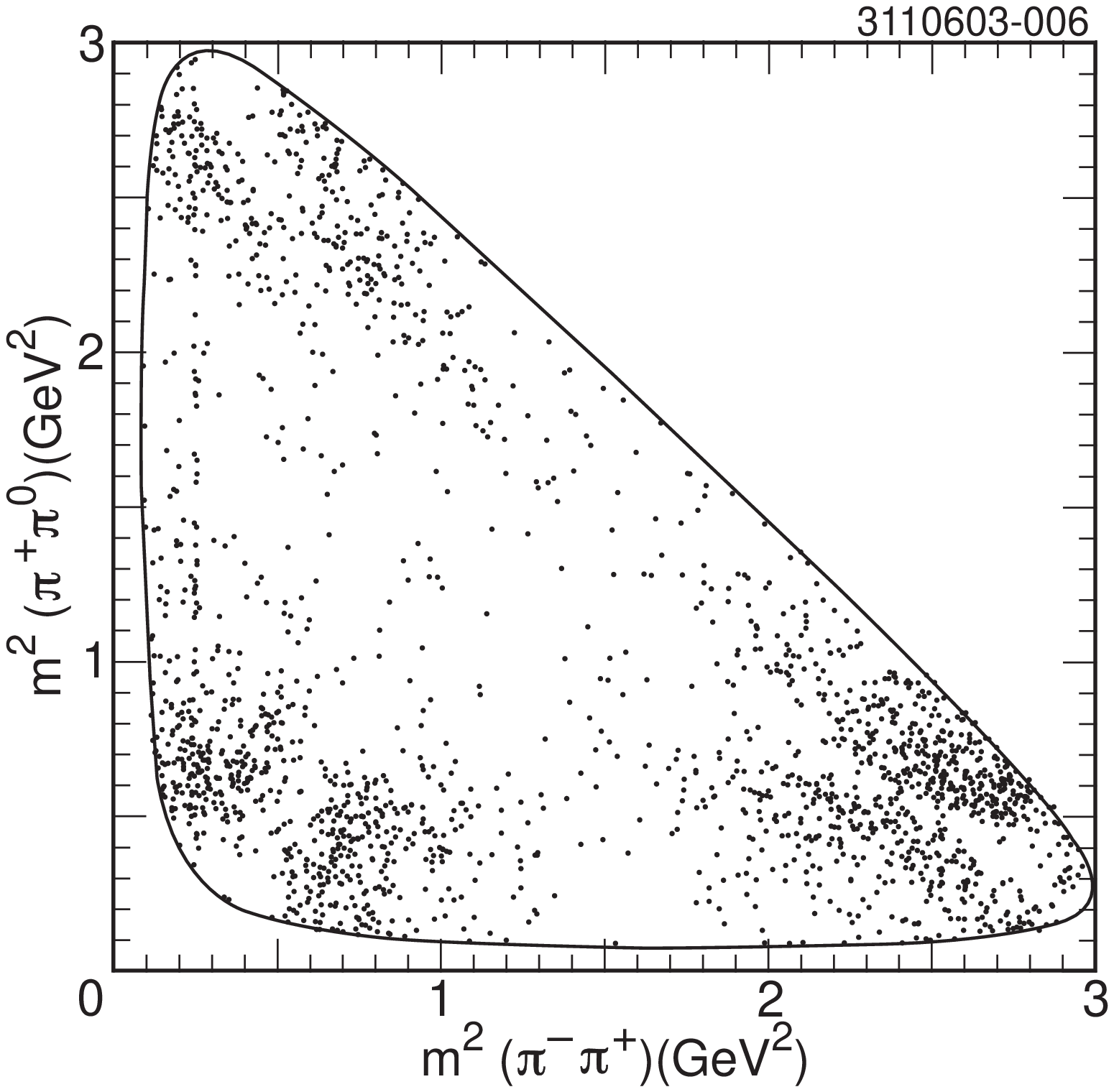}&
\hspace*{-2mm}\includegraphics[width=4.2cm]{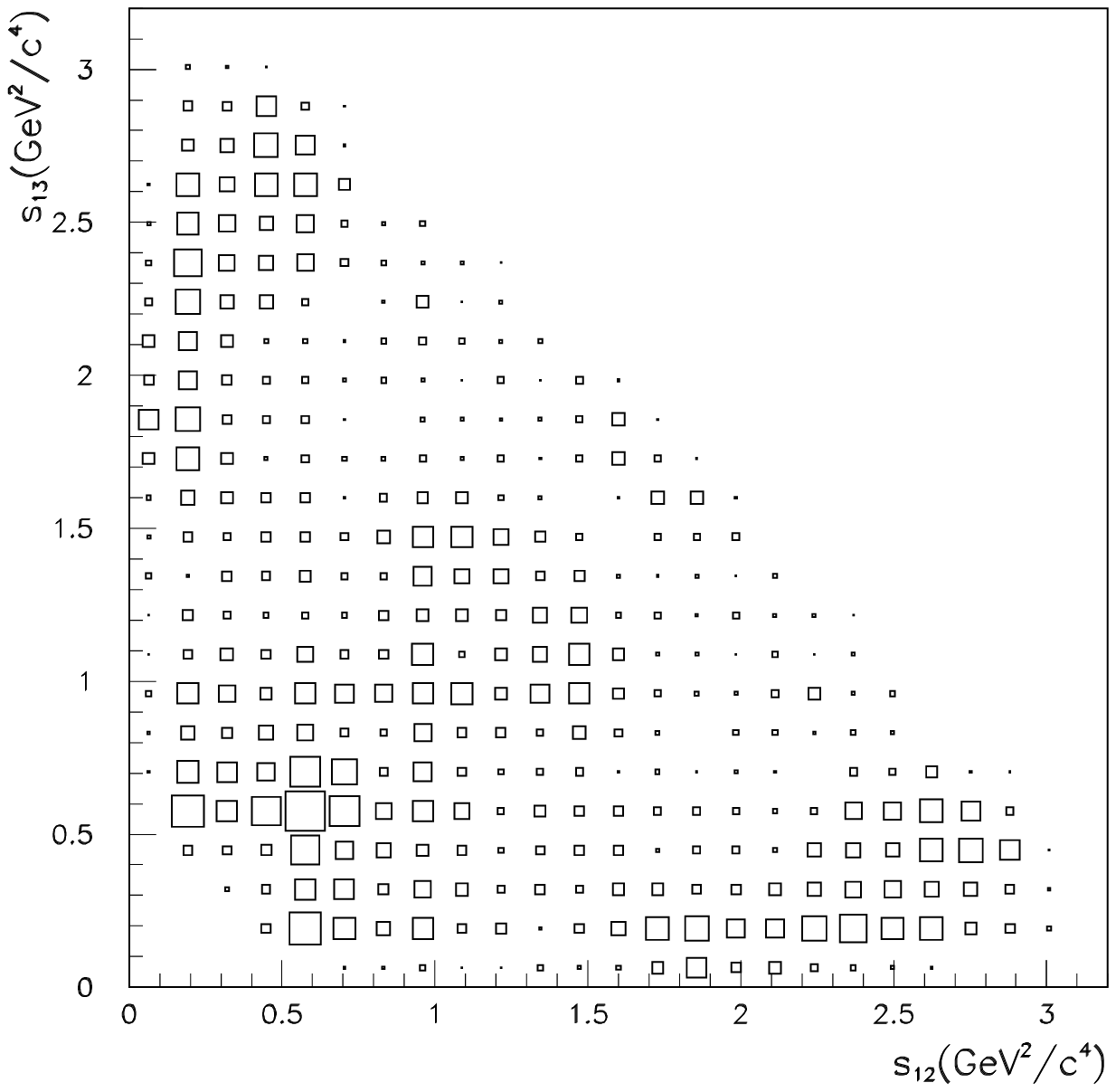}&
\hspace*{-2mm}\includegraphics[width=4.0cm]{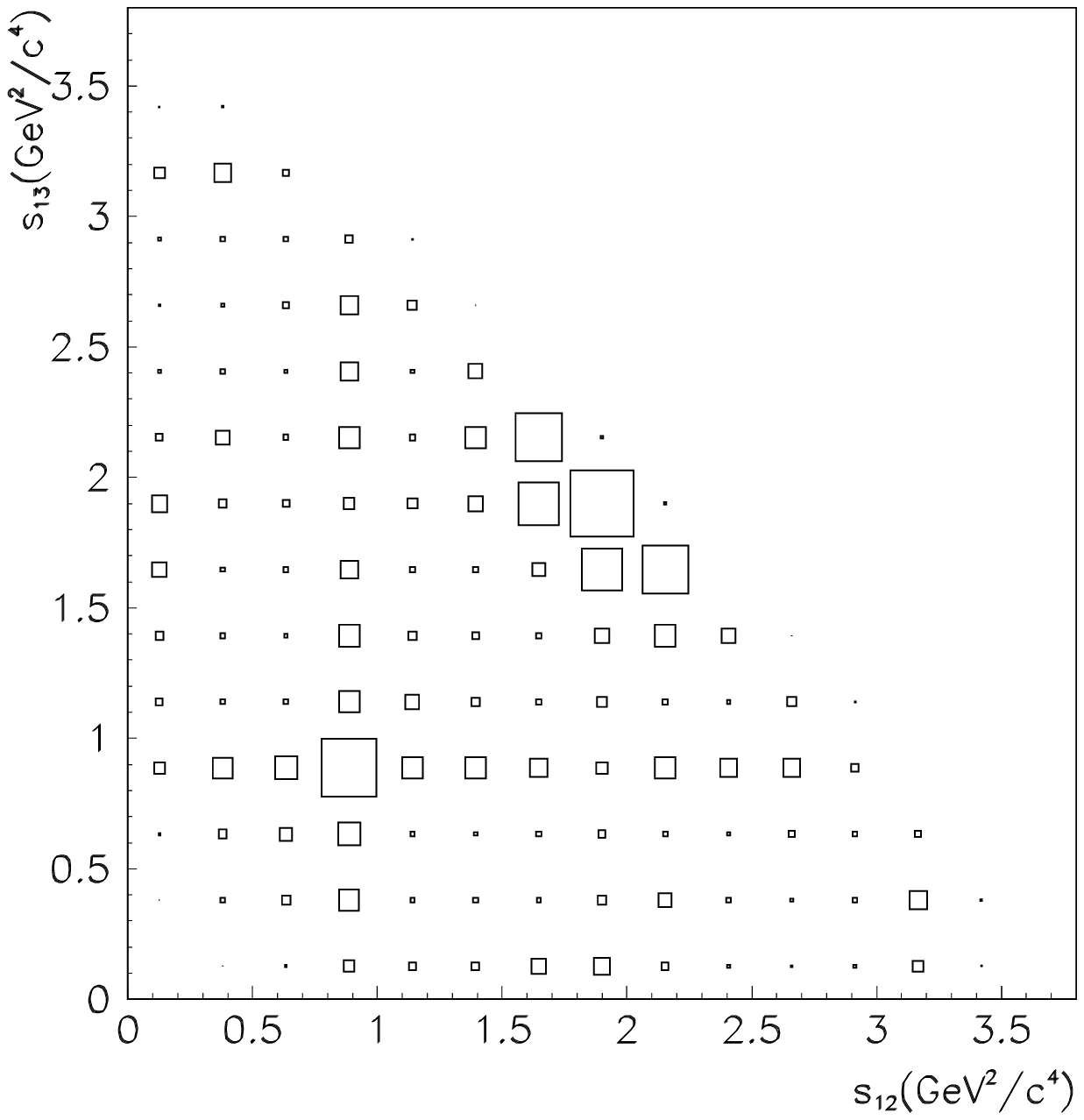}&
\hspace*{-2mm}\includegraphics[width=4.2cm]{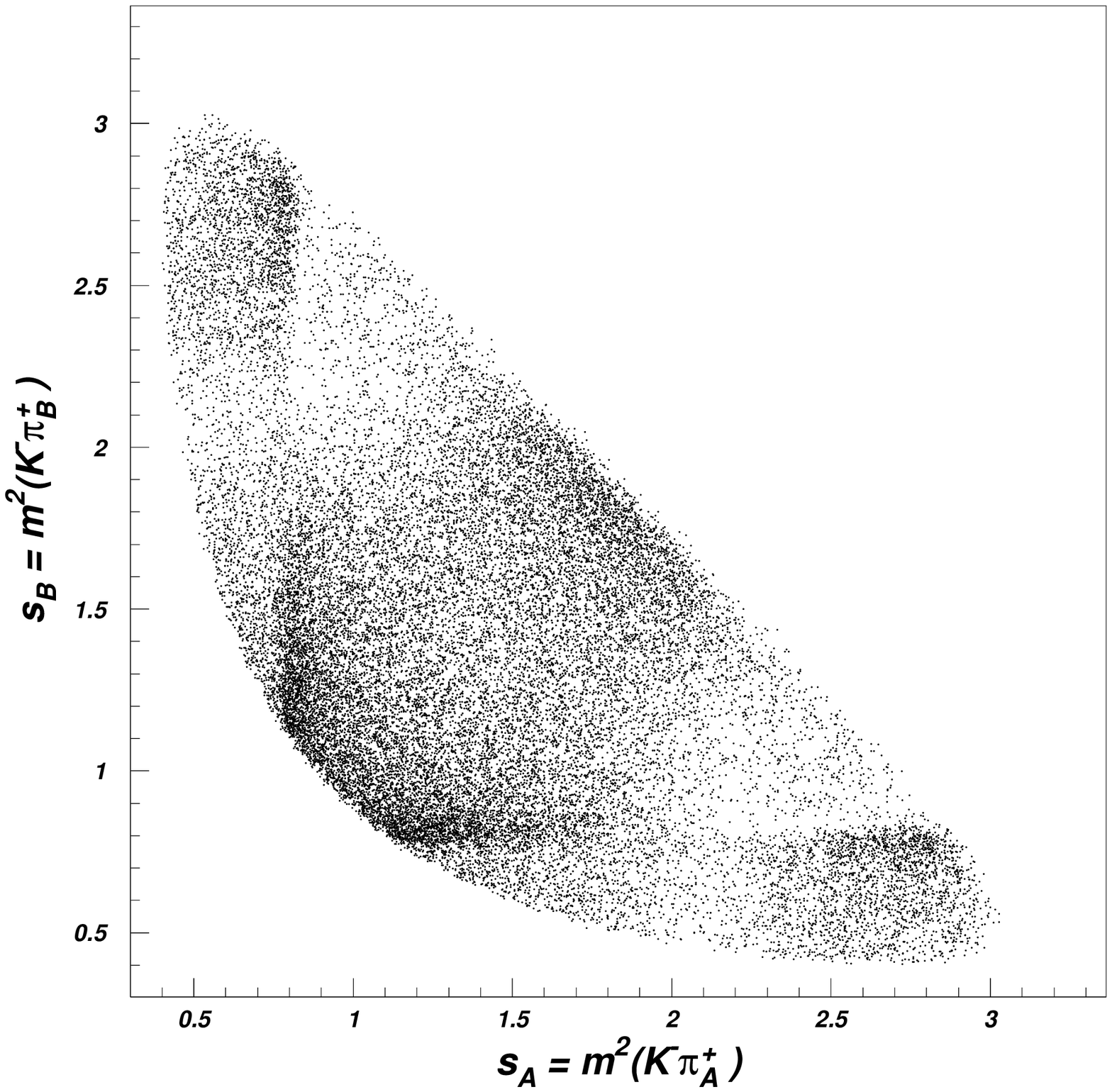}
\vspace{1mm}\\
\hspace*{-6mm}\includegraphics[width=4.1cm]{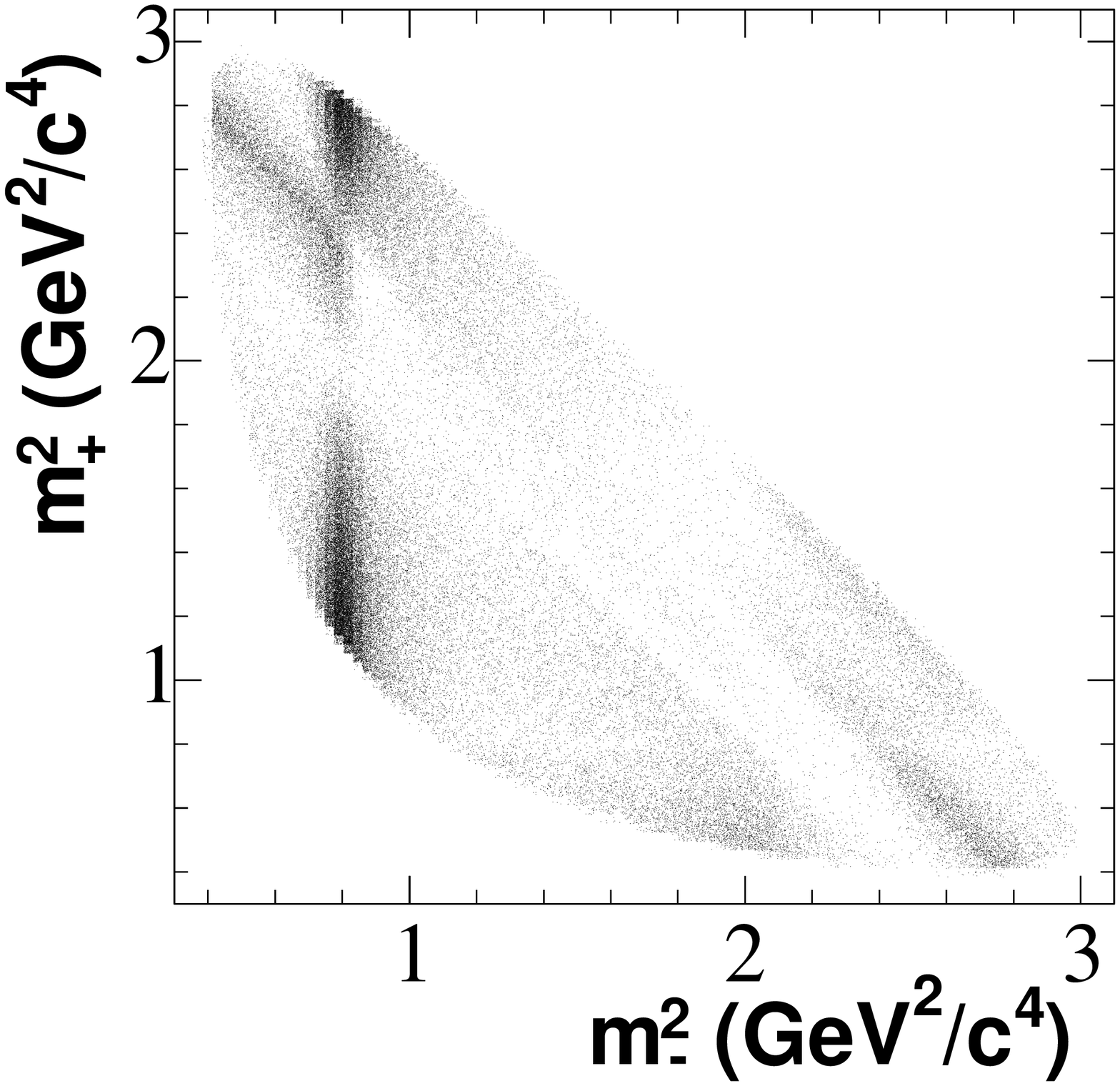}&
\hspace*{-2mm}\includegraphics[width=4.3cm]{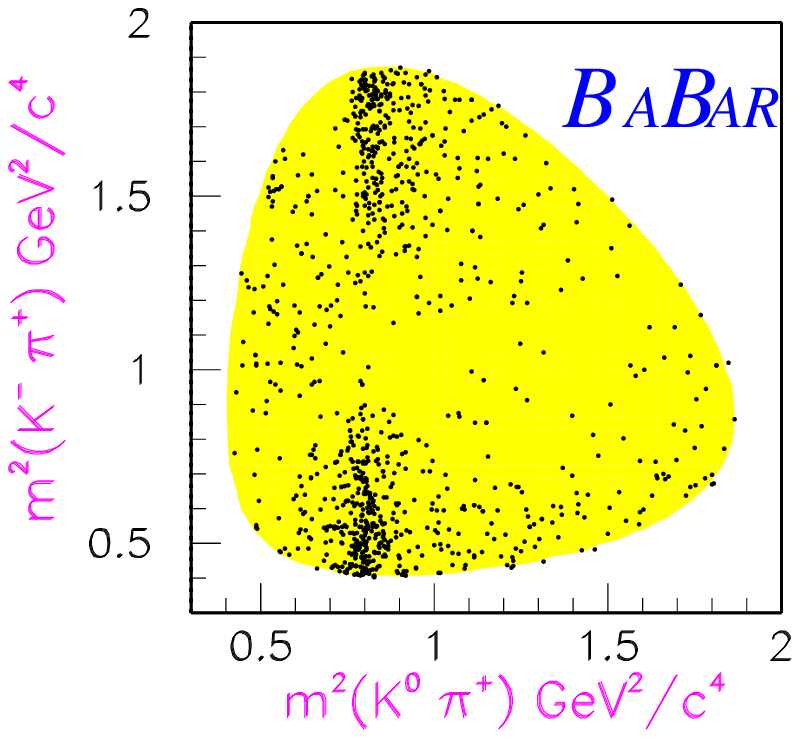}&
\hspace*{-2mm}\includegraphics[width=4.1cm,height=4.0cm]{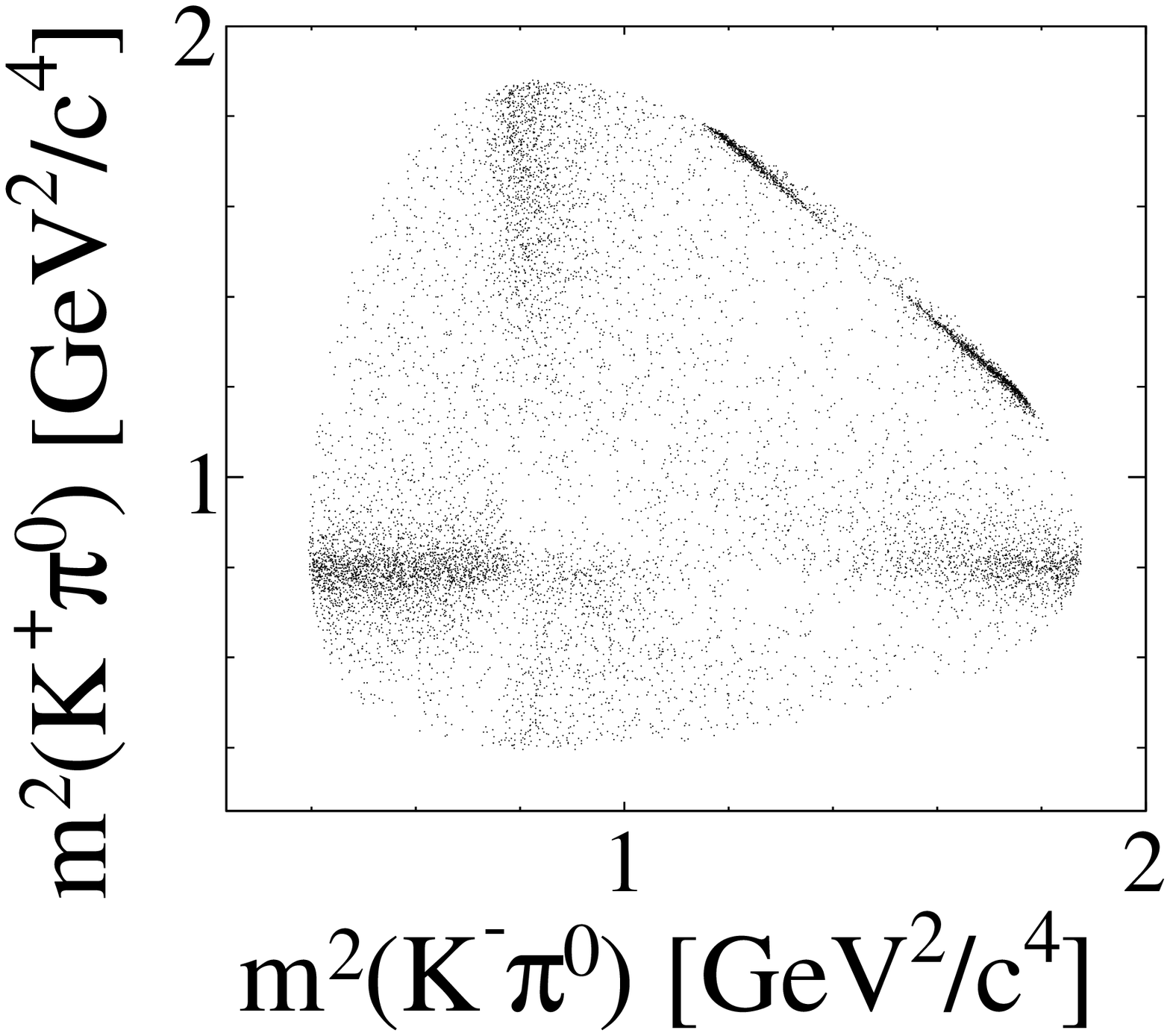}&
\hspace*{-2mm}\includegraphics[width=4.1cm]{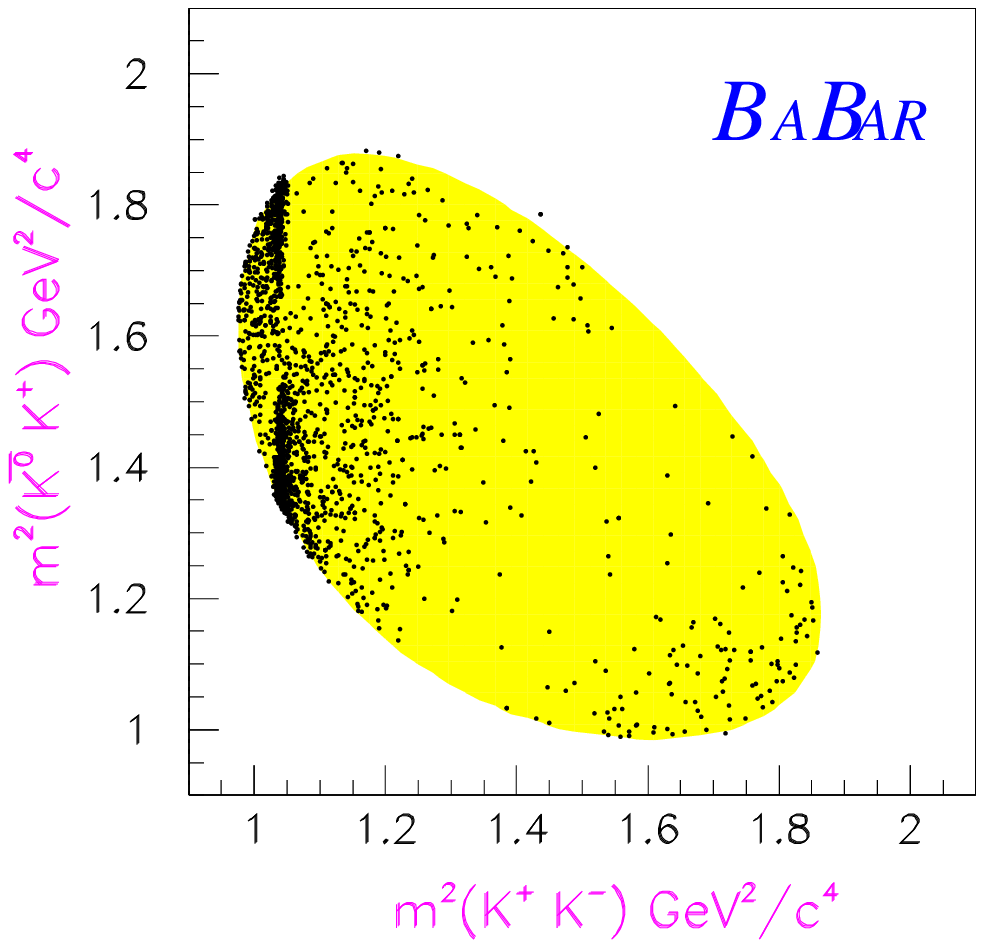}
\end{tabular}\vspace{-85mm}\\
\hspace{34mm}a\hspace{43mm}b\hspace{42mm}c\hspace{43mm}d\vspace{40mm}\\
\hspace{33mm}e\hspace{44mm}f\hspace{42mm}g\hspace{44mm}h\vspace{33mm}
\ec
\caption{\label{D-decays}Dalitz plots for decays.
$ D^0$ (a), $ D^+$ (b) and $ D_s^+$ (c) decays into $3\pi$;
\protect\cite{Frolov:2003jf,Aitala:2000xu,Aitala:2000xt};
$ D^+ \to  K^- \pi^+ \pi^+$ (d), \protect\cite{Aitala:2005yh}.
$ D^0\to K^0_S \pi^+\pi^-$ (e) \protect\cite{Aubert:2005iz};
$ D^0 \to K^0 K^- \pi^+$ (f) \protect\cite{Aubert:2002yc};
$ D^0 \to \overline{K}^0 K^+ \pi^-$
(g) \protect\cite{Collaboration:2007dc}; $ D^0 \to \overline{K}^0 K^+
K^-$ (h) \protect\cite{Aubert:2005sm}. } \end{figure} Fig.
\ref{D-decays} shows a series of $D^0$, $D^+$, and $ D_s^+$ Dalitz
plots for different decay modes. The comparison of  $D^0$, $D^+$, and $
D_s^+$ decays into three pions reveals an interesting but not yet fully
understood phenomenology. A summary of Dalitz plot analyses is given by
Asner in the Review of Particle Properties \cite{Eidelman:2004wy}.
Analyses of the same Dalitz plot made in different collaborations
mostly arrive at similar conclusions concerning contributing resonances
and their fractional contributions except for some -- partly important
-- details. We give here a description of a few Dalitz plots to
underline their potential for meson spectroscopy.

In $ D^0\to 3\pi$ decays, a $c$ quark converts into a $d$ quark, the
transition is Cabibbo suppressed. The CLEO Dalitz plot
\cite{Frolov:2003jf} shows significant structures due to $\rho^+\pi^-$
production and weaker contributions from $\rho^0\pi^0$ and
$\rho^-\pi^+$. There are depletions along the three diagonal lines
separating the three charge states of the $\rho$ indicating destructive
interference between the three amplitudes at appropriate mass values.
The dominance of $\rho^+\pi^-$ is unexpected; the leading diagram (see
Fig. \ref{fig:d-decay}) produces $\rho^-\pi^+$, the annihilation
diagram cannot contribute, colour suppressed diagrams lead to equal
production of all three charge states. Possibly, constructive and
destructive interference between leading and non-leading diagrams are
responsible for the inversion of the expected yields. So far, CLEO is
the only experiment which reported results from this reaction, and only
in conference proceedings.

The leading diagram for $ D^+\to 2\pi^+\pi^-$ decays produces a $\pi^+$
and a neutral meson. In the Dalitz plot of E791 at Fermilab
\cite{Aitala:2000xu}, bands due to $\rho(770), f_0(980)$ and
$f_2(1270)$ can be recognised. The $\rho$ band is depleted in the
centre due to the $\cos^2\theta$ angular distribution expected for a
transition from a pseudoscalar initial state to a vector plus
pseudoscalar meson. The asymmetry along the angular distribution points
to a substantial interference between even and odd partial waves. The
partial wave analysis found strong evidence for the $\sigma(485)$ with
mass $478^{+24}_{-23} \pm 17$ MeV/$c^2$ and width $324^{+42}_{-40} \pm
21$ MeV/$c^2$ accounting for approximately half of all decays.
Recently, the CLEO collaboration performed a Dalitz plot study of
$D^+\to \pi^+\pi^+\pi^-$ decays \cite{Bonvicini:2007tc} using different
S-wave parametrisations. They confirm the large $\sigma(485)\pi^+$
yield ($41.8\pm2.9$\%). The isobar fit used in addition $\rho^0\pi^+$
($20.0\pm2.5$\%), $f_2(1270)\pi^+$ ($18.2\pm2.7$\%), and
$f_0(980)\pi^+$ ($4.1\pm0.9$\%). The region above 1\,GeV/c$^2$ was
described by two Breit-Wigner amplitudes for $f_0(1370)$ and
$f_0(1500)$ contributing with $2.6\pm1.9$\% and $3.4\pm1.3$\%,
respectively. The existence of $f_0(1370)$ was not questioned; when its
mass was introduced as a free fit parameter, $M=1260$\,MeV/c$^2$ was
found.

A surprise in the $D^+_s\to\pi^+\pi^+\pi^-$ Dalitz plot shown in Fig.
\ref{D-decays}c is the large yield and the large scalar intensity
\cite{Frabetti:1997sx}. In $D_s^+ =c\bar s$ decays, the leading
contribution stems from the process shown in Fig. \ref{fig:d-decay}a in
which the $c$ quark converts -- by emission of a $W^+\to\pi^+$ -- into
an $s$ quark forming an $s\bar s$ state. However, the $\pi^+$ recoils
against a $\pi^+\pi^-$ (and not the expected $K^+K^-$) pair.
Obviously, some resonances offer a bridge from primary $s\bar s$ states
to pionic final states. Such a behaviour is known from pseudoscalar
mesons which have large $n\bar n$ and $s\bar s$ components. Likely,
scalar mesons have flavour wave functions which do not correspond to an
ideal mixing angle. A more subtle interpretation of the $s\bar s\to
\pi\pi$ transition is discussed in section \ref{Our own interpretation}
where we give our own interpretation of the scalar mesons. The
$\sigma(485)$ is not required in fits to $ D^+_s$ decays; it is
generated dynamically from $\pi\pi$ interactions and seems to have weak
coupling to $s\bar s$; $\rho$ production is nearly absent.

The largest contribution stems visibly from $f_0(980)\pi^+$. There is a
second structure at about 2\,GeV$^2$/c$^4$ which partial wave analyses
assign to $\rho(1450)\pi^+$, $f_2(1270)\pi^+$, and $f_0\pi^+$. A very
important aspect is the mass of the scalar resonance. E687
\cite{Frabetti:1997sx} determined the scalar meson mass to
1475\,MeV/c$^2$ with 100\,MeV/c$^2$ width, Focus \cite{Malvezzi:2003jp}
found $M=1475\pm 10$\,MeV/c$^2$ and $\Gamma=112\pm 24$\,MeV/c$^2$.
These results are not incompatible with the standard $f_0(1500)$. E791
\cite{Aitala:2000xt} imposed the $f_0(1370)$ mass; when left free, the
parameters optimised for $M=1434\pm 18\pm 9$\,MeV/c$^2$ and
$\Gamma=172\pm 32\pm 6$\,MeV/c$^2$. In \cite{Link:2003gb}, four scalar
mesons above 1\,GeV/c$^2$ were used to describe the data, $f_0(1370)$,
$f_0(1500)$, $f_0(1750)$, and a wide $f_0(1200-1600)$. It is thus an
open question if the enhancement at $M^2\sim 2$\,GeV$^2$ in $ D^+_s\to
2\pi^+\pi^-$ is due to $f_0(1370)$ or $f_0(1500)$ production, to
production of both, or due to some dynamical effect. The reason why
$\rho(1450)$ makes a large contribution to $D_s$ decays (and
$\rho(770)$ at most weakly) while $\rho(770)$ is observed in $D$ decays
(and $\rho(1450)$ not) is not understood.

The Focus collaboration reported an interesting analysis of
both data sets, $D^+$ and $ D_s^+$ decays into $2\pi^+\pi^-$ using the
same resonances \cite{Malvezzi:2003jp}. An equivalent fit using a $K$
matrix for the S-wave was also used, but this approach does not give
easy access to the fractional yields of S-wave resonances. The Focus
data were also fitted \cite{Link:2003gb} using a highly flexible set of
amplitudes proposed by Anisovich and Sarantsev \cite{Anisovich:2002ij};
the latter results will be discussed in section \ref{The
Anisovich-Sarantsev picture}.

The $D^+_s\to \pi^- \pi^+ \pi^+$ Dalitz plot of the E791 collaboration
\cite{Aitala:2000xt} was reanalysed using $K$-matrices in $P$-vector
approach, or Breit-Wigner amplitudes and a Flatt\'e parametrisation to
describe $f_0(980)$ \cite{Klempt:2007ms}. As in other analyses,
$f_0(980)$ resonance provided the most significant contribution. The
emphasis was laid on the question if $f_0(1370)$ is required to
describe the data. The answer was no. When $f_0(1500)$ and $f_0(1710)$
were used in the fit, the uncertainty in the mass increased and
$M=1458\pm 37$\,MeV/c$^2$ was found which is $1.2\sigma$ compatible
with the PDG mass.

In Table~\ref{D-and-D-s} results are collected of the analysis
\cite{Malvezzi:2003jp} using Breit-Wigner amplitudes to describe the
dynamics of the process. The physical content of the results are
similar to those obtained by E687 \cite{Frabetti:1997sx} even though
some numbers are different (in particular a larger fraction of the data
was assigned to a nonresonant (NR) amplitude in \cite{Frabetti:1997sx}.
The results of \cite{Aitala:2000xt} and \cite{Aitala:2000xu} are
presented as small numbers in Table~\ref{D-and-D-s}. Recently, CLEO
reported a preliminary analysis of $D^+\to 3\pi$
\cite{Bonvicini:2006ge} with different numbers but similar results. In
an analysis of  $D^0\to K^+ K^- \pi^0$ decays \cite{Cawlfield:2006hm},
the existence of both, $f_0(1370)$ and $f_0(1500)$ is assumed but not
explored further.

\begin{table}[pt]
\caption{\label{D-and-D-s}
$ D^+$ and $ D^+_s$ fit fractions and phases with the isobar model
\cite{Malvezzi:2003jp}. The results from
\cite{Aitala:2000xu,Aitala:2000xt} are shown as small numbers for
comparison. } \vspace{2mm} \bc \renewcommand{\arraystretch}{1.8}
{\footnotesize \begin{tabular}{ccccc} \hline\hline
   &  \multicolumn{2}{c}{$ D^+\to 2\pi^+\pi^-$} &
\multicolumn{2}{c}{$ D^+_s\to 2\pi^+\pi^-$} \\
\hline
resonance & fit  fraction(\%) & phase (deg) & fit fraction(\%) & phase (deg) \\
\hline
NR&9.8 $\pm$ 4.3 &145.9 $\pm$ 17.7 & 25.5 $\pm$ 4.6 & 246.5 $\pm$
4.7\vspace{-2mm} \\
&{\scriptsize   7.8 $\pm$ 6.6 }&{\scriptsize 246.5 $\pm$ 20.3 }&{\scriptsize 0.5 $\pm$
2.2}&{\scriptsize 181 $\pm$ 107} \vspace{-2mm} \\
$\rho(770)\pi^+$ & 32.8 $\pm$ 3.8 & 208.8 $\pm$ 16.8   & - \vspace{-3mm} \\
&{\scriptsize  33.6 $\pm$ 3.9 }& 0 (fixed)&{\scriptsize 5.8$\pm$ 4.4}&{\scriptsize 109 $\pm$ 25}\vspace{-2mm} \\
$\rho(1450)\pi^+$ & -  &  &
4.1 $\pm$ 1.0 & 187.3 $\pm$ 15.3\vspace{-3mm}\\ &{\scriptsize  - }&
&{\scriptsize 4.4$\pm$ 2.1}&{\scriptsize 162 $\pm$ 31}\vspace{-2mm} \\
$f_2(1275)\pi^+$&12.3 $\pm$2.1&146.7$\pm$17.7&9.8$\pm$ 1.3&140.2 $\pm$
9.2\vspace{-3mm} \\ &{\scriptsize 19.4$\pm$ 2.6}&{\scriptsize 246.5
$\pm$ 6.0}&{\scriptsize 19.7$\pm$ 3.4} &{\scriptsize  133 $\pm$ 31 }
\vspace{-2mm} \\ $f_0(485)\pi^+$ & 18.9 $\pm$ 5.3 &  49.0 $\pm$ 30.7  & -
\vspace{-3mm} \\
 &{\scriptsize 46.3 $\pm$ 9.3} &{\scriptsize
40.7 $\pm$ 9.5 } & - \vspace{-2mm} \\ $f_0(980)\pi^+$   & 6.7 $\pm$ 1.5 &
0(fixed) & 94.4 $\pm$ 3.8&0(fixed) \vspace{-3mm} \\ &{\scriptsize 6.2
$\pm$ 1.4} &{\scriptsize 0(fixed)} & {\scriptsize 56.5 $\pm$
6.4}&{\scriptsize 0(fixed)} \vspace{-2mm} \\ $f_0(1475)\pi^+$   & 1.8 $\pm$
1.2 &  28.2 $\pm$ 25.8  & 17.4 $\pm$  3.1 & 249.7 $\pm$ 6.4
\vspace{-3mm} \\ {\scriptsize$f_0(1370)\pi^+$}&{\scriptsize  2.3 $\pm$ 1.8}&
{\scriptsize 270.4 $\pm$ 17.8 } & {\scriptsize 32.4 $\pm$ 7.9} &
{\scriptsize 198 $\pm$ 33 } \\ \hline\hline \end{tabular}}
\renewcommand{\arraystretch}{1.5}
\ec
\end{table}

The data on $ D^+ \to  K^- \pi^+ \pi^+$ \cite{Aitala:2005yh} shown in
Fig. \ref{D-decays}d were used to develop a model-independent
partial-wave analysis of the S-wave component of the $K\pi$ system. The
amplitudes were determined for ranges of $K^-\pi^+$ invariant mass down
to threshold. This was an essential extension of the LASS data and thus
helped to extract the $K_0^*(750)$ discussed in section
\ref{Kpi scattering} (see also \cite{Aubert:2005yj}).

The four BaBar Dalitz plots on $D^0$ decays in the lower row evidence
the importance of $S$-$P$ interference and the potential of charm
decays for light meson spectroscopy. The $ D^0\to K^0\pi^+\pi^-$ Dalitz
plot reveals strong interference effects between $K^*(892)$ and
$(K\pi)_{S-{\rm wave}}$ interfering with $\rho$ and the $
(\pi\pi)_{S-{\rm wave}}$ \cite{Aubert:2005iz}. The decays $D^0\to
K^0K^-\pi^+$ were shown at a conference \cite{Aubert:2002yc}; no visible
structure except $K^*(892)$ production is observed. There are hints for
interference with the $K\pi$ S-wave. The $ D^0\to K^+K^-\pi^0$ Dalitz
plot \cite{Collaboration:2007dc} shows significant contributions from
$K^*(892)$ and $\phi(1020)$.  The fits assigned the largest fraction to
the $(K\pi)_{S-wave}$. The LASS parametrisation and a fit using the
$(K\pi)_{S-wave}$ from \cite{Aitala:2005yh} both resulted in
acceptable fits. Finally, the reaction $D^0\to K^0K^-K^+$
\cite{Aubert:2005sm} shows strong $K^0\phi(1020)$ production
interfering with $K^0a_0(980)^0$. The yield of $K^-a_0(980)^+$ is
significant, the two doubly-Cabibbo suppressed decay modes $K^+a_0
^-(980)$ and $\bar K^0f_0^0(980)$ make no significant contributions.
There is some additional contribution, which can be described by the
tail of a broad resonance peaking well outside the phase space.

\begin{figure}[pt]
\begin{center}
\includegraphics[width=17cm]{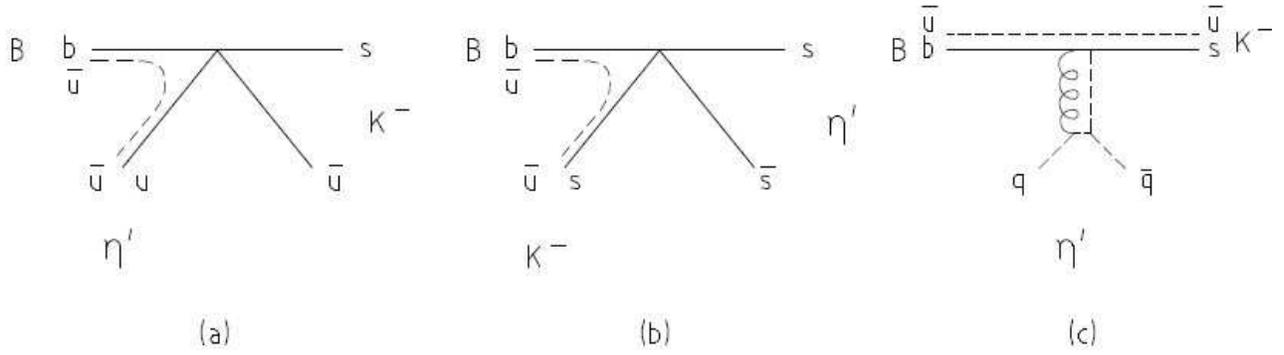}
\vspace{-2mm}\end{center}
\caption{\label{Ketaprime}
Two-body decay $ B^-\to K^-\eta'$ with amplitudes in which
transitions take place with (a) $s\to K^-$, (b) $s\to \eta'$ and (c)
$ K^-\eta'$ production via intermediate gluons. The dashed lines
represent 'soft' QCD \cite{Minkowski:2004xf}. }
\end{figure}

Decays of $ B$ mesons offer a wide phase space at the expense of small
event numbers per MeV.  Charmless decays may proceed via
the Penguin diagram in which the $b$ quark converts under $W$ emission
into a $u,c$, or $t$ quark. The latter quark reabsorbs the $W$ turning
into an $s$ (or $d$) quark. Based on a comparison of data on $ B$
decays into two pseudoscalar mesons, Minkowski and Ochs argue that a
large fraction of $\eta^{\prime}$ production has to proceed via a
gluonic amplitude (see Fig.~\ref{Ketaprime}c). This amplitude is very
similar to the one needed in radiative J/$\psi$ production.

The BABAR and BELLE collaboration have analysed the Dalitz plots for the
three-body charmless decays  $ B^+\to K^+K^+K^-$ and  $ B^+\to
K^+\pi^+\pi^-$ \cite{Garmash:2004wa}, and for $ B^0\to K^+ K^-K^0_S$
and $ B^0\to K^0_S \pi^+ \pi^-$
\cite{Aubert:2005kd,Aubert:2005wb}. These data, even though low in
statistics, provide an important key to the dynamics of scalar mesons.
The BELLE data are based on 152 million $B\bar{B}$ pairs produced on
the $\Upsilon(4S)$ resonance. Three significant structures were
observed, the $f_0(980)$ and a small enhancement in the $\pi\pi$ mass
distribution at about $1.3$~GeV/c$^2$, denoted as $f_X(1300)$, and resonant
structure at $1.5$~GeV/c$^2$ in the $ K\bar K$ mass distribution. The
latter is best described as a scalar resonance; mass and width are
compatible with standard values for $f_0(1500)$.

Now, there is a problem: if the $f_0(1500)$ were the same particle
as observed in other experiments, it would decay to $\pi^+\pi^-$ about
five times more frequently than to $ K^+K^-$. This is not observed. In
the $\pi^+\pi^-$ mass distribution there is, instead, a peak at
1300\,MeV/c$^2$. Both these aspects were confirmed in the BABAR data.
The BABAR collaboration concludes that the nature of the $f_0(1500)$
contribution remains unclear. Identification of this state as the
$f_0(1500)$ leads to inconsistency with the measurement of the $ B^0
\to f_0(1500) K^0_S, f_0(1500) \to \pip\pim$ decay.

\begin{figure}[pb]
\begin{center}
\includegraphics[width=0.48\textwidth]{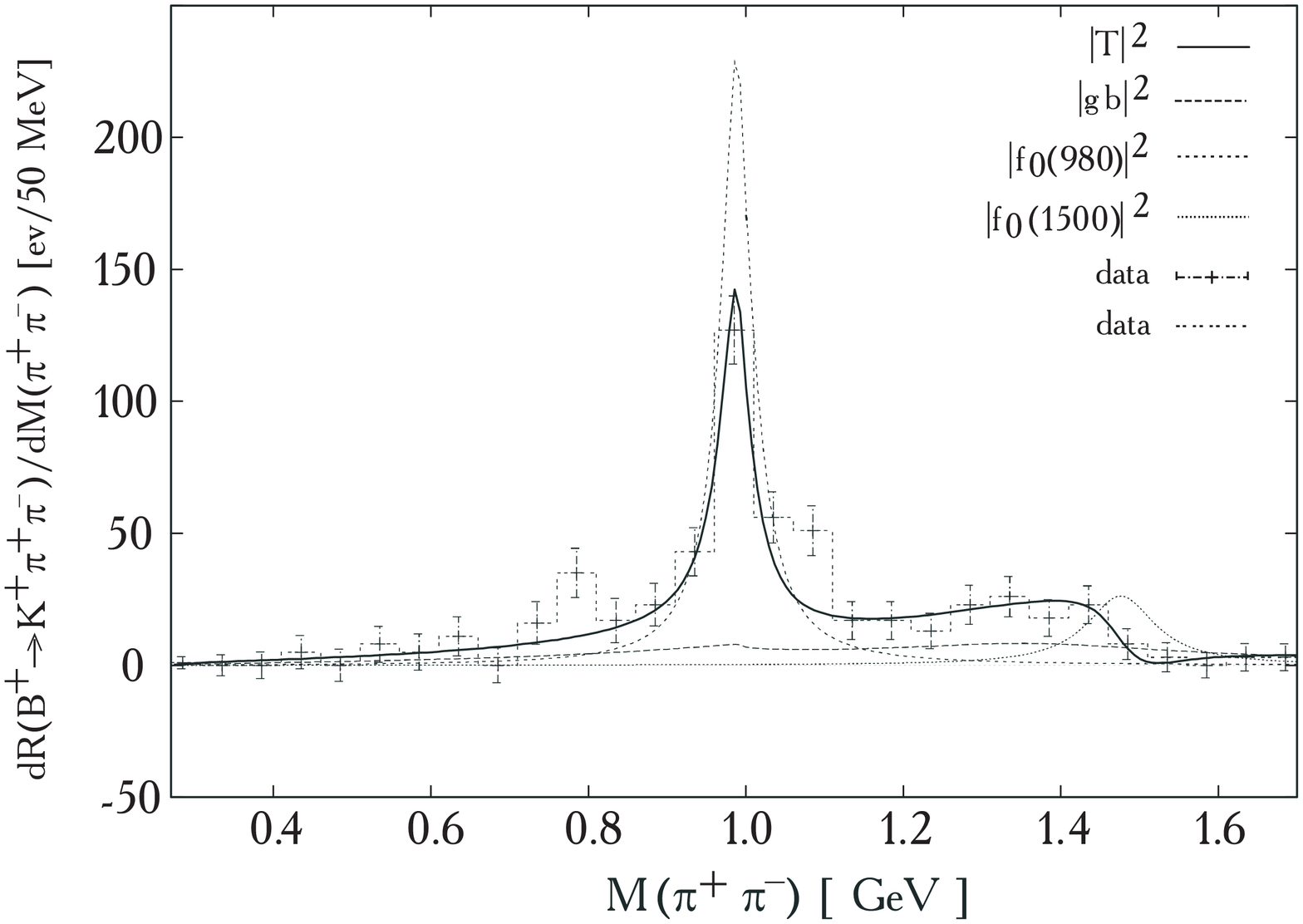}
\includegraphics[width=0.48\textwidth]{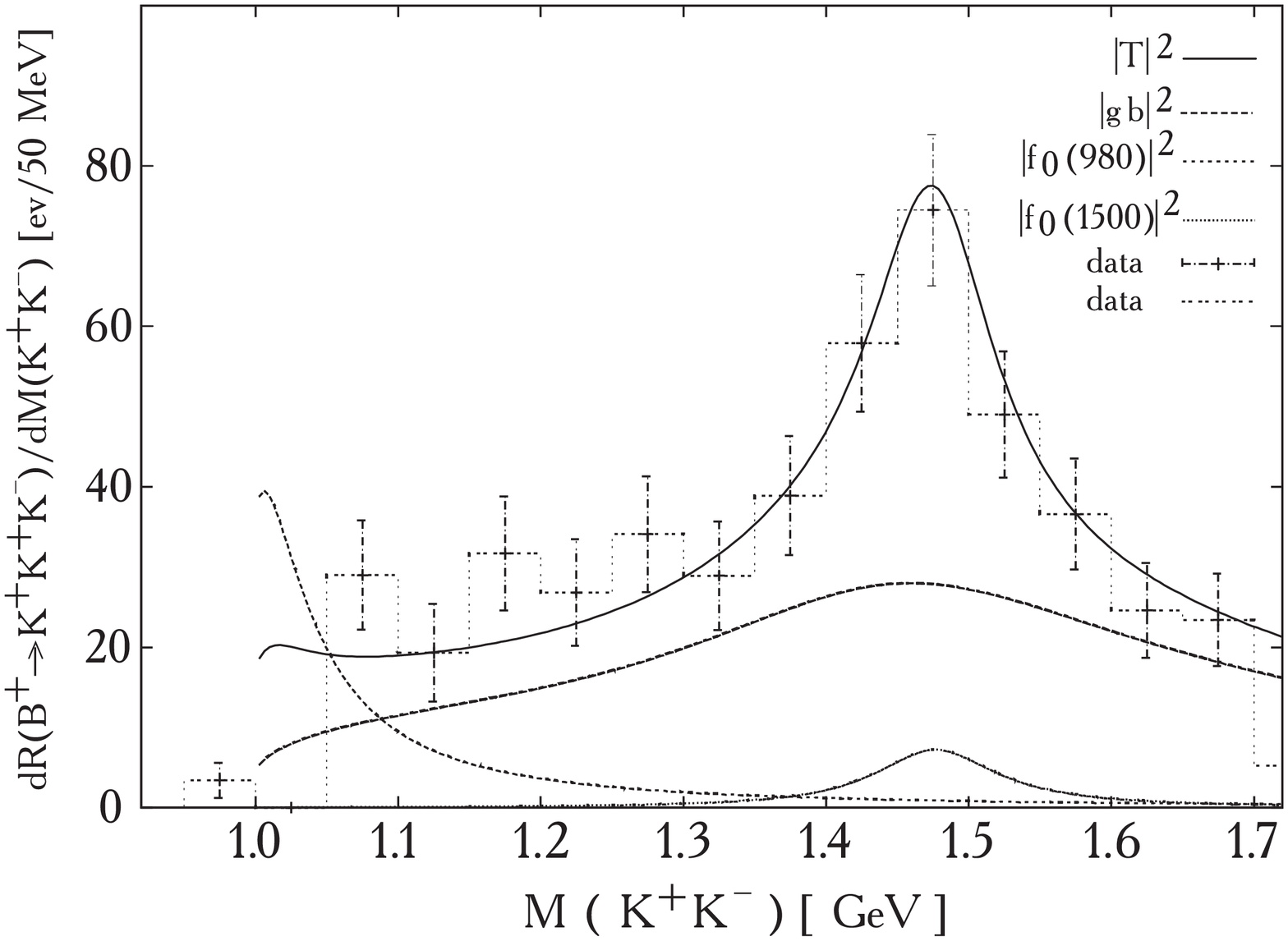}
\vspace{-62mm}\\
\hspace{-40mm}a\hspace{85mm}b\vspace{50mm}\\
\end{center}
\caption{\label{fig:minkochs}
$\pi^+\pi^-$ (a) and $ K^+K^-$ (b) mass spectra in $ B$-decays as
\cite{Garmash:2004wa} measured by BELLE  in comparison with a model
amplitude $|T|^2$ of the coherent superposition of $f_0(980),\
f_0(1500)$ and a broad background amplitude. Also shown are the
individual resonance terms $|T_R|^2$. The background
interferes destructively with both $f_0(980)$ and $f_0(1500)$ in their
$\pi^+\pi^-$ decays while there is  constructive  interference between
background and $f_0(1500)$ in case of $f_0(1500)\to  K^+K^-$
\cite{Minkowski:2004xf}.
}
\end{figure}

Fig.~\ref{fig:minkochs} shows the $\pi^+\pi^-$ and $ K^+K^-$ mass
spectra from $ B\to K\,  \pi^+\pi^-$ and $ B\to K\,  K^+K^-$,
respectively, and a fit to the data by Minkowski and Ochs
\cite{Minkowski:2004xf}. The $\pi^+\pi^-$ exhibits a clear $f_0(980)$
signal and some additional intensity which can be described by
$f_0(1370)$. However, this interpretation is very problematic: the $
K^+K^-$ peak at 1500\,MeV/c$^2$, if assigned to $f_0(1500)$, must be
accompanied by a peak in $\pi^+\pi^-$. This is not the case. Hence a
new scalar state would need to be introduced which couples strongly to
$ K^+K^-$ and weakly to $\pi^+\pi^-$. (This very argument is used to
identify two scalar mesons $f_0(1710)$ and $f_0(1790)$, see section
\ref{Scalar mesons in hadronic J/psi decays}). Minkowski and Ochs have
shown that both spectra can be explained by assuming a broad SU(3)
isoscalar background amplitude (which they identify with a scalar
glueball) and $f_0(980)$ plus $f_0(1500)$ with standard properties
\cite{Minkowski:2004xf}. Fig.~\ref{fig:minkochs} shows how the
destructive interference creates a fall off of intensity at 1.5\,GeV/c$^2$ in
the $\pi^+\pi^-$ invariant mass distributions and a peak in the  $
K^+K^-$ due to constructive interference. A $f_0(1370)$ is not needed
in the fit. The main difference in this analysis compared to the
analyses of the BABAR and BELLE collaborations is the treatment of
unitarity: the $f_0(1370)$ resonance decaying to $\pi\pi$ and a second
(new) $f_0(1500)$ resonance decaying into $ K\bar K$ is not needed
if there is a background amplitude close to the unitarity limit plus
resonant amplitudes which are added in a unitarity conserving way, in a
$K$-matrix or using the Dalitz-Tuan formalism. Even in a production
experiment, unitarity constraints play a decisive $\rm r\hat{o}le$. This may be
unexpected. The strong dip in the centrally produced $\rho\rho$
$J^{PC}=0^{++}$ wave in Fig.~\ref{wadip}b at exactly 1.5\,GeV/c$^2$
underlines the importance of unitarity constraints.

The change of the $f_0(1500)$ interference phase from constructive
(in $ K\bar K$) to destructive (in $\pi\pi$) suggests that the
$f_0(1500)$ wave function has a minus sign between its $n\bar n$ and
$s\bar s$ component. The background amplitude is assumed to be
flavour singlet. The minus sign was already found in
\cite{Minkowski:1998mf} from a comparison of the Argand phase motion
for $ \pi\pi\to \eta\eta$ and to $ K\bar K$. Obviously, $f_0(1500)$ is
flavour-octet like. This is in agreement with its large coupling to
$\eta\eta^{\prime}$.

\subsection{\label{Scalar mesons in hadronic J/psi decays}
Scalar mesons in hadronic J/$\psi$ decays}

\begin{figure}[pt]
\centering{
\includegraphics[width=0.65\textwidth,height=0.75\textheight]{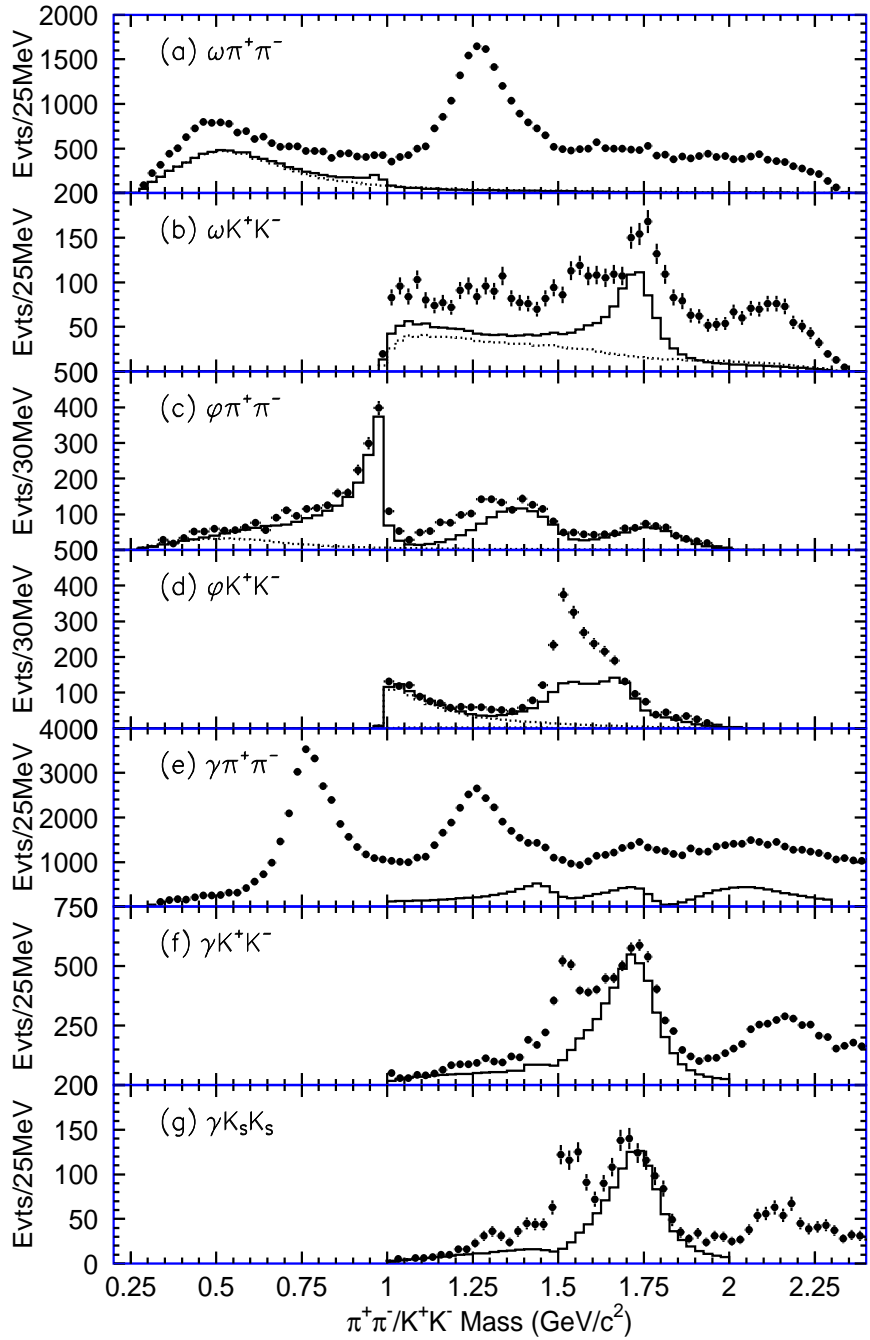}
}
\caption{\label{scalars}
The invariant mass distributions of the pseudoscalar meson
pairs recoiling against $\omega$, $\phi$, or $\gamma$ in J/$\psi$
decays measured at BES-II. The dots with error bars are data, the
solid histograms are the scalar contribution from PWA, and the
dashed lines in (a) through (c) are contributions of $\sigma(485)$ from
the fits, while the dashed line in (d) is the $f_0(980)$. Not the full
mass spectra are analyzed in (e), (f), and (g). The figure is taken
from \cite{Li:2006ni}. }  \end{figure}

Hadronic decays of J/$\psi$ into $\omega$ and $\phi$ mesons recoiling
against a meson $X$ give access to the flavour content of $X$. Due to
the OZI rule, J/$\psi\to\omega X$ couples to the $n\bar n$ component of
$X$ while J/$\psi\to\phi X$ couples to $s\bar s$. The BESII detector
$-$ described briefly in section \ref{BES} $-$ has recorded
$58\cdot 10^6$ J/$\psi$ decays. The results on J/$\psi$ decays to
$\omega$ or $\phi$ recoiling against two pseudoscalar mesons are
summarised in Table \ref{tab:recoil}; important mass spectra are
collected in Fig. \ref{scalars}. Fig. \ref{scalars}a,b show $\pi\pi$
and $ K\bar K$ systems recoiling against an $\omega$, Fig.
\ref{scalars}c,d  those recoiling against a $\phi$. Data on radiative
J/$\psi$ decays into $\pi^+\pi^-$ (e) and $ K\bar K$ (f,g) are also
included in the figure.

\begin {table}[pb]
\caption{\label{tab:recoil}
Masses, widths and branching fractions (in $10^{-4}$) of
scalar resonances recoiling against $\omega$ and $\phi$ mesons in
J/$\psi$ decays. Masses and width are given from the data with the
largest branching fraction. The yields given as $\leq$ were derived
from fits with imposed masses and widths with no forcing evidence for
presence of the contribution.  \vspace{2mm}
 }
\begin {center}
\renewcommand{\arraystretch}{1.4}
\begin{tabular}{ccccccc}                                 \hline\hline
Channel &Mass&Width&$\mathcal
B(J/\psi\to\omega X,$&$\mathcal B(J/\psi\to\omega X,$
        &$\mathcal B(J/\psi\to\phi X,$&$\mathcal B(J/\psi\to\phi X,$ \\
        & (MeV/c$^2$)& (MeV/c$^2$) &  $X\to\pi\pi)$ & $X\to K \bar{K})$
        &$X\to\pi\pi)$ & $X\to K \bar{K})$ \\                  \hline
$\sigma$ & $541\pm39$ & $504\pm84$&$\sim 20$&&$1.6\pm0.6$&$0.2\pm0.1$\\
$f_0(980)$   & \multicolumn{2}{c}{Flatt\'e formula}&$\leq1.2$&fixed &
$5.4\pm0.9$ & $4.5\pm0.8$                                          \\
$f_0(1370)$ & $1350\pm 50$& $265\pm40$&&& $4.3\pm1.1$ & $0.3\pm0.3$  \\
$f_0(1500)$  & PDG & PDG  &$\leq1.2$&& $1.7\pm0.8$ & $0.8\pm0.5$  \\
$f_0(1710)$ &$M = 1738 \pm 30$ & $\Gamma = 125 \pm 20$            &
$\leq0.5$&$13.2\pm2.6$& & $2.0\pm0.7$                            \\
$f_0(1790)$ & $1790^{+40}_{-30}$ & $270^{+60}_{-30}$ &&& $6.2\pm1.4$
& $1.6\pm 0.8$                                      \\ \hline\hline
\end {tabular} \renewcommand{\arraystretch}{1.0} \end
{center} \end {table}

The following observations can be made:
\begin{itemize}
\item  The region below 1\,GeV/c$^2$:
\end{itemize}
The $\pi\pi$ mass distribution from J/$\psi\to\omega\pi^+\pi^-$ shows a
significant peak at about 600\,MeV/c$^2$ (discussed as $\sigma(485)$ in
section \ref{Scalar mesons}), the $\pi\pi$ mass
distribution recoiling against the $\phi$ not. This apparent puzzle
was explained recently \cite{Lahde:2006wr} by the difference of the
strange and non-strange formfactors of the pion: the
J/$\psi\to\phi\pi^+\pi^-$ process is mainly driven by the strange
scalar pion formfactor which vanishes in leading order in chiral
perturbation theory. The non-strange formfactor is not suppressed at
small energies, and the $\sigma(485)$ appears. A second remarkable
feature is the high visibility of $f_0(980)$ in the $\pi\pi$ mass
spectrum recoiling against a $\phi$. The corresponding $ K\bar K$
spectrum shows, up to the $f_0(1500)$, no significant features. In the
recoil against an $\omega$, $f_0(980)$ gives only small contributions.
Careful inspection of the $\pi\pi$ spectrum in the latter reaction
reveals a small dip (Fig. 4 in \cite{Lahde:2006wr}); the fit in
\cite{Lahde:2006wr} demonstrates compatibility of $f_0(980)$ $-$
generated by meson-meson interactions $-$ with all four reactions. The
singlet/octet fraction varies as a function of the mass in the $K\bar
K$ threshold region, and we refrain from assigning a singlet or octet
flavour wave function to it.

\begin{itemize}
\item The 1.3\,$-$\,1.6\,GeV/c$^2$ region:
\end{itemize}

Above 1\,GeV/c$^2$, the $\pi\pi$ mass distribution from
J/$\psi\to\omega\pi^+\pi^-$ is dominated by a strong $f_2(1270)$ peak.
The partial wave analysis finds nearly no scalar contribution.
In $\phi\pi\pi$ the scalar contribution is substantial and peaks at
about 1400\,MeV/c$^2$ followed by a dip at about 1520\,MeV/c$^2$. In both
$ K\bar K$ spectra, there are peaks due to $f_2'(1525)$. In
$K\bar K$ recoiling against $\phi$, a significant scalar
component is found beneath the $f_2'(1525)$ which is assigned to $\phi
f_0(1500)$. The best evidence for $f_0(1370)$ (if any) stems from
J/$\psi\to\phi\pi^+\pi^-$.

\begin{itemize}
\item The 1.6\,$-$\,1.9\,GeV/c$^2$ region:
\end{itemize}

The 1700-1800\,MeV/c$^2$ region is the most complicated part of the spectrum.
In the $ K\bar K$ mass distributions recoiling against $\omega$
mesons, a striking peak is observed at 1710\,MeV/c$^2$ which is absent in the
spectrum recoiling against $\phi$'s. On the contrary, a $\pi^+\pi^-$
peak at about 1770\,MeV/c$^2$ is seen in $\phi\pi^+\pi^-$ which is absent in
$\phi K\bar K$. This is very puzzling. The BES collaboration
suggests that the phenomena originate from production of two different
resonances, a $f_0(1710)$ mainly $s\bar s$ state decaying into $
\bar KK$, and a $f_0(1790)$ mainly $n\bar n$ state decaying into
$\pi\pi$. At the present stage, it is a mystery why the mainly $s\bar s$
$f_0(1710)$ state is produced recoiling against an $\omega$, and the
mainly $n\bar n$ $f_0(1770)$ state is seen in its recoil against a
$\phi$.

\subsection{\label{Scalar mesons in radiative J/psi decays}
Scalar mesons in radiative J/$\psi$ decays}

Radiative J/$\psi$ decays bare the promise to reveal unambiguously the
presence of glueballs. In this process, the $\bar cc$ system supposedly
decays into one photon and two interacting gluons. If their interaction
resonates, a glueball is formed. Because of the importance of J/$\psi$
radiative decays for the discussion of scalar mesons and their
interpretation, its seems appropriate to highlight some important
historical aspects.

We have already discussed the $\eta(1440)$ or $\iota(1440)$, which was
interpreted in section~\ref{Isoscalar resonances revisited} as $\eta$
radial excitation but was believed to be a glueball when first seen in
radiative J/$\psi$ decay. With a pseudoscalar glueball at 1.4\,GeV/c$^2$, a
tensor glueball was expected below 2\,GeV/c$^2$, and there was great
excitement when a broad bump in the radiatively produced $\eta\eta$
mass spectrum was discovered by the Crystal Ball collaboration
\cite{Edwards:1981ex}. The bump extended from 1450 to 1800\,MeV/c$^2$, was
found with a yield of $(4.9\pm1.4\pm1.0)\cdot 10^{-4}$, and was called
$\Theta(1640)$. The $2^{++}$ quantum numbers originated from an excess
of four events above $|\cos\theta|>0.95$. If these events are
neglected, scalar quantum numbers follow. We now believe that the bump
comprises contributions from $f_0(1500)$ and $f_0(1760)$ which we
estimate to be in the order of 1:4. Then

\begin{eqnarray}
\label{br1780_1}
\mathcal B_{J/\psi\to\gamma
f_0(1500)\to\gamma\eta\eta}&=&(1.0\pm0.3\pm0.2\pm0.3)\cdot 10^{-4}
\nonumber\\
\mathcal B_{J/\psi\to\gamma f_0(1760)\to\gamma\eta\eta}&=&
(3.9\pm1.1\pm0.8\pm0.3)\cdot 10^{-4}\,,
\end{eqnarray}

where the first error is of statistical nature, the second the
systematic error, and the third one (introduced here) estimates the
uncertainty in dividing the signal into two components.
 The $\eta\eta$ peak was confirmed
by the GAMS collaboration in central production, another glue-rich
environment. GAMS reported $M=1755\pm 8$\,MeV/c$^2$ \cite{Alde:1986xr}
and, with higher statistics, $M=1744\pm 15, \ \Gamma < 80$\,MeV/c$^2$
\cite{Alde:1991ns}. The branching ratio (\ref{br1780_1}) and other
ratios are collected in Table \ref{scalardecay}.

\begin{table}[pb]
\caption{\label{scalardecay}
Branching fraction  $\times
10^4$ for production of scalar mesons in J/$\psi$
radiative decays. The $f_0(1500)$ yield is calculated from
the $\eta\eta$ and $\sigma\sigma$ and the known $f_0(1500)$ decay
branching ratios. \vspace{2mm}}
\begin {center}
\renewcommand{\arraystretch}{1.4}
\begin{tabular}{c|c|cc|c|c|c}
\hline                                                        \hline
Decay channel &$f_0(980)$ & $f_0(1500)$ && $f_0(1710)$, $f_0(1790)$,
$f_0(1810)$ & $f_0(2100)$& reference \\ \hline
$\omega\omega$&&-&&$3.1\pm0.6$&&\cite{Ablikim:2006ca} \\
$\omega\phi$&&-&&\qquad\qquad\qquad $\mid 2.61 \pm 0.27 \pm 0.65$ &&
\cite{Ablikim:2006dw} \\ $\sigma\sigma$ (or $\rho\rho$)&& $8.0 \pm
3.5$&& $9.0 \pm 1.3$ & $13 \pm 5$&
\cite{Bugg:2006uk} \\ $\pi\pi$&
$\leq4$&$1.00\pm0.03\pm0.42$&&$3.96\pm0.06$ & &
\cite{Ablikim:2006db} \\ $\eta\eta$&&$1.0\pm0.3\pm0.2\pm0.3$&&
$3.9\pm1.1\pm0.8\pm0.3$       & &
Eq. (\ref{br1780_1}) \\ $ K\bar
K$&&-&&$9.62\pm0.29^{+2.11+2.81}_{-1.86-0.00}\mid$ \qquad\qquad\qquad &
& \cite{Bai:2003ww}\\ \hline  total &$\leq 5$&$16\pm7$&& $31.6 \pm 4.1$
& $>13 \pm 5$     \\ \hline\hline \end{tabular}
\renewcommand{\arraystretch}{1.0} \end {center} \end{table}

MARKIII \cite{Becker:1986zt} observed the $\Theta$ at 1720\,MeV/c$^2$ in
J/$\psi\to\gamma K^+K^-$ (and possibly $\gamma\pi^+\pi^-$),
found $2^{++}$ quantum numbers and a yield $ \mathcal
B_{J/\psi\to\gamma f_J(1710)\to\gamma K^{+} K^{-}}=
(4.8\pm0.6\pm0.9)\, 10^{-4}\,.$  A reanalysis by Dunwoodie
\cite{Dunwoodie:1997an} identified the $\Theta$ as scalar
resonance and reported contributions from two scalar states

\renewcommand{\arraystretch}{1.5}
\begin{center}
\small
\begin{tabular}{ccr}
$M = 1429^{+43}_{-37}$ & $\Gamma = 169^{+111}_{-76}$  & $\mathcal
B_{J/\psi\to\gamma f_0(1430)}=(4.3^{+2.7}_{-1.3})\cdot 10^{-4}$ \\
\multicolumn{3}{c}{\scriptsize$ \Gamma\left[f_{0}(1430)\to
K\bar{K} \right]/ \Gamma_{f_{0}(1430)\to \pi\pi}=
0.15^{+0.22}_{-0.13}$} \\
$M = 1704^{+16}_{-23}$ & $\Gamma = 124^{+53}_{-44}$  & $\mathcal
B_{J/\psi\to\gamma f_0(1710)}=(9.5^{+2.5}_{-2.0})\cdot 10^{-4}$ \\
\multicolumn{3}{c}{\scriptsize$ \Gamma_{f_{0}(1710)\to \pi\pi}/
\Gamma_{f_{0}(1710)\to K\bar{K}}= 0.27^{+0.17}_{-0.12}$}
\\ \end{tabular} \end{center} \renewcommand{\arraystretch}{1.0}

where the yields $\mathcal B$ refer to the sum of the two decay modes.

BES I \cite{Bai:1996dc} decomposed the structure at about 1750\,MeV/c$^2$
into a (small) scalar component at 1781\,MeV/c$^2$ plus a large tensor
component at 1696\,MeV/c$^2$. Hence the situation remained unclear.

Data on radiative J/$\psi$ decays into $\pi\pi$ \cite{Ablikim:2006db}
and $ K\bar K$ \cite{Bai:2003ww} from BESII were already shown in
Fig. \ref{scalars}e,f,g. In the reaction J/$\psi\to\gamma\pi^+\pi^-$,
the recoiling photon has a large energy when the $\pi^+\pi^-$ masses
are small; a large background due to three-pion production (with one
$\pi^0\to\gamma\gamma$ and one $\gamma$ lost) cannot be avoided, and
the authors have limited the partial wave analysis to the region above
1\,GeV/c$^2$. The reaction reveals a strong $f_2(1270)$ contribution.
The scalar part rises slowly, peaks at about 1430\,MeV/c$^2$ followed
by a faster fall-off with minimum at 1520\,MeV/c$^2$, rises again to a
bump at 1710\,MeV/c$^2$ and has a second dip at 1820\,MeV/c$^2$. It is
legitimate to ask whether the bumps or the dips should be identified as
resonances. The BES Collaboration uses Breit-Wigner amplitudes without
a coherent background amplitude; hence the bumps are identified with
scalar resonances.  The first resonance was found at a mass of $1466\pm
6\pm 20$\,MeV/c$^2$ and a width of $108{^{+14}_{-11}}\pm
21$\,MeV/c$^2$. The branching fraction is given in Table
\ref{scalardecay}. The second $0^{++}$ resonance parameters are
determined to $M=1765^{+4}_{-3}\pm 12$\,MeV/c$^2$ and $\Gamma=
145\pm8\pm69$\,MeV/c$^2$. A joint analysis of the two reactions
J/$\psi\to\gamma  K^+K^-$ and into $\gamma K^0_SK^0_S$ did not require
introduction of $f_0(1500)$, and gave mass and width
$M=1740\pm4$\,MeV/c$^2$ and $\Gamma = 166^{+5}_{-8}$\,MeV/c$^2$,
respectively, for the $f_0(1710)$.

\begin{figure}[pt]
\bc
\begin{tabular}{cc}
\includegraphics[width=0.44\textwidth,height=0.33\textwidth,clip=on]{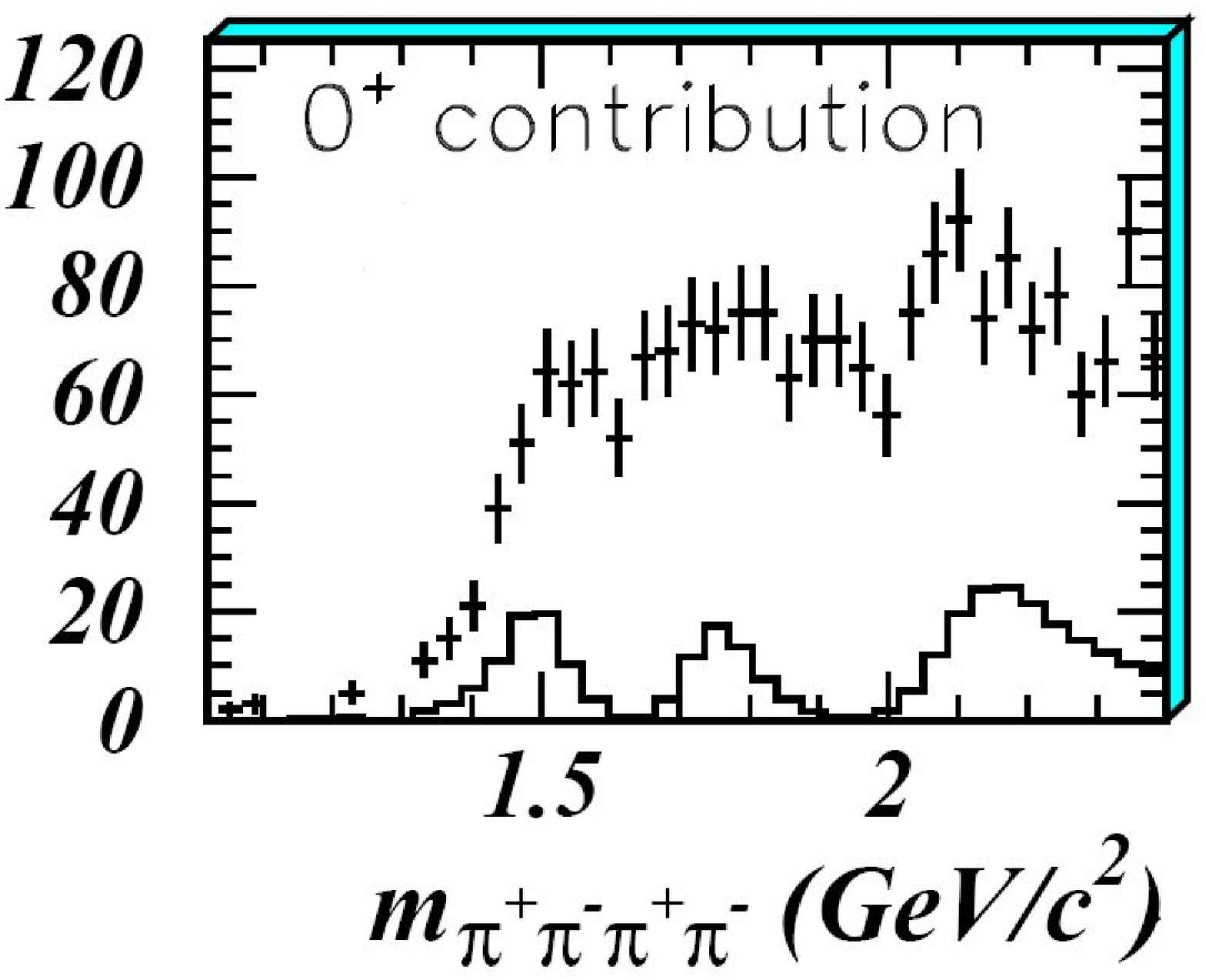}&
\includegraphics[width=0.44\textwidth,height=0.35\textwidth,clip=on]{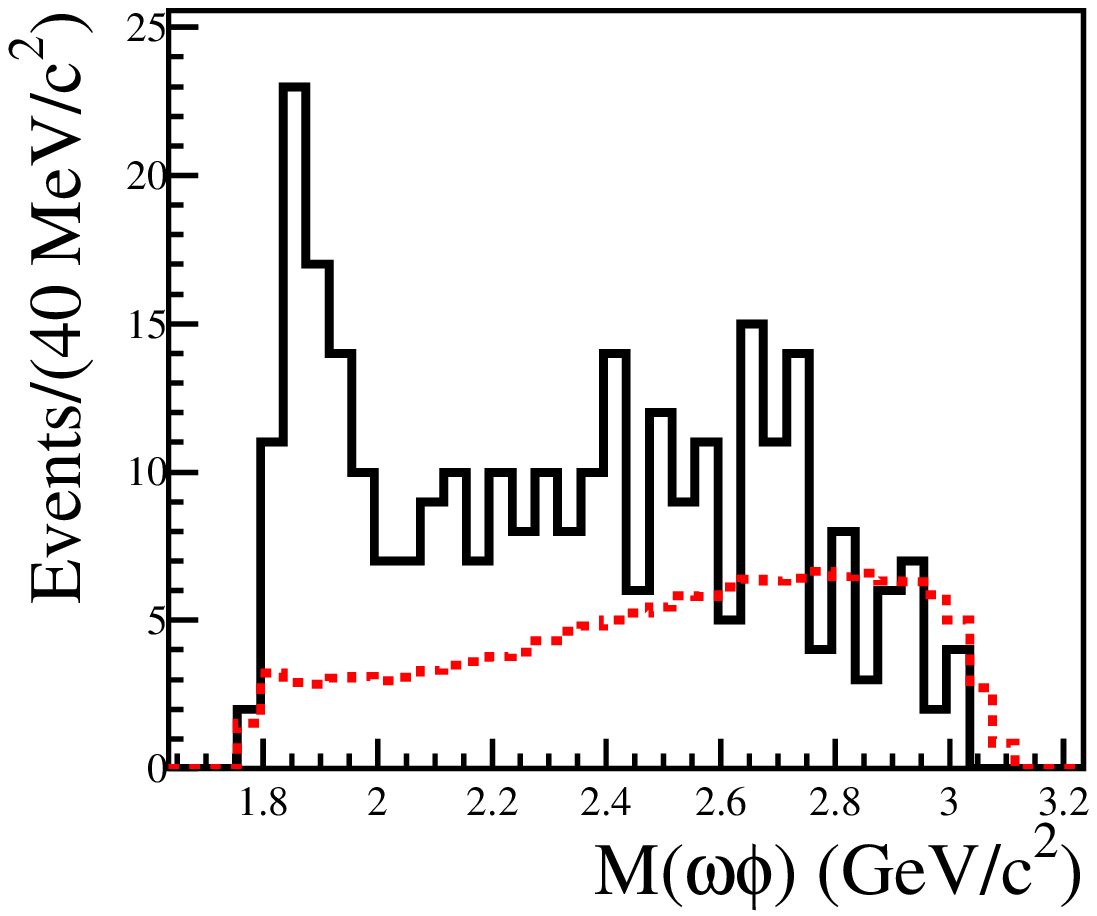}
\end{tabular}
\ec
\caption{\label{clok}
Left: $\pi^+\pi^-\pi^+\pi^-$ invariant mass for
J/$\psi\to\gamma\pi^+\pi^-\pi^+\pi^-$ and the scalar contribution (from
an isobar fit). Right: The $ K^+K^-\pi^+\pi^-\pi^0$ invariant
mass distribution in radiative J/$\psi$ decays for events in which the
$\pi^+\pi^-\pi^0$ mass is compatible with the $\omega$ mass and the
$ K^+K^-$ mass with the $\phi$ mass. The dashed curve represents the
$\omega\phi$ acceptance. }
\end{figure}

Radiative decay J/$\psi\to\gamma\pi^+\pi^-\pi^+\pi^-$ had been
observed by DM2 and MARKIII \cite{Bisello:1988as,Baltrusaitis:1985nd}.
The data were interpreted as a series of pseudoscalar resonances
at about 1.5, 1.75 and 2.1\,GeV/c$^2$. Bugg noticed the similarity of the
pattern of $\eta\eta$ resonances observed in antiproton-proton
annihilation at $2950 < s^{1/2} < 3620$\,MeV/c$^2$ \cite{Armstrong:1993fh},
the data are reproduced in Fig.~\ref{e760_etaeta}. The $\eta\eta$
states cannot have pseudoscalar quantum numbers. This observation
motivated a reanalysis \cite{Bugg:1995jq} suggesting scalar quantum
numbers for the three structures in
J/$\psi\to\gamma\pi^+\pi^-\pi^+\pi^-$ \cite{Bisello:1988as}. This
conjecture was confirmed by BES \cite{Bai:1999mm}. Fig. \ref{clok}a
shows the resulting $4\pi$ invariant mass distribution and the fraction
assigned to the scalar partial wave which was found to originate from
$\sigma\sigma$ decays. A study of radiative J/$\psi$ decays into
$\omega\omega$ \cite{Ablikim:2006ca} found a peak at 1.76\,GeV/c$^2$,
just above the $\omega\omega$ threshold. Analysis of angular
correlations assigned predominantly pseudoscalar quantum numbers to the
$\omega\omega$ system below 2\,GeV/c$^2$. A partial wave found small
$0^{++}$ and $2^{++}$ contributions in addition, with a product
branching fraction $\mathcal B (J/\psi\to\gamma f_0(1710)) \cdot
\mathcal B(f_0(1710)\to\omega\omega) = (3.1 \pm 0.6)\cdot 10^{-3}$.
In a recent discussion of $f_0(1790)$, Bugg included a small fraction of
$\rho\rho$ decays and estimated the $\omega\omega$ contribution
\cite{Bugg:2006uk}. The decays into $\pi^+\pi^-\pi^+\pi^-$ via
$\rho\rho$ or $\sigma\sigma$ lead to different Clebsch-Gordan
corrections, 2.25 and 3 respectively, for the total four-pion yield. We
use $2.6\pm 0.4$ to determine the total 4$\pi$ branching ratio from the
$\pi^+\pi^-\pi^+\pi^-$ contribution.

For $f_0(1500)$, $\omega\omega$ is not open. For the four-pion state,
the Crystal Barrel Collaboration found a mixture of $\rho\rho$,
$\sigma\sigma$ and other decays. Again, we use $2.6\pm 0.4$ to
calculate the total $f_0(1500)\to 4\pi$ contribution from  the
J/$\psi\to\gamma\pi^+\pi^-\pi^+\pi^-$ yield.

A threshold enhancement was observed in the doubly OZI suppressed
decay J/$\psi\to\gamma\omega\phi$ \cite{Ablikim:2006dw}, see Fig.
\ref{clok}b. The partial wave analysis favoured scalar quantum numbers
with the $\omega\phi$ system in S-wave. Mass and width of the
enhancement were determined to be $M = 1812^{+19}_{-26}\pm
18$\,MeV/c$^2$ and $\Gamma = 105\pm20\pm28$\,MeV/c$^2$, respectively,
the product branching fraction was $\mathcal{B}(J/\psi\to\gamma
X)\cdot\mathcal{B}(X\to\omega\phi) = (2.61 \pm 0.27\pm0.65)\,10^{-4}$.
It is suggested to be an 'unconventional' state and stimulated a
variety of different interpretations. These will be discussed in
section  \ref{Scalar mesons and their interpretation}.

Is $f_0(1810)$ an additional state\,? First we note that, in the limit
of SU(3) invariance, decays into $\phi\omega$ are forbidden from an
isoscalar-singlet state; $f_0(1810)$ must have a strong octet
component. The nominal $\phi\omega$ threshold mass is 1802\,MeV/c$^2$, the
$ K^*K^*$ system has a (mean) threshold at 1788\,MeV/c$^2$. Thus both
thresholds fall into the $f_0(1760)$ natural widths, and $\phi$
formation by $K^*\bar K^*$ rescattering could be a natural
explanation. A study of J/$\psi\to\gamma K^*K^*$ \cite{Bai:1999mk}
found however a broad pseudoscalar state
($M,\Gamma=1800,500$\,MeV/c$^2$). Zou, Dong and Bugg \cite{Bugg:1999jc}
reanalysed the data using the Flatt\'e formula and dispersive
corrections for the opening of decay channels. Changing the formalism
shifted the resonance mass to 2190\,MeV and the width to 800\,MeV. In
both analyses, there was little scalar intensity. If $f_0(1760)$ does
not decay into $K^*K^*$ with a significant yield, the interpretation of
$f_0(1810)\to\phi\omega$ decays by a rescattering mechanism seems to be
excluded.

The process J/$\psi\to\gamma f_0(980)$ was not yet observed. There is
a small threshold enhancement in the $ K^+K^-$ mass distribution
from which we estimate (using $\Gamma_{f_0(980)\to  K\bar
K}/\Gamma_{tot}=0.16$) an order of magnitude of \be \mathcal
B_{J/\psi\to\gamma f_0(980)} \leq  5\, 10^{-4}\,.
\ee

A summary of results on radiative production of scalar mesons  in
J/$\psi$ decays is given in Table~\ref{scalardecay}. The radiative
yield of scalar isoscalar mesons increases with increasing mass. This
is an important observation. Naively, one might expect an increase of
radiative yields with the third power of the photon momentum. However,
if the reaction is viewed as J/$\psi\to \gamma gg$, the chance of two
gluons to be produced with a low mass vanishes with $m^{-3}$.

\begin{figure}[pt]
\hspace{5mm}\includegraphics[width=0.9\textwidth,clip=on]{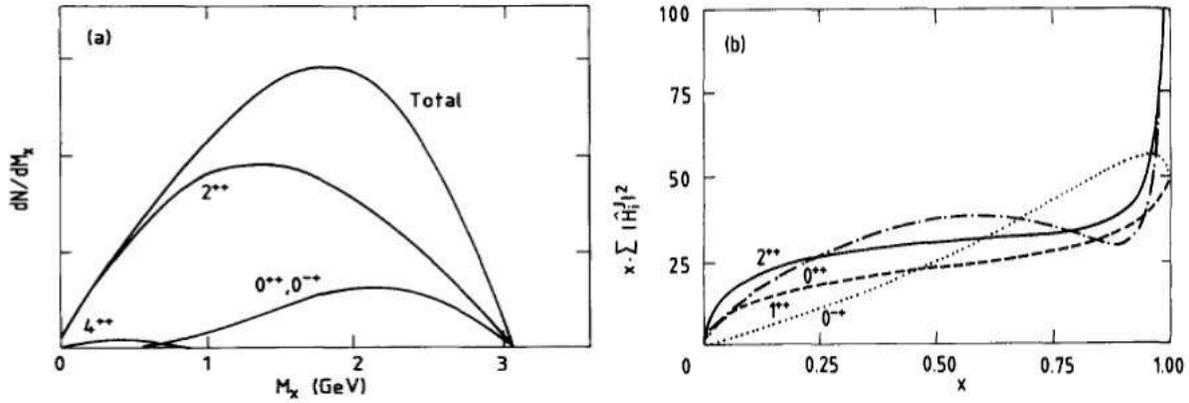}
\caption{\label{gg-jpc}
Energy and spin-parity distributions of a two-gluon system recoiling
against a photon in radiative decays of heavy quarkonia as functions of
the produced mass (a) or as function of $x=M_{gg}/M_{Q\bar Q}$. For
$x\to 1$, a soft photon is radiated. a) is calculated for massless
gluons \cite{Billoire:1978xt}, b) for off-shell gluons
\cite{Korner:1982vg}. } \end{figure}

The energy and
spin-parity distribution of a two-gluon system recoiling against the
photon were calculated by Billoire {\it et al.} \cite{Billoire:1978xt}
assuming that the two gluons are massless. The distributions are shown
in Fig. \ref{gg-jpc}a. For small $M_X$, the scalar intensity vanishes.
Assuming gluon-hadron duality we can expect that these predictions are
valid for hadrons as well.

K\"orner {\it et al.} \cite{Korner:1982vg} consider virtual gluons as
well. Quark loop integrals are calculated in nonrelativistic
approximation. Their distributions, shown in Fig. \ref{gg-jpc}b, does
not predict the reduction of the scalar intensity at $M_x\to 0, \ (x\to
1)$. On the other hand, Billoire {\it et al.} predict absence of the
$1^{++}$ wave which is forbidden in their model due to the Landau-Yang
theorem. The model of K\"orner {\it et al.} predicts sizable $f_1$
radiative yields, in agreement with experiment. Both calculations use
perturbative methods, and the limit of applicability seems to be
reached.

We have assigned flavour octet wave functions to both $f_0(1500)$ and
$f_0(1760)$; we conjecture that $f_0(2100)$ is octet as well. Is this
assignment compatible with them being produced in radiative J/$\psi$
decays? First, also octet mesons are produced in radiative J/$\psi$
decays, see Table \ref{bes1750}. The summation of all seen radiative
yields into isoscalar $0^{-+}$ mesons gives 2.25\%. All isoscalar
$0^{++}$ mesons give a yield of 0.63\% (see Table \ref{scalardecay}).
Of course, this is an estimate since we do not know what fractions are
missing. From the calculation of \cite{Billoire:1978xt} shown in Fig.
\ref{gg-jpc}, scalar and pseudoscalar mesons are expected with similar
yields. SU(3) symmetry breaking, final state radiation and/or
singlet-octet mixing may contribute to the production of the observed
scalar mesons.

\subsection{\label{Scalar mesons in radiative Upsilon decays}
Scalar mesons in radiative $\Upsilon$ decays}

The CLEOIII collaboration studied $21\cdot 10^6$ $\Upsilon$
annihilations. As in the case of J/$\psi$, radiative decays give access
to a final states which are supposedly glueball-rich. Compared to
J/$\psi$, radiative decays of the $\Upsilon$ are a challenge. The ratio
\begin{equation}
 \frac{\Gamma_{\Upsilon} ^{rad}}{\Gamma_{\Upsilon} ^{tot}}=
\left(\frac{\alpha_S(\Upsilon)}{\alpha_S(J/\psi)}\right)^4
\left(\frac{\Gamma_{\Upsilon\to e^+e^-}}{\Gamma_{\Upsilon} ^{tot}}
\cdot \frac{\Gamma_{J/\psi} ^{tot}}{\Gamma_{J/\psi\to e^+e^-}}
\cdot \frac{\Gamma_{J/\psi} ^{rad}}{\Gamma_{J/psi} ^{tot}} \right)
            \approx 0.013,
\end{equation}
instead of 8\% for J/$\psi$'s and we expect a suppression of about a
factor $\sim6$. However, there is a second important reduction factor
from the energy and spin-parity distribution of the two-gluon system
recoiling against the photon, see Fig. \ref{gg-jpc}a. The $2^{++}$
contribution peaks at half the quarkonium mass; for tensor mesons a
factor 2 reduction is predicted at the mass of the $f_2(1270)$. For
scalar mesons the factor is about 20 for the $f_0(1760)$ and increases
rapidly for smaller masses. With $\mathcal
B_{J/\psi\to\gamma f_2(1270)}=1.38\cdot 10^{-3}$ and the scalar
radiative yields from Table \ref{scalardecay} we expect yields which
are compared in Table \ref{compcleo3} to the CLEOIII results
\cite{Besson:2005ud}.

\begin{table}[pt]
\caption{\label{compcleo3}Comparison of observed \cite{Besson:2005ud}
and expected yields (in units of $10^{-5}$) in radiative decays into
tensor and scalar mesons. \vspace{2mm}} \bc
\renewcommand{\arraystretch}{1.4} \begin{tabular}{lcc} \hline\hline
Reaction & observed yield & expected yield \\           \hline
$\mathcal B_{\Upsilon\to\gamma f_2(1270)}$& $10.5 ~\pm
1.6~(\mathrm{stat})^{+ 1.9}_{-1.8}~(\mathrm{syst})$ & $\approx 10$ \\
$\mathcal B_{\Upsilon(1S) \rightarrow \gamma f_0(1500)}$ & $< 1.17$&
$\approx 0.5$ \\ $\mathcal B_{\Upsilon\to\gamma f_2(1760), f_2(1760)\to
2\pi^0}$   & $< 1.2 $ &           $\approx 1$   \\
\hline\hline \end{tabular} \renewcommand{\arraystretch}{1.4}
\ec \end{table}

The comparison shows that searches for a scalar glueball are much more
sensitive in radiative J/$\psi$ decays. In $\Upsilon$ radiative decays,
the scalar mass spectrum is dominated by the tensor contributions. A
tensor glueball, expected at about 2.4\,GeV/c$^2$, might be detectable in
$\Upsilon$ radiative decays if it exists with a reasonably small
width, and if the statistics can be increased significantly.

\subsection{\label{Scalar mesons in chi decays}
Scalar mesons in $\chi_{cJ}$ decays}

A first study of $\chi_{c0}$, $\chi_{c1}$, and $\chi_{c2}$ decays into
three mesons was reported in \cite{Athar:2006gh}. For $\chi_{c1}$ to
$\pi^+\pi^-\eta$, $K^+ K^- pi^0$, and $\pi^+ K^- K^0_S$, Dalitz plot
analysis were presented showing the potential of these decays for
light-meson spectroscopy.

\subsection{\label{Scalar mesons in two-photon fusion}
Scalar mesons in two-photon fusion}

The ALEPH collaboration at LEP searched for $\gamma\gamma$ production
of the glueball candidates $f_0(1500)$ and $f_J(1710)$ via their decay
to $\pi^+\pi^-$. No signal was observed; upper limits to the product
of $\gamma\gamma$ width and $\pi^+\pi^-$ branching ratio of the
$f_0(1500)$ and the $f_J(1710)$ were reported to be
$\Gamma_{\gamma\gamma\to f_0(1500)}\cdot BR(f_0(1500)\to\pi^+\pi^-) <
0.31$\,keV/c$^2$ and $\Gamma_{(\gamma\gamma\to f_J(1710)}. BR(f_J(1710)\to
\pi^+\pi^-) < 0.55$\,keV/c$^2$ at 95\% confidence level \cite{Barate:1999ze}.
From the decay branching ratios given above we estimate
$\Gamma_{\gamma\gamma\to f_0(1500)} < 1.6$\,KeV/c$^2$ and
$\Gamma_{(\gamma\gamma\to f_J(1710)} < 6$\,KeV/c$^2$ assuming that the 1700
to 1800\,MeV/c$^2$ region is dominated by one scalar meson (no $f_0(1790)$
and no $f_2(1720)$). The expected width is zero for a pure glueball
not mixing with $q\bar q$ mesons, 4.5\,keV/c$^2$ for a $n\bar n$, and
0.4\,keV/c$^2$ for a $s\bar s$ meson  \cite{Barnes:1985cy}.

\markboth{\sl Meson spectroscopy} {\sl Scalar mesons and their
interpretation} \clearpage\setcounter{equation}{0}\section{\label{Scalar mesons and their interpretation}
Scalar mesons and their interpretation}

\subsection{\label{Global views of scalar mesons and the scalar
glueball}
Global views of scalar mesons and the scalar glueball}

\begin{figure}[pt]
\vspace{-0.6cm}
\begin{center}
\includegraphics[width=0.4\textwidth,height=0.3\textwidth,clip=on]{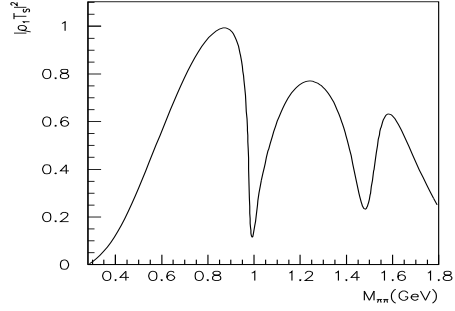}
\vspace{-0.8cm}\caption{\label{sampl} The
$I=0$ $\pi\pi$ S-wave amplitude squared~\cite{Li:2000jq}.}
\end{center}
\end{figure}

The interpretation of the spectrum of scalar mesons is still highly
controversial. The foundations of the different interpretations are
not well enough established to allow us to decide which scenario is
closest to the physical truth. The main differences arise because of
ambiguities in the interpretation of the following parts:

\begin{enumerate}

\item Are $f_0(980)$ and $a_0(980)$ genuine
$q\bar q$ resonances, or are they additional states? If the latter is
the case, they may be $qq\bar q\bar q$ states (Jaffe), $ K\bar K$
molecules (Isgur), or generated by a doubling of states due to strong
coupling of scalar states to the meson-meson continua (Beveren,
Tornqvist)\,? Do $\sigma(485)$, $\kappa(700)$, $f_0(980)$, and
$a_0(980)$ form a nonet?

\vspace{2mm}\item Is $f_0(1370)$ a true $q\bar q$ resonance or
generated by $\rho\rho$ molecular dynamics\,? The resonance is needed
to fit very different data sets and is accepted by the Particle Data
Group as established particle. However, the evidence is much weaker
when individual channels are discussed. A proper $f_0(1370)$ phase
motion has never been established. On the contrary, wherever the scalar
phase motion was tested, no evidence for an additional resonance
between $f_0(980)$ and $f_0(1500)$ was found.

\vspace{2mm}\item Are $f_0(1710)$ and $f_0(1790)$ distinct objects? In
mixing scenarios discussed below there must be a scalar resonance in
this mass range with a sizable $s\bar s$ component. The $f_0(1710)$ has
a large coupling to $K\bar K$, the $f_0(1790)$ not. At the first
glance, it is plausible to assign a large $s\bar s$ component to
$f_0(1710)$ while $f_0(1790)$ could be the radial excitation of
$f_0(1370)$. However, $f_0(1710)$ is observed recoiling against
$\omega$ in  J/$\psi$ decays. Thus, $f_0(1710)$ is produced via its
$n\bar n$ component and decays via its $s\bar s$ component. $f_0(1790)$
decaying into $\pi\pi$ is produced in recoil against $\phi$, via its
$s\bar s$ component. There is a clear conflict.

\vspace{2mm}\item The $f_0(1810)$ is intriguingly close to $f_0(1790)$
and to two thresholds, $K^*\bar K^*$ and $\phi\omega$. It is observed
in the latter decay mode. The yield in radiative J/$\psi$ decays is too
large to assign it to a conventional $f_0(1790)$. Its production and
decay is doubly OZI rule violating. The most obvious interpretation is
that of a tetraquark resonance. Is it separate from $f_0(1790)$?
Possibly, the three observations at 1710, 1790, and 1810\,MeV/c$^2$ are
traces of one resonance, $f_0(1760)$.

 \vspace{2mm}\item Is $f_0(2100)$ a respectable resonance? It is seen
as a dip in the GAMS data on $\pi\pi$ scattering, shown in Fig.
\ref{fig:gams-scalar}, as a $\eta\eta$ peak of unknown (even) spin in
$\bar pp$ annihilation in flight into $\pi^0\eta\eta$ (Fig.
\ref{e760_etaeta}), and is suggested to make a significant contribution
to J/$\psi$ radiative decay into four pions, see Fig. \ref{clok}.

\vspace{2mm}\item Does the $f_0(2100)$ region house two components,
$f_0(2000)$ reported in central production \cite{Barberis:2000em} and
possibly confirmed in $\bar pp$ annihilation in flight
\cite{Anisovich:2000ae,Uman:2006xb} and $f_0(2100)$ seen in
\cite{Anisovich:2000ae,Uman:2006xb} and observed in J/$\psi\to \gamma
4\pi$ \cite{Bai:1999mm}\,?

\vspace{2mm}\item What is the dynamical origin of the broad scalar
component shown in Fig.~\ref{sampl}\,?

\end{enumerate}

The low mass part (up to $\sim$ 1.2\,GeV/c$^2$) can well be described
by $t$-channel $\rho$ and some $f_2(1270)$ exchange. Thus it seems
plausible that it is generated by $t$-channel exchange dynamics. At
high energies, cross sections can be described by a series of
$s$-channel resonances; $t$-channel amplitudes reproduce the mean
cross section averaging over the many individual resonances. Possibly,
the broad $\pi\pi$ S-wave scattering background amplitude  -- let it be
called $f_0(1000)$ -- is equivalent to a series of $t$-channel exchange
processes. Duality arguments then assign a $q\bar q$ nature to
$f_0(1000)$. On the other hand, it is also very attractive to assume
that the broad component is the scalar glueball.

\begin{figure}[ph] \begin{tabular}{cccc}
\hspace{-3mm}\includegraphics*[width=.24\textwidth]{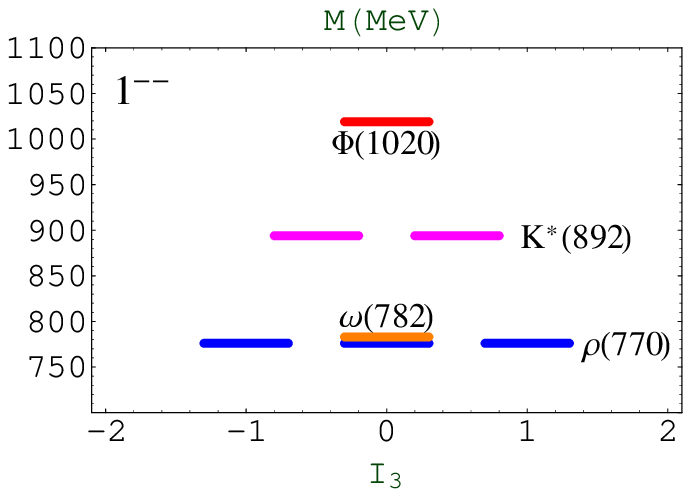}&
\hspace{-9mm}\includegraphics*[width=.24\textwidth]{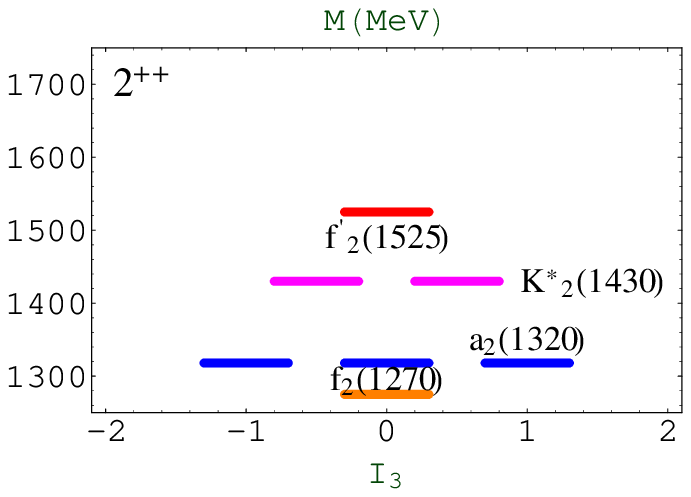}&
\hspace{-4mm}\includegraphics*[width=.24\textwidth]{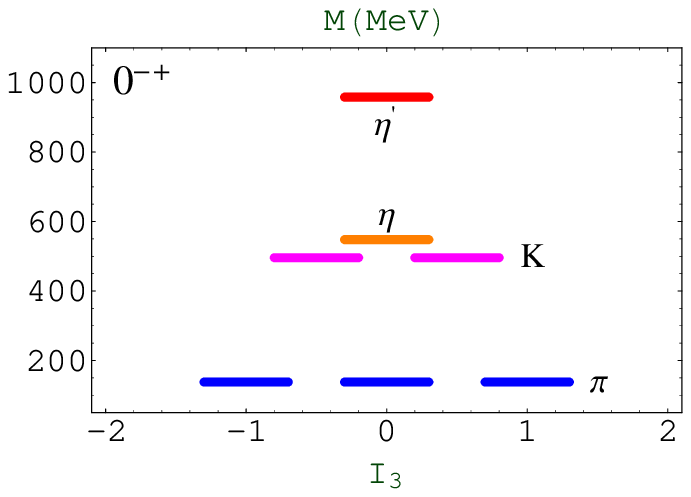}&
\hspace{-3mm}\includegraphics*[width=.24\textwidth]{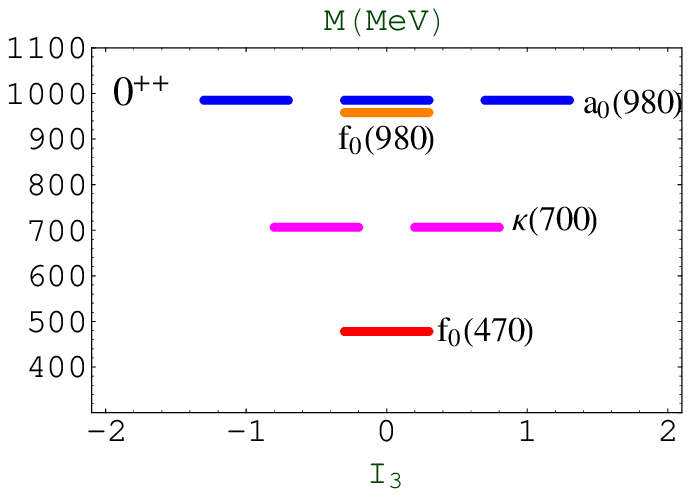}\\
\hspace{7mm}\includegraphics*[width=.24\textwidth]{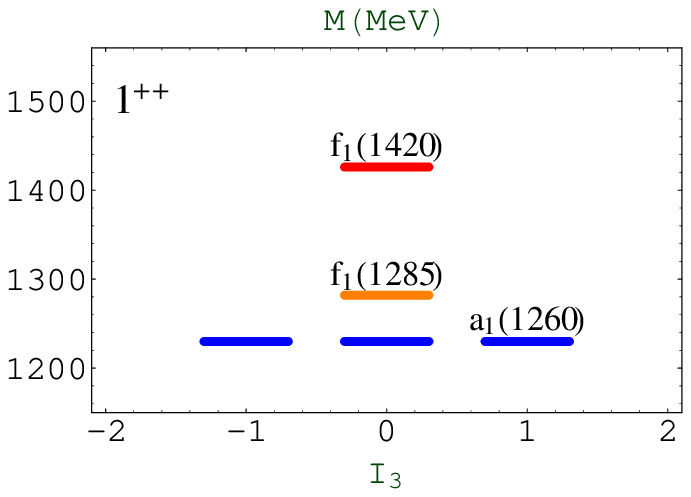}&
\hspace{-10mm}\includegraphics*[width=.24\textwidth]{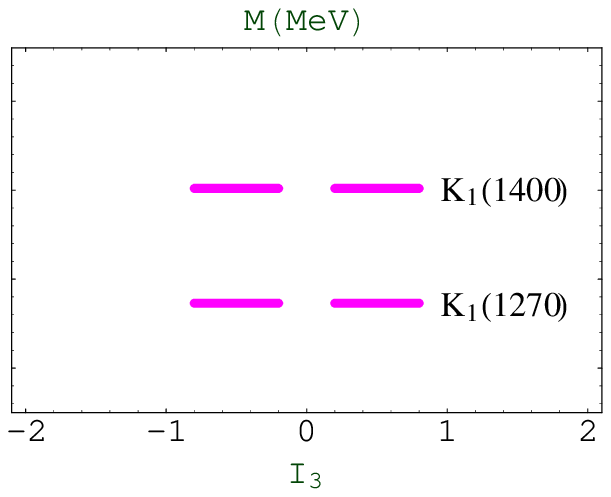}&
\hspace{-16mm}\includegraphics*[width=.24\textwidth]{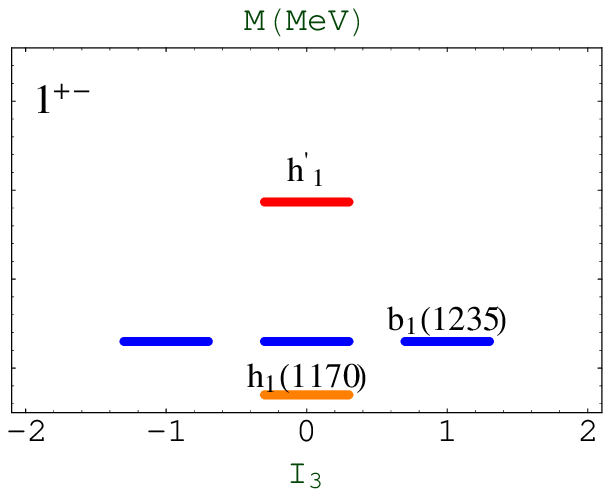}&
\hspace{-3mm}\includegraphics*[width=.24\textwidth]{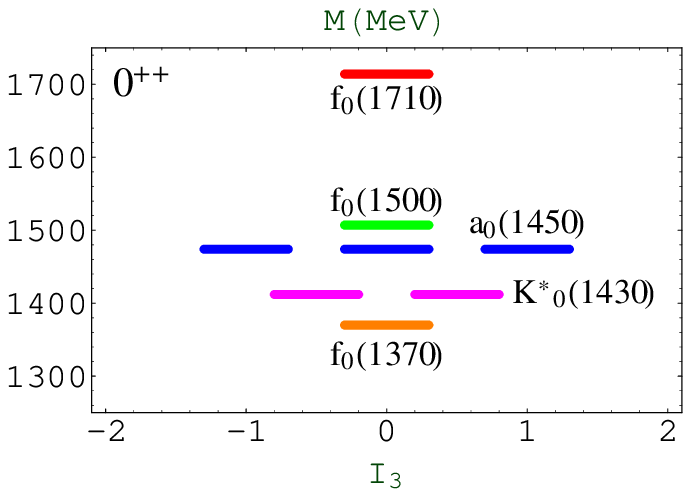}\\
\end{tabular}
\caption{\label{maiani}Mass spectra of the lowest lying meson
multiplets. Most multiplets show a high-mass $s\bar s$ state, an
isospin  doublet of strange mesons (plus their antiparticles), and an
isotriplet which is approximately mass degenerate with the $n\bar n$
state. The strange $K_1$ mesons are mixtures of the two $K_{1A}$ and
$K_{1B}$ states (belonging to the $J^{PC}=1^{+\pm}$ nonets,
respectively). The pseudoscalar mesons have no ideal mixing, the
$\eta$ is much heavier than the $\pi$. The lower-mass scalar nonet has
a structure like vector, tensor, and axialvector mesons have, but it is
inverted. The other scalar mesons may form a decuplet possibly
evidencing the intrusion of a glueball into a scalar meson nonet. The
figure is adapted from \cite{Maiani:2006rq}. } \vspace*{-2mm}
\end{figure}

Figure \ref{maiani} compares the scalar mass spectrum with those of
other nonets. Vector and tensor nonets show ideal mixing and a very
regular pattern. Axial vector mesons exhibit a similar pattern even
though modified due to $K_{1A}-K_{1B}$ mixing (see section
\ref{Mixing}). The pseudoscalar and low-mass scalar spectra are similar
but inverted. In all these cases, nine mesons with identical $J^{PC}$
fall into a limited mass gap which identifies the nine mesons as one
single nonet. The scalar mesons at higher mass resist such a simple
pattern. In a reasonable mass gap of 400\,MeV/c$^2$, ten scalar states
are reported \cite{Eidelman:2004wy}. A popular interpretation assigns
these ten states to a scalar nonet plus a scalar glueball. The three
isoscalar states are supposed to mix forming the three observed states.

\subsection{\label{Glueball-qbar q mixing scenarios}
Glueball-$q\bar q$ mixing scenarios}

The most popular view of
the scalar mesons assumes that $\sigma(485)$, $\kappa(700)$,
$f_0(980)$, and $a_0(980)$ are bound by molecular forces or are $qq\bar
q\bar q$ states, unrelated to $q\bar q$ spectroscopy. In the approach
of Beveren and Tornqvist and their collaborators, these mesons acquire
a large $q\bar q$ fraction, but the number of the lowest-lying scalar
states is still doubled due to the molecular forces between the decay
products of scalar mesons. Thus the lowest mass scalar nonet can still
be discarded in as far as the abundance of scalar mesons is in the
focus of interest.

Then, there are three states, $f_0(1370)$, $f_0(1500)$, and $f_0(1760)$
which need to be interpreted. One of them, $f_0(1760)$, couples
strongly to $ K\bar K$ and could be the $1^3P_0$ $s\bar s$ state.
The two remaining states, $f_0(1370)$ and $f_0(1500)$, cannot both be
$1^3P_0$ $n\bar n$ states; hence there must be one non-$q\bar q$ state.
Since the lowest mass scalar glueball is predicted to fall into this
mass range, it is natural to assume that a glueball has intruded the
spectrum of scalar mesons, mixes with them thus forming the 3 observed
states $f_0(1370)$, $f_0(1500)$, and $f_0(1760)$. An extension is
offered when $f_0(1760)$ is split into a $1^3P_0$ $s\bar s$ $f_0(1710)$
and a $2^3P_0$ $n\bar n$ $f_0(1790)$ state.

Different mixing scenarios have been developed in which the primordial
glueball has a mass above or in between the two $q\bar q$ states. The
differences between the models lie in the assumed mass of the glueball
-- which can be taken from lattice gauge calculation -- and in imposing
(or not) the known decay branching ratios of scalar mesons. These
mixing scenarios were pioneered by two papers by Amsler and Close
\cite{Amsler:1995tu,Amsler:1995td}. In these papers, it was shown that
the properties of $f_0(1370)$ and $f_0(1500)$ are incompatible with
them both being $\qqbar$ mesons. In contrast,  $f_0(1500)$ mass and
decays are compatible  with it being the ground state glueball mixed
with the nearby states of the $0^{++}$ $q\bar{q}$ nonet. These
conclusion were confirmed when data from central production were
included in the analysis \cite{Close:2000yk,Close:2001ga}. The $s\bar
s$ state predicted in \cite{Amsler:1995tu,Amsler:1995td} was identified
with the $f_0(1710)$. Further support for a glueball nature of
$f_0(1500)$ was presented in \cite{Amsler:2002ey}. The upper limits on
the two-photon widths of $f_0(1500)$ and $f_0(1710)$ suggest that
these two mesons cannot have sizable $n\bar n$ components. Hence they
must have glueball and $s\bar s$ wave functions, the decay pattern
assigns the larger glueball content to $f_0(1500)$. The production
rates of $f_0^i$ ($i=1$, 2, 3 corresponding to $f_0(1710)$,
$f_0(1500)$, and $f_0(1370)$) in J/$\psi\to V f_0 \to V PP$
($V$=vector, $P$=pseudoscalar meson) were also exploited to test
glueball and $q\bar{q}$ nonet mixing scenarios. In \cite{Close:2005vf}
it is shown that the peculiar patterns in these branching ratios can be
understood by assuming the presence of significant glueball fractions in
the scalar wavefunctions. Recently, branching ratios from J/$\psi$
decays into a vector and a scalar meson were used to determine the
mixing \cite{Close:2005vf,Zhao:2005nv}; OZI rule violating is required
to play a significant $\rm r\hat{o}le$ \cite{Zhao:2005ip}. J/$\psi$
decays were found to be very sensitive to the structure of scalar
mesons. The analysis confirmed the claim for a glueball in the
$1.5-1.7$ GeV/c$^2$ region, in line with Lattice QCD. The mixing
parameters were compatible with previous analyses.

Starting from lattice results on the scalar glueball mass, Weingarten
and Lee \cite{Weingarten:1996pp,Lee:1999kv} see the main glueball
component in $f_0(1710)$ while $f_0(1500)$ is believed to be dominantly
$s\bar s$. The decay pattern of scalar mesons was not included in the
analysis. De-Min Li {\it et al.} studied the influence of the position
of the primordial glueball mass on the mixing parameters
\cite{Li:2000yn}. Crystal Barrel results on 4$\pi$ decays of scalar
mesons were included in the analysis by F\"assler and collaborators
\cite{Strohmeier-Presicek:1998fi,Strohmeier-Presicek:1999yv}. In this
analysis, slightly more than 50\% of the $f_0(1500)$ wave function is
of gluonic nature, the $f_0(1710)$ comprises 32\% and the $f_0(1370)$
12\% of glue. Refined variants of this approach can be found in
\cite{Giacosa:2004ug,Giacosa:2005qr,Giacosa:2005zt}. A viable mass
matrix can however also be constructed with the largest glueball
fraction assigned to $f_0(1370)$  \cite{Li:2000cj}.

In a relativistic quark model Celenza {\it et al.}
\cite{Celenza:2000uk} find a good fit to the scalar mass spectrum by
adding a glueball with mass of about 1700 MeV/c$^2$. The scalar meson
nonet is nearly ideally mixed, with the $f_0(980)$ interpreted as
having a 10\% $s\bar s$ component. The $f_0(1370)$ is the dominantly
$s\bar s$ state, while $f_0(1500)$ is a $n\bar n$ state having a
single node. The next radial excitation is expected at 1843 MeV/c$^2$.
Thus, $f_0(1760)$ was identified as the state with the largest glueball
component \cite{Celenza:2000uk}. The scheme can be extended to contain
not only $q\bar q$ mesons and glueballs but also tetraquarks
\cite{Vijande:2005jd,Maiani:2006rq} or hybrids \cite{Zhao:2005ip}. The
possible discovery of a further scalar meson at 1810\,MeV/c$^2$ may
require the extension into one of these directions.

Based on QCD spectral sum rules, Narison interprets  $\sigma, \kappa,
\delta$, S$^*$ as nonet of $1^3P_0$ ground state $q\bar q$ mesons.
The properties of the isoscalar states are strongly influenced by their
coupling with a low-mass glueball. He suggests that $f_0(1370)$ is
mostly $n\bar n$ while $f_0(1500)$, $f_0(1710)$, and $f_0(1790)$ should
have significant gluonium components in their wave functions
\cite{Narison:2005wc}. Invoking sum rules as well, Zhang, and Steele
\cite{Zhang:2006xp} interpret the light scalar meson nonet as nonet of
tetraquarks.

\begin{figure}[ph]
\bc
{\tiny $\sigma(485)$\quad.98\quad1.37\quad1.5\quad1.75 GeV/c$^2$\hspace{3mm}
 $\sigma(485)$\quad .98\quad 1.37\quad 1.5\quad 1.75 GeV/c$^2$\hspace{3mm}
$\sigma(485)$\quad .98\quad 1.37\quad 1.5\quad 1.75 GeV/c$^2$
\phantom{tttttrrrrrrrrr}\hfill 1.79\,GeV/c$^2$ }
\includegraphics[width=1.09\textwidth,height=5cm,clip=on]{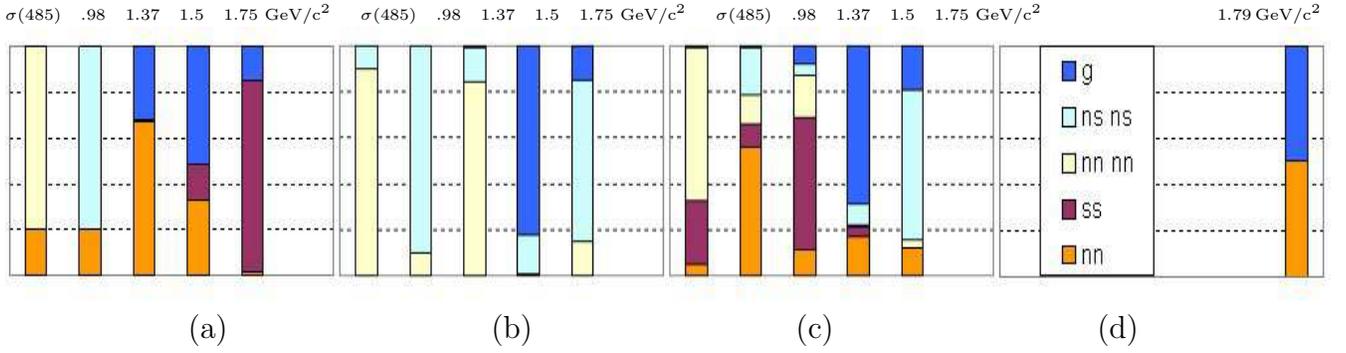}
\vspace{-18mm}
\bc
(a)\hspace{35mm}(b)\hspace{35mm}(c)\hspace{35mm}(d)
 \ec
\vspace{-3mm}
\caption{\label{fig:mixing} Decomposition of scalar isoscalar states
into different components. a) $n\bar n, s\bar s$ and glueball
\cite{Close:2001ga}. b) $qq\bar q, \bar q, qq\bar q\bar q$ and glueball
\cite{Maiani:2006rq}. c) $n\bar q, s\bar s, qq\bar q\bar q$ and
glueball \cite{Fariborz:2006xq}. d) The $f_0(1790)$ and $f_0(1810)$ are
treated as one state with very large glueball content
\cite{Bugg:2006uk}. Notation: $n\bar n={\tiny\frac{1}{\sqrt 2}}(u\bar u
+d\bar d)$, $q\bar q=8_F\oplus 1_F$. }
\ec
 \end{figure}

A few recently suggested interpretations of the scalar mesons below
1\,GeV/c$^2$ and decompositions of $f_0(1370)$, $f_0(1500)$, and
$f_0(1710)$ into their $n\bar n, s\bar s$ and glueball components are
collected in Fig.~\ref{fig:mixing}. Close and Kirk
(Fig.~\ref{fig:mixing}a) interpret $\sigma(485)$ and S$^*$ as molecules
with small $q\bar q$ admixture; the largest glueball component is
contained in $f_0(1500)$ \cite{Close:2001ga}. Maiani, Piccinini,
Polosa, and Riquer interpret $\sigma, \kappa, \delta$, S$^*$ as nonet
of tetraquark states. A decuplet consisting of $f_0(1370)$,
$K^*_0(1430)$, $a_0(1475)$, $f_0(1500)$, and $f_0(1710)$ is suggested
to originate from a nonet of tetraquark states plus a glueball. The
isoscalar mesons mix, the mixing fractions are shown in
Fig.~\ref{fig:mixing}b; $f_0(1500)$ is nearly a pure glueball
\cite{Maiani:2006rq}.  Fariborz starts from a nonlinear chiral
Lagrangian, developed in \cite{Black:1998wt,Fariborz:2005gm}, and
studies mixing of two scalar meson nonets, a two-quark nonet and a
tetraquark nonet, and a scalar glueball \cite{Fariborz:2006xq}. The
observed isoscalar states receive contributions from $gg$, $q\bar q$,
and $qq\bar q\bar q$ with fractions displayed in Fig.~\ref{fig:mixing}c.
Bugg underlines the unusual $\phi\omega$ decay mode in the scalar wave,
having a peak position at 1810\,MeV/c$^2$ \cite{Bugg:2006uk}. He
combines it with the 1790\,MeV/c$^2$ peak and derives decay properties
of an scalar resonance which require a large $q\bar q$ - glueball
mixing. Bicudo, Cotanch, Llanes-Estrada, and Robertson argue in favour
of a glueball interpretation \cite{Bicudo:2006sd}.
Fig.~\ref{fig:mixing}d reminds of this possibility. None of the papers
mentions that glueball decays into $\phi\omega$ violate SU(3) symmetry.

\subsection{\label{The Anisovich-Sarantsev picture}
The Anisovich-Sarantsev picture}

 \def\wh{\widehat}

The Gatchina group fitted a large set of different reactions and final
states as outlined in section \ref{Systematic of qbarq mesons in
planes}. Here we discuss the S-wave which is parameterised by a
K-matrix. In a simplified version, the scalar partial wave amplitude is
written in the form

\begin{tabular}{ccc} $ \wh A(s)\ =\ \wh K(s) \left[1-i\hat\rho\wh
K(s)\right]^{-1}, $ & $ \wh K=\left | \begin{array}{c} K_{11}, K_{12},
... \\ K_{21}, K_{22}, ... \\
 ... \\
\end{array}
\right |,
$
&
$
\wh \rho=\left | \begin{array}{c}
\rho_1, 0, ... \\
0\;\;\;, \rho_2, ... \\
 ... \\
\end{array}
\right |
$
\\
\end{tabular}

with a $5\otimes5$ matrix which includes the decay modes 1,2,3,4 and 5
and $1=\pi\pi$, $2=K\bar K$, $3=\eta\eta$, $4=\eta\eta'$,
$5=\pi\pi\pi\pi$. Threshold singularities are taken into account
correctly. The phase space elements were defined as $$
\rho_{\pi\pi}=\sqrt{\frac{s-4m_\pi^2}{s}}, \quad
\rho_{KK}=\sqrt{\frac{s-4m_K^2}{s}}, \quad ...
$$
The definition of the amplitudes became more complex in the course of
time and with an increasing number of reactions studied.

Fit parameters are the K-matrix elements, represented by a sum
of pole terms $g^{(n)}_ag^{(n)}_b/(\mu^2_n-s)$, and a smooth
$s$-dependent term $f_{ab}(s)$:
$$ K_{ab}\ =\ \sum \limits_n
\frac{g^{(n)}_ag^{(n)}_b}{\mu^2_n-s}+f_{ab}(s)\ ,
$$
$M_n$,  $g_a$, $g_b$ are K-matrix parameters. The full amplitudes
used recently in fitting procedure are documented in
\cite{Anisovich:2002ij}. From the fits, positions of poles (masses and
total widths of the resonances) and pole residues and hence partial
widths to meson channels $\pi\pi , K\bar K , \eta\eta , \eta\eta',
\pi\pi \pi\pi $ were deduced. Masses and widths of scalar mesons are
given in Table \ref{tab:as}, the decay branching ratios in Table
\ref{andreyscalar}. The $\sigma(485)$ was not required in the fits but
its existence was not excluded.  Possibly, its absence is due to the
threshold behaviour of the amplitudes which were not constrained by
chiral symmetry. A small modification of the threshold behaviour of
the amplitudes would likely accommodate $\sigma(485)$ and $\kappa(700)$
as additional resonances. In \cite{Anisovich:2005jn} it is suggested
that the $\sigma(485)$ resonance, if it exists, may originate from the
singularity of a flavour singlet Coulomb-like potential at $r\to 0$.
The $\kappa(700)$ was not mentioned.

\begin{table}[pb]
\caption{\label{tab:as}Pole positions of scalar mesons from a fit to a
large number of data sets \cite{Anisovich:2002ij}.\vspace{2mm}
 }
\bc
\renewcommand{\arraystretch}{1.4}
\begin{tabular}{cccc}
\hline\hline
$f_0(980)$ & $a_0(980)$ &  $ K_0^*(1430)$ & $f_0(1370)$\\
 $(1015\pm 15)- i(43\pm 8)$    &  $988\pm 5^1$ &
 $(1415\pm 25)- i(165\pm 25)$  & $(1310\pm 20)- i(160\pm 20)$ \\
$f_0(1500)$ & $a_0(1450)$ &  $ K_0^*(1800)$ & $f_0(1760)$ \\
$(1496\pm  8)- i( 58\pm 10)$  &
 $(1490\pm 5)-i(70\pm10)$      &
$(1820\pm 40)- i(125\pm 50)$  &
 $(1780\pm 30)- i(140\pm 20)^2$ \\
&Glueball: $f_0(1530)$ &
\multicolumn{2}{c}{$(1530^{+90}_{-250}) - i(560\pm 140)$}
 \vspace{5mm}\\
\hline\hline
\end{tabular}
\renewcommand{\arraystretch}{1.0}
{\scriptsize
$^1g_{\pi\eta}=0.436\pm 0.02$, $\left(\frac{g_{K\bar
K}}{g_{\pi\eta}}\right)^2=1.23\pm 0.10$ \qquad\ $^2$Alternative
solution: $(1780\pm 50)- i(220\pm 50)$ } \ec
\caption{\label{andreyscalar}
Partial decay widths of scalar mesons \cite{Anisovich:2002ij}.
\vspace{2mm}}
\bc
\renewcommand{\arraystretch}{1.4}
\begin{tabular}{lcccccc}
\hline\hline
Resonance &$\Gamma_{\pi\pi}$&$\Gamma_{K\bar K}$&
$\Gamma_{\eta\eta}$&$\Gamma_{\eta\eta'}$&$\Gamma_{\pi\pi\pi\pi}$& $\Gamma_{tot}$\\
\hline
$f_0(980):$ &$55\pm 5$ & $10\pm 1$ & -- & -- & $2\pm 1$ & $68\pm 10$ \\
$f_0(1300):$&$66\pm 10$ & $6\pm 4$ & $5\pm 2$  & -- & $230\pm 50$ &
$300\pm 40$\\
$f_0(1500):$&$23\pm 5$ & $6\pm 2$ & $3\pm 1$ & $0.1\pm
0.1$ & $80\pm 10$ & $110\pm 10$\\
$f_0(1760):$&$30\pm 5$ &  $100\pm 10$ &
$5\pm 3$ & $7\pm 3$ & $5\pm 5$ &$140\pm 40$\\
   &$105\pm 10$ &  $2\pm 1$ & $8\pm 1$ & $5\pm 3$ & $170\pm 40$
&$300\pm 50$\\ \hline\hline \end{tabular}
\renewcommand{\arraystretch}{1.0}
\ec
\end{table}

\begin{figure}[pt] \bc
\begin{tabular}{cccc}
\hspace{1mm}\includegraphics[width=0.23\textwidth,height=0.23\textwidth]{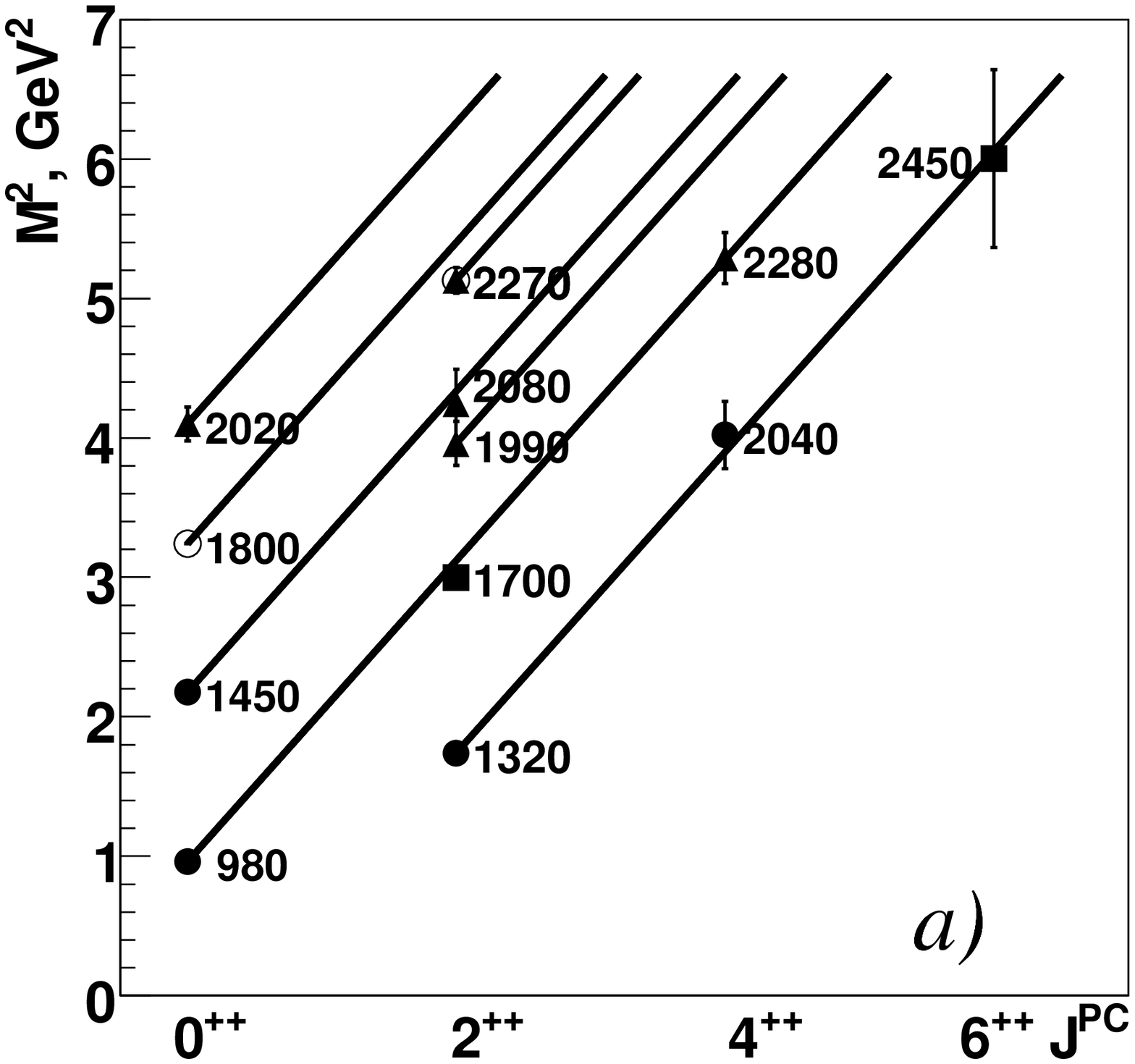}&
\hspace{1mm}\includegraphics[width=0.23\textwidth,height=0.23\textwidth]{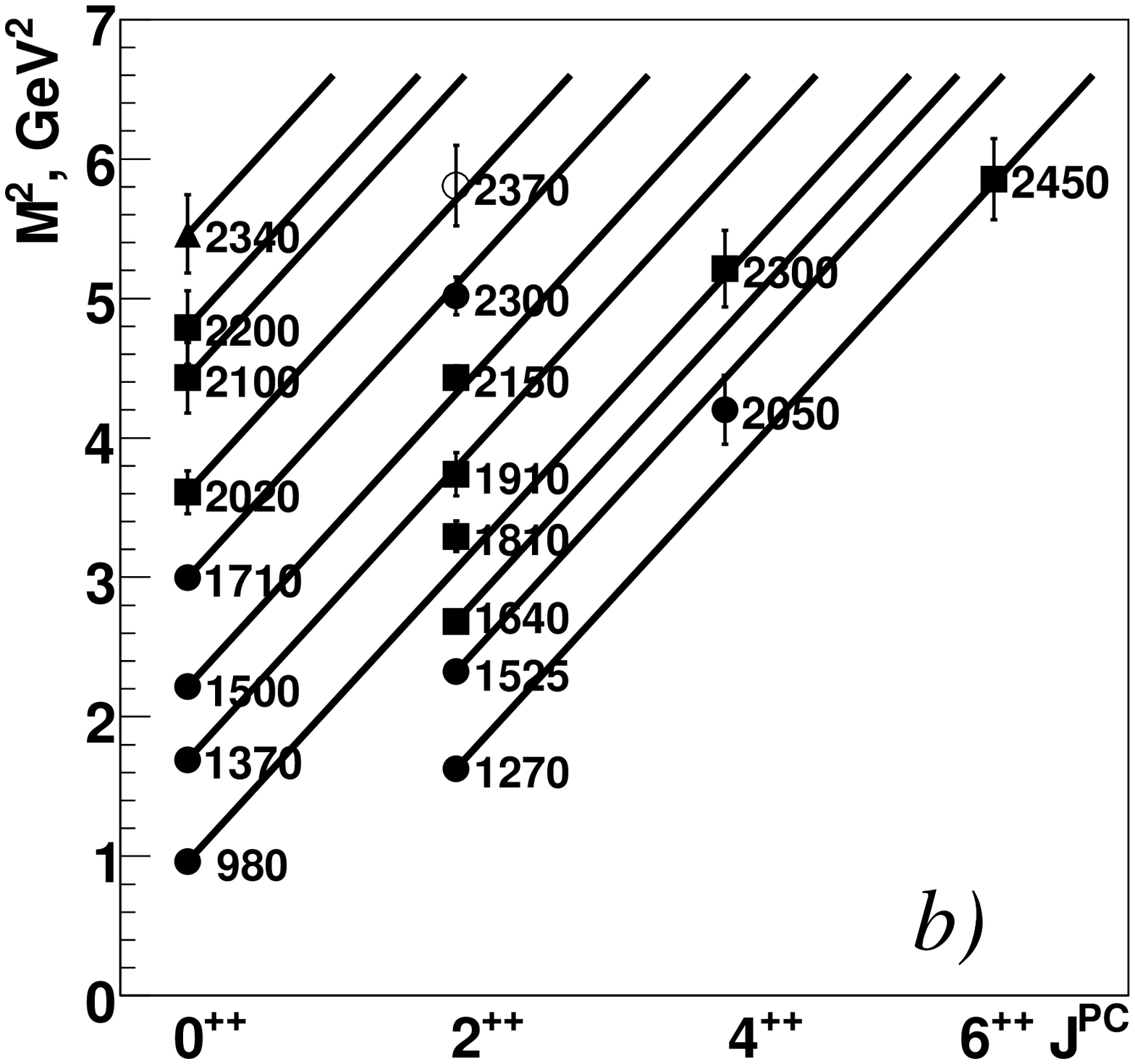}&
\hspace{1mm}\includegraphics[width=0.23\textwidth,height=0.23\textwidth]{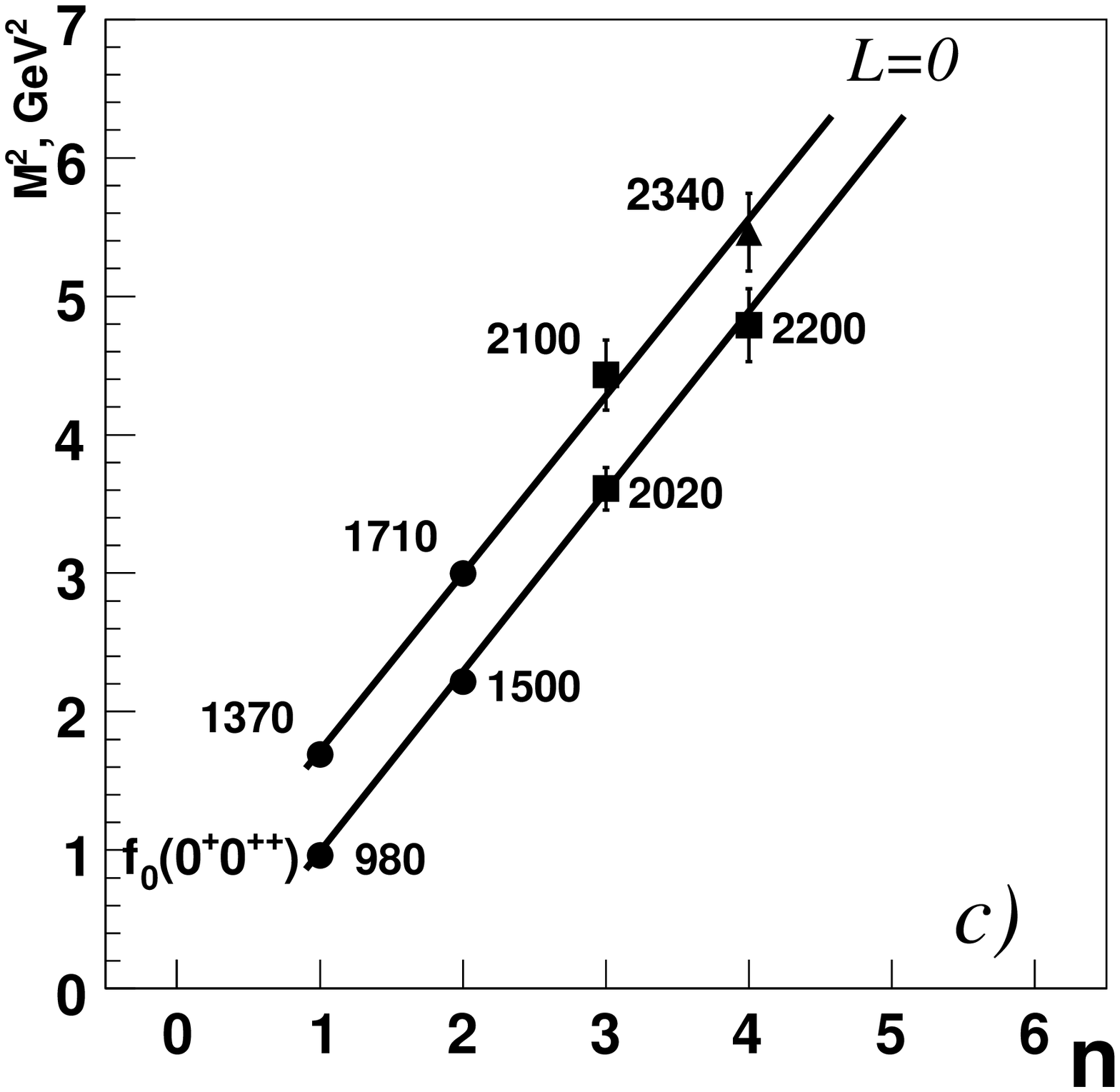}&
\hspace{1mm}\includegraphics[width=0.23\textwidth,height=0.23\textwidth]{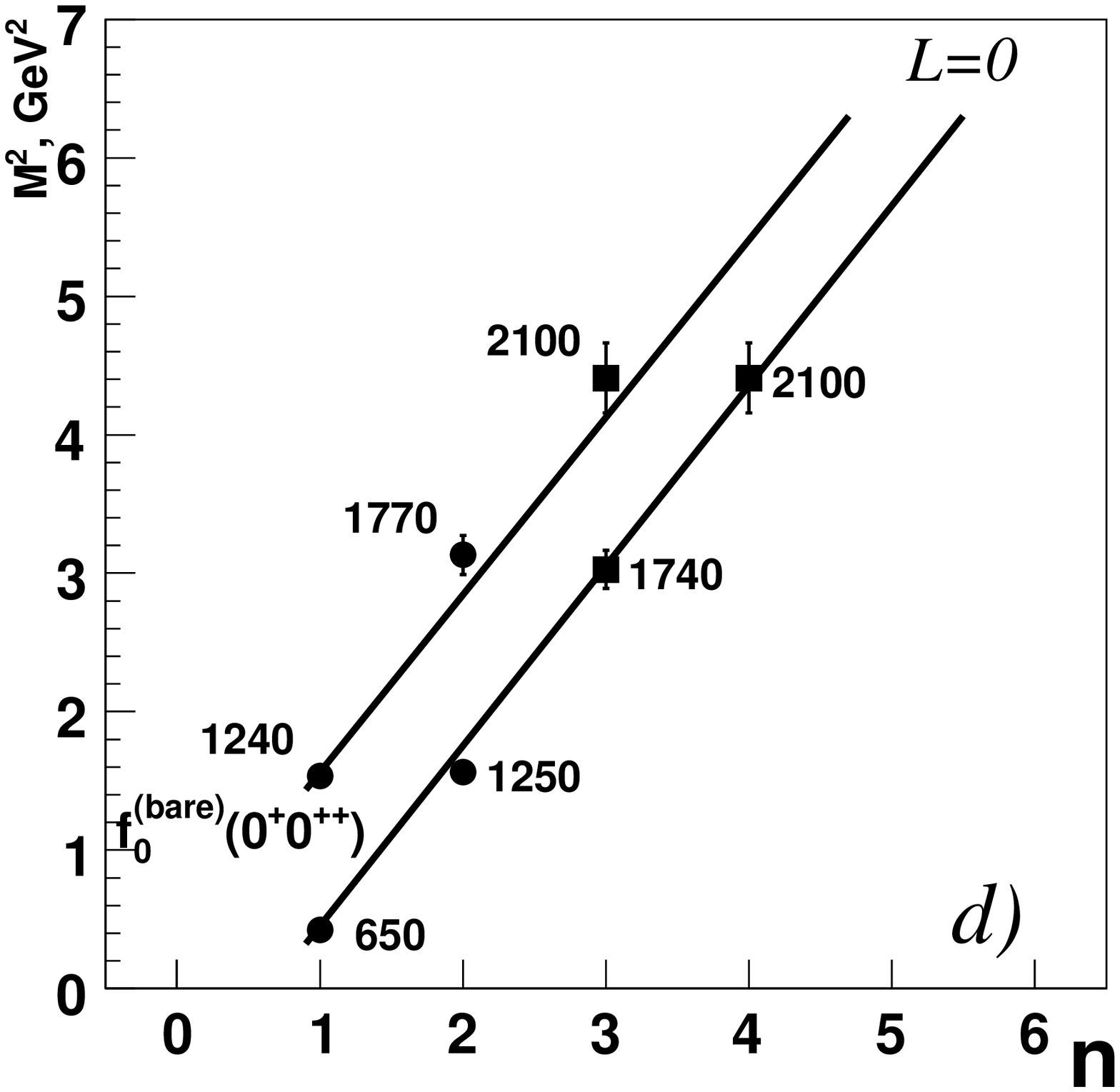}\\
\end{tabular}\ec
\caption{\label{scalarplot}
Leading and daughter Regge trajectories of isovector (a) and isocsalar
(b) positive-parity mesons with even $J$. Different symbols are used to
identify states belonging to the same trajectory. Fig. (b) contains
separate straight lines for mesons with a larger and a smaller $s\bar
s$ component. In particular the isoscalar scalar states reveal a rich
spectrum. The $f_0(00^{++})$-trajectories for the physical states (c)
and for the K-matrix pole positions (d) \cite{Anisovich:2000kx}.}
\end{figure}

The interpretation of the Gatchina group \cite{Anisovich:2002ij}
started from the quark model assignments presented in Figs.
\ref{AAS_planes} and \ref{psaas} -- shown in sections \ref{Systematic
of qbarq mesons in planes} and \ref{Other interpretations},
respectively -- and in Fig. \ref{scalarplot}. The two scalar mesons
$f_0(980), a_0(980)$ were interpreted as $q\bar q$ mesons. The main
reasons for this assignment were found in the consistency of their
masses with trajectories on  $(n,M^2)$ and $(J,M^2)$ planes,
in the high yield of $f_0(980)$ in $D_s$ decays, in the fact that
$\phi$ radiative yields of  $f_0(980)$ and  $a_0(980)$ agree with
calculations assuming a $q\bar q$ structure of these mesons, and in the
need of their presence in K-matrix fits and a consistent interpretation
of K- and T-matrix poles (see below) \cite{Anisovich:2005kt}.

With $f_0(980)$ and  $a_0(980)$ as $q\bar q$ states, the following
spectroscopic assigments were suggested:
$$
\{~a_0(980) , f_0(1300) , f_0(980) ,  K^*_0(1430)~\} \qquad\
\{~a_0(1450) , f_0(1760) , f_0(1500) ,  K^*_0(1430)~\}
$$
for the ground state scalar nonet and its first radial excitation.
These are the pole positions of the scattering matrix or T-matrix. They
correspond to the experimentally observed states. These poles did not
suffice to get a good description of the data, a wide background
amplitude was needed in addition. When parametrised as $s$-channel
resonance,  its mass was found to be \bc $0^{++}$ \ glueball: \qquad
$M=1530^{\ +90}_{-250}\quad;\quad\Gamma=560\pm140$ \ .
\ec
Its decay couplings to the channels $\pi\pi,K\bar K, \eta\eta,
\eta\eta'$ were found to be flavour neutral; the state is a flavour
singlet state. This gave the reason to consider this state as the
lightest scalar glueball.

The K-matrix poles are direct fit parameters but do not correspond to
observables quantities. However, when the coupling of T-matrix poles
are reduced (and then set to zero), the T-matrix poles approach the
K-matrix pole (and coincide with them for vanishing couplings). It is
thus tempting to identify the K-matrix poles with the quark model
states in the absence of decays. A full dynamical model of strong
interactions would include meson decay couplings but such calculations
are not available. To remedy this situation the Gatchina group
suggested that the K-matrix pole may represent the bare states and that
dressing of bare $q\bar q$ states (which results in observable meson
resonances) can be mimicked by switching on the couplings of K-matrix
poles to the respective final states. This is an untested but very
intuitive assumption. Hence the positions of the K-matrix poles and
their couplings to meson channels were identified as masses and
couplings of 'bare' states. For the first nonet, the K-matrix poles
were found to have, within 20\%, the same SU(3) coupling, for second
nonet members the coupling was nearly identical. (Of course, the SU(3)
couplings had to be multiplied with the appropriate isoscalar
coefficients.)

The bare states were interpreted as $q\bar q$
states before dressing, before the couplings to meson-meson final
states are turned on, and these are the states which should be compared
to quark model calculations. It is only after coupling to the
meson-meson continuum that the quark-model states acquire finite
widths. When the couplings to the
meson continua were turned on, for the fitted values of the coupling
constants $g_{\alpha}$, the physical states evolve. Their masses can
be determined by inspecting the T-matrix; their poles positions give
the physical masses and the total widths, the residues give the partial
decay widths. These are the quantities which are independent of the
production process.

On this basis, the bare states can be classified into quark-antiquark
nonets. In particular it can be imposed that (or checked if) two scalar
isoscalar mesons assigned to the same multiplet have orthogonal flavour
wave functions and that a $5^{\rm th}$ scalar isoscalar state has
couplings of a SU(3) isosinglet state, of a glueball candidate.
Its position differs in different solutions and the authors quote
$f_0(1200-1600)$. It is a rather broad state, with a width between 1
and 2\,GeV/c$^2$. It is interpreted as the descendant of a primary
glueball (even though mixing with SU(3) isoscalar mesons cannot be
excluded).

The strongest argument in favour of the Anisovich-Sarantsev approach is
the beautiful agreement between their K-matrix poles and the meson
mass spectrum calculated in the Bonn quark model using
instanton-induced interactions (see section \ref{Instanton-induced
interactions}). The comparison of K-matrix poles and of the
calculated masses is shown in Table~\ref{scalar_asmp}.

\begin{table}[ph]
\caption{\label{scalar_asmp}Comparison of K-matrix poles \cite{Anisovich:2002ij}
and meson masses calculated in a relativistic quark model with
instanton-induced forces \cite{Klempt:1995ku}. \vspace{2mm}}
\renewcommand{\arraystretch}{1.6}
\bc
\begin{tabular}{ccccc}
\hline\hline
$1^3P_0q\bar q$: \quad&\quad $f_0^{bare}(720\pm 100)$ \quad&\quad $a_0^{bare}(960\pm
30)$ \quad&\quad
     $ K_0^{bare}(1220^{+\; 50}_{-150})$ \quad&\quad $f_0^{bare}(1260\pm 30)$\\
Model            \quad&\quad  $f_0(665)$              \quad&\quad   $a_0(1057)$           \quad&\quad
     $ K_0^*(1187)$                      \quad&\quad  $f_0(1262)$            \\
$2^3P_0q\bar q$: \quad&\quad $f_0^{bare}(1230^{+150}_{-\; 30})$ \quad&\quad$a_0^{bare}(1650\pm 50)$\quad&\quad
     $ K_0^{bare}(1885^{+\; 50}_{-100})$ \quad&\quad $f_0^{bare}(1810\pm 30)$\\
Model:           \quad&\quad $f_0(1262)$              \quad&\quad $a_0(1665)$             \quad&\quad
     $ K_0^*(1788)$                      \quad&\quad $f_0(1870)$            \\
\hline\hline
\end{tabular}
\ec
\renewcommand{\arraystretch}{1.6}
\end{table}

The agreement is very impressive. The identification of K-matrix poles
and quark model states (which are calculated by neglecting their
coupling to the meson-meson continuum) seems very plausible even though
one has to admit that K-matrix poles are not observable quantities.
Even worse, there is no one-to-one correspondence of K- and T-matrix
poles. The position of K-matrix poles can be changed by a large amount,
and the T-matrix poles can still remain at the same position. Only in
the limit of narrow resonances, K-matrix and T-matrix poles remain at
similar values. Quantitative details on the spread of K-matrix poles
for a given T-matrix pole can be found in a recent paper by Jaffe
\cite{Jaffe:2007id}.

The mass and width of the wide state are not well defined, the mass
were found in the range 1200 to 1600\,MeV/c$^2$, the width between 1000
and 2000\,MeV/c$^2$. The large width is accumulated from its mixing
with $q\bar q$ states. Anisovich, Bugg, and Sarantsev developed the
idea of width accumulation of an exotic state which overlaps with $\bar
qq$ resonances \cite{Anisovich:1997nk}. The idea can best be followed
in a concrete example. Let $Re\;M_1^2$ and $Re\;M_2^2$ be different but
$M_1\Gamma_1=M_2\Gamma_2=M\Gamma$. Then,
\begin{eqnarray}
 M_1^2=M_{R1}^2-iM\Gamma\,,\qquad M_2^2=M_{R2}^2-iM\Gamma\,.
\label{v48}
\end{eqnarray}
The diagonalisation of the mass matrix yields:
\begin{eqnarray}
M_{A,B}^2=\frac 12(M_{R1}^2+M_{R2}^2)-iM\Gamma
\pm\sqrt{\frac 14(M_{R1}^2-M_{R2}^2)^2-M^2\Gamma^2}.
\label{v49}
\end{eqnarray}
In the limit  $2M\Gamma\ll |M^2_{R1}-M^2_{R2}|$, there is no mixing and
the location of the two poles remains unchanged. For
$|M^2_{R1}-M^2_{R2}|\ll 2M\Gamma$ the two poles coincide in mass, with
one state being almost stable while the width of the other resonance is
close to $2\Gamma$. Of course, this is a simplified example. The two
quark model states $ K_{1A}$ and  $ K_{1B}$ mix to form $K_{1}(1270)$
and $ K_{1}(1400)$. Their total widths are similar, 90 and
174\,MeV/c$^2$, respectively. But they acquire different partial
widths: $ K_{1}(1270)$ couples to $ K^*\pi$ with $(16\pm 5)$\%
branching ratio and $(53\pm 6)$\% to $ K\rho$  and $ K\omega$;  $
K_{1}(1400)$ decays with $(94\pm 6)$\% to $ K^*\pi$ and with $(5\pm
4)$\% to $ K\rho$  and $ K\omega$. In this example, none of the
states acquires a broad total width; rather they decay into two
orthogonal final states.

\subsection{\label{The red dragon}
The red dragon of Minkowski and Ochs}

Minkowski and Ochs \cite{Minkowski:1998mf} interpret the isoscalar
states $f_0(980)$ and $f_0(1500)$ together with $a_0(980)$ and
$K_0^*(1430)$ as members of the $q\overline q$ nonet with
$J^{PC}=0^{++}$ of lowest mass. They show that the $f_0(980)$
and $\eta^{\prime}$ have very similar production patterns in J/$\psi$
decay when recoiling against an $\omega$ or $\phi$, and that $f_0(1500)$
has a flavour wave function in which the $n\bar n$ and $s\bar s$
components have opposite signs. The data are compatible with $f_0(980)$
as isoscalar singlet and $f_0(1500)$ as isoscalar octet member of the
scalar nonet. Hence flavour mixing in the scalar sector resembles more
the one observed in the pseudoscalar nonet even though with inverted
masses. Scalar mesons do not follow the ideal mixing pattern of vector
and tensor mesons and as most other meson nonets do. The pair
$f_0(980)$ and $\eta^{\prime}$ forms a (nearly) mass-degenerate parity
doublet of scalar isoscalar states.

In the mass range from 400 MeV/c$^2$ up to about 1700 MeV/c$^2$, the
$\pi\pi$ amplitude describes a full loop in the Argand diagram after
the $f_0(980)$ and $f_0(1500)$ states are subtracted. This background
amplitude is called {\it red dragon} and identified with the expected
glueball. The $\sigma(485)$ and $f_0(1370)$ are considered to be part of
the red dragon and not to be genuine resonances. It should be noted that
the authors interpret $\eta(1440)$ as pseudoscalar glueball (see
however section \ref{Pseudoscalar mesons}), and $f_J((1710)$ with $J=2$
as tensor glueball.

The glueball scenario suggested by Minkowski and Ochs has some
similarity with the Anisovich-Sarantsev picture even though the scalar
$q\bar q$ nonets are completely different. Minkowski and Ochs interpret
the observed T-matrix poles at 980\,MeV/c$^2$ and at 1500\,MeV/c$^2$ as
$q\bar q$ states; the $f_0(1370)$ is put into question, the $f_0(1760)$
is argued to have tensor quantum numbers and suggested to be the tensor
glueball. Jointly with $a_0(980)$ and $ K^*_0(1430)$, one scalar nonet
is suggested. Anisovich and Sarantsev identify two full scalar nonets.
Their T-matrix poles, the observable resonances, are given by
$f_0(980), f_0(1370), f_0(1500)$, $f_0(1760)$, $a_0(980), a_0(1450)$, $
K^*(1430)$ and $ K^*_0(1760)$. Common in both pictures is the glueball:
it is a broad about 1\,GeV/c$^2$ wide object with a mass of 1 to
1.5\,GeV/c$^2$.

\subsection{\label{Scalar mesons with instanton-induced interactions}
Scalar mesons with instanton-induced interactions}

In 1995, a quark model was suggested \cite{Klempt:1995ku} which is
based on quarks having an effective constituent quark mass, a linear
interaction modelling confinement and a residual interaction based on
instanton effects~\cite{'tHooft:1976fv,Shifman:1979nz}. The model had
the virtue of reproducing near exactly the mass spectrum of the
ground-state pseudoscalar mesons and their flavour mixing. The
instanton-induced interaction inverts their sign when going from the
pseudoscalar to the scalar world. Thus they lower the mass of states
whose SU(3) flavour structure is dominantly flavour singlet and pushes
the dominantly flavour octet states to higher masses, see Fig.
\ref{instanton-for-mesons} in section \ref{Instanton-induced
interactions}. The calculated masses of the ground state scalar mesons
were predicted to be
\begin{equation}  \qquad a_0(1370),\;\;\;\;\;\; K_0^*(1430),\;\;\;\;\;\;
                         f_0(980),\;\;\;\;\;\; f_0(1470)
\end{equation} which can be compared to the observed states
\begin{equation}a_0(980)/a_0(1450),\;\;\;\;\;\; K_0^*(1430),\;\;\;\;\;\;
                          f_0(980),\;\;\;\;\;\; f_0(1500)\,.
\end{equation}
The $f_0(980)$ is calculated to have a nearly pure flavour singlet wave
function and $f_0(1500)$ to be a flavour octet state, in accordance
with its large coupling to $\eta\eta^{\prime}$\footnote{\footnotesize
The ratio $0.38\pm0.16$ of partial widths $f_0(1500)\to
\eta\eta^{\prime}$ to $f_0(1500)\to \eta\eta$ is rather large in spite
of the much more favourable phase space for $\eta\eta$ decays. A
isoscalar singlet can of course not decay into $\eta\eta^{\prime}$
while an octet is allowed to decay into $\eta\eta$.}.

Preserving the molecular picture of the $a_0(980), f_0(980)$ doublet
yields a nonet of a scalar $q\bar q$ nonet
\begin{equation}
a_0(1450),\;\;\;\;\;\; K_0^*(1430),\;\;\;\;\;\;
f_0(1000),\;\;\;\;\;\; f_0(1500)\,.
\end{equation}
In this case, the broad background component is identified with the
ground-state scalar meson. In the dual Regge picture, the scattering
amplitude can be described by a sum of $s$-channel resonances or
by a sequence of $t$-channel exchanges. The Pomeron belongs to the
allowed $t$-channel exchanges, it is dual to a `background' in the
direct $s$-channel and can possibly be identified with a contribution
of a glueball. From $\pi\pi$ scattering, amplitudes for $t$-channel
exchanges with defined isospin $I_t$ was constructed by Quigg
\cite{Quigg:1974je}. The $I_t=0$ amplitude rises with energy and
is sizable already below 1 GeV. This amplitude could be due to Pomeron
exchange which would suggest that the broad background component could
be a glueball. Of course, exchanges of isoscalar $q\bar q$ mesons like
$f_2(1270)$ contribute to this amplitude, hence the broad isoscalar
scalar background could also be of $q\bar q$ nature. Its width may
speak against this interpretation, but the analysis of Ritter, Metsch,
M\"unz, and Petry  \cite{Ritter:1996xh} has shown that instanton
induced interaction can make significant contributions to the decay
amplitude for the decay of scalar mesons into two pseudoscalar mesons.
The solution space for interference of this amplitude and a
conventional $^3P_0$ decay amplitude is not yet explored. Hence very
different widths within a scalar nonet are not ruled out.

 If $f_0(1370)$ does not exist as $q\bar q$ resonance, if $f_0(1000)$
is of $q\bar q$ nature, and if the lowest mass scalar meson form a
separate nonet there is, below 1.7\,GeV/c$^2$, no room  left for the
scalar glueball \cite{Klempt:2000ud}.

The eminent $\rm r\hat{o}le$ played by the $U_A(1)$ symmetry-breaking
t' Hooft interaction in scalar mesons was underlined by Dmitrasinovic
\cite{Dmitrasinovic:1996fi} in the Nambu-Jona-Lasinio model. He derived
an approximate sum rule
\be
m^2 _{\eta^{\prime}}+m^2 _{\eta}-2m_K ^2 = m^2 _{f_0 ^{\prime}}+m^2
_{f_0} -2m^2 _{K_0} \label{dmitra} \ee linking the mass splittings
between the pseudoscalar and the scalar singlet and octet mesons. The
splitting must have the same size but opposite signs. A scalar nonet
was suggested to comprise \begin{equation} \qquad
a_0(1320),\;\;\;\;\;\; K_0^*(1430),\;\;\;\;\;\;
f_0(1000),\;\;\;\;\;\; f_0(1590) \end{equation}
 Burakovsky, partly with Goldman, calculated the mass spectra of
P-wave mesons using a non-relativistic quark model
\cite{Burakovsky:1997ci,Burakovsky:1997cj}. The calculation assumed
degeneracy of tensor and scalar mesons, except for the scalar
isosinglet state. Its mass is put in 'by hand' using eq. (\ref{dmitra}).
In the Nambu-Jona-Lasinio model with instanton-induced interactions, a
SU(3) multiplet mass formula was obtained consistent with this
assignment \cite{Dmitrasinovic:1996fi}.

\subsection{\label{Our own interpretation}
Our own interpretation}

\subsubsection{\label{Light scalar nonets}
Light scalar nonets}

The starting point of our interpretation is $f_0(1500)$. It is a mainly
SU(3) octet state; this follows from its strong $\eta\eta^{\prime}$
decay mode \cite{Amsler:1994ah}, from the phase motion of the scalar
wave in $\pi^- p\to p\,\eta\eta$ and $p\,K^0_SK^0_S$
\cite{Minkowski:1998mf}, and from the constructive interference of
$B^{\pm}\to K^{\pm}f_0, f_0\to (K\bar K)$ and destructive interference
in $B^{\pm}\to K^{\pm}f_0, f_0\to (\pi\pi)$ \cite{Minkowski:2004xf}.
The different interference patterns in these two reactions would lead,
if not properly taken into account, to the necessity to introduce three
scalar states, a scalar state at about 1350\,MeV/c$^2$ coupling to
$\pi\pi$, the standard $f_0(1500)$, and a new state at 1500\,MeV/c$^2$
with strong coupling to $K\bar K$ and weak coupling to $\pi\pi$
\cite{Garmash:2004wa,Aubert:2005kd,Aubert:2005wb}. The latter state
would have properties incompatible with the standard Crystal Barrel
state which decays into $\pi\pi$ four times more frequent than into
$K\bar K$. We refuse this interpretation. Instead, we enforce the
interpretation of Minkowski and Ochs \cite{Minkowski:2004xf} that the
$f_0(1500)$ with standard properties interferes with a scalar isoscalar
background.

The $f_0(1500)\to (K\bar K)$ peak would require introduction of a new
resonance if interference effects were not properly taken into account.
With this warning, we examine the BES puzzle on J/$\psi$ decays into a
scalar mesons recoiling against an $\omega$ and a $\phi$ meson. The
1700-1800\,MeV/c$^2$ region is often supposed to be split into a
low-mass $f_0(1710)$ and a high-mass $f_0(1790)$. The problem is that
the low mass state $f_0(1710)$ is produced in J/$\psi\to \omega
f_0(1710)$ as a mainly $n\bar n$ state. But is only observed decay mode
is $K\bar K$. The $f_0(1790)$ is produced in J/$\psi\to \phi
f_0(1790)$ as a mainly $s\bar s$ state. But the latter state decays
into $\pi\pi$. At the first glance, this is completely contradictory.
To increase the complexity of the pattern, a scalar structure was
observed in radiative J/$\psi$ decays which decays into $\phi\omega$.
Decays into $\phi\omega$ are OZI rule violating; radiative production
of such a state violates the OZI rule as well. The BES collaboration
announced the observation as doubly OZI rule violating effect.

We start with the assumption that the three observations $f_0(1710)$,
$f_0(1790)$, and $f_0(1810)$ represent one single state with unusual
properties which we call $f_0(1760)$. In the SU(3) limit, $\phi\omega$
is flavour octet. The radiative yield of $f_0(1760)$ is large, hence
it does not have only a small octet component, its flavour structure
must be octet-like (with a negative sign between \nnb\ and \ssb ).
Production of flavour-octet mesons in radiative J/$\psi$ decays is
unexpected. Initial state radiation in J/$\psi$ decays produces a
system of pure glue (with minimum of two gluons in a perturbative
picture). These cannot couple to one flavour-octet meson. However,
final state radiation could play a significant $\rm r\hat{o}le$ as
well. Here it is worthwhile to recall that bremsstrahlung is produced
proportional to $m^{-3}$ of the radiating system. These factors
counterbalance arguments based on the number of gluons. Production of
octet mesons seems therefore not implausible.

A $q\bar q$ octet state can decay into $\rho\rho$, $\omega\omega$, and
$K^*K^*$ but not into $\phi\omega$. However, a tetraquark S-wave
configuration with the two antisymmetric quark pairs may be
energetically favoured compared to $q\bar q$ in P-wave. An
octet-state tetraquark wave function is derived by adding a second
$q\bar q$ pair to the $q\bar q$ wave function in a way suggested by
Jaffe. In this way one obtains

 \begin{equation}
\label{acht}
f_0^{8}=\frac{1}{\sqrt
6}|\uub + \ddb -2\ssb> \qquad \Rightarrow \qquad \frac{1}{\sqrt
6}|2\uub\ddb - \uub\ssb - \ddb\ssb> \end{equation}

which decays naturally, by falling apart, into $\pi\pi$, $K\bar K$,
$\rho\rho$, $\omega\omega$, $K^*K^*$, and into $\phi\omega$. The
tetraquark flavour singlet wave function

\be
\label{eins}
f_0^{1} = \frac{1}{\sqrt 3}|u\bar
u+d\bar d+s\bar s> \qquad \Rightarrow \qquad \frac{1}{\sqrt 3}|u\bar
ud\bar d+u\bar us\bar s +d\bar ds\bar s>.
\ee

has the same decay modes as the octet state. It is important to
emphasize that a small tetraquark fraction in the flavour wave function
and a large decay probability of this component is sufficient to result
in a significant impact on the observed decay pattern.

Likely, SU(3) singlet and octet amplitudes contribute to the final
state as in the case of $B$ decays discussed above (see Fig.
\ref{fig:minkochs} and their discussion in the text). Interference
between singlet and octet amplitudes could be responsible for the
weird production and decay pattern in J/$\psi$ decays into scalar
mesons recoiling against an $\omega$ or $\phi$. Note that adding or
subtracting eqs. (\ref{acht}) and (\ref{eins}) with appropriate factors
yields pure $\uub\ddb$ and $\uub\ssb - \ddb\ssb$ states. The
former state does not couple to $K\bar K$, the latter not to $\pi\pi$.

The $a_0(1475)$ mass is larger than that of $K^*_0(1430)$ even though
the latter state carries strangeness, the former one not.  This
`inverse' mass pattern has been discussed often and used as an argument
against $a_0(1475)$ and $K^*_0(1430)$ belonging to the same SU(3)
nonet. In Jaffe's scalar tetraquark nonet, $a_0$ masses are heavier
than $K^*_0$ masses by 150\,MeV/c$^2$, see Fig. \ref{mes:fig:four-q}.
This for tetraquark states, the inverse mass pattern emerges naturally.
Mixing with $q\bar q$ can obviously reproduce the observed masses.

When $f_0(980)$, $f_0(1500)$, and $f_0(1760)$ are interpreted as SU(3)
octet states, where are the associated singlet resonances? Scalar
isosinglet resonances couple directly to the QCD vacuum, in contrast to
the octet states. Thus isosinglet states can become very broad. We
suggest that all flavour-singlet scalar mesons get absorbed into a wide
structure; the transition from the flavour-singlet $^3P_0$ mesons to
the vacuum by emission of two pseudoscalar mesons, of two Goldstone
bosons, is too fast to support the existence of individual resonances.
The scalar glueball could be part of this broad background. The
background is thus `gluish' in the same sense as $\eta^{\prime}$ mesons
are gluish. But the coupling to the QCD vacuum is much stronger for
$^3P_0$ quantum numbers than for the flavour-singlet pseudoscalar
$\eta^{\prime}$. The isoscalar scalar amplitude manifests itself only
by interference with the narrow octet scalar mesons.

\begin{figure}[pt] \bc
\includegraphics[width=0.5\textwidth]{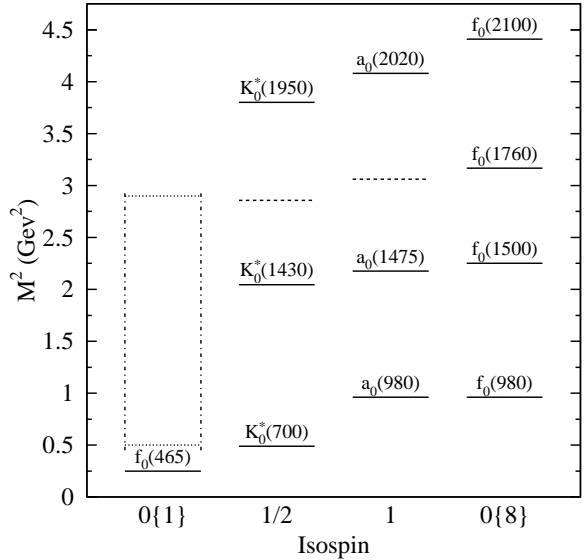}
\caption{\label{ourscalar}
Scalar mesons with 4 octets: \{$a_0(980), K^*_0(700), f_0(980)$\},
\{$a_0(1475), K^*_0(1430), f_0(1500)$\}, \{$a_0(xxx), K^*_0(xxx),
f_0(1760)$\}, \{$a_0(2020), K^*_0(1950), f_0(2100)$\}. The ninth
members are flavour SU(3) singlets. They are not separated into
individual resonances. A glueball component is possible but unproven.
}\ec
\end{figure}

In the schematic view of  Fig. \ref{sampl}, the wide scalar background
amplitude is assumed to be isoscalar, the interspersed narrow
resonances octet states. The $\sigma(485)$ is partner of $f_0(980)$, the
other isosinglet partners are not easily identified due to their large
widths. Fig. \ref{ourscalar} sketches this scenario.

We see immediately one problem: interpreted as octet state, $f_0(1760)$
needs an isovector and isodoublet partner. Hence either $f_0(1760)$ is
isosinglet (and the problematic production and decay pattern observed
in  J/$\psi$ decays remains an unsolved issue), or we need to postulate
the existence of a $a_0(1720)$ and a $K^*_0(1680)$ resonance, where the
masses are estimates. It is not unrealistic that these two states have
escaped detection so far. The two isovector states $a_0(1475)$ and
$a_0(2020)$ were both discovered in $\bar pp$ annihilation, the
$a_0(1475)$ in $\bar pp$ annihilation at rest, the $a_0(2020)$ in $\bar
pp$ annihilation in flight, in formation. Both techniques do not give
access to an eventual $a_0(1720)$. The $K^*_0(1720)$ may have escaped
observation due to dominant contributions of $K^*_0(1680)$ and
$K_1(1720)$. In the discussion of tensor mesons
\cite{Anisovich:2005xd,Anisovich:2005iv,Anisovich:2006ma} (see section
\ref{JPC=2++}), an additional $K^*_2(1650)$ was predicted. Hence, the
non-observation of these two missing states may not be a decisive
argument against our view.

The mass of isoscalar scalar mesons is speculative. Nevertheless we
suggest a scenario which may have some plausibility even though it is
not enforced by data. Instanton induced interactions increase the mass
of the pseudoscalar isoscalar $q\bar q$ mesons, of $\eta^{\prime}$ and
$\eta(1835)$. Their masses lie well above other all S-wave ground-state
mesons and above all other first radial excitations, respectively. The
same forces lower the masses of scalar isoscalar $q\bar q$ mesons, by
about the same amount. However, scalar mesons are not dominantly $q\bar
q$ and the action of instanton induced forces is likely reduced for the
latter component. Instead of a mass shift of 400\,MeV/c$^2$
($m_{\eta^{\prime} } - m_{\eta}$), we estimate the mass shift to $\sim
200$\,MeV/c$^2$. Thus we estimate isoscalar masses to be 1300\,MeV/c$^2$,
1560\,MeV/c$^2$, 1900\,MeV/c$^2$. The lower two masses are close to
values found in partial wave analyses: $f_0(1370)$ is considered as
established resonance; we believe it to exist but to be narrowed
artificially by the $f_0(980)$ and $f_0(1500)$ resonances. Anisovich
and Sarantsev find a scalar isoscalar resonance with a pole at
$(1530^{+90}_{-250}) - i(560\pm 140)$. Possibly the same pole was found
at a somewhat larger mass, at $(1.67-0.26i)$\,GeV in \cite{Li:2000jq}.

Why is there so little phase motion for $f_0(1300)$? A likely answer
can be seen looking at the $\sigma(485)$ phase motion. The $90^{\circ}$
phase is reached at 750\,MeV/c$^2$, a further $45^{\circ}$ shift
(corresponding to a half width) is reached (subtracting $f_0(980)$ at
1200\,MeV/c$^2$. A pole far from the real axis has a slow phase motion
and a $180^{\circ}$ phase shift may never be reached.

Where is the scalar glueball? We believe it to be dissolved as well into
the wide scalar background amplitude. We recall that the wide scalar
component of low-energy $\pi\pi$ scattering can be described
quantitatively by molecular forces between the pions, by $t$-channel
exchange of $\rho$ and $f_2(1270)$. Duality arguments link amplitudes
for  $t$-channel exchange of $\rho$ and $f_2(1270)$ mesons to
$s$-channel $q\bar q$ resonances. A sum of broad S-wave resonances can
thus be represented by a sum of a few meson $t$-channel exchanges. The
$t$-channel exchange may contain a contribution from Pomeron exchange
which in $s$-channel corresponds to a glueball. This scenario is a
possibility how the glueball may hide; at present there is no evidence
for this is reality.

The question arises how we can understand the absence of a narrow scalar
glueball which is so firmly predicted by QCD inspired models and by QCD
on the lattice. The reasons have to be the approximations made on the
lattice. QCD on the lattice neglects the coupling of the gluon field to
$q\bar q$ pairs. This is called quenched approximation. Recent glueball
mass calculations on the lattice include couplings to fermion loops
\cite{Bali:2000vr} but pions are still too heavy to represent the true
chiral limit. It is hence not excluded that the scalar glueball is very
broad.

To clarify this question, high precision data on J/$\psi$ decays into
a vector meson ($\gamma$, $\omega$ and $\phi$) and two
pseudoscalar mesons ($\pi\pi$, $K\bar K$, $\eta\eta$, $\eta\eta'$)
are needed, and detailed studies of the decay of scalar mesons including
radiative decays \cite{Pacetti:2007eg}. Ultimately, a series of Rosner
plots as functions of the scalar mass will be required. A Rosner plot
for pseudoscalar mesons was shown in Fig. \ref{plotetap} in section
\ref{Introduction}.

\subsubsection{\label{Dynamical generation of resonances and flavour
exotics}
Dynamical generation of resonances and flavour exotics}

We have seen that there are severe arguments which justify interpreting
some mesons as being generated from molecular forces between the mesons
into which they decay. For the isospin quartet ($a_0(980)$, $f_0(980)$)
the discussion\cite{Achasov:1987ts,Baru:2003qq,Anisovich:2004qr,%
Baru:2004xg,Kalashnikova:2004ta,Achasov:2005ub} concentrates, with
refutation \cite{Achasov:2006cq} and counter-arguments
\cite{Kalashnikova:2006vp}, and a recent detailed analysis
\cite{Hanhart:2006nr,Hanhart:2007wa}, on the question if data are
compatible with or enforce one of the favoured interpretations, if
$a_0(980)$ and $f_0(980)$ are dynamically generated, $q\bar q$, or
tetraquark states.

The mesonic flavour wave function of $a_0^+(980)$ can be expanded into
its Fock components \be |a_0(980)^+> = \alpha |u\bar d> + \beta |u\bar
ds\bar s> + \gamma |K^+\bar K^0> \cdots
\ee plus higher
order configurations. In this expansion, quark and antiquark in $u\bar
d$ are in P-wave while the $us\bar d\bar s$ and $K^+\bar K^0$ systems
are in S-wave. The Pauli principle forbids to add a $|u\bar du\bar
u>$ or $|u\bar dd\bar d>$ component. The two components $|u\bar ds\bar
s>$ and $|K^+\bar K^0>$ differ in their colour configuration: in the
latter wave function, $u\bar s$ and $s\bar d$ are colour singlets, in
the former this is not imposed. The tetraquark wave function contains
the molecular picture. The dynamical assumptions are of course
different in these two cases.

Jaffe has reminded us recently that tetraquark configurations and the
molecular picture cannot be meaningfully distinguished when the state
is near or above important thresholds \cite{Jaffe:2007id}. He
considered a toy model in which the four quarks are very weakly
bound inside of a strong interaction volume; outside, the $K\bar K$
system is supposed not to interact. Then he assumed the colour forces
are strong enough to bind the system but with a small binding energy.
The small binding energy forces the system to become large. Is that
system of tetraquark nature or of molecular character? The origin of
that state lies certainly in the colour forces. The wave function
exhibits molecular character.

Similar arguments have been raised in baryon spectroscopy. In the
quark model, the $N(1535)S_{11}$ resonance with $J=1/2$ is the twin
brother of $N(1520)D_{13}$ with $J=3/2$. They both have intrinsic
orbital angular momentum $\ell=1$ and the 3 quarks spins couple to
$s=1/2$ (some $s=3/2$ component can be mixed into the wave function.
On the other hand, the decay properties of the $N(1535)S_{11}$
resonance can very well be understood from the S-wave
meson-baryon interaction in the energy range around the $\eta N$
threshold using a coupled channel chiral Lagrangian with $\pi N$, $\eta
N$, $K \Lambda$, $K \Sigma$ in the isospin-$1/2$, $l=0$ partial
wave. A fit to scattering data then yielded parameters, supporting a
quasi-bound $K \Sigma$-state which can be identified with the
$N(1535)S_{11}$ resonance \cite{Kaiser:1995cy}. Does this observation
require two states, a quark model state {\it and} a quasi-bound
$K \Sigma$-state? We do not believe so. Meson-nucleon scattering is
partly driven by resonances which in turn can be reconstructed from the
scattering amplitude. The quark model and dynamically generated
resonances are, as we believe, different views of the same object.

Coming back to the low-lying scalar mesons.
The main topic of this discussion is if we can expect two states,
a $|u\bar d>$ and a $|us\bar d\bar s>$ meson (or $K\bar K$ molecule),
possibly mixing but creating two (or even three) states. This is a
dynamical question to which answers can be given within a model or by
looking at the observed pattern of states. The two orthogonal
components $\alpha |u\bar d> + \beta |us\bar d\bar s>$ and $- \beta
|u\bar d> + \alpha|us\bar d\bar s>$ may both form resonances. Yet it is
possible as well that one of them disappears as scattering state. In
SU(3), the number of tetraquark wave functions is 9, see Fig.
\ref{mes:fig:four-q}. This is just the number of $q\bar q$ states.
This situation does not repeat in the case of SU(4). Charm provides an
additional degree of freedom, and it is rewarding to include $u,d,s,c$
quarks into the discussion. In SU(4), there are 16 $q\bar q$ states but
36 tetraquark configurations which have quark pairs which are
antisymmetric in spin and flavour: 9 tetraquark states with $u,d,s$
quarks only, 20 tetraquark states with open charm or anti-charm, and 9
tetraquark states with hidden charm. The 16 $q\bar q$ states are the
light-quark nonet, a $D^+, D^0, \bar D^0, D^-$ quartet, a $D_s^+$, and
$D_s^-$, and the $c\bar c$ state; three of them carry charm (and three
anticharm). The 10 tetraquark states with open charm have wave
functions: \vspace{-5mm} \begin{eqnarray} c\bar d\,u\bar u\quad;\quad
c\bar d\,s\bar s\quad;\quad c\bar d\,d\bar s\quad;\quad c\bar d\,s\bar
u\quad;\quad c\bar d\,u\bar s\quad \vspace{-8mm} \nonumber \\ c\bar
u\,d\bar d\quad;\quad c\bar u\,s\bar s\quad;\quad c\bar u\,u\bar
s\quad;\quad c\bar u\,s\bar d\quad;\quad c\bar u\,d\bar s\quad
\label{fullfourq} \vspace{-2mm} \end{eqnarray} These tetraquark
combinations can all have $qq$ in colour {$\bf 3$} and spin $^1S_0$.
Thus mesons with wave functions $1/\sqrt 2|c\bar s(\uub+\ddb)>$ and
$1/\sqrt 2|c\bar s(\uub-\ddb)>$ could exist. However, the last two wave
functions in both rows of (\ref{fullfourq}) are flavour exotic, and
such states have never been observed.

Out of the 10 wave functions in (\ref{fullfourq}), three can mix with
$q\bar q$, the other seven tetraquark wave functions cannot.
Phenomenologically, none of the seven extra states has been found.
We conjecture that the forces in the tetraquark system and the
molecular forces are not strong enough to provide binding, they fall
apart as suggested by Jaffe \cite{Jaffe:1976ig,Jaffe:1976ih}. Only
those tetraquark configurations exist which can mix with $q\bar q$. The
tetraquark component plays a significant $\rm r\hat{o}le$ for scalar mesons with
masses close to or above their respective thresholds for OZI rule
allowed decays. The tetraquark flavour wave functions of scalar charmed
states are given by $c\bar d$(\uub+\ssb), $c\bar u$(\ddb+\ssb), $c\bar
s$(\uub+\ddb). We identify $c\bar s$(\uub+\ddb) with $D_s^*(2317)$. The
$c\bar d$(\uub+\ssb) and $c\bar u$(\ddb+\ssb) ground states might be
$D^{*+,0}_0(2350)$. We believe that these are the first excited scalar
states, and that the ground state $D^{*+,0}(1980)$ is not yet found. At
such a low mass, SU(3) symmetry breaking will reduce the \ssb\
component; in high excitations, the \ssb\ component may play a $\rm
r\hat{o}le$.

The conjecture of just three tetraquark scalar states with open charm
can be tested at the $D_sK^0_S$ threshold. A $(c\bar s)(s\bar d)=(c\bar
d)(s\bar s)$ tetraquark state could exist and should be observed as
narrow peak in the $D^+\pi^0$ mass distribution. Its decay would be OZI
rule violating. Fig.  \ref{scalarview} would suggest it to have a mass
$\sim 50$\,MeV/c$^2$ below the $D_s^*(2317)K^0$ threshold, i.e. about
2750\,MeV/c$^2$. It should show up as peak in Fig. \ref{fig:d2p-d3p}b.
Clearly, considerably more statistics is needed to exclude its
existence. We predict that it will not be found.

It should be noted that the existence or not of flavour-exotic
tetraquark states is of considerable importance. If tetraquark states
not allowed to mix with $q\bar q$ states are unstable -- as we claim --
then any resonance with exotic quantum numbers must be of hybrid
nature. In the light-quark sector, all non-exotic tetraquark states  in
Jaffe configuration can mix with $q\bar q$ and arguments in favour of
tetraquark states have to rely on their abundance. In the charm sector,
flavour-exotic tetraquark states might still exist; their discovery or
quantitative upper limits are urgently needed. Even if they should not
exist, QCD may support tetraquark bound states in very heavy plus light
quark systems, like $bc\bar u\bar d$ (Richard, private communication).

\subsubsection{\label{Scalar states from the sigma to chib0(1P)}
Scalar states from the $\sigma$ to $\chi_{b0}(1P)$}

\begin{figure}[ph]
\vspace{-5mm}
\bc
\begin{tabular}{cc}
\includegraphics[width=0.49\textwidth]{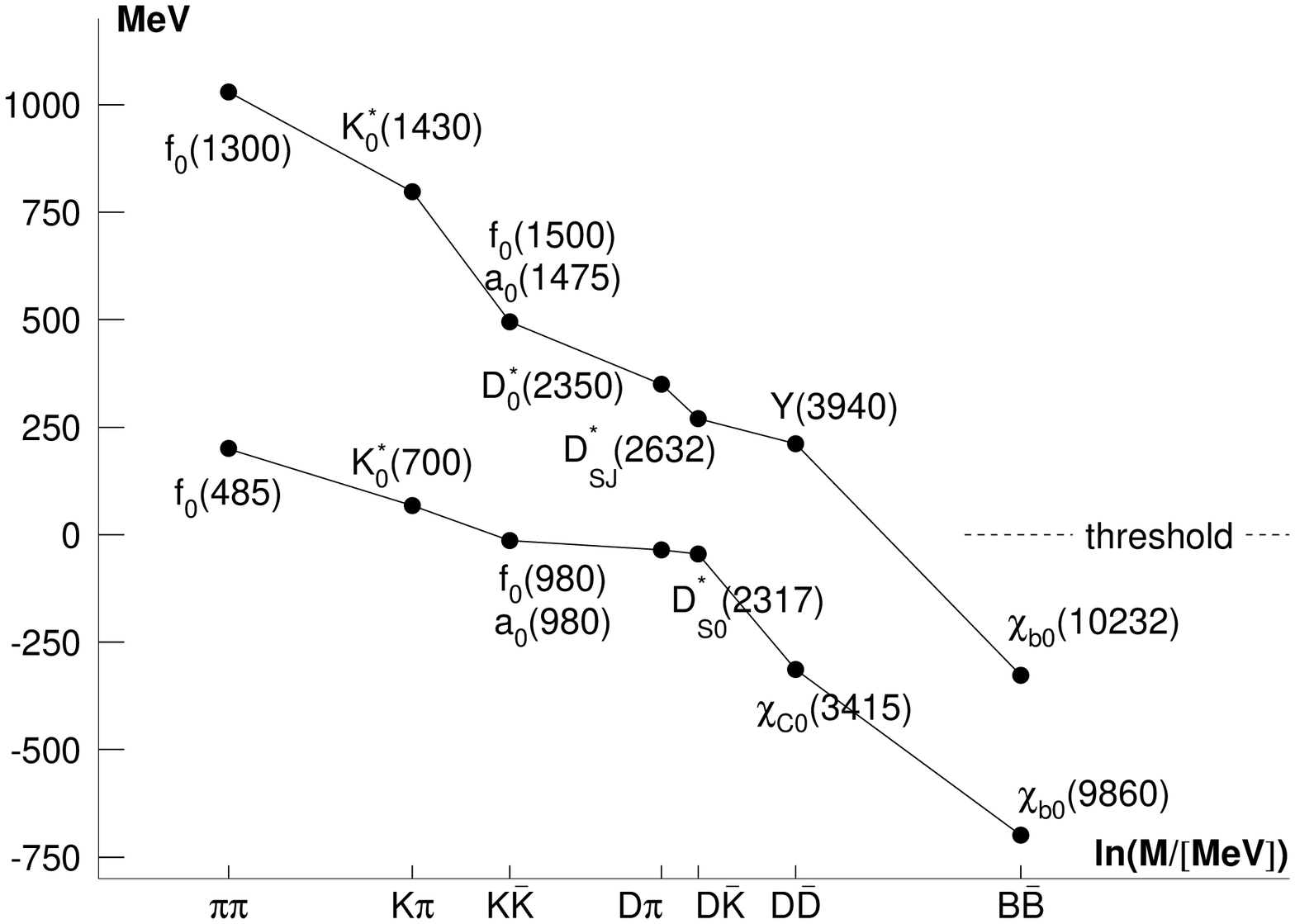}&
\includegraphics[width=0.49\textwidth]{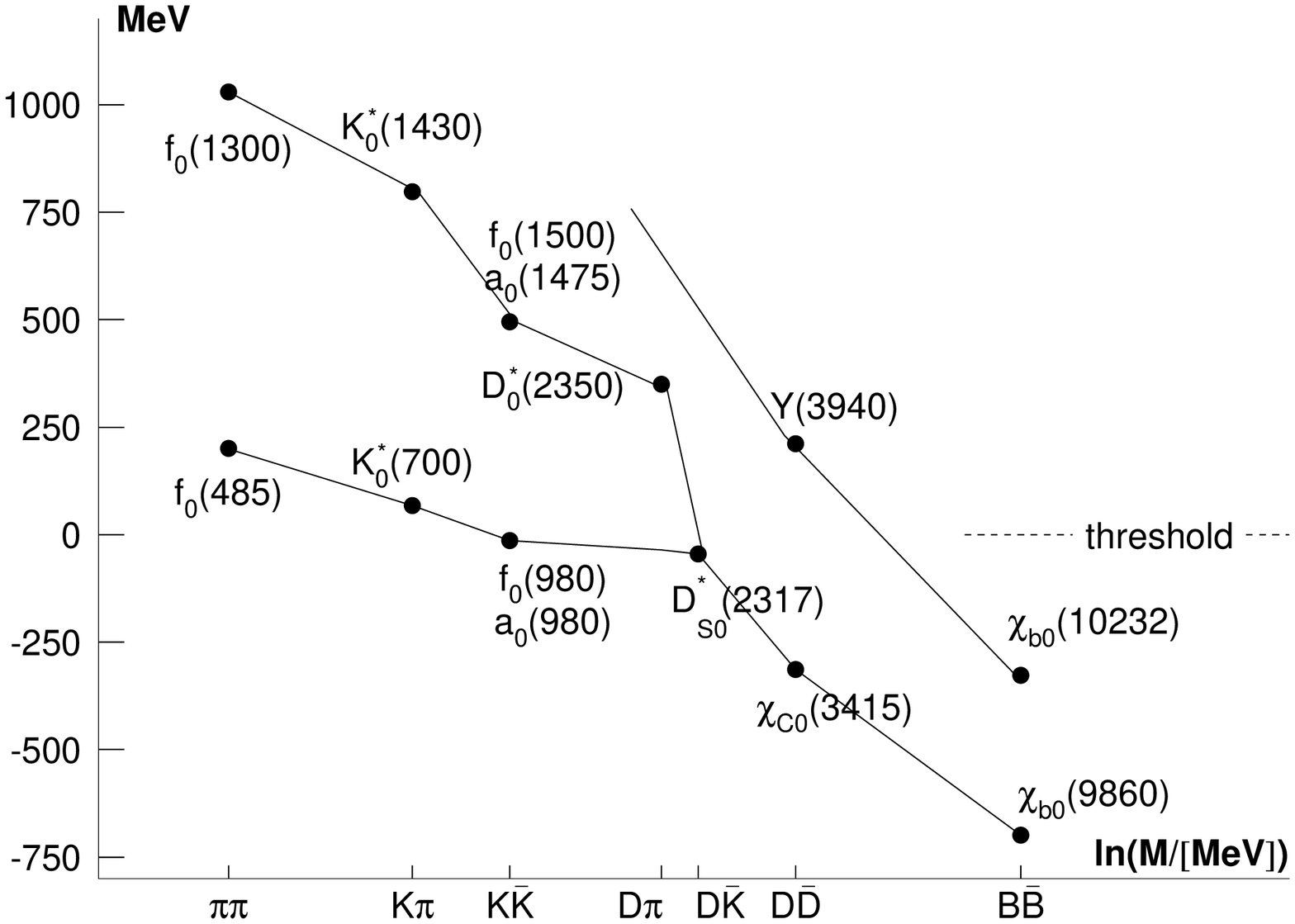}
\end{tabular}
\vspace{-2mm}
\ec
\caption{\label{scalarview}
Excitation energy of scalar mesons above their respective thresholds
for OZI allowed decays. The diagram suggests that the $\chi_{b0}(1P)$
would change its mass continuously down to the $\sigma(485)$ mass, and the
$\chi_{b0}(2P)$ to the $f_0(1300)$ mass, when the quark masses are
reduced continuously from $b\bar b$ to $n\bar n$. The horizontal scale
is given by the logarithmic threshold masses. }
\end{figure}

In this last section we summarise our view of the mass spectrum
of scalar ground states and of their first radial excitation. In Fig.
\ref{scalarview}a we display our view.

The nine mesons \{$a_0(980), K^*_0(700), f_0(465), f_0(980)$\} form a
low-mass scalar nonet. Their Fock space expansion contains $q\bar q$,
$qq\bar q\bar q$, and meson-meson components. In this sense we
interpret \{$a_0(980), K^*_0(700), f_0(465), f_0(980)$\} as nonet of
$q\bar q$ mesons, with diquark (and antidiquark) having colour wave
functions in \{$\bf{\bar 3}$\} (and \{$\bf 3$\}) representation and
spins in $^1S_0$, and with the two $q\bar q$ subsystems spending most
of the time as $K$ or $\bar K$. So far, we have presented an accepted
(even though not undisputed) view. However, we interpret $f_0(980)$ as
flavour octet and $f_0(465)$ as flavour singlet state even though
flavour SU(3) is broken at such a low mass, and $f_0(465)$ is expected
to have a small \ssb\ component only. With $K^*_0(700)$ as lowest
scalar state carrying strangeness, there is consensus that
$K^*_0(1430)$ is the first excitation above it. Controversies will
arise from our assignment of $f_0(1500)$ and $a_0(1475)$ as octet
states and $K^*_0(1430)$ partners. We omit the disputed narrow
$f_0(1370)$ and interpret the broad scalar isoscalar background
$f_0(1300)$ -- the old $\epsilon(1300)$ -- as $f_0(465)$ excitation.
The puzzling mass pattern of $D^*_0(2350)$ and $D^*_{s0}(2317)$ -- with
the $c\bar n$ state heavier than its $c\bar s$ companion -- is resolved
by interpreting the former state as excited, the latter one as ground
state. A $D^*_0$ ground state is then expected at about
1980\,MeV/c$^2$, below the $D\pi$ threshold. In the limit of chiral
symmetry, the $D\pi$ interaction vanishes when pion momenta approach
zero. This may shift the mass of the hypothesised $D^*_0$ state to
higher masses. Beveren, Costa, Kleefeld and Rupp
\cite{vanBeveren:2005ha} propose it to have 2114\,MeV.  The
$D^*_{s0}(2317)$ radial excitation is expected at about
2650\,MeV/c$^2$. The Selex state $D^*_{sJ}(2632)$ has the right mass to
fill this slot; however there is no obvious reason why it is so narrow.
Alternatively, the $D_{sJ}(2860)$ might play this $\rm r\hat{o}le$
\cite{vanBeveren:2006st}. The $Y(3940)$ fits well as radial $\chi_{c0}$
excitation.

Bugg (private communication) pointed out that an inclusion of the
rapidly growing phase space might shift the $D^*_0(1980)$ pole position
up to 2350\,MeV; if this is true, an additional $D^*_0(1980)$ would, of
course, not be needed. The Selex state $D^*_{sJ}(2632)$ could be fake.
Hanhart (private communication) insisted that $f_0(980)$ is of
molecular character and has no relation to $q\bar q$ or $qq\bar q\bar
q$, in the same way as the deuteron is a proton-neutron bound state and
no six-quark state. If these two views are correct, the low-mass scalar
meson nonet decouples from $q\bar q$ physics becoming an independent
branch, and the ground state scalar mesons are all above 1\,GeV/c$^2$.
This more conventional scenario is shown in Fig. \ref{scalarview}b.

Our preferred interpretation is contained in Fig. \ref{scalarview}a.
Its most striking aspect is the smooth change of the excitation
energy with the mass of the two pseudoscalar mesons to which the scalar
mesons couple strongly. The plot suggests that the $\chi_{b0}(1P)$
state gradually changes into $f_0(465)$ when the quark masses are
lowered continuously. Of course, this does not imply that $f_0(465)$
and $\chi_{b0}(1P)$ have the same structure. The $\chi_{b0}(1P)$ is
well understood as $q\bar q$ state; when the quark masses get smaller,
the mass approaches the threshold for OZI allowed decays and tetraquark
and molecular components become important. But still, $f_0(465)$
occupies the slot of the quark model ground state, and no light $q\bar
q$ scalar isoscalar meson which resembles the $\chi_{b0}(1P)$. Quantum
chromodynamics forms states of $q\bar q$ structure when both quark
masses exceed one GeV; for light quarks, QCD forms bound states of a
more complicated dynamical structure.

\markboth{\sl Meson spectroscopy} {\sl Summary}
\clearpage\setcounter{equation}{0}\section{\label{Outlook}
Outlook\quad$\cdots$\quad and a look back}
We recall some novel features which we developed while writing this
review. A summary of facts and ideas was already presented at the
beginning of this review.

The most fascinating aspect is the new view of scalar mesons, from the
$\chi_{b0}(9860)$ to the $\sigma$. The scalar mesons have brought with
them many puzzles: from the long-debated $\sigma(485)$ and the twin
brothers $f_0(980)/a_0(980)$ to the hotly disputed
$f_0(1370)/f_0(1500)/f_0(1760)$ complex and its strange production and
decay pattern in J/$\psi$, $B$ and $D$ decays. It may look surprising
that a view can be developed with embraces all these different
observations with a minimum set of assumptions. In this view, the
$\chi_{b0}(9860)$ is, as expected, a $b\bar b$ bound state and
$\chi_{c0}(3415)$ a $c\bar c$ state: the $h_{1c}(3542)$ mass coincides
with the centre of gravity of the spin-triplet states. The summation
includes $\chi_{c0}(3415)$. If $\chi_{c0}(3415)$ would have non-$c\bar
c$ components, the coincidence would be extremely fortuitous. With
decreasing quark masses, the scalar states approach their respective
thresholds for OZI allowed decays and become strongly coupled to the
meson-meson continuum. The $q\bar q$ seed develops a tetraquark
component and, for very low mass, a molecular object is formed. The
tetraquark and molecular components govern the meson masses and their
decay properties. While $\chi_{b0}(9860)$ and $\sigma(465)$ are of very
different nature, their roots are the same. But when when a light
quark-antiquark pair with scalar quantum numbers is created out of the
vacuum, QCD transforms the initial $q\bar q$ seed into a complex object
which is efficiently interpreted using methods of chiral perturbation
theory instead of the conventional quark model. The light-quark
isoscalar mesons seem to organise themselves into comparatively narrow
octet states and into broad flavour singlet states. The latter states
have the quantum numbers of the QCD vacuum and couple strongly to
pions, to the lightest Goldstone bosons. The isoscalar scalar mesons
are so broad that they do not show up as individual resonances, they
merge into a wide background. The scalar glueball may be part of this
general background.  The strange production and decay patterns observed
for scalar mesons can be understood by interference between the
wide SU(3) singlet states and the narrow octet states and their
expansion from primarily formed $q\bar q$ seeds into tetraquark
clusters.

Many new questions have emerged. Even though our belief in glueballs
is shaken, there will be undoubtedly new experimental searches for
glueballs and new attempts to interpret the scalar mesons as $q\bar q$
mesons mixing with a scalar glueball. Also the search for spin-parity
exotics will go on; arguments have been put forward that charmonium
hybrids could be narrower and easier to detect. And optimists hope for
narrow high-mass glueballs. The new excitingly narrow charmonium states
lead to the expectation of flavour-exotic states, of tetraquark states
which cannot be reduced to $c\bar c$. So far there is no experimental
evidence for such states but it is important to clarify these
conjectures.

We expect that the `boring' spectrum of high-mass $q\bar q$ mesons will
be completed in future experiments and that some important questions
will find answers. We still do not know the basic mechanisms
responsible for high orbital and radial excitations. Are the agents in
a $a_6^+(2450)$ meson still a constituent $u$ and a $\bar d$ quark? Do
the masses of the constituent quarks change when mesons are excited?
What connects the quark and the antiquark? A gluonic string? What kind
of nonperturbative fluctuations are important for small and for large
separations? Do condensates play a $\rm r\hat{o}le$ in the `empty'
space between quark and antiquark? Is there really something like an
effective one-gluon exchange? Are there instanton-induced interactions?
We believe, these are fundamental questions which need to be solved
before we can claim to have understood QCD.

It appears from this review that answers often come from unexpected
directions. The best evidence for the octet character of $f_0(1500)$
comes from $B$ decays. The tetraquark nature of the low-mass scalar
mesons was suggested 30 year ago. We believe this to be a much broader
concept and found support for this view from J/$\psi$ decays and from
$B^+\to \omega J/\psi K^+$ decays. The new narrow charmonium and
open-charm states have emphasized the $\rm r\hat{o}le$ of the opening
of new thresholds. The systematic of the mesons with open charm taught
us that scalar mesons are likely not be just tetraquark states; a
$q\bar q$ component may be needed to bind the system. One major
conclusion of the review can be drawn: hadron spectroscopy is still
full of surprises, a fact which teaches us modesty. And we have learned
that hadron spectroscopy does not advance by performing one experiment
only. Different approaches stimulate each other and shed light from
different directions on a subject which continues to be fascinating and
intellectually demanding.

Physics does not deal only with objects. Physics cannot be done without
the enthusiasm of colleagues and collaborators. Here we would like to
render our deepest thanks to all with whom we had the priviledge to
cooperate in one of the experiments. Our views developed in a multitude
of discussions; here we mention a few but in the conviction to have
forgotten enlightening discussions with people not listed below.
Nevertheless we mention here N. Achasov, R. Alkofer, C. Amsler, A.V.
Anisovich$^*$, V.V. Anisovich$^*$, R. Armenteros, D. Aston, T. Barnes,
Chr. Batty, K. Beck, E. van Beveren, F. Bradamante, S.J. Brodsky, D.
Bugg, S. Capstick, S.U. Chung$^*$, F.E. Close, G.F. De Teramond, D.
Diakonov, A. Donnachie, W. Dunwoodie, W. D\"unnweber, St. Dytman, A.
Dzierba, W. D\"unnweber, P. Eugenio, A. F\"assler, M. F\"assler, H.
Fritzsch, U. Gastaldi, K. G\"oke, C. Guaraldo, H. Hammer$^*$, C.
Hanhart$^*$, D. von Harrach, D. Herzog, N. Isgur, S. Ishida, B.
Jaffe$^*$, A. Kirk, K. K\"onigsmann, F. Klein,  K. Kleinknecht$^*$, H.
Koch, K. K\"onigsmann, S. Krewald$^*$, B. Kubis$^*$, G.D. Lafferty, R.
Landua, L.G. Landsberg, S. Lange, Weiguo Li, D.B. Lichtenberg, M.
Locher, R.S. Longacre, V.E. Markushin, A. Masoni, A. Martin, B.
Meadows$^*$, U.-G. Mei\ss ner$^*$,  V. Metag, B. Metsch$^*$, C.A.
Meyer, P. Minkowski, L. Montanet, S. Narison, V.A. Nikonov, W.
Ochs$^*$, L.B. Okun, S.L. Olsen, E. Oset, M. Ostrick, Ph. Page, A.
Palano, E. Paul, J.R. Pelaez, M. Pennington$^*$, K. Peters, H. Petry,
M.V. Polyakov, J. Pretz$^*$, Yu.D. Prokoshkin, J.M. Richard$^*$, A.
Sarantsev$^*$, A. Sch\"afer, F. Scheck, H. Schmieden, B. Schoch, K.
Seth, Jin Shan$^*$, Xiaoyan Shen, E.S. Swanson, W. Schwille, A.P.
Szczepaniak, J. Speth, K. Takamatsu, L. Tiator, P. Tru\"ol, U.
Thoma$^*$, T. Tsuru, Th. Walcher, Chr. Weinheimer, W. Weise, N. Wermes,
U. Wiedner$^*$, H. Wittig, H. Willutzki, B.S. Zou$^*$, and C. Zupancic.
Those marked with a $^*$ had read critically parts of the manuscript
and made thoughtful suggestions. D. Bugg and B. Schoch read the full
manuscript, to them our deepest thanks. Early contributions by Bernard
Metsch and Mike Pennington in the initial phase of the project are
kindly acknowledged. Of course, the responsibility for any remaining
inaccuracies rests with the authors.

We as authors have collaborated closely with our direct colleagues. We
both would like to thank them for the wonderful time we have spend
with them. A.Z. would like to appreciate the collaboration with D.V.
Amelin, E.B. Berdnikov, S.I. Bityukov, G.V. Borisov, V.A. Dorofeev,
R.I. Dzhelyadin, Yu.P. Gouz, Yu.M. Ivanyushenkov, A.V. Ivashin, V.V.
Kabachenko, I.A. Kachaev, A.N. Karyukhin, Yu.A. Khokhlov, G.A.
Klyuchnikov, V.F. Konstantinov, S.V. Kopikov,  V.V. Kostyukhin, A.A.
Kriushin, M.A. Kulagin, S.A. Likhoded, V.D. Matveev, A.P. Ostankov,
D.I. Ryabchikov, A.A. Solodkov, A.V. Solodkov, O.V. Solovianov, E.A.
Starchenko, N.K. Vishnevsky, and E.A. Vlasov in common experimental
efforts.

E.K. has had the privilege to work in the course of time with a large
number of PhD students. The results were certainly not achieved without
their unflagging enthusiasm for physics. I would like to mention O.
Bartholomy, J. Brose, V. Crede, K.D. Duch, A. Ehmanns, I. Fabry, M.
Fuchs, G. G\"artner, J. Junkersfeld, J. Haas, R. Hackmann, M. Heel,
Chr. Heimann, M. Herz, G. Hilkert, I. Horn, B. Kalteyer, F. Kayser, R.
Landua, J. Link, J. Lotz, M. Matveev, K. Neubecker, H. v.Pee, K.
Peters, B. Pick, W. Povkov, J. Reifenr\"other, G. Reifenr\"other, J.
Reinnarth, St. Resag, E. Sch\"afer, C. Schmidt, R. Schulze, R.
Schneider, O. Schreiber,  S. Spanier, Chr. Stra\ss burger, J.S. Suh, T.
Sczcepanek, U. Thoma, F. Walter, K. Wittmack, H. Wolf, R.W. Wodrich, M.
Ziegler. Very special thanks go to my colleague Hartmut Kalinowsky with
whom I have worked with jointly for more than 35 years. Anyone familiar
with one of the experiments I was working on knows and admires his
competence and endurance and his invaluable contributions to our common
efforts. To him my very personal thanks.

This work (A.Z.) was supported in part by the Presidential Grant
5911.206.2.

Nathan Isgur and Lucien Montanet were exceptional personalities, as
individuals and as physicists, and have made outstanding contributions
in many areas of physics. We are grateful for their friendship and will
have them in our memories, ever.

\markboth{\sl Meson spectroscopy} {\sl Bibliography  }

\clearpage
\cleardoublepage
\bibliographystyle{elsart-num}
\bibliography{ref-intro,ref-qcd,ref-meth,ref-hq,ref-ex,ref-ps,ref-exp,ref-s}
\end{document}